\documentclass[prd,12pt,tightenlines,aps,letterpaper,amsmath,amssymb,superscriptaddress]{revtex4-2}
\usepackage[utf8]{inputenc}
\usepackage{amsmath}
\usepackage{amssymb}
\usepackage{graphicx}
\usepackage{xcolor}
\usepackage{xspace}
\usepackage[inline]{enumitem}
\usepackage[nowatermark]{fixmetodonotes}

\usepackage{natbib}
\usepackage{hyperref}
\hypersetup{
    colorlinks=true,        
    linkcolor=blue,         
    citecolor=blue,        
    filecolor=blue,      
    urlcolor=blue,       
    pdfstartview=           
}                           
\usepackage{mciteplus}

\mciteErrorOnUnknownfalse

\cftsetindents{section}{1em}{2.5em}
\cftsetindents{subsection}{3.7em}{2.5em}

\newcommand\snowmass{\begin{center} \rule[-0.2in]{\hsize}{0.01in}\\\rule{\hsize}{0.01in}\\
\vskip 0.1in Submitted to the  Proceedings of the US Community Study\\
on the Future of Particle Physics (Snowmass 2021)\\
\rule{\hsize}{0.01in}\\\rule[+0.2in]{\hsize}{0.01in} \end{center}}

\providecommand{\nubar}{\ensuremath{\bar{\nu}}}

\newcommand{\matrixel}[3]{\left\langle #1 \right| #2 \left| #3 \right\rangle}  

\newcommand{\str}[1]{}

\newcommand{\CoLeader}{Co-leader.}
\newcommand{\FNAL}{Fermi National Accelerator Laboratory, Batavia, IL, USA}
\newcommand{\ANL}{Argonne National Laboratory, Lemont, IL, USA}
\newcommand{\MSU}{Michigan State University, East Lansing, MI, USA}
\newcommand{\SLAC}{SLAC National Accelerator Laboratory, Stanford University, Menlo Park, CA, USA}
\newcommand{\TAMU}{Mitchell Institute for Fundamental Physics and Astronomy, Department of Physics and Astronomy, Texas A\&M University, College Station, TX, 77843}
\newcommand{\WashU}{Department of Physics, Washington University in St. Louis, MO, 63130, USA}
\newcommand{\MCSS}{McDonnell Center for the Space Sciences at the Washington University in St. Louis, MO, 63130, USA}
\newcommand{\UGhent}{Ghent University, Department of Physics and Astronomy, B-9000 Ghent, Belgium}
\newcommand{\TIPFA}{INFN-TIFPA Trento Institute of Fundamental Physics and Applications, Trento, Italy}
\newcommand{\WISC}{Department of Physics, University of Wisconsin, Madison, WI 53706 USA}
\newcommand{\BERN}{Albert Einstein Center for Fundamental Physics, Institute for Theoretical Physics, University of Bern, Sidlerstrasse 5, 3012 Bern, Switzerland}
\newcommand{\DESY}{Deutsches Elektronen-Synchrotron DESY, Notkestr. 85, 22607 Hamburg, Germany}
\newcommand{\TUD}{Department of Physics, Technische Universit\"at Darmstadt, 64289 Darmstadt, Germany}
\newcommand{\HELM}{ExtreMe Matter Institute EMMI, GSI Helmholtzzentrum f\"ur Schwerionenforschung GmbH, 64291 Darmstadt, Germany}
\newcommand{\MPIK}{Max-Planck-Institut f\"ur Kernphysik, 69117 Heidelberg, Germany}
\newcommand{\BAR}{Department of Quantum Physics and Astrophysics and Institute of Cosmos Sciences, University of Barcelona, Spain}
\newcommand{\LANL}{Los Alamos National Laboratory, Los Alamos, NM 87545}
\newcommand{\MIT}{Massachusetts Institute of Technology, Cambridge, MA, USA 02139}
\newcommand{\Mainz}{Helmholtz-Institut Mainz, Johannes Gutenberg-Universit\"{a}t Mainz, D-55099 Mainz, Germany}
\newcommand{\UCBerkeley}{Department of Physics, University of California, Berkeley, CA, 94720, USA}
\newcommand{\LBNL}{Lawrence Berkeley National Laboratory, Berkeley,
CA, 94720, USA}
\newcommand{\UKY}{Department of Physics and Astronomy, University of Kentucky, Lexington, KY 40506, USA}
\newcommand{\Aligarch}{Department of Physics, Aligarh Muslim University, Aligarh-202002, India}
\newcommand{\Osaka}{Research Center for Nuclear Physics, Osaka University, 567-0047 Osaka, Japan}
\newcommand{\IFC}{Instituto de F\'{i}sica Corpuscular, Consejo Superior de Investigaciones Cient\'{i}ficas  and Universidad de Valencia,  E-46980 Paterna, Valencia, Spain}
\newcommand{\Madrid}{Grupo de Física Nuclear, Departamento de Estructura de la Materia, Física Térmica y Electrónica and IPARCOS, Facultad de Ciencias Físicas, Universidad Complutense de Madrid, CEI Moncloa, Madrid 28040, Spain}
\newcommand{\Kings}{Department of Physics, King's College London, London WC2R 2LS, UK}
\newcommand{\FRIB}{Facility for Rare Isotope Beams, Michigan State University, MI 48824, USA}
\newcommand{\IIT}{Department of Physics, Illinois Institute of Technology, Chicago, IL 60616, USA}
\newcommand{\JLAB}{Jefferson Lab, Newport News, Virginia 23606, USA}
\newcommand{\WM}{Department of Physics, William \& Mary, Williamsburg, Virginia 23185, USA}
\newcommand{\Liverpool}{University of Liverpool, Department of Physics,Liverpool L69 7ZE, UK}
\newcommand{\ORNL}{Oak Ridge National Laboratory, Oak Ridge, TN 37831, USA}
\newcommand{\NW}{Northwestern University, Dept.~of Physics \& Astronomy, Evanston, IL 60208, USA}
\newcommand{\UCI}{Department of Physics and Astronomy, University of California, Irvine, CA 92697-4575, USA}
\newcommand{\Granada}{Department of Atomic, Molecular and Nuclear Physics, University of Granada, Granada-E18071, Spain}
\newcommand{\UFL}{Department of Physics, University of Florida, Gainesville, FL 32611, USA}
\newcommand{\BNL}{Brookhaven National Laboratory, Upton, NY 11973, USA}
\newcommand{\TelAviv}{Tel Aviv University, School of Physics and Astronomy, Tel Aviv, Israel}
\newcommand{\UWr}{University of Wrocław, Institute of Theoretical Physics, 50-204 Wrocław, Poland}
\newcommand{\IPSA}{IPSA-DRII and Sorbonne Universit\'e, Universit\'e Paris Diderot, CNRS/IN2P3, Laboratoire de Physique Nucl\'eaire et de Hautes Energies (LPNHE), Paris, France}
\newcommand{\Beijing}{Institute of High Energy Physics, Chinese Academy of Sciences and School of Physical Sciences, University of Chinese Academy of Sciences, Beijing 100049, China}
\newcommand{\STFC}{STFC Rutherford Appleton Laboratory, Particle Physics Department  Oxfordshire OX11 0QX, UK}
\newcommand{\Pavia}{Dipartimento di Fisica, Universit\`{a} degli Studi di Pavia and INFN, Sezione di Pavia,  I-27100 Pavia, Italy}
\newcommand{\Kamioka}{Kamioka Observatory, ICRR, The University of Tokyo, 5061205, Gifu, Japan}
\newcommand{\UMD}{Maryland Center for Fundamental Physics and Department of Physics,
University of Maryland, College Park, MD 20742, USA}

\begin{document}

\title{Theoretical tools for neutrino scattering: \\ interplay between lattice QCD, EFTs, nuclear physics, phenomenology, and neutrino event generators}

\snowmass

\author{L.~Alvarez~Ruso}\affiliation{\IFC}

\author{A.~M.~Ankowski}\affiliation{\SLAC}

\author{S.~Bacca}\affiliation{\Mainz} 

\author{A.~B.~Balantekin}\affiliation{\WISC} 

\author{J.~Carlson}\altaffiliation{\CoLeader}\affiliation{\LANL} 

\author{S.~Gardiner}\altaffiliation{\CoLeader}\affiliation{\FNAL}

\author{R.~Gonz\'alez-Jim\'enez}\affiliation{\Madrid}

\author{R.~Gupta}\affiliation{\LANL}

\author{T.~J.~Hobbs}\affiliation{\FNAL}\affiliation{\IIT}

\author{M.~Hoferichter}\affiliation{\BERN}

\author{J.~Isaacson}\affiliation{\FNAL}

\author{N.~Jachowicz}\altaffiliation{\CoLeader}\affiliation{\UGhent}

\author{W.I.~Jay}\affiliation{\MIT}

\author{T.~Katori}\affiliation{\Kings}

\author{F.~Kling}\affiliation{\DESY}

\author{A.~S.~Kronfeld}\affiliation{\FNAL}

\author{S.~W.~Li}\affiliation{\FNAL}

\author{H.-W.~Lin}\affiliation{\MSU}

\author{K.-F.~Liu}\affiliation{\UKY}

\author{A.~Lovato}\affiliation{\ANL}\affiliation{\TIPFA}

\author{K.~Mahn}\altaffiliation{\CoLeader}\affiliation{\MSU}

\author{J.~Men\'endez}\affiliation{\BAR}

\author{A.~S.~Meyer}\affiliation{\UCBerkeley}\affiliation{\LBNL}

\author{J.~Morfin}\affiliation{\FNAL}

\author{S.~Pastore}\affiliation{\WashU}\affiliation{\MCSS}

\author{N.~Rocco}\affiliation{\FNAL} 

\author{M.~Sajjad~Athar}\affiliation{\Aligarch}

\author{T.~Sato}\affiliation{\Osaka}

\author{A.~Schwenk}\affiliation{\TUD}\affiliation{\HELM}\affiliation{\MPIK}

\author{P.~E.~Shanahan}\affiliation{\MIT}

\author{L.~E.~Strigari}\altaffiliation{\CoLeader}\affiliation{\TAMU} 

\author{M.~Wagman}\altaffiliation{\CoLeader}\affiliation{\FNAL}

\author{X.~Zhang}\affiliation{\FRIB}

\author{Y.~Zhao}\affiliation{\ANL}

\author{\\\vspace{16pt} \textbf{Endorsers}:  B.~Acharya}\affiliation{\ORNL}
\author{L.~Andreoli}\affiliation{\WashU}
\author{C.~Andreopoulos}\affiliation{\Liverpool}\affiliation{\STFC}
\author{J.~L.~Barrow}\affiliation{\MIT}\affiliation{\TelAviv}
\author{T.~Bhattacharya}\affiliation{\LANL}
\author{V.~Brdar}\affiliation{\FNAL}\affiliation{\NW}
\author{Z.~Davoudi}\affiliation{\UMD}
\author{C.~Giusti} \affiliation{\Pavia}
\author{Y.~Hayato}\affiliation{\Kamioka}
\author{A.~N.~Khan}\affiliation{\MPIK}
\author{D.~Kim}\affiliation{\TAMU}
\author{Y.~F.~Li}\affiliation{\Beijing}
\author{M.~Lin}\affiliation{\BNL}
\author{P.~Machado}\affiliation{\FNAL}
\author{M.~Martini}\affiliation{\IPSA}
\author{K. Niewczas}\affiliation{\UWr}\affiliation{\UGhent}
\author{V.~Pandey}\affiliation{\UFL}
\author{A.~Papadopoulou}\affiliation{\MIT}
\author{R.~Plestid}\affiliation{\UKY}\affiliation{\FNAL}
\author{M.~Roda}\affiliation{\Liverpool}
\author{I.~Ruiz~Simo}\affiliation{\Granada}
\author{J.~N.~Simone}\affiliation{\FNAL}
\author{R.~S.~Sufian}\affiliation{\WM}\affiliation{\JLAB}
\author{J.~Tena-Vidal}\affiliation{\Liverpool}
\author{O.~Tomalak}\affiliation{\LANL}
\author{Y.-D.~Tsai}\affiliation{\UCI}
\author{J.~M.~Ud\'{\i}as}\affiliation{\Madrid}

\maketitle

\clearpage
\tableofcontents

\clearpage

\section{Executive summary}
\label{sec:ES}

Neutrino physics is entering a precision era in which measurements of neutrino oscillations, astrophysical neutrinos from supernovae and other sources, and coherent neutrino scattering will provide insight on the nature of neutrino masses, the presence of CP violation, and perhaps more exotic new physics in the neutrino sector.
Maximizing the discovery potential of increasingly precise neutrino experiments will require an improved theoretical understanding of neutrino-nucleus cross sections over a wide range of energies that uses a combination of lattice QCD, nuclear effective theories, phenomenological models, and neutrino event generators to make reliable theory predictions for experimentally relevant nuclei.

At low energies, calculations of MeV-scale exclusive scattering involving nuclear ground and excited states will be needed to reconstruct the energies of astrophysical neutrinos from supernovae bursts, and more precise determinations of neutron nuclear structure factors will be needed to search for new physics using increasingly precise measurement of coherent elastic neutrino scattering. Such precision determinations of neutron distributions require data input, and since, besides parity-violating electron scattering,  coherent neutrino nucleus scattering is the most promising source thereof, measurements at different momentum transfers and off different nuclear targets are paramount to disentangle nuclear effects and potential new-physics contributions. 
Current and future accelerator neutrino experiments involve higher energies, and achieving predictions of GeV-scale neutrino-nucleus cross sections with few-percent uncertainties will be essential for DUNE in particular.
Predictions for experimentally observable hadronic final states are required to reconstruct the incident neutrino energies used in oscillation analyses, but several different reaction channels including quasi-elastic, multi-nucleon, resonant processes, and deep inelastic scattering can contribute to particular final-state event rates.
Cross section contributions from different channels have different energy dependence, and theoretical input on this decomposition is required to accurately extrapolate cross section results between different energies.

Quasi-elastic scattering can be accurately described in nuclear effective theories in which the distributions of nucleons within a nucleus are calculated using quantum Monte Carlo methods for light nuclei, up to carbon, and more approximate nuclear many-body methods based in coupled-cluster theory or factorization of the hadronic final state for heavier nuclei.
Electron scattering experiments provide crucial data on aspects of nuclear structure and vector-current form factors, while nucleon elastic axial form factors that also enter neutrino scattering are known less precisely from experiment.
Lattice QCD can be used to calculate axial as well as vector nucleon form factors and is beginning to provide predictions with complete error budgets and few-percent precision for these observables.
Over the next five to ten years, both lattice QCD determinations of nucleon elastic axial form factors and nuclear many-body calculations of quasi-elastic scattering could achieve few-percent uncertainties using existing theoretical methods and available computing resources.

A large fraction of the DUNE neutrino flux is above the pion-production threshold, and both resonant and non-resonant pion production processes must be theoretically understood at the ten-percent level in order to achieve few-percent overall cross-section uncertainties.
The dominant role of the $\Delta(1232)$ resonance makes accurately modeling in this energy region a high priority for theory investigations.
Although this energy region is relatively well-studied, resonant axial current responses are poorly constrained from experiment.
New data for neutrino scattering on proton or deuteron targets would be very valuable for constraining these responses.
Lattice QCD can also provide valuable information about resonant and non-resonant nucleon axial transition form factors; current studies are more exploratory than elastic form factor calculations and further theoretical and computational efforts will be required to deliver results with complete error budgets.
Nuclear many-body theory investigations of nuclear modifications to resonant and non-resonant pion production and absorption processes will also be crucial.
It will also be essential to consistently implement reliable models of resonant scattering in neutrino event generators.
Besides the $\Delta$, higher-energy nucleon resonances must also be included in cross section predictions, and theoretical studies of the nuclear modifications to vector and axial current responses in the high-energy part of the resonance region will also be indispensable.

An important challenge for achieving precise neutrino-nucleus cross-section predictions for the energy range relevant for DUNE will be reliably bridging the transition regions between low- and high-energy theories, which use different degrees of freedom to describe neutrino-nucleus interactions.
Extrapolations of vector-current structure functions from the dynamical coupled-channel model of the resonance region to the DIS region approach the corresponding structure function results obtained from partonic descriptions valid at high energies, but analogous extrapolations of axial-current structure functions between resonance and DIS regions do not agree.
There is a strong need for new experimental data  of neutrino scattering on nucleons and nuclei as well as theoretical studies of how to consistently model this transition region. 
New neutrino-hydrogen/deuterium DIS measurements would greatly inform theoretical models and help more precisely determine the combinations of parton distribution functions (PDFs) relevant to neutrino scattering as well as benchmarks for validating phenomenological models of the transition region.
Lattice QCD studies of PDFs are rapidly maturing and can also provide insight on aspects of nucleon and nuclear structure functions relevant to neutrino scattering.
Detailed phenomenological studies of the shallow inelastic scattering (SIS), DIS, and transition regions will be needed to obtain consistent models of neutrino scattering, and dedicated efforts to consistently model the transition region in neutrino event generators will be essential.

Simulations of neutrino scattering physics play a crucial role in experimental analyses. Creating a stronger and more efficient pipeline for neutrino event generator development will be necessary for the experimental community to fully benefit from these anticipated theory improvements. Significant organizational barriers to that goal currently exist, but they can be largely overcome through enhanced support for inter-disciplinary collaboration (across theory, experiment, and computation, as well as high-energy and nuclear physics), improved career incentives for physicists working on generators, and leadership to establish and pursue community priorities.
Support for theoretical efforts on neutrino scattering at the interface of high-energy and nuclear physics will be critical for achieving reliable cross-section predictions across the range of energies relevant to DUNE.
Sustained support for event generator development will further be essential in order to ensure that all relevant theoretical models are combined consistently in experimental analyses.

This WP was informed by the many LOIs received in the first stage of the Snowmass community planning exercise and by the participants to the \href{https://indico.fnal.gov/event/45039/}{Snowmass workshops Mini-Workshop on Neutrino Theory} (Sept 21-23, 2020), and \href{https://indico.fnal.gov/event/50335/}{ Mini-workshop in preparation for the white paper “Theoretical tools for neutrino scattering: the interplay between lattice QCD, EFTs, nuclear physics, phenomenology, and neutrino event generators”}
(Aug 23-25, 2021). This is a cross-frontier white paper solicited by the following Snowmass topical groups: \href{https://snowmass21.org/theory/lattice}{TF05} (Lattice gauge theory); \href{https://snowmass21.org/theory/neutrino}{TF11/NF08} (Neutrino theory); and \href{https://snowmass21.org/neutrino/cross_sections/start}{NF06} (Neutrino interaction cross sections).

\clearpage

\section{The needs of the neutrino experimental program}
\label{sec:exp_needs}

The planned neutrino experimental program probes a wide range of open physics questions. Broadly, there are three energy regimes of interest. At low energies are precision measurements of coherent elastic neutrino scattering (CE$\nu$NS) and astrophysical sources (supernova neutrino bursts, SNB). At energies around 1 GeV, measurements of neutrino oscillation, searches for exotic physics (sterile neutrinos, light dark matter), and searches for beyond-Standard-Model (BSM) processes, including proton decay, are made with atmospheric and accelerator based neutrino sources. At very high energies, are astrophysical neutrino searches and precision tests of Standard Model processes; detector response, resolution and statistical precision are the limiting factors in very high energy measurements so we will not discuss them further here. 

{\bf\boldmath Low Energy Nuclear Processes: $E_\nu\sim1$--100~MeV:}  CE$\nu$NS is a neutral-current process in which a neutrino elastically scatters off the whole nucleus. The first detection of CE$\nu$NS was published by the COHERENT collaboration~\cite{COHERENT:2017ipa} in 2017, and opened an exciting chapter of using CE$\nu$NS to test the Standard Model and search for new physics~\cite{Patton:2012jr,Coloma:2017ncl,Liao:2017uzy,Cadeddu:2017etk,Papoulias:2017qdn,Farzan:2018gtr,Abdullah:2018ykz,Denton:2018xmq,Canas:2018rng,Esteban:2018ppq,AristizabalSierra:2018eqm,Billard:2018jnl,Dutta:2019eml,Dutta:2019nbn,Cadeddu:2020lky,Miranda:2020tif}. The interaction rates of CE$\nu$NS are sensitive to neutron distributions in the nuclear targets (neutron nuclear structure factors), which dominate the theoretical uncertainties~\cite{Payne:2019wvy,AristizabalSierra:2019zmy,Hoferichter:2020osn}. As CE$\nu$NS experiments continue to improve their experimental precision~\cite{CONNIE:2019swq,XENON:2020gfr,COHERENT:2020iec,CONUS:2020skt,COHERENT:2021xhx}, more precise theory calculations of these structure factors are needed. In the same way as the proton responses are derived from electron scattering data, such precision calculations require data input, with CE$\nu$NS the most promising source probing the neutron distribution besides parity-violating electron scattering. To disentangle a potential new-physics contribution, measurements for different momentum transfers and a variety of nuclear targets are thus mandatory. 

In the same energy window, neutrinos can also inelastically scatter off nuclei via charged-current or neutral-current interactions~\cite{Raghavan:1986fg,Haxton:1987kc,Fukugita:1988hg,Engel:1996zt,KARMEN:1998xmo,Kolbe:1999au,Hayes:1999ew,Volpe:2000zn,LSND:2001fbw,Kolbe:2002gk}. This is a crucial detection channel for supernova neutrino bursts~\cite{Super-Kamiokande:2007zsl,Duba:2008zz,Scholberg:2012id,Laha:2014yua,JUNO:2015zny,Lu:2016ipr,Li:2020ujl,DUNE:2020zfm}. In particular, DUNE will enable a high-statistics detection of MeV electron neutrinos via $\nu_e +$Ar$\rightarrow e^- + {}^{40}$K$^*$. This channel is also important for solar-neutrino studies, where impressive sensitivity is possible~\cite{Capozzi:2018dat}. To reconstruct the energy of the incoming neutrinos, we need to know the exclusive cross sections to each individual excited state in $^{40}$K. At slightly higher energies, e.g., $E_\nu\gtrsim 50$~MeV, there could be nucleon knockout in the final state~\cite{KOLBE1992599,PhysRevLett.76.2629,Gardiner:2018zfg}, which could significantly bias the energy reconstruction if the outgoing nucleons are not detected, so the need for theoretical cross section predictions with accurate final states is even more pressing.

{\bf\boldmath Intermediate Energy Cross Sections: $E_\nu \sim 0.1$--20~GeV:} 
In general, for oscillation physics and rare or exotic searches, neutrino interaction cross sections are important~\cite{Mosel:2016cwa,NuSTEC:2017hzk}. In this energy regime, charged current quasi-elastic, multi-nucleon, 
resonant processes, deep-inelastic scattering, and the transition region play an increasingly important role for future oscillation measurements. Between 100 GeV to 1 TeV, neutrino experiments use deep inelastic scattering for Standard Model cross section physics and for searches for new physics beyond the Standard Model. Accelerator-based (anti)neutrino sources and atmospheric neutrinos have a broad energy spectrum, so multiple channels contribute to event rates; the energy dependence of each process is important as oscillation depends on energy. Rare charged or neutral current processes~\cite{Ankowski:2015lma} may also be important as signal or background as well, especially for exotics searches~\cite{Altmannshofer:2014pba,Magill:2016hgc,Magill:2017mps,Coloma:2017ppo,deGouvea:2018cfv,Bertuzzo:2018itn,Ballett:2019xoj,Berryman:2019dme,Altmannshofer:2019zhy,Schwetz:2020xra,Atkinson:2021rnp}. 
For each process, well-grounded theoretical predictions are needed to assess event rates and uncertainties. This is complicated by the nuclear dynamics of the target medium (commonly, carbon, oxygen or argon). Furthermore, neutrino experiments also need predictions for all relevant flavors of neutrinos ($\nu_e$, $\nu_\mu$, $\nu_\tau$) and antineutrinos, to perform appearance searches (e.g., $\nu_\mu \rightarrow \nu_e$) or measure CP violation ($\nu_\mu \rightarrow \nu_e$ vs.\ $\overline{\nu}_\mu \rightarrow \overline{\nu}_e$).

Furthermore, the signal selection  may depend on the composition and kinematics of exclusive final states. The unprecedented increases to beam exposure and detector size also enable explorations of final states in increasing detail; DUNE for example will have a highly capable near detector complex, which includes improved particle detection thresholds and sensitivity to the energy evolution of the cross section~\cite{DUNE:2021tad}. Theory needs to provide experimental programs with semi-inclusive or exclusive predictions as well as inclusive to ensure a robust implementation of interaction models in experiments, which may need to make approximations to be used.

In addition to model development and robust prediction, theory is needed on  uncertainty quantification~\cite{Ankowski:2014yfa}. A positive example of this is the assessment of uncertainties for models used in accelerator experiments~\cite{Barbieri:2019ual,Andreoli:2021cxo,Chakrani:2022tey}. 
Theory may be refined from existing or new measurements from electron scattering~\cite{Ankowski:2022thw},
and the uncertainties on those measurements also need to be quantified and propagated to experiments.
Multiple theoretical efforts are also critical to 
discuss different approximations or assumptions within models or within experimental implementation of those models.

A critical need for the future will be close collaboration between theory and experimental groups.  Historically, the expertise needed to carry out theoretical work in neutrino scattering may come from high energy physics (HEP) or nuclear physics (NP). Programs like the Neutrino Theory Network supports work by theorists working on specific topics across NP and HEP, but this does not address broader issues of problems beyond individual groups nor prioritization of experimental needs.
The (theory) needs of experiments evolve with time, as the experiment data sample size increases and includes more complex techniques, so dedicated iteration between theory and experiment is needed. Community wide efforts were started with the ECT* and FNAL~\cite{Barrow:2020gzb} workshops on creating a detailed discussion between theorists and experimentalists. 

Learning from the successes of the LHC can help to accelerate and strengthen the theory improvements in neutrino physics in the next 5--10 years. The Les Houches studies have been vital in developing precision theory calculations for the LHC~\cite{Amoroso:2020lgh,Brooijmans:2020yij}. Workshops that produce meaningful and controlled studies around open problems could be very helpful.  Experiments need to provide qualitative and quantitative information about which sources of systematic uncertainty or types of effects are important for a given analysis. Theorists need to provide what effects are not considered in a given model, or which important assumptions should be revisited. Given the complexity and importance of near detector information, experiments may need to carry out sensitivity estimates with coordinated input from theory groups.  
The formation of a topical group around specific issues would accelerate and focus effort across the theoretical and experimental communities. 

In the last decade, there has been success incorporating theory into oscillation analyses. For example, initial models of the dominant reaction channel, charged current quasi-elastic scattering (CCQE), for the Tokai-to-Kamioka (T2K) experiment, were very crude~\cite{PhysRevLett.26.445,Smith:1972xh}. With the help of multiple theoretical efforts~\cite{Nieves:2014lpa,Benhar:1994hw,Benhar:2005dj}, T2K's uncertainty treatment now includes multiple CCQE models and information from electron scattering data~\cite{T2K:2021xwb}. 
Furthermore, models also have been developed with both one body and two body currents and incorporated into the analysis and/or uncertainty treatment~\cite{Gran:2013kda,Martini:2009uj,RuizSimo:2016rtu,Gallmeister:2016dnq}.
This was an iterative process, where models were compared to external data, key features discussed, impact tested within the T2K oscillation analysis, and finally incorporated in the systematic uncertainty estimation and/or baseline model used; with the theory effort done with dedicated involvement of individuals, guest lectures in T2K meetings and at conferences and workshops. While this one example comes from T2K and a subset of theoretical efforts, it is worth noting that there is a wide community of  experimental and theoretical efforts focused on developing models and uncertainties. The inclusion of theoretical models in experimental analyses also requires essential work on dedicated event simulations to interface to experiments, called generators; the role of generators is described in Section~\ref{sec:generator}.

Current and future oscillation experiments are anticipated to need detailed predictions 
of the reactions summarized in Table~\ref{tab:neutrino_interactions}, inclusive, semi-inclusive, or exclusive results for a range of relevant targets materials (especially oxygen, carbon, and argon).  Each of those channels will need robust predictions, incorporating electron scattering, single nucleon neutrino scattering, or photoproduction data where appropriate. An experimental program is underway to measure precise electron scattering data covering the broad kinematics of long baseline experiments. Theory historically is necessary for a complete set of uncertainties to be assessed. A critical part of appearance searches like DUNE will be predictions of differences between neutrinos and antineutrino cross sections ($\nu_e$ vs.\ $\overline{\nu}_e$) and  differences between flavors ($\nu_e$, $\nu_\tau$ and $\nu_\mu$). The transition region, described in detail in Section~\ref{sec:sis-dis}, is anticipated to play a more important role in oscillation analyses than previously. The priority among these important theoretical topics needs to be determined by experiments and shared with the community.

\renewcommand{\arraystretch}{1.4}
\begin{table}
\centering
\begin{tabular}{||c | c | c||} 
 \hline
 Process & Neutrino Energy Range & Example Final State \\ \hline\hline
 Coherent Elastic Scattering & $\lesssim 50$~MeV & $\nu +A$ \\ \hline 
 Inelastic Scattering & $\lesssim 100$~MeV & $e+ {}^\mathrm{A}(Z+1)^*(\rightarrow{}^\mathrm{A}(Z+1)+n\gamma)$\\ \hline 
 Quasi-Elastic Scattering & 100~MeV--1~GeV & $l+p + X$ \\ \hline 
 Two-Nucleon Emission   & 1 GeV & $ l+2N+X$ \\ \hline 
 Resonance Production  & 1--3 GeV & $l + \Delta(\rightarrow N+\pi)+X$ \\ \hline
 Shallow Inelastic Scattering  & 3--5 GeV & $ l + n \pi + X$ \\ \hline   
 Deep Inelastic Scattering  & $\gtrsim 5$~GeV & $ l + n \pi + X$  \\
 \hline
\end{tabular}
\caption{Main neutrino interaction channels in different energy ranges. 
}
\label{tab:neutrino_interactions}
\end{table}

\newcommand{\cevns}{\protect{CE$\nu$NS}\xspace}

\section{Coherent elastic neutrino-nucleus scattering }
\label{sec:SM} 

Coherent elastic neutrino-nucleus scattering (CE$\nu$NS) is a neutral-current process that arises when the momentum transfer in the neutrino-nucleus interaction is less than the inverse of the size of the nucleus. For typical nuclei, this corresponds to neutrinos with energies $E_\nu\lesssim 50$ MeV. In the SM, the interaction  is mediated by the $Z$-boson, with its vector component leading to the coherent enhancement~\cite{Freedman:1973yd}. As reference point, we first write the cross section in the form
\begin{align}
\frac{\text{d}\sigma}{\text{d}T}=\frac{G_F^2M}{4\pi}\bigg(1-\frac{M T}{2E_\nu^2}\bigg)Q_\text{w}^2\big[F_\text{w}(q^2)\big]^2\,,
\label{eq:SMcrosssection} 
\end{align}
where $G_F$ is the Fermi constant, $T=E_R=q^2/(2M)=E_\nu-E_\nu'$ is the nuclear recoil energy (taking values in $[0,2E_\nu^2/(M+2E_\nu)]$), $F_\text{w}(q^2)$ is the weak form factor, $M$ is the mass of the target nucleus, and $E_\nu$ ($E_\nu'$) is the energy of the incoming (outgoing) neutrino. The tree-level weak charge is defined by 
\begin{align}
Q_\text{w}=Z \big(1-4\sin^2\theta_W\big)-N\,,
\end{align}
with proton number $Z$, neutron number $N$, and weak mixing angle $\sin^2 \theta_W$. To first approximation, the weak form factor $F_\text{w}(q^2)$ depends on the nuclear density distribution of protons and neutrons. In the coherence limit $q^2\to 0$ it is normalized to $F_\text{w}(0)=1$, with the coherent enhancement of the cross section reflected by the scaling with $N^2$ via the weak charge, given the accidental suppression of the proton weak charge $Q_\text{w}^p\ll1$ (see Eq.~\eqref{eq:weakcharges} below). Consequently, this implies that \cevns is mainly sensitive to the neutron distribution in the nucleus.     

In writing the cross section as in Eq.~\eqref{eq:SMcrosssection} a number of subtleties have been ignored: subleading kinematic effects, axial-vector contributions, form-factor effects besides the density distributions, and radiative corrections. In the subsequent sections, each of these effects is addressed in more detail, providing an extended presentation of the related discussion in Ref.~\cite{CEvNS_WP}.

\subsection{Structure of the Standard-Model contribution}

The quark-level interactions in the SM are
\begin{equation}
\label{Lagr_SM}
{\mathcal{L}}^\text{SM}=-\sqrt{2}G_F\sum_{q=u,d,s}\Big(C_q^V\bar\nu\gamma^\mu P_L\nu \,\bar q\gamma_\mu q
+C_q^A\bar\nu\gamma^\mu P_L\nu \,\bar q\gamma_\mu\gamma_5 q\Big)\,,\\
\end{equation}
with $P_L=(1-\gamma_5)/2$ and tree-level Wilson coefficients
\begin{align}
\label{Wilson_SM}
C_u^V&=\frac{1}{2}\bigg(1-\frac{8}{3}\sin^2\theta_W\bigg)\,,\qquad
C_d^V=C_s^V=-\frac{1}{2}\bigg(1-\frac{4}{3}\sin^2\theta_W\bigg)\,,\notag\\
C_u^A&=-C_d^A=-C_s^A=-\frac{1}{2}\,.
\end{align} 
The vector operator gives rise to the coherent contribution quoted in Eq.~\eqref{eq:SMcrosssection}, while the axial-vector operator adds an additional contribution that is not coherently enhanced. Including the dominant kinematic corrections, the cross section can be written in the form
\begin{equation}
\label{CEvNS_SM}
\frac{\text{d} \sigma}{\text{d} T}=\frac{G_F^2 M}{4\pi}\bigg(1-\frac{M T}{2E_\nu^2}-\frac{T}{E_\nu}\bigg)Q_\text{w}^2\big[F_\text{w}(q^2)\big]^2
+\frac{G_F^2M}{4\pi}\bigg(1+\frac{M T}{2E_\nu^2}-\frac{T}{E_\nu}\bigg)F_A(q^2)\,,
\end{equation}
with an axial-vector form factor $F_A(q^2)$ discussed in detail in Sec.~\ref{sec:Nuclear}. This contribution vanishes for nuclei with even number of protons and neutrons, which have spin-zero ground states. 

Moving from the quark-level interactions in Eq.~\eqref{Lagr_SM} to the neutrino-nucleus cross section in Eq.~\eqref{CEvNS_SM} involves a two-step process~\cite{Hoferichter:2020osn}. In the first step, hadronic matrix elements are required to obtain the matching to single-nucleon operators, i.e., vector and axial-vector form factors of the nucleon, respectively. For the vector operators, the normalization is determined via the valence-quark content, leading to the relations
\begin{equation}
\label{eq:weakcharges}
Q_\text{w}^p=2(2C_u^V+C_d^V)=1-4\sin^2\theta_W\,,\qquad 
Q_\text{w}^n=2(C_u^V+2C_d^V)=-1\,,
\end{equation}
while the $q$-dependent corrections, expressed in terms of radii and magnetic moments, are subsumed into the weak form factor $F_\text{w}(q^2)$. Similarly, $F_A(q^2)$ depends on the axial charges and radii of the nucleon. 
In the second step, the nuclear responses need to be derived from a multipole expansion~\cite{Serot:1978vj,Donnelly:1978tz,Donnelly:1979ezn,Serot:1979yk,Walecka:1995mi}, in which the leading contribution can be interpreted in terms of the proton and neutron density distributions. The detailed breakdown of the form factors $F_\text{w}(q^2)$ and $F_A(q^2)$ is provided in Sec.~\ref{sec:Nuclear}. In addition, the relations presented here hold at tree level in the SM, with radiative corrections discussed in Sec.~\ref{sec:radiative}.

\subsection{Nuclear and hadronic physics} 
\label{sec:Nuclear}

Due to the suppression of the weak charge of the proton, the most important nuclear response required for the interpretation of \cevns~experiments is related to the neutron distribution. While the charge density of nuclei has been probed extensively in elastic electron scattering experiments~\cite{Hofstadter:1956qs,DeJager:1987qc, Fricke:1995zz, Angeli:2013epw}, the neutron density distributions are hard to determine. While precise experimental data exist for observables that are sensitive to the neutron density distribution or the neutron skin, such as the nuclear dipole polarizability~\cite{Tamii:2011pv,Rossi:2013xha,Hashimoto:2015ema,Birkhan:2016qkr}, efforts using hadronic probes require a careful analysis of model-dependent uncertainties (see, e.g., Ref.~\cite{Thiel:2019tkm}). In contrast, electroweak processes such as parity-violating electron scattering (PVES)~\cite{Donnelly:1989qs} and \cevns~
have long been considered as clean probes of the neutron densities. Both of which, though long considered experimentally challenging, have become a reality in recent years: 
\cevns on CsI~\cite{COHERENT:2017ipa} and Ar~\cite{COHERENT:2020iec} has been observed by the COHERENT collaboration, and the PREX and PREX II experiments at Jefferson Lab (JLab) have determined the weak form factor of $^{208}$Pb at two values of the momentum transfer~\cite{Abrahamyan:2012gp, Horowitz:2012tj,PREX:2021umo}, with results from the CREX experiment on  $^{48}$Ca~\cite{Kumar:2020ejz} are expected soon. In the future, 
the MREX experiment at MESA aims at a yet higher precision in $^{208}$Pb~\cite{Becker:2018ggl}. 

The observation of \cevns~can therefore further provide important nuclear structure information through the the determination of the weak form factor, which constrains the neutron density distribution and thus the neutron radius and the neutron skin, at least at low momentum transfers where the process remains coherent~\cite{Horowitz:2003cz,Patton:2012jr,Cadeddu:2017etk,Ciuffoli:2018qem,Payne:2019wvy,Yang:2019pbx,AristizabalSierra:2019zmy,Papoulias:2019lfi,Hoferichter:2020osn,Co:2020gwl,Coloma:2020nhf,VanDessel:2020epd}. 
These measurements complement PVES experiments not only due to 
additional data, but also due to different energy ranges and nuclear targets, which could be used to calibrate nuclear-structure calculations. Furthermore, improved measurements of the neutron skin would have important consequences for the equation of state of neutron-rich matter, which plays an essential role in understanding the structure and evolution of neutron stars~\cite{RocaMaza:2011pm,Tsang:2012se,Lattimer:2012xj,Hebeler:2013nza,Hagen:2015yea}.

However, arguably the most intricate aspect of nuclear-structure input concerns searches for BSM physics. Without independent experimental information for the neutron responses, which, potentially apart from PVES, is difficult to obtain, \cevns~cross sections provide constraints on the combination of nuclear responses and BSM effects. In fact, in order to derive BSM constraints beyond the level at which current nuclear-structure calculations constrain the neutron distribution, a combined analysis of multiple targets and momentum transfers is required to distinguish between nuclear structure and potential BSM contributions. To do so, a detailed understanding of the nuclear responses is prerequisite, as we discuss in the remainder of this section.  

Traditionally, the weak form factor 
\begin{equation}
\label{Eq:Fw}
F_\text{w}(q^2) = \frac{1}{Q_\text{w}} \left[ Z Q_\text{w}^p F_{p}(q^2) +N Q_\text{w}^n F_n(q^2)\right]
\end{equation}
has been modeled in terms of proton and neutron densities 
\begin{equation}
\label{Eq:Fn}
F_n(q^2) = \frac{4\pi}{N} \int \text{d}r~r^2~\frac{\sin(qr)}{qr}~\rho_n(r)\,,\qquad 
F_p(q^2) = \frac{4\pi}{Z} \int \text{d}r~r^2~\frac{\sin(qr)}{qr}~\rho_p(r)\,,
\end{equation}
where $\rho_n(r)$ and $\rho_p(r)$ are neutron and proton density distributions normalized to the neutron and proton numbers. 
Phenomenological form factors, such as Helm~\cite{Helm:1956zz} and Klein-Nystrand~\cite{Klein:1999qj}, are 
based on empirical fits to elastic electron scattering data, and similar parameterizations are assumed for the neutron form factor.
In the Helm approach~\cite{Helm:1956zz}, the nucleon distribution is given by the convolution of a uniform density with radius $R_0$ and a Gaussian profile with width $s$, the surface thickness. The resulting form factor is
\begin{equation}
F_{\text{Helm}}(q^2) = \frac{3 j_1(qR_0)}{qR_0} e^{-q^2s^2/2}\,,
\end{equation}
where $j_1(x)$ is the spherical Bessel function of order one. The Klein-Nystrand approach~\cite{Klein:1999qj} relies on a surface-diffuse distribution that results from folding a short-range Yukawa potential with range $a_k$ over a hard sphere distribution with radius $R_A$. The resulting form factor becomes
\begin{equation}
F_{\text{KN}}(q^2) = \frac{3 j_1(qR_A)}{qR_A} \left[\frac{1}{1+q^2a_k^2} \right]\,.
\end{equation}
In both cases, it should be stressed that these parameterizations need to assume a value for the neutron radius---related to $R_0$ or $R_A$---and only try to capture the leading nuclear responses, with the neutron distribution largely unconstrained. Actual nuclear-structure calculations of the nuclear responses are based on relativistic mean-field methods~\cite{Horowitz:2003cz,Yang:2019pbx},
nonrelativistic energy-density functionals~\cite{Patton:2012jr,Co:2020gwl,VanDessel:2020epd}, shell-model calculations~\cite{Hoferichter:2016nvd,Hoferichter:2018acd,Hoferichter:2020osn}, and, for argon, 
a first-principles calculation using coupled-cluster theory~\cite{Payne:2019wvy}.

\begin{figure}[t]
	\begin{center}
		\includegraphics[width=0.48\linewidth,bb=0 0 300 300]{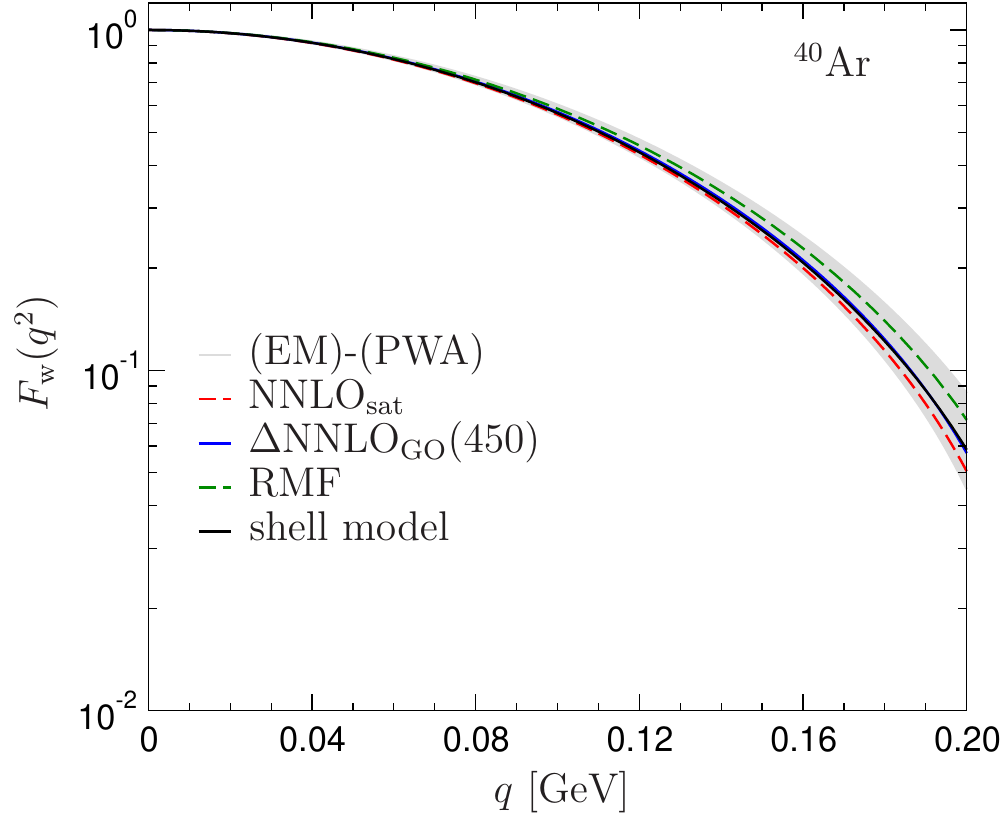}
	\end{center}
	\caption{Theoretical predictions for the weak form factor of $^{40}$Ar, from relativistic mean-field methods~\cite{Yang:2019pbx}, coupled-cluster~\cite{Payne:2019wvy}, and shell-model~\cite{Hoferichter:2020osn} calculations. The curves/bands labeled (EM)-(PWA), NNLO$_\text{sat}$, and $\Delta$NNLO$_\text{GO}$(450) refer to the chiral interactions considered in Ref.~\cite{Payne:2019wvy}. Figure adapted from Ref.~\cite{Hoferichter:2020osn}.}
	\label{fig:Fweak_Ar}
\end{figure}

Retaining all responses that at least display some degree of coherent enhancement, the weak form factor can be expressed as~\cite{Hoferichter:2020osn}
\begin{align}
\label{Fw_definition}
F_\text{w}(q^2)&=\frac{1}{Q_\text{w}}\bigg[\bigg(Q_\text{w}^p\Big(1-\frac{\langle r_E^2\rangle^p}{6} q^2 -\frac{1}{8{m_N}^2} q^2\Big)
-Q_\text{w}^n \frac{\langle r_E^2\rangle^n+\langle r_{E,s}^2\rangle^N}{6} q^2\bigg) {\mathcal{F}}^M_p(q^2)\notag\\
&+\bigg(Q_\text{w}^n\Big(1-\frac{\langle r_E^2\rangle^p+\langle r_{E,s}^2\rangle^N}{6} q^2 -\frac{1}{8{m_N}^2} q^2\Big)-Q_\text{w}^p \frac{\langle r_E^2\rangle^n}{6} q^2\bigg){\mathcal{F}}^M_n(q^2)\notag\\
&+ \frac{Q_\text{w}^p(1+2\kappa^p)+2Q_\text{w}^n(\kappa^n+\kappa_s^N)}{4{m_N}^2} q^2 {\mathcal{F}}^{\Phi''}_p(q^2)\notag\\
&+ \frac{Q_\text{w}^n(1+2\kappa^p+2\kappa_s^N)+2Q_\text{w}^p\kappa^n}{4{m_N}^2} q^2 {\mathcal{F}}^{\Phi''}_n(q^2)\bigg]\,,
\end{align}
where $m_N$ denotes the nucleon mass. The $M$ responses ${\mathcal{F}}_{N}^M$, $N=\{p,n\}$, correspond to the charge distribution and the $\Phi''$ responses ${\mathcal{F}}_{N}^{\Phi''}$ to spin-orbit corrections, which add coherently for nucleons with spin aligned with the orbital angular momentum. The coefficients are determined by the weak charges $Q_\text{w}^{N}$, but, in addition, involve the nucleon matrix elements of the vector current, expressed here in terms of the charge radii $\langle r_E^2\rangle^N$, strangeness radii $\langle r_{E,s}^2\rangle^N$, and magnetic moments $\kappa^N$, $\kappa_s^N$. In consequence, since $F_\text{w}(q^2)$ originates from a linear combination of weak charges, hadronic matrix elements, and nuclear responses, its shape will change if BSM contributions modify the weak charges.  
Further corrections could be expected from two-body currents, but for the relevant responses such contributions only start at loop level in the chiral expansion~\cite{Hoferichter:2020osn}. Figure~\ref{fig:Fweak_Ar} compares several predictions for argon's $F_\text{w}(q^2)$. The theoretical spread indicates the accuracy with which nuclear responses can currently be calculated.

\begin{figure}[t]
	\begin{center}
		\includegraphics[width=0.48\linewidth,bb=0 0 300 300]{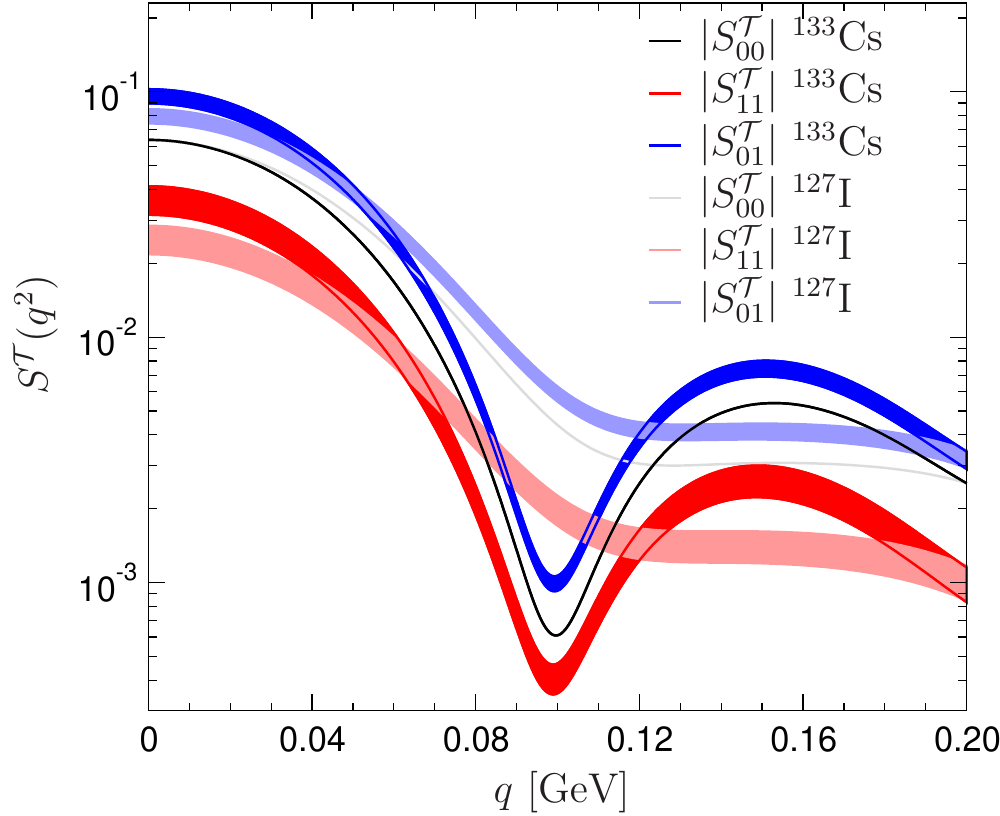}
	\end{center}
	\caption{Calculated tranverse spin-dependent structure factors for \cevns, see Eq.~\eqref{FA}, exemplarily shown for cesium and iodine~\cite{Hoferichter:2020osn}. Figure taken from Ref.~\cite{Hoferichter:2020osn}.}
	\label{fig:SijT}
\end{figure}

The decomposition of the subleading axial-vector form factor reads~\cite{Hoferichter:2020osn}
\begin{equation}
\label{FA}
F_A(q^2)=\frac{8\pi}{2J+1}
\Big(\big(g_A^{s,N}\big)^2 S_{00}^\mathcal{T}(q^2)-g_A g_A^{s,N} S_{01}^\mathcal{T}(q^2)+ g_A^2 S_{11}^\mathcal{T}(q^2)\Big)\,,
\end{equation}
where $J>0$ is the nuclear spin, $g_A$, $g_A^{s,N}$ refer to the appropriate nucleon axial-vector matrix elements, and $S_{ij}^\mathcal{T}(q^2)$ are the nuclear responses in isospin basis (with $i,j=0/1$ for the isoscalar/isovector parts). The dominant contribution arises from the isovector component, with normalization
\begin{equation}
F_A(0)=\frac{4}{3}g_A^2\frac{J+1}{J}\big(\langle \mathbf{S}_p\rangle-\langle \mathbf{S}_n\rangle\big)^2\,,
\end{equation}
when strangeness and two-body corrections are neglected. $\langle \mathbf{S}_N\rangle$ refers to the neutron and proton spin expectation values in the nucleus. The calculation of $S_{ij}^\mathcal{T}(q^2)$ requires a careful multipole decomposition, which shows that only the transverse part contributes to \cevns, with additional corrections from two-body currents and the axial radius, see Fig.~\ref{fig:SijT} for recent results~\cite{Hoferichter:2020osn}. 

\subsection{Radiative corrections}
\label{sec:radiative}

The relation~\eqref{eq:weakcharges} for the weak charges holds true at tree-level, in which case $Q_\text{w}^{p,n}$ are flavor universal and apply both to neutrino and electron scattering. Once including radiative corrections, process- and flavor-dependent contributions arise, in such a way that separate weak charges need to be defined. 
For \cevns, the corresponding radiative corrective have been studied in Refs.~\cite{Barranco:2005yy,Erler:2013xha,Tomalak:2020zfh}, see also Ref.~\cite{Crivellin:2021bkd} for a comparison. Keeping the decomposition $Q_\text{w}=Z Q_\text{w}^p+N Q_\text{w}^n$, one has from Ref.~\cite{Erler:2013xha}
\begin{align}
 Q_\text{w}^{\nu_e, p}&=0.0766\,, & Q_\text{w}^{\nu_\mu, p}&=0.0601\,, &
 Q_\text{w}^{\nu_\tau, p}&=0.0513\,,\notag\\
 Q_\text{w}^{\nu_\ell, n}&=-1.0233\,,
\end{align}
i.e., only $Q_\text{w}^{\nu_\ell, p}$ becomes flavor dependent. These values are in agreement with Ref.~\cite{Tomalak:2020zfh}
\begin{align}
    Q_\text{w}^{\nu_e, p}&=0.0747(34)\,, & Q_\text{w}^{\nu_e, p}-Q_\text{w}^{\nu_\mu, p}&=0.01654\,, & Q_\text{w}^{\nu_\mu, p}-Q_\text{w}^{\nu_\tau, p}&=0.00876\,,\notag\\
    Q_\text{w}^{\nu_\ell, n}&=-1.02352(25)\,. 
\end{align}
The main difference between Refs.~\cite{Erler:2013xha,Tomalak:2020zfh} concerns the treatment of the light-quark loops in $\gamma$--$Z$ mixing diagrams, which lead to non-perturbative effects that have been absorbed into $Q_\text{w}^{\nu_\ell, p}$. 

The consequences of process-dependent corrections become apparent when comparing to the SM values for the weak charges probed in electron scattering~\cite{Erler:2013xha,Zyla:2020zbs}
\begin{equation}
  Q_\text{w}^{e, p}=0.0710\,,\qquad   Q_\text{w}^{e, n}=-0.9891\,,
\end{equation}
which include further corrections ($\gamma Z$ box diagrams and axial-current renormalization) that do not play a role in \cevns. 

\subsection{Neutrino magnetic moment} 
Since oscillation experiments demonstrated that neutrinos have mass, they should at least carry magnetic dipole moments. If the CP-invariance is violated, they can have electric dipole moments as well. The value of the magnetic moment is very small in the SM, but they may be larger than the SM prediction if new physics beyond SM contributes. The SM value  for a Dirac neutrino is as low as of the order of $10^{-20} \mu_B$ for the inverted hierarchy and lower for the normal hierarchy \cite{Balantekin:2013sda}. 

Best laboratory limits on the neutrino magnetic moment comes from neutrino-electron scattering experiments using either reactor or solar neutrinos as neutrino source. Smallest possible limits come from smallest electron recoil energy which can be measured since electromagnetic contribution to the cross section would exceed the weak contribution at lower electron recoil energies. Current experimental limits are of the order of $2.9 \times 10^{- 11} \mu_B$ \cite{Beda:2013mta,Borexino:2017fbd}. 
Mass eigenstates of neutrinos have well-defined magnetic moments. Hence the measured value of the neutrino magnetic moment in a given experiment also depends on the proportion of different mass eigenstates present in the signal. At \cevns this proportion is very different than that at solar or reactor neutrino experiments. Astrophysical limits on neutrino magnetic moments are somewhat tighter than laboratory limits, but they are subject to systematic errors. They typically explore the consequences of energy losses due to the creation of neutrino-antineutrino pairs from stellar plasmas. Significant energy loss would prevent Cepheid stars from being formed if $\mu_{\nu} >  4 \times 10^{-11} \mu_B$ \cite{Mori:2020qqd}. The presence of stars at the tip of the red giant branch in globular clusters require $\mu_{\nu} < 1.5 \times 10^{-12} \mu_B$ \cite{Capozzi:2020cbu}. 

A~\cevns~experiment has a dominant contribution from $Z$-boson exchange, but also a much smaller subdominant contribution coming from the presence of neutrino magnetic moment which exchanges a photon with the nucleus. Unlike the dominant contribution which probes the neutron distribution in the target, the subdominant contribution is sensitive to the proton distribution. 

Majorana neutrinos cannot have diagonal magnetic moments. This makes theoretical constraints on Majorana neutrino magnetic moments somewhat weaker \cite{Davidson:2005cs,Bell:2006wi}, suggesting that if a magnetic moment with a value slightly below the above limits is experimentally observed, neutrinos are likely to be Majorana particles.

\section{Lattice QCD and inputs for neutrino scattering}\label{sec:lqcd}

Neutrino-nucleus scattering is described in the SM by the exchange of a $W^\pm$ or $Z^0$ boson between a neutrino and a quark that is bound in a nucleus. 
Low-energy scattering can be accurately described using nuclear effective theories, while high-energy scattering can be factorized into hard scattering amplitudes calculable in perturbative QCD and nonperturbative PDFs. 
For intermediate neutrino energies around 1 GeV, nonperturbative QCD processes such as pion and other resonance production make significant cross-section contributions.
Lattice QCD (LQCD) provides a first principal  framework for numerically calculating the QCD path integral with systematically improvable control over systematics. It can be used to determine nonperturbative inputs to nuclear effective theories, nucleon and nuclear PDFs, and benchmark phenomenological models of resonant scattering in the transition region between the low- and high-energy expansions provided by nuclear effective theories and perturbative QCD. 

The information provided by LQCD is often complementary to that provided by neutrino and electron scattering experiments, in part because some systems that are relatively simple to study in LQCD are challenging to study experimentally, such as free neutrons, while other systems such as large nuclei are challenging to study directly with LQCD.
Further, it is straightforward to study both the axial and vector components of the electroweak currents relevant for neutrino scattering using LQCD.
For elastic form factors and other observables, vector-current LQCD results can be compared with precise results from electron scattering experiments, see Ref.~\cite{Ankowski:2022thw}, and used to validate LQCD methods, while precise axial-current  LQCD results will provide predictions that can be used to inform and validate nuclear effective theories.
Moreover, we note that, complementary to direct lattice calculations, there is also an extensive
literature~\cite{Lu:1997sd,Zhang:2019iyx,Hobbs:2014lea,Perdrisat:2006hj,A1:2013fsc,Ye:2017gyb,Bernard:1998gv,Schindler:2006it,Chung:1991st,Cardarelli:1995dc,Miller:2002ig,Ma:2002ir,Ma:2002xu,Punjabi:2015bba,Hill:2017wgb,Bernard:2001rs,Kelly:2004hm,Bhattacharya:2011ah,Bodek:2007ym,Alvarez-Ruso:2018rdx}
on determinations of the electromagnetic and axial form factors of the nucleon based on various phenomenological fits and theoretical models. In this whitepaper, we concentrate primarily on lattice-based approaches, which, as discussed above, are capable of accessing kinematics and flavor currents that can otherwise be challenging to constrain empirically in an {\it ab initio} fashion immediately related to the QCD Lagrangian. Still, we stress that there are valuable synergies between the available lattice and phenomenological/model-based methods that can help extend or benchmark one approach off the other.
This section discusses the status and outlook for LQCD calculations relevant for neutrino scattering from low to high energies, see also Ref.~\cite{Kronfeld:2019nfb}.
The use of LQCD results to constrain inputs to nuclear effective theories and validate phenomenological models of the transition region is discussed further in the following sections.

\subsection{Nucleon form factors \label{ssec:nucleon_form_factors}}

The scattering amplitude for elastic charged-current neutrino-nucleon scattering in the isospin limit can be expressed as a linear combination of four form factors~\cite{LlewellynSmith:1971uhs}: the Dirac $F_1$, Pauli $F_2$, axial $G_A$, and induced pseudoscalar $\widetilde G_P$ form factors defined by
\begin{eqnarray}
  \left\langle N(\vec{p}+\vec{q}) | A_\mu^a (\vec{q}) | N(\vec{p})\right\rangle  &=&
  {\overline u}_N(\vec{p}+\vec{q}) \left(
     G_A(Q^2) \gamma_\mu + {\widetilde G_P} (Q^2) \frac{q_\mu}{M_N}
 \right) \gamma_5 \tau^a u_N(\vec{p}),
 \label{eq:Axial_FF} \\
  \left\langle N(\vec{p}+\vec{q}) | V_\mu^a (\vec{q}) | N(\vec{p})\right\rangle
  &=& {\overline u}_N(\vec{p}+\vec{q})\ \left(F_1(Q^2)\gamma^\mu
        + F_2(Q^2)\frac{i\sigma^{\mu\nu}q_\nu}{2M_N} \right)\tau^a  u_N(\vec{p})\,,
\label{eq:EM_FF}
\end{eqnarray}
where the isovector axial current is $A_\mu^a = \overline{q}\gamma_\mu \gamma_5 \tau^a q$ with $\tau^a$ a Pauli matrix in isospin space, the isovector vector current is $V_\mu^a(x) = \overline{q}\gamma_\mu \tau^a q$, the $u_N(\vec{p})$ are Dirac spinors, and $M_N$ is the nucleon mass.
Form factors with spacelike momentum transfers can be computed using Euclidean matrix elements accessible to LQCD where  $q = (E_{\vec{p} +\vec{q}} - E_{\vec{p}}
,\vec{q})$ and $Q^2 = \vec{q}^2 - (E_{\vec{p} +\vec{q}} - E_{\vec{p}})^2$ and the form factors can then be used to compute scattering cross sections where $Q^2 = -(q^0)^2 + \vec{q}^2 > 0$.

The electric $G_E = F_1 - \frac{Q^2}{4M_N^2} F_2$ and magnetic $G_M = F_1 + F_2$ form factors are often used in place of $F_1$ and $F_2$ to provide a complete set of isovector  nucleon elastic form factors $\{G_E, G_M, G_A, \widetilde{G}_P\}$.
The same four nucleon form factors and their flavor-singlet analogs are used to compute quasi-elastic charged- and neutral-current neutrino-nucleus scattering cross sections in nuclear many-body models and neutrino event generators, as discussed below.
$G_E$ and $G_M$ are precisely known from electron scattering experiments and provide validation for LQCD methods and uncertainty quantification, while LQCD predictions of $G_A$ and $\widetilde{G}_P$ with robust control of systematic uncertainties will provide valuable input for nuclear effective theory and neutrino event generator predictions of quasi-elastic scattering.
The current status of lattice QCD calculations of the electromagnetic form factors $G_E$ and $ G_M$ 
is discussed in Ref.~\cite{Park:2021ypf}. The lattice results are now in good 
agreement  with the Kelly~\cite{Kelly:2004hm} or the rational fraction~\cite{Xiong:2019umf} parameterization of the experimental data. 
Calculations of the axial form factors $G_A$ and $\widetilde G_P$ are still maturing. The current main uncertainty  comes from correctly identifying all the excited states that contribute to the 3-point correlation functions and including them in the extraction of the ground-state matrix elements.

\begin{figure}[t]
    \includegraphics[width=0.49\linewidth]{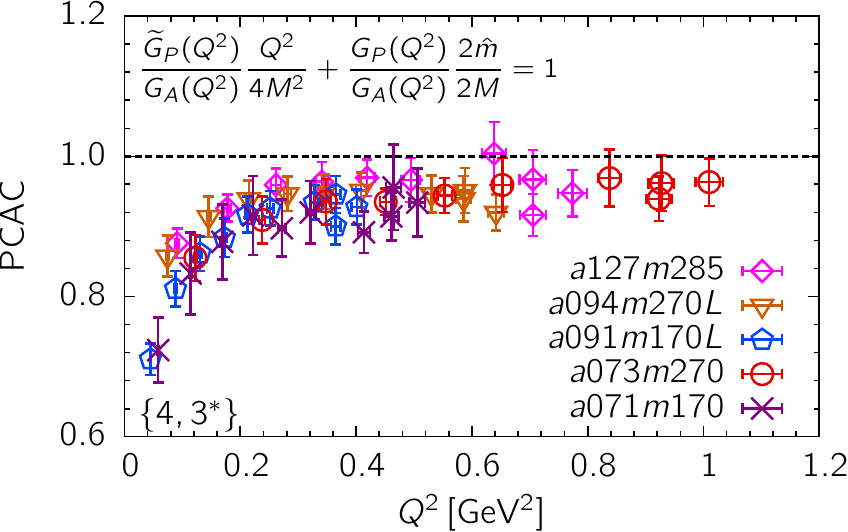}
    \includegraphics[width=0.49\linewidth]{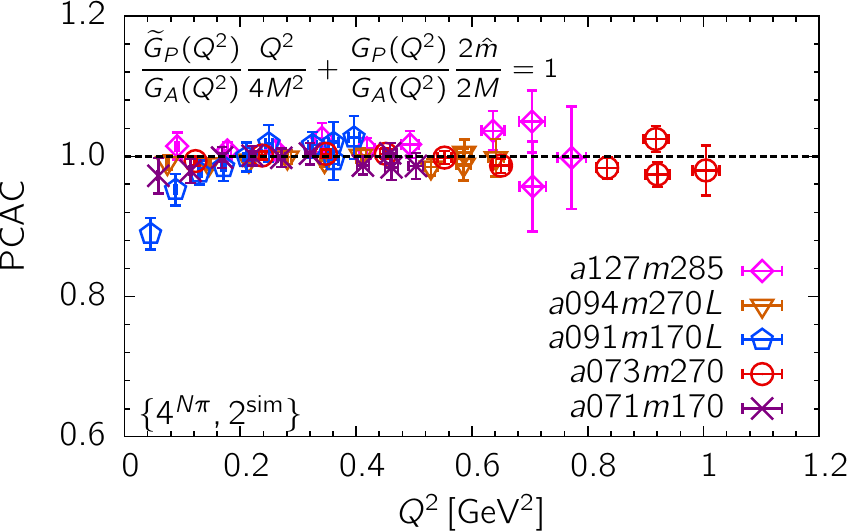}
\vspace{-0.1in}
  \caption{The axial $G_A$, induced pseudoscalar $\widetilde G_P$, and pseudoscalar $G_P$ nucleon form factors do not accurately satisfy Eq.~\eqref{eq:PCAC} using a standard method of determining the form factor in LQCD, left, but do satisfy this relation at the expected accuracy when the lowest $N\pi$ excited state is explicitly included in the analysis. Figure reproduced from Ref.~\protect\cite{Park:2021ypf}.  
  \label{fig:PCAC}
  }
\end{figure}

\begin{figure}
    \centering
    \includegraphics[width=0.6\textwidth]{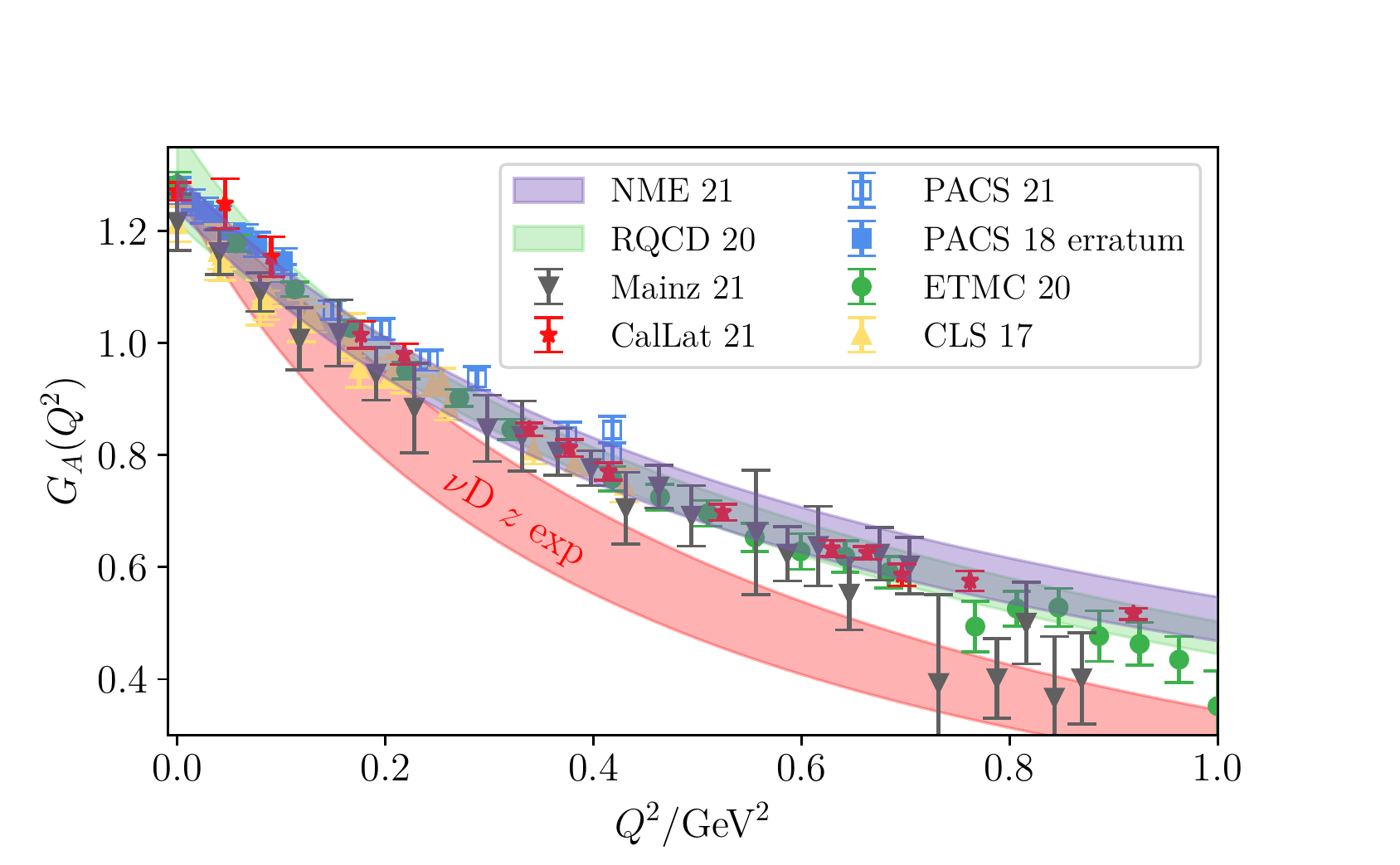}
    \caption{
    Summary of current LQCD calculations of the nucleon axial form factor,
    compared to the $z$ expansion deuterium bubble chamber parameterization
    from Ref.~\cite{Meyer:2016oeg} (red band).
    Lattice results from two collaborations,
    RQCD~\cite{RQCD:2019jai} (green) and NME~\cite{Park:2021ypf} (purple),
    are plotted as bands.
    These account for systematics due to extrapolation in parameters such as
    the lattice spacing, finite volume, and pion mass.
    The results of RQCD are obtained by enforcing the constraint that the
    form factors satisfy Eq.~\ref{eq:PCAC}.
    The NME result is taken from Eq.(55) of Ref.~\cite{Park:2021ypf}
    with an inflation of the uncertainty on $g_A$ and $b_0$ by a factor of 3
    to account for neglected variations due to lattice spacing and quark mass artifacts.
    Other LQCD
    results~\cite{Hasan:2017wwt,Shintani:2018ozy,Alexandrou:2020okk,Ishikawa:2021eut,Meyer:2021vfq,Djukanovic:2021yqg}, shown as scatter points, are each from a single ensemble and subject to similar level of unaccounted systematic effects since simulations have been done away from the physical point.
    Figure reproduced from Ref.~\cite{Meyer:2022mix}.
    }
    \label{fig:gaq2_summary}
\end{figure}

The pseudoscalar current $P^a = \overline{q} \gamma_5 \tau^a q$ and the associated nucleon form factor $G_P$ defined by
\begin{eqnarray}
  \left\langle N(\vec{p}+\vec{q}) | P (\vec{q}) | N(\vec{p})\right\rangle  &=&
{\overline u}_N(\vec{p}+\vec{q})\ G_P(q^2)\ \gamma_5 u_N(\vec{p}) \,, \label{eq:PS_FF}
\end{eqnarray}
enters
the chiral Ward identity $\partial_\mu A_\mu^a = 2 m P^a$ (also called the PCAC relation), which holds in QCD for all quark mass values and is only violated in LQCD by discretization errors.
The chiral ward identity can be used to derive an important check on the systematic uncertainties of LQCD form factor calculations, i.e., they satisfy the generalized Goldberger-Treiman relation 
\begin{equation}
2 { m_{ud}} G_P(Q^2) = 2 M_N G_A(Q^2) - \frac{Q^2}{2M_N} {\widetilde G}_P(Q^2),
\label{eq:PCAC}
\end{equation}
where $m_{ud} = (m_u + m_d)/2$. This relation is valid under the assumption that the matrix elements on the left hand side of Eqs.~\eqref{eq:Axial_FF}--\eqref{eq:EM_FF} are determined within the nucleon ground-state. 
Under the further assumption that $\widetilde{G}_P$ and $G_P$ are dominated by pion pole contributions, which is valid at leading order (LO) in chiral perturbation theory, the $\pi NN$ form factor $g_{\pi {\rm NN}}$ can also be related to these form factors~\cite{Liu:1994dr}.
Contributions from excited states can also arise because the interpolating operators used to create and annihilate nucleon states in LQCD couple not only to the nucleon but to all excitations  with the same quantum numbers, including 
multihadron states such as $N\pi$, $N\pi \pi$, $\dots$.
The failure of many early LQCD calculations to satisfy Eq.~\eqref{eq:PCAC} within expected discretization errors (see Ref.~\cite{Park:2021ypf} for a summary) can be regarded as an indication that excited-state contributions (ESC) are present in these calculations and were not fully removed in the extraction of matrix elements.
In chiral perturbation theory, these contributions are expected to be large in the axial and pseudoscalar channels~\cite{Bar:2018xyi,bar:2019gfx}.
Note that for quantitative agreement subleading loop effects, see Ref.~\cite{Gupta:2021ahb}, are important due to enhancements of low-energy constants that describe the effects of the $\Delta(1232)$ resonance.
In LQCD, the importance of multi-hadron excited-state contamination in axial form factor calculations was established clearly in Ref.~\cite{Jang:2019vkm}, where it was demonstrated that explicit inclusion of $N\pi$ excited-state effects using information from a different correlation function where excited-state effects are more pronounced reduced violations of Eq.~\eqref{eq:PCAC} from tens of percent to a few percent as shown in Fig.~\ref{fig:PCAC}.
This work also demonstrated that a relationship between  $\widetilde G_P$ and $G_A$ arising from the pion-pole dominance hypothesis~\cite{Goldberger:1958vp}, which is valid at LO in chiral perturbation theory, is highly correlated with Eq.~\eqref{eq:PCAC} and both relations are approximately satisfied if and only if excited-state effects are adequately removed in the extraction of $G_A$, $\widetilde G_P$ and $G_P$.
The current status of results for $G_A$ in calculations including extrapolations to the continuum limit and physical quark masses, as well as demonstration that excited-state effects do not spoil Eq.~\eqref{eq:PCAC},  is shown in Fig.~\ref{fig:gaq2_summary}.
 A rough estimate, based 
on these calculations, of the time required to reduce all systematics other than 
excited state effects to $\lesssim 2\%$ and ESC to $\lesssim 5\%$ is 5 million node hours on Summit supercomputer 
at ORNL and its follow on.  This size of resource should become available within the next five years.  

A complete systematic uncertainty budget for LQCD calculations of form factors must account for the size of multi-hadron excited-state effects, as well as other systematic uncertainties such as discretization and finite-volume effects.
There has been recent progress towards the application of variational methods to $N\pi$ scattering~\cite{Andersen:2017una,Silvi:2021uya} and pion-production amplitudes~\cite{Barca:2021iak} that will shed further light on the size of $N\pi$ and other excited-state effects and enable future calculations to explicitly isolate and remove $N\pi$ and other excited-state contributions to axial form factors  that may be relevant for the goal of achieving form-factor predictions with few-percent uncertainties.
Such variational calculations would also simultaneously enable calculations of nucleon pion-production amplitudes and are discussed further in Sec.~\ref{sec:lqcd_res} below.

\subsection{Two-body currents}\label{sec:lqcd_mec}

Effective theories with nucleon degrees of freedom describe electroweak interactions of nuclei as a sum of electroweak current interactions with constituent nucleons, parameterized by the nucleon form factors above, plus corrections dominantly arising from two-body current interactions with correlated pairs of nucleons (sometimes called ``meson-exchange currents'').
LQCD calculations of electroweak matrix elements of light nuclei include all multi-nucleon correlations arising from strong interactions, and two-body currents can be determined by matching the results of LQCD calculations of multi-nucleon electroweak matrix elements to nuclear effective theory calculations of the same quantities.
The computational costs of LQCD calculations of nuclei grow exponentially as the nucleon number is increased and as the quark masses are reduced, consequently LQCD calculations of electroweak nuclear matrix elements have so far been limited to two- and three-nucleon systems with heavier-than-physical quark masses, see Refs.~\cite{Beane:2010em,Drischler:2019xuo,Davoudi:2020ngi} for reviews.

For energies much below the pion-production threshold, pionless EFT can be used to parameterize the neutrino-deuteron cross section and other electroweak observables in terms of an isovector two-body axial current parameter $L_{1A}$ and other comparatively well-known quantities.
The parameter $L_{1A}$ has been phenomenologically calculated using  reactor neutrinos \cite{Butler:2002cw}, SNO and SK data \cite{Chen:2002pv,Balantekin:2003ep}, and helioseismology \cite{Brown:2002ih}.
Calculations of the axial-current matrix element $\left< pp | A_\mu^+ | d \right>$ relevant for the $pp$ fusion process $p + p \rightarrow d  + e^+ + \overline{n}_e$ have been performed in LQCD using a single unphysically-heavy value of the $u$ and $d$ quark masses and matched to pionless EFT by calculating the same background-field correlation functions in both theories in order to determine $L_{1A}$~\cite{Savage:2016kon}.
Isoscalar and strange-quark axial matrix elements relevant for neutral-current interactions have also been computed with the same quark mass values~\cite{Chang:2017eiq}.
LQCD calculations of the axial-current matrix element $\left< {}^3\text{He} | A_\mu^+ | {}^3\text{H} \right>$ relevant for triton $\beta$-decay have been performed at two unphysically large quark mass values~\cite{Savage:2016kon,Parreno:2021ovq} and used to constrain $L_{1A}$ using techniques for directly matching finite-volume LQCD and pionless EFT observables~\cite{Barnea:2013uqa,Eliyahu:2019nkz,Detmold:2021oro}.
Although systematic uncertainties from quark-mass dependence, discretization effects, and excited-state effects are not fully controlled in these exploratory LQCD calculations, there is good agreement between these LQCD results for $L_{1A}$ and results  extracted directly from experimental data.

Excited-state effects arising from unbound multi-nucleon states can lead to significant systematic uncertainties in LQCD calculations of multi-nucleon observables analogous to the $N\pi$ excited-state effects in nucleons discussed above.
Results for two-baryon energy spectra obtained using variational methods~\cite{Francis:2018qch,Horz:2020zvv,Green:2021qol,Amarasinghe:2021lqa} show tensions with non-variational results obtained by multiple groups that used asymmetric correlation functions with similar interpolating operators~\cite{Beane:2012vq,Yamazaki:2012hi,Beane:2013br,Berkowitz:2015eaa,Orginos:2015aya,Yamazaki:2015asa,Wagman:2017tmp,Illa:2020nsi}.
Direct comparisons of results using different interpolating operators on the same gauge-field ensemble for $\pi\pi$~\cite{Dudek:2012xn,Wilson:2015dqa}, $N\pi$~\cite{Lang:2012db,Kiratidis:2015vpa}, and $NN$~\cite{Amarasinghe:2021lqa} systems highlight the importance of including operators that have significant overlap with all low-energy states  in order to obtain a complete description of the low-energy spectrum.
If such a sufficiently complete operator set can be identified, then variational methods can be used to robustly remove excited-state effects from ground-state energy and matrix element determinations.
LQCD calculations of few-nucleon electroweak matrix elements using variational methods to control excited states, as well as using approximately physical quark masses and robust continuum extrapolations, could be achieved within the next five to ten years given sufficient computing resources.
In addition to determinations of $L_{1A}$, such LQCD calculations of two-nucleon axial form factors could provide systematically controlled predictions for the more poorly known momentum-dependence of two-body current effects.

\subsection{The resonance region}\label{sec:lqcd_res}

Reaction channels involving pion and other resonance production provide the dominant contributions to neutrino-nucleus scattering cross sections for few-GeV energies~\cite{Formaggio:2013kya}.
Robust theory predictions with quantified uncertainties for resonant scattering cross sections will be particularly important for DUNE, where the neutrino flux will be peaked around 2 GeV energies~\cite{DUNE:2016evb}.
Achieving systematic uncertainty targets~\cite{DUNE:2015lol} of $2-3\%$ precision on the total cross section in this energy region will require accurate predictions of $\Delta$ production and other resonant scattering cross sections with few-percent precision.
Electron scattering experiments can be used to determine vector-current contributions to resonance effects directly,
but axial-current contributions from electron scattering are only measured indirectly by applying low-energy theorems to decays of the produced resonances~\cite{Bernard:2001rs}.
Determinations of axial $N \rightarrow \Delta$ transition form factors often use leading-order chiral perturbation theory and phenomenological model results~\cite{Adler:1968tw} in order to reduce the number of transition form factors because there is not enough data to constrain the full form factor parameterization~\cite{LlewellynSmith:1971uhs,Hernandez:2007qq,Hernandez:2010bx}. 
It will be important to determine all of the independent vector and axial $N\rightarrow \Delta$ transition form factors, conventionally denoted $C_3^V,\ldots,C_6^V$ and $C_{3}^A,\ldots,C_6^A$, in order to make resonant cross-section predictions with few-percent accuracy and in particular to determine the angular dependence of $\Delta$ resonance production and decay.
Accurate determinations of these transition form factors as well as non-resonant nucleon pion production amplitudes are required to provide reliable predictions of neutrino-nucleus cross sections in the resonance region.

LQCD can be used to calculate resonant and non-resonant nucleon pion-production amplitudes as well as the elastic form factors discussion above.
In general, LQCD calculations determine correlation functions that encode the spectrum and matrix elements of the energy eigenstates of QCD.
Variational methods can be applied to symmetric correlation-function matrices built from a set of interpolating operators in order to construct approximate energy eigenstates and obtain upper bounds on the true energy levels of QCD~\cite{Fox:1981xz,Michael:1982gb,Luscher:1990ck}.
Future calculations of nucleon form factors applying variational methods to a set of interpolating operators including both local $N$ operators and products of plane-wave $N\pi$ operators can be used not only to obtain $N\rightarrow N$ elastic form factors with $N\pi$ excited-state effects explicitly removed but also $N \rightarrow N\pi$ transition form factors for resonances with the nucleon quantum numbers within the same calculation.
Calculations that further include interpolating operators with the $\Delta$ quantum numbers that strongly overlap with states in the vicinity of the $\Delta(1232)$ resonance can be used to determine $N \rightarrow \Delta$ transition form factors.

It is essential for variational methods to include operators overlapping with all low-energy states present in the spectrum in order to reliably control excited-state effects~\cite{Dudek:2012xn,Wilson:2015dqa,Lang:2012db,Kiratidis:2015vpa,Amarasinghe:2021lqa}.
This complicates the determination of resonant pion-production amplitudes, because in large volumes there are multiple excited states associated with unbound $N\pi$ and $N\pi\pi$ systems that are below or comparable to the energy of the $\Delta(1232)$ resonance.
Large sets of interpolating operators are required in order to describe the low-energy states associated with $N\pi$ and $N\pi\pi$ systems with different relative momenta, and calculations will require large computational resources and state-of-the-art algorithms for approximating ``all-to-all'' quark propagators~\cite{HadronSpectrum:2009krc,Morningstar:2011ka,Detmold:2019fbk,Li:2020hbj}.
Further complications arise from the fact that the finite-volume matrix elements determined by LQCD calculations are not simply related to infinite-volume resonant form factors.
It is possible to relate LQCD finite-volume matrix elements to infinite-volume form factors either using generalizations of L{\"u}scher's quantization condition~\cite{Luscher:1986pf,Luscher:1990ux,Luscher:1991cf,Rummukainen:1995vs,Lellouch:2000pv,Briceno:2014uqa,Briceno:2015csa,Briceno:2015tza,Baroni:2018iau} or by directly matching between LQCD and effective theory results for finite-volume energies and matrix elements and subsequently using the effective theory to predict infinite-volume form factors and other physical observables.
Further studies are needed to determine the most efficient way to match effective theory descriptions of resonant form factors to LQCD results for finite-volume matrix elements. 

There is a long history of LQCD studies of the $\Delta$ resonance, 
 including determinations of vector- and axial-current $N \rightarrow \Delta$ transition form factors using unphysically heavy quark masses that result in a stable $\Delta$ baryon~\cite{Alexandrou:2006mc,Alexandrou:2007dt,Alexandrou:2010uk}.
Although it is difficult to estimate the systematic uncertainties associated with quark-mass effects for these form factors, LQCD calculations of $N\rightarrow \Delta$ form factors with stable $\Delta$ baryons still provide useful information about experimentally poorly known observables such as the angular distributions of $\Delta$ production and decay amplitudes.
More recent calculations
with pion masses below about $300~$MeV find the $\Delta$ to be an unstable $N\pi$ resonance~\cite{Alexandrou:2013ata,Alexandrou:2015hxa}.
Recent calculations have applied variational methods and included both $\Delta \sim qqq$ operators and products of plane-wave $N \pi$ operators that overlap with states corresponding to $N\pi$ scattering states~\cite{Andersen:2017una,Silvi:2021uya}.
These calculations have enable determinations of $N\pi$ phase shifts that have been fit to a Breit-Wigner form to predict the mass and width of the $\Delta$ resonance.
Exploratory calculations of $N \rightarrow N \pi$ axial matrix-element including multi-hadron $N\pi$ operators have also recently been performed~\cite{Barca:2021iak}.
For such transition form factors involving unstable resonances, amplitudes are computed for real-valued invariant masses and must be analytically continued to the resonance pole position.
To get a complete error budget, both lattice spacing
 and finite volume uncertainties must be taken into account
 and control over the parameterization of the scattering phase shift
 used to obtain the resonance properties is needed.
This necessitates the use of multiple lattice ensembles
 to perform the relevant extrapolations to the physical point.
 Although challenging, a pilot study of the methodology exists in the meson sector
 for the $\rho\to\pi\gamma^\ast$ transition~\cite{Owen:2015fra}.

Besides the  $\Delta(1232)$, the low-energy nucleon resonances that are the most accessible to LQCD and among the most important for the neutrino oscillation program are the $N^\ast(1440)$ (Roper)
 and the negative parity $N^\ast(1535)$ resonances~\cite{Zyla:2020zbs}; see Sec.~\ref{inelastic} for further details. 
Several LQCD studies have investigated whether the Roper resonance more closely resembles a 3-quark state or a bound state of two or more hadrons, including variational calculations employing $N\pi$ as well as $N\sigma$ interpolating operators~\cite{Lang:2016hnn}.
Many calculations 
appear to prefer larger masses for the Roper resonance
\cite{Lang:2016hnn,Wu:2017qve,Roberts:2013ipa,Engel:2013ig,Mahbub:2013ala,Alexandrou:2013fsu,Alexandrou:2014mka,Kiratidis:2016hda,Edwards:2011jj},
 while others prefer a mass that extrapolates to the experimental Roper mass at the physical point~\cite{Chen:2004gp,xQCD:2019jke}.
This was attributed in Refs.~\cite{Liu:2016rwa,xQCD:2019jke}
 to a sensitivity to the chiral properties of the action.
The negative-parity $N^\ast(1535)$ (or $S_{11}$) resonance is also relevant for neutrino scattering and shares the same quantum numbers as an S-wave $N\pi$ system.
This resonance has been less studied in LQCD than the $\Delta$ or Roper, but calculations have been performed using both single-hadron $q^3$~\cite{Edwards:2011jj,Kiratidis:2015vpa} and multi-hadron $N\pi$ ~\cite{Lang:2012db,Verduci:2014csa,Verduci:2014btc} interpolating operators with pion masses ranging down to 260~MeV.
More precise determinations of Roper and other $N^{\ast}$ resonance properties and transition form factors would both illuminate the structures of these states and inform models of resonant neutrino scattering by providing exclusive transition form factor results complementary to experimental pion-nucleon data. 
The same LQCD methodology can be applied to higher-energy $N^{\ast}$ resonances, but it is presently computationally unfeasible to include the multitude of $N\pi$, $N\pi\pi$, $N\pi\pi\pi$, and other multi-hadron scattering states with energies below larger resonance masses.

Over the next five to ten years, calculations of $N\rightarrow \Delta$ and $N\rightarrow N^{\ast}$ transition form factors for the resonances discussed above provide suitable targets for LQCD calculations.
Achieving controlled systematic uncertainties in calculations involving excited-state matrix elements is challenging, but encouraging steps towards this goal have been made in recent years. Further, the same variational methods that enable controlled calculations of resonant and non-resonant $N\rightarrow N\pi$ transition form factors can help quantify and reduce systematic uncertainties for the quasi-elastic region by  providing determinations of nucleon elastic form factors with $N\pi$ excited-state contamination explicitly removed.

\subsection{The hadron tensor}{\label{sec:hadr_lqcd}}

For energies above two- and three-pion production thresholds, it is difficult to account explicitly for all resonant and non-resonant scattering channels, and so it is simpler to consider inclusive or semi-inclusive processes.
Cross-sections for $\nu A$ scattering processes involving a final-state lepton and a fully inclusive sum over hadronic final-states $X$ can be written as linear combinations of the elements of the hadron tensor $W^{\mu\nu}$ of the nucleus $A$ with initial-state four-momentum $p$, defined as
\begin{equation}
  W_A^{\mu\nu}(p,q) = \frac{1}{4\pi}\int d^{4}ze^{iq\cdot z}\left\langle A,p+q\left|J_{\mu}^{\dagger}(z)J_{\nu}(0)\right|A,p\right\rangle. \label{eq:Wmunudef}
\end{equation}
For scattering events with $|\mathbf{q}| \gg 1/d$, where $d$ is the typical inter-nucleon separation in a nucleus, the nuclear hadron tensor can be factorized into products of nuclear spectral functions and nucleon hadron tensors using the impulse approximation~\cite{Benhar:2006wy}:
\begin{equation}
  W_A^{\mu\nu}(\mathbf{q},\omega) = 
\int d^3k\,dE\left(\frac{M_N}{E}\right)
\left[
  Z \, S_p(\mathbf{k}, E) \, W_p^{\mu\nu}
    + (A-Z) \, S_n(\mathbf{k}, E) \, W_n^{\mu\nu}
\right],
\end{equation}
where $q = (\omega, \mathbf{q})$ and the nucleus is taken to be initially at rest.
The nuclear spectral functions $S_N(\mathbf{k}, E)$ (with $N=n,p$) describe the probability of finding a nucleon $N$ with energy $E$ and momentum $\mathbf{k}$ inside the nucleus.
Nuclear spectral functions can be computed to high precision for nuclei as large as $^{12}$C using non-relativistic nuclear many-body theory, see Refs.~\cite{Lynn:2019rdt,Carlson:2014vla} for reviews and Sec.~\ref{sec:many_body} below for further discussion.
The nucleon hadron tensors $W_N^{\mu\nu}$ (with $N=p,n$) encode the hadronic structure of individual nucleons interacting with the external current $J$. They are defined by Eq.~\eqref{eq:Wmunudef} with $A$ replaced by $N$, which is equivalent to~\cite{Benhar:2006wy}
\begin{equation}
  W_N^{\mu\nu}(\mathbf{k},E,\mathbf{q},\omega) =  \sum_X
    \matrixel{N,\mathbf{k}}{J^\mu}{X,\mathbf{k}+\mathbf{q}}
    \matrixel{X,\mathbf{k}+\mathbf{q}}{J^\nu}{N, \mathbf{k}}
    \delta(\omega + M_N - E - E_X ).
\end{equation}
Within the impulse approximation, it is therefore consistent to describe $\nu A$ scattering using the free nucleon spectral functions except that the nucleon energy is that of a bound nucleon, $M_N - E$, rather than the nucleon mass.
Corrections to the impulse approximation including two-body currents~\cite{Benhar:2015ula,Rocco:2015cil,Rocco:2018mwt} and pion production~\cite{Rocco:2019gfb} can be included within an extended factorization scheme that therefore connects determinations of the nucleon hadron tensor to predictions for $\nu A$ cross sections for experimentally relevant nuclei.
 
The nucleon hadron tensor is not directly accessible in the Euclidean spacetime. However,  its Laplace transform
\begin{equation}
  W_N^{\mu\nu}(\mathbf{k}, E, \mathbf{q}, \tau) = \int d\omega\, e^{-\omega \tau} W_N^{\mu\nu}(\mathbf{k}, E, \mathbf{q}, \omega). \label{eq:WE}
\end{equation}
can be formulated as a Euclidean path integral that can be calculated using LQCD~\cite{Liu:1993cv,Liu:1999ak,Aglietti:1998mz,Detmold:2005gg,Can:2020sxc}. 
Inverting this relation to extract the nucleon hadron tensor from its Laplace transform 
is a numerically delicate and challenging problem.
Developing robust methods for solving this inverse problem numerically is the subject of current research: the Maximum Entropy Method (MEM)~\cite{Bryan:1990,Jarrell:1996rrw}, Bayesian Reconstruction (BR)~\cite{Burnier:2013nla}, and the Backus-Gilbert Method (BG)~\cite{Backus:1968,Hansen:2017mnd} have all been investigated in connection with the nucleon hadron tensor~\cite{Liang:2019frk}.
Another approach that exploits the analytic structure of Euclidean Green functions was recently explored in the context of condensed matter physics~\cite{PhysRevLett.126.056402}.
Extensions of these ideas may also prove useful for LQCD calculations of the hadron tensor.
The hadronic tensor in Eq.~\eqref{eq:WE} may also be related to the total
scattering cross section by using its $\tau$ dependence to perform an
weighted integral over $\omega$~\cite{Fukaya:2020wpp}.

LQCD calculations of the nucleon hadron tensor provide a valuable window on $\nu A$ scattering cross sections at energies where QCD is nonperturbative and decompositions into exclusive channels are unfeasible.
Hadron tensor calculations can be extended to large momentum transfers in order to study the transition to the DIS region, although very fine lattice spacings are required to probe energies relevant to DIS~\cite{Liang:2019frk}.
For low energies below inelastic thresholds, the nucleon hadron tensor can be written as a sum of products of the elastic nucleon form factors discussed in Sec.~\ref{ssec:nucleon_form_factors}.
Agreement between form factor and hadron tensor results provides a non-trivial cross-check on the LQCD methodology of these calculations,
which has been satisfied by direct comparisons of hadron tensor and vector form factor results in Ref.~\cite{Liang:2019frk}.
During the next five years, calculations of the nucleon hadron tensor will continue to mature and begin to provide reliable nonperturbative predictions for $\nu A$ scattering cross sections at energies above inelastic thresholds.

\subsection{DIS structure functions}{\label{sec:lqcd_dis}}

In the high-energy DIS region, hadronic cross sections factorize into partonic cross sections calculable with perturbative QCD and light-cone structure functions such as PDFs that must be determined through global fits to experimental data and/or nonperturbative calculations.
LQCD calculations are performed in Euclidean spacetime where path integrals include positive-definite factors of $\propto e^{-S}$ that can be used for importance sampling, and light-cone structure functions cannot be directly calculated using LQCD.
Despite this obstruction, LQCD can provide useful nonperturbative input that can be used alongside experimental data in global fits to determine PDFs and other structure functions.
There is a great deal of complementarity between LQCD and experimental results for PDFs, with for example parton flavor separations of polarized PDFs relatively straightforward for LQCD but determinations of unpolarized nucleon and nuclear PDFs obtained more precisely from high-energy scattering experiments.

By performing an operator product expansion, Mellin moments of PDFs and other structure functions can be related to matrix elements of local operators, which have been targets of LQCD calculations for a long time~\cite{Kronfeld:1984zv,Martinelli:1988rr,Gockeler:1995wg}. The first Mellin moments with the insertion of a vector current describe the momentum fractions carried by quarks and gluons within an asymptotically high-energy hadron. The calculations of these quantities have been performed using physical quark masses~\cite{Alexandrou:2017oeh,Fan:2018dxu,Mondal:2020cmt}. Other low-order Mellin moments (helicity, transversity) have been calculated with LQCD. Although the lowest moments of unpolarized PDFs are known more precisely from experiment than LQCD calculations, LQCD results for moments of transversity PDFs have already been demonstrated to improve the precision of global PDF extractions~\cite{Lin:2017snn,Lin:2017stx,Constantinou:2020hdm}. LQCD calculations of higher moments of PDFs are challenging because the relevant operators mix with lower-dimensional operators under renormalization. Novel methods to circumvent this difficulty have been proposed~\cite{Detmold:2005gg,Braun:2007wv,Davoudi:2012ya,Monahan:2015lha} and are being actively explored, see Refs.~\cite{Lin:2017snn,Constantinou:2020hdm} for reviews.

A method for directly calculating the $x$-dependence of PDFs using LQCD is provided by the quasi PDF approach~\cite{Ji:2013dva} and related techniques~\cite{Ma:2014jla,Radyushkin:2017cyf,Chambers:2017dov}. See Refs.~\cite{Lin:2017snn,Cichy:2018mum,Ji:2020ect,Constantinou:2020hdm} for reviews. 
In this approach, nonlocal matrix elements describing Euclidean analogs of the light-cone-separated operators defining PDFs are computed in boosted hadron states, for example the unpolarized quark quasi PDF for a hadron $h$ with momentum in $z$-direction $P_z$ is defined by
\begin{equation}
\label{eq:lamet}
    \tilde{q}(x,P_z) = \int_{-\infty}^\infty \frac{dz}{4\pi} \, e^{-i x z P_z} \left< h(P_z) | \overline{q}(z) \Gamma W(z,0) q(0) | h(P_z) \right>,
\end{equation}
where $W(z,0)$ is a Wilson line for the interval $[0,z]$ and $\Gamma=\gamma_4$. Quasi PDFs can be related to light-cone PDFs using perturbation theory up to power corrections that vanish for $P_z\rightarrow \infty$. The need to extrapolate LQCD results to $P_z\rightarrow \infty$, truncation effects arising from approximating the Fourier transform using finite Wilson displacement results, and nonlocal operator renormalization lead to challenging systematic uncertainties, but there has been significant recent progress in understanding these issues~\cite{Lin:2017ani,Alexandrou:2019lfo,Lin:2019ocg,Ji:2020brr,LatticePartonCollaborationLPC:2021xdx}. LQCD results are most reliable for intermediate $x$ away from the endpoints of the physical region $x \in [0,1]$. 
Current LQCD results are able to significantly improve the precision of global PDF analyses of isovector polarized (with $\Gamma=\gamma_5\gamma_z$ in Eq.~\ref{eq:lamet} ) PDFs~\cite{Bringewatt:2020ixn}. These PDFs are directly relevant for neutrino-nucleon DIS, and over the next five years LQCD calculations of quasi PDFs with full lattice systematic errors  can improve predictions for neutrino DIS by providing reliable determinations of nucleon isovector unpolarized and polarized PDFs at intermediate $x$. 
There has been a LQCD study investigating  the 
$s(x)-\bar{s}(x)$ asymmetry~\cite{Zhang:2020dkn}, an important step towards LQCD calculations of a complete flavor decomposition of nucleon PDFs that can be used for neutrino DIS calculations. 
A joint LQCD and global PDF fitting community whitepaper~\cite{Lin:2017snn} estimates that LQCD determinations of strange PDFs with 10\% precision will significantly improve global fits.
We note that developing and benchmarking the above-mentioned relations between the lattice-calculable quasi PDFs and the exact light-cone PDFs would be aided by precise data for the latter, which are determined phenomenologically; detailed studies of the PDF sensitivity of high-energy data for quasi-PDF and lattice-calculable Mellin moments are available, {\it e.g.}, Ref.~\cite{Hobbs:2019gob}.
Additional light-cone structure functions such as TMDPDFs may be relevant for precisely predicting neutrino semi-inclusive DIS (SIDIS) processes such as high-energy pion production.
LQCD can be used to constrain the nonperturbative evolution of TMDPDFs~\cite{Ebert:2018gzl,Shanahan:2020zxr,Schlemmer:2021aij,LatticeParton:2020uhz,Li:2021wvl,Shanahan:2021tst} as well as quasi TMDPDFs that can be nonperturbatively related to light-cone TMDPDFs~\cite{Ji:2014hxa,Ji:2019sxk,LatticeParton:2020uhz,Li:2021wvl,Ebert:2022fmh}, including the spin and flavor combinations of TMDPDFs relevant for neutrino scattering. 
More details on the future prospects for lattice PDFs can be found in a parallel Snowmass whitepaper~\cite{Constantinou:2022yye}. 

Differences between nuclear and nucleon PDFs, including effects connected through EFT to the famous EMC effect, can be calculated directly using LQCD for light nuclei~\cite{Chen:2004zx}.
Phenomenological determinations of nuclear PDFs (nPDFs)~\cite{Segarra:2020gtj,Kovarik:2015cma,AbdulKhalek:2020yuc,Walt:2019slu,Eskola:2016oht,deFlorian:2011fp,Hirai:2007sx} have long attempted to constrain nuclear modifications to the free-nucleon PDFs, generally by fitting a smooth parametrization of the $A$ dependence in combined QCD analyses. While such analyses have made steady progress in recent years, continuously extending $A$-dependent nPDFs to describe light nuclei has been challenging, such that lattice QCD input on these systems --- in the form of moments and quasi PDFs --- would be very informative. In addition, nuclear data used to constrain nPDFs is frequently expressed in the form of ratios with respect to deuterium; as such, lattice simulations of the deuteron would be very helpful for the purpose of unraveling any potential systematic effects. 
Nuclear effects on isovector quark PDF constraints, which can be obtained relatively simply and precisely using LQCD, are relevant for charged-current processes in neutrino-nucleus scattering and, in particular, have been suggested as one potential source of the NuTeV anomaly~\cite{Cloet:2009qs}, in conjunction with nucleon-level effects from the $s\!-\!\bar{s}$ asymmetry~\cite{Davidson:2001ji,Kretzer:2003wy} and parton-level charge-symmetry breaking.
Exploratory calculations have been performed of isovector quark as well as gluon momentum fractions of two- and three-nucleon systems using a single gauge-field ensemble with unphysically heavy quark masses~\cite{Winter:2017bfs,Detmold:2020snb}.
Calculations of nuclear quasi PDFs are also possible in principle but would require considerable computational resources because increasing baryon number and including large Wilson lines both lead to exponential signal-to-noise degradation.
LQCD calculations of PDF moments of light nuclei including physical quark masses, continuum extrapolations, and systematic control of unbound multi-nucleon excited-state effects will be challenging but could be achieved within the next five years.

\section{Nuclear many-body theory approaches}\label{sec:many_body}

\subsection{Introduction}

A detailed understanding of neutrino scattering from nuclei is required to extract information on neutrino properties from the accelerator-neutrino program. These properties include the neutrino mass differences, mixing angles, and particularly the neutrino mass hierarchy and the CP-violating mixing angle. In particular, in neutrino experiments, the neutrino energy distribution, a critical ingredient in neutrino oscillation measurements, is a-priori unknown and must be inferred from the final state charged leptons and the emitted nucleons and pions.

This is a particularly challenging problem because of
the wide range of energies and momenta involved in these experiments, from quasielastic scattering dominated by single-nucleon knockout process, to the pion production region eventually to the deep inelastic region at high $Q^2$. Each of these regimes requires knowledge of the nuclear ground state and the electroweak coupling and propagation of the struck nucleons, hadrons, or partons. The range of challenges is extreme; quasielastic scattering and deep inelastic scattering are conceptually the easiest to understand, but ultimately we would like to be able to predict both inclusive and exclusive cross sections across a wide range of kinematics.  In particular, the CP violating phase is expected to have a significant impact in the quasi-elastic regime and at much higher energies and momenta. A consistent extraction of the CP-violating phase in different regimes is required to make a convincing high-precision measurement.

In this section, we summarize the current state of the art theory of electron- and neutrino-scattering scattering from nuclei, a brief comparison to selected experimental results, and prospects of dramatically improving the theory over the next 5-10 years and connecting to experiment through improvements to the generators used in neutrino experiments. We also describe connections to lattice QCD calculations of one- and potentially two-nucleon electroweak couplings and other neutrino experiments including double-beta decay and coherent neutrino scattering.

\subsection{Theory}
\label{sec:theory}
 
Microscopic nuclear many-body approaches aim at describing the structure and dynamics of atomic nuclei in terms of the individual interactions among protons and neutrons, which are treated as ``fundamental'' degrees of freedom. The nucleus is modeled as a collection of $A$ non-relativistic point-like nucleons whose dynamics is dictated by the Hamiltonian
\begin{equation}
 H = \sum_i K_i + \sum_{i<j} v_{ij} + \sum_{i<j<k} V_{ijk}\ . 
\end{equation}
In the above equation, $K_i$ is the non-relativistic single-nucleon kinetic energy, while $v_{ij}$ and $V_{ijk}$ are two-nucleon (NN) and three-nucleon (3N) potentials; four- and higher-body potentials are assumed to be suppressed. The interactions of nuclei with external electroweak probes is mediated by charge ($\rho$) and current ({\bf j}) operators that are consistent with the nuclear interactions. As such, they are also expanded in a series of many-body operators as
\begin{eqnarray}
\rho    &=& \sum_i {\rho}_i({\bf q}) + \sum_{i<j} {\rho}_{ij}({\bf q}) + \dots\ , \\
\nonumber
{\bf j} &=&  \sum_i {\bf j}_i({\bf q}) + \sum_{i<j} {\bf j}_{ij}({\bf q}) +\dots  \ ,
\end{eqnarray}
where ${\bf q}$ is the momentum transferred to the nucleus. For example, in the Impulse Approximation (IA), that is retaining only leading one-body operators in the equations above, nuclear electromagnetic charge and current distributions reduce to the sums of those associated with individual protons and neutrons. The electromagnetic single-nucleon couplings are given by the proton or neutron charge form factor, while the nucleon vector current includes convection and magnetization terms. For neutrino scattering we also need to include the axial and pseudoscalar form factors. The electromagnetic form factors can be measured with electron scattering and have been determined with significant precision over the range of momenta relevant to neutrino experiments. The experimental determination of the axial form factor require experiments on the lightest nuclei, such as the deuteron.  All the electroweak form factors can also be calculated from lattice QCD, and recently there has been significant progress in this area. Comparison with the measured electromagnetic form factors provide excellent tests lending confidence to the calculations of the remaining axial and pseudoscalar form factors, as discussed at length in  Sec.~\ref{ssec:nucleon_form_factors}.

The IA picture of the nucleus is, however, incomplete as it fails to explain, {\it e.g.}, the excess in the electromagnetic transverse nuclear response induced by electrons~\cite{Benhar:2006wy,Carlson:2001mp}. Corrections that account  for processes in which external probes couple to pairs of interacting nucleons  need to be incorporated in the theoretical {\it ab initio} description. 

Traditionally, phenomenological NN interactions have been constructed by including the long-range one-pion exchange interaction, while different  schemes are implemented to account for intermediate and short range effects, including multiple-pion-exchange, contact terms, heavy-meson-exchange, or excitation of nucleons into virtual $\Delta$-isobars. Highly-realistic interactions~\cite{Wiringa:1994wb,Bonnr,machleidt2001} of this kind, such as the Argonne $v_{18}$ (AV18) potential~\cite{Wiringa:1994wb}, involve a number of parameters that are determined by fitting experimental data; the AV18 can fit to the Nijmegen NN scattering database with $\chi^2$ per datum of about 1. Phenomenological 3N interactions, consistent with the NN ones, have been developed. They are generally expressed as a sum of a two-pion-exchange P-wave term, a two-pion-exchange S-wave contribution, a three-pion-exchange contribution, plus a contact interaction. Their inclusion is essential for reproducing the energy spectrum of atomic nuclei. For instance, the Illinois-7 3N force~\cite{Pieper:2008}, when used in conjunction with AV18, can reproduce the spectrum of nuclei up to C$^{12}$ with percent-level accuracy.  Meson-exchange currents~\cite{Villars47,Chemtob71,Friar77,rho79,Towner84,Riska84,Carlson:1997,Marcucci98,Marcucci05,Shen:2012xz,Bacca_Pastore_2014} (MEC) follow naturally once meson-exchange mechanism are invoked  to describe interactions between individual nucleons. They  account for processes in which  the external probe couples with mesons being exchanged between nucleons and are found to be essential to explain the data. 

Recent years have witnessed the tremendous development and success of chiral Effective Field Theory~\cite{Weinberg:1979sa,Weinberg:1990rz,Weinberg:1991um,vanKolck93,Ordonez:1992xp,Ordonez96,Bernard95,Epelbaum:2008ga,Epelbaum12,Epelbaum:2014efa,Entem:2003ft,Machleidt:2011zz,ekstrom2015accurate} ($\chi$EFTs) that grounds the achievements of more phenomenological theoretical approaches into the broken symmetry pattern of QCD, the fundamental theory of strong interactions. The relevant degrees of freedom of $\chi$EFTs are again are bounds states of QCD, {\it i.e.}, pions, nucleons, and $\Delta$'s, $\dots$. On this basis, their dynamics is completely determined by that associated with the underlying degrees of freedom of quarks and gluons, that is QCD. However, at low energies, QCD does not have a simple solution because the strong coupling constant becomes too large and perturbative techniques cannot be applied to solve it. $\chi$EFT is a low-energy approximation of QCD  valid in the energy regime where the typical momenta involved, generically indicated by $Q$, are such that $Q\ll\Lambda_\chi \sim 1$ GeV, where $\Lambda_\chi$ is the chiral-symmetry breaking scale. $\chi$EFT provides us with effective Lagrangians describing the interactions between pions, nucleons, and $\Delta$'s---as well
as the interactions of these hadrons with electroweak field---that preserve all the symmetries, in particular chiral symmetry, exhibited by the underlying theory of QCD at low-energy. These effective interactions, and the transition amplitudes derived from them,  can be expanded in powers of the small expansion parameter $Q/\Lambda_\chi$. 

\begin{figure}[htp]
    \centering
    \includegraphics[width=0.95\textwidth]{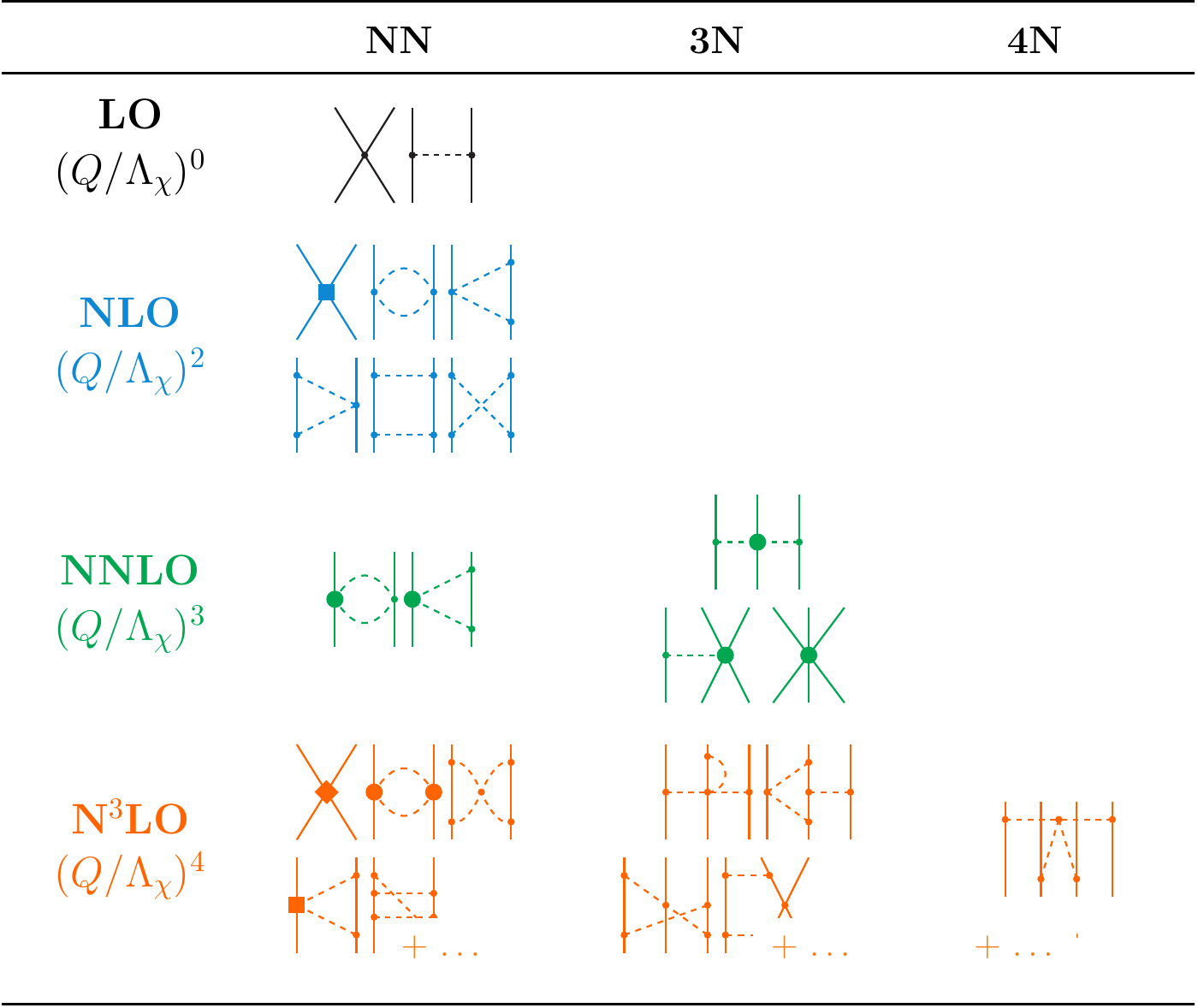}
   \caption{Figure~\cite{10.3389/fphy.2020.00379} courtesy of H. Hergert.  Chiral two-, three-, and four-nucleon forces through next-to-next-to-next-to-leading order (N3LO) in the chiral expansion. Dashed lines represent pion exchanged between nucleons. The large solid circles, boxes and diamonds represent vertices that are proportional to LECs of the theory (see text).  }
\label{fig:chipot}
\end{figure}

It is then possible in principle to evaluate nuclear observables to any degree $\nu$ of desired accuracy, 
with an associated theoretical error roughly given by $(Q/\Lambda_\chi)^{(\nu+1)}$. 
This scheme has been widely utilized to study both nuclear forces~\cite{vanKolck93,Ordonez:1992xp,Ordonez96,Bernard95,Epelbaum:2008ga,Epelbaum12,Epelbaum:2014efa,Entem:2003ft,Machleidt:2011zz,ekstrom2015accurate} and nuclear
electroweak currents~\cite{Park93,Park96,Phillips07,Pastore:2008ui,Pastore:2009is,Pastore:2011ip,Kolling:2009iq,Kolling:2011mt,Kolling:2012cs,Krebs:2016rqz}. The many-body operators emerging from direct evaluations
of the transitions amplitudes with interactions provided by $\chi$EFT Lagrangians  involve multiple-pion exchange operators, as well as contact-like interaction terms. 
As an example, the consistent two-, three- and many-nucleon chiral interactions up to next-to-next-to-next-to leading order (N3LO) in the chiral expansion 
are represented in Fig.~\ref{fig:chipot}, where the empirical suppression of three-nucleon
interactions with respect to two-nucleon interactions, and so on is explained by 
the adopted power counting scheme. Additionally, within the $\chi$EFT formulation, many-body electroweak currents are by construction consistent with the associated nuclear forces. 
In practice, chiral EFT introduces a set
of low-energy constants (LECs) that, in principle,
can be calculated from QCD, but are in practice fit
to experimental data. The LECs related to the short-range two-nucleon interactions are typically fit to the deuteron and nucleon-nucleon scattering data, and the analog ones related to the three-nucleon interaction are fit to properties of light nuclei. In both cases, the LECs describing the long-range interactions can be determined independently from pion-nucleon scattering~\cite{Hoferichter:2015tha,Siemens:2016jwj,Hoferichter:2015hva}, and thus, as a prediction of chiral EFT, do not lead to new parameters that would need to be determined in nuclear systems.

Many-body nuclear interactions have been over the years developed up to N5LO in the chiral expansion~\cite{entem2015dominant,epelbaum2015precision,epelbaum2015precision,reinert2018semilocal}. Most many-body calculations are still at much lower order, however, and often at lower cutoff scales $\Lambda$.
A variety of quantum many-body approaches~\cite{ barrett2013ab,jurgenson2013structure,hagen2014coupled,hagen2014coupled,bogner2010low,Carlson:2014vla,hergert2013medium} are used for
these calculations, including coupled cluster (CC),
the no-core shell model (NCSM), and Variational and Green's function Monte Carlo (VMC and GFMC) and Auxiliary Field
Diffusion Monte Carlo (AFDMC). Each of these approaches
has different strengths and weaknesses depending upon
the system size and the momentum cutoff  of the interaction.

The community is vigorously exploring the importance of including  $\Delta$'s as explicit degrees of freedom to improve the convergence of the chiral expansion~\cite{10.3389/fphy.2019.00245,krebs2007nuclear,piarulli2018light,piarulli2015minimally,Piarulli:2016vel}. 
Ground-state properties can be calculated within
the typical convergence pattern of $\chi$EFTs, including both bulk properties like charge and radii as well as intermediate quantities like momentum distributions
and spectral functions that are important ingredients
in model calculations of lepton-nucleon scattering.
Accelerator neutrino experiments are likely to require
high order calculations as well as calculations with
higher cutoffs due to the larger energy and momenta involved.
 
Elementary amplitudes, including elastic and transition nucleonic form-factors, as well as LECs entering the chiral many-body interactions and currents, are the main inputs to the nuclear models. Nucleonic electromagnetic form factors are, in most cases, well-known from electron scattering experiments. In neutrino scattering, in addition to probing vector currents, one probes also axial couplings along with different quark flavor structures. The data, in this case, are scarce or poorly known which makes theoretical LQCD calculations extremely valuable to constrain and ground both nuclear EFTs and phenomenological models, as recently outlined in a whitepaper by the USQCD Collaboration~\cite{Kronfeld:2019nfb}. Calculated elastic form factors are already achieving a precision that is competing with that of experimental data~\cite{Kallidonis:2018cas,Alexandrou:2018sjm,Sufian:2018qtw,Jang:2019vkm,Jang:2019jkn}, and with increased control of statistical and systematic uncertainties in the future nucleonic form factor calculations will provide solid inputs to nuclear EFT studies of electroweak interactions as discussed in Sec.~\ref{ssec:nucleon_form_factors}.

Inelastic electroweak transition amplitudes involving $\pi$ or other meson production, or hadronic resonances, such as the $\Delta$, are also required as inputs to EFT descriptions of nuclei involving two-body currents and explicit $\pi$ and $\Delta$ degrees of freedom relevant for multi-hundred-GeV incident neutrinos and are less well-known experimentally than elastic nucleon form factors~\cite{Hernandez:2007qq,Piarulli:2016vel}.
Although LQCD calculations are limited to finite-volume Euclidean correlation functions---see Secs.~\ref{sec:lqcd_mec} and ~\ref{sec:lqcd_res} for a detailed discussion, there has been significant progress in extracting resonance physics from finite-volume observables~\cite{Briceno:2017max} and, in particular,  a formalism has been developed to relate multi-hadron finite-volume matrix elements to infinite-volume resonant electroweak transition amplitudes~\cite{Lellouch:2000pv,Hansen:2012tf,Briceno:2015tza,Baroni:2018iau}.
LQCD results for finite-volume energy levels and matrix elements can also be matched directly to the corresponding EFT results with the goal of constraining the parameters governing resonance production~\cite{Eliyahu:2019nkz}.
The derivation of the nucleon hadron tensor governing inclusive neutrino-nucleus scattering based spectral reconstruction techniques to relate Euclidean and Minkowski correlation functions~\cite{Liang:2019frk} is discussed in Sec.~\ref{sec:hadr_lqcd}. The latter could be readily implemented in many-body approaches that rely on a factorization scheme, see Sec.~\ref{subsec:QE}. Finally, Sec.~\ref{sec:lqcd_dis} focuses on how LQCD can be used to constrain the required PDFs by computing PDF moments related to nucleon and nuclear matrix elements of local operators. Using these LQCD constraints on PDFs in one- and few-nucleon systems, EFT can be used to extrapolate LQCD constraints to larger nuclei of experimental relevance~\cite{Chen:2016bde,Lynn:2019vwp}.

\subsection{Low-energy neutrino processes}

At neutrino energies below  a few tens of MeV,  the dominant mode of neutrinos-nucleus interactions is  
coherent elastic neutrino scattering (CE$\nu$NS) and
its cross sections is directly proportional to the weak form factor $F_\text{w}$ of the nucleus.
 
\begin{figure}[t]
	\begin{center}
		\includegraphics[width=0.48\textwidth,clip]{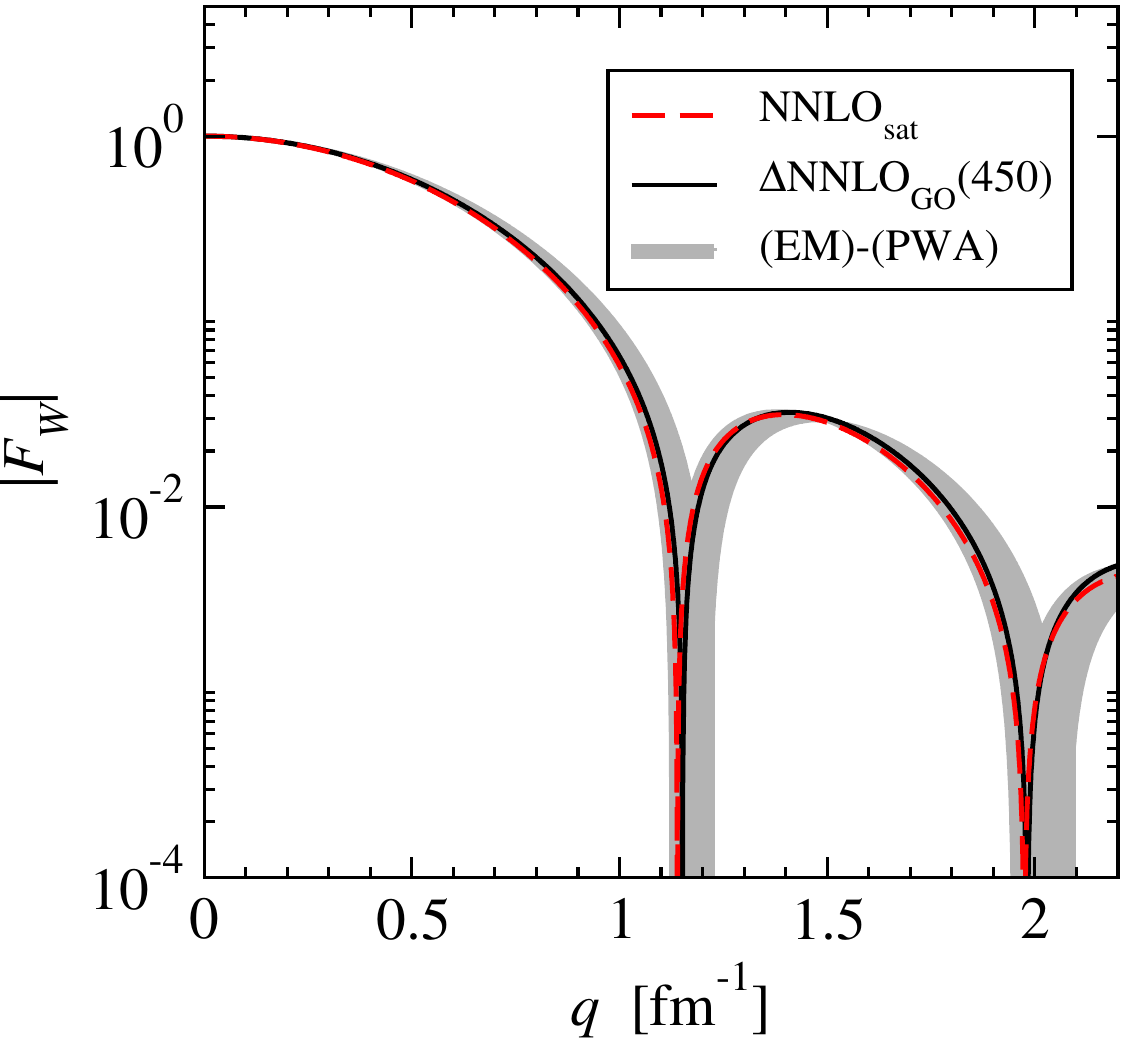}
	\end{center}
	\caption{Coupled-cluster theory predictions for the weak form factor of $^{40}$Ar using different $\chi$EFT interactions, labeled with NNLO$_{\rm sat}$, NNLO$_{\rm GO}(450)$ and (EM)-(PWA). The range in momentum transfer $q$ is extended with respect to Fig.~\ref{fig:Fweak_Ar}, where the coupled-cluster results are compared to relativistic mean-field and shell-model calculations. Figure adapted from Ref.~\cite{Payne:2019wvy}.}
	\label{fig:Fweak_Ar_CC}
\end{figure}
On the one hand, because $F_\text{w}$  is primarily determined by the distribution of the neutrons within the nucleus,
 CE$\nu$NS  offers  an opportunity to expand our understanding of nuclear structure, given that from a precision measurement of the cross section one could extract the neutron radius~\cite{Cadeddu:2017etk}.
 On the other hand, if  the neutron-distribution and the weak form factor are constrained either from another experiment of from accurate theory, one can use CE$\nu$NS 
to search for signature of non-standard neutrino interactions with high sensitivity (see Section~\ref{sec:SM} for more details). 

Many-body theory plays an important role in supporting the experimental CE$\nu$NS program.
First of all, several targets are planned to be used in the CE$\nu$NS investigations that span from medium to heavy nuclei. Nuclear theory can be used to bridge the various targets. So far, the only nucleus for which CE$\nu$NS was investigated from the ab initio point of view is $^{40}$Ar~\cite{Payne:2019wvy}, see Figure~\ref{fig:Fweak_Ar_CC}. There, Hamiltonian inspired from $\chi$EFT were used together with one-body currents to compute the weak form factor. Below  momentum transfers of $q=50$ MeV/c, results were found to be quite stable with respect to variations of the Hamiltonian (namely the parameterization of the $\chi$EFT and corresponding low-energy constants), leading to a nominal 2$\%$ uncertainty. Meson exchange currents affect  form factors only at higher momenta.
Other interesting  CE$\nu$NS targets are $^{23}$Na to $^{127}$I and $^{133}$Cs. In particular for the  heavy-mass nuclei, the challenge will be to extend the ab initio methods to that mass range preserving both the connection to QCD and the accuracy. 

The same chiral effective field theory that governs neutrino scattering from nuclei governs nuclear beta decay and double beta decay. Beta decay processes serve as a valuable check on the effective field theory at low energies and momenta~\cite{King:2020wmp,Gysbers:2019uyb}, and can be used for studies of BSM physis. Neutrinoless double beta decay, if observed, indicates lepton number violation and is often interpreted as evidence for Majorana neutrinos. From an observation of neutrinoless double beta decay, calculated nuclear matrix elements allow extraction of the absolute mass scale of neutrinos (see Snowmass WP~\cite{ndbdWP}.) 

At low neutrino energies, neutrinos can also inelastically scatter from nuclei,  exciting low-lying nuclear states and at a bit higher energies ejecting nucleons from the nucleus. Many studies of these
rates have been undertaken,~\cite{Raghavan:1986fg,Haxton:1987kc,Fukugita:1988hg,Engel:1996zt,KARMEN:1998xmo,Kolbe:1999au,Hayes:1999ew,Volpe:2000zn,LSND:2001fbw,Kolbe:2002gk} but modern calculations using many-body advances and consistent  interaction and currents across a range of kinematics would be valuable.  These inelastic  processes play a key role in setting the nuclear environment in core-collapse supernovae and neutron star mergers, for example. Measuring these processes in terrestrial detectors enables one  to obtain the flavor- and energy-dependent neutrino flux from supernovae, which can inform us about the internal dynamics of the astrophysical site~\cite{Super-Kamiokande:2007zsl,Duba:2008zz,Scholberg:2012id,Laha:2014yua,JUNO:2015zny,Lu:2016ipr,Li:2020ujl,DUNE:2020zfm}.

DUNE will enable a high-statistics detection of MeV $\nu_e$ via $\nu_e +$Ar$\rightarrow e^- + {}^{40}$K$^*$. This channel is also important for solar-neutrino studies.~\cite{Capozzi:2018dat}. For these reasons, new calculations with reliable error bands are needed.
To reconstruct the energy of the incoming neutrinos, we need to know the exclusive cross sections to each individual excited state in $^{40}$K. At slightly higher energies, e.g., $\gtrsim 50$~MeV, there could be nucleons knockout in the final state~\cite{KOLBE1992599,PhysRevLett.76.2629,Gardiner:2018zfg}.
 
At higher excitation energies, also collective modes in the nucleus can be excited and eventually the quasi-elastic regime is reached. 
Modeling these inelastic processes from the theoretical point of view is more challenging than calculating ground state properties and often more challenging than calculations of inclusive neutrino cross sections.

\begin{figure}[htp]
    \centering
    \includegraphics[width=0.95\textwidth]{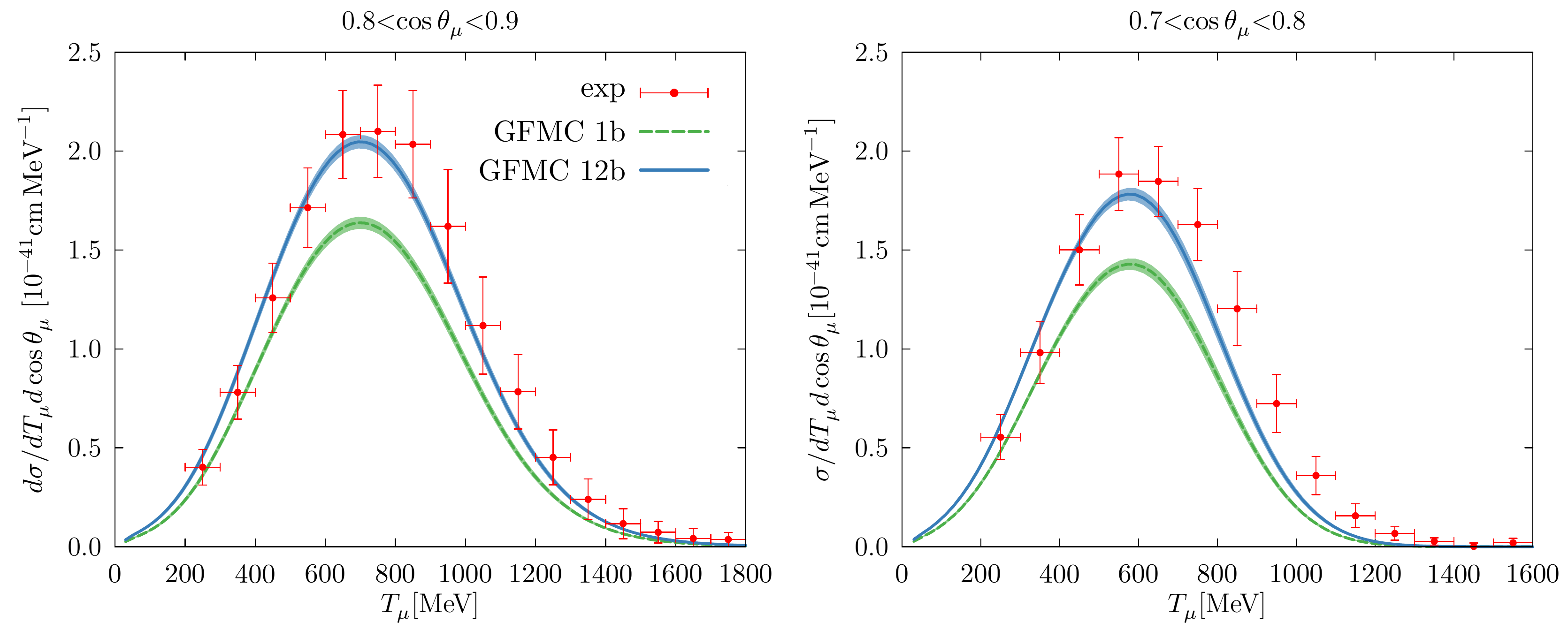}
   \caption{Adapted from Ref.~\cite{Lovato:2020kba}. MiniBooNE flux-folded double differential cross sections per target neutron for $\nu_\mu$-CCQE scattering on $^{12}$C, displayed as a function of the muon kinetic energy (T$_\mu$) for different ranges of $\cos\theta_\mu$. The experimental data and their shape uncertainties are from Ref. [46]. The additional 10.7\% normalization uncertainty is not shown here. Calculated cross sections are obtained with a dipole axial form factor with $\Lambda_A = 1.0$ GeV.}
\label{fig:gfmc_inclusive}
\end{figure}

\subsection{Quasi-elastic processes }
\label{subsec:QE}
 
\noindent 

\noindent 
The cross section of inclusive scattering of electrons or neutrinos by nuclei is given  by a sum over the individual response functions, that depend on energy and momentum transfer, times kinematical factors as
\begin{equation}
\frac{d\sigma}{d\Omega} ({\bf q},\omega) = \sum_i K_i ({\bf q, \omega}) R_i ({\bf q, \omega})\,.
\end{equation}
In particular for neutrino-nucleus scattering, five response functions $R_i$ are required. They can be obtained through the calculation of the relevant two-point functions
\begin{eqnarray}
    \label{resp}
    R_i ({\bf q} ,\omega)&  = & \int
    dt e^{ i \omega t} \  \langle 0 | {\bf j}^\dagger ({\bf q}) \ e^{- i H t}\  {\bf j}^\dagger ({\bf q}) | 0 \rangle \nonumber \\
    & = & \sum_f \langle 0 | {\bf j}^\dagger ({\bf q}) | f \rangle \langle f | {\bf j} ({\bf q}) |0 \rangle \delta (E_f - E_0 - \omega)\,,
\end{eqnarray}
where $|0\rangle$ is the nuclear ground state and $|f\rangle$ are all possible final states of the $A$-nucleon system.  The first expression in Eq.~(\ref{resp}) is written in the time domain, while the second is in the energy domain with the time integration yielding the energy conserving delta function.
While these two-point functions cannot be calculated
exactly except for extremely simple nuclei, they can
be used to obtain very accurate calculations of the
inclusive responses~\cite{leidemann2013,Bacca_Pastore_2014}. 
In principle, the same quantum many-body methods that are used to
determine ground-state properties can be used to
calculate inclusive response functions when integral transforms are used.
Below we discuss the various approaches  to compute response functions, which provide a wealth of information on
the inclusive cross section including a full treatment
of initial-state two-nucleon correlations, two-nucleon
 currents consistent with the nuclear potentials as discussed in Sec.~\ref{sec:theory}, and final-state interactions.
\\

\noindent{\bf Continuum quantum Monte Carlo approaches}\\
Among microscopic methods, the variational Monte Carlo (VMC) and Green's function Monte Carlo (GFMC) approaches utilizes quantum Monte Carlo (QMC) techniques to fully retain the complexity of many-body correlations and associated electroweak currents. QMC methods have been extensively applied to study the structure and electroweak properties of light nuclei, including electromagnetic moments and form factors, low-energy transitions and beta decays~\citep{Carlson:2014vla,King:2020wmp}. Exploiting the fact that quasielastic responses are smooth functions of energy and momentum transfer, one can compute  the imaginary-time response, where the factor $e^{-iHt}$  in Eq.~(\ref{resp}) is replaced by $e^{- H \tau}$. The corresponding Laplace transform, dubbed as Euclidean response functions, can essentially be evaluated exactly. Bayesian techniques, most notably Maximum Entropy~\cite{Bryan:1990,Jarrell:1996rrw}, are then used to retrieve the energy dependence of the response functions from their Euclidean counterparts. Recently, algorithms based on artificial neural networks have been developed to solve this problem. They have proven to be more accurate than Maximum Entropy in the low-energy region, which is relevant for detecting supernovae neutrinos, and more robust against high noise levels in the Euclidean response functions~\cite{Raghavan:2020bze}.

\begin{figure*}[thb]
	\includegraphics[width=0.6\textwidth]{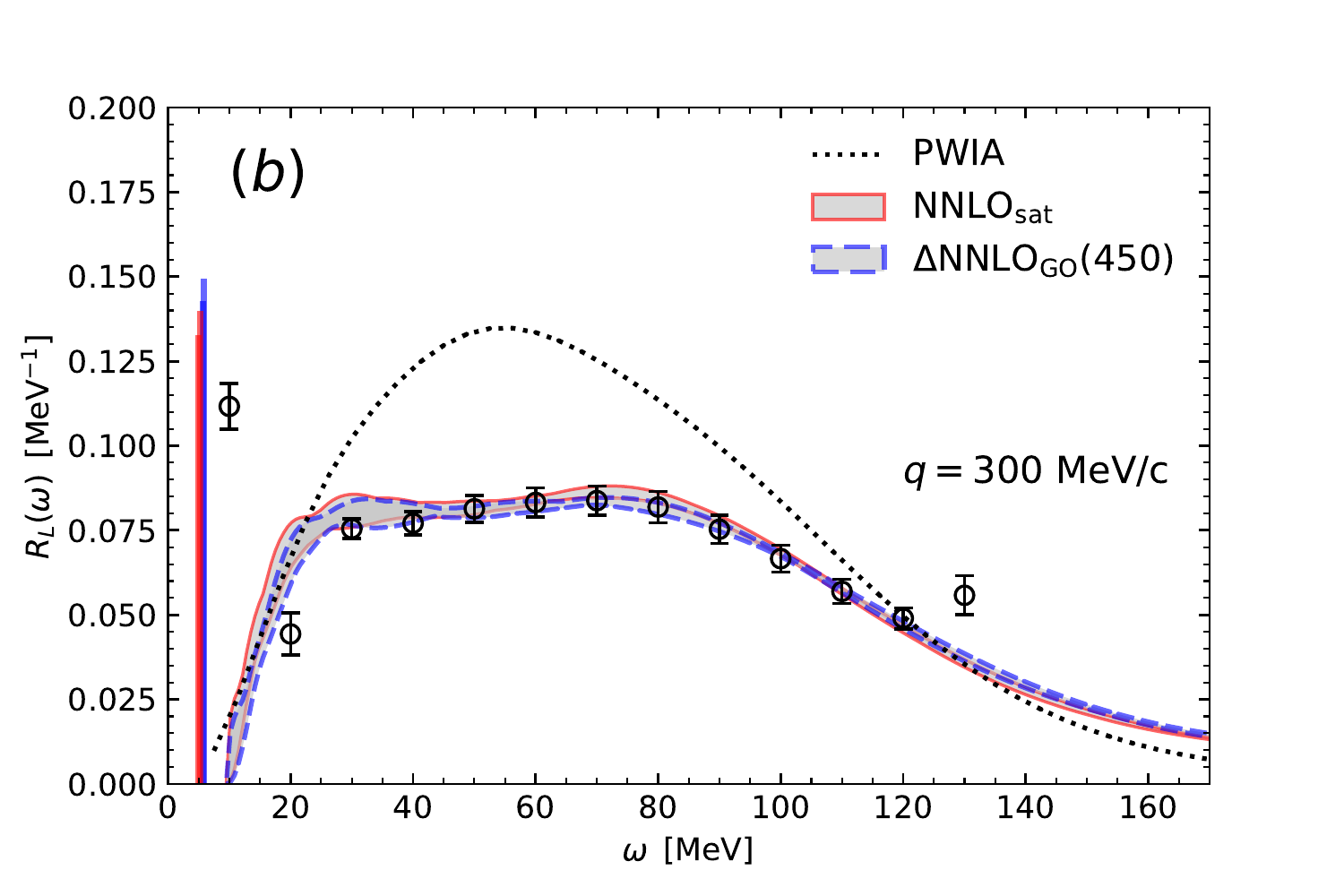}
	\caption{Electromagnetic longitudinal response function for $^{40}$Ca for $q=300 MeV/c $
	calculated with two $\chi$EFT forces in comparison to experimental data~\cite{Williamson:1997zz}. Figure adapted from Ref.~\cite{Sobczyk2021}.}
	\label{40Ca}
\end{figure*}

The GFMC has been employed to perform {\it virtually exact} calculations of inclusive electron- and  neutrino-scattering~\citep{Lovato:2016gkq,Lovato:2020kba} on $^4$He and $^{12}$C, which turned out to be in excellent agreement with experiments in the quasielastic region, see Fig.~\ref{fig:gfmc_inclusive}. The GFMC method retains all of the  spin-isospin components of the nuclear wave function which causes an exponential scaling with the number of nucleons. The computational cost currently limits its applicability to light nuclei, up to $^{12}$C. The AFDMC~\cite{Schmidt:1999lik} reduces the computational cost from exponential to polynomial in $A$ by representing the spin-isospin degrees of freedom in terms of products of single-particle states. A promising avenue to be pursued in the future includes tackling the electroweak responses of medium mass nuclei, including $^{16}$O and $^{40}$Ar within AFDMC. In this regard, an importart role is expected to be played by artificial neural-network representations of the AFDMC wave function~\cite{Adams:2020aax,Gnech:2021wfn}. Preliminary AFDMC calculations of the density response functions of $^{4}$He are in excellent agreement with the GFMC ones. Both the GFMC and the AFDMC method, while being extremely accurate, suffer some limitations that hamper their direct applicability to the forthcoming DUNE data analysis. In particular, they can not address exclusive reactions and include fully-realistic kinematics and currents. However, they will provide invaluable benchmarks on inclusive observables up to the moderate momentum transfer regime, for the more approximate methods discussed below. \\
     
\noindent{\bf Coupled-cluster  approach}\\ 
\noindent Another many-body method that can be used to compute response functions and lepton-nucleus cross sections is coupled-cluster theory. In this theory, one imprints correlations onto a starting  Slater determinant  using an exponential ansatz~\cite{hagen2013c}.
Response functions can be computed within the Lorentz integral transform method~\cite{efros1994}, leading to the solution of  a coupled-cluster equation of motion~\cite{bacca2013}. 
Recently, the longitudinal response function of $^{40}$Ca was investigated using  $\chi$EFT potentials and one-body currents. As very good description of the electron scattering experimental data was obtained, as shown in Figure~\ref{40Ca}.
 This approach can be extended to neutrino scattering in the quasi-elastic region. Nuclei such as $^{16}$O and $^{40}$Ar, which are typical targets in neutrino long-baseline experiments, are within the reach of this many-body method. More effort will need to be devoted into the inclusion of two-body currents, which have to be expanded into multipoles, and higher order correlations, which might be important at intermediate momentum transfer. \\

\noindent {\bf Microscopic factorization approaches}\\
\noindent 
Methods based on the factorization of the final hadronic state, such as those relying on the spectral function (SF) of the nucleus~\cite{Rocco:2020jlx,Barbieri:2019ual,Rocco:2019gfb} and the short-time approximation (STA)~\cite{Pastore:2019urn} are suitable to study  larger nuclear systems ($A>12$) relevant to the experimental program, while retaining most of the important effects coming from multi-nucleon physics. 

\begin{figure}[!htbp]
\centering
\includegraphics[height=2.4in]{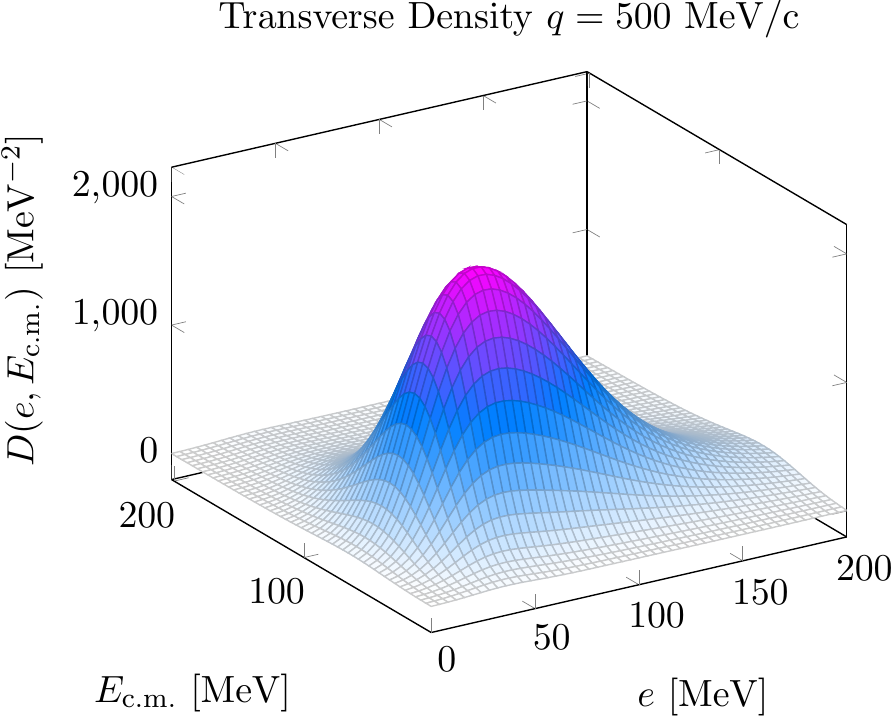}
\caption{\small Alpha particle transverse response densities at  $q\,$=$\,500$ MeV/c.  
The surface plots
show the response densities as functions of relative energy $e$ and center-of-mass energy $E_{\rm c.m.}$.
}
\label{fig:sta}
\end{figure}

The STA~\cite{Pastore:2019urn} algorithm has been developed to calculate nuclear responses in nuclei with $A>12$ within a QMC framework. At present, is has been tested within the VMC method to study electron scattering from the alpha particle and the trinucleon systems~\cite{Andreoli:2021cxo}. It is, however, exportable to other QMC approaches~\cite{Carlson:2014vla}  that are applicable to study larger nuclear systems, {\it e.g.}, the AFDMC.
The computational algorithm exploits a factorization scheme to consistently retain two-body physics, namely two-body correlations and associated two-body currents. Despite limiting the description of the scattering process to interactions of the probe with pairs
of correlated nucleons, the STA is found to be in good agreement with both GFMC predictions and experimental data 
for electron scattering from the alpha particle and the trinucleon systems~\cite{Pastore:2019urn,Barrow:2020mfy,Andreoli:2021cxo}. Importantly, the STA can account for interference effects between one- and two-body current contributions that are found to be essential to explain, {\it e.g.}, the observed excess in the transverse electromagnetic nuclear response~\cite{Benhar:2006wy,Carlson:2001mp}. Moreover, 
due to the factorization scheme, the STA provides us with additional information at the vertex where the probe interacts with the pair of correlated nucleons via one- and two-body electroweak currents. This information is cast in nuclear response densities, which are expressed in terms of the relative and center of mass energies of the struck nucleon pair. Upon integration of the response
densities, one recovers the nuclear response functions. In Fig.~\ref{fig:sta}, we show the electromagnetic transverse response density of $^4$He for external momentum transfer $q=500$ MeV/c. Response densities provide with valuable information for the event generators, as discussed in more detail in Sec.~\ref{sec:generator}. The STA method can accommodate fully-relativistic kinematic and currents, as well as pion production mechanisms (as already demonstrated within the spectral function formalism) and provide detailed information on the kinematic variables associated with the hadronic final states.\\

The framework based on the factorization of the hadronic final state and realistic SFs has been extensively utilized to describe electron-nucleus scattering data in the limit of moderate and high momentum transfer~\cite{Benhar:2006wy,Ankowski:2014yfa}. Within this approach, the hadronic final state is factorized in terms of a free nucleon state and $A-1$ spectator nucleons --- which can either be in a bound or an unbound state --- and all nuclear-structure information is encoded by the SF. The SF of finite nuclei is written as a sum of two terms: the single-nucleon mean field and the two-body correlation contribution. 
The first term is associated to the low momentum and removal-energy region. On the other hand, the correlation contribution includes unbound states of the $A-1$ spectator system in which at least one of the spectator nucleons is in the continuum, and it provides strength in the high momentum and energy region. The nuclear SF has been computed within different semi-phenomenological~\cite{Benhar:1994hw,Ivanov:2018nlm} and ab-initio many-body approaches~\cite{Rocco:2018vbf,Barbieri:2019ual}.  
More recently, the SF of $A=3$ and $A=4$ nuclei has been obtained from VMC and GFMC calculations~\cite{Andreoli:2021cxo}. This is particularly relevant for studying the dependence of the different observables from the nuclear interactions adopted in the calculation.  

\begin{figure}[h!]
    \centering
    \includegraphics[width=0.52\linewidth]{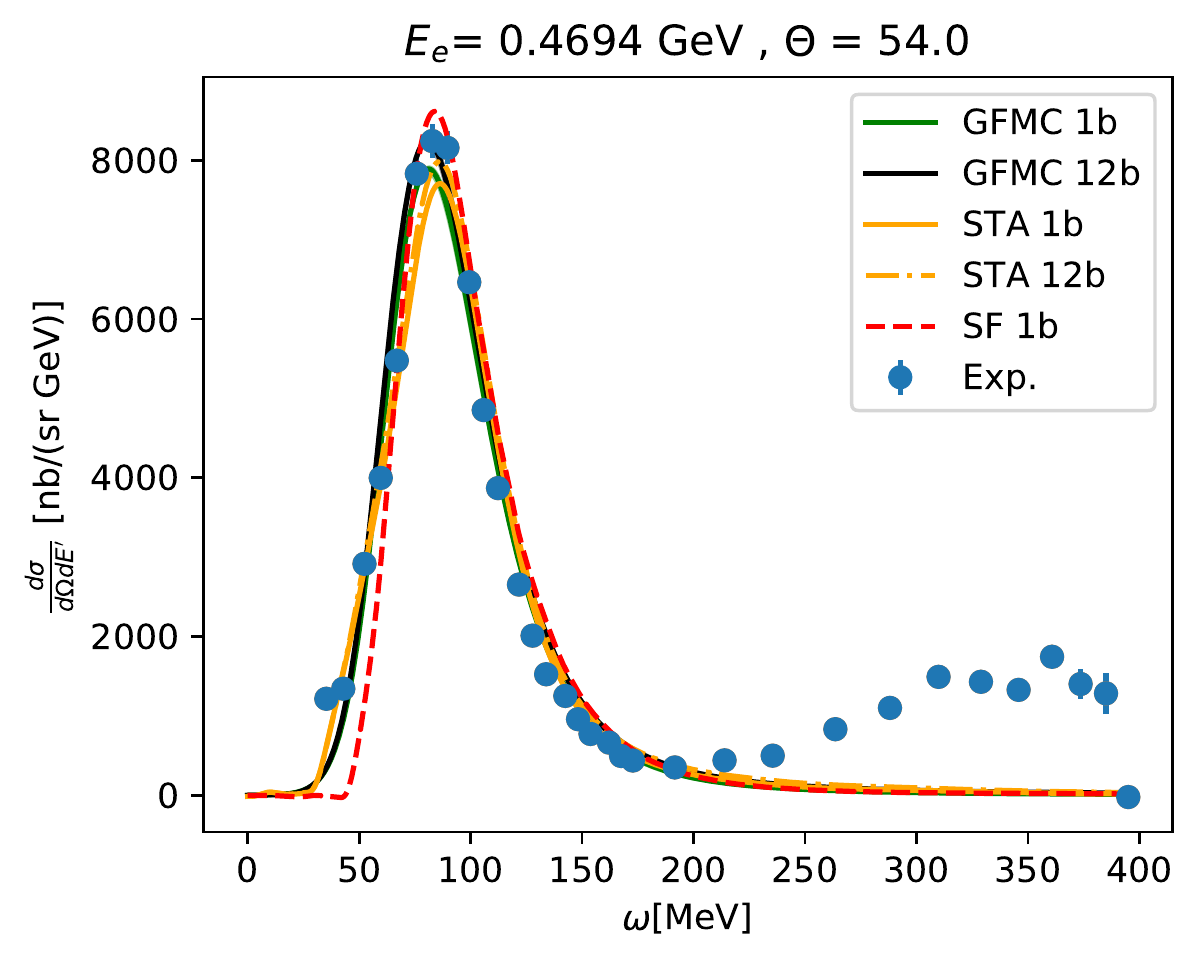}
    \hspace{0.5cm}
    \includegraphics[width=0.42\linewidth]{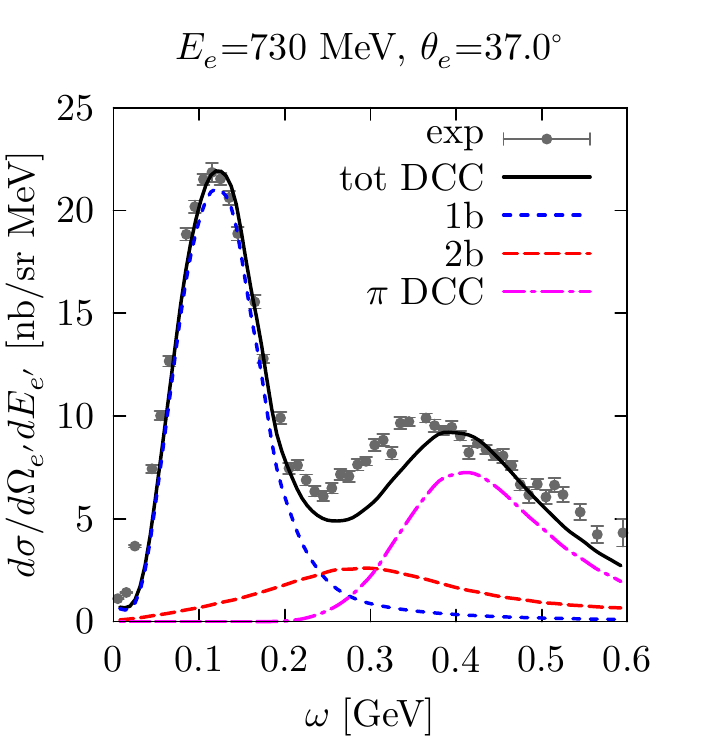}
    \caption{Left Panel: from Ref.~\cite{Andreoli:2021cxo}, inclusive double-differential cross sections for electron scattering on $^3$He at $469$ MeV and $54^\circ$ scattering angle. The blue points represent the experimental data of Ref.~\cite{Carlson:2001mp}. The black and green curves correspond to the GFMC one- and one- plus two-body current contributions. The yellow solid and dashed curves display the STA one- and one- plus two-body current calculations and the red dashed line show the SF results, where only the one-body current operator has been included. Right Panel: Inclusive $^{12}$C(e,e’) cross sections at 730 MeV and $37^\circ$ scattering angle. The theoretical calculations have been obtained within the SF approach using an extended factorization scheme. The short- dashed (blue) line and dashed (red) line correspond to one- and two-body current contributions, respectively. The dash-dotted (magenta) lines represent $\pi$-production contributions. The solid (black) line is the total result. The figure is adapted from Ref.~\cite{Rocco:2019gfb}.}
    \label{fig:SF-STA}
\end{figure}

Comparing the results obtained within three approaches based on the same description of nuclear dynamics of the initial target state--SF, Short Time Approximation and GFMC-- enables a precise quantification of the uncertainties inherent to factorization schemes, as shown in the left panel of Fig.~\ref{fig:SF-STA}. The SF framework has been already extended and generalized to include two-nucleon emission processes induced by relativistic meson-exchange currents ~\cite{Rocco:2015cil,Rocco:2018mwt} and applied to calculate the electroweak inclusive cross sections of carbon and oxygen~\cite{Rocco:2018mwt}. To tackle the resonance production region, the electroweak pion production amplitudes generated within the dynamical coupled-channel (DCC) model~\cite{Kamano:2013iva,Nakamura:2015rta,Kamano:2016bgm} have been included in the factorization scheme (see also Sec.~\ref{sec:pi_dis}). The results obtained using the semi-phenomenological SF of Ref.~\cite{Benhar:1994hw} are displayed in the right panel of Fig.~\ref{fig:SF-STA}.

In the future, we plan on leveraging AFDMC techniques in the STA and SF approaches to tackle larger nuclei, including $^{16}$O and $^{40}$Ar, that are relevant for the neutrino oscillation program. Finally, using different nuclear interactions and currents derived from EFTs will enable us to provide an estimate of the theoretical uncertainty in the neutrino-nucleus scattering cross sections. Inputs from LQCD calculations such as nucleon form factors, elementary nucleon matrix elements, and inelastic transition amplitudes involving $\pi$ mesons and $\Delta$ resonances--see Secs.~\ref{ssec:nucleon_form_factors}, \ref{sec:lqcd_mec}, and \ref{sec:lqcd_res}--will be readily implemented in the STA and SF as they become available. 

Factorization approaches appear quite accurate in describing inclusive scattering, but they also provide important information on the state of the nucleus at the electroweak vertex. This information includes
one- and two-nucleons momenta and energies at the vertex. In principle these states could be further time-evolved to explicit final states of the system. At present this could only be done in very small systems, $A=3$ and $4$,
because of the huge dimensionality of the relevant Hilbert space. For larger systems this time evolution is evaluated through the semi-classical approaches used in event generators. Explicit real-time calculations in $A=3$ and $4$ could be used to test these semi-classical approximations and perhaps improve them.

In the future, it may be possible to perform real-time evolution of the nuclear many-body state through quantum computers~\cite{Roggero:2019,Roggero:2020,Hall:2021}.  These could in principle provide very accurate inclusive cross sections through evaluation of the real-time two point functions. Even a short time evolution would give valuable information on the response. More intriguing, though more difficult, would be to follow the time evolution to larger times where one could isolate the contributions to explicit final states.  While this is well beyond the capabilities of current hardware, the two-point functions are expected to be a relatively near term application of quantum hardware. For quasi-elastic scattering the number of qubits required and the evolution time, corresponding to the circuit depth, are modest compared to many other quantum many-body properties at low energies.

\begin{figure}[h!]
\begin{center}
\includegraphics[width=\textwidth]{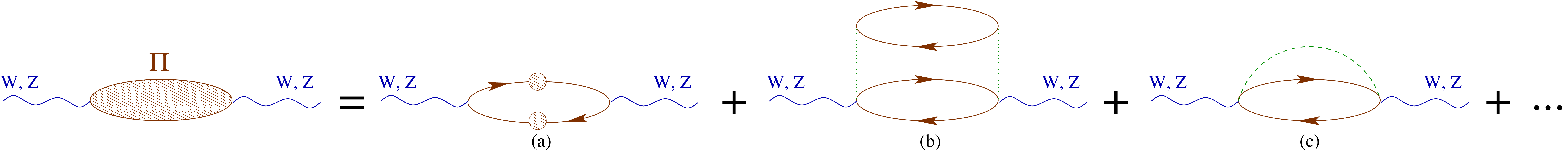}
\caption{\label{fig:polar}
  Diagrammatic representation of many-body contributions to the polarization propagator. Solid (dashed) lines correspond to
free nucleon (pion) propagators; dotted lines stand for effective nucleon-nucleon interactions. The solid lines with a blob represent full (dressed) nucleon propagators. For nucleons, the lines pointing to the right (left) denote particle (hole) states.}
\end{center}
\end{figure}

\vspace{0.5cm}
\noindent
{\bf Polarization propagator approach}\\
For leptons and, in particular,  (anti)neutrinos scattering off an extended system such as a nuclear target the inclusive cross section per unit volume in the Laboratory frame is given by
\begin{equation}
\label{eq:inclcs}
\frac{d}{d^3r} \left(\frac{d \sigma}{d \Omega(k') dk'^0} \right) = \frac{C}{4 \pi^2} \frac{| \vec{k}' |}{| \vec{k} |} \, L_{\alpha \beta} W^{\alpha \beta} \,.
\end{equation}
Constant $C$ is process specific. For example in the case of neutrino-induced charge current interactions involving only light quarks (such as quasielastic scattering or pion production), $C=(G_F V_{ud})^2$ while the leptonic tensor is    
\begin{equation}
\label{eq:leptensor}
L_{\alpha \beta} = k_\alpha k'_\beta  + k'_\alpha k_\beta - g_{\alpha \beta} k\cdot k' \pm i \epsilon_{\alpha \beta \sigma \delta} k'^\sigma k^\delta \,
\end{equation}
where $k (k')$ are the initial (final) lepton momenta. The hadronic tensor $W^{\alpha \beta}$, introduced in Eq.~\ref{eq:Wmunudef} can be expressed in terms of the so-called polarization propagator~\cite{fetterwalecka}
\begin{equation}
\label{eq:polar}
W^{\alpha \beta} = -\frac{1}{\pi} \, \mathrm{Im}\, \Pi^{\alpha \beta}  \,.
\end{equation}
Some examples of different pieces of the polarization propagator are diagrammatically represented in Fig~\ref{fig:polar}. Internal lines denote in-medium propagators of particle and hole states. The imaginary parts of these diagrams that contribute to the hadronic tensor can be obtained using Cutkosky rules. The blobs in diagram (a) denote selfenergy insertions that account for nucleon-nucleon interactions, both mean-field and short-range correlations. In fact, the contribution of diagram (a) to quasielastic scattering can be cast as
\begin{equation}
\label{eq:polar(a)}
\mathrm{Im}\, \Pi^{\alpha \beta} = - 2 \pi^2 \int \frac{d^4 p}{(2 \pi)^4} H^{\beta \alpha}  \mathcal{A}_p(p+q) \, \mathcal{A}_h(p)
\end{equation}
where $\mathcal{A}_{p,h}$ are nothing but the particle and hole spectral functions discussed above. Tensor $H^{\beta \alpha}$ is the free nucleon counterpart of $W^{\alpha \beta}$. Diagram (b) is an example of a MEC contribution  involving two nucleons, which cannot be reduced to (a). Diagram (c) is a pion production contribution to the (semi)inclusive cross section.

\begin{figure}[h!]
    \centering
    \includegraphics[width=0.42\linewidth]{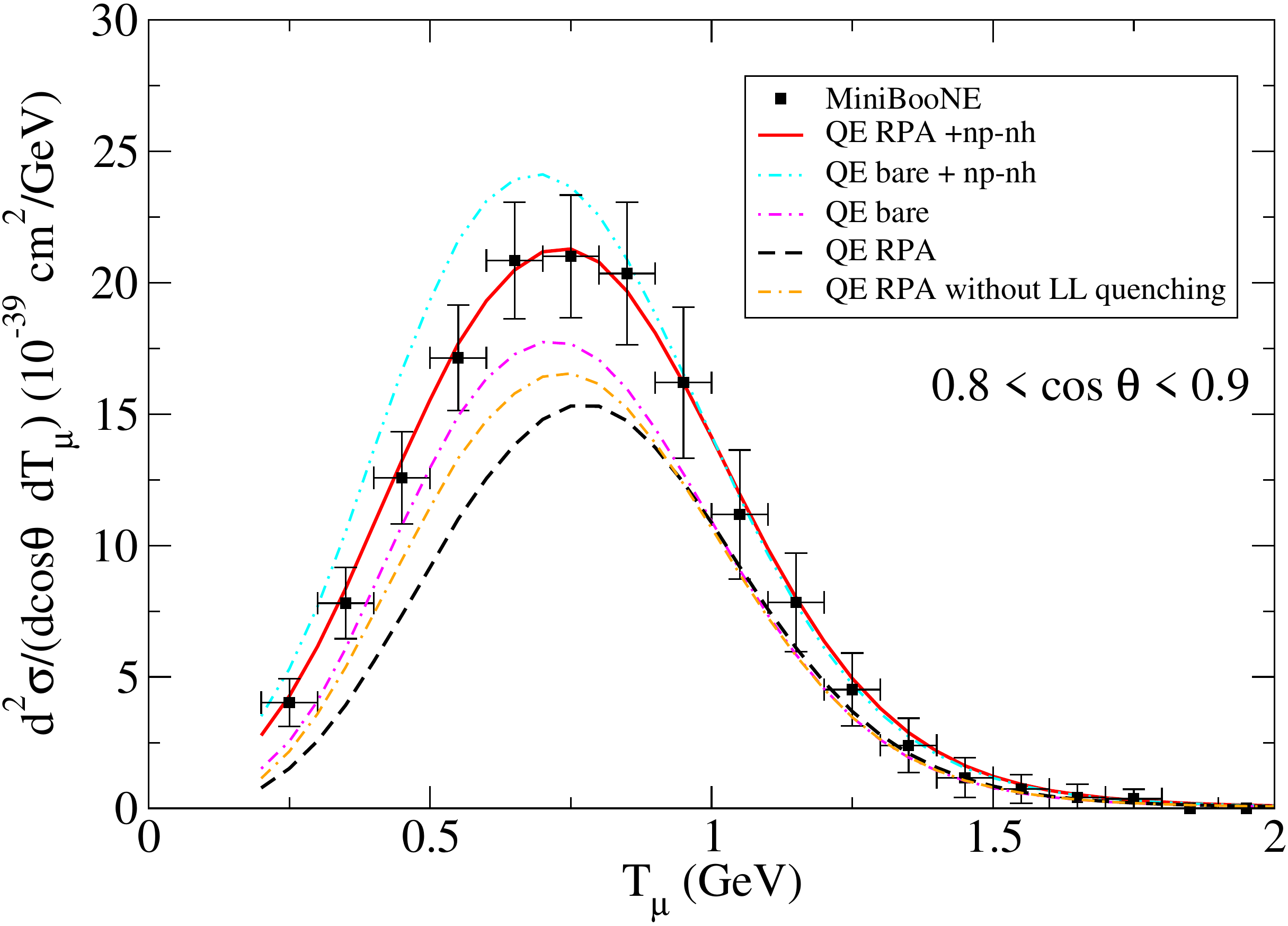}
    \includegraphics[width=0.42\linewidth]{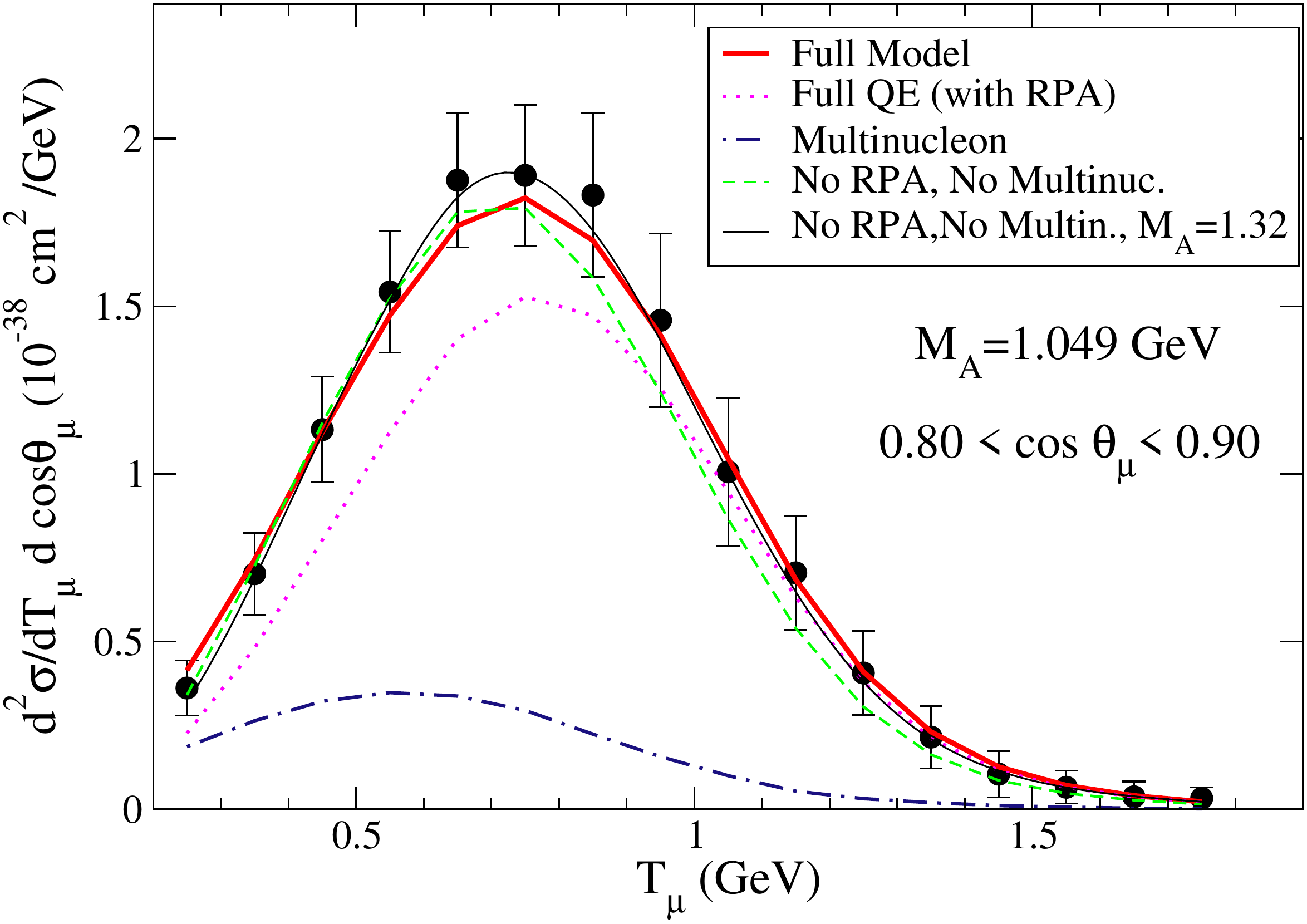}
    \caption{Quasielastic-like $\nu$-$^{12}$C double differential cross section  averaged over the MiniBooNE flux as a function of the muon kinetic energy and for the  $0.80 < \cos\theta_\mu < 0.90$ angular bin calculated in Ref.~\cite{Martini:2011wp} (left) and \cite{Nieves:2011yp} (right) compared to the MiniBooNE data~\cite{MiniBooNE:2010bsu}. In the right panel the data has been rescaled by a factor 0.9 (compatible with flux uncertainties).}
    \label{fig:MiniBooNEcomp}
\end{figure}

The different terms of the polarization propagator can be calculated following various strategies. A possibility is to compute them for infinite nuclear matter and adapted to finite nuclei using the local density approximation. The fact that plain waves provide a convenient basis in infinite nuclear matter considerably simplifies the calculations, making it easy to account for relativistic effects and hadronic degrees of freedom beyond pions and nucleons. It is also straightforward to predict the nuclear mass dependence of the observables. This approximation is more realistic for heavy nuclei but has been used for carbon and oxygen isotopes. This approach has been extensively applied to the study of a variety of particle-nucleus interactions. It is however not valid at low momentum transfers as it cannot describe discrete transitions or the excitation of collective states. In the case of neutrino scattering, pioneering studies focused on quasielastic scattering~\cite{Singh:1992dc,Kim:1994zea} where the RPA equations were solved in the ring approximation, with the above mentioned assumptions. More recently, explicit $\Delta$s, pion production and MEC have been incorporated~\cite{Martini:2009uj,Nieves:2011pp}. These studies first showed the importance of two-nucleon mechanisms for few-GeV neutrino interactions with nuclei. Comparisons of these models to MiniBooNE quasielastic-like ({\it i.e.} without produced or absorbed pions) cross sections are shown in Fig.~\ref{fig:MiniBooNEcomp}. A good agreement has also been obtained with T2K data~\cite{T2K:2016jor}. 
On the other hand discrepancies with these models have been found at the higher energy and momentum transfers probed at MINERvA and NOvA as can be noticed in Refs.~\cite{MINERvA:2015ydy,NOvA:2020rbg}. These discrepancies have been attributed to an underestimation of the MEC contribution, although it is unclear if the deficiency should be actually ascribed to the theoretical model itself or to its implementation in the event generator used to analyse and compare to data. Nevertheless, MEC at these kinematics and the role of heavier mesonic and baryonic degrees of freedom should be further investigated.

While the described approach is in principle not suitable for exclusive final states, the local density approximation allows to obtain a reaction probability at a given spacial coordinate, making it possible to propagate the final state particles using semiclassical methods (cascade or transport). In this way single~\cite{Nieves:2005rq,Leitner:2006ww} and multiple~\cite{Sobczyk:2020dkn} nucleon knockout but also pion production~\cite{Leitner:2006ww,Lalakulich:2010ss} have been investigated. By using structure functions integrated over space as input, event generators do not take into account correlations of dynamical origin predicted by theory. Progress in this direction should be pursued. \\  

\noindent
{\bf Mean-field approaches}\\
\noindent 
Mean-field or shell-model approaches are able to capture a good part of the nuclear dynamics by describing the ground state nucleus as a set of independent-particle nucleon wave functions that are solutions of the mean-field equations. 
The quasielastic cross section can be efficiently modeled by describing the knocked-out nucleon as a scattering solution of the wave equation. In the inclusive case, the flux has to be conserved, so one can use mean-field potentials with only the real part \cite{Maieron03,Meucci09,Kim07,Butkevich07,Pandey16,Gonzalez-Jimenez20} or full complex optical potentials, in which the flux lost (transferred to inelastic channels) is recovered by a summation over those channels, as done in the Relativistic Green Function model \cite{Capuzzi91,Meucci09,Ivanov16b}. For exclusive scenarios, one needs to account for the flux moved to the inelastic channels (absorption, multi-particle emission, charge exchange, etc). This is done by using phenomenological complex optical potentials \cite{Udias93,Udias01} which are usually fitted to elastic nucleon-nucleus scattering data~\cite{Cooper93,Cooper09}.

In neutrino experiments, fully exclusive conditions are never satisfied because the neutrino energy is unknown and the limited acceptances of the detectors make it impossible to detect the complete final state. Therefore, Monte Carlo (MC) neutrino event generators have to deal with inclusive and semi-inclusive scenarios. In MC generators, due to the factorization ‘elementary vertex $\times$ hadron propagation’, the inclusive cross section will not be affected by the cascade process. Therefore, the primary model (the one that describes the elementary vertex) should be able to provide a good inclusive response. But, additionally, it is preferable to use primary models that provide information on the hadrons, so that they could be used as the ‘seed’ for the cascade in a more consistent way. The mean-field models discussed here satisfy these requirement, i.e., full hadronic information and good inclusive results.

\begin{figure}[h!]
    \centering
    \includegraphics[width=.4\linewidth]{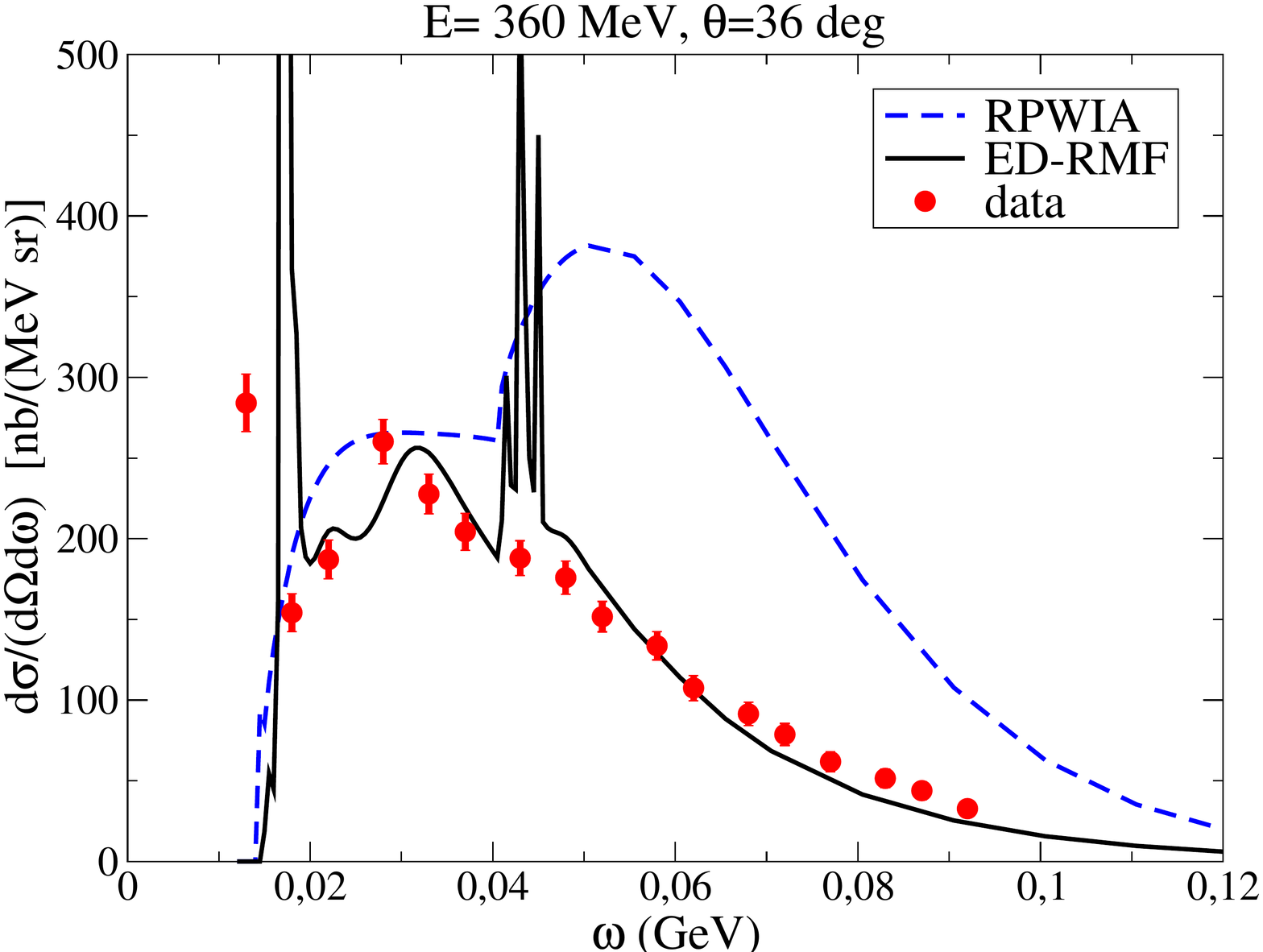}
    \includegraphics[width=.4\linewidth]{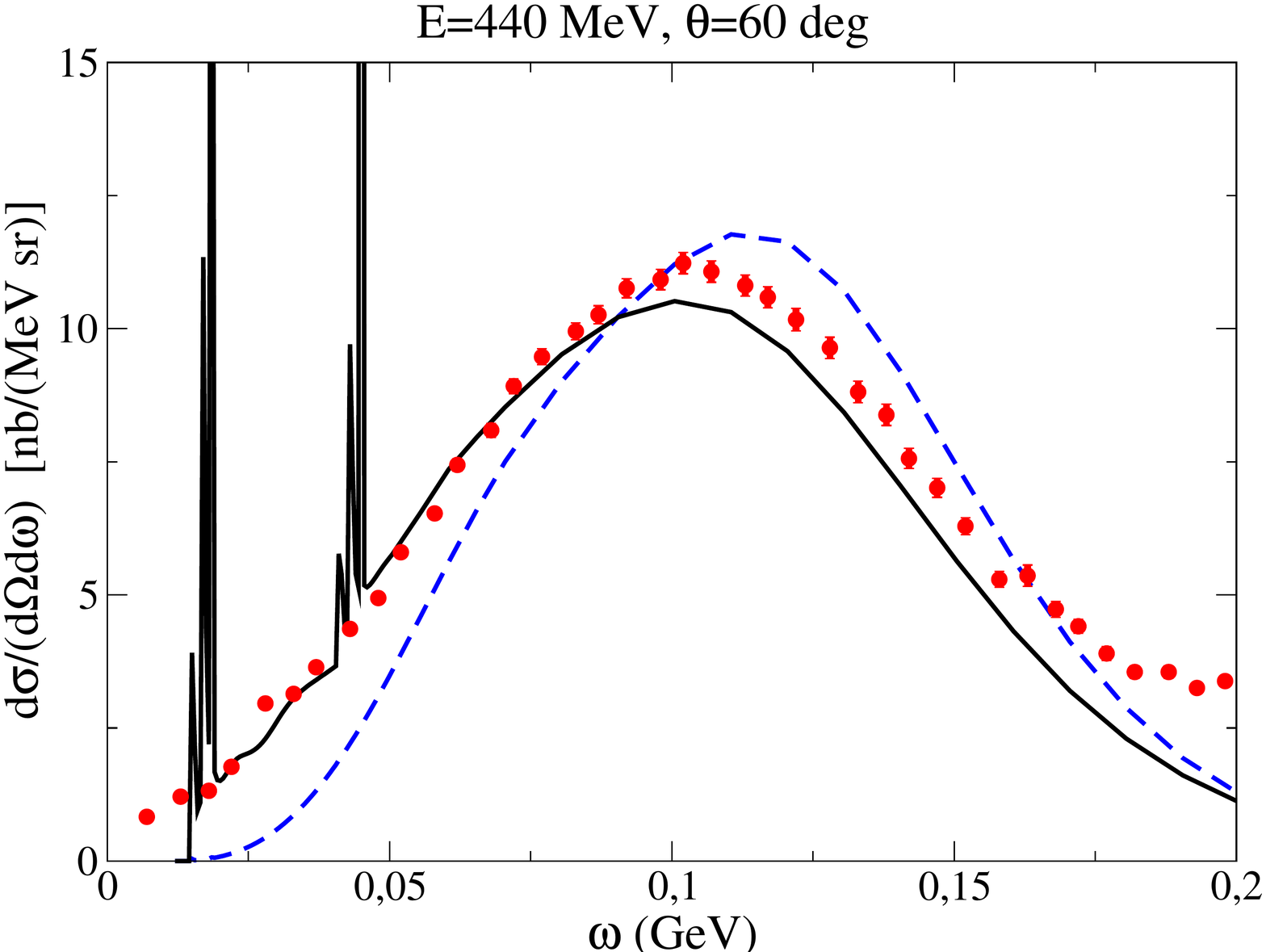}
    \caption{Inclusive $^{12}$C(e,e') cross section data compared with different model predictions: RPWIA
(distortion and Pauli blocking are neglected) and ED-RMF (full model). $E$, $q$ and $\omega$ represent the electron incident energy, the scattering angle and the energy transfer, respectively. Figure adapted from \cite{Gonzalez-Jimenez19}.}
    \label{fig:EDRMF-vs-RPWIA}
\end{figure}

In the shell-model approaches, nuclear effects like Pauli blocking, binding energy and distortion (or final-state interactions) are consistently incorporated. In Fig.~\ref{fig:EDRMF-vs-RPWIA}, we show the results of a plane-wave model (labeled as RPWIA) and the 'full' model (labeled as ED-RMF), that uses a distorted wave for the final nucleon. We compare both approaches with inclusive electron scattering data. A detail discussion can be found in \cite{Gonzalez-Jimenez19,Nikolakopoulos19}.

Of course, the complexity has a cost, both relativity and distorted waves break the factorization scheme that appears in plane-wave based models, this means that these non-factorized models demand important computational resources and, also, it is not easy to incorporate them in the MC generators. Though, some work is currently being done in this direction \cite{Gonzalez-Jimenez:2021ohu,Nikolakopoulos22}.

The CRPA model ~\cite{Jachowicz:2021ieb} is based on a Hartree-Fock mean-field description supplemented with a random phase approximation approach to include long-range correlations, important for the description of processes at low energy transfers.  The CRPA approach has been successfully validated against available electron scattering data, and is especially suited to describe the cross section at incoming energies below 100 MeV or for  small lepton scattering angles, where nuclear structure effects are known to be important, in an efficient way.            The RPA allows correlations to be present in the ground state of the nuclear system and additionally allows the particles to interact by means of the residual two-body force. The random phase approximation hence goes a step beyond the zeroth-order mean-field approach and describes a nuclear state as the coherent superposition of particle-hole and hole-particle contributions out of a correlated ground state. This approach some of the collectivity present in the nucleus to be accounted for.  The consistency of the implementation of nuclear correlations can be guaranteed by using the same effective force to build the mean field and  residual interaction. Solving the equations for the RPA propagator in coordinate space allows the continuum to be taken into account. The formalism has been extended to higher energies, allowing for a uniform description from threshold up to the QE peak regime. One of the advantages of this approach is that it allows for a straightforward extension to the description of exclusive processes.

When the momentum transfer involved in the electroweak reaction is high, typically $q>400$ MeV, relativistic effects, not only on the kinematics but also on the dynamics, are relevant. The relativistic mean-field model is a fully relativistic approach, and as so, it can make predictions over the entire energy region of interest for neutrino experiments. It has been used for years to describe inclusive and exclusive electron-nucleus scattering and neutrino scattering; a recent review of the formalism and results can be found in \cite{Amaro20}.\\

\noindent
{\bf Semi-phenomenological factorization approaches}\\
\noindent 
Over the years a considerable number of effective models has been developed to describe neutrino-nucleus cross sections, in the quasi-elastic regime and beyond.
Whereas these models are based on approximate schemes and are as a matter of fact not as rigorous as ab-initio approaches, 
they provide powerful and flexible tools to study various aspects of neutrino-nucleus cross sections.
They often originate from equivalent electron-scattering efforts and have been benchmarked extensively against the plethora of 
high-quality data that is available in the electromagnetic sector. In the axial weak sector, detailed comparisons with ab-initio results can further corroborate the validity of these models.

Confronted with the rigorousness of ab-initio approaches, an important advantage of these descriptions is their computational efficiency, and in a number of cases their ability to be extended to the description of more exclusive processes in a straightforward way. Other advantages are a full inclusion of relativistic nuclear dynamics, as well as to the treatment of  heavier nuclear targets without exaggerate numerical cost.
Without aiming to be exhaustive, often adopted  approaches include Superscaling methods as SuSA(v2), exploiting equivalences between electron- and neutrino scattering processes, based on the scaling behaviour exhibited by electroweak scattering data and ported to relativistic mean-field modeling \cite{Gonzalez-Jimenez:2014eqa} and extended to include meson-exchange contributions to the cross-section \cite{Megias16a}. See \cite{Amaro21} for a recent review. \\

\noindent
{\bf Fast emulators for complex theoretical models}\\
\noindent
An emulator~\cite{Melendez:2022kid} can be considered as an efficient and accurate tool to interpolate and extrapolate the solutions of a complex theoretical model in the model parameter space. As a piece of computer code, the emulator runs extremely fast and can be easily transported between physicists. It can be readily incorporated into the experimental event generators as well. Such emulators have been rapidly developed for both many-body calculations of nuclear properties~\cite{Frame:2017fah, Ekstrom:2019lss,Konig:2019adq,Yoshida:2021jbl,Hu:2021trw,Bonilla:2022rph} and few-body scattering and reaction calculations~\cite{Furnstahl:2020abp,Bai:2021xok,Drischler:2021qoy,Melendez:2021lyq,Zhang:2021jmi}. It will be valuable to generalize the former studies to emulate  nuclear response function calculations. The latter ones could also be expanded to emulate hadronic coupled-channel models~\cite{Nakamura2022}. In addition, when the emulators for different theoretical models become available, it is natural to explore the so-called model mixing method from Bayesian statistics~\cite{Phillips:2020dmw}. The method provides a systematic framework to combine these models (emulators) to form a mixture model and estimate the theoretical uncertainty of the collective knowledge~\cite{Phillips:2020dmw,BAND_Framework}. The mixture model can also be implemented in the event generators.

\subsection{Pion production and DIS}{\label{sec:pi_dis}}

\begin{figure}[h!]
    \centering
    \includegraphics[width=0.8\linewidth]{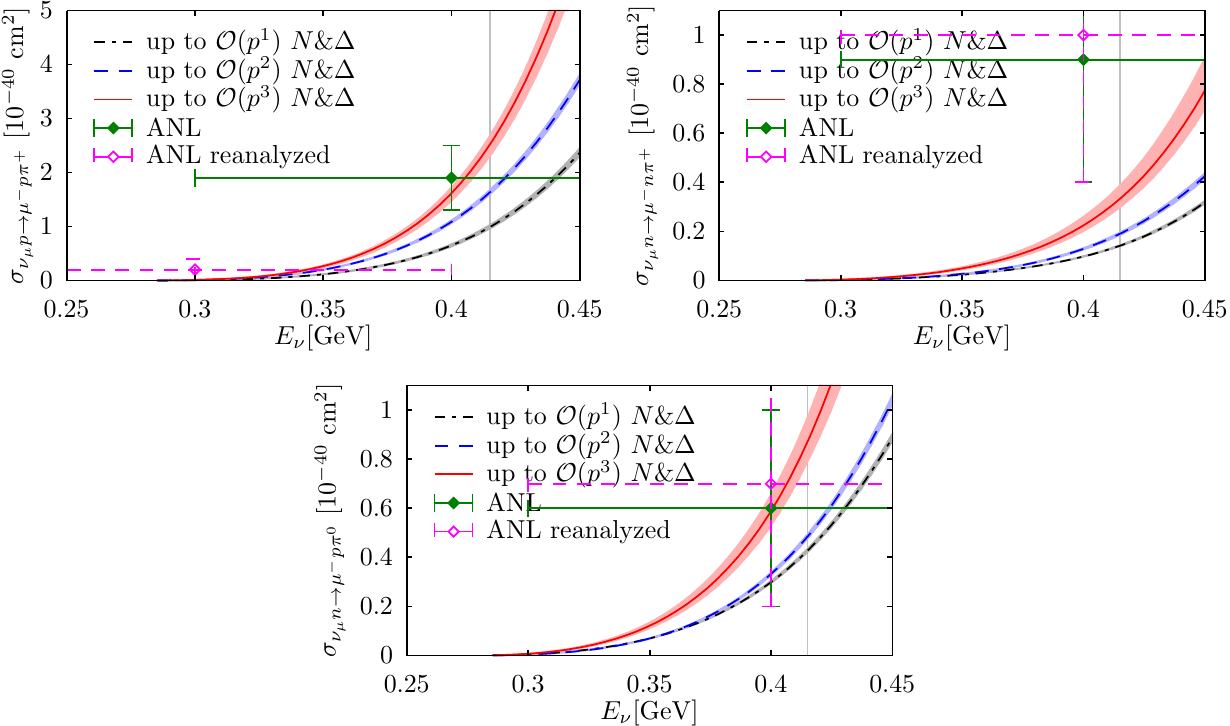}
    \caption{Charged-current neutrino-induced pion production at different chiral orders. Plots taken from Ref.~\cite{Yao:2018pzc}.}
    \label{fig:pionChPT}
\end{figure}

Neutrino-induced inelastic scattering and, in particular, pion production can be investigated using effective field theory. For this purpose the chiral Lagrangian with mesons and baryons is coupled to weak bosons as external fields. Close to threshold,  systematic treatment of quantum corrections is possible using chiral perturbation theory. Weak pion production has been studied in Refs.~\cite{Yao:2018pzc,Yao:2019avf} up to next-to-next-to-leading order using the covariant formulation of the theory and with explicit $\Delta(1232)$ degrees of freedom. For charged-current interactions, amplitudes are expressed in terms of 22 LECs, most of which have been determined in non-neutrino processes such as pion-nucleon scattering,  pion photo and electroproduction data~\cite{GuerreroNavarro:2020kwb}. The remaining LECs could be fixed using neutrino-nucleon data with full kinematic reconstruction and high enough statistics in the kinematic region where a perturbative treatment is applicable. The total cross section calculated at different chiral orders is shown in Fig.~\ref{fig:pionChPT}. The error bars come from the LEC uncertainties.  
LEC determination is not only interesting but would open the possibility to use pion production as a standard candle for neutrino-flux determination with controlled theoretical errors. One should bear in mind that this approach is limited to low energy and momentum transfers, nevertheless effective field theory calculations provide a well-founded low-energy benchmark for phenomenological models aimed at the description of weak pion production processes in the broad kinematic range of interest for current and future neutrino-oscillation experiments.

The production of real pions in the final state will be crucial for the correct understanding of the DUNE results. 
This contribution has been recently included within the SF formalism by generalizing the factorization of the hadronic final state to include pion state (see right panel of Fig.~\ref{fig:SF-STA}). 
In analogy with the one-body case discussed in the previous Section, the one-body one-pion (1b1$\pi$) incoherent contribution to the hadron tensor
is written in terms of the one-nucleon SF and the elementary matrix elements has been obtained within the sophisticated DCC model~\cite{Kamano:2013iva,Nakamura:2015rta,Kamano:2016bgm} able to describe the  
$\pi N \rightarrow \pi N$, $\gamma N \rightarrow \pi N$, and $N(e,e'\pi)N$ reactions accounting for meson-baryon channels  and nucleon resonances up to an invariant of $W=$ 2 GeV. When they become available, LQCD calculations of $N\rightarrow N\pi$ and $N \pi \rightarrow N\pi $ amplitudes will provide valuable constraints on the phenomenological models currently employed to describe these processes. 
How to correctly describe the transition between the pion production and the region dominated by the DIS is an open question that will need to be carefully addressed in the future; for a more detailed discussion see Sec.~\ref{sec:sis-dis} below.

\section{Neutrino-induced 
shallow and deep inelastic scattering}
\label{sec:sis-dis}

\subsection{Introduction}

\begin{figure}[h!]
    \centering
    \includegraphics[width=0.85\textwidth]{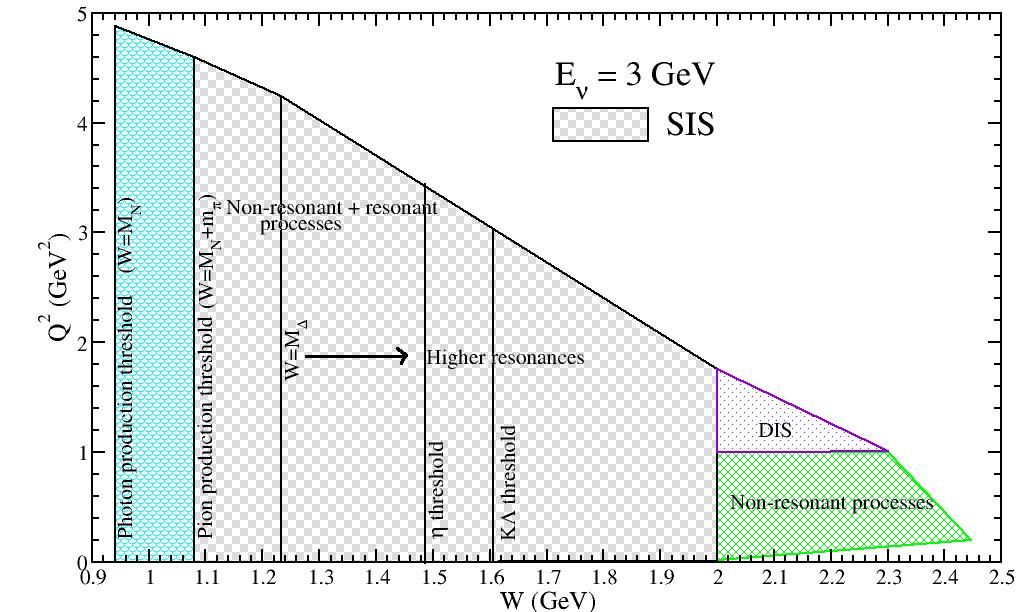}
    \includegraphics[width=0.85\textwidth]{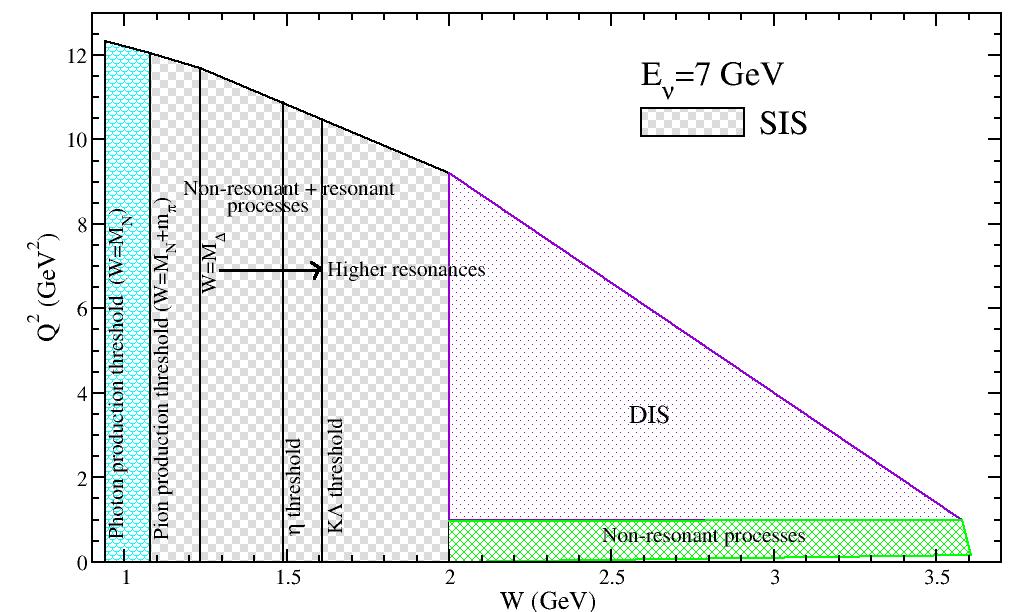}
    \caption{$(W,Q^2)$ {\it landscape}. for neutrino-nucleon scattering at two representative laboratory neutrino energies.}
    \label{fig:landscape}
\end{figure}

Neutrino and antineutrino scattering off nucleons 
exhibits a very rich phenomenology in a broad kinematic range, which can be defined in terms of the invariant mass of the final hadronic system, $W$ and minus the four-momentum transferred to the nucleon squared, $Q^2$. As illustrated in Fig.~\ref{fig:landscape} for two incident neutrino energies, different $(W,Q^2)$ regions show the prevalence of distinct degrees of freedom and dynamics. Above the pion production threshold $W \approx 1080$~MeV the excitation of the $\Delta(1232)$ dominates, but at higher $W$ the dynamics results from a non-trivial interplay of overlaping baryon resonances, non-resonant amplitudes and their interference. It is this region of $W$ above the $\Delta(1232)$ and at moderate $Q^2 \lesssim 1$~GeV$^2$ that we refer to as Shallow Inelastic Scattering (SIS).  Figure~\ref{fig:landscape} shows its prevalence at $E_\nu \sim 3$~GeV.
As $Q^2$ grows, one approaches the onset of Deep Inleastic Scattering (DIS).  The science of this complex region, poorly understood both theoretically and experimentally~\cite{SajjadAthar:2020nvy,NuSTEC:2017hzk,Andreopoulos:2019gvw}, encompasses the \emph{transition} from strong interactions described in terms of hadronic degrees of freedom to  those among quarks and gluons described by perturbative QCD. 

Neutrino-scattering simulations often describe this transition using parton distribution functions empirically extrapolated from the DIS region to lower $W$ and $Q^2$ by Bodek {\it et. al.}~\cite{Bodek:2002ps,Bodek:2004pc,Bodek:2010km}. Duality arguments constrain the inclusive cross section but do not predict the specific particle content of the final state. Therefore, efforts to extend the description in terms of quarks and gluons towards lower $W$ and $Q^2$ by including higher-twist corrections~\cite{Dasgupta:1996hh} should be complemented with a realistic modeling of the SIS region using hadronic degrees of freedom. Progress in this direction has been significant (see for instance Ref.~\cite{Nakamura:2015rta}) but is hindered by the lack of experimental information about the axial current for inelastic processes at non-zero $Q^2$. 

Modern experiments with (heavy) nuclear targets have provided and will keep providing valuable information on these issues, but the presence of nuclear effects 
such as Fermi motion, Pauli blocking, long- and short-range correlations, two- and three-body currents and, very significantly, final-state interactions tends to blur the information required to refine the hadronic description in the way outlined above.  Since a large fraction of events in NOvA~\cite{Acero:2019ksn} and DUNE~\cite{Abi:2020wmh}, and in atmospheric neutrino measurements at IceCube-Upgrade~\cite{IceCube-Gen2:2020qha}, KM3NeT~\cite{Adrian-Martinez:2016fdl}, Super- and Hyper-Kamiokande~\cite{Fukuda:2002uc,Abe:2018uyc}, are from the SIS and DIS regions, there is a definite need to improve our knowledge of this physics.

 \subsection{Inelastic processes} \label{inelastic}

As indicated in the introduction, the onset of the  inelastic regime is marked by single pion production. Neutrino-nucleon 
inelastic scattering predominantly leads to single pion ($\pi N$) but also to $\gamma N$, $\pi \pi N$, $\eta N$, $\rho N$, $K N$, $\pi \Sigma$, $\bar{K} N$, $K Y$, $\ldots$ final states. At small energy and momentum transfers (or, equivalently, close to threshold in $W$ and small $Q^2$), the amplitudes for these processes are constrained by the approximate chiral symmetry of QCD~\cite{Hernandez:2007qq,Hernandez:2007ej,RafiAlam:2010kf, Alam:2012zz,Wang:2013wva}. In this regime, Chiral Perturbation Theory allows for a systematic improvement by computing higher-order corrections, as done in Refs.~\cite{Yao:2018pzc,Yao:2019avf} for single-pion production, but a theoretical description covering the whole kinematics available with few-GeV neutrinos demands  phenomenological modeling using external (non neutrino) information. Indeed, thanks to approximate flavor symmetries and the partial conservation of the axial current (PCAC), electron- and meson-nucleon scattering  provide very valuable input for the description of weak inelastic processes. The axial current contribution however remains largely unconstrained, which calls for new measurements on elementary (hydrogen, deuterium) targets.

\begin{figure}[h!]
    \centering
    \includegraphics[width=0.48\textwidth]{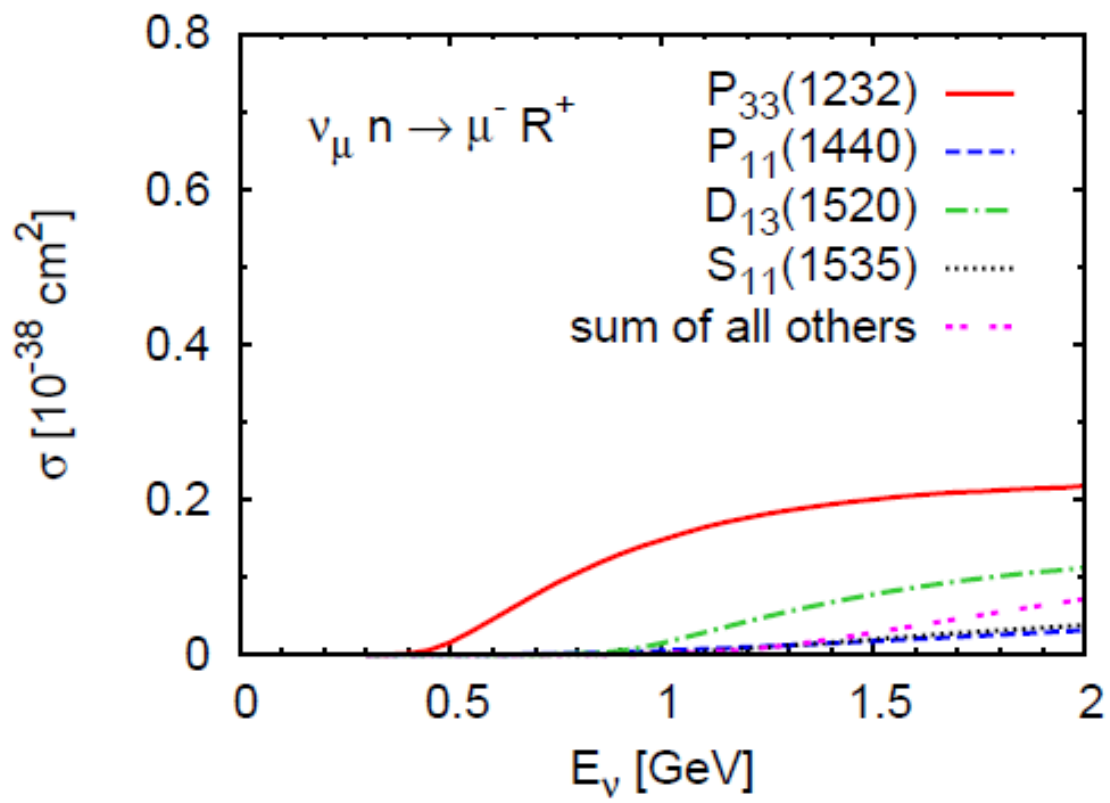}
    \includegraphics[width=0.5\textwidth]{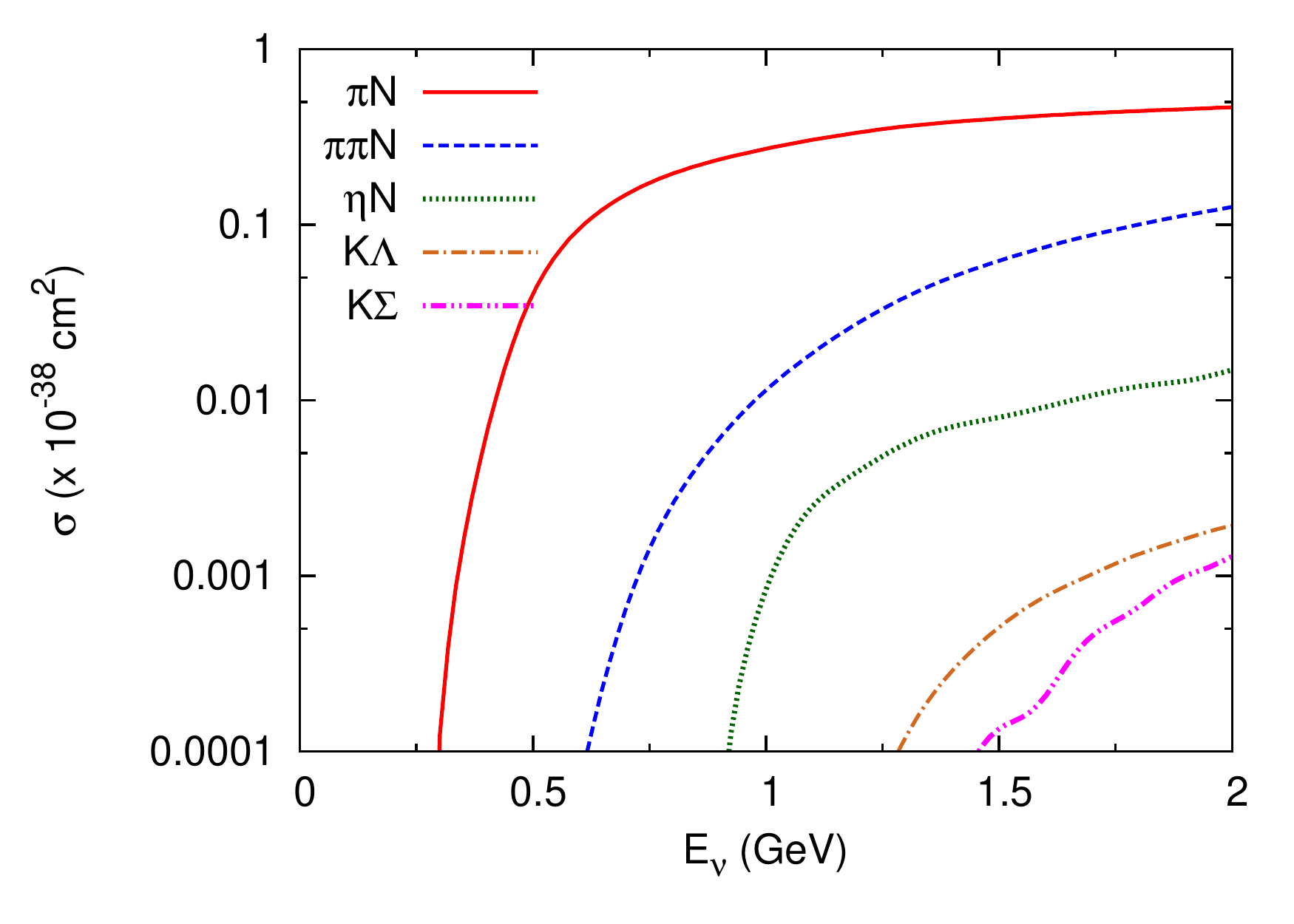}
    \caption{Charged-current $\nu_\mu - n$ inelastic cross section. Left panel: contribution from different baryon resonances according to GiBUU (adapted from Ref. \cite{Leitner:2008ue}). Right panel: cross section for different final states as predicted by the DCC model \cite{Nakamura:2015rta}.}
    \label{fig:CCnu-n}
\end{figure}

Away from threshold, most of these reactions are dominated by baryon resonances, albeit with sizable contributions from non-resonant amplitudes and their interference with the resonant counterpart~\cite{Leitner:2008ue,RafiAlam:2015fcw}. In the case of $\pi N$, but also $\gamma N$ final states, $\Delta(1232)$ excitation is dominant. Among heavier baryonic resonances, the $N(1520)$ has been identified as the most relevant one in Ref.~\cite{Leitner:2008ue} (left panel of Fig. \ref{fig:CCnu-n}), while the $N(1440)$ has the largest contribution in most $\pi N$ channels according to Ref.~\cite{RafiAlam:2015fcw}. As it is well known from $\pi N$ scattering, $N(1535)$ prevails for the $\eta N$ final state. Different models for neutrino-nucleon inelastic scattering in the resonance region have been developed. The Rein-Sehgal approach~\cite{Rein:1980wg} relies on unrealistic transition form factors calculated with a constituent quark model. It is, nonetheless, extensively adopted by neutrino event generators. Implementations therein have updates some resonance properties such as masses, decay widths and branching ratios but interferences are neglected; the non-resonant background is treated in an effective manner as a smooth $P_{11}$-like term or as a downward extrapolation of DIS contributions for the higher energy region. This model has been recently updated with a more realistic (close to threshold) non-resonant part of the amplitude and empirical input for the vector part of the transition~\cite{Kabirnezhad:2017jmf,Kabirnezhad:2020wtp} current.  
The Giessen Boltzmann-Uehling-Uhlenbeckmodel (GiBUU) model~\cite{Buss:2011mx} relies on the Mainz Unitary Isobar Model (MAID)  analysis~\cite{Drechsel:2007if} of electron-nucleon pion production as input for the vector part of both resonant and non-resonant amplitudes. Owing to PCAC, Goldberger-Treiman relations are derived to relate the leading axial nucleon-to-resonance couplings to the $\pi N$ partial decay width of the resonances. The axial part of the non-resonant current is built from the vector part in a purely phenomenological way. The dynamical coupled channel (DCC) approach~\cite{Nakamura:2015rta,Kamano:2013iva,Kamano:2016bgm} is consistently constrained using the $eN$ and $\pi N$ vast amount of  data to predict not only weak single but also double pion production and other meson-baryon final states up to $W$ of about 2.2 GeV (see right panel of Fig. \ref{fig:CCnu-n})

As stated above, at low energy and momentum transfers, the non-resonant part of the amplitude is constrained by QCD symmetries. It is common that model builders extend the validity of chiral amplitudes towards high $Q^2$ by introducing phenomenological form factors~\cite{Hernandez:2007qq,Hernandez:2007ej,RafiAlam:2010kf, Alam:2012zz,Wang:2013wva}. Unitarization, which becomes important as $W$ increases, is absent in tree-level amplitudes but can be partially restored by imposing Watson's theorem in relevant multipoles~\cite{Alvarez-Ruso:2015eva,Saul-Sala:2021swb}. For neutrino interactions, a complete unitarization in coupled channels has only been implemented by the DCC model~\cite{Nakamura:2015rta,Kamano:2013iva,Kamano:2016bgm} and in the strangeness $S=-1$ sector within the chiral-unitary approach ~\cite{Ren:2015bsa}. High invariant masses $W \gtrsim 2$~GeV, above the resonance region and low $Q^2$ is the realm of diffractive scattering where non-resonant amplitudes need to be improved within the Regge approach~\cite{Gonzalez-Jimenez:2016qqq,Gonzalez-Jimenez:2017fea,Nikolakopoulos:2018gtf}.

When neutrinos scatter off heavy nuclear targets, the presence of the nuclear medium poses additional challenges for the reaction modeling. The initial nucleon is often assumed to be free, with a Fermi momentum according to the global or local Fermi gas models, or interacting with a nuclear mean field. The description of the initial state in terms of spectral functions for interacting nucleons has also been applied to weak pion production \cite{Rocco:2019gfb}. Given the prevalent role of the $\Delta(1232)$ excitation in pion production, it is not surprising that the in-medium modification of the $\Delta$ propagator~\cite{Oset:1987re} is very important: the main effect is the increase of the resonance width (broadening) by many-body processes. The role of two-nucleon currents, partially considered for pion production \cite{Hernandez:2013jka,Rocco:2019gfb} and photon emission \cite{Chanfray:2021wie} remains largely unexplored for inelastic processes. In their way out of the nucleus, pions undergo final state interactions (FSI). They can be absorbed, change their energy, angle and charge. FSI are often simulated using semiclassical methods in which particles move freely between collisions (cascades) or follow classical trajectories in a mean field \cite{Buss:2011mx}.

The next generation of experiments will hence require considerable efforts towards a more precise modeling of neutrino interactions in the resonance regime and beyond.  A prerequisite for the understanding of nuclear cross sections as they appear in the detectors is a thorough understanding of the elementary process on 
the nucleon, that necessarily has to be appropriately constrained by data.  
Even for processes on the nucleon our knowledge is limited  in the axial sector, and the influence of the nuclear medium is largely unexplored especially for more convoluted reaction mechanisms. The strong convolution of interaction mechanisms in data brings along the need to address a considerable  number of issues with strong priority, as will be outlined in detail in section \ref{pathforward}.

\subsection{Quark-Hadron duality}

The transition from resonant/non-resonant production to DIS is marked by increasing $W$, which in turn corresponds to growing $Q^2$, and naturally evolves into scattering off the quark in the nucleon that can be described by perturbative QCD.  On the way to this QCD-described scattering region there is a significant contribution from the  non-perturbative QCD regime. This is a very complex kinematic transition region, encompassing interactions that can be described in terms of hadrons as well as quarks, that should be well-described by the application of quark-hadron duality~\cite{Bloom:1970xb} where baryonic resonant and non-resonant processes behave on average like DIS in similar $Q^2$ and $W$ regions.  

To further define the concept of duality, consider that perturbative QCD is well defined and calculable in terms of asymptotically free quarks and gluons, yet the process of confinement ensures that it is hadrons, pions and protons, that are observed.    One speaks the language of quarks/gluons in the DIS region and, as $W$ decreases, transitions to speak the language of hadrons in the SIS region that includes both resonant and non-resonant pion production.  Duality can then be considered as a conceptual experimental bridge between free and confined partons.   The resonances can  be considered as a continuing part of the behavior observed in DIS, which would suggest there is a connection between the behavior of resonances and QCD, perhaps even a common origin in terms of a point-like structure for both resonance and DIS interactions.  

More formally, in the 1970's, Bloom and Gilman~\cite{Bloom:1970xb} defined duality by comparing the structure functions obtained from inclusive electron-nucleon DIS scattering with resonance production in similar experiments and the observation that the average over resonances is approximately equal to the leading twist contribution measured in the DIS region. 
That is the DIS scaling curve extrapolated down into the resonance region passes through the average of the "peaks and valleys" of the resonant structure.   It is important to recall that the understanding of this SIS region is critical for long-baseline oscillation experiments where, for example, in the future DUNE experiment
 around 50\% of the interactions will be in these SIS and DIS regions with $W$ above the mass of the $\Delta$ resonance.

\begin{figure}[h]
\begin{center}
\includegraphics[width=0.55\textwidth]{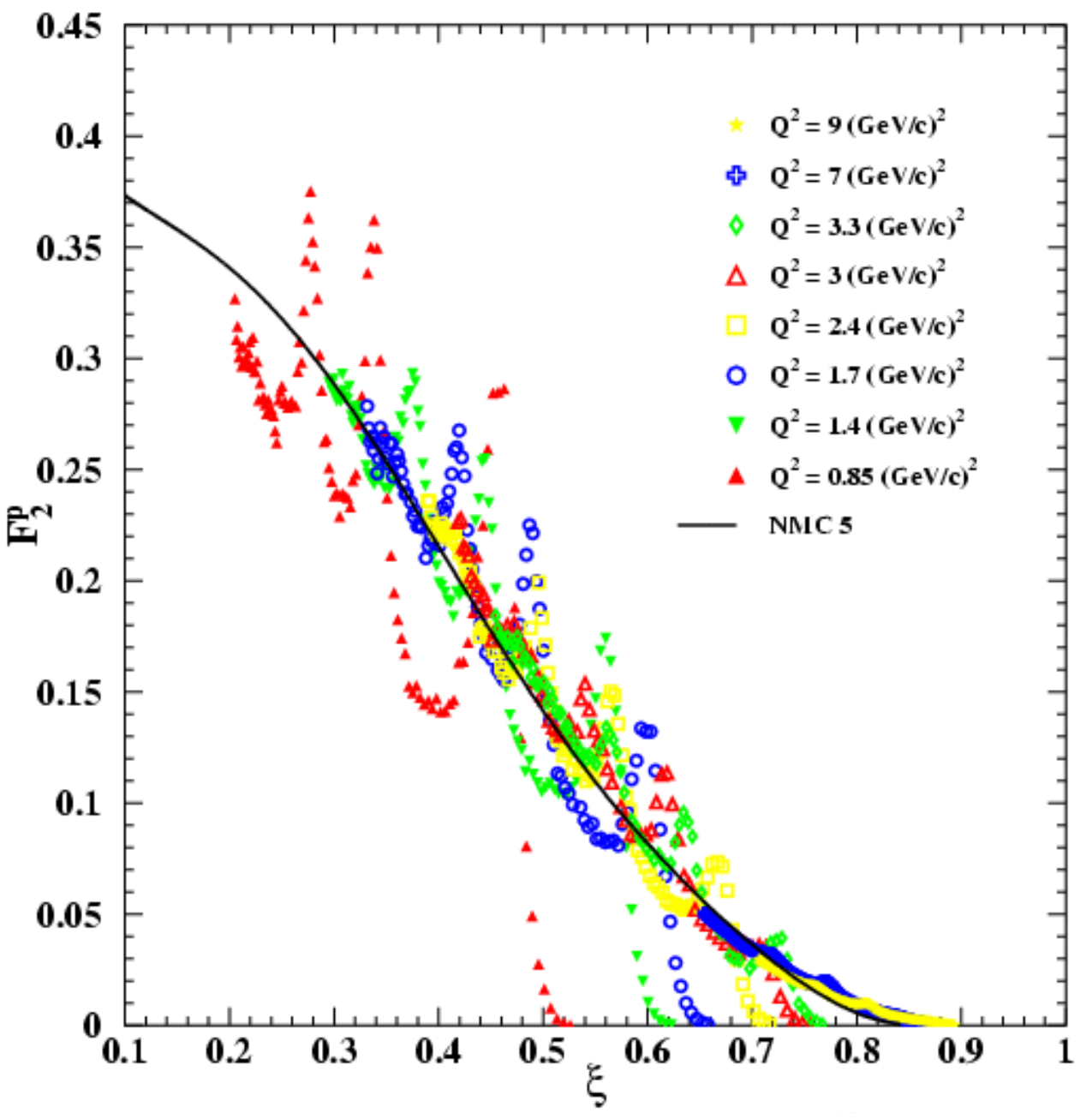}
\caption{ Comparison of $F_2^p$ from the series of resonances measured by E94-110 vs the Nachtmann variable $\xi$ at the indicated $Q^2$ compared to the extrapolated DIS measurement from the NMC collaboration at 5 GeV$^2$ }
\label{eN_duality94110}
\end{center}
\end{figure}

This higher-$W$ SIS region between the $\Delta$ resonance and DIS has been quite intensively  studied experimentally in electron/muon-nucleon (e/$\mu$-N) interactions and somewhat less thoroughly in electron/muon-nucleus (e/$\mu$-A) scattering. The studies of e/$\mu$-N interactions in this kinematic region have been used to test this hypothesis of quark-hadron duality.   An early Jefferson Lab measurement (E94-110)  
showed that global duality was clearly observed for $Q^2 \ge 0.5$~GeV$^2$, as can be seen in Fig.~\ref{eN_duality94110}, with resonances following the extrapolated DIS curve.  

A quantitative description of how well duality is satisfied can be accomplished by defining the ratio of integrals of structure functions, over the same  interval in the Nachtmann variable $\xi (x, Q^2) = {2x}/{[1+\sqrt{1+4x^2M_N^2/Q^2}]}$, from the resonance (RES) region and DIS region.  To keep the same $\xi$ interval in the higher $W$ DIS region compared to the lower $W$ RES region requires a different $Q^2$ for the RES and DIS regions, thus the indexing of $Q^2$ in the ratios.  This method tests local duality within the integrals limits and for perfect quark-hadron local duality the value of the ratio would be 1.0.
\begin{equation}
{\cal I}_j (Q^2_{RES}, Q^2_{DIS}) = \frac{\int_{\xi_{min}}^{\xi_{max}} d\xi F_j^{RES}(\xi, Q^2_{RES})}
{\int_{\xi_{min}}^{\xi_{max}} d\xi F_j^{DIS}(\xi, Q^2_{DIS})}\
\label{dratio}
\end{equation}

Unfortunately, the experimental study of duality with neutrinos is very restricted since the measurement of resonance production by $\nu$-N interactions is confined to rather low-statistics data obtained in hydrogen and deuterium bubble chamber experiments from the 70's and 80's.  Attempting to study duality with experimental $\nu$-A scattering is also restricted due to very limited results above the $\Delta$ resonance in the SIS region. A recent NuSTEC workshop (\href{https://indico.cern.ch/event/727283/overview}{NuSTEC SIS/DIS Workshop})~\cite{Andreopoulos:2019gvw} concentrating on this SIS region with neutrino-nucleus interactions emphasized the considerable problems facing the neutrino community in this transition region.  Since there are no high-statistics experimental data available across the SIS region, $\nu$-N and $\nu$-A scattering duality studies are by necessity limited to theoretical models. Yet even the theoretical study of $\nu$-N/A duality is sparse with only  several full studies in the literature \cite{Lalakulich:2006yn,Lalakulich:2009zza} and references therein.

For lepton-A interactions, the GiBUU and Ghent groups have used their respective resonance models to evaluate duality.  The main difference in the models is that GiBUU~\cite{Kaskulov:2011pr}
 uses a resonance model that includes single- and multi-$\pi$ decays plus heavier decay states while the Ghent model~\cite{Gonzalez-Jimenez:2016qqq} concentrates on 1$\pi$ decays.

\begin{figure}[h]
\begin{center}
\includegraphics[width=0.95\textwidth]{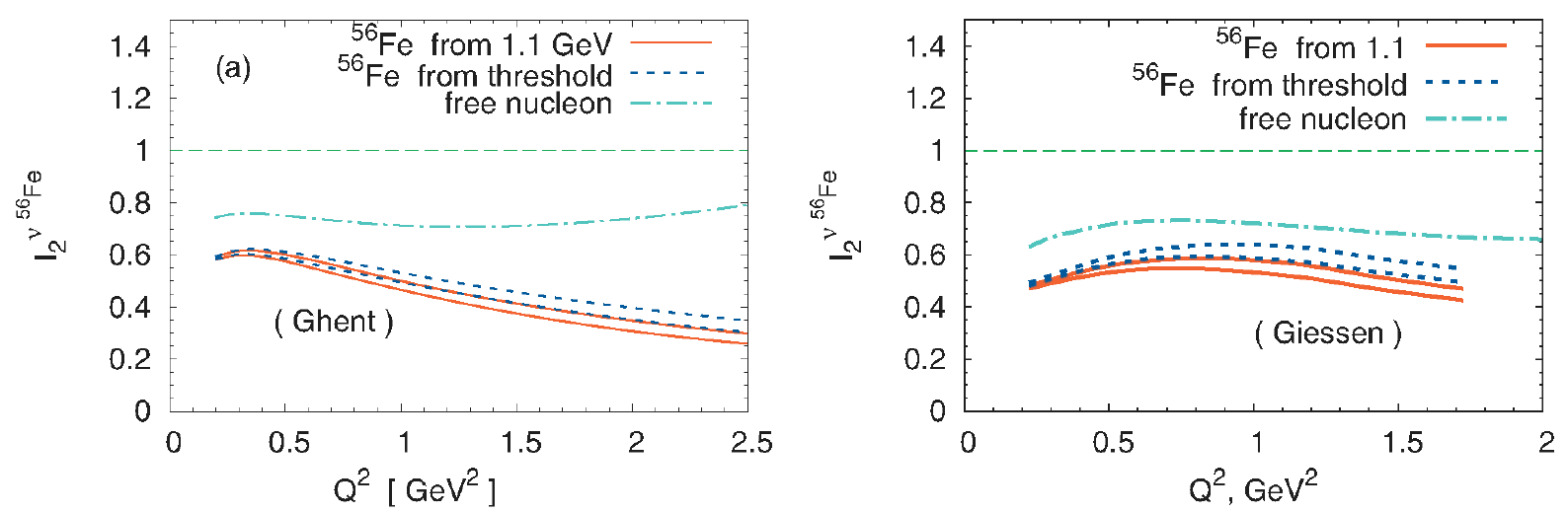}
\caption{Figure from~\cite{Lalakulich:2009zza}: Ratio I$_2^{\nu Fe}$ for iron calculated within the Ghent~\cite{Praet:2008zz} (left) and Giessen~\cite{Mosel:2017nzk} (right) models. For Fe the results are displayed for two choices of the lower limit of the numerator in the integral $I_j$: $W$ = 1.1 GeV (solid line) and "threshold" that takes into account the Fermi motion within the Fe nucleus (dotted line). For each of these two choices they used two sets of DIS data in determining the denominator of the integral I, one at $Q^2_{DIS}$ = 12.59 GeV$^2$ and the other at 19.95 GeV$^2$.  The ratio I$_2^{\nu N}$ for the free nucleon (dash-dotted line) is shown for comparison}
\label{fig_I2nuFe}
\end{center}
\end{figure}

 They observed as in Fig.~\ref{fig_I2nuFe} 
 that the\emph{ computed} integrated resonance strength is about half of the measured DIS one. 
 Specifically, they found for nuclei such as Fe ratios of 0.6 for electro-production and 0.4 for neutrino production. This could point towards a scale dependence in the role of the nuclear effects suggesting that nuclear effects act differently at lower $Q^2$ (resonance regime) than at higher $Q^2$ (DIS regime).  This would not be surprising.
 
On the other hand, the contributions of the non-resonant background was ignored in these analyses. It was stressed that 
a theoretical or phenomenological model for the non-resonant background across the entire resonance region will be required.  
 
As more data with better precision become available on inclusive lepton scattering from nucleons and nuclei a verification of quark-hadron duality with sufficient accuracy would provide a way to describe lepton-nucleon and lepton-nucleus scattering over the entire SIS region  
and give an indication of how well current event simulators are modeling the SIS region.  If the application of duality to our event generators can help us with this understanding, it should be explored.

\subsection{Deep Inelastic Scattering}
The shallow inelastic region covers resonance excitation on the nucleon that, together with a non-resonant continuum and the interference between them yields the total inelastic cross section.
 As $Q^2$ grows and surpasses  $\sim 1$ GeV$^2$, non-resonant interactions begin to take place with quarks within the nucleon indicating the start of deep inelastic scattering (DIS).
In this kinematical region, the cross section may be written in terms of DIS structure functions, which are described via QCD factorization theorems. These theorems permit an explicit separation into short-distance kernels that are perturbatively calculable and long-distance correlations quantified by the nonperturbative PDFs. Presently, there is no sharp kinematic boundary on $W$ and $Q^2$ for the onset of deep inelastic scattering in literature. Generally $Q^2 > 1 $GeV$^2$ is chosen for the onset of DIS and a kinematic
constraint of $W >$ 2 GeV is also applied to describe the DIS region by minimizing contributions from resonant states.

 For DIS from nuclear targets, the cross sections are expressed in terms of nuclear structure functions. In the weak sector, theoretically these nuclear structure functions have been studied by mainly two groups, one Kulagin and Petti~\cite{Kulagin:2007ju, Kulagin:2004ie}, and the  Aligarh-Valencia group~\cite{Zaidi:2019asc,Zaidi:2019mfd, Haider:2016zrk, Haider:2015vea, Ansari:2021cao} the other. Significantly, the nucleon structure functions are the basic inputs in the determination of nuclear structure functions and the scattering cross section. Therefore, a proper understanding of the nucleon structure functions becomes quite important~\cite{Zaidi:2019asc, Ansari:2020xne}.

 Especially in the low- and moderate-$Q^2$ region(s), there can be a nontrivial interply between perturbative and nonperturbative QCD effects. In the case of the former, the ultimate cross section can exhibit strong dependence on the chosen perturbative order, which for contemporary DIS calculations, runs from leading-order accuracy to NLO, and, more recently, up to NNLO; this theoretical accuracy applies at the level of the perturbatively-calculable matrix elements as well as the computed PDFs and DGLAP evolution kernels. At the same time, nonperturbative effects are also important in this region, including target mass corrections (TMCs)~\cite{Brady:2011uy,Schienbein:2007gr,Georgi:1976ve}, which arise due to the inherently nonperturbative nature of the hadronic bound-state mass, and dynamical higher twist effects (HTs)~\cite{Ellis:1982cd}, which stem from multi-parton correlations within the target hadron. These nonperturbative effects are particularly important at high $x$ and low $Q^2$. The HT effects have been constrained phenomenologically through attempts to fit, {\it e.g.}, twist-4 PDFs at fixed perturbative order, as in Ref.~\cite{Accardi:2016qay}. As such, disentangling higher-order perturbative corrections from HT corrections is a serious and nontrivial undertaking. We also stress that further development of the perturbative QCD aspects of charged-current DIS is needed; these include studies of the consistent implementation of heavy-quark effects into perturbative structure-function calculation --- a consideration which has been shown to enhance perturbative stability~\cite{Gao:2021fle}. In the end, it will be imperative to carry out further consistency studies relating the SIS-region nonperturbative background to formulations of the charged-current DIS structure functions based on rigorous (non)perturbative QCD. This activity is a priority for achieving target precision(s) at DUNE/LBNF for $E_\nu\! \sim\! \mathrm{few\, GeV}$.

\begin{figure}
    \centering
    \includegraphics[height=12 cm, width=0.95\textwidth]{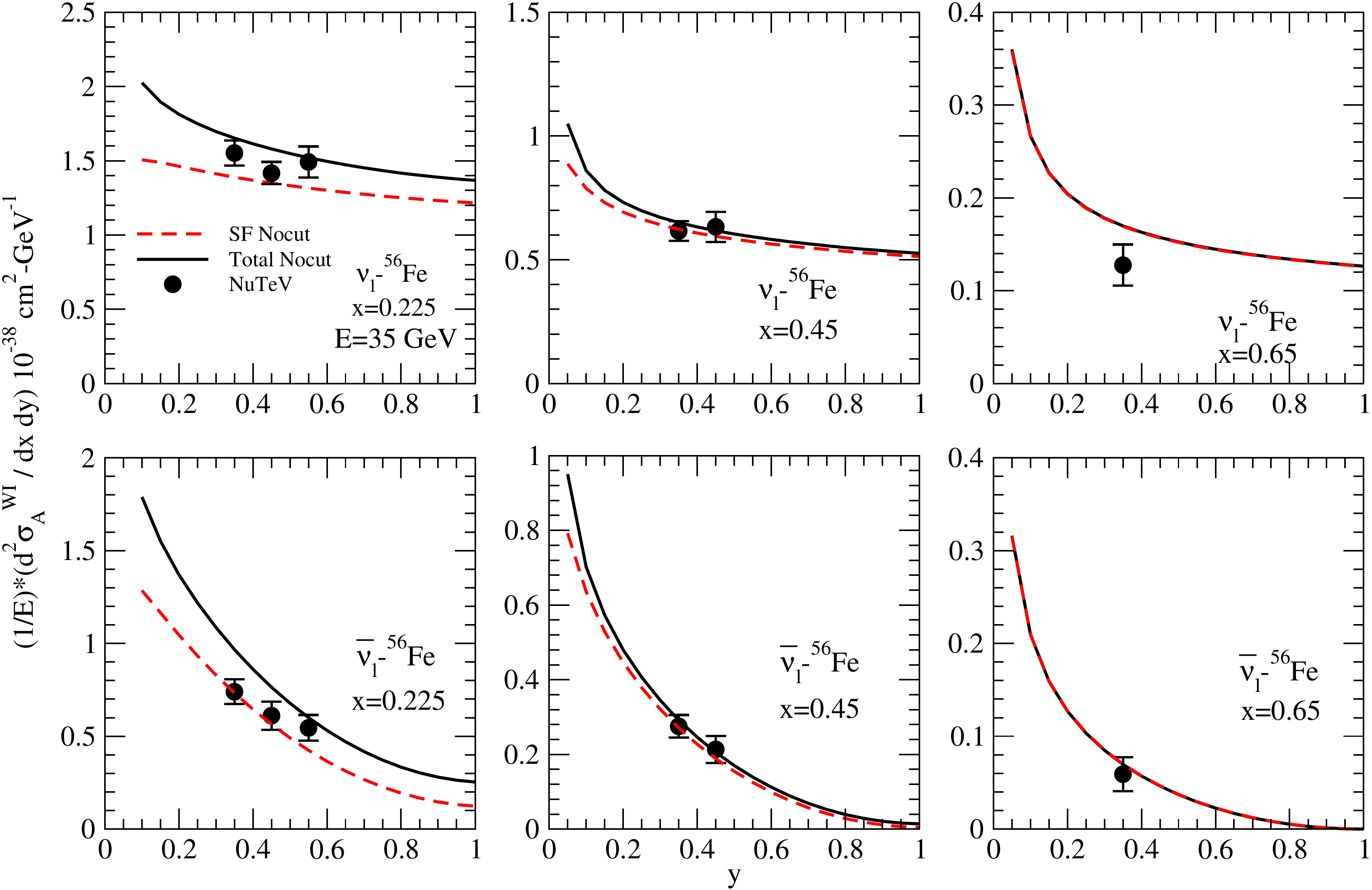}
    \caption{Differential cross section vs $y$ for the different values of $x$ for the incoming beam energy of $E_\nu$=35GeV for $\nu-^{56}$Fe DIS (Top panel) and  $\bar\nu-^{56}$Fe DIS (Bottom panel) processes. Theoretical results are shown with the spectral function only(dashed line) and with the full model by including also mesonic effect and (anti)shadowing corrections (solid line) at NNLO. Solid circles with the error bars are limited experimental data points of NuTeV. }
    \label{fig:fig8}
\end{figure}

 The Aligarh-Valencia group has studied nuclear medium effects in the structure functions in a microscopic model which uses relativistic nucleon spectral functions to describe the target nucleon momentum distribution incorporating the effects of Fermi motion, binding energy and nucleon correlations in a field theoretical model. In Fig.\ref{fig:fig8}, the theoretical results of the Aligarh-Valencia group for the (anti)neutrino differential cross sections at $E_\nu$=35 GeV are presented. These results are shown with the spectral function only and with the full model by including also meson-cloud corrections from a specific ansatz and (anti)shadowing corrections at NNLO, where it can be observed that the mesonic contributions play important role in the region of $0.2 \le x \le 0.5$. 

 Neutrino scattering plays an important role in the phenomenological QCD analysis of DIS since the weak current has the unique ability to probe specific quark currents within the target nucleon or nucleus, thus helping to resolve the flavor dependence of the nucleon’s constituents.
 This significantly enhances the study of parton distribution functions and complements studies with electromagnetic probes. However, as helpful as this ability of the weak-interaction may be, it should be emphasized that all high-statistic neutrino experiments have had to use heavier nuclear targets. This means the PDFs extracted from these experiments are for nucleons in the nuclear environment and are thus \textit{nuclear} parton distribution functions (nPDF). There is considerable difference between these A-dependent nPDFs and the free nucleon PDFs. Furthermore, since the relevant nuclear effects could involve multiple nucleon scattering as in shadowing or scattering from correlated nucleon pairs as possibly in the EMC effect these nPDFs might better be considered {\it effective} nPDFs and not necessarily the PDFs of single bound nucleons.  

Although it has been emphasized that neutrino DIS scattering could be a particularly rich source for flavor separation in determining free proton parton distribution functions, a serious problem  is the very poor state of knowledge of $\nu$-\emph{free nucleon} interactions.  There are presently only low-statistics bubble chamber results from the 1970's and 1980's that have relatively large statistical and systematic errors.  This severely limits the impact of neutrino scattering in free nucleon PDFs.  That these rather imprecise results are then used as the start of neutrino interaction simulations by the current community's event generators is also a matter of real concern.

\begin{figure*}[t]
    \centering
    \includegraphics[width=0.55\textwidth]{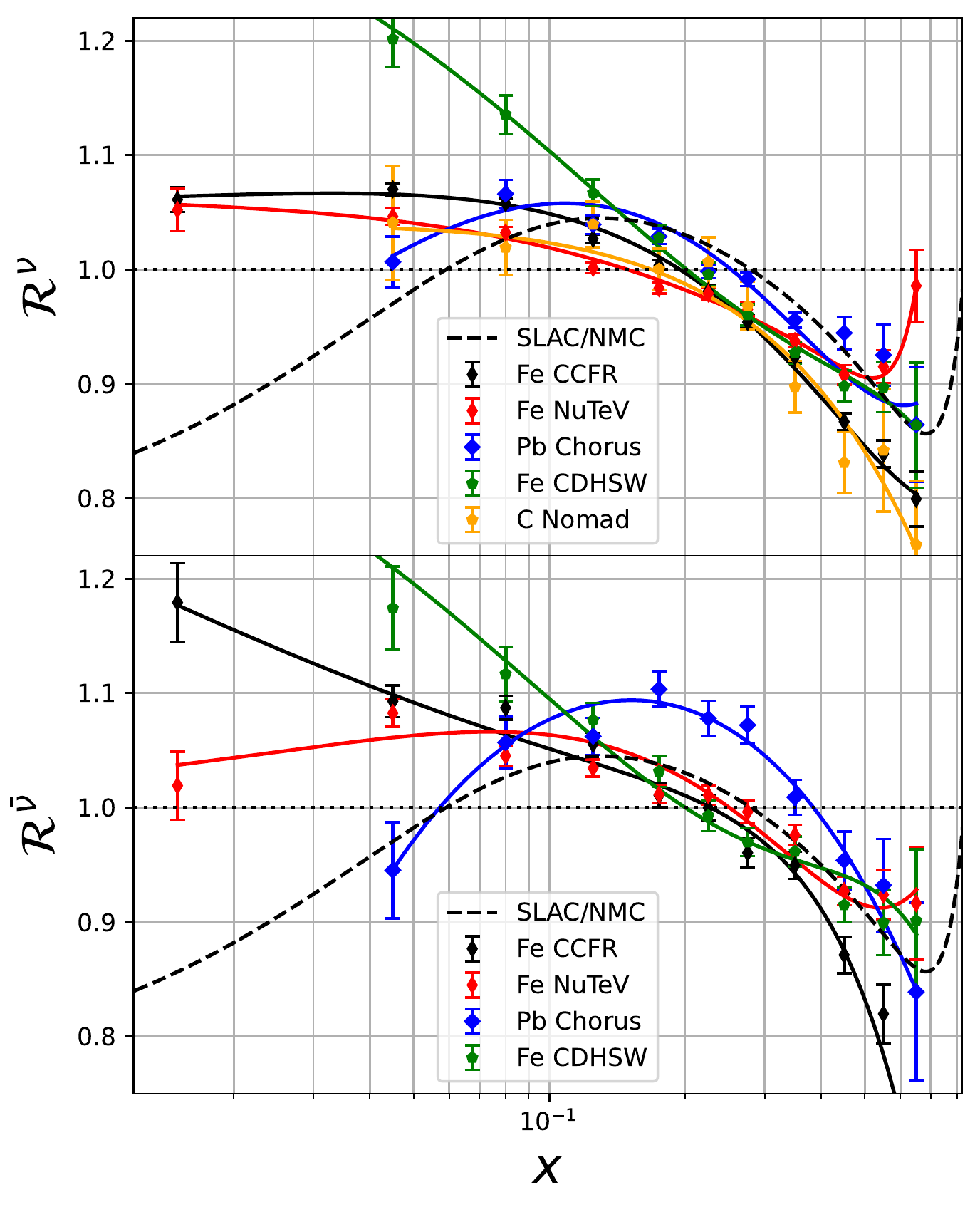}
    \caption{ PRELIMINARY figure from an nCTEQ paper in preparation. The weighted average of the cross section ratio for $4\leq Q^2<30$ GeV$^2$ and $W^2>12.25$ GeV$^2$ from NuTeV, CCFR, Chorus, CDHSW, and NOMAD. The denominator ($\sigma_{free}$) are computed using CTEQ6M nucleon PDFs.
    }
    \label{fig:Rnunubar}
\end{figure*}

The NuTeV, CCFR, CDHSW $\nu/\nubar$-Fe and CHORUS $\nu/\nubar$-Pb experiments are the most recent high-statistics DIS experiments that  have published double-differential $\nu/\nubar$-A scattering cross sections as well as very detailed studies of systematic errors.   
Using the results from these experiments, nuclear effects of charged current deep inelastic $\nu/\nubar$-A scattering were studied by the nuclear CTEQ (nCTEQ) collaboration \footnote{Refer to \url{https://ncteq.hepforge.org} for details of the nCTEQ collaboration} in the framework of a $\chi^2$ analysis and, in particular, a  set of iron and lead nuclear correction factors for the structure functions were extracted as in Fig.\ref{fig:Rnunubar}.
Comparing these results with structure function correction factors for $\ell^\pm$-Fe scattering it was determined that the neutrino correction factors differ in both shape and magnitude, particularly at small $x$, from the correction factors for $\ell^\pm$-Fe scattering. 

This difference, although not unexpected theoretically especially in the shadowing and antishadowing regions, is not universally seen by all groups examining nPDFs of neutrinos.  It is imperative that we carefully consider these contrasting results and gain an understanding of the $\nu$-A nuclear correction factors.  

However the results from a much wider variety of nuclear targets in a neutrino beam, able to access DIS kinematics, will be needed to definitively answer this question. 
Steps in this direction could be achieved at the upcoming neutrino experiments at the LHC, which will utilize the intense and strongly collimated beam of TeV energy neutrinos of all three flavors that are produced in LHC collisions to study neutrino interactions at even human-made energies. With FASER$\nu$ and SND@LHC, two emulsion-based detectors with tungsten targets will already start their operation in 2022~\cite{FASER:2019dxq, SHiP:2020sos}. A continuation of this program during the HL-LHC era with significantly increased event rates has been proposed in the context of the Forward Physics Facility~\cite{Anchordoqui:2021ghd, Feng:2022inv}. This proposal includes three dedicated neutrino detectors: a liquid argon based detector, FLArE, an electronic neutrino detector, AdvSND, and an emulsion based neutrino detector, FASER$\nu$2. While FLArE would be able to test the structure functions for argon that are relevant for DUNE, both FASER$\nu$2 and AdvSND would have the ability to change the target material and collect data for a variety of nuclear targets. In addition, the emulsion based experiments will be sensitive to heavy quark flavors, providing the opportunity to study the strange quark content of the proton via charm associated neutrino interactions.

\subsection{Hadronization}
Hadronization is not described by a fundamental theory such as perturbative QCD, but it is based on phenomenological models~\cite{Webber:1983if,Andersson:1983ia}. Usually in scattering experiments, the energy and direction of  the incoming probe particle are known. By measuring the outgoing  particle's energy and scattering angle, the kinematics of interaction, namely energy and three-momentum transfer $|{\bf q}|$ are then determined and the interaction kinematics is fixed. Similarly, other kinematic variables, including  $Q^2$, $W$, Bjorken variable $x$, and inelasticity $y$ are determined. This is however not the case for neutrino scattering experiments. The incoming neutrino's direction is known, but the neutrino beam is often wideband and the energy is not precisely known. Thus,  measuring the  outgoing charged lepton's energy and scattering angle does not suffice to determine the  neutrino's energy and the energy transfer unless all outgoing particles, including hadrons, are measured. For heavy nuclear target experiments, which include all neutrino scattering experiments, FSI  prevent direct connections between the observed hadrons and the  hadrons produced by the primary neutrino interaction. In this situation, experimentalists heavily rely on simulations to interpret the available  hadron information and reconstruct the vertex kinematics. Thus,  hadronization models constitute an essential input for current and future neutrino experiments  in the SIS and DIS regions, including DUNE and all other atmospheric and high-energy astrophysical neutrino experiments. A precise hadron measurement is the target of next generation neutrino experiments~\cite{Chukanov:2016lra,Adams:2018fud,NINJA:2020gbg,NINJA:2020bvx}. 

The PYTHIA hadronization package~\cite{Sjostrand:2006za,Sjostrand:2007gs} is based on the Lund string model~\cite{Andersson:1983ia} and is adopted by all neutrino oscillation experiments. Here, confinement of partons is modeled by a relativistic one-dimensional string which represents a color flux between a quark and an anti-quark. The hadronization process is described by breaking up these strings to produce more quark-antiquark pairs. PYTHIA has many parameters to be tuned, and the default scale for PYTHIA6 is $\sim 35$~GeV, whereas the default scale of PYTHIA8 is even higher. In fact, basic assumptions in these hadronization models break down at low energy, and neutrino experiments around 1-10 GeV must rely on other methods to produce hadrons. In event generators, PYTHIA is used for $W\gtrsim 2$ GeV, but often extended to even lower energies. 

\begin{figure*}[t]
    \centering
    \includegraphics[width=0.45\textwidth]{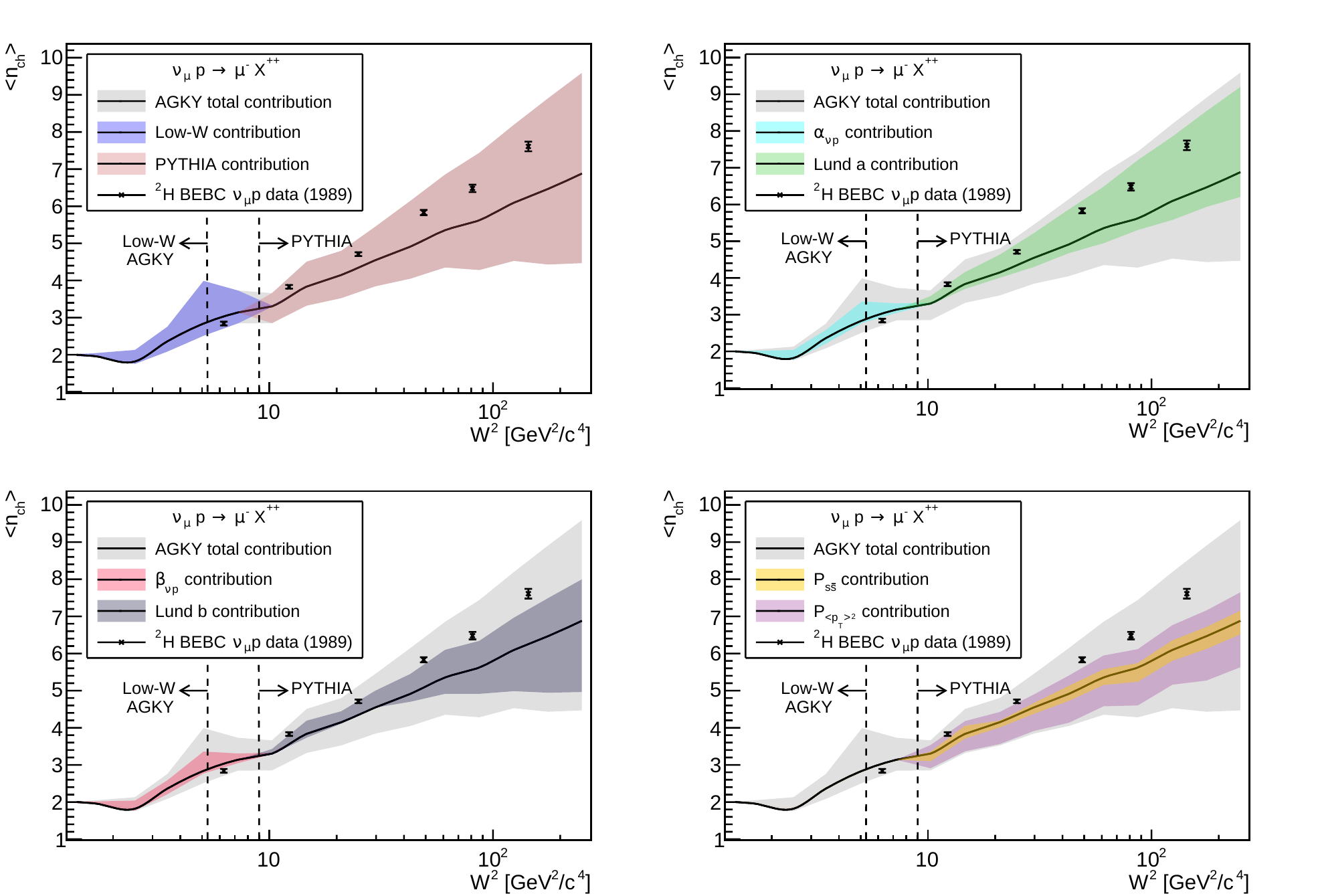}
    \includegraphics[width=0.49\textwidth]{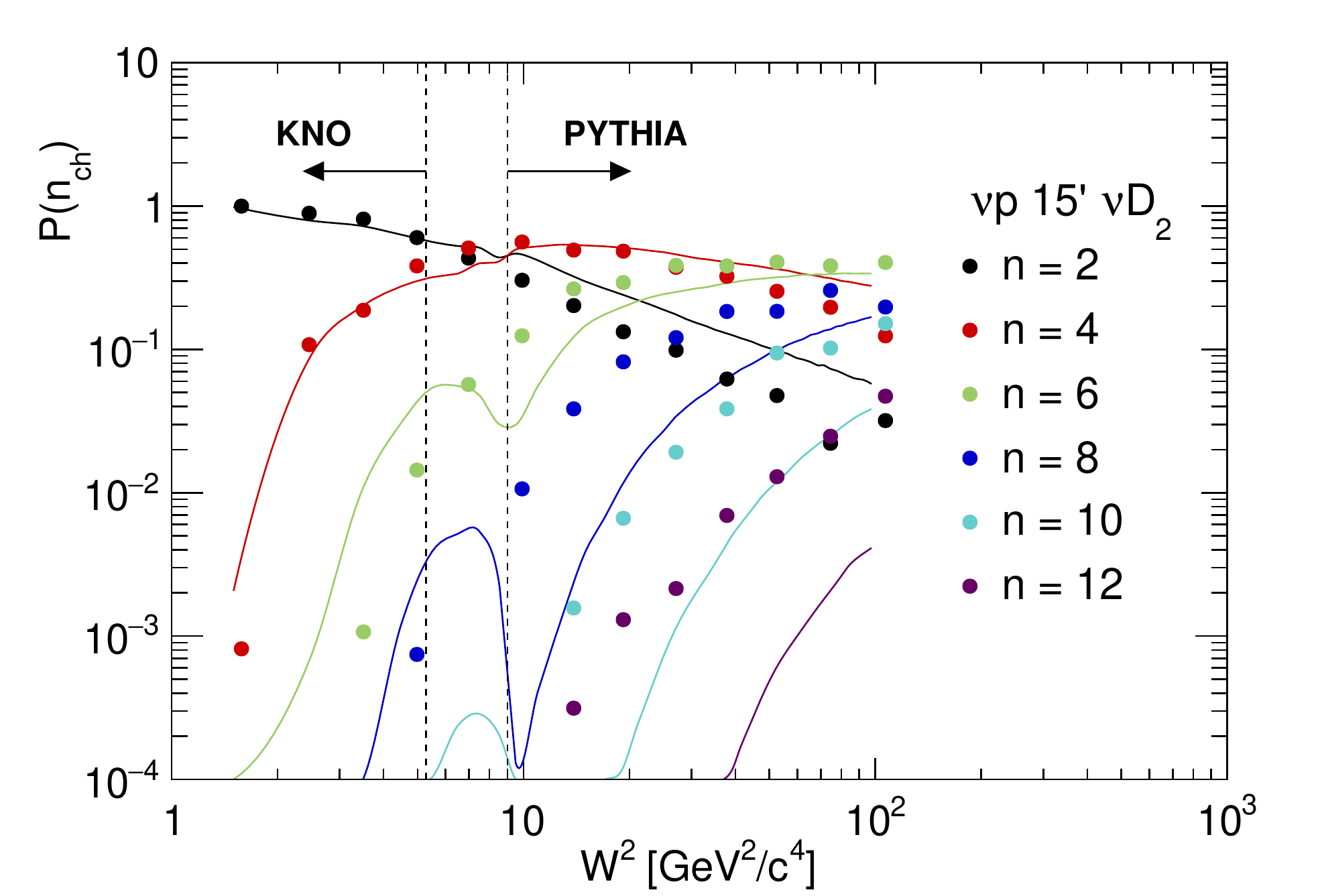}
    \caption{GENIE hadron multiplicity simulation (v3.0.6, tune \texttt{G18\_02a\_02\_11a}). The left figure shows an averaged charged hadron multiplicity predicted by GENIE including systematic errors. The right figure shows normalized topological cross sections. The GENIE predictions are compared with  data from the Fermilab 15' deuteron bubble chamber with $\nu_\mu$ beam~\cite{Zieminska:1983bs}.
    }
    \label{fig:had}
\end{figure*}

In the SIS region ($W\lesssim 2$~GeV) neutrino interaction generators have to use custom hadronization models. A popular approach is to extract averaged charged hadron multiplicities from external bubble chamber data ~\cite{Yang:2009zx,Bronner:2016gmz,GENIE:2021wox}. Isospin symmetry is used to produce the averaged neutral pion multiplicity. Then this model is smoothly connected to PYTHIA at given $W$.  The left panel of  Fig.~\ref{fig:had}  shows an example of such an averaged multiplicity prediction from GENIE~\cite{GENIE:2021wox}. The transition from the low-$W$ to the high-$W$ model is governed by the AGKY model~\cite{Yang:2009zx}. The problem is seen in the high-$W$ region predicted by PYTHIA6. First, PYTHIA6's default prediction cannot describe the bubble chamber data. Second, bubble chamber data often lack systematic uncertainties,  leading to many tensions~\cite{Kuzmin:2013tza,Katori:2014fxa}. To accommodate this, the hadronization model needs to have a large systematic error~\cite{GENIE:2021wox}. 
It is also important to simulate the dispersion of the hadron multiplicities. In the low-$W$ region, the dispersion is extracted from bubble chamber data with the empirical KNO scaling law~\cite{KOBA1972317}. This allows one to make accurate event-by-event hadron multiplicity simulations. At high-$W$, the dispersion is also simulated by PYTHIA. However, as can be seen in the right panel of Fig.~\ref{fig:had}, topological cross sections are not smoothly connected in the low-$W$ and high-$W$ transition region. This indicates that the  dispersion provided by the low-$W$ and high-$W$ hadronization models is different. Since the low-$W$ hadron multiplicity dispersion is extracted from bubble chamber data, this discontinuity also means that the dispersion predicted by default PYTHIA6 around $\le 10$ GeV is incompatible with neutrino bubble chamber data. At this moment, event-by-event hadron simulation is difficult in this energy region even at nucleon level. 

\section{Neutrino event generators}
\label{sec:generator}

Across the broad range of energies of interest for current and future
investigations of neutrino physics, realistic simulations of neutrino
interactions are a critical ingredient for the design, execution, and
interpretation of experimental analyses. These simulations are generally
carried out using Monte Carlo techniques implemented within computer programs
known as \textit{event generators}. Advances in our collective theoretical
understanding of neutrino interactions, such as those described in the previous
sections, will be essential to the progress of the field. However, unless these
advances are appropriately translated into improvements to neutrino event
generators, the benefit of theory efforts on experimental precision will be
severely limited at best.

Current capabilities and technical needs for neutrino event generators,
including physics content and computational methods, are considered at length
in a separate white paper~\cite{HEPGeneratorWP}. This section supplements
that material with a discussion of
\begin{enumerate*}[label=(\Alph*)]
\item the need to better organize and strengthen support for generator
development and related theoretical work, and 
\item experimental needs in light of how generator predictions enter into
analyses. 
\end{enumerate*}

\subsection{Organizational needs}

While the development of neutrino event generators is technically demanding,
sociological and organizational challenges are currently the greatest
hindrances to progress. Participation in generator-related activities is poorly
incentivized for both theorists and experimentalists, and opportunities to
pursue neutrino generator development as one's primary research activity are rare. As a result, new work on neutrino event generators is
largely driven by idiosyncratic short-term needs of individual experiments and
interests of small theory groups. A need for greater coordination and
prioritization of such activities is widely recognized in the neutrino
scattering community. Despite promising initial discussions that have taken
place in a series of recent
workshops~\cite{ECTStarWorkshop2018,ECTStarWorkshop2019,FNALWorkshop2020,
Barrow:2020gzb}; however, neither a clear leadership structure nor significant
institutional support to carry out the related work have yet emerged. The
present reliance on piecemeal solutions emerging spontaneously from the
community is unsustainable in light of the accelerating demands of precision
experiments.

A notable challenge for neutrino event generator development work is the wide
range of required expertise, which is cross-cutting along multiple dimensions.
Open questions in the field of high-energy physics, such as those related to
neutrino mass, are a main motivation for improving the quality of neutrino
interaction simulations. Achieving those improvements, however, requires a deep
knowledge of nuclear physics. The cooperation needed for success in neutrino
generator development cannot occur to the extent that funding agencies impose
rigid boundaries between these two domains. Participation is needed from
theorists, experimentalists, and computing experts to ensure that neutrino
generators
\begin{enumerate*}[label=(\arabic*)]
\item reflect our best understanding of the underlying scattering physics (and
associated uncertainties),
\item are responsive to experimental needs and new data sets, and
\item adopt best practices for scientific software development and
user support.
\end{enumerate*}
Experience and tools from other subfields, notably simulation efforts for
collider physics, are currently underused and should be explored more
thoroughly by the neutrino generator community.

While the big-picture need to improve neutrino event generators has been
appreciated for some time, there is not yet clear consensus on the specific
theory improvements that should be prioritized. This is in part due to the
complexity of the theoretical problem that must be solved: neutrino experiments
are sensitive to many details of generator predictions, but a unified,
first-principles description of all relevant neutrino scattering phenomena is
unavailable. Competing theoretical approaches make different approximations and
vary in their domain of validity. Generator authors are thus obliged to stitch
together multiple models to achieve complete simulations. This is ideally done
in collaboration with theorists to minimize inconsistencies, but incentives for
their direct involvement are currently poor.

It is also presently difficult to fully quantify the level of improvement
needed in various aspects of the theoretical models implemented in generators.
In future accelerator-based neutrino oscillation experiments, for example,
percent-level control of all systematic uncertainties (including those related
to interaction modeling) will be needed to obtain definitive measurements of
neutrino properties~\cite{NuSTEC:2017hzk}. Cross-section calculations for
neutrino scattering on complex nuclei typically have theoretical uncertainties
of $O(10\%)$ or larger, but modern experiments routinely apply data-driven
constraints to achieve higher precision. Novel, highly sophisticated techniques
for constraining neutrino cross-section uncertainties will be enabled by the
DUNE near detector complex, but deficiencies in the interaction model cannot be
completely removed~\cite{DUNE:2021tad}.
On a practical level, there is also a need to further develop global analyses of the relevant neutrino and charged-lepton scattering data, including with more comprehensive model and parametric uncertainties. In this respect, expertise in the larger HEP community related to QCD analyses will be instructive.
Support for closer collaboration between theorists and experimentalists will allow the best return on investment
of effort in both spheres: experimental constraint strategies can be designed
to be robust against ``known unknowns'' in our understanding of the relevant
nuclear physics, and theoretical investigations can become more targeted
towards those modeling details which are most poorly constrained.

While there is widespread agreement in the neutrino generator community on the
need for specific technical improvements (e.g., a standardized output
format~\cite{Barrow:2020gzb}) and on the overall importance of achieving
higher-quality simulations, many other issues remain controversial. Some
long-standing disagreements involve tradeoffs between competing models and/or
between theoretical rigor and computational simplicity (e.g., GiBUU transport
versus intranuclear cascade treatments of hadronic final-state
interactions~\cite[Sec.~4]{Mosel:2019vhx}). These tradeoffs will continue to be
explored in light of new neutrino cross-section measurements and theoretical
progress.

Differing perspectives also exist within the neutrino generator community on
the best strategies for organizing, pursuing, and supporting the needed
development work. Generators created by theory groups emphasize consistency and
the quality of the underlying physics models. While the generator's role as an
aid to theoretical investigation can easily motivate implementation of new
models, there is little incentive to build interfaces with beam and detector
simulations and provide other essential infrastructure used in experimental
workflows. Generators developed primarily by experimentalists provide such
tools out of necessity, but making them compatible with other codes is
labor-intensive, requires maintenance, and potentially dilutes already meager
rewards. Generator development is regarded as service work and must be balanced
against pursuing a physics analysis in order for junior experimentalists to
have a good chance of career advancement. This motivates underinvestment and
pursuit of short-term solutions by experiments. If junior collaborators contributing to essential generator work are unable to find long-term employment in the field (or are dissuaded from doing so), then this situation also poses a significant risk to continuity of expertise.


\subsection{Experimental use cases}
\label{sec:generator_exp_use_cases}

The neutrino interaction models provided by event generators are used in the
context of experimental analyses for a number of vital tasks. Chief among these
for accelerator-based oscillation measurements (but also important for other
applications) is neutrino energy estimation: since neutrino beams are not
monoenergetic, extraction of oscillation probabilities from experimental data
requires the incident neutrino energy to be reconstructed on an event-by-event
basis. For the complex nuclear targets used in contemporary experiments, the
necessary corrections for missing energy are large and highly sensitive to many
aspects of the underlying theoretical calculations. A recent
study~\cite{CLAS:2021neh} that benchmarked standard neutrino energy
reconstruction techniques against electron scattering measurements revealed
major discrepancies, even for a generator-based model that provided a good
description of inclusive electron-nucleus cross sections in the quasielastic
region.

Another key experimental application of neutrino event generators is to
calculate expected event rates which are used to interpret data. In searches
for new physics processes, the expected event rates serve as a reference
prediction from the Standard Model (and effective nuclear theory based upon
it). For measurements of oscillation parameters, the expected event rates
provide the unoscillated spectrum. In both cases, a priori generator
predictions are typically refined by experiments via empirical tuning of model
parameters~\cite{NOvA:2020rbg,uBooNEGENIE,T2K:2021xwb} and data-driven
constraints between separate parts of the apparatus (e.g., detectors at
different distances from the neutrino source) or measurements performed using
different event selections (e.g., $\nu_\mu$ versus $\nu_e$ data). While such
techniques represent a powerful means of detecting and mitigating interaction
mismodeling, the need to properly relate the reference data (used for model
constraints) to a distinct generator prediction (used to interpret the ultimate
result) leads to some degree of unavoidable model dependence even in ideal
circumstances. There is also some risk of ``tuning away'' evidence of new
physics processes that would be noticed if a generator model that required less
tuning were used. Deficiencies in neutrino event generator models are already a
leading source of systematic uncertainty in current oscillation
analyses~\cite{T2K:2021xwb,NOvA:2021nfi,Brdar:2021cgb}, and the needed improvements will only
become more urgent as the size of experimental data sets continues to
grow~\cite{NuSTEC:2017hzk}.

A third class of experimental tasks for which neutrino event generators are
crucial involves corrections for imperfect detector performance. While
\textit{in situ} measurements are routinely used to characterize backgrounds due
to cosmic rays and natural radioactivity, the contribution of neutrino-induced
backgrounds (involving event topologies which are not of interest for a
particular analysis) must be estimated using an event generator prediction.
While simulations of the detector response are obviously crucial for applying
corrections related to inefficiency and finite resolution, these corrections can
also be sensitive to details of the neutrino interaction model. For instance,
the efficiency of an inclusive charged-current event selection (which attempts
to identify all neutrino interactions that produce a particular charged lepton)
may depend to some extent on the expected multiplicity and kinematics of the
hadronic final-state particles. High-quality event generator predictions of
complete final states for all significant interaction modes (and all relevant
target nuclei, including inactive detector components) are thus essential for an
accurate interpretation of neutrino data.

When considering the various experimental use cases for neutrino event
generators, it should be noted that a thorough assessment of theoretical
uncertainties on all aspects of a generator's interaction model is an
indispensable requirement. Neutrino generators developed primarily as an aid to
theoretical studies (e.g., GiBUU) typically do not include software tools to
calculate these uncertainties. This is a major reason why such generators,
despite their strengths in other areas, have not seen widespread adoption in
experimental simulation workflows. In the absence of theoretical guidance
and/or built-in generator support for uncertainty quantification, experimental
collaborations must resort to ad hoc approaches which may not be well-grounded
in theory and which typically require significant investment (and often
duplication) of effort.

\section{Summary and path forward}\label{pathforward}

Current and future oscillation experiments need a better understanding and realistic modeling of neutrino-nucleus scattering. 
To meet these challenges, we need coordinated work by both nuclear
physics and particle physics communities; in theory, experiment, and simulation. Such a commitment is beneficial to both communities to achieve broader scientific goals in multidisciplinary topics.  

Efforts to improve theoretical modeling bring along a strongly growing demand for additional experimental constraints on inputs to theoretical calculations: 
 \begin{itemize}
     \item 
Neutrino-hydrogen/deuteron scattering experiments — Even if photon- electron- and meson-nucleon scattering data provide a priceless input to model neutrino interactions on nucleons, the properties of the axial current at finite $Q^2$ remain largely unknown and experimentally unconstrained.  Lattice QCD may be able to partially fill this gap, but there is also a strong need for new $\nu$H\/D experiments to remove systematic uncertainties and a complete understanding of neutrino-nucleon interactions 
(see $\nu$-H\/D LoI~\cite{Snowmass2021:nu-H} and WP~\cite{HDWP}).

\item Electron-nucleus scattering experiments — Modern neutrino-nucleus models rely on the experience gathered in the description of electron scattering data. 
Precision measurements of inclusive electron-nucleus scattering at a wide variety of kinematics are important for validating nuclear models~\cite{Ankowski:2020qbe}. 
Recent electron scattering measurements~\cite{Benhar:2014nca,JeffersonLabHallA:2018zyx,Dai:2018gch,Murphy:2019wed,JeffersonLabHallA:2020rcp,JeffersonLabHallA:2022cit,adi_ashkenazi_2020_3959538} on various targets (including Ar) indicate sizable discrepancies in the generator models beyond the quasielastic peak~\cite{Ankowski:2020qbe,electronsforneutrinos:2020tbf}. The essential first step is to incorporate the information on the nuclear ground state~\cite{Benhar:2014nca,JeffersonLabHallA:2020rcp,JeffersonLabHallA:2022cit}. Coverage must be extended into the SIS kinematic region and information on the final-state mesons and nucleons should be added~\cite{e4nu17,e4nu18,Ankowski:2019mfd} so that FSI models can be tested. This topic is discussed at length in the Snowmass WP~\cite{Ankowski:2022thw}.
\item
Neutrino-nucleus scattering experiments — Neutrino experiments 
such as MiniBooNE, T2K, NOvA, and MINERvA have published cross-section data mainly for $\text{CH}_n$ and $\text{H}_2\text{O}$ targets in QE region. Limited data on heavier targets (Ar, Fe, Pb) and higher energy processes are also available~\cite{Tice:2014pgu,Mousseau:2016snl,Betancourt:2017uso,Abe:2015biq,NINJA:2020bvx,Adamson:2016hyz,Wu:2007ab,Adamson:2009ju,Lyubushkin:2008pe}. These data offer an opportunity to test nuclear dependent DIS models in neutrinos. 
The SBN program (MicroBooNE, SBND, ICARUS)~\cite{Antonello:2015lea} and ArgonCube~\cite{Abi:2020wmh} can provide Ar cross-section data relevant for the SIS region.  
More extensive experimental studies focusing on meson final states in a broad kinematic range can test our understanding of the neutrino SIS physics as well as FSIs~\cite{PhysRevD.100.072005}. 

\item
  Following the first observation of neutrinos at the LHC~\cite{FASER:2021mtu}, a novel LHC neutrino program is being established with the construction the FASER$\nu$~\cite{FASER:2019dxq} and SND@LHC~\cite{SHiP:2020sos} detectors and its continuation through a dedicated Forward Physics Facility is being proposed~\cite{Anchordoqui:2021ghd}. These experiments will study neutrino-nucleus interactions at TeV energies on different targets (including Ar and W) and provide input in a novel kinematic regime. This will, for example, offer the opportunity to constrain nuclear PDFs, to test the modeling of hadronization inside cold nuclear matter, to probe heavy charm and bottom quark mass effects, and to study properties of tau neutrino interactions with high statistics. This topic is discussed at length in the Snowmass WP~\cite{Feng:2022inv}.
\end{itemize}

Many theoretical topics in neutrino scattering require further study in order to meet the needs of neutrino experiment. We identify the following high-priority topics as essential for study over the next 5-10 years, ordered roughly in the order that they are discussed above:
\begin{itemize}

    \item For CE$\nu$NS, the cross section should be mapped as a function of the neutron number. The cross section is now measured by COHERENT using CsI and Ar targets, these should be extended to both heavy and light nuclei, for example Ne, Ge, and Xe. Understanding this behavior as a function of neutron number will be important in calibrating the prediction to the Standard Model and for measuring the neutron nuclear structure factors. The experimental CE$\nu$NS program is described in more detail in Ref.~\cite{CEvNS_WP}. 
    
    \item It will be additionally important to measure the CE$\nu$NS cross section as a function of neutrino energy. This will require detection with neutrinos from nuclear reactors as well as astrophysical sources. Measuring the cross section as a function of neutrino energy will test the (tree-level) Standard Model result that the cross section is independent of neutrino flavor, and at the same time provide additional information that, together with the dependence on the nuclear target, can be used to disentangle a potential new-physics contribution from nuclear effects. Any differences in the cross section between flavor components that go beyond the expected small radiative corrections would represent a hint of physics beyond the Standard Model.  
    
    \item Measurement of the angular distribution of the CE$\nu$NS cross section may also provide information on physics beyond the Standard Model. For example, scalar and vector-like interactions predict differences in the outgoing angular nuclear recoil distribution. 

  \item{Lattice QCD calculations of nucleon elastic axial form factors with few-percent precision are achievable with present techniques and computing resources and will provide valuable input to nuclear many-body calculations. Providing results with complete error budgets including physical quark mass, continuum, and infinite-volume extrapolations is a near-term priority.}

  \item{Extend lattice QCD calculations of few-nucleon electroweak matrix elements to control the systematic uncertainties present in exploratory calculations and determine vector and axial two-nucleon form factors that can be used to determine LECs for two-body currents in nuclear effective theories.}

  \item{Extend exploratory lattice QCD calculations of $N\pi$ scattering as well as resonant and non-resonant $N\rightarrow N\pi$ and $N\rightarrow \Delta$ transition form factors. Vector-current form factors can be used for validation while axial-current form factors would provide valuable predictions.}

  \item{Lattice QCD can predict aspects of nucleon and nuclear PDFs, the hadron tensor, and other structure functions relevant to neutrino DIS. As calculations with controlled systematic uncertainties become available, they should be used to augment experimental data in global fits.}

\item{Inclusive quasi-elastic scattering  (along with the deep inelastic scattering) 
is comparatively simple. It can be
    well understood with realistic nuclear interactions and currents. Though much progress has been made, significant further effort is needed to take full advantage of the experimental regime and connect it to others at lower and higher neutrino energies.}

\item{Extend quantum Monte Carlo calculations of inclusive electroweak response functions to $^{16}$O and $^{40}$Ar nuclei, which are relevant to the accelerator-neutrino program. Leveraging machine-learning methods will be particularly useful in both representing the wave function and in reconstructing the energy dependence of the response functions from imaginary-time correlators.}

\item{Incorporate modern evaluations of nucleon form factors into nuclear many-body approaches, including both experimental and lattice QCD results where available. As further calculations of inelastic processes and two-nucleon currents become available, incorporate them into the chiral EFT.}

\item{Test many-body calculations across a wide range of energies and momenta including both electron and neutrino data.  The same interactions/currents should be able to describe low-energy inclusive neutrino scattering, astrophysical inelastic processes on nuclei, and quasi-elastic scattering.}

\item{ Incorporate relativistic kinematics and currents directly into the many-body approaches. This may be easier in the factorization schemes since one only has to treat a modest number of degrees of freedom, ideally one would also treat relativistic corrections to the final state interactions.}

\item{Extend many-body factorized approaches to include effective field theory and related models of pion production and $\Delta(1232)$ resonances, and their propagation in the nuclear medium.}

\item{Use factorization algorithms to gain information about exclusive final states. 
At present the one- and two-nucleon vertex can be treated quantum mechanically, but propagation through the medium is treated semiclassically in generators. 
Classical vs. exact quantum evolution can be tested in the very simplest ($A=3,4$) nuclei, and advances can be incorporated into generators.}

\item
Although relatively well-studied, the dominant role of 
  the $\Delta(1232)$ resonance region in GeV neutrino
  reactions makes a precise description of this energy region a first
  priority task.  In particular the axial response of the nucleon at higher $Q^2$ needs to be better constrained. LQCD calculations could provide important input here, and new data on neutrino-proton and neutrino-deuteron pion production would prove extremely valuable in this respect.  Another possible source of information is provided  by  parity-violating  electron scattering~\cite{Wang:2014guo} where backward electron scattering enhances the effect of 
   interference and  provides information about axial form factors that can be  highlighted investigating  parity-violating asymmetry data.

 \item
 As mentioned in Sec \ref{inelastic}, the reaction dynamics becomes more complicated
  beyond the Delta region.
  It is of special interest to hadron physics  to investigate higher nucleon resonances and their structure. Neutrino models currently implemented in generators require the non-resonant part of the amplitude to be included in a consistent way.
The axial vector current  response of the nucleon and its  $Q^2$ and $W$ running when approaching the boundary of the resonance region, constitutes an interesting problem. In a naive parton model, the inclusive strength of the vector and axial vector current is the same, while in a hadron picture, vector and axial vector transition form factors are expected to reflect the structure of baryon resonances. The electromagnetic structure functions of the DCC model indeed approach the  partonic picture for large $Q^2$ and $W$, while that the of axial vector current does not.  Again, this is mainly due to our poor knowledge of axial vector current.

\item
Whereas currently pion-nucleus interactions and medium
  modifications of the $\Delta$ in the  nucleus
  are taken into account in descriptions of  inclusive cross sections, this work needs to be extended  towards the description of 
  semi-inclusive reactions like $(l,l'\pi)$, $(l,l' N)$ or QE-like Delta production.
  Meson-exchange current contributions in the pion-production region, only studied
  in the past for exclusive pion photoproduction reactions, also need to be better explored. In the higher resonance region, there are clear indications that 
  the vacuum properties of $N^*$ and $\Delta$ states are unsuitable for nuclear pion production processes, 
  as supported by   photo-nuclear reaction data showing  a disappearing  $N^*$ signal.
  Theoretical work  for neutrino reactions along the lines of Ref.~\cite{Hirata:2001sw}, taking into account a combination of nuclear effects in a consistent way  will be hence indispensable.

\item
 With the lack of a coherent picture of the SIS region, 
the models presently used in generators are either smoothed descriptions of inclusive data or often inconsistent mixtures of models \cite{Mosel:2019vhx}. 
Recently, a fairly complete group of generator experts started a new initiative to
improve structural issues~\cite{Barrow:2020gzb}. The present task to develop a consistent and accurate SIS model is a very interesting and challenging physics problem that requires proficiency in both nuclear physics and particle physics.  One of the sources of the present inconsistency is the different framing in different sub-fields.  A more complete picture is needed to achieve a coherent model.

\item
In the transition from SIS to DIS, differences between Monte Carlo generators often yield inconsistent predictions, as shown graphically in, e.g., \cite{SajjadAthar:2020nvy} and by Bronner in \cite{Andreopoulos:2019gvw}. The pioneering PDF-based approach of Bodek-Yang \cite{Yang:1998zb,Bodek:2002ps,Bodek:2004pc,Bodek:2010km} and more phenomenological, theory-guided, structure function approaches that do not rely on a parton decomposition (see, e.g., \cite{Capella:1994cr,Reno:2006hj}), merit study in view of the availability of more recent PDFs, studies of target mass and higher twist corrections, and next-to-next-to-leader order \cite{Kataev:1999bp,Vermaseren:2005qc,Moch:2004xu,Moch:2008fj} perturbative treatments of DIS \cite{Zaidi:2019asc}. 

\item 
For the study of quark-hadron duality with neutrinos there is a strong need for $\nu/\nubar$ data on both nucleons and nuclei covering the transition region running in $W$ from 1.5 to 2.0 GeV off nucleons and nuclei. Even without data and studying quark-hadron duality using available models, the need of a much improved 
theoretical or phenomenological model for the non-resonant background across the entire resonance region will be required.  

\item
The study of deep-inelastic scattering with neutrinos would be significantly improved with the neutrino-hydrogen/deuterium mentioned at the top of this section. In addition there is a lack of DIS off a range of nuclear targets, particularly the lower A nuclei, that is limiting the extraction of nuclear parton distributions.

\item Hadronization model tuning suffers from tensions in old bubble chamber data~\cite{Yang:2009zx,Kuzmin:2013tza,Katori:2014fxa,Bronner:2016gmz,GENIE:2021wox}. Currently, the tuning of the neutrino hadronization models is mainly relying on  $\nu-$H/D experimental data. A modern $\nu-$H/D experiment is necessary to remove systematic errors, and to confirm multiplicity predictions used in experimental analyses. It may be possible to tune hadron multiplicities from heavy target neutrino data. In this case, hadronization models and FSI models may be tuned together. 

\item Event generators are critical in connecting theoretical calculations to neutrino data for the determination of oscillation parameters and a variety of other analysis topics. To achieve the required experimental precision going forward, these theory improvements must be incorporated correctly and efficiently into simulations.

\item Experiments rely upon event generators to estimate signal and backgrounds, efficiency corrections, and systematic uncertainties. Full final-state predictions must be provided for all relevant neutrino energies, target nuclei, and scattering processes. Meeting these needs requires theoretical models to be combined in an approximate but self-consistent way with thorough uncertainty quantification. Greater theory guidance on the best strategies for meeting these needs (and support for providing such guidance) can help to improve upon the existing solutions.

\item An optimal development model for improving neutrino event generators will likely involve both increased support and new technical strategies for implementing models. An example of the latter has been the use of tables of pre-computed inclusive response functions to evaluate neutrino cross sections~\cite{Schwehr:2016pvn,Dolan:2019bxf,Barrow:2020mfy,Dolan:2021rdd}. This strategy enables a straightforward implementation of multiple models, including those which would be computationally impractical otherwise. However, without further extensions (e.g., additional tables to describe the hadronic final state), it leads to an incorrect treatment of exclusive observables.
Alternatively, developing interfaces able to directly incorporate part of a theory code into the event generator will be an alternative strategy to be further explored in the future (see Snowmass WP~\cite{HEPGeneratorWP}.)

\item Neutrino event generator development currently focuses on the intermediate energy regime of interest for accelerator-based oscillation experiments, although some tools exist for both lower and higher energies. Increased support for generator work should be coupled with a consideration of possible needs from the wider neutrino community.

\end{itemize}

A thorough understanding of neutrino scattering is still in need of extensive theoretical and experimental efforts. Realistic theoretical modeling of scattering should provide accurate predictions of neutrino-nucleus interactions, as well as meaningful theoretical uncertainties. New neutrino cross-section measurements to guide and benchmark model improvements will be essential, as will be sustained support for event generator development and theoretical and computational efforts at the interface of HEP and NP.
Achieving accurate and precise theoretical descriptions of neutrino scattering anchored in the SM and consistently incorporated into event generators will maximize the potential for discovery as the field moves into the precision era.

\acknowledgments{
L.A.R. acknowledges the support from the Spanish Ministerio de Ciencia e Innovaci\'on under contract PID2020-112777GB-I00, the EU STRONG-2020 project under the program H2020-INFRAIA-2018-1, grant agreement no. 824093 and by Generalitat Valenciana under contract PROMETEO/2020/023.
A.M.A. is supported by the U.S. Department of Energy, Office of Science (DOE) under Award No. DEAC02-76SF00515.
A.B.B. is supported by U.S.~Department of Energy, Office of Science, Office of High Energy Physics, under Award No.~DE-SC0019465 and by the U.S. National Science Foundation Grants No.~PHY-2020275 and PHY-2108339.
R.G. is supported by the U.S. Department of Energy, Office of Science, Office of High Energy Physics under Contract No.~DE-AC52-06NA25396. 
R.G.J. is supported by the government of Madrid and Complutense University under Project PR65/19-22430.
M.H. is supported by the Swiss National Science Foundation, Project No.~PCEFP2\_181117.
N.J. acknowledges support by the Research Foundation Flanders (FWO-Flanders).
W.J. is supported by the U.S. Department of Energy, Office of Science, Office of Nuclear Physics under grant Contract Numbers DE-SC0011090 and DE-SC0021006. 
F.K. is supported by the Deutsche Forschungsgemeinschaft under Germany’s Excellence Strategy -- EXC 2121 Quantum Universe -- 390833306. 
T.K.  acknowledges the support from the Science and Technology Council Facilities, UK. 
H.W.L. is partially supported by the U. S.  National Science Foundation under grant PHY 1653405 and  and by the  Research  Corporation  for  Science  Advancement through the Cottrell Scholar Award. 
K.F.L. is supported in part by the U.S. DOE
Grant No. DE-SC0013065  and DOE Grant No. DE-AC05-06OR23177.
A.L. is supported by the U.S. Department of Energy, Office of Science, Office of Nuclear Physics, under contract DE-AC02-06CH11357 and the NUCLEI SciDAC program.
K.M. is supported by U.S. Department of Energy, Office of Science, under grant  DE-SC0015903.
J.M. is supported by the ``Ram\'on y Cajal'' program with grant RYC-2017-22781, and grants CEX2019-000918-M and PID2020-118758GB-I00 funded by MCIN/AEI/10.13039/501100011033 and, as appropriate, by "ESF Investing in your future".
A.S.M. is supported by the Department of Energy, Office of Nuclear Physics, under Contract No. DE-SC00046548.
S.P. is supported by the U.S.~Department of Energy under contract DE-SC0021027, through the Neutrino Theory Network and the FRIB Theory Alliance award DE-SC0013617.
T.S. is supported by JSPS KAKENHI Grant JP19H05104.
A.S. is supported by the European Research Council (ERC) under the European Union's Horizon 2020 research and innovation programme (Grant Agreement No.~101020842).
P.E.S. is supported in part by the U.S.~Department of Energy, Office of Science, Office of Nuclear Physics under grant Contract Number DE-SC0011090 and by the U.S. Department of Energy Early Career Award DE-SC0021006, and by the National Science Foundation under EAGER grant 2035015 and under Cooperative Agreement PHY-2019786 (The NSF AI Institute for Artificial Intelligence and Fundamental Interactions, http://iaifi.org/).
L.E.S. acknowledges support from DOE Grant de-sc0010813. 
X.Z. is  supported by the U.S. Department of Energy, Office of Science, Office of Nuclear Physics, under the FRIB Theory Alliance award DE-SC0013617.
Y.Z. is supported by the U.S. Department of Energy, Office of Science, Office of Nuclear Physics through Contract No.~DE-AC02-06CH11357, and partially supported by an LDRD initiative at Argonne National Laboratory under Project~No.~2020-0020.
This manuscript has been authored by Fermi Research Alliance, LLC under Contract No. DEAC02-07CH11359 with the U.S. Department of Energy, Office of Science, Office of High Energy Physics.
}

\bibliographystyle{apsrev4-1}
\bibliography{bib}

\begin{thebibliography}{573}%
\makeatletter
\providecommand \@ifxundefined [1]{%
 \@ifx{#1\undefined}
}%
\providecommand \@ifnum [1]{%
 \ifnum #1\expandafter \@firstoftwo
 \else \expandafter \@secondoftwo
 \fi
}%
\providecommand \@ifx [1]{%
 \ifx #1\expandafter \@firstoftwo
 \else \expandafter \@secondoftwo
 \fi
}%
\providecommand \natexlab [1]{#1}%
\providecommand \enquote  [1]{``#1''}%
\providecommand \bibnamefont  [1]{#1}%
\providecommand \bibfnamefont [1]{#1}%
\providecommand \citenamefont [1]{#1}%
\providecommand \href@noop [0]{\@secondoftwo}%
\providecommand \href [0]{\begingroup \@sanitize@url \@href}%
\providecommand \@href[1]{\@@startlink{#1}\@@href}%
\providecommand \@@href[1]{\endgroup#1\@@endlink}%
\providecommand \@sanitize@url [0]{\catcode `\\12\catcode `\$12\catcode
  `\&12\catcode `\#12\catcode `\^12\catcode `\_12\catcode `\%12\relax}%
\providecommand \@@startlink[1]{}%
\providecommand \@@endlink[0]{}%
\providecommand \url  [0]{\begingroup\@sanitize@url \@url }%
\providecommand \@url [1]{\endgroup\@href {#1}{\urlprefix }}%
\providecommand \urlprefix  [0]{URL }%
\providecommand \Eprint [0]{\href }%
\providecommand \doibase [0]{http://dx.doi.org/}%
\providecommand \selectlanguage [0]{\@gobble}%
\providecommand \bibinfo  [0]{\@secondoftwo}%
\providecommand \bibfield  [0]{\@secondoftwo}%
\providecommand \translation [1]{[#1]}%
\providecommand \BibitemOpen [0]{}%
\providecommand \bibitemStop [0]{}%
\providecommand \bibitemNoStop [0]{.\EOS\space}%
\providecommand \EOS [0]{\spacefactor3000\relax}%
\providecommand \BibitemShut  [1]{\csname bibitem#1\endcsname}%
\let\auto@bib@innerbib\@empty
\bibitem [{\citenamefont {Akimov}\ \emph {et~al.}(2017)\citenamefont {Akimov}
  \emph {et~al.}}]{COHERENT:2017ipa}%
  \BibitemOpen
  \bibfield  {author} {\bibinfo {author} {\bibfnamefont {D.}~\bibnamefont
  {Akimov}} \emph {et~al.} (\bibinfo {collaboration} {COHERENT}),\ }\href
  {\doibase 10.1126/science.aao0990} {\bibfield  {journal} {\bibinfo  {journal}
  {Science}\ }\textbf {\bibinfo {volume} {357}},\ \bibinfo {pages} {1123}
  (\bibinfo {year} {2017})},\ \Eprint {http://arxiv.org/abs/1708.01294}
  {arXiv:1708.01294 [nucl-ex]} \BibitemShut {NoStop}%
\bibitem [{\citenamefont {Patton}\ \emph {et~al.}(2012)\citenamefont {Patton},
  \citenamefont {Engel}, \citenamefont {McLaughlin},\ and\ \citenamefont
  {Schunck}}]{Patton:2012jr}%
  \BibitemOpen
  \bibfield  {author} {\bibinfo {author} {\bibfnamefont {K.}~\bibnamefont
  {Patton}}, \bibinfo {author} {\bibfnamefont {J.}~\bibnamefont {Engel}},
  \bibinfo {author} {\bibfnamefont {G.~C.}\ \bibnamefont {McLaughlin}}, \ and\
  \bibinfo {author} {\bibfnamefont {N.}~\bibnamefont {Schunck}},\ }\href
  {\doibase 10.1103/PhysRevC.86.024612} {\bibfield  {journal} {\bibinfo
  {journal} {Phys. Rev. C}\ }\textbf {\bibinfo {volume} {86}},\ \bibinfo
  {pages} {024612} (\bibinfo {year} {2012})},\ \Eprint
  {http://arxiv.org/abs/1207.0693} {arXiv:1207.0693 [nucl-th]} \BibitemShut
  {NoStop}%
\bibitem [{\citenamefont {Coloma}\ \emph
  {et~al.}(2017{\natexlab{a}})\citenamefont {Coloma}, \citenamefont
  {Gonzalez-Garcia}, \citenamefont {Maltoni},\ and\ \citenamefont
  {Schwetz}}]{Coloma:2017ncl}%
  \BibitemOpen
  \bibfield  {author} {\bibinfo {author} {\bibfnamefont {P.}~\bibnamefont
  {Coloma}}, \bibinfo {author} {\bibfnamefont {M.~C.}\ \bibnamefont
  {Gonzalez-Garcia}}, \bibinfo {author} {\bibfnamefont {M.}~\bibnamefont
  {Maltoni}}, \ and\ \bibinfo {author} {\bibfnamefont {T.}~\bibnamefont
  {Schwetz}},\ }\href {\doibase 10.1103/PhysRevD.96.115007} {\bibfield
  {journal} {\bibinfo  {journal} {Phys. Rev. D}\ }\textbf {\bibinfo {volume}
  {96}},\ \bibinfo {pages} {115007} (\bibinfo {year} {2017}{\natexlab{a}})},\
  \Eprint {http://arxiv.org/abs/1708.02899} {arXiv:1708.02899 [hep-ph]}
  \BibitemShut {NoStop}%
\bibitem [{\citenamefont {Liao}\ and\ \citenamefont
  {Marfatia}(2017)}]{Liao:2017uzy}%
  \BibitemOpen
  \bibfield  {author} {\bibinfo {author} {\bibfnamefont {J.}~\bibnamefont
  {Liao}}\ and\ \bibinfo {author} {\bibfnamefont {D.}~\bibnamefont
  {Marfatia}},\ }\href {\doibase 10.1016/j.physletb.2017.10.046} {\bibfield
  {journal} {\bibinfo  {journal} {Phys. Lett. B}\ }\textbf {\bibinfo {volume}
  {775}},\ \bibinfo {pages} {54} (\bibinfo {year} {2017})},\ \Eprint
  {http://arxiv.org/abs/1708.04255} {arXiv:1708.04255 [hep-ph]} \BibitemShut
  {NoStop}%
\bibitem [{\citenamefont {Cadeddu}\ \emph {et~al.}(2018)\citenamefont
  {Cadeddu}, \citenamefont {Giunti}, \citenamefont {Li},\ and\ \citenamefont
  {Zhang}}]{Cadeddu:2017etk}%
  \BibitemOpen
  \bibfield  {author} {\bibinfo {author} {\bibfnamefont {M.}~\bibnamefont
  {Cadeddu}}, \bibinfo {author} {\bibfnamefont {C.}~\bibnamefont {Giunti}},
  \bibinfo {author} {\bibfnamefont {Y.~F.}\ \bibnamefont {Li}}, \ and\ \bibinfo
  {author} {\bibfnamefont {Y.~Y.}\ \bibnamefont {Zhang}},\ }\href {\doibase
  10.1103/PhysRevLett.120.072501} {\bibfield  {journal} {\bibinfo  {journal}
  {Phys. Rev. Lett.}\ }\textbf {\bibinfo {volume} {120}},\ \bibinfo {pages}
  {072501} (\bibinfo {year} {2018})},\ \Eprint
  {http://arxiv.org/abs/1710.02730} {arXiv:1710.02730 [hep-ph]} \BibitemShut
  {NoStop}%
\bibitem [{\citenamefont {Papoulias}\ and\ \citenamefont
  {Kosmas}(2018)}]{Papoulias:2017qdn}%
  \BibitemOpen
  \bibfield  {author} {\bibinfo {author} {\bibfnamefont {D.~K.}\ \bibnamefont
  {Papoulias}}\ and\ \bibinfo {author} {\bibfnamefont {T.~S.}\ \bibnamefont
  {Kosmas}},\ }\href {\doibase 10.1103/PhysRevD.97.033003} {\bibfield
  {journal} {\bibinfo  {journal} {Phys. Rev. D}\ }\textbf {\bibinfo {volume}
  {97}},\ \bibinfo {pages} {033003} (\bibinfo {year} {2018})},\ \Eprint
  {http://arxiv.org/abs/1711.09773} {arXiv:1711.09773 [hep-ph]} \BibitemShut
  {NoStop}%
\bibitem [{\citenamefont {Farzan}\ \emph {et~al.}(2018)\citenamefont {Farzan},
  \citenamefont {Lindner}, \citenamefont {Rodejohann},\ and\ \citenamefont
  {Xu}}]{Farzan:2018gtr}%
  \BibitemOpen
  \bibfield  {author} {\bibinfo {author} {\bibfnamefont {Y.}~\bibnamefont
  {Farzan}}, \bibinfo {author} {\bibfnamefont {M.}~\bibnamefont {Lindner}},
  \bibinfo {author} {\bibfnamefont {W.}~\bibnamefont {Rodejohann}}, \ and\
  \bibinfo {author} {\bibfnamefont {X.-J.}\ \bibnamefont {Xu}},\ }\href
  {\doibase 10.1007/JHEP05(2018)066} {\bibfield  {journal} {\bibinfo  {journal}
  {JHEP}\ }\textbf {\bibinfo {volume} {05}},\ \bibinfo {pages} {066} (\bibinfo
  {year} {2018})},\ \Eprint {http://arxiv.org/abs/1802.05171} {arXiv:1802.05171
  [hep-ph]} \BibitemShut {NoStop}%
\bibitem [{\citenamefont {Abdullah}\ \emph {et~al.}(2018)\citenamefont
  {Abdullah}, \citenamefont {Dent}, \citenamefont {Dutta}, \citenamefont
  {Kane}, \citenamefont {Liao},\ and\ \citenamefont
  {Strigari}}]{Abdullah:2018ykz}%
  \BibitemOpen
  \bibfield  {author} {\bibinfo {author} {\bibfnamefont {M.}~\bibnamefont
  {Abdullah}}, \bibinfo {author} {\bibfnamefont {J.~B.}\ \bibnamefont {Dent}},
  \bibinfo {author} {\bibfnamefont {B.}~\bibnamefont {Dutta}}, \bibinfo
  {author} {\bibfnamefont {G.~L.}\ \bibnamefont {Kane}}, \bibinfo {author}
  {\bibfnamefont {S.}~\bibnamefont {Liao}}, \ and\ \bibinfo {author}
  {\bibfnamefont {L.~E.}\ \bibnamefont {Strigari}},\ }\href {\doibase
  10.1103/PhysRevD.98.015005} {\bibfield  {journal} {\bibinfo  {journal} {Phys.
  Rev. D}\ }\textbf {\bibinfo {volume} {98}},\ \bibinfo {pages} {015005}
  (\bibinfo {year} {2018})},\ \Eprint {http://arxiv.org/abs/1803.01224}
  {arXiv:1803.01224 [hep-ph]} \BibitemShut {NoStop}%
\bibitem [{\citenamefont {Denton}\ \emph {et~al.}(2018)\citenamefont {Denton},
  \citenamefont {Farzan},\ and\ \citenamefont {Shoemaker}}]{Denton:2018xmq}%
  \BibitemOpen
  \bibfield  {author} {\bibinfo {author} {\bibfnamefont {P.~B.}\ \bibnamefont
  {Denton}}, \bibinfo {author} {\bibfnamefont {Y.}~\bibnamefont {Farzan}}, \
  and\ \bibinfo {author} {\bibfnamefont {I.~M.}\ \bibnamefont {Shoemaker}},\
  }\href {\doibase 10.1007/JHEP07(2018)037} {\bibfield  {journal} {\bibinfo
  {journal} {JHEP}\ }\textbf {\bibinfo {volume} {07}},\ \bibinfo {pages} {037}
  (\bibinfo {year} {2018})},\ \Eprint {http://arxiv.org/abs/1804.03660}
  {arXiv:1804.03660 [hep-ph]} \BibitemShut {NoStop}%
\bibitem [{\citenamefont {Ca\~nas}\ \emph {et~al.}(2018)\citenamefont
  {Ca\~nas}, \citenamefont {Garc\'es}, \citenamefont {Miranda},\ and\
  \citenamefont {Parada}}]{Canas:2018rng}%
  \BibitemOpen
  \bibfield  {author} {\bibinfo {author} {\bibfnamefont {B.~C.}\ \bibnamefont
  {Ca\~nas}}, \bibinfo {author} {\bibfnamefont {E.~A.}\ \bibnamefont
  {Garc\'es}}, \bibinfo {author} {\bibfnamefont {O.~G.}\ \bibnamefont
  {Miranda}}, \ and\ \bibinfo {author} {\bibfnamefont {A.}~\bibnamefont
  {Parada}},\ }\href {\doibase 10.1016/j.physletb.2018.07.049} {\bibfield
  {journal} {\bibinfo  {journal} {Phys. Lett. B}\ }\textbf {\bibinfo {volume}
  {784}},\ \bibinfo {pages} {159} (\bibinfo {year} {2018})},\ \Eprint
  {http://arxiv.org/abs/1806.01310} {arXiv:1806.01310 [hep-ph]} \BibitemShut
  {NoStop}%
\bibitem [{\citenamefont {Esteban}\ \emph {et~al.}(2018)\citenamefont
  {Esteban}, \citenamefont {Gonzalez-Garcia}, \citenamefont {Maltoni},
  \citenamefont {Martinez-Soler},\ and\ \citenamefont
  {Salvado}}]{Esteban:2018ppq}%
  \BibitemOpen
  \bibfield  {author} {\bibinfo {author} {\bibfnamefont {I.}~\bibnamefont
  {Esteban}}, \bibinfo {author} {\bibfnamefont {M.~C.}\ \bibnamefont
  {Gonzalez-Garcia}}, \bibinfo {author} {\bibfnamefont {M.}~\bibnamefont
  {Maltoni}}, \bibinfo {author} {\bibfnamefont {I.}~\bibnamefont
  {Martinez-Soler}}, \ and\ \bibinfo {author} {\bibfnamefont {J.}~\bibnamefont
  {Salvado}},\ }\href {\doibase 10.1007/JHEP08(2018)180} {\bibfield  {journal}
  {\bibinfo  {journal} {JHEP}\ }\textbf {\bibinfo {volume} {08}},\ \bibinfo
  {pages} {180} (\bibinfo {year} {2018})},\ \bibinfo {note} {[Addendum: JHEP
  {\bf 12}, 152 (2020)]},\ \Eprint {http://arxiv.org/abs/1805.04530}
  {arXiv:1805.04530 [hep-ph]} \BibitemShut {NoStop}%
\bibitem [{\citenamefont {Aristizabal~Sierra}\ \emph
  {et~al.}(2018)\citenamefont {Aristizabal~Sierra}, \citenamefont {De~Romeri},\
  and\ \citenamefont {Rojas}}]{AristizabalSierra:2018eqm}%
  \BibitemOpen
  \bibfield  {author} {\bibinfo {author} {\bibfnamefont {D.}~\bibnamefont
  {Aristizabal~Sierra}}, \bibinfo {author} {\bibfnamefont {V.}~\bibnamefont
  {De~Romeri}}, \ and\ \bibinfo {author} {\bibfnamefont {N.}~\bibnamefont
  {Rojas}},\ }\href {\doibase 10.1103/PhysRevD.98.075018} {\bibfield  {journal}
  {\bibinfo  {journal} {Phys. Rev. D}\ }\textbf {\bibinfo {volume} {98}},\
  \bibinfo {pages} {075018} (\bibinfo {year} {2018})},\ \Eprint
  {http://arxiv.org/abs/1806.07424} {arXiv:1806.07424 [hep-ph]} \BibitemShut
  {NoStop}%
\bibitem [{\citenamefont {Billard}\ \emph {et~al.}(2018)\citenamefont
  {Billard}, \citenamefont {Johnston},\ and\ \citenamefont
  {Kavanagh}}]{Billard:2018jnl}%
  \BibitemOpen
  \bibfield  {author} {\bibinfo {author} {\bibfnamefont {J.}~\bibnamefont
  {Billard}}, \bibinfo {author} {\bibfnamefont {J.}~\bibnamefont {Johnston}}, \
  and\ \bibinfo {author} {\bibfnamefont {B.~J.}\ \bibnamefont {Kavanagh}},\
  }\href {\doibase 10.1088/1475-7516/2018/11/016} {\bibfield  {journal}
  {\bibinfo  {journal} {JCAP}\ }\textbf {\bibinfo {volume} {11}},\ \bibinfo
  {pages} {016} (\bibinfo {year} {2018})},\ \Eprint
  {http://arxiv.org/abs/1805.01798} {arXiv:1805.01798 [hep-ph]} \BibitemShut
  {NoStop}%
\bibitem [{\citenamefont {Dutta}\ \emph {et~al.}(2019)\citenamefont {Dutta},
  \citenamefont {Liao}, \citenamefont {Sinha},\ and\ \citenamefont
  {Strigari}}]{Dutta:2019eml}%
  \BibitemOpen
  \bibfield  {author} {\bibinfo {author} {\bibfnamefont {B.}~\bibnamefont
  {Dutta}}, \bibinfo {author} {\bibfnamefont {S.}~\bibnamefont {Liao}},
  \bibinfo {author} {\bibfnamefont {S.}~\bibnamefont {Sinha}}, \ and\ \bibinfo
  {author} {\bibfnamefont {L.~E.}\ \bibnamefont {Strigari}},\ }\href {\doibase
  10.1103/PhysRevLett.123.061801} {\bibfield  {journal} {\bibinfo  {journal}
  {Phys. Rev. Lett.}\ }\textbf {\bibinfo {volume} {123}},\ \bibinfo {pages}
  {061801} (\bibinfo {year} {2019})},\ \Eprint
  {http://arxiv.org/abs/1903.10666} {arXiv:1903.10666 [hep-ph]} \BibitemShut
  {NoStop}%
\bibitem [{\citenamefont {Dutta}\ \emph {et~al.}(2020)\citenamefont {Dutta},
  \citenamefont {Kim}, \citenamefont {Liao}, \citenamefont {Park},
  \citenamefont {Shin},\ and\ \citenamefont {Strigari}}]{Dutta:2019nbn}%
  \BibitemOpen
  \bibfield  {author} {\bibinfo {author} {\bibfnamefont {B.}~\bibnamefont
  {Dutta}}, \bibinfo {author} {\bibfnamefont {D.}~\bibnamefont {Kim}}, \bibinfo
  {author} {\bibfnamefont {S.}~\bibnamefont {Liao}}, \bibinfo {author}
  {\bibfnamefont {J.-C.}\ \bibnamefont {Park}}, \bibinfo {author}
  {\bibfnamefont {S.}~\bibnamefont {Shin}}, \ and\ \bibinfo {author}
  {\bibfnamefont {L.~E.}\ \bibnamefont {Strigari}},\ }\href {\doibase
  10.1103/PhysRevLett.124.121802} {\bibfield  {journal} {\bibinfo  {journal}
  {Phys. Rev. Lett.}\ }\textbf {\bibinfo {volume} {124}},\ \bibinfo {pages}
  {121802} (\bibinfo {year} {2020})},\ \Eprint
  {http://arxiv.org/abs/1906.10745} {arXiv:1906.10745 [hep-ph]} \BibitemShut
  {NoStop}%
\bibitem [{\citenamefont {Cadeddu}\ \emph {et~al.}(2020)\citenamefont
  {Cadeddu}, \citenamefont {Dordei}, \citenamefont {Giunti}, \citenamefont
  {Li}, \citenamefont {Picciau},\ and\ \citenamefont
  {Zhang}}]{Cadeddu:2020lky}%
  \BibitemOpen
  \bibfield  {author} {\bibinfo {author} {\bibfnamefont {M.}~\bibnamefont
  {Cadeddu}}, \bibinfo {author} {\bibfnamefont {F.}~\bibnamefont {Dordei}},
  \bibinfo {author} {\bibfnamefont {C.}~\bibnamefont {Giunti}}, \bibinfo
  {author} {\bibfnamefont {Y.~F.}\ \bibnamefont {Li}}, \bibinfo {author}
  {\bibfnamefont {E.}~\bibnamefont {Picciau}}, \ and\ \bibinfo {author}
  {\bibfnamefont {Y.~Y.}\ \bibnamefont {Zhang}},\ }\href {\doibase
  10.1103/PhysRevD.102.015030} {\bibfield  {journal} {\bibinfo  {journal}
  {Phys. Rev. D}\ }\textbf {\bibinfo {volume} {102}},\ \bibinfo {pages}
  {015030} (\bibinfo {year} {2020})},\ \Eprint
  {http://arxiv.org/abs/2005.01645} {arXiv:2005.01645 [hep-ph]} \BibitemShut
  {NoStop}%
\bibitem [{\citenamefont {Miranda}\ \emph {et~al.}(2020)\citenamefont
  {Miranda}, \citenamefont {Papoulias}, \citenamefont {Sanchez~Garcia},
  \citenamefont {Sanders}, \citenamefont {T\'ortola},\ and\ \citenamefont
  {Valle}}]{Miranda:2020tif}%
  \BibitemOpen
  \bibfield  {author} {\bibinfo {author} {\bibfnamefont {O.~G.}\ \bibnamefont
  {Miranda}}, \bibinfo {author} {\bibfnamefont {D.~K.}\ \bibnamefont
  {Papoulias}}, \bibinfo {author} {\bibfnamefont {G.}~\bibnamefont
  {Sanchez~Garcia}}, \bibinfo {author} {\bibfnamefont {O.}~\bibnamefont
  {Sanders}}, \bibinfo {author} {\bibfnamefont {M.}~\bibnamefont {T\'ortola}},
  \ and\ \bibinfo {author} {\bibfnamefont {J.~W.~F.}\ \bibnamefont {Valle}},\
  }\href {\doibase 10.1007/JHEP05(2020)130} {\bibfield  {journal} {\bibinfo
  {journal} {JHEP}\ }\textbf {\bibinfo {volume} {05}},\ \bibinfo {pages} {130}
  (\bibinfo {year} {2020})},\ \bibinfo {note} {[Erratum: JHEP {\bf 01}, 067
  (2021)]},\ \Eprint {http://arxiv.org/abs/2003.12050} {arXiv:2003.12050
  [hep-ph]} \BibitemShut {NoStop}%
\bibitem [{\citenamefont {Payne}\ \emph {et~al.}(2019)\citenamefont {Payne},
  \citenamefont {Bacca}, \citenamefont {Hagen}, \citenamefont {Jiang},\ and\
  \citenamefont {Papenbrock}}]{Payne:2019wvy}%
  \BibitemOpen
  \bibfield  {author} {\bibinfo {author} {\bibfnamefont {C.~G.}\ \bibnamefont
  {Payne}}, \bibinfo {author} {\bibfnamefont {S.}~\bibnamefont {Bacca}},
  \bibinfo {author} {\bibfnamefont {G.}~\bibnamefont {Hagen}}, \bibinfo
  {author} {\bibfnamefont {W.}~\bibnamefont {Jiang}}, \ and\ \bibinfo {author}
  {\bibfnamefont {T.}~\bibnamefont {Papenbrock}},\ }\href {\doibase
  10.1103/PhysRevC.100.061304} {\bibfield  {journal} {\bibinfo  {journal}
  {Phys. Rev. C}\ }\textbf {\bibinfo {volume} {100}},\ \bibinfo {pages}
  {061304} (\bibinfo {year} {2019})},\ \Eprint
  {http://arxiv.org/abs/1908.09739} {arXiv:1908.09739 [nucl-th]} \BibitemShut
  {NoStop}%
\bibitem [{\citenamefont {Aristizabal~Sierra}\ \emph
  {et~al.}(2019)\citenamefont {Aristizabal~Sierra}, \citenamefont {Liao},\ and\
  \citenamefont {Marfatia}}]{AristizabalSierra:2019zmy}%
  \BibitemOpen
  \bibfield  {author} {\bibinfo {author} {\bibfnamefont {D.}~\bibnamefont
  {Aristizabal~Sierra}}, \bibinfo {author} {\bibfnamefont {J.}~\bibnamefont
  {Liao}}, \ and\ \bibinfo {author} {\bibfnamefont {D.}~\bibnamefont
  {Marfatia}},\ }\href {\doibase 10.1007/JHEP06(2019)141} {\bibfield  {journal}
  {\bibinfo  {journal} {JHEP}\ }\textbf {\bibinfo {volume} {06}},\ \bibinfo
  {pages} {141} (\bibinfo {year} {2019})},\ \Eprint
  {http://arxiv.org/abs/1902.07398} {arXiv:1902.07398 [hep-ph]} \BibitemShut
  {NoStop}%
\bibitem [{\citenamefont {Hoferichter}\ \emph {et~al.}(2020)\citenamefont
  {Hoferichter}, \citenamefont {Men\'endez},\ and\ \citenamefont
  {Schwenk}}]{Hoferichter:2020osn}%
  \BibitemOpen
  \bibfield  {author} {\bibinfo {author} {\bibfnamefont {M.}~\bibnamefont
  {Hoferichter}}, \bibinfo {author} {\bibfnamefont {J.}~\bibnamefont
  {Men\'endez}}, \ and\ \bibinfo {author} {\bibfnamefont {A.}~\bibnamefont
  {Schwenk}},\ }\href {\doibase 10.1103/PhysRevD.102.074018} {\bibfield
  {journal} {\bibinfo  {journal} {Phys. Rev. D}\ }\textbf {\bibinfo {volume}
  {102}},\ \bibinfo {pages} {074018} (\bibinfo {year} {2020})},\ \Eprint
  {http://arxiv.org/abs/2007.08529} {arXiv:2007.08529 [hep-ph]} \BibitemShut
  {NoStop}%
\bibitem [{\citenamefont {Aguilar-Arevalo}\ \emph {et~al.}(2019)\citenamefont
  {Aguilar-Arevalo} \emph {et~al.}}]{CONNIE:2019swq}%
  \BibitemOpen
  \bibfield  {author} {\bibinfo {author} {\bibfnamefont {A.}~\bibnamefont
  {Aguilar-Arevalo}} \emph {et~al.} (\bibinfo {collaboration} {CONNIE}),\
  }\href {\doibase 10.1103/PhysRevD.100.092005} {\bibfield  {journal} {\bibinfo
   {journal} {Phys. Rev. D}\ }\textbf {\bibinfo {volume} {100}},\ \bibinfo
  {pages} {092005} (\bibinfo {year} {2019})},\ \Eprint
  {http://arxiv.org/abs/1906.02200} {arXiv:1906.02200 [physics.ins-det]}
  \BibitemShut {NoStop}%
\bibitem [{\citenamefont {Aprile}\ \emph {et~al.}(2021)\citenamefont {Aprile}
  \emph {et~al.}}]{XENON:2020gfr}%
  \BibitemOpen
  \bibfield  {author} {\bibinfo {author} {\bibfnamefont {E.}~\bibnamefont
  {Aprile}} \emph {et~al.} (\bibinfo {collaboration} {XENON}),\ }\href
  {\doibase 10.1103/PhysRevLett.126.091301} {\bibfield  {journal} {\bibinfo
  {journal} {Phys. Rev. Lett.}\ }\textbf {\bibinfo {volume} {126}},\ \bibinfo
  {pages} {091301} (\bibinfo {year} {2021})},\ \Eprint
  {http://arxiv.org/abs/2012.02846} {arXiv:2012.02846 [hep-ex]} \BibitemShut
  {NoStop}%
\bibitem [{\citenamefont {Akimov}\ \emph
  {et~al.}(2021{\natexlab{a}})\citenamefont {Akimov} \emph
  {et~al.}}]{COHERENT:2020iec}%
  \BibitemOpen
  \bibfield  {author} {\bibinfo {author} {\bibfnamefont {D.}~\bibnamefont
  {Akimov}} \emph {et~al.} (\bibinfo {collaboration} {COHERENT}),\ }\href
  {\doibase 10.1103/PhysRevLett.126.012002} {\bibfield  {journal} {\bibinfo
  {journal} {Phys. Rev. Lett.}\ }\textbf {\bibinfo {volume} {126}},\ \bibinfo
  {pages} {012002} (\bibinfo {year} {2021}{\natexlab{a}})},\ \Eprint
  {http://arxiv.org/abs/2003.10630} {arXiv:2003.10630 [nucl-ex]} \BibitemShut
  {NoStop}%
\bibitem [{\citenamefont {Bonet}\ \emph {et~al.}(2021)\citenamefont {Bonet}
  \emph {et~al.}}]{CONUS:2020skt}%
  \BibitemOpen
  \bibfield  {author} {\bibinfo {author} {\bibfnamefont {H.}~\bibnamefont
  {Bonet}} \emph {et~al.} (\bibinfo {collaboration} {CONUS}),\ }\href {\doibase
  10.1103/PhysRevLett.126.041804} {\bibfield  {journal} {\bibinfo  {journal}
  {Phys. Rev. Lett.}\ }\textbf {\bibinfo {volume} {126}},\ \bibinfo {pages}
  {041804} (\bibinfo {year} {2021})},\ \Eprint
  {http://arxiv.org/abs/2011.00210} {arXiv:2011.00210 [hep-ex]} \BibitemShut
  {NoStop}%
\bibitem [{\citenamefont {Akimov}\ \emph
  {et~al.}(2021{\natexlab{b}})\citenamefont {Akimov} \emph
  {et~al.}}]{COHERENT:2021xhx}%
  \BibitemOpen
  \bibfield  {author} {\bibinfo {author} {\bibfnamefont {D.}~\bibnamefont
  {Akimov}} \emph {et~al.} (\bibinfo {collaboration} {COHERENT}),\ }\href
  {\doibase 10.1088/1748-0221/16/08/P08048} {\bibfield  {journal} {\bibinfo
  {journal} {JINST}\ }\textbf {\bibinfo {volume} {16}},\ \bibinfo {pages}
  {P08048} (\bibinfo {year} {2021}{\natexlab{b}})},\ \Eprint
  {http://arxiv.org/abs/2104.09605} {arXiv:2104.09605 [physics.ins-det]}
  \BibitemShut {NoStop}%
\bibitem [{\citenamefont {Raghavan}\ \emph {et~al.}(1986)\citenamefont
  {Raghavan}, \citenamefont {Pakvasa},\ and\ \citenamefont
  {Brown}}]{Raghavan:1986fg}%
  \BibitemOpen
  \bibfield  {author} {\bibinfo {author} {\bibfnamefont {R.~S.}\ \bibnamefont
  {Raghavan}}, \bibinfo {author} {\bibfnamefont {S.}~\bibnamefont {Pakvasa}}, \
  and\ \bibinfo {author} {\bibfnamefont {B.~A.}\ \bibnamefont {Brown}},\ }\href
  {\doibase 10.1103/PhysRevLett.57.1801} {\bibfield  {journal} {\bibinfo
  {journal} {Phys. Rev. Lett.}\ }\textbf {\bibinfo {volume} {57}},\ \bibinfo
  {pages} {1801} (\bibinfo {year} {1986})}\BibitemShut {NoStop}%
\bibitem [{\citenamefont {Haxton}(1987)}]{Haxton:1987kc}%
  \BibitemOpen
  \bibfield  {author} {\bibinfo {author} {\bibfnamefont {W.~C.}\ \bibnamefont
  {Haxton}},\ }\href {\doibase 10.1103/PhysRevD.36.2283} {\bibfield  {journal}
  {\bibinfo  {journal} {Phys. Rev. D}\ }\textbf {\bibinfo {volume} {36}},\
  \bibinfo {pages} {2283} (\bibinfo {year} {1987})}\BibitemShut {NoStop}%
\bibitem [{\citenamefont {Fukugita}\ \emph {et~al.}(1988)\citenamefont
  {Fukugita}, \citenamefont {Kohyama},\ and\ \citenamefont
  {Kubodera}}]{Fukugita:1988hg}%
  \BibitemOpen
  \bibfield  {author} {\bibinfo {author} {\bibfnamefont {M.}~\bibnamefont
  {Fukugita}}, \bibinfo {author} {\bibfnamefont {Y.}~\bibnamefont {Kohyama}}, \
  and\ \bibinfo {author} {\bibfnamefont {K.}~\bibnamefont {Kubodera}},\ }\href
  {\doibase 10.1016/0370-2693(88)90513-8} {\bibfield  {journal} {\bibinfo
  {journal} {Phys. Lett. B}\ }\textbf {\bibinfo {volume} {212}},\ \bibinfo
  {pages} {139} (\bibinfo {year} {1988})}\BibitemShut {NoStop}%
\bibitem [{\citenamefont {Engel}\ \emph {et~al.}(1996)\citenamefont {Engel},
  \citenamefont {Kolbe}, \citenamefont {Langanke},\ and\ \citenamefont
  {Vogel}}]{Engel:1996zt}%
  \BibitemOpen
  \bibfield  {author} {\bibinfo {author} {\bibfnamefont {J.}~\bibnamefont
  {Engel}}, \bibinfo {author} {\bibfnamefont {E.}~\bibnamefont {Kolbe}},
  \bibinfo {author} {\bibfnamefont {K.}~\bibnamefont {Langanke}}, \ and\
  \bibinfo {author} {\bibfnamefont {P.}~\bibnamefont {Vogel}},\ }\href
  {\doibase 10.1103/PhysRevC.54.2740} {\bibfield  {journal} {\bibinfo
  {journal} {Phys. Rev. C}\ }\textbf {\bibinfo {volume} {54}},\ \bibinfo
  {pages} {2740} (\bibinfo {year} {1996})},\ \Eprint
  {http://arxiv.org/abs/nucl-th/9606031} {arXiv:nucl-th/9606031} \BibitemShut
  {NoStop}%
\bibitem [{\citenamefont {Armbruster}\ \emph {et~al.}(1998)\citenamefont
  {Armbruster} \emph {et~al.}}]{KARMEN:1998xmo}%
  \BibitemOpen
  \bibfield  {author} {\bibinfo {author} {\bibfnamefont {B.}~\bibnamefont
  {Armbruster}} \emph {et~al.} (\bibinfo {collaboration} {KARMEN}),\ }\href
  {\doibase 10.1016/S0370-2693(98)00087-2} {\bibfield  {journal} {\bibinfo
  {journal} {Phys. Lett. B}\ }\textbf {\bibinfo {volume} {423}},\ \bibinfo
  {pages} {15} (\bibinfo {year} {1998})}\BibitemShut {NoStop}%
\bibitem [{\citenamefont {Kolbe}\ \emph {et~al.}(1999)\citenamefont {Kolbe},
  \citenamefont {Langanke},\ and\ \citenamefont {Vogel}}]{Kolbe:1999au}%
  \BibitemOpen
  \bibfield  {author} {\bibinfo {author} {\bibfnamefont {E.}~\bibnamefont
  {Kolbe}}, \bibinfo {author} {\bibfnamefont {K.}~\bibnamefont {Langanke}}, \
  and\ \bibinfo {author} {\bibfnamefont {P.}~\bibnamefont {Vogel}},\ }\href
  {\doibase 10.1016/S0375-9474(99)00152-9} {\bibfield  {journal} {\bibinfo
  {journal} {Nucl. Phys. A}\ }\textbf {\bibinfo {volume} {652}},\ \bibinfo
  {pages} {91} (\bibinfo {year} {1999})},\ \Eprint
  {http://arxiv.org/abs/nucl-th/9903022} {arXiv:nucl-th/9903022} \BibitemShut
  {NoStop}%
\bibitem [{\citenamefont {Hayes}\ and\ \citenamefont
  {Towner}(2000)}]{Hayes:1999ew}%
  \BibitemOpen
  \bibfield  {author} {\bibinfo {author} {\bibfnamefont {A.~C.}\ \bibnamefont
  {Hayes}}\ and\ \bibinfo {author} {\bibfnamefont {I.~S.}\ \bibnamefont
  {Towner}},\ }\href {\doibase 10.1103/PhysRevC.61.044603} {\bibfield
  {journal} {\bibinfo  {journal} {Phys. Rev. C}\ }\textbf {\bibinfo {volume}
  {61}},\ \bibinfo {pages} {044603} (\bibinfo {year} {2000})},\ \Eprint
  {http://arxiv.org/abs/nucl-th/9907049} {arXiv:nucl-th/9907049} \BibitemShut
  {NoStop}%
\bibitem [{\citenamefont {Volpe}\ \emph {et~al.}(2000)\citenamefont {Volpe},
  \citenamefont {Auerbach}, \citenamefont {Colo}, \citenamefont {Suzuki},\ and\
  \citenamefont {Van~Giai}}]{Volpe:2000zn}%
  \BibitemOpen
  \bibfield  {author} {\bibinfo {author} {\bibfnamefont {C.}~\bibnamefont
  {Volpe}}, \bibinfo {author} {\bibfnamefont {N.}~\bibnamefont {Auerbach}},
  \bibinfo {author} {\bibfnamefont {G.}~\bibnamefont {Colo}}, \bibinfo {author}
  {\bibfnamefont {T.}~\bibnamefont {Suzuki}}, \ and\ \bibinfo {author}
  {\bibfnamefont {N.}~\bibnamefont {Van~Giai}},\ }\href {\doibase
  10.1103/PhysRevC.62.015501} {\bibfield  {journal} {\bibinfo  {journal} {Phys.
  Rev. C}\ }\textbf {\bibinfo {volume} {62}},\ \bibinfo {pages} {015501}
  (\bibinfo {year} {2000})},\ \Eprint {http://arxiv.org/abs/nucl-th/0001050}
  {arXiv:nucl-th/0001050} \BibitemShut {NoStop}%
\bibitem [{\citenamefont {Auerbach}\ \emph {et~al.}(2001)\citenamefont
  {Auerbach} \emph {et~al.}}]{LSND:2001fbw}%
  \BibitemOpen
  \bibfield  {author} {\bibinfo {author} {\bibfnamefont {L.~B.}\ \bibnamefont
  {Auerbach}} \emph {et~al.} (\bibinfo {collaboration} {LSND}),\ }\href
  {\doibase 10.1103/PhysRevC.64.065501} {\bibfield  {journal} {\bibinfo
  {journal} {Phys. Rev. C}\ }\textbf {\bibinfo {volume} {64}},\ \bibinfo
  {pages} {065501} (\bibinfo {year} {2001})},\ \Eprint
  {http://arxiv.org/abs/hep-ex/0105068} {arXiv:hep-ex/0105068} \BibitemShut
  {NoStop}%
\bibitem [{\citenamefont {Kolbe}\ \emph {et~al.}(2002)\citenamefont {Kolbe},
  \citenamefont {Langanke},\ and\ \citenamefont {Vogel}}]{Kolbe:2002gk}%
  \BibitemOpen
  \bibfield  {author} {\bibinfo {author} {\bibfnamefont {E.}~\bibnamefont
  {Kolbe}}, \bibinfo {author} {\bibfnamefont {K.}~\bibnamefont {Langanke}}, \
  and\ \bibinfo {author} {\bibfnamefont {P.}~\bibnamefont {Vogel}},\ }\href
  {\doibase 10.1103/PhysRevD.66.013007} {\bibfield  {journal} {\bibinfo
  {journal} {Phys. Rev. D}\ }\textbf {\bibinfo {volume} {66}},\ \bibinfo
  {pages} {013007} (\bibinfo {year} {2002})}\BibitemShut {NoStop}%
\bibitem [{\citenamefont {Ikeda}\ \emph {et~al.}(2007)\citenamefont {Ikeda}
  \emph {et~al.}}]{Super-Kamiokande:2007zsl}%
  \BibitemOpen
  \bibfield  {author} {\bibinfo {author} {\bibfnamefont {M.}~\bibnamefont
  {Ikeda}} \emph {et~al.} (\bibinfo {collaboration} {Super-Kamiokande}),\
  }\href {\doibase 10.1086/521547} {\bibfield  {journal} {\bibinfo  {journal}
  {Astrophys. J.}\ }\textbf {\bibinfo {volume} {669}},\ \bibinfo {pages} {519}
  (\bibinfo {year} {2007})},\ \Eprint {http://arxiv.org/abs/0706.2283}
  {arXiv:0706.2283 [astro-ph]} \BibitemShut {NoStop}%
\bibitem [{\citenamefont {Duba}\ \emph {et~al.}(2008)\citenamefont {Duba} \emph
  {et~al.}}]{Duba:2008zz}%
  \BibitemOpen
  \bibfield  {author} {\bibinfo {author} {\bibfnamefont {C.~A.}\ \bibnamefont
  {Duba}} \emph {et~al.},\ }\href {\doibase 10.1088/1742-6596/136/4/042077}
  {\bibfield  {journal} {\bibinfo  {journal} {J. Phys. Conf. Ser.}\ }\textbf
  {\bibinfo {volume} {136}},\ \bibinfo {pages} {042077} (\bibinfo {year}
  {2008})}\BibitemShut {NoStop}%
\bibitem [{\citenamefont {Scholberg}(2012)}]{Scholberg:2012id}%
  \BibitemOpen
  \bibfield  {author} {\bibinfo {author} {\bibfnamefont {K.}~\bibnamefont
  {Scholberg}},\ }\href {\doibase 10.1146/annurev-nucl-102711-095006}
  {\bibfield  {journal} {\bibinfo  {journal} {Ann. Rev. Nucl. Part. Sci.}\
  }\textbf {\bibinfo {volume} {62}},\ \bibinfo {pages} {81} (\bibinfo {year}
  {2012})},\ \Eprint {http://arxiv.org/abs/1205.6003} {arXiv:1205.6003
  [astro-ph.IM]} \BibitemShut {NoStop}%
\bibitem [{\citenamefont {Laha}\ \emph {et~al.}(2014)\citenamefont {Laha},
  \citenamefont {Beacom},\ and\ \citenamefont {Agarwalla}}]{Laha:2014yua}%
  \BibitemOpen
  \bibfield  {author} {\bibinfo {author} {\bibfnamefont {R.}~\bibnamefont
  {Laha}}, \bibinfo {author} {\bibfnamefont {J.~F.}\ \bibnamefont {Beacom}}, \
  and\ \bibinfo {author} {\bibfnamefont {S.~K.}\ \bibnamefont {Agarwalla}},\
  }\href@noop {} {\  (\bibinfo {year} {2014})},\ \Eprint
  {http://arxiv.org/abs/1412.8425} {arXiv:1412.8425 [hep-ph]} \BibitemShut
  {NoStop}%
\bibitem [{\citenamefont {An}\ \emph {et~al.}(2016)\citenamefont {An} \emph
  {et~al.}}]{JUNO:2015zny}%
  \BibitemOpen
  \bibfield  {author} {\bibinfo {author} {\bibfnamefont {F.}~\bibnamefont {An}}
  \emph {et~al.} (\bibinfo {collaboration} {JUNO}),\ }\href {\doibase
  10.1088/0954-3899/43/3/030401} {\bibfield  {journal} {\bibinfo  {journal} {J.
  Phys. G}\ }\textbf {\bibinfo {volume} {43}},\ \bibinfo {pages} {030401}
  (\bibinfo {year} {2016})},\ \Eprint {http://arxiv.org/abs/1507.05613}
  {arXiv:1507.05613 [physics.ins-det]} \BibitemShut {NoStop}%
\bibitem [{\citenamefont {Lu}\ \emph {et~al.}(2016)\citenamefont {Lu},
  \citenamefont {Li},\ and\ \citenamefont {Zhou}}]{Lu:2016ipr}%
  \BibitemOpen
  \bibfield  {author} {\bibinfo {author} {\bibfnamefont {J.-S.}\ \bibnamefont
  {Lu}}, \bibinfo {author} {\bibfnamefont {Y.-F.}\ \bibnamefont {Li}}, \ and\
  \bibinfo {author} {\bibfnamefont {S.}~\bibnamefont {Zhou}},\ }\href {\doibase
  10.1103/PhysRevD.94.023006} {\bibfield  {journal} {\bibinfo  {journal} {Phys.
  Rev. D}\ }\textbf {\bibinfo {volume} {94}},\ \bibinfo {pages} {023006}
  (\bibinfo {year} {2016})},\ \Eprint {http://arxiv.org/abs/1605.07803}
  {arXiv:1605.07803 [hep-ph]} \BibitemShut {NoStop}%
\bibitem [{\citenamefont {Li}\ \emph {et~al.}(2021{\natexlab{a}})\citenamefont
  {Li}, \citenamefont {Roberts},\ and\ \citenamefont {Beacom}}]{Li:2020ujl}%
  \BibitemOpen
  \bibfield  {author} {\bibinfo {author} {\bibfnamefont {S.~W.}\ \bibnamefont
  {Li}}, \bibinfo {author} {\bibfnamefont {L.~F.}\ \bibnamefont {Roberts}}, \
  and\ \bibinfo {author} {\bibfnamefont {J.~F.}\ \bibnamefont {Beacom}},\
  }\href {\doibase 10.1103/PhysRevD.103.023016} {\bibfield  {journal} {\bibinfo
   {journal} {Phys. Rev. D}\ }\textbf {\bibinfo {volume} {103}},\ \bibinfo
  {pages} {023016} (\bibinfo {year} {2021}{\natexlab{a}})},\ \Eprint
  {http://arxiv.org/abs/2008.04340} {arXiv:2008.04340 [astro-ph.HE]}
  \BibitemShut {NoStop}%
\bibitem [{\citenamefont {Abi}\ \emph {et~al.}(2021)\citenamefont {Abi} \emph
  {et~al.}}]{DUNE:2020zfm}%
  \BibitemOpen
  \bibfield  {author} {\bibinfo {author} {\bibfnamefont {B.}~\bibnamefont
  {Abi}} \emph {et~al.} (\bibinfo {collaboration} {DUNE}),\ }\href {\doibase
  10.1140/epjc/s10052-021-09166-w} {\bibfield  {journal} {\bibinfo  {journal}
  {Eur. Phys. J. C}\ }\textbf {\bibinfo {volume} {81}},\ \bibinfo {pages} {423}
  (\bibinfo {year} {2021})},\ \Eprint {http://arxiv.org/abs/2008.06647}
  {arXiv:2008.06647 [hep-ex]} \BibitemShut {NoStop}%
\bibitem [{\citenamefont {Capozzi}\ \emph {et~al.}(2019)\citenamefont
  {Capozzi}, \citenamefont {Li}, \citenamefont {Zhu},\ and\ \citenamefont
  {Beacom}}]{Capozzi:2018dat}%
  \BibitemOpen
  \bibfield  {author} {\bibinfo {author} {\bibfnamefont {F.}~\bibnamefont
  {Capozzi}}, \bibinfo {author} {\bibfnamefont {S.~W.}\ \bibnamefont {Li}},
  \bibinfo {author} {\bibfnamefont {G.}~\bibnamefont {Zhu}}, \ and\ \bibinfo
  {author} {\bibfnamefont {J.~F.}\ \bibnamefont {Beacom}},\ }\href {\doibase
  10.1103/PhysRevLett.123.131803} {\bibfield  {journal} {\bibinfo  {journal}
  {Phys. Rev. Lett.}\ }\textbf {\bibinfo {volume} {123}},\ \bibinfo {pages}
  {131803} (\bibinfo {year} {2019})},\ \Eprint
  {http://arxiv.org/abs/1808.08232} {arXiv:1808.08232 [hep-ph]} \BibitemShut
  {NoStop}%
\bibitem [{\citenamefont {Kolbe}\ \emph {et~al.}(1992)\citenamefont {Kolbe},
  \citenamefont {Langanke}, \citenamefont {Krewald},\ and\ \citenamefont
  {Thielemann}}]{KOLBE1992599}%
  \BibitemOpen
  \bibfield  {author} {\bibinfo {author} {\bibfnamefont {E.}~\bibnamefont
  {Kolbe}}, \bibinfo {author} {\bibfnamefont {K.}~\bibnamefont {Langanke}},
  \bibinfo {author} {\bibfnamefont {S.}~\bibnamefont {Krewald}}, \ and\
  \bibinfo {author} {\bibfnamefont {F.-K.}\ \bibnamefont {Thielemann}},\ }\href
  {\doibase https://doi.org/10.1016/0375-9474(92)90175-J} {\bibfield  {journal}
  {\bibinfo  {journal} {Nuclear Physics A}\ }\textbf {\bibinfo {volume}
  {540}},\ \bibinfo {pages} {599} (\bibinfo {year} {1992})}\BibitemShut
  {NoStop}%
\bibitem [{\citenamefont {Langanke}\ \emph {et~al.}(1996)\citenamefont
  {Langanke}, \citenamefont {Vogel},\ and\ \citenamefont
  {Kolbe}}]{PhysRevLett.76.2629}%
  \BibitemOpen
  \bibfield  {author} {\bibinfo {author} {\bibfnamefont {K.}~\bibnamefont
  {Langanke}}, \bibinfo {author} {\bibfnamefont {P.}~\bibnamefont {Vogel}}, \
  and\ \bibinfo {author} {\bibfnamefont {E.}~\bibnamefont {Kolbe}},\ }\href
  {\doibase 10.1103/PhysRevLett.76.2629} {\bibfield  {journal} {\bibinfo
  {journal} {Phys. Rev. Lett.}\ }\textbf {\bibinfo {volume} {76}},\ \bibinfo
  {pages} {2629} (\bibinfo {year} {1996})}\BibitemShut {NoStop}%
\bibitem [{\citenamefont {Gardiner}(2018)}]{Gardiner:2018zfg}%
  \BibitemOpen
  \bibfield  {author} {\bibinfo {author} {\bibfnamefont {S.~J.}\ \bibnamefont
  {Gardiner}},\ }\emph {\bibinfo {title} {{Nuclear Effects in Neutrino
  Detection}}},\ \href {\doibase 10.2172/1637626} {Ph.D. thesis},\ \bibinfo
  {school} {UC, Davis} (\bibinfo {year} {2018})\BibitemShut {NoStop}%
\bibitem [{\citenamefont {Mosel}(2016)}]{Mosel:2016cwa}%
  \BibitemOpen
  \bibfield  {author} {\bibinfo {author} {\bibfnamefont {U.}~\bibnamefont
  {Mosel}},\ }\href {\doibase 10.1146/annurev-nucl-102115-044720} {\bibfield
  {journal} {\bibinfo  {journal} {Ann. Rev. Nucl. Part. Sci.}\ }\textbf
  {\bibinfo {volume} {66}},\ \bibinfo {pages} {171} (\bibinfo {year} {2016})},\
  \Eprint {http://arxiv.org/abs/1602.00696} {arXiv:1602.00696 [nucl-th]}
  \BibitemShut {NoStop}%
\bibitem [{\citenamefont {Alvarez-Ruso}\ \emph {et~al.}(2018)\citenamefont
  {Alvarez-Ruso} \emph {et~al.}}]{NuSTEC:2017hzk}%
  \BibitemOpen
  \bibfield  {author} {\bibinfo {author} {\bibfnamefont {L.}~\bibnamefont
  {Alvarez-Ruso}} \emph {et~al.} (\bibinfo {collaboration} {NuSTEC}),\ }\href
  {\doibase 10.1016/j.ppnp.2018.01.006} {\bibfield  {journal} {\bibinfo
  {journal} {Prog. Part. Nucl. Phys.}\ }\textbf {\bibinfo {volume} {100}},\
  \bibinfo {pages} {1} (\bibinfo {year} {2018})},\ \Eprint
  {http://arxiv.org/abs/1706.03621} {arXiv:1706.03621 [hep-ph]} \BibitemShut
  {NoStop}%
\bibitem [{\citenamefont {Ankowski}\ \emph
  {et~al.}(2015{\natexlab{a}})\citenamefont {Ankowski}, \citenamefont
  {Barbaro}, \citenamefont {Benhar}, \citenamefont {Caballero}, \citenamefont
  {Giusti}, \citenamefont {Gonz\'alez-Jim\'enez}, \citenamefont {Megias},\ and\
  \citenamefont {Meucci}}]{Ankowski:2015lma}%
  \BibitemOpen
  \bibfield  {author} {\bibinfo {author} {\bibfnamefont {A.~M.}\ \bibnamefont
  {Ankowski}}, \bibinfo {author} {\bibfnamefont {M.~B.}\ \bibnamefont
  {Barbaro}}, \bibinfo {author} {\bibfnamefont {O.}~\bibnamefont {Benhar}},
  \bibinfo {author} {\bibfnamefont {J.~A.}\ \bibnamefont {Caballero}}, \bibinfo
  {author} {\bibfnamefont {C.}~\bibnamefont {Giusti}}, \bibinfo {author}
  {\bibfnamefont {R.}~\bibnamefont {Gonz\'alez-Jim\'enez}}, \bibinfo {author}
  {\bibfnamefont {G.~D.}\ \bibnamefont {Megias}}, \ and\ \bibinfo {author}
  {\bibfnamefont {A.}~\bibnamefont {Meucci}},\ }\href {\doibase
  10.1103/PhysRevC.92.025501} {\bibfield  {journal} {\bibinfo  {journal} {Phys.
  Rev. C}\ }\textbf {\bibinfo {volume} {92}},\ \bibinfo {pages} {025501}
  (\bibinfo {year} {2015}{\natexlab{a}})},\ \Eprint
  {http://arxiv.org/abs/1506.02673} {arXiv:1506.02673 [nucl-th]} \BibitemShut
  {NoStop}%
\bibitem [{\citenamefont {Altmannshofer}\ \emph {et~al.}(2014)\citenamefont
  {Altmannshofer}, \citenamefont {Gori}, \citenamefont {Pospelov},\ and\
  \citenamefont {Yavin}}]{Altmannshofer:2014pba}%
  \BibitemOpen
  \bibfield  {author} {\bibinfo {author} {\bibfnamefont {W.}~\bibnamefont
  {Altmannshofer}}, \bibinfo {author} {\bibfnamefont {S.}~\bibnamefont {Gori}},
  \bibinfo {author} {\bibfnamefont {M.}~\bibnamefont {Pospelov}}, \ and\
  \bibinfo {author} {\bibfnamefont {I.}~\bibnamefont {Yavin}},\ }\href
  {\doibase 10.1103/PhysRevLett.113.091801} {\bibfield  {journal} {\bibinfo
  {journal} {Phys. Rev. Lett.}\ }\textbf {\bibinfo {volume} {113}},\ \bibinfo
  {pages} {091801} (\bibinfo {year} {2014})},\ \Eprint
  {http://arxiv.org/abs/1406.2332} {arXiv:1406.2332 [hep-ph]} \BibitemShut
  {NoStop}%
\bibitem [{\citenamefont {Magill}\ and\ \citenamefont
  {Plestid}(2017)}]{Magill:2016hgc}%
  \BibitemOpen
  \bibfield  {author} {\bibinfo {author} {\bibfnamefont {G.}~\bibnamefont
  {Magill}}\ and\ \bibinfo {author} {\bibfnamefont {R.}~\bibnamefont
  {Plestid}},\ }\href {\doibase 10.1103/PhysRevD.95.073004} {\bibfield
  {journal} {\bibinfo  {journal} {Phys. Rev. D}\ }\textbf {\bibinfo {volume}
  {95}},\ \bibinfo {pages} {073004} (\bibinfo {year} {2017})},\ \Eprint
  {http://arxiv.org/abs/1612.05642} {arXiv:1612.05642 [hep-ph]} \BibitemShut
  {NoStop}%
\bibitem [{\citenamefont {Magill}\ and\ \citenamefont
  {Plestid}(2018)}]{Magill:2017mps}%
  \BibitemOpen
  \bibfield  {author} {\bibinfo {author} {\bibfnamefont {G.}~\bibnamefont
  {Magill}}\ and\ \bibinfo {author} {\bibfnamefont {R.}~\bibnamefont
  {Plestid}},\ }\href {\doibase 10.1103/PhysRevD.97.055003} {\bibfield
  {journal} {\bibinfo  {journal} {Phys. Rev. D}\ }\textbf {\bibinfo {volume}
  {97}},\ \bibinfo {pages} {055003} (\bibinfo {year} {2018})},\ \Eprint
  {http://arxiv.org/abs/1710.08431} {arXiv:1710.08431 [hep-ph]} \BibitemShut
  {NoStop}%
\bibitem [{\citenamefont {Coloma}\ \emph
  {et~al.}(2017{\natexlab{b}})\citenamefont {Coloma}, \citenamefont {Machado},
  \citenamefont {Martinez-Soler},\ and\ \citenamefont
  {Shoemaker}}]{Coloma:2017ppo}%
  \BibitemOpen
  \bibfield  {author} {\bibinfo {author} {\bibfnamefont {P.}~\bibnamefont
  {Coloma}}, \bibinfo {author} {\bibfnamefont {P.~A.~N.}\ \bibnamefont
  {Machado}}, \bibinfo {author} {\bibfnamefont {I.}~\bibnamefont
  {Martinez-Soler}}, \ and\ \bibinfo {author} {\bibfnamefont {I.~M.}\
  \bibnamefont {Shoemaker}},\ }\href {\doibase 10.1103/PhysRevLett.119.201804}
  {\bibfield  {journal} {\bibinfo  {journal} {Phys. Rev. Lett.}\ }\textbf
  {\bibinfo {volume} {119}},\ \bibinfo {pages} {201804} (\bibinfo {year}
  {2017}{\natexlab{b}})},\ \Eprint {http://arxiv.org/abs/1707.08573}
  {arXiv:1707.08573 [hep-ph]} \BibitemShut {NoStop}%
\bibitem [{\citenamefont {de~Gouv\^ea}\ \emph {et~al.}(2019)\citenamefont
  {de~Gouv\^ea}, \citenamefont {Fox}, \citenamefont {Harnik}, \citenamefont
  {Kelly},\ and\ \citenamefont {Zhang}}]{deGouvea:2018cfv}%
  \BibitemOpen
  \bibfield  {author} {\bibinfo {author} {\bibfnamefont {A.}~\bibnamefont
  {de~Gouv\^ea}}, \bibinfo {author} {\bibfnamefont {P.~J.}\ \bibnamefont
  {Fox}}, \bibinfo {author} {\bibfnamefont {R.}~\bibnamefont {Harnik}},
  \bibinfo {author} {\bibfnamefont {K.~J.}\ \bibnamefont {Kelly}}, \ and\
  \bibinfo {author} {\bibfnamefont {Y.}~\bibnamefont {Zhang}},\ }\href
  {\doibase 10.1007/JHEP01(2019)001} {\bibfield  {journal} {\bibinfo  {journal}
  {JHEP}\ }\textbf {\bibinfo {volume} {01}},\ \bibinfo {pages} {001} (\bibinfo
  {year} {2019})},\ \Eprint {http://arxiv.org/abs/1809.06388} {arXiv:1809.06388
  [hep-ph]} \BibitemShut {NoStop}%
\bibitem [{\citenamefont {Bertuzzo}\ \emph {et~al.}(2018)\citenamefont
  {Bertuzzo}, \citenamefont {Jana}, \citenamefont {Machado},\ and\
  \citenamefont {Zukanovich~Funchal}}]{Bertuzzo:2018itn}%
  \BibitemOpen
  \bibfield  {author} {\bibinfo {author} {\bibfnamefont {E.}~\bibnamefont
  {Bertuzzo}}, \bibinfo {author} {\bibfnamefont {S.}~\bibnamefont {Jana}},
  \bibinfo {author} {\bibfnamefont {P.~A.~N.}\ \bibnamefont {Machado}}, \ and\
  \bibinfo {author} {\bibfnamefont {R.}~\bibnamefont {Zukanovich~Funchal}},\
  }\href {\doibase 10.1103/PhysRevLett.121.241801} {\bibfield  {journal}
  {\bibinfo  {journal} {Phys. Rev. Lett.}\ }\textbf {\bibinfo {volume} {121}},\
  \bibinfo {pages} {241801} (\bibinfo {year} {2018})},\ \Eprint
  {http://arxiv.org/abs/1807.09877} {arXiv:1807.09877 [hep-ph]} \BibitemShut
  {NoStop}%
\bibitem [{\citenamefont {Ballett}\ \emph {et~al.}(2019)\citenamefont
  {Ballett}, \citenamefont {Hostert}, \citenamefont {Pascoli}, \citenamefont
  {Perez-Gonzalez}, \citenamefont {Tabrizi},\ and\ \citenamefont
  {Zukanovich~Funchal}}]{Ballett:2019xoj}%
  \BibitemOpen
  \bibfield  {author} {\bibinfo {author} {\bibfnamefont {P.}~\bibnamefont
  {Ballett}}, \bibinfo {author} {\bibfnamefont {M.}~\bibnamefont {Hostert}},
  \bibinfo {author} {\bibfnamefont {S.}~\bibnamefont {Pascoli}}, \bibinfo
  {author} {\bibfnamefont {Y.~F.}\ \bibnamefont {Perez-Gonzalez}}, \bibinfo
  {author} {\bibfnamefont {Z.}~\bibnamefont {Tabrizi}}, \ and\ \bibinfo
  {author} {\bibfnamefont {R.}~\bibnamefont {Zukanovich~Funchal}},\ }\href
  {\doibase 10.1103/PhysRevD.100.055012} {\bibfield  {journal} {\bibinfo
  {journal} {Phys. Rev. D}\ }\textbf {\bibinfo {volume} {100}},\ \bibinfo
  {pages} {055012} (\bibinfo {year} {2019})},\ \Eprint
  {http://arxiv.org/abs/1902.08579} {arXiv:1902.08579 [hep-ph]} \BibitemShut
  {NoStop}%
\bibitem [{\citenamefont {Berryman}\ \emph {et~al.}(2020)\citenamefont
  {Berryman}, \citenamefont {de~Gouv\^ea}, \citenamefont {Fox}, \citenamefont
  {Kayser}, \citenamefont {Kelly},\ and\ \citenamefont
  {Raaf}}]{Berryman:2019dme}%
  \BibitemOpen
  \bibfield  {author} {\bibinfo {author} {\bibfnamefont {J.~M.}\ \bibnamefont
  {Berryman}}, \bibinfo {author} {\bibfnamefont {A.}~\bibnamefont
  {de~Gouv\^ea}}, \bibinfo {author} {\bibfnamefont {P.~J.}\ \bibnamefont
  {Fox}}, \bibinfo {author} {\bibfnamefont {B.~J.}\ \bibnamefont {Kayser}},
  \bibinfo {author} {\bibfnamefont {K.~J.}\ \bibnamefont {Kelly}}, \ and\
  \bibinfo {author} {\bibfnamefont {J.~L.}\ \bibnamefont {Raaf}},\ }\href
  {\doibase 10.1007/JHEP02(2020)174} {\bibfield  {journal} {\bibinfo  {journal}
  {JHEP}\ }\textbf {\bibinfo {volume} {02}},\ \bibinfo {pages} {174} (\bibinfo
  {year} {2020})},\ \Eprint {http://arxiv.org/abs/1912.07622} {arXiv:1912.07622
  [hep-ph]} \BibitemShut {NoStop}%
\bibitem [{\citenamefont {Altmannshofer}\ \emph {et~al.}(2019)\citenamefont
  {Altmannshofer}, \citenamefont {Gori}, \citenamefont {Mart\'\i{}n-Albo},
  \citenamefont {Sousa},\ and\ \citenamefont
  {Wallbank}}]{Altmannshofer:2019zhy}%
  \BibitemOpen
  \bibfield  {author} {\bibinfo {author} {\bibfnamefont {W.}~\bibnamefont
  {Altmannshofer}}, \bibinfo {author} {\bibfnamefont {S.}~\bibnamefont {Gori}},
  \bibinfo {author} {\bibfnamefont {J.}~\bibnamefont {Mart\'\i{}n-Albo}},
  \bibinfo {author} {\bibfnamefont {A.}~\bibnamefont {Sousa}}, \ and\ \bibinfo
  {author} {\bibfnamefont {M.}~\bibnamefont {Wallbank}},\ }\href {\doibase
  10.1103/PhysRevD.100.115029} {\bibfield  {journal} {\bibinfo  {journal}
  {Phys. Rev. D}\ }\textbf {\bibinfo {volume} {100}},\ \bibinfo {pages}
  {115029} (\bibinfo {year} {2019})},\ \Eprint
  {http://arxiv.org/abs/1902.06765} {arXiv:1902.06765 [hep-ph]} \BibitemShut
  {NoStop}%
\bibitem [{\citenamefont {Schwetz}\ \emph {et~al.}(2020)\citenamefont
  {Schwetz}, \citenamefont {Zhou},\ and\ \citenamefont
  {Zhu}}]{Schwetz:2020xra}%
  \BibitemOpen
  \bibfield  {author} {\bibinfo {author} {\bibfnamefont {T.}~\bibnamefont
  {Schwetz}}, \bibinfo {author} {\bibfnamefont {A.}~\bibnamefont {Zhou}}, \
  and\ \bibinfo {author} {\bibfnamefont {J.-Y.}\ \bibnamefont {Zhu}},\ }\href
  {\doibase 10.1007/JHEP07(2021)200} {\bibfield  {journal} {\bibinfo  {journal}
  {JHEP}\ }\textbf {\bibinfo {volume} {21}},\ \bibinfo {pages} {200} (\bibinfo
  {year} {2020})},\ \Eprint {http://arxiv.org/abs/2105.09699} {arXiv:2105.09699
  [hep-ph]} \BibitemShut {NoStop}%
\bibitem [{\citenamefont {Atkinson}\ \emph {et~al.}(2021)\citenamefont
  {Atkinson}, \citenamefont {Coloma}, \citenamefont {Martinez-Soler},
  \citenamefont {Rocco},\ and\ \citenamefont {Shoemaker}}]{Atkinson:2021rnp}%
  \BibitemOpen
  \bibfield  {author} {\bibinfo {author} {\bibfnamefont {M.}~\bibnamefont
  {Atkinson}}, \bibinfo {author} {\bibfnamefont {P.}~\bibnamefont {Coloma}},
  \bibinfo {author} {\bibfnamefont {I.}~\bibnamefont {Martinez-Soler}},
  \bibinfo {author} {\bibfnamefont {N.}~\bibnamefont {Rocco}}, \ and\ \bibinfo
  {author} {\bibfnamefont {I.~M.}\ \bibnamefont {Shoemaker}},\ }\href@noop {}
  {\  (\bibinfo {year} {2021})},\ \Eprint {http://arxiv.org/abs/2105.09357}
  {arXiv:2105.09357 [hep-ph]} \BibitemShut {NoStop}%
\bibitem [{\citenamefont {Abed~Abud}\ \emph {et~al.}(2021)\citenamefont
  {Abed~Abud} \emph {et~al.}}]{DUNE:2021tad}%
  \BibitemOpen
  \bibfield  {author} {\bibinfo {author} {\bibfnamefont {A.}~\bibnamefont
  {Abed~Abud}} \emph {et~al.} (\bibinfo {collaboration} {DUNE}),\ }\href
  {\doibase 10.3390/instruments5040031} {\bibfield  {journal} {\bibinfo
  {journal} {Instruments}\ }\textbf {\bibinfo {volume} {5}},\ \bibinfo {pages}
  {31} (\bibinfo {year} {2021})},\ \Eprint {http://arxiv.org/abs/2103.13910}
  {arXiv:2103.13910 [physics.ins-det]} \BibitemShut {NoStop}%
\bibitem [{\citenamefont {Ankowski}\ \emph
  {et~al.}(2015{\natexlab{b}})\citenamefont {Ankowski}, \citenamefont
  {Benhar},\ and\ \citenamefont {Sakuda}}]{Ankowski:2014yfa}%
  \BibitemOpen
  \bibfield  {author} {\bibinfo {author} {\bibfnamefont {A.~M.}\ \bibnamefont
  {Ankowski}}, \bibinfo {author} {\bibfnamefont {O.}~\bibnamefont {Benhar}}, \
  and\ \bibinfo {author} {\bibfnamefont {M.}~\bibnamefont {Sakuda}},\ }\href
  {\doibase 10.1103/PhysRevD.91.033005} {\bibfield  {journal} {\bibinfo
  {journal} {Phys. Rev. D}\ }\textbf {\bibinfo {volume} {91}},\ \bibinfo
  {pages} {033005} (\bibinfo {year} {2015}{\natexlab{b}})},\ \Eprint
  {http://arxiv.org/abs/1404.5687} {arXiv:1404.5687 [nucl-th]} \BibitemShut
  {NoStop}%
\bibitem [{\citenamefont {Barbieri}\ \emph {et~al.}(2019)\citenamefont
  {Barbieri}, \citenamefont {Rocco},\ and\ \citenamefont
  {Som\`a}}]{Barbieri:2019ual}%
  \BibitemOpen
  \bibfield  {author} {\bibinfo {author} {\bibfnamefont {C.}~\bibnamefont
  {Barbieri}}, \bibinfo {author} {\bibfnamefont {N.}~\bibnamefont {Rocco}}, \
  and\ \bibinfo {author} {\bibfnamefont {V.}~\bibnamefont {Som\`a}},\ }\href
  {\doibase 10.1103/PhysRevC.100.062501} {\bibfield  {journal} {\bibinfo
  {journal} {Phys. Rev. C}\ }\textbf {\bibinfo {volume} {100}},\ \bibinfo
  {pages} {062501} (\bibinfo {year} {2019})},\ \Eprint
  {http://arxiv.org/abs/1907.01122} {arXiv:1907.01122 [nucl-th]} \BibitemShut
  {NoStop}%
\bibitem [{\citenamefont {Andreoli}\ \emph {et~al.}(2022)\citenamefont
  {Andreoli}, \citenamefont {Carlson}, \citenamefont {Lovato}, \citenamefont
  {Pastore}, \citenamefont {Rocco},\ and\ \citenamefont
  {Wiringa}}]{Andreoli:2021cxo}%
  \BibitemOpen
  \bibfield  {author} {\bibinfo {author} {\bibfnamefont {L.}~\bibnamefont
  {Andreoli}}, \bibinfo {author} {\bibfnamefont {J.}~\bibnamefont {Carlson}},
  \bibinfo {author} {\bibfnamefont {A.}~\bibnamefont {Lovato}}, \bibinfo
  {author} {\bibfnamefont {S.}~\bibnamefont {Pastore}}, \bibinfo {author}
  {\bibfnamefont {N.}~\bibnamefont {Rocco}}, \ and\ \bibinfo {author}
  {\bibfnamefont {R.~B.}\ \bibnamefont {Wiringa}},\ }\href {\doibase
  10.1103/PhysRevC.105.014002} {\bibfield  {journal} {\bibinfo  {journal}
  {Phys. Rev. C}\ }\textbf {\bibinfo {volume} {105}},\ \bibinfo {pages}
  {014002} (\bibinfo {year} {2022})},\ \Eprint
  {http://arxiv.org/abs/2108.10824} {arXiv:2108.10824 [nucl-th]} \BibitemShut
  {NoStop}%
\bibitem [{\citenamefont {Chakrani}\ \emph {et~al.}(2022)\citenamefont
  {Chakrani}, \citenamefont {Buizza~Avanzini},\ and\ \citenamefont
  {Dolan}}]{Chakrani:2022tey}%
  \BibitemOpen
  \bibfield  {author} {\bibinfo {author} {\bibfnamefont {J.}~\bibnamefont
  {Chakrani}}, \bibinfo {author} {\bibfnamefont {M.}~\bibnamefont
  {Buizza~Avanzini}}, \ and\ \bibinfo {author} {\bibfnamefont {S.}~\bibnamefont
  {Dolan}},\ }in\ \href@noop {} {\emph {\bibinfo {booktitle} {{22nd
  International Workshop on Neutrinos from Accelerators}}}}\ (\bibinfo {year}
  {2022})\ \Eprint {http://arxiv.org/abs/2202.03219} {arXiv:2202.03219
  [hep-ph]} \BibitemShut {NoStop}%
\bibitem [{\citenamefont {Ankowski}\ \emph {et~al.}(2022)\citenamefont
  {Ankowski} \emph {et~al.}}]{Ankowski:2022thw}%
  \BibitemOpen
  \bibfield  {author} {\bibinfo {author} {\bibfnamefont {A.~M.}\ \bibnamefont
  {Ankowski}} \emph {et~al.}\ }(\bibinfo {year} {2022})\ \Eprint
  {http://arxiv.org/abs/2203.06853} {arXiv:2203.06853 [hep-ex]} \BibitemShut
  {NoStop}%
\bibitem [{\citenamefont {Barrow}\ \emph {et~al.}(2020)\citenamefont {Barrow}
  \emph {et~al.}}]{Barrow:2020gzb}%
  \BibitemOpen
  \bibfield  {author} {\bibinfo {author} {\bibfnamefont {J.}~\bibnamefont
  {Barrow}} \emph {et~al.},\ }\href@noop {} {\  (\bibinfo {year} {2020})},\
  \bibinfo {note} {arXiv:2008.06566},\ \Eprint
  {http://arxiv.org/abs/2008.06566} {arXiv:2008.06566 [hep-ex]} \BibitemShut
  {NoStop}%
\bibitem [{\citenamefont {Amoroso}\ \emph {et~al.}(2020)\citenamefont {Amoroso}
  \emph {et~al.}}]{Amoroso:2020lgh}%
  \BibitemOpen
  \bibfield  {author} {\bibinfo {author} {\bibfnamefont {S.}~\bibnamefont
  {Amoroso}} \emph {et~al.},\ }in\ \href@noop {} {\emph {\bibinfo {booktitle}
  {{11th Les Houches Workshop on Physics at TeV Colliders}: {PhysTeV Les
  Houches}}}}\ (\bibinfo {year} {2020})\ \Eprint
  {http://arxiv.org/abs/2003.01700} {arXiv:2003.01700 [hep-ph]} \BibitemShut
  {NoStop}%
\bibitem [{\citenamefont {Brooijmans}\ \emph {et~al.}(2020)\citenamefont
  {Brooijmans} \emph {et~al.}}]{Brooijmans:2020yij}%
  \BibitemOpen
  \bibfield  {author} {\bibinfo {author} {\bibfnamefont {G.}~\bibnamefont
  {Brooijmans}} \emph {et~al.},\ }in\ \href@noop {} {\emph {\bibinfo
  {booktitle} {{11th Les Houches Workshop on Physics at TeV Colliders}:
  {PhysTeV Les Houches}}}}\ (\bibinfo {year} {2020})\ \Eprint
  {http://arxiv.org/abs/2002.12220} {arXiv:2002.12220 [hep-ph]} \BibitemShut
  {NoStop}%
\bibitem [{\citenamefont {Moniz}\ \emph {et~al.}(1971)\citenamefont {Moniz},
  \citenamefont {Sick}, \citenamefont {Whitney}, \citenamefont {Ficenec},
  \citenamefont {Kephart},\ and\ \citenamefont {Trower}}]{PhysRevLett.26.445}%
  \BibitemOpen
  \bibfield  {author} {\bibinfo {author} {\bibfnamefont {E.~J.}\ \bibnamefont
  {Moniz}}, \bibinfo {author} {\bibfnamefont {I.}~\bibnamefont {Sick}},
  \bibinfo {author} {\bibfnamefont {R.~R.}\ \bibnamefont {Whitney}}, \bibinfo
  {author} {\bibfnamefont {J.~R.}\ \bibnamefont {Ficenec}}, \bibinfo {author}
  {\bibfnamefont {R.~D.}\ \bibnamefont {Kephart}}, \ and\ \bibinfo {author}
  {\bibfnamefont {W.~P.}\ \bibnamefont {Trower}},\ }\href {\doibase
  10.1103/PhysRevLett.26.445} {\bibfield  {journal} {\bibinfo  {journal} {Phys.
  Rev. Lett.}\ }\textbf {\bibinfo {volume} {26}},\ \bibinfo {pages} {445}
  (\bibinfo {year} {1971})}\BibitemShut {NoStop}%
\bibitem [{\citenamefont {Smith}\ and\ \citenamefont
  {Moniz}(1972)}]{Smith:1972xh}%
  \BibitemOpen
  \bibfield  {author} {\bibinfo {author} {\bibfnamefont {R.~A.}\ \bibnamefont
  {Smith}}\ and\ \bibinfo {author} {\bibfnamefont {E.~J.}\ \bibnamefont
  {Moniz}},\ }\href {\doibase 10.1016/0550-3213(75)90612-4} {\bibfield
  {journal} {\bibinfo  {journal} {Nucl. Phys. B}\ }\textbf {\bibinfo {volume}
  {43}},\ \bibinfo {pages} {605} (\bibinfo {year} {1972})},\ \bibinfo {note}
  {[Erratum: Nucl. Phys. B {\bf 101}, 547 (1975)]}\BibitemShut {NoStop}%
\bibitem [{\citenamefont {Nieves}\ \emph {et~al.}(2016)\citenamefont {Nieves},
  \citenamefont {Gran}, \citenamefont {Ruiz~Simo}, \citenamefont {S\'anchez},\
  and\ \citenamefont {Vicente~Vacas}}]{Nieves:2014lpa}%
  \BibitemOpen
  \bibfield  {author} {\bibinfo {author} {\bibfnamefont {J.}~\bibnamefont
  {Nieves}}, \bibinfo {author} {\bibfnamefont {R.}~\bibnamefont {Gran}},
  \bibinfo {author} {\bibfnamefont {I.}~\bibnamefont {Ruiz~Simo}}, \bibinfo
  {author} {\bibfnamefont {F.}~\bibnamefont {S\'anchez}}, \ and\ \bibinfo
  {author} {\bibfnamefont {M.~J.}\ \bibnamefont {Vicente~Vacas}},\ }\href
  {\doibase 10.1016/j.nuclphysbps.2015.09.295} {\bibfield  {journal} {\bibinfo
  {journal} {Nucl. Part. Phys. Proc.}\ }\textbf {\bibinfo {volume} {273-275}},\
  \bibinfo {pages} {1830} (\bibinfo {year} {2016})},\ \Eprint
  {http://arxiv.org/abs/1411.7821} {arXiv:1411.7821 [hep-ph]} \BibitemShut
  {NoStop}%
\bibitem [{\citenamefont {Benhar}\ \emph {et~al.}(1994)\citenamefont {Benhar},
  \citenamefont {Fabrocini}, \citenamefont {Fantoni},\ and\ \citenamefont
  {Sick}}]{Benhar:1994hw}%
  \BibitemOpen
  \bibfield  {author} {\bibinfo {author} {\bibfnamefont {O.}~\bibnamefont
  {Benhar}}, \bibinfo {author} {\bibfnamefont {A.}~\bibnamefont {Fabrocini}},
  \bibinfo {author} {\bibfnamefont {S.}~\bibnamefont {Fantoni}}, \ and\
  \bibinfo {author} {\bibfnamefont {I.}~\bibnamefont {Sick}},\ }\href {\doibase
  10.1016/0375-9474(94)90920-2} {\bibfield  {journal} {\bibinfo  {journal}
  {Nucl. Phys. A}\ }\textbf {\bibinfo {volume} {579}},\ \bibinfo {pages} {493}
  (\bibinfo {year} {1994})}\BibitemShut {NoStop}%
\bibitem [{\citenamefont {Benhar}\ \emph {et~al.}(2005)\citenamefont {Benhar},
  \citenamefont {Farina}, \citenamefont {Nakamura}, \citenamefont {Sakuda},\
  and\ \citenamefont {Seki}}]{Benhar:2005dj}%
  \BibitemOpen
  \bibfield  {author} {\bibinfo {author} {\bibfnamefont {O.}~\bibnamefont
  {Benhar}}, \bibinfo {author} {\bibfnamefont {N.}~\bibnamefont {Farina}},
  \bibinfo {author} {\bibfnamefont {H.}~\bibnamefont {Nakamura}}, \bibinfo
  {author} {\bibfnamefont {M.}~\bibnamefont {Sakuda}}, \ and\ \bibinfo {author}
  {\bibfnamefont {R.}~\bibnamefont {Seki}},\ }\href {\doibase
  10.1103/PhysRevD.72.053005} {\bibfield  {journal} {\bibinfo  {journal} {Phys.
  Rev. D}\ }\textbf {\bibinfo {volume} {72}},\ \bibinfo {pages} {053005}
  (\bibinfo {year} {2005})},\ \Eprint {http://arxiv.org/abs/hep-ph/0506116}
  {arXiv:hep-ph/0506116} \BibitemShut {NoStop}%
\bibitem [{\citenamefont {Abe}\ \emph {et~al.}(2021)\citenamefont {Abe} \emph
  {et~al.}}]{T2K:2021xwb}%
  \BibitemOpen
  \bibfield  {author} {\bibinfo {author} {\bibfnamefont {K.}~\bibnamefont
  {Abe}} \emph {et~al.} (\bibinfo {collaboration} {T2K}),\ }\href {\doibase
  10.1103/PhysRevD.103.112008} {\bibfield  {journal} {\bibinfo  {journal}
  {Phys. Rev. D}\ }\textbf {\bibinfo {volume} {103}},\ \bibinfo {pages}
  {112008} (\bibinfo {year} {2021})},\ \Eprint
  {http://arxiv.org/abs/2101.03779} {arXiv:2101.03779 [hep-ex]} \BibitemShut
  {NoStop}%
\bibitem [{\citenamefont {Gran}\ \emph {et~al.}(2013)\citenamefont {Gran},
  \citenamefont {Nieves}, \citenamefont {Sanchez},\ and\ \citenamefont
  {Vicente~Vacas}}]{Gran:2013kda}%
  \BibitemOpen
  \bibfield  {author} {\bibinfo {author} {\bibfnamefont {R.}~\bibnamefont
  {Gran}}, \bibinfo {author} {\bibfnamefont {J.}~\bibnamefont {Nieves}},
  \bibinfo {author} {\bibfnamefont {F.}~\bibnamefont {Sanchez}}, \ and\
  \bibinfo {author} {\bibfnamefont {M.~J.}\ \bibnamefont {Vicente~Vacas}},\
  }\href {\doibase 10.1103/PhysRevD.88.113007} {\bibfield  {journal} {\bibinfo
  {journal} {Phys. Rev. D}\ }\textbf {\bibinfo {volume} {88}},\ \bibinfo
  {pages} {113007} (\bibinfo {year} {2013})},\ \Eprint
  {http://arxiv.org/abs/1307.8105} {arXiv:1307.8105 [hep-ph]} \BibitemShut
  {NoStop}%
\bibitem [{\citenamefont {Martini}\ \emph {et~al.}(2009)\citenamefont
  {Martini}, \citenamefont {Ericson}, \citenamefont {Chanfray},\ and\
  \citenamefont {Marteau}}]{Martini:2009uj}%
  \BibitemOpen
  \bibfield  {author} {\bibinfo {author} {\bibfnamefont {M.}~\bibnamefont
  {Martini}}, \bibinfo {author} {\bibfnamefont {M.}~\bibnamefont {Ericson}},
  \bibinfo {author} {\bibfnamefont {G.}~\bibnamefont {Chanfray}}, \ and\
  \bibinfo {author} {\bibfnamefont {J.}~\bibnamefont {Marteau}},\ }\href
  {\doibase 10.1103/PhysRevC.80.065501} {\bibfield  {journal} {\bibinfo
  {journal} {Phys. Rev. C}\ }\textbf {\bibinfo {volume} {80}},\ \bibinfo
  {pages} {065501} (\bibinfo {year} {2009})},\ \Eprint
  {http://arxiv.org/abs/0910.2622} {arXiv:0910.2622 [nucl-th]} \BibitemShut
  {NoStop}%
\bibitem [{\citenamefont {Ruiz~Simo}\ \emph {et~al.}(2017)\citenamefont
  {Ruiz~Simo}, \citenamefont {Amaro}, \citenamefont {Barbaro}, \citenamefont
  {De~Pace}, \citenamefont {Caballero},\ and\ \citenamefont
  {Donnelly}}]{RuizSimo:2016rtu}%
  \BibitemOpen
  \bibfield  {author} {\bibinfo {author} {\bibfnamefont {I.}~\bibnamefont
  {Ruiz~Simo}}, \bibinfo {author} {\bibfnamefont {J.~E.}\ \bibnamefont
  {Amaro}}, \bibinfo {author} {\bibfnamefont {M.~B.}\ \bibnamefont {Barbaro}},
  \bibinfo {author} {\bibfnamefont {A.}~\bibnamefont {De~Pace}}, \bibinfo
  {author} {\bibfnamefont {J.~A.}\ \bibnamefont {Caballero}}, \ and\ \bibinfo
  {author} {\bibfnamefont {T.~W.}\ \bibnamefont {Donnelly}},\ }\href {\doibase
  10.1088/1361-6471/aa6a06} {\bibfield  {journal} {\bibinfo  {journal} {J.
  Phys. G}\ }\textbf {\bibinfo {volume} {44}},\ \bibinfo {pages} {065105}
  (\bibinfo {year} {2017})},\ \Eprint {http://arxiv.org/abs/1604.08423}
  {arXiv:1604.08423 [nucl-th]} \BibitemShut {NoStop}%
\bibitem [{\citenamefont {Gallmeister}\ \emph {et~al.}(2016)\citenamefont
  {Gallmeister}, \citenamefont {Mosel},\ and\ \citenamefont
  {Weil}}]{Gallmeister:2016dnq}%
  \BibitemOpen
  \bibfield  {author} {\bibinfo {author} {\bibfnamefont {K.}~\bibnamefont
  {Gallmeister}}, \bibinfo {author} {\bibfnamefont {U.}~\bibnamefont {Mosel}},
  \ and\ \bibinfo {author} {\bibfnamefont {J.}~\bibnamefont {Weil}},\ }\href
  {\doibase 10.1103/PhysRevC.94.035502} {\bibfield  {journal} {\bibinfo
  {journal} {Phys. Rev. C}\ }\textbf {\bibinfo {volume} {94}},\ \bibinfo
  {pages} {035502} (\bibinfo {year} {2016})},\ \Eprint
  {http://arxiv.org/abs/1605.09391} {arXiv:1605.09391 [nucl-th]} \BibitemShut
  {NoStop}%
\bibitem [{\citenamefont {Freedman}(1974)}]{Freedman:1973yd}%
  \BibitemOpen
  \bibfield  {author} {\bibinfo {author} {\bibfnamefont {D.~Z.}\ \bibnamefont
  {Freedman}},\ }\href {\doibase 10.1103/PhysRevD.9.1389} {\bibfield  {journal}
  {\bibinfo  {journal} {Phys.\ Rev.\ D}\ }\textbf {\bibinfo {volume} {9}},\
  \bibinfo {pages} {1389} (\bibinfo {year} {1974})}\BibitemShut {NoStop}%
\bibitem [{\citenamefont {Abdullah}\ \emph {et~al.}(2022)\citenamefont
  {Abdullah} \emph {et~al.}}]{CEvNS_WP}%
  \BibitemOpen
  \bibfield  {author} {\bibinfo {author} {\bibfnamefont {M.}~\bibnamefont
  {Abdullah}} \emph {et~al.}\ }(\bibinfo {year} {2022})\ \Eprint
  {http://arxiv.org/abs/2203.07361} {arXiv:2203.07361 [hep-ph]} \BibitemShut
  {NoStop}%
\bibitem [{\citenamefont {Serot}(1978)}]{Serot:1978vj}%
  \BibitemOpen
  \bibfield  {author} {\bibinfo {author} {\bibfnamefont {B.}~\bibnamefont
  {Serot}},\ }\href {\doibase 10.1016/0375-9474(78)90561-4} {\bibfield
  {journal} {\bibinfo  {journal} {Nucl. Phys. A}\ }\textbf {\bibinfo {volume}
  {308}},\ \bibinfo {pages} {457} (\bibinfo {year} {1978})}\BibitemShut
  {NoStop}%
\bibitem [{\citenamefont {Donnelly}\ and\ \citenamefont
  {Peccei}(1979)}]{Donnelly:1978tz}%
  \BibitemOpen
  \bibfield  {author} {\bibinfo {author} {\bibfnamefont {T.}~\bibnamefont
  {Donnelly}}\ and\ \bibinfo {author} {\bibfnamefont {R.}~\bibnamefont
  {Peccei}},\ }\href {\doibase 10.1016/0370-1573(79)90010-3} {\bibfield
  {journal} {\bibinfo  {journal} {Phys. Rept.}\ }\textbf {\bibinfo {volume}
  {50}},\ \bibinfo {pages} {1} (\bibinfo {year} {1979})}\BibitemShut {NoStop}%
\bibitem [{\citenamefont {Donnelly}\ and\ \citenamefont
  {Haxton}(1979)}]{Donnelly:1979ezn}%
  \BibitemOpen
  \bibfield  {author} {\bibinfo {author} {\bibfnamefont {T.}~\bibnamefont
  {Donnelly}}\ and\ \bibinfo {author} {\bibfnamefont {W.}~\bibnamefont
  {Haxton}},\ }\href {\doibase 10.1016/0092-640X(79)90003-2} {\bibfield
  {journal} {\bibinfo  {journal} {Atom. Data Nucl. Data Tabl.}\ }\textbf
  {\bibinfo {volume} {23}},\ \bibinfo {pages} {103} (\bibinfo {year}
  {1979})}\BibitemShut {NoStop}%
\bibitem [{\citenamefont {Serot}(1979)}]{Serot:1979yk}%
  \BibitemOpen
  \bibfield  {author} {\bibinfo {author} {\bibfnamefont {B.}~\bibnamefont
  {Serot}},\ }\href {\doibase 10.1016/0375-9474(79)90435-4} {\bibfield
  {journal} {\bibinfo  {journal} {Nucl. Phys. A}\ }\textbf {\bibinfo {volume}
  {322}},\ \bibinfo {pages} {408} (\bibinfo {year} {1979})}\BibitemShut
  {NoStop}%
\bibitem [{\citenamefont {Walecka}(1995)}]{Walecka:1995mi}%
  \BibitemOpen
  \bibfield  {author} {\bibinfo {author} {\bibfnamefont {J.}~\bibnamefont
  {Walecka}},\ }\href@noop {} {\emph {\bibinfo {title} {{Theoretical nuclear
  and subnuclear physics}}}},\ Vol.~\bibinfo {volume} {16}\ (\bibinfo {year}
  {1995})\BibitemShut {NoStop}%
\bibitem [{\citenamefont {Hofstadter}(1956)}]{Hofstadter:1956qs}%
  \BibitemOpen
  \bibfield  {author} {\bibinfo {author} {\bibfnamefont {R.}~\bibnamefont
  {Hofstadter}},\ }\href {\doibase 10.1103/RevModPhys.28.214} {\bibfield
  {journal} {\bibinfo  {journal} {Rev. Mod. Phys.}\ }\textbf {\bibinfo {volume}
  {28}},\ \bibinfo {pages} {214} (\bibinfo {year} {1956})}\BibitemShut
  {NoStop}%
\bibitem [{\citenamefont {De~Vries}\ \emph {et~al.}(1987)\citenamefont
  {De~Vries}, \citenamefont {De~Jager},\ and\ \citenamefont
  {De~Vries}}]{DeJager:1987qc}%
  \BibitemOpen
  \bibfield  {author} {\bibinfo {author} {\bibfnamefont {H.}~\bibnamefont
  {De~Vries}}, \bibinfo {author} {\bibfnamefont {C.}~\bibnamefont {De~Jager}},
  \ and\ \bibinfo {author} {\bibfnamefont {C.}~\bibnamefont {De~Vries}},\
  }\href {\doibase 10.1016/0092-640X(87)90013-1} {\bibfield  {journal}
  {\bibinfo  {journal} {Atom.\ Data Nucl.\ Data Tabl.}\ }\textbf {\bibinfo
  {volume} {36}},\ \bibinfo {pages} {495} (\bibinfo {year} {1987})}\BibitemShut
  {NoStop}%
\bibitem [{\citenamefont {Fricke}\ \emph {et~al.}(1995)\citenamefont {Fricke},
  \citenamefont {Bernhardt}, \citenamefont {Heilig}, \citenamefont {Schaller},
  \citenamefont {Schellenberg}, \citenamefont {Shera},\ and\ \citenamefont
  {de~Jager}}]{Fricke:1995zz}%
  \BibitemOpen
  \bibfield  {author} {\bibinfo {author} {\bibfnamefont {G.}~\bibnamefont
  {Fricke}}, \bibinfo {author} {\bibfnamefont {C.}~\bibnamefont {Bernhardt}},
  \bibinfo {author} {\bibfnamefont {K.}~\bibnamefont {Heilig}}, \bibinfo
  {author} {\bibfnamefont {L.}~\bibnamefont {Schaller}}, \bibinfo {author}
  {\bibfnamefont {L.}~\bibnamefont {Schellenberg}}, \bibinfo {author}
  {\bibfnamefont {E.}~\bibnamefont {Shera}}, \ and\ \bibinfo {author}
  {\bibfnamefont {C.}~\bibnamefont {de~Jager}},\ }\href {\doibase
  10.1006/adnd.1995.1007} {\bibfield  {journal} {\bibinfo  {journal} {Atom.
  Data Nucl. Data Tabl.}\ }\textbf {\bibinfo {volume} {60}},\ \bibinfo {pages}
  {177} (\bibinfo {year} {1995})}\BibitemShut {NoStop}%
\bibitem [{\citenamefont {Angeli}\ and\ \citenamefont
  {Marinova}(2013)}]{Angeli:2013epw}%
  \BibitemOpen
  \bibfield  {author} {\bibinfo {author} {\bibfnamefont {I.}~\bibnamefont
  {Angeli}}\ and\ \bibinfo {author} {\bibfnamefont {K.}~\bibnamefont
  {Marinova}},\ }\href {\doibase 10.1016/j.adt.2011.12.006} {\bibfield
  {journal} {\bibinfo  {journal} {Atom. Data Nucl. Data Tabl.}\ }\textbf
  {\bibinfo {volume} {99}},\ \bibinfo {pages} {69} (\bibinfo {year}
  {2013})}\BibitemShut {NoStop}%
\bibitem [{\citenamefont {Tamii}\ \emph {et~al.}(2011)\citenamefont {Tamii}
  \emph {et~al.}}]{Tamii:2011pv}%
  \BibitemOpen
  \bibfield  {author} {\bibinfo {author} {\bibfnamefont {A.}~\bibnamefont
  {Tamii}} \emph {et~al.},\ }\href {\doibase 10.1103/PhysRevLett.107.062502}
  {\bibfield  {journal} {\bibinfo  {journal} {Phys. Rev. Lett.}\ }\textbf
  {\bibinfo {volume} {107}},\ \bibinfo {pages} {062502} (\bibinfo {year}
  {2011})},\ \Eprint {http://arxiv.org/abs/1104.5431} {arXiv:1104.5431
  [nucl-ex]} \BibitemShut {NoStop}%
\bibitem [{\citenamefont {Rossi}\ \emph {et~al.}(2013)\citenamefont {Rossi}
  \emph {et~al.}}]{Rossi:2013xha}%
  \BibitemOpen
  \bibfield  {author} {\bibinfo {author} {\bibfnamefont {D.}~\bibnamefont
  {Rossi}} \emph {et~al.},\ }\href {\doibase 10.1103/PhysRevLett.111.242503}
  {\bibfield  {journal} {\bibinfo  {journal} {Phys. Rev. Lett.}\ }\textbf
  {\bibinfo {volume} {111}},\ \bibinfo {pages} {242503} (\bibinfo {year}
  {2013})}\BibitemShut {NoStop}%
\bibitem [{\citenamefont {Hashimoto}\ \emph {et~al.}(2015)\citenamefont
  {Hashimoto} \emph {et~al.}}]{Hashimoto:2015ema}%
  \BibitemOpen
  \bibfield  {author} {\bibinfo {author} {\bibfnamefont {T.}~\bibnamefont
  {Hashimoto}} \emph {et~al.},\ }\href {\doibase 10.1103/PhysRevC.92.031305}
  {\bibfield  {journal} {\bibinfo  {journal} {Phys. Rev. C}\ }\textbf {\bibinfo
  {volume} {92}},\ \bibinfo {pages} {031305} (\bibinfo {year} {2015})},\
  \Eprint {http://arxiv.org/abs/1503.08321} {arXiv:1503.08321 [nucl-ex]}
  \BibitemShut {NoStop}%
\bibitem [{\citenamefont {Birkhan}\ \emph {et~al.}(2017)\citenamefont {Birkhan}
  \emph {et~al.}}]{Birkhan:2016qkr}%
  \BibitemOpen
  \bibfield  {author} {\bibinfo {author} {\bibfnamefont {J.}~\bibnamefont
  {Birkhan}} \emph {et~al.},\ }\href {\doibase 10.1103/PhysRevLett.118.252501}
  {\bibfield  {journal} {\bibinfo  {journal} {Phys. Rev. Lett.}\ }\textbf
  {\bibinfo {volume} {118}},\ \bibinfo {pages} {252501} (\bibinfo {year}
  {2017})},\ \Eprint {http://arxiv.org/abs/1611.07072} {arXiv:1611.07072
  [nucl-ex]} \BibitemShut {NoStop}%
\bibitem [{\citenamefont {Thiel}\ \emph {et~al.}(2019)\citenamefont {Thiel},
  \citenamefont {Sfienti}, \citenamefont {Piekarewicz}, \citenamefont
  {Horowitz},\ and\ \citenamefont {Vanderhaeghen}}]{Thiel:2019tkm}%
  \BibitemOpen
  \bibfield  {author} {\bibinfo {author} {\bibfnamefont {M.}~\bibnamefont
  {Thiel}}, \bibinfo {author} {\bibfnamefont {C.}~\bibnamefont {Sfienti}},
  \bibinfo {author} {\bibfnamefont {J.}~\bibnamefont {Piekarewicz}}, \bibinfo
  {author} {\bibfnamefont {C.}~\bibnamefont {Horowitz}}, \ and\ \bibinfo
  {author} {\bibfnamefont {M.}~\bibnamefont {Vanderhaeghen}},\ }\href {\doibase
  10.1088/1361-6471/ab2c6d} {\bibfield  {journal} {\bibinfo  {journal} {J.
  Phys. G}\ }\textbf {\bibinfo {volume} {46}},\ \bibinfo {pages} {093003}
  (\bibinfo {year} {2019})},\ \Eprint {http://arxiv.org/abs/1904.12269}
  {arXiv:1904.12269 [nucl-ex]} \BibitemShut {NoStop}%
\bibitem [{\citenamefont {Donnelly}\ \emph {et~al.}(1989)\citenamefont
  {Donnelly}, \citenamefont {Dubach},\ and\ \citenamefont
  {Sick}}]{Donnelly:1989qs}%
  \BibitemOpen
  \bibfield  {author} {\bibinfo {author} {\bibfnamefont {T.}~\bibnamefont
  {Donnelly}}, \bibinfo {author} {\bibfnamefont {J.}~\bibnamefont {Dubach}}, \
  and\ \bibinfo {author} {\bibfnamefont {I.}~\bibnamefont {Sick}},\ }\href
  {\doibase 10.1016/0375-9474(89)90432-6} {\bibfield  {journal} {\bibinfo
  {journal} {Nucl. Phys. A}\ }\textbf {\bibinfo {volume} {503}},\ \bibinfo
  {pages} {589} (\bibinfo {year} {1989})}\BibitemShut {NoStop}%
\bibitem [{\citenamefont {Abrahamyan}\ \emph {et~al.}(2012)\citenamefont
  {Abrahamyan} \emph {et~al.}}]{Abrahamyan:2012gp}%
  \BibitemOpen
  \bibfield  {author} {\bibinfo {author} {\bibfnamefont {S.}~\bibnamefont
  {Abrahamyan}} \emph {et~al.},\ }\href {\doibase
  10.1103/PhysRevLett.108.112502} {\bibfield  {journal} {\bibinfo  {journal}
  {Phys. Rev. Lett.}\ }\textbf {\bibinfo {volume} {108}},\ \bibinfo {pages}
  {112502} (\bibinfo {year} {2012})},\ \Eprint {http://arxiv.org/abs/1201.2568}
  {arXiv:1201.2568 [nucl-ex]} \BibitemShut {NoStop}%
\bibitem [{\citenamefont {Horowitz}\ \emph {et~al.}(2012)\citenamefont
  {Horowitz} \emph {et~al.}}]{Horowitz:2012tj}%
  \BibitemOpen
  \bibfield  {author} {\bibinfo {author} {\bibfnamefont {C.}~\bibnamefont
  {Horowitz}} \emph {et~al.},\ }\href {\doibase 10.1103/PhysRevC.85.032501}
  {\bibfield  {journal} {\bibinfo  {journal} {Phys. Rev. C}\ }\textbf {\bibinfo
  {volume} {85}},\ \bibinfo {pages} {032501} (\bibinfo {year} {2012})},\
  \Eprint {http://arxiv.org/abs/1202.1468} {arXiv:1202.1468 [nucl-ex]}
  \BibitemShut {NoStop}%
\bibitem [{\citenamefont {Adhikari}\ \emph {et~al.}(2021)\citenamefont
  {Adhikari} \emph {et~al.}}]{PREX:2021umo}%
  \BibitemOpen
  \bibfield  {author} {\bibinfo {author} {\bibfnamefont {D.}~\bibnamefont
  {Adhikari}} \emph {et~al.} (\bibinfo {collaboration} {PREX}),\ }\href
  {\doibase 10.1103/PhysRevLett.126.172502} {\bibfield  {journal} {\bibinfo
  {journal} {Phys. Rev. Lett.}\ }\textbf {\bibinfo {volume} {126}},\ \bibinfo
  {pages} {172502} (\bibinfo {year} {2021})},\ \Eprint
  {http://arxiv.org/abs/2102.10767} {arXiv:2102.10767 [nucl-ex]} \BibitemShut
  {NoStop}%
\bibitem [{\citenamefont {Kumar}(2020)}]{Kumar:2020ejz}%
  \BibitemOpen
  \bibfield  {author} {\bibinfo {author} {\bibfnamefont {K.~S.}\ \bibnamefont
  {Kumar}} (\bibinfo {collaboration} {PREX, CREX}),\ }\href {\doibase
  10.1016/j.aop.2019.168012} {\bibfield  {journal} {\bibinfo  {journal} {Annals
  Phys.}\ }\textbf {\bibinfo {volume} {412}},\ \bibinfo {pages} {168012}
  (\bibinfo {year} {2020})}\BibitemShut {NoStop}%
\bibitem [{\citenamefont {Becker}\ \emph {et~al.}(2018)\citenamefont {Becker}
  \emph {et~al.}}]{Becker:2018ggl}%
  \BibitemOpen
  \bibfield  {author} {\bibinfo {author} {\bibfnamefont {D.}~\bibnamefont
  {Becker}} \emph {et~al.},\ }\href {\doibase 10.1140/epja/i2018-12611-6} {\
  (\bibinfo {year} {2018}),\ 10.1140/epja/i2018-12611-6},\ \Eprint
  {http://arxiv.org/abs/1802.04759} {arXiv:1802.04759 [nucl-ex]} \BibitemShut
  {NoStop}%
\bibitem [{\citenamefont {Horowitz}\ \emph {et~al.}(2003)\citenamefont
  {Horowitz}, \citenamefont {Coakley},\ and\ \citenamefont
  {McKinsey}}]{Horowitz:2003cz}%
  \BibitemOpen
  \bibfield  {author} {\bibinfo {author} {\bibfnamefont {C.~J.}\ \bibnamefont
  {Horowitz}}, \bibinfo {author} {\bibfnamefont {K.~J.}\ \bibnamefont
  {Coakley}}, \ and\ \bibinfo {author} {\bibfnamefont {D.~N.}\ \bibnamefont
  {McKinsey}},\ }\href {\doibase 10.1103/PhysRevD.68.023005} {\bibfield
  {journal} {\bibinfo  {journal} {Phys. Rev. D}\ }\textbf {\bibinfo {volume}
  {68}},\ \bibinfo {pages} {023005} (\bibinfo {year} {2003})},\ \Eprint
  {http://arxiv.org/abs/astro-ph/0302071} {arXiv:astro-ph/0302071 [astro-ph]}
  \BibitemShut {NoStop}%
\bibitem [{\citenamefont {Ciuffoli}\ \emph {et~al.}(2018)\citenamefont
  {Ciuffoli}, \citenamefont {Evslin}, \citenamefont {Fu},\ and\ \citenamefont
  {Tang}}]{Ciuffoli:2018qem}%
  \BibitemOpen
  \bibfield  {author} {\bibinfo {author} {\bibfnamefont {E.}~\bibnamefont
  {Ciuffoli}}, \bibinfo {author} {\bibfnamefont {J.}~\bibnamefont {Evslin}},
  \bibinfo {author} {\bibfnamefont {Q.}~\bibnamefont {Fu}}, \ and\ \bibinfo
  {author} {\bibfnamefont {J.}~\bibnamefont {Tang}},\ }\href {\doibase
  10.1103/PhysRevD.97.113003} {\bibfield  {journal} {\bibinfo  {journal} {Phys.
  Rev. D}\ }\textbf {\bibinfo {volume} {97}},\ \bibinfo {pages} {113003}
  (\bibinfo {year} {2018})},\ \Eprint {http://arxiv.org/abs/1801.02166}
  {arXiv:1801.02166 [physics.ins-det]} \BibitemShut {NoStop}%
\bibitem [{\citenamefont {Yang}\ \emph {et~al.}(2019)\citenamefont {Yang},
  \citenamefont {Hernandez},\ and\ \citenamefont {Piekarewicz}}]{Yang:2019pbx}%
  \BibitemOpen
  \bibfield  {author} {\bibinfo {author} {\bibfnamefont {J.}~\bibnamefont
  {Yang}}, \bibinfo {author} {\bibfnamefont {J.~A.}\ \bibnamefont {Hernandez}},
  \ and\ \bibinfo {author} {\bibfnamefont {J.}~\bibnamefont {Piekarewicz}},\
  }\href {\doibase 10.1103/PhysRevC.100.054301} {\bibfield  {journal} {\bibinfo
   {journal} {Phys. Rev. C}\ }\textbf {\bibinfo {volume} {100}},\ \bibinfo
  {pages} {054301} (\bibinfo {year} {2019})},\ \Eprint
  {http://arxiv.org/abs/1908.10939} {arXiv:1908.10939 [nucl-th]} \BibitemShut
  {NoStop}%
\bibitem [{\citenamefont {Papoulias}\ \emph {et~al.}(2020)\citenamefont
  {Papoulias}, \citenamefont {Kosmas}, \citenamefont {Sahu}, \citenamefont
  {Kota},\ and\ \citenamefont {Hota}}]{Papoulias:2019lfi}%
  \BibitemOpen
  \bibfield  {author} {\bibinfo {author} {\bibfnamefont {D.~K.}\ \bibnamefont
  {Papoulias}}, \bibinfo {author} {\bibfnamefont {T.~S.}\ \bibnamefont
  {Kosmas}}, \bibinfo {author} {\bibfnamefont {R.}~\bibnamefont {Sahu}},
  \bibinfo {author} {\bibfnamefont {V.~K.~B.}\ \bibnamefont {Kota}}, \ and\
  \bibinfo {author} {\bibfnamefont {M.}~\bibnamefont {Hota}},\ }\href {\doibase
  10.1016/j.physletb.2019.135133} {\bibfield  {journal} {\bibinfo  {journal}
  {Phys. Lett. B}\ }\textbf {\bibinfo {volume} {800}},\ \bibinfo {pages}
  {135133} (\bibinfo {year} {2020})},\ \Eprint
  {http://arxiv.org/abs/1903.03722} {arXiv:1903.03722 [hep-ph]} \BibitemShut
  {NoStop}%
\bibitem [{\citenamefont {Co'}\ \emph {et~al.}(2020)\citenamefont {Co'},
  \citenamefont {Anguiano},\ and\ \citenamefont {Lallena}}]{Co:2020gwl}%
  \BibitemOpen
  \bibfield  {author} {\bibinfo {author} {\bibfnamefont {G.}~\bibnamefont
  {Co'}}, \bibinfo {author} {\bibfnamefont {M.}~\bibnamefont {Anguiano}}, \
  and\ \bibinfo {author} {\bibfnamefont {A.}~\bibnamefont {Lallena}},\ }\href
  {\doibase 10.1088/1475-7516/2020/04/044} {\bibfield  {journal} {\bibinfo
  {journal} {JCAP}\ }\textbf {\bibinfo {volume} {04}},\ \bibinfo {pages} {044}
  (\bibinfo {year} {2020})},\ \Eprint {http://arxiv.org/abs/2001.04684}
  {arXiv:2001.04684 [nucl-th]} \BibitemShut {NoStop}%
\bibitem [{\citenamefont {Coloma}\ \emph {et~al.}(2020)\citenamefont {Coloma},
  \citenamefont {Esteban}, \citenamefont {Gonzalez-Garcia},\ and\ \citenamefont
  {Men\'endez}}]{Coloma:2020nhf}%
  \BibitemOpen
  \bibfield  {author} {\bibinfo {author} {\bibfnamefont {P.}~\bibnamefont
  {Coloma}}, \bibinfo {author} {\bibfnamefont {I.}~\bibnamefont {Esteban}},
  \bibinfo {author} {\bibfnamefont {M.~C.}\ \bibnamefont {Gonzalez-Garcia}}, \
  and\ \bibinfo {author} {\bibfnamefont {J.}~\bibnamefont {Men\'endez}},\
  }\href {\doibase 10.1007/JHEP08(2020)030} {\bibfield  {journal} {\bibinfo
  {journal} {JHEP}\ }\textbf {\bibinfo {volume} {08}},\ \bibinfo {pages} {030}
  (\bibinfo {year} {2020})},\ \Eprint {http://arxiv.org/abs/2006.08624}
  {arXiv:2006.08624 [hep-ph]} \BibitemShut {NoStop}%
\bibitem [{\citenamefont {Van~Dessel}\ \emph {et~al.}(2020)\citenamefont
  {Van~Dessel}, \citenamefont {Pandey}, \citenamefont {Ray},\ and\
  \citenamefont {Jachowicz}}]{VanDessel:2020epd}%
  \BibitemOpen
  \bibfield  {author} {\bibinfo {author} {\bibfnamefont {N.}~\bibnamefont
  {Van~Dessel}}, \bibinfo {author} {\bibfnamefont {V.}~\bibnamefont {Pandey}},
  \bibinfo {author} {\bibfnamefont {H.}~\bibnamefont {Ray}}, \ and\ \bibinfo
  {author} {\bibfnamefont {N.}~\bibnamefont {Jachowicz}},\ }\href@noop {} {\
  (\bibinfo {year} {2020})},\ \Eprint {http://arxiv.org/abs/2007.03658}
  {arXiv:2007.03658 [nucl-th]} \BibitemShut {NoStop}%
\bibitem [{\citenamefont {Roca-Maza}\ \emph {et~al.}(2011)\citenamefont
  {Roca-Maza}, \citenamefont {Centelles}, \citenamefont {Vinas},\ and\
  \citenamefont {Warda}}]{RocaMaza:2011pm}%
  \BibitemOpen
  \bibfield  {author} {\bibinfo {author} {\bibfnamefont {X.}~\bibnamefont
  {Roca-Maza}}, \bibinfo {author} {\bibfnamefont {M.}~\bibnamefont
  {Centelles}}, \bibinfo {author} {\bibfnamefont {X.}~\bibnamefont {Vinas}}, \
  and\ \bibinfo {author} {\bibfnamefont {M.}~\bibnamefont {Warda}},\ }\href
  {\doibase 10.1103/PhysRevLett.106.252501} {\bibfield  {journal} {\bibinfo
  {journal} {Phys. Rev. Lett.}\ }\textbf {\bibinfo {volume} {106}},\ \bibinfo
  {pages} {252501} (\bibinfo {year} {2011})},\ \Eprint
  {http://arxiv.org/abs/1103.1762} {arXiv:1103.1762 [nucl-th]} \BibitemShut
  {NoStop}%
\bibitem [{\citenamefont {Tsang}\ \emph {et~al.}(2012)\citenamefont {Tsang}
  \emph {et~al.}}]{Tsang:2012se}%
  \BibitemOpen
  \bibfield  {author} {\bibinfo {author} {\bibfnamefont {M.}~\bibnamefont
  {Tsang}} \emph {et~al.},\ }\href {\doibase 10.1103/PhysRevC.86.015803}
  {\bibfield  {journal} {\bibinfo  {journal} {Phys. Rev. C}\ }\textbf {\bibinfo
  {volume} {86}},\ \bibinfo {pages} {015803} (\bibinfo {year} {2012})},\
  \Eprint {http://arxiv.org/abs/1204.0466} {arXiv:1204.0466 [nucl-ex]}
  \BibitemShut {NoStop}%
\bibitem [{\citenamefont {Lattimer}\ and\ \citenamefont
  {Lim}(2013)}]{Lattimer:2012xj}%
  \BibitemOpen
  \bibfield  {author} {\bibinfo {author} {\bibfnamefont {J.~M.}\ \bibnamefont
  {Lattimer}}\ and\ \bibinfo {author} {\bibfnamefont {Y.}~\bibnamefont {Lim}},\
  }\href {\doibase 10.1088/0004-637X/771/1/51} {\bibfield  {journal} {\bibinfo
  {journal} {Astrophys. J.}\ }\textbf {\bibinfo {volume} {771}},\ \bibinfo
  {pages} {51} (\bibinfo {year} {2013})},\ \Eprint
  {http://arxiv.org/abs/1203.4286} {arXiv:1203.4286 [nucl-th]} \BibitemShut
  {NoStop}%
\bibitem [{\citenamefont {Hebeler}\ \emph {et~al.}(2013)\citenamefont
  {Hebeler}, \citenamefont {Lattimer}, \citenamefont {Pethick},\ and\
  \citenamefont {Schwenk}}]{Hebeler:2013nza}%
  \BibitemOpen
  \bibfield  {author} {\bibinfo {author} {\bibfnamefont {K.}~\bibnamefont
  {Hebeler}}, \bibinfo {author} {\bibfnamefont {J.~M.}\ \bibnamefont
  {Lattimer}}, \bibinfo {author} {\bibfnamefont {C.~J.}\ \bibnamefont
  {Pethick}}, \ and\ \bibinfo {author} {\bibfnamefont {A.}~\bibnamefont
  {Schwenk}},\ }\href {\doibase 10.1088/0004-637X/773/1/11} {\bibfield
  {journal} {\bibinfo  {journal} {Astrophys. J.}\ }\textbf {\bibinfo {volume}
  {773}},\ \bibinfo {pages} {11} (\bibinfo {year} {2013})},\ \Eprint
  {http://arxiv.org/abs/1303.4662} {arXiv:1303.4662 [astro-ph.SR]} \BibitemShut
  {NoStop}%
\bibitem [{\citenamefont {Hagen}\ \emph {et~al.}(2015)\citenamefont {Hagen}
  \emph {et~al.}}]{Hagen:2015yea}%
  \BibitemOpen
  \bibfield  {author} {\bibinfo {author} {\bibfnamefont {G.}~\bibnamefont
  {Hagen}} \emph {et~al.},\ }\href {\doibase 10.1038/nphys3529} {\bibfield
  {journal} {\bibinfo  {journal} {Nature Phys.}\ }\textbf {\bibinfo {volume}
  {12}},\ \bibinfo {pages} {186} (\bibinfo {year} {2015})},\ \Eprint
  {http://arxiv.org/abs/1509.07169} {arXiv:1509.07169 [nucl-th]} \BibitemShut
  {NoStop}%
\bibitem [{\citenamefont {Helm}(1956)}]{Helm:1956zz}%
  \BibitemOpen
  \bibfield  {author} {\bibinfo {author} {\bibfnamefont {R.~H.}\ \bibnamefont
  {Helm}},\ }\href {\doibase 10.1103/PhysRev.104.1466} {\bibfield  {journal}
  {\bibinfo  {journal} {Phys. Rev.}\ }\textbf {\bibinfo {volume} {104}},\
  \bibinfo {pages} {1466} (\bibinfo {year} {1956})}\BibitemShut {NoStop}%
\bibitem [{\citenamefont {Klein}\ and\ \citenamefont
  {Nystrand}(1999)}]{Klein:1999qj}%
  \BibitemOpen
  \bibfield  {author} {\bibinfo {author} {\bibfnamefont {S.}~\bibnamefont
  {Klein}}\ and\ \bibinfo {author} {\bibfnamefont {J.}~\bibnamefont
  {Nystrand}},\ }\href {\doibase 10.1103/PhysRevC.60.014903} {\bibfield
  {journal} {\bibinfo  {journal} {Phys. Rev. C}\ }\textbf {\bibinfo {volume}
  {60}},\ \bibinfo {pages} {014903} (\bibinfo {year} {1999})},\ \Eprint
  {http://arxiv.org/abs/hep-ph/9902259} {arXiv:hep-ph/9902259} \BibitemShut
  {NoStop}%
\bibitem [{\citenamefont {Hoferichter}\ \emph
  {et~al.}(2016{\natexlab{a}})\citenamefont {Hoferichter}, \citenamefont
  {Klos}, \citenamefont {Men\'endez},\ and\ \citenamefont
  {Schwenk}}]{Hoferichter:2016nvd}%
  \BibitemOpen
  \bibfield  {author} {\bibinfo {author} {\bibfnamefont {M.}~\bibnamefont
  {Hoferichter}}, \bibinfo {author} {\bibfnamefont {P.}~\bibnamefont {Klos}},
  \bibinfo {author} {\bibfnamefont {J.}~\bibnamefont {Men\'endez}}, \ and\
  \bibinfo {author} {\bibfnamefont {A.}~\bibnamefont {Schwenk}},\ }\href
  {\doibase 10.1103/PhysRevD.94.063505} {\bibfield  {journal} {\bibinfo
  {journal} {Phys. Rev. D}\ }\textbf {\bibinfo {volume} {94}},\ \bibinfo
  {pages} {063505} (\bibinfo {year} {2016}{\natexlab{a}})},\ \Eprint
  {http://arxiv.org/abs/1605.08043} {arXiv:1605.08043 [hep-ph]} \BibitemShut
  {NoStop}%
\bibitem [{\citenamefont {Hoferichter}\ \emph {et~al.}(2019)\citenamefont
  {Hoferichter}, \citenamefont {Klos}, \citenamefont {Men{\'e}ndez},\ and\
  \citenamefont {Schwenk}}]{Hoferichter:2018acd}%
  \BibitemOpen
  \bibfield  {author} {\bibinfo {author} {\bibfnamefont {M.}~\bibnamefont
  {Hoferichter}}, \bibinfo {author} {\bibfnamefont {P.}~\bibnamefont {Klos}},
  \bibinfo {author} {\bibfnamefont {J.}~\bibnamefont {Men{\'e}ndez}}, \ and\
  \bibinfo {author} {\bibfnamefont {A.}~\bibnamefont {Schwenk}},\ }\href
  {\doibase 10.1103/PhysRevD.99.055031} {\bibfield  {journal} {\bibinfo
  {journal} {Phys. Rev. D}\ }\textbf {\bibinfo {volume} {99}},\ \bibinfo
  {pages} {055031} (\bibinfo {year} {2019})},\ \Eprint
  {http://arxiv.org/abs/1812.05617} {arXiv:1812.05617 [hep-ph]} \BibitemShut
  {NoStop}%
\bibitem [{\citenamefont {Barranco}\ \emph {et~al.}(2005)\citenamefont
  {Barranco}, \citenamefont {Miranda},\ and\ \citenamefont
  {Rashba}}]{Barranco:2005yy}%
  \BibitemOpen
  \bibfield  {author} {\bibinfo {author} {\bibfnamefont {J.}~\bibnamefont
  {Barranco}}, \bibinfo {author} {\bibfnamefont {O.~G.}\ \bibnamefont
  {Miranda}}, \ and\ \bibinfo {author} {\bibfnamefont {T.~I.}\ \bibnamefont
  {Rashba}},\ }\href {\doibase 10.1088/1126-6708/2005/12/021} {\bibfield
  {journal} {\bibinfo  {journal} {JHEP}\ }\textbf {\bibinfo {volume} {12}},\
  \bibinfo {pages} {021} (\bibinfo {year} {2005})},\ \Eprint
  {http://arxiv.org/abs/hep-ph/0508299} {arXiv:hep-ph/0508299 [hep-ph]}
  \BibitemShut {NoStop}%
\bibitem [{\citenamefont {Erler}\ and\ \citenamefont
  {Su}(2013)}]{Erler:2013xha}%
  \BibitemOpen
  \bibfield  {author} {\bibinfo {author} {\bibfnamefont {J.}~\bibnamefont
  {Erler}}\ and\ \bibinfo {author} {\bibfnamefont {S.}~\bibnamefont {Su}},\
  }\href {\doibase 10.1016/j.ppnp.2013.03.004} {\bibfield  {journal} {\bibinfo
  {journal} {Prog. Part. Nucl. Phys.}\ }\textbf {\bibinfo {volume} {71}},\
  \bibinfo {pages} {119} (\bibinfo {year} {2013})},\ \Eprint
  {http://arxiv.org/abs/1303.5522} {arXiv:1303.5522 [hep-ph]} \BibitemShut
  {NoStop}%
\bibitem [{\citenamefont {Tomalak}\ \emph {et~al.}(2021)\citenamefont
  {Tomalak}, \citenamefont {Machado}, \citenamefont {Pandey},\ and\
  \citenamefont {Plestid}}]{Tomalak:2020zfh}%
  \BibitemOpen
  \bibfield  {author} {\bibinfo {author} {\bibfnamefont {O.}~\bibnamefont
  {Tomalak}}, \bibinfo {author} {\bibfnamefont {P.}~\bibnamefont {Machado}},
  \bibinfo {author} {\bibfnamefont {V.}~\bibnamefont {Pandey}}, \ and\ \bibinfo
  {author} {\bibfnamefont {R.}~\bibnamefont {Plestid}},\ }\href {\doibase
  10.1007/JHEP02(2021)097} {\bibfield  {journal} {\bibinfo  {journal} {JHEP}\
  }\textbf {\bibinfo {volume} {02}},\ \bibinfo {pages} {097} (\bibinfo {year}
  {2021})},\ \Eprint {http://arxiv.org/abs/2011.05960} {arXiv:2011.05960
  [hep-ph]} \BibitemShut {NoStop}%
\bibitem [{\citenamefont {Crivellin}\ \emph {et~al.}(2021)\citenamefont
  {Crivellin}, \citenamefont {Hoferichter}, \citenamefont {Kirk}, \citenamefont
  {Manzari},\ and\ \citenamefont {Schnell}}]{Crivellin:2021bkd}%
  \BibitemOpen
  \bibfield  {author} {\bibinfo {author} {\bibfnamefont {A.}~\bibnamefont
  {Crivellin}}, \bibinfo {author} {\bibfnamefont {M.}~\bibnamefont
  {Hoferichter}}, \bibinfo {author} {\bibfnamefont {M.}~\bibnamefont {Kirk}},
  \bibinfo {author} {\bibfnamefont {C.~A.}\ \bibnamefont {Manzari}}, \ and\
  \bibinfo {author} {\bibfnamefont {L.}~\bibnamefont {Schnell}},\ }\href
  {\doibase 10.1007/JHEP10(2021)221} {\bibfield  {journal} {\bibinfo  {journal}
  {JHEP}\ }\textbf {\bibinfo {volume} {10}},\ \bibinfo {pages} {221} (\bibinfo
  {year} {2021})},\ \Eprint {http://arxiv.org/abs/2107.13569} {arXiv:2107.13569
  [hep-ph]} \BibitemShut {NoStop}%
\bibitem [{\citenamefont {Zyla}\ \emph {et~al.}(2020)\citenamefont {Zyla} \emph
  {et~al.}}]{Zyla:2020zbs}%
  \BibitemOpen
  \bibfield  {author} {\bibinfo {author} {\bibfnamefont {P.~A.}\ \bibnamefont
  {Zyla}} \emph {et~al.} (\bibinfo {collaboration} {Particle Data Group}),\
  }\href {\doibase 10.1093/ptep/ptaa104} {\bibfield  {journal} {\bibinfo
  {journal} {PTEP}\ }\textbf {\bibinfo {volume} {2020}},\ \bibinfo {pages}
  {083C01} (\bibinfo {year} {2020})}\BibitemShut {NoStop}%
\bibitem [{\citenamefont {Balantekin}\ and\ \citenamefont
  {Vassh}(2014)}]{Balantekin:2013sda}%
  \BibitemOpen
  \bibfield  {author} {\bibinfo {author} {\bibfnamefont {A.~B.}\ \bibnamefont
  {Balantekin}}\ and\ \bibinfo {author} {\bibfnamefont {N.}~\bibnamefont
  {Vassh}},\ }\href {\doibase 10.1103/PhysRevD.89.073013} {\bibfield  {journal}
  {\bibinfo  {journal} {Phys. Rev. D}\ }\textbf {\bibinfo {volume} {89}},\
  \bibinfo {pages} {073013} (\bibinfo {year} {2014})},\ \Eprint
  {http://arxiv.org/abs/1312.6858} {arXiv:1312.6858 [hep-ph]} \BibitemShut
  {NoStop}%
\bibitem [{\citenamefont {Beda}\ \emph {et~al.}(2013)\citenamefont {Beda},
  \citenamefont {Brudanin}, \citenamefont {Egorov}, \citenamefont {Medvedev},
  \citenamefont {Pogosov}, \citenamefont {Shevchik}, \citenamefont
  {Shirchenko}, \citenamefont {Starostin},\ and\ \citenamefont
  {Zhitnikov}}]{Beda:2013mta}%
  \BibitemOpen
  \bibfield  {author} {\bibinfo {author} {\bibfnamefont {A.~G.}\ \bibnamefont
  {Beda}}, \bibinfo {author} {\bibfnamefont {V.~B.}\ \bibnamefont {Brudanin}},
  \bibinfo {author} {\bibfnamefont {V.~G.}\ \bibnamefont {Egorov}}, \bibinfo
  {author} {\bibfnamefont {D.~V.}\ \bibnamefont {Medvedev}}, \bibinfo {author}
  {\bibfnamefont {V.~S.}\ \bibnamefont {Pogosov}}, \bibinfo {author}
  {\bibfnamefont {E.~A.}\ \bibnamefont {Shevchik}}, \bibinfo {author}
  {\bibfnamefont {M.~V.}\ \bibnamefont {Shirchenko}}, \bibinfo {author}
  {\bibfnamefont {A.~S.}\ \bibnamefont {Starostin}}, \ and\ \bibinfo {author}
  {\bibfnamefont {I.~V.}\ \bibnamefont {Zhitnikov}},\ }\href {\doibase
  10.1134/S1547477113020027} {\bibfield  {journal} {\bibinfo  {journal} {Phys.
  Part. Nucl. Lett.}\ }\textbf {\bibinfo {volume} {10}},\ \bibinfo {pages}
  {139} (\bibinfo {year} {2013})}\BibitemShut {NoStop}%
\bibitem [{\citenamefont {Agostini}\ \emph {et~al.}(2017)\citenamefont
  {Agostini} \emph {et~al.}}]{Borexino:2017fbd}%
  \BibitemOpen
  \bibfield  {author} {\bibinfo {author} {\bibfnamefont {M.}~\bibnamefont
  {Agostini}} \emph {et~al.} (\bibinfo {collaboration} {Borexino}),\ }\href
  {\doibase 10.1103/PhysRevD.96.091103} {\bibfield  {journal} {\bibinfo
  {journal} {Phys. Rev. D}\ }\textbf {\bibinfo {volume} {96}},\ \bibinfo
  {pages} {091103} (\bibinfo {year} {2017})},\ \Eprint
  {http://arxiv.org/abs/1707.09355} {arXiv:1707.09355 [hep-ex]} \BibitemShut
  {NoStop}%
\bibitem [{\citenamefont {Mori}\ \emph {et~al.}(2021)\citenamefont {Mori},
  \citenamefont {Kusakabe}, \citenamefont {Balantekin}, \citenamefont
  {Kajino},\ and\ \citenamefont {Famiano}}]{Mori:2020qqd}%
  \BibitemOpen
  \bibfield  {author} {\bibinfo {author} {\bibfnamefont {K.}~\bibnamefont
  {Mori}}, \bibinfo {author} {\bibfnamefont {M.}~\bibnamefont {Kusakabe}},
  \bibinfo {author} {\bibfnamefont {A.~B.}\ \bibnamefont {Balantekin}},
  \bibinfo {author} {\bibfnamefont {T.}~\bibnamefont {Kajino}}, \ and\ \bibinfo
  {author} {\bibfnamefont {M.~A.}\ \bibnamefont {Famiano}},\ }\href {\doibase
  10.1093/mnras/stab595} {\bibfield  {journal} {\bibinfo  {journal} {Mon. Not.
  Roy. Astron. Soc.}\ }\textbf {\bibinfo {volume} {503}},\ \bibinfo {pages}
  {2746} (\bibinfo {year} {2021})},\ \Eprint {http://arxiv.org/abs/2009.00293}
  {arXiv:2009.00293 [astro-ph.SR]} \BibitemShut {NoStop}%
\bibitem [{\citenamefont {Capozzi}\ and\ \citenamefont
  {Raffelt}(2020)}]{Capozzi:2020cbu}%
  \BibitemOpen
  \bibfield  {author} {\bibinfo {author} {\bibfnamefont {F.}~\bibnamefont
  {Capozzi}}\ and\ \bibinfo {author} {\bibfnamefont {G.}~\bibnamefont
  {Raffelt}},\ }\href {\doibase 10.1103/PhysRevD.102.083007} {\bibfield
  {journal} {\bibinfo  {journal} {Phys. Rev. D}\ }\textbf {\bibinfo {volume}
  {102}},\ \bibinfo {pages} {083007} (\bibinfo {year} {2020})},\ \Eprint
  {http://arxiv.org/abs/2007.03694} {arXiv:2007.03694 [astro-ph.SR]}
  \BibitemShut {NoStop}%
\bibitem [{\citenamefont {Davidson}\ \emph {et~al.}(2005)\citenamefont
  {Davidson}, \citenamefont {Gorbahn},\ and\ \citenamefont
  {Santamaria}}]{Davidson:2005cs}%
  \BibitemOpen
  \bibfield  {author} {\bibinfo {author} {\bibfnamefont {S.}~\bibnamefont
  {Davidson}}, \bibinfo {author} {\bibfnamefont {M.}~\bibnamefont {Gorbahn}}, \
  and\ \bibinfo {author} {\bibfnamefont {A.}~\bibnamefont {Santamaria}},\
  }\href {\doibase 10.1016/j.physletb.2005.08.086} {\bibfield  {journal}
  {\bibinfo  {journal} {Phys. Lett. B}\ }\textbf {\bibinfo {volume} {626}},\
  \bibinfo {pages} {151} (\bibinfo {year} {2005})},\ \Eprint
  {http://arxiv.org/abs/hep-ph/0506085} {arXiv:hep-ph/0506085} \BibitemShut
  {NoStop}%
\bibitem [{\citenamefont {Bell}\ \emph {et~al.}(2006)\citenamefont {Bell},
  \citenamefont {Gorchtein}, \citenamefont {Ramsey-Musolf}, \citenamefont
  {Vogel},\ and\ \citenamefont {Wang}}]{Bell:2006wi}%
  \BibitemOpen
  \bibfield  {author} {\bibinfo {author} {\bibfnamefont {N.~F.}\ \bibnamefont
  {Bell}}, \bibinfo {author} {\bibfnamefont {M.}~\bibnamefont {Gorchtein}},
  \bibinfo {author} {\bibfnamefont {M.~J.}\ \bibnamefont {Ramsey-Musolf}},
  \bibinfo {author} {\bibfnamefont {P.}~\bibnamefont {Vogel}}, \ and\ \bibinfo
  {author} {\bibfnamefont {P.}~\bibnamefont {Wang}},\ }\href {\doibase
  10.1016/j.physletb.2006.09.055} {\bibfield  {journal} {\bibinfo  {journal}
  {Phys. Lett. B}\ }\textbf {\bibinfo {volume} {642}},\ \bibinfo {pages} {377}
  (\bibinfo {year} {2006})},\ \Eprint {http://arxiv.org/abs/hep-ph/0606248}
  {arXiv:hep-ph/0606248} \BibitemShut {NoStop}%
\bibitem [{\citenamefont {Lu}\ \emph {et~al.}(1998)\citenamefont {Lu},
  \citenamefont {Thomas},\ and\ \citenamefont {Williams}}]{Lu:1997sd}%
  \BibitemOpen
  \bibfield  {author} {\bibinfo {author} {\bibfnamefont {D.-H.}\ \bibnamefont
  {Lu}}, \bibinfo {author} {\bibfnamefont {A.~W.}\ \bibnamefont {Thomas}}, \
  and\ \bibinfo {author} {\bibfnamefont {A.~G.}\ \bibnamefont {Williams}},\
  }\href {\doibase 10.1103/PhysRevC.57.2628} {\bibfield  {journal} {\bibinfo
  {journal} {Phys. Rev. C}\ }\textbf {\bibinfo {volume} {57}},\ \bibinfo
  {pages} {2628} (\bibinfo {year} {1998})},\ \Eprint
  {http://arxiv.org/abs/nucl-th/9706019} {arXiv:nucl-th/9706019} \BibitemShut
  {NoStop}%
\bibitem [{\citenamefont {Zhang}\ \emph
  {et~al.}(2020{\natexlab{a}})\citenamefont {Zhang}, \citenamefont {Hobbs},\
  and\ \citenamefont {Miller}}]{Zhang:2019iyx}%
  \BibitemOpen
  \bibfield  {author} {\bibinfo {author} {\bibfnamefont {X.}~\bibnamefont
  {Zhang}}, \bibinfo {author} {\bibfnamefont {T.~J.}\ \bibnamefont {Hobbs}}, \
  and\ \bibinfo {author} {\bibfnamefont {G.~A.}\ \bibnamefont {Miller}},\
  }\href {\doibase 10.1103/PhysRevD.102.074026} {\bibfield  {journal} {\bibinfo
   {journal} {Phys. Rev. D}\ }\textbf {\bibinfo {volume} {102}},\ \bibinfo
  {pages} {074026} (\bibinfo {year} {2020}{\natexlab{a}})},\ \Eprint
  {http://arxiv.org/abs/1912.07797} {arXiv:1912.07797 [nucl-th]} \BibitemShut
  {NoStop}%
\bibitem [{\citenamefont {Hobbs}\ \emph {et~al.}(2015)\citenamefont {Hobbs},
  \citenamefont {Alberg},\ and\ \citenamefont {Miller}}]{Hobbs:2014lea}%
  \BibitemOpen
  \bibfield  {author} {\bibinfo {author} {\bibfnamefont {T.~J.}\ \bibnamefont
  {Hobbs}}, \bibinfo {author} {\bibfnamefont {M.}~\bibnamefont {Alberg}}, \
  and\ \bibinfo {author} {\bibfnamefont {G.~A.}\ \bibnamefont {Miller}},\
  }\href {\doibase 10.1103/PhysRevC.91.035205} {\bibfield  {journal} {\bibinfo
  {journal} {Phys. Rev. C}\ }\textbf {\bibinfo {volume} {91}},\ \bibinfo
  {pages} {035205} (\bibinfo {year} {2015})},\ \Eprint
  {http://arxiv.org/abs/1412.4871} {arXiv:1412.4871 [nucl-th]} \BibitemShut
  {NoStop}%
\bibitem [{\citenamefont {Perdrisat}\ \emph {et~al.}(2007)\citenamefont
  {Perdrisat}, \citenamefont {Punjabi},\ and\ \citenamefont
  {Vanderhaeghen}}]{Perdrisat:2006hj}%
  \BibitemOpen
  \bibfield  {author} {\bibinfo {author} {\bibfnamefont {C.~F.}\ \bibnamefont
  {Perdrisat}}, \bibinfo {author} {\bibfnamefont {V.}~\bibnamefont {Punjabi}},
  \ and\ \bibinfo {author} {\bibfnamefont {M.}~\bibnamefont {Vanderhaeghen}},\
  }\href {\doibase 10.1016/j.ppnp.2007.05.001} {\bibfield  {journal} {\bibinfo
  {journal} {Prog. Part. Nucl. Phys.}\ }\textbf {\bibinfo {volume} {59}},\
  \bibinfo {pages} {694} (\bibinfo {year} {2007})},\ \Eprint
  {http://arxiv.org/abs/hep-ph/0612014} {arXiv:hep-ph/0612014} \BibitemShut
  {NoStop}%
\bibitem [{\citenamefont {Bernauer}\ \emph {et~al.}(2014)\citenamefont
  {Bernauer} \emph {et~al.}}]{A1:2013fsc}%
  \BibitemOpen
  \bibfield  {author} {\bibinfo {author} {\bibfnamefont {J.~C.}\ \bibnamefont
  {Bernauer}} \emph {et~al.} (\bibinfo {collaboration} {A1}),\ }\href {\doibase
  10.1103/PhysRevC.90.015206} {\bibfield  {journal} {\bibinfo  {journal} {Phys.
  Rev. C}\ }\textbf {\bibinfo {volume} {90}},\ \bibinfo {pages} {015206}
  (\bibinfo {year} {2014})},\ \Eprint {http://arxiv.org/abs/1307.6227}
  {arXiv:1307.6227 [nucl-ex]} \BibitemShut {NoStop}%
\bibitem [{\citenamefont {Ye}\ \emph {et~al.}(2018)\citenamefont {Ye},
  \citenamefont {Arrington}, \citenamefont {Hill},\ and\ \citenamefont
  {Lee}}]{Ye:2017gyb}%
  \BibitemOpen
  \bibfield  {author} {\bibinfo {author} {\bibfnamefont {Z.}~\bibnamefont
  {Ye}}, \bibinfo {author} {\bibfnamefont {J.}~\bibnamefont {Arrington}},
  \bibinfo {author} {\bibfnamefont {R.~J.}\ \bibnamefont {Hill}}, \ and\
  \bibinfo {author} {\bibfnamefont {G.}~\bibnamefont {Lee}},\ }\href {\doibase
  10.1016/j.physletb.2017.11.023} {\bibfield  {journal} {\bibinfo  {journal}
  {Phys. Lett. B}\ }\textbf {\bibinfo {volume} {777}},\ \bibinfo {pages} {8}
  (\bibinfo {year} {2018})},\ \Eprint {http://arxiv.org/abs/1707.09063}
  {arXiv:1707.09063 [nucl-ex]} \BibitemShut {NoStop}%
\bibitem [{\citenamefont {Bernard}\ \emph {et~al.}(1998)\citenamefont
  {Bernard}, \citenamefont {Fearing}, \citenamefont {Hemmert},\ and\
  \citenamefont {Meissner}}]{Bernard:1998gv}%
  \BibitemOpen
  \bibfield  {author} {\bibinfo {author} {\bibfnamefont {V.}~\bibnamefont
  {Bernard}}, \bibinfo {author} {\bibfnamefont {H.~W.}\ \bibnamefont
  {Fearing}}, \bibinfo {author} {\bibfnamefont {T.~R.}\ \bibnamefont
  {Hemmert}}, \ and\ \bibinfo {author} {\bibfnamefont {U.~G.}\ \bibnamefont
  {Meissner}},\ }\href {\doibase 10.1016/S0375-9474(98)00175-4} {\bibfield
  {journal} {\bibinfo  {journal} {Nucl. Phys. A}\ }\textbf {\bibinfo {volume}
  {635}},\ \bibinfo {pages} {121} (\bibinfo {year} {1998})},\ \bibinfo {note}
  {[Erratum: Nucl.Phys.A 642, 563--563 (1998)]},\ \Eprint
  {http://arxiv.org/abs/hep-ph/9801297} {arXiv:hep-ph/9801297} \BibitemShut
  {NoStop}%
\bibitem [{\citenamefont {Schindler}\ \emph {et~al.}(2007)\citenamefont
  {Schindler}, \citenamefont {Fuchs}, \citenamefont {Gegelia},\ and\
  \citenamefont {Scherer}}]{Schindler:2006it}%
  \BibitemOpen
  \bibfield  {author} {\bibinfo {author} {\bibfnamefont {M.~R.}\ \bibnamefont
  {Schindler}}, \bibinfo {author} {\bibfnamefont {T.}~\bibnamefont {Fuchs}},
  \bibinfo {author} {\bibfnamefont {J.}~\bibnamefont {Gegelia}}, \ and\
  \bibinfo {author} {\bibfnamefont {S.}~\bibnamefont {Scherer}},\ }\href
  {\doibase 10.1103/PhysRevC.75.025202} {\bibfield  {journal} {\bibinfo
  {journal} {Phys. Rev. C}\ }\textbf {\bibinfo {volume} {75}},\ \bibinfo
  {pages} {025202} (\bibinfo {year} {2007})},\ \Eprint
  {http://arxiv.org/abs/nucl-th/0611083} {arXiv:nucl-th/0611083} \BibitemShut
  {NoStop}%
\bibitem [{\citenamefont {Chung}\ and\ \citenamefont
  {Coester}(1991)}]{Chung:1991st}%
  \BibitemOpen
  \bibfield  {author} {\bibinfo {author} {\bibfnamefont {P.~L.}\ \bibnamefont
  {Chung}}\ and\ \bibinfo {author} {\bibfnamefont {F.}~\bibnamefont
  {Coester}},\ }\href {\doibase 10.1103/PhysRevD.44.229} {\bibfield  {journal}
  {\bibinfo  {journal} {Phys. Rev. D}\ }\textbf {\bibinfo {volume} {44}},\
  \bibinfo {pages} {229} (\bibinfo {year} {1991})}\BibitemShut {NoStop}%
\bibitem [{\citenamefont {Cardarelli}\ \emph {et~al.}(1995)\citenamefont
  {Cardarelli}, \citenamefont {Pace}, \citenamefont {Salme},\ and\
  \citenamefont {Simula}}]{Cardarelli:1995dc}%
  \BibitemOpen
  \bibfield  {author} {\bibinfo {author} {\bibfnamefont {F.}~\bibnamefont
  {Cardarelli}}, \bibinfo {author} {\bibfnamefont {E.}~\bibnamefont {Pace}},
  \bibinfo {author} {\bibfnamefont {G.}~\bibnamefont {Salme}}, \ and\ \bibinfo
  {author} {\bibfnamefont {S.}~\bibnamefont {Simula}},\ }\href {\doibase
  10.1016/0370-2693(95)00921-7} {\bibfield  {journal} {\bibinfo  {journal}
  {Phys. Lett. B}\ }\textbf {\bibinfo {volume} {357}},\ \bibinfo {pages} {267}
  (\bibinfo {year} {1995})},\ \Eprint {http://arxiv.org/abs/nucl-th/9507037}
  {arXiv:nucl-th/9507037} \BibitemShut {NoStop}%
\bibitem [{\citenamefont {Miller}(2002)}]{Miller:2002ig}%
  \BibitemOpen
  \bibfield  {author} {\bibinfo {author} {\bibfnamefont {G.~A.}\ \bibnamefont
  {Miller}},\ }\href {\doibase 10.1103/PhysRevC.66.032201} {\bibfield
  {journal} {\bibinfo  {journal} {Phys. Rev. C}\ }\textbf {\bibinfo {volume}
  {66}},\ \bibinfo {pages} {032201} (\bibinfo {year} {2002})},\ \Eprint
  {http://arxiv.org/abs/nucl-th/0207007} {arXiv:nucl-th/0207007} \BibitemShut
  {NoStop}%
\bibitem [{\citenamefont {Ma}\ \emph {et~al.}(2002{\natexlab{a}})\citenamefont
  {Ma}, \citenamefont {Qing},\ and\ \citenamefont {Schmidt}}]{Ma:2002ir}%
  \BibitemOpen
  \bibfield  {author} {\bibinfo {author} {\bibfnamefont {B.-Q.}\ \bibnamefont
  {Ma}}, \bibinfo {author} {\bibfnamefont {D.}~\bibnamefont {Qing}}, \ and\
  \bibinfo {author} {\bibfnamefont {I.}~\bibnamefont {Schmidt}},\ }\href
  {\doibase 10.1103/PhysRevC.65.035205} {\bibfield  {journal} {\bibinfo
  {journal} {Phys. Rev. C}\ }\textbf {\bibinfo {volume} {65}},\ \bibinfo
  {pages} {035205} (\bibinfo {year} {2002}{\natexlab{a}})},\ \Eprint
  {http://arxiv.org/abs/hep-ph/0202015} {arXiv:hep-ph/0202015} \BibitemShut
  {NoStop}%
\bibitem [{\citenamefont {Ma}\ \emph {et~al.}(2002{\natexlab{b}})\citenamefont
  {Ma}, \citenamefont {Qing},\ and\ \citenamefont {Schmidt}}]{Ma:2002xu}%
  \BibitemOpen
  \bibfield  {author} {\bibinfo {author} {\bibfnamefont {B.-Q.}\ \bibnamefont
  {Ma}}, \bibinfo {author} {\bibfnamefont {D.}~\bibnamefont {Qing}}, \ and\
  \bibinfo {author} {\bibfnamefont {I.}~\bibnamefont {Schmidt}},\ }\href
  {\doibase 10.1103/PhysRevC.66.048201} {\bibfield  {journal} {\bibinfo
  {journal} {Phys. Rev. C}\ }\textbf {\bibinfo {volume} {66}},\ \bibinfo
  {pages} {048201} (\bibinfo {year} {2002}{\natexlab{b}})},\ \Eprint
  {http://arxiv.org/abs/hep-ph/0204082} {arXiv:hep-ph/0204082} \BibitemShut
  {NoStop}%
\bibitem [{\citenamefont {Punjabi}\ \emph {et~al.}(2015)\citenamefont
  {Punjabi}, \citenamefont {Perdrisat}, \citenamefont {Jones}, \citenamefont
  {Brash},\ and\ \citenamefont {Carlson}}]{Punjabi:2015bba}%
  \BibitemOpen
  \bibfield  {author} {\bibinfo {author} {\bibfnamefont {V.}~\bibnamefont
  {Punjabi}}, \bibinfo {author} {\bibfnamefont {C.~F.}\ \bibnamefont
  {Perdrisat}}, \bibinfo {author} {\bibfnamefont {M.~K.}\ \bibnamefont
  {Jones}}, \bibinfo {author} {\bibfnamefont {E.~J.}\ \bibnamefont {Brash}}, \
  and\ \bibinfo {author} {\bibfnamefont {C.~E.}\ \bibnamefont {Carlson}},\
  }\href {\doibase 10.1140/epja/i2015-15079-x} {\bibfield  {journal} {\bibinfo
  {journal} {Eur. Phys. J. A}\ }\textbf {\bibinfo {volume} {51}},\ \bibinfo
  {pages} {79} (\bibinfo {year} {2015})},\ \Eprint
  {http://arxiv.org/abs/1503.01452} {arXiv:1503.01452 [nucl-ex]} \BibitemShut
  {NoStop}%
\bibitem [{\citenamefont {Hill}\ \emph {et~al.}(2018)\citenamefont {Hill},
  \citenamefont {Kammel}, \citenamefont {Marciano},\ and\ \citenamefont
  {Sirlin}}]{Hill:2017wgb}%
  \BibitemOpen
  \bibfield  {author} {\bibinfo {author} {\bibfnamefont {R.~J.}\ \bibnamefont
  {Hill}}, \bibinfo {author} {\bibfnamefont {P.}~\bibnamefont {Kammel}},
  \bibinfo {author} {\bibfnamefont {W.~J.}\ \bibnamefont {Marciano}}, \ and\
  \bibinfo {author} {\bibfnamefont {A.}~\bibnamefont {Sirlin}},\ }\href
  {\doibase 10.1088/1361-6633/aac190} {\bibfield  {journal} {\bibinfo
  {journal} {Rept. Prog. Phys.}\ }\textbf {\bibinfo {volume} {81}},\ \bibinfo
  {pages} {096301} (\bibinfo {year} {2018})},\ \Eprint
  {http://arxiv.org/abs/1708.08462} {arXiv:1708.08462 [hep-ph]} \BibitemShut
  {NoStop}%
\bibitem [{\citenamefont {Bernard}\ \emph {et~al.}(2002)\citenamefont
  {Bernard}, \citenamefont {Elouadrhiri},\ and\ \citenamefont
  {Mei{\ss}ner}}]{Bernard:2001rs}%
  \BibitemOpen
  \bibfield  {author} {\bibinfo {author} {\bibfnamefont {V.}~\bibnamefont
  {Bernard}}, \bibinfo {author} {\bibfnamefont {L.}~\bibnamefont
  {Elouadrhiri}}, \ and\ \bibinfo {author} {\bibfnamefont {U.-G.}\ \bibnamefont
  {Mei{\ss}ner}},\ }\href {\doibase 10.1088/0954-3899/28/1/201} {\bibfield
  {journal} {\bibinfo  {journal} {J. Phys. G}\ }\textbf {\bibinfo {volume}
  {28}},\ \bibinfo {pages} {R1} (\bibinfo {year} {2002})},\ \Eprint
  {http://arxiv.org/abs/hep-ph/0107088} {arXiv:hep-ph/0107088} \BibitemShut
  {NoStop}%
\bibitem [{\citenamefont {Kelly}(2004)}]{Kelly:2004hm}%
  \BibitemOpen
  \bibfield  {author} {\bibinfo {author} {\bibfnamefont {J.~J.}\ \bibnamefont
  {Kelly}},\ }\href {\doibase 10.1103/PhysRevC.70.068202} {\bibfield  {journal}
  {\bibinfo  {journal} {Phys. Rev. C}\ }\textbf {\bibinfo {volume} {70}},\
  \bibinfo {pages} {068202} (\bibinfo {year} {2004})}\BibitemShut {NoStop}%
\bibitem [{\citenamefont {Bhattacharya}\ \emph {et~al.}(2011)\citenamefont
  {Bhattacharya}, \citenamefont {Hill},\ and\ \citenamefont
  {Paz}}]{Bhattacharya:2011ah}%
  \BibitemOpen
  \bibfield  {author} {\bibinfo {author} {\bibfnamefont {B.}~\bibnamefont
  {Bhattacharya}}, \bibinfo {author} {\bibfnamefont {R.~J.}\ \bibnamefont
  {Hill}}, \ and\ \bibinfo {author} {\bibfnamefont {G.}~\bibnamefont {Paz}},\
  }\href {\doibase 10.1103/PhysRevD.84.073006} {\bibfield  {journal} {\bibinfo
  {journal} {Phys. Rev. D}\ }\textbf {\bibinfo {volume} {84}},\ \bibinfo
  {pages} {073006} (\bibinfo {year} {2011})},\ \Eprint
  {http://arxiv.org/abs/1108.0423} {arXiv:1108.0423 [hep-ph]} \BibitemShut
  {NoStop}%
\bibitem [{\citenamefont {Bodek}\ \emph {et~al.}(2008)\citenamefont {Bodek},
  \citenamefont {Avvakumov}, \citenamefont {Bradford},\ and\ \citenamefont
  {Budd}}]{Bodek:2007ym}%
  \BibitemOpen
  \bibfield  {author} {\bibinfo {author} {\bibfnamefont {A.}~\bibnamefont
  {Bodek}}, \bibinfo {author} {\bibfnamefont {S.}~\bibnamefont {Avvakumov}},
  \bibinfo {author} {\bibfnamefont {R.}~\bibnamefont {Bradford}}, \ and\
  \bibinfo {author} {\bibfnamefont {H.~S.}\ \bibnamefont {Budd}},\ }\href
  {\doibase 10.1140/epjc/s10052-007-0491-4} {\bibfield  {journal} {\bibinfo
  {journal} {Eur. Phys. J. C}\ }\textbf {\bibinfo {volume} {53}},\ \bibinfo
  {pages} {349} (\bibinfo {year} {2008})},\ \Eprint
  {http://arxiv.org/abs/0708.1946} {arXiv:0708.1946 [hep-ex]} \BibitemShut
  {NoStop}%
\bibitem [{\citenamefont {Alvarez-Ruso}\ \emph {et~al.}(2019)\citenamefont
  {Alvarez-Ruso}, \citenamefont {Graczyk},\ and\ \citenamefont
  {Saul-Sala}}]{Alvarez-Ruso:2018rdx}%
  \BibitemOpen
  \bibfield  {author} {\bibinfo {author} {\bibfnamefont {L.}~\bibnamefont
  {Alvarez-Ruso}}, \bibinfo {author} {\bibfnamefont {K.~M.}\ \bibnamefont
  {Graczyk}}, \ and\ \bibinfo {author} {\bibfnamefont {E.}~\bibnamefont
  {Saul-Sala}},\ }\href {\doibase 10.1103/PhysRevC.99.025204} {\bibfield
  {journal} {\bibinfo  {journal} {Phys. Rev. C}\ }\textbf {\bibinfo {volume}
  {99}},\ \bibinfo {pages} {025204} (\bibinfo {year} {2019})},\ \Eprint
  {http://arxiv.org/abs/1805.00905} {arXiv:1805.00905 [hep-ph]} \BibitemShut
  {NoStop}%
\bibitem [{\citenamefont {Kronfeld}\ \emph {et~al.}(2019)\citenamefont
  {Kronfeld}, \citenamefont {Richards}, \citenamefont {Detmold}, \citenamefont
  {Gupta}, \citenamefont {Lin}, \citenamefont {Liu}, \citenamefont {Meyer},
  \citenamefont {Sufian},\ and\ \citenamefont {Syritsyn}}]{Kronfeld:2019nfb}%
  \BibitemOpen
  \bibfield  {author} {\bibinfo {author} {\bibfnamefont {A.~S.}\ \bibnamefont
  {Kronfeld}}, \bibinfo {author} {\bibfnamefont {D.~G.}\ \bibnamefont
  {Richards}}, \bibinfo {author} {\bibfnamefont {W.}~\bibnamefont {Detmold}},
  \bibinfo {author} {\bibfnamefont {R.}~\bibnamefont {Gupta}}, \bibinfo
  {author} {\bibfnamefont {H.-W.}\ \bibnamefont {Lin}}, \bibinfo {author}
  {\bibfnamefont {K.-F.}\ \bibnamefont {Liu}}, \bibinfo {author} {\bibfnamefont
  {A.~S.}\ \bibnamefont {Meyer}}, \bibinfo {author} {\bibfnamefont
  {R.}~\bibnamefont {Sufian}}, \ and\ \bibinfo {author} {\bibfnamefont
  {S.}~\bibnamefont {Syritsyn}} (\bibinfo {collaboration} {USQCD}),\ }\href
  {\doibase 10.1140/epja/i2019-12916-x} {\bibfield  {journal} {\bibinfo
  {journal} {Eur. Phys. J. A}\ }\textbf {\bibinfo {volume} {55}},\ \bibinfo
  {pages} {196} (\bibinfo {year} {2019})},\ \Eprint
  {http://arxiv.org/abs/1904.09931} {arXiv:1904.09931 [hep-lat]} \BibitemShut
  {NoStop}%
\bibitem [{\citenamefont {Llewellyn~Smith}(1972)}]{LlewellynSmith:1971uhs}%
  \BibitemOpen
  \bibfield  {author} {\bibinfo {author} {\bibfnamefont {C.~H.}\ \bibnamefont
  {Llewellyn~Smith}},\ }\href {\doibase 10.1016/0370-1573(72)90010-5}
  {\bibfield  {journal} {\bibinfo  {journal} {Phys. Rept.}\ }\textbf {\bibinfo
  {volume} {3}},\ \bibinfo {pages} {261} (\bibinfo {year} {1972})}\BibitemShut
  {NoStop}%
\bibitem [{\citenamefont {Park}\ \emph {et~al.}(2022)\citenamefont {Park},
  \citenamefont {Gupta}, \citenamefont {Yoon}, \citenamefont {Mondal},
  \citenamefont {Bhattacharya}, \citenamefont {Jang}, \citenamefont {Jo\'o},\
  and\ \citenamefont {Winter}}]{Park:2021ypf}%
  \BibitemOpen
  \bibfield  {author} {\bibinfo {author} {\bibfnamefont {S.}~\bibnamefont
  {Park}}, \bibinfo {author} {\bibfnamefont {R.}~\bibnamefont {Gupta}},
  \bibinfo {author} {\bibfnamefont {B.}~\bibnamefont {Yoon}}, \bibinfo {author}
  {\bibfnamefont {S.}~\bibnamefont {Mondal}}, \bibinfo {author} {\bibfnamefont
  {T.}~\bibnamefont {Bhattacharya}}, \bibinfo {author} {\bibfnamefont {Y.-C.}\
  \bibnamefont {Jang}}, \bibinfo {author} {\bibfnamefont {B.}~\bibnamefont
  {Jo\'o}}, \ and\ \bibinfo {author} {\bibfnamefont {F.}~\bibnamefont {Winter}}
  (\bibinfo {collaboration} {Nucleon Matrix Elements (NME)}),\ }\href {\doibase
  10.1103/PhysRevD.105.054505} {\bibfield  {journal} {\bibinfo  {journal}
  {Phys. Rev. D}\ }\textbf {\bibinfo {volume} {105}},\ \bibinfo {pages}
  {054505} (\bibinfo {year} {2022})},\ \Eprint
  {http://arxiv.org/abs/2103.05599} {arXiv:2103.05599 [hep-lat]} \BibitemShut
  {NoStop}%
\bibitem [{\citenamefont {Xiong}\ \emph {et~al.}(2019)\citenamefont {Xiong}
  \emph {et~al.}}]{Xiong:2019umf}%
  \BibitemOpen
  \bibfield  {author} {\bibinfo {author} {\bibfnamefont {W.}~\bibnamefont
  {Xiong}} \emph {et~al.},\ }\href {\doibase 10.1038/s41586-019-1721-2}
  {\bibfield  {journal} {\bibinfo  {journal} {Nature}\ }\textbf {\bibinfo
  {volume} {575}},\ \bibinfo {pages} {147} (\bibinfo {year}
  {2019})}\BibitemShut {NoStop}%
\bibitem [{\citenamefont {Liu}\ \emph {et~al.}(1995)\citenamefont {Liu},
  \citenamefont {Dong}, \citenamefont {Draper},\ and\ \citenamefont
  {Wilcox}}]{Liu:1994dr}%
  \BibitemOpen
  \bibfield  {author} {\bibinfo {author} {\bibfnamefont {K.~F.}\ \bibnamefont
  {Liu}}, \bibinfo {author} {\bibfnamefont {S.~J.}\ \bibnamefont {Dong}},
  \bibinfo {author} {\bibfnamefont {T.}~\bibnamefont {Draper}}, \ and\ \bibinfo
  {author} {\bibfnamefont {W.}~\bibnamefont {Wilcox}},\ }\href {\doibase
  10.1103/PhysRevLett.74.2172} {\bibfield  {journal} {\bibinfo  {journal}
  {Phys. Rev. Lett.}\ }\textbf {\bibinfo {volume} {74}},\ \bibinfo {pages}
  {2172} (\bibinfo {year} {1995})},\ \Eprint
  {http://arxiv.org/abs/hep-lat/9406007} {arXiv:hep-lat/9406007} \BibitemShut
  {NoStop}%
\bibitem [{\citenamefont {B{\"a}r}(2019{\natexlab{a}})}]{Bar:2018xyi}%
  \BibitemOpen
  \bibfield  {author} {\bibinfo {author} {\bibfnamefont {O.}~\bibnamefont
  {B{\"a}r}},\ }\href {\doibase 10.1103/PhysRevD.99.054506} {\bibfield
  {journal} {\bibinfo  {journal} {Phys. Rev. D}\ }\textbf {\bibinfo {volume}
  {99}},\ \bibinfo {pages} {054506} (\bibinfo {year} {2019}{\natexlab{a}})},\
  \Eprint {http://arxiv.org/abs/1812.09191} {arXiv:1812.09191 [hep-lat]}
  \BibitemShut {NoStop}%
\bibitem [{\citenamefont {B{\"a}r}(2019{\natexlab{b}})}]{bar:2019gfx}%
  \BibitemOpen
  \bibfield  {author} {\bibinfo {author} {\bibfnamefont {O.}~\bibnamefont
  {B{\"a}r}},\ }\href {\doibase 10.1103/PhysRevD.100.054507} {\bibfield
  {journal} {\bibinfo  {journal} {Phys. Rev. D}\ }\textbf {\bibinfo {volume}
  {100}},\ \bibinfo {pages} {054507} (\bibinfo {year} {2019}{\natexlab{b}})},\
  \Eprint {http://arxiv.org/abs/1906.03652} {arXiv:1906.03652 [hep-lat]}
  \BibitemShut {NoStop}%
\bibitem [{\citenamefont {Gupta}\ \emph {et~al.}(2021)\citenamefont {Gupta},
  \citenamefont {Park}, \citenamefont {Hoferichter}, \citenamefont
  {Mereghetti}, \citenamefont {Yoon},\ and\ \citenamefont
  {Bhattacharya}}]{Gupta:2021ahb}%
  \BibitemOpen
  \bibfield  {author} {\bibinfo {author} {\bibfnamefont {R.}~\bibnamefont
  {Gupta}}, \bibinfo {author} {\bibfnamefont {S.}~\bibnamefont {Park}},
  \bibinfo {author} {\bibfnamefont {M.}~\bibnamefont {Hoferichter}}, \bibinfo
  {author} {\bibfnamefont {E.}~\bibnamefont {Mereghetti}}, \bibinfo {author}
  {\bibfnamefont {B.}~\bibnamefont {Yoon}}, \ and\ \bibinfo {author}
  {\bibfnamefont {T.}~\bibnamefont {Bhattacharya}},\ }\href {\doibase
  10.1103/PhysRevLett.127.242002} {\bibfield  {journal} {\bibinfo  {journal}
  {Phys. Rev. Lett.}\ }\textbf {\bibinfo {volume} {127}},\ \bibinfo {pages}
  {242002} (\bibinfo {year} {2021})},\ \Eprint
  {http://arxiv.org/abs/2105.12095} {arXiv:2105.12095 [hep-lat]} \BibitemShut
  {NoStop}%
\bibitem [{\citenamefont {Jang}\ \emph
  {et~al.}(2020{\natexlab{a}})\citenamefont {Jang}, \citenamefont {Gupta},
  \citenamefont {Yoon},\ and\ \citenamefont {Bhattacharya}}]{Jang:2019vkm}%
  \BibitemOpen
  \bibfield  {author} {\bibinfo {author} {\bibfnamefont {Y.-C.}\ \bibnamefont
  {Jang}}, \bibinfo {author} {\bibfnamefont {R.}~\bibnamefont {Gupta}},
  \bibinfo {author} {\bibfnamefont {B.}~\bibnamefont {Yoon}}, \ and\ \bibinfo
  {author} {\bibfnamefont {T.}~\bibnamefont {Bhattacharya}},\ }\href {\doibase
  10.1103/PhysRevLett.124.072002} {\bibfield  {journal} {\bibinfo  {journal}
  {Phys. Rev. Lett.}\ }\textbf {\bibinfo {volume} {124}},\ \bibinfo {pages}
  {072002} (\bibinfo {year} {2020}{\natexlab{a}})},\ \Eprint
  {http://arxiv.org/abs/1905.06470} {arXiv:1905.06470 [hep-lat]} \BibitemShut
  {NoStop}%
\bibitem [{\citenamefont {Goldberger}\ and\ \citenamefont
  {Treiman}(1958)}]{Goldberger:1958vp}%
  \BibitemOpen
  \bibfield  {author} {\bibinfo {author} {\bibfnamefont {M.~L.}\ \bibnamefont
  {Goldberger}}\ and\ \bibinfo {author} {\bibfnamefont {S.~B.}\ \bibnamefont
  {Treiman}},\ }\href {\doibase 10.1103/PhysRev.111.354} {\bibfield  {journal}
  {\bibinfo  {journal} {Phys. Rev.}\ }\textbf {\bibinfo {volume} {111}},\
  \bibinfo {pages} {354} (\bibinfo {year} {1958})}\BibitemShut {NoStop}%
\bibitem [{\citenamefont {Andersen}\ \emph {et~al.}(2018)\citenamefont
  {Andersen}, \citenamefont {Bulava}, \citenamefont {H\"orz},\ and\
  \citenamefont {Morningstar}}]{Andersen:2017una}%
  \BibitemOpen
  \bibfield  {author} {\bibinfo {author} {\bibfnamefont {C.~W.}\ \bibnamefont
  {Andersen}}, \bibinfo {author} {\bibfnamefont {J.}~\bibnamefont {Bulava}},
  \bibinfo {author} {\bibfnamefont {B.}~\bibnamefont {H\"orz}}, \ and\ \bibinfo
  {author} {\bibfnamefont {C.}~\bibnamefont {Morningstar}},\ }\href {\doibase
  10.1103/PhysRevD.97.014506} {\bibfield  {journal} {\bibinfo  {journal} {Phys.
  Rev. D}\ }\textbf {\bibinfo {volume} {97}},\ \bibinfo {pages} {014506}
  (\bibinfo {year} {2018})},\ \Eprint {http://arxiv.org/abs/1710.01557}
  {arXiv:1710.01557 [hep-lat]} \BibitemShut {NoStop}%
\bibitem [{\citenamefont {Silvi}\ \emph {et~al.}(2021)\citenamefont {Silvi}
  \emph {et~al.}}]{Silvi:2021uya}%
  \BibitemOpen
  \bibfield  {author} {\bibinfo {author} {\bibfnamefont {G.}~\bibnamefont
  {Silvi}} \emph {et~al.},\ }\href {\doibase 10.1103/PhysRevD.103.094508}
  {\bibfield  {journal} {\bibinfo  {journal} {Phys. Rev. D}\ }\textbf {\bibinfo
  {volume} {103}},\ \bibinfo {pages} {094508} (\bibinfo {year} {2021})},\
  \Eprint {http://arxiv.org/abs/2101.00689} {arXiv:2101.00689 [hep-lat]}
  \BibitemShut {NoStop}%
\bibitem [{\citenamefont {Barca}\ \emph {et~al.}(2021)\citenamefont {Barca},
  \citenamefont {Bali},\ and\ \citenamefont {Collins}}]{Barca:2021iak}%
  \BibitemOpen
  \bibfield  {author} {\bibinfo {author} {\bibfnamefont {L.}~\bibnamefont
  {Barca}}, \bibinfo {author} {\bibfnamefont {G.~S.}\ \bibnamefont {Bali}}, \
  and\ \bibinfo {author} {\bibfnamefont {S.}~\bibnamefont {Collins}},\ }in\
  \href@noop {} {\emph {\bibinfo {booktitle} {{38th International Symposium on
  Lattice Field Theory}}}}\ (\bibinfo {year} {2021})\ \Eprint
  {http://arxiv.org/abs/2110.11908} {arXiv:2110.11908 [hep-lat]} \BibitemShut
  {NoStop}%
\bibitem [{\citenamefont {Meyer}\ \emph {et~al.}(2016)\citenamefont {Meyer},
  \citenamefont {Betancourt}, \citenamefont {Gran},\ and\ \citenamefont
  {Hill}}]{Meyer:2016oeg}%
  \BibitemOpen
  \bibfield  {author} {\bibinfo {author} {\bibfnamefont {A.~S.}\ \bibnamefont
  {Meyer}}, \bibinfo {author} {\bibfnamefont {M.}~\bibnamefont {Betancourt}},
  \bibinfo {author} {\bibfnamefont {R.}~\bibnamefont {Gran}}, \ and\ \bibinfo
  {author} {\bibfnamefont {R.~J.}\ \bibnamefont {Hill}},\ }\href {\doibase
  10.1103/PhysRevD.93.113015} {\bibfield  {journal} {\bibinfo  {journal} {Phys.
  Rev. D}\ }\textbf {\bibinfo {volume} {93}},\ \bibinfo {pages} {113015}
  (\bibinfo {year} {2016})},\ \Eprint {http://arxiv.org/abs/1603.03048}
  {arXiv:1603.03048 [hep-ph]} \BibitemShut {NoStop}%
\bibitem [{\citenamefont {Bali}\ \emph {et~al.}(2020)\citenamefont {Bali},
  \citenamefont {Barca}, \citenamefont {Collins}, \citenamefont {Gruber},
  \citenamefont {L\"offler}, \citenamefont {Sch\"afer}, \citenamefont
  {S\"oldner}, \citenamefont {Wein}, \citenamefont {Weish\"aupl},\ and\
  \citenamefont {Wurm}}]{RQCD:2019jai}%
  \BibitemOpen
  \bibfield  {author} {\bibinfo {author} {\bibfnamefont {G.~S.}\ \bibnamefont
  {Bali}}, \bibinfo {author} {\bibfnamefont {L.}~\bibnamefont {Barca}},
  \bibinfo {author} {\bibfnamefont {S.}~\bibnamefont {Collins}}, \bibinfo
  {author} {\bibfnamefont {M.}~\bibnamefont {Gruber}}, \bibinfo {author}
  {\bibfnamefont {M.}~\bibnamefont {L\"offler}}, \bibinfo {author}
  {\bibfnamefont {A.}~\bibnamefont {Sch\"afer}}, \bibinfo {author}
  {\bibfnamefont {W.}~\bibnamefont {S\"oldner}}, \bibinfo {author}
  {\bibfnamefont {P.}~\bibnamefont {Wein}}, \bibinfo {author} {\bibfnamefont
  {S.}~\bibnamefont {Weish\"aupl}}, \ and\ \bibinfo {author} {\bibfnamefont
  {T.}~\bibnamefont {Wurm}} (\bibinfo {collaboration} {RQCD}),\ }\href
  {\doibase 10.1007/JHEP05(2020)126} {\bibfield  {journal} {\bibinfo  {journal}
  {JHEP}\ }\textbf {\bibinfo {volume} {05}},\ \bibinfo {pages} {126} (\bibinfo
  {year} {2020})},\ \Eprint {http://arxiv.org/abs/1911.13150} {arXiv:1911.13150
  [hep-lat]} \BibitemShut {NoStop}%
\bibitem [{\citenamefont {Hasan}\ \emph {et~al.}(2018)\citenamefont {Hasan},
  \citenamefont {Green}, \citenamefont {Meinel}, \citenamefont {Engelhardt},
  \citenamefont {Krieg}, \citenamefont {Negele}, \citenamefont {Pochinsky},\
  and\ \citenamefont {Syritsyn}}]{Hasan:2017wwt}%
  \BibitemOpen
  \bibfield  {author} {\bibinfo {author} {\bibfnamefont {N.}~\bibnamefont
  {Hasan}}, \bibinfo {author} {\bibfnamefont {J.}~\bibnamefont {Green}},
  \bibinfo {author} {\bibfnamefont {S.}~\bibnamefont {Meinel}}, \bibinfo
  {author} {\bibfnamefont {M.}~\bibnamefont {Engelhardt}}, \bibinfo {author}
  {\bibfnamefont {S.}~\bibnamefont {Krieg}}, \bibinfo {author} {\bibfnamefont
  {J.}~\bibnamefont {Negele}}, \bibinfo {author} {\bibfnamefont
  {A.}~\bibnamefont {Pochinsky}}, \ and\ \bibinfo {author} {\bibfnamefont
  {S.}~\bibnamefont {Syritsyn}},\ }\href {\doibase 10.1103/PhysRevD.97.034504}
  {\bibfield  {journal} {\bibinfo  {journal} {Phys. Rev. D}\ }\textbf {\bibinfo
  {volume} {97}},\ \bibinfo {pages} {034504} (\bibinfo {year} {2018})},\
  \Eprint {http://arxiv.org/abs/1711.11385} {arXiv:1711.11385 [hep-lat]}
  \BibitemShut {NoStop}%
\bibitem [{\citenamefont {Shintani}\ \emph {et~al.}(2019)\citenamefont
  {Shintani}, \citenamefont {Ishikawa}, \citenamefont {Kuramashi},
  \citenamefont {Sasaki},\ and\ \citenamefont {Yamazaki}}]{Shintani:2018ozy}%
  \BibitemOpen
  \bibfield  {author} {\bibinfo {author} {\bibfnamefont {E.}~\bibnamefont
  {Shintani}}, \bibinfo {author} {\bibfnamefont {K.-I.}\ \bibnamefont
  {Ishikawa}}, \bibinfo {author} {\bibfnamefont {Y.}~\bibnamefont {Kuramashi}},
  \bibinfo {author} {\bibfnamefont {S.}~\bibnamefont {Sasaki}}, \ and\ \bibinfo
  {author} {\bibfnamefont {T.}~\bibnamefont {Yamazaki}},\ }\href {\doibase
  10.1103/PhysRevD.99.014510} {\bibfield  {journal} {\bibinfo  {journal} {Phys.
  Rev. D}\ }\textbf {\bibinfo {volume} {99}},\ \bibinfo {pages} {014510}
  (\bibinfo {year} {2019})},\ \bibinfo {note} {[Erratum: Phys. Rev. D {\bf
  102}, 019902 (2020)]},\ \Eprint {http://arxiv.org/abs/1811.07292}
  {arXiv:1811.07292 [hep-lat]} \BibitemShut {NoStop}%
\bibitem [{\citenamefont {Alexandrou}\ \emph {et~al.}(2021)\citenamefont
  {Alexandrou} \emph {et~al.}}]{Alexandrou:2020okk}%
  \BibitemOpen
  \bibfield  {author} {\bibinfo {author} {\bibfnamefont {C.}~\bibnamefont
  {Alexandrou}} \emph {et~al.},\ }\href {\doibase 10.1103/PhysRevD.103.034509}
  {\bibfield  {journal} {\bibinfo  {journal} {Phys. Rev. D}\ }\textbf {\bibinfo
  {volume} {103}},\ \bibinfo {pages} {034509} (\bibinfo {year} {2021})},\
  \Eprint {http://arxiv.org/abs/2011.13342} {arXiv:2011.13342 [hep-lat]}
  \BibitemShut {NoStop}%
\bibitem [{\citenamefont {Ishikawa}\ \emph {et~al.}(2021)\citenamefont
  {Ishikawa}, \citenamefont {Kuramashi}, \citenamefont {Sasaki}, \citenamefont
  {Shintani},\ and\ \citenamefont {Yamazaki}}]{Ishikawa:2021eut}%
  \BibitemOpen
  \bibfield  {author} {\bibinfo {author} {\bibfnamefont {K.-I.}\ \bibnamefont
  {Ishikawa}}, \bibinfo {author} {\bibfnamefont {Y.}~\bibnamefont {Kuramashi}},
  \bibinfo {author} {\bibfnamefont {S.}~\bibnamefont {Sasaki}}, \bibinfo
  {author} {\bibfnamefont {E.}~\bibnamefont {Shintani}}, \ and\ \bibinfo
  {author} {\bibfnamefont {T.}~\bibnamefont {Yamazaki}} (\bibinfo
  {collaboration} {PACS}),\ }\href {\doibase 10.1103/PhysRevD.104.074514}
  {\bibfield  {journal} {\bibinfo  {journal} {Phys. Rev. D}\ }\textbf {\bibinfo
  {volume} {104}},\ \bibinfo {pages} {074514} (\bibinfo {year} {2021})},\
  \Eprint {http://arxiv.org/abs/2107.07085} {arXiv:2107.07085 [hep-lat]}
  \BibitemShut {NoStop}%
\bibitem [{\citenamefont {Meyer}\ \emph {et~al.}(2021)\citenamefont {Meyer}
  \emph {et~al.}}]{Meyer:2021vfq}%
  \BibitemOpen
  \bibfield  {author} {\bibinfo {author} {\bibfnamefont {A.~S.}\ \bibnamefont
  {Meyer}} \emph {et~al.},\ }in\ \href@noop {} {\emph {\bibinfo {booktitle}
  {{38th International Symposium on Lattice Field Theory}}}}\ (\bibinfo {year}
  {2021})\ \Eprint {http://arxiv.org/abs/2111.06333} {arXiv:2111.06333
  [hep-lat]} \BibitemShut {NoStop}%
\bibitem [{\citenamefont {Djukanovic}\ \emph {et~al.}(2021)\citenamefont
  {Djukanovic}, \citenamefont {von Hippel}, \citenamefont {Koponen},
  \citenamefont {Meyer}, \citenamefont {Ottnad}, \citenamefont {Schulz},\ and\
  \citenamefont {Wittig}}]{Djukanovic:2021yqg}%
  \BibitemOpen
  \bibfield  {author} {\bibinfo {author} {\bibfnamefont {D.}~\bibnamefont
  {Djukanovic}}, \bibinfo {author} {\bibfnamefont {G.}~\bibnamefont {von
  Hippel}}, \bibinfo {author} {\bibfnamefont {J.}~\bibnamefont {Koponen}},
  \bibinfo {author} {\bibfnamefont {H.~B.}\ \bibnamefont {Meyer}}, \bibinfo
  {author} {\bibfnamefont {K.}~\bibnamefont {Ottnad}}, \bibinfo {author}
  {\bibfnamefont {T.}~\bibnamefont {Schulz}}, \ and\ \bibinfo {author}
  {\bibfnamefont {H.}~\bibnamefont {Wittig}},\ }in\ \href@noop {} {\emph
  {\bibinfo {booktitle} {{38th International Symposium on Lattice Field
  Theory}}}}\ (\bibinfo {year} {2021})\ \Eprint
  {http://arxiv.org/abs/2112.00127} {arXiv:2112.00127 [hep-lat]} \BibitemShut
  {NoStop}%
\bibitem [{\citenamefont {Meyer}\ \emph {et~al.}(2022)\citenamefont {Meyer},
  \citenamefont {Walker-Loud},\ and\ \citenamefont
  {Wilkinson}}]{Meyer:2022mix}%
  \BibitemOpen
  \bibfield  {author} {\bibinfo {author} {\bibfnamefont {A.~S.}\ \bibnamefont
  {Meyer}}, \bibinfo {author} {\bibfnamefont {A.}~\bibnamefont {Walker-Loud}},
  \ and\ \bibinfo {author} {\bibfnamefont {C.}~\bibnamefont {Wilkinson}},\
  }\href@noop {} {\  (\bibinfo {year} {2022})},\ \Eprint
  {http://arxiv.org/abs/2201.01839} {arXiv:2201.01839 [hep-lat]} \BibitemShut
  {NoStop}%
\bibitem [{\citenamefont {Beane}\ \emph {et~al.}(2011)\citenamefont {Beane},
  \citenamefont {Detmold}, \citenamefont {Orginos},\ and\ \citenamefont
  {Savage}}]{Beane:2010em}%
  \BibitemOpen
  \bibfield  {author} {\bibinfo {author} {\bibfnamefont {S.~R.}\ \bibnamefont
  {Beane}}, \bibinfo {author} {\bibfnamefont {W.}~\bibnamefont {Detmold}},
  \bibinfo {author} {\bibfnamefont {K.}~\bibnamefont {Orginos}}, \ and\
  \bibinfo {author} {\bibfnamefont {M.~J.}\ \bibnamefont {Savage}},\ }\href
  {\doibase 10.1016/j.ppnp.2010.08.002} {\bibfield  {journal} {\bibinfo
  {journal} {Prog. Part. Nucl. Phys.}\ }\textbf {\bibinfo {volume} {66}},\
  \bibinfo {pages} {1} (\bibinfo {year} {2011})},\ \Eprint
  {http://arxiv.org/abs/1004.2935} {arXiv:1004.2935 [hep-lat]} \BibitemShut
  {NoStop}%
\bibitem [{\citenamefont {Drischler}\ \emph
  {et~al.}(2021{\natexlab{a}})\citenamefont {Drischler}, \citenamefont
  {Haxton}, \citenamefont {McElvain}, \citenamefont {Mereghetti}, \citenamefont
  {Nicholson}, \citenamefont {Vranas},\ and\ \citenamefont
  {Walker-Loud}}]{Drischler:2019xuo}%
  \BibitemOpen
  \bibfield  {author} {\bibinfo {author} {\bibfnamefont {C.}~\bibnamefont
  {Drischler}}, \bibinfo {author} {\bibfnamefont {W.}~\bibnamefont {Haxton}},
  \bibinfo {author} {\bibfnamefont {K.}~\bibnamefont {McElvain}}, \bibinfo
  {author} {\bibfnamefont {E.}~\bibnamefont {Mereghetti}}, \bibinfo {author}
  {\bibfnamefont {A.}~\bibnamefont {Nicholson}}, \bibinfo {author}
  {\bibfnamefont {P.}~\bibnamefont {Vranas}}, \ and\ \bibinfo {author}
  {\bibfnamefont {A.}~\bibnamefont {Walker-Loud}},\ }\href {\doibase
  10.1016/j.ppnp.2021.103888} {\bibfield  {journal} {\bibinfo  {journal} {Prog.
  Part. Nucl. Phys.}\ }\textbf {\bibinfo {volume} {121}},\ \bibinfo {pages}
  {103888} (\bibinfo {year} {2021}{\natexlab{a}})},\ \Eprint
  {http://arxiv.org/abs/1910.07961} {arXiv:1910.07961 [nucl-th]} \BibitemShut
  {NoStop}%
\bibitem [{\citenamefont {Davoudi}\ \emph {et~al.}(2021)\citenamefont
  {Davoudi}, \citenamefont {Detmold}, \citenamefont {Orginos}, \citenamefont
  {Parre\~no}, \citenamefont {Savage}, \citenamefont {Shanahan},\ and\
  \citenamefont {Wagman}}]{Davoudi:2020ngi}%
  \BibitemOpen
  \bibfield  {author} {\bibinfo {author} {\bibfnamefont {Z.}~\bibnamefont
  {Davoudi}}, \bibinfo {author} {\bibfnamefont {W.}~\bibnamefont {Detmold}},
  \bibinfo {author} {\bibfnamefont {K.}~\bibnamefont {Orginos}}, \bibinfo
  {author} {\bibfnamefont {A.}~\bibnamefont {Parre\~no}}, \bibinfo {author}
  {\bibfnamefont {M.~J.}\ \bibnamefont {Savage}}, \bibinfo {author}
  {\bibfnamefont {P.}~\bibnamefont {Shanahan}}, \ and\ \bibinfo {author}
  {\bibfnamefont {M.~L.}\ \bibnamefont {Wagman}},\ }\href {\doibase
  10.1016/j.physrep.2020.10.004} {\bibfield  {journal} {\bibinfo  {journal}
  {Phys. Rept.}\ }\textbf {\bibinfo {volume} {900}},\ \bibinfo {pages} {1}
  (\bibinfo {year} {2021})},\ \Eprint {http://arxiv.org/abs/2008.11160}
  {arXiv:2008.11160 [hep-lat]} \BibitemShut {NoStop}%
\bibitem [{\citenamefont {Butler}\ \emph {et~al.}(2002)\citenamefont {Butler},
  \citenamefont {Chen},\ and\ \citenamefont {Vogel}}]{Butler:2002cw}%
  \BibitemOpen
  \bibfield  {author} {\bibinfo {author} {\bibfnamefont {M.}~\bibnamefont
  {Butler}}, \bibinfo {author} {\bibfnamefont {J.-W.}\ \bibnamefont {Chen}}, \
  and\ \bibinfo {author} {\bibfnamefont {P.}~\bibnamefont {Vogel}},\ }\href
  {\doibase 10.1016/S0370-2693(02)02868-X} {\bibfield  {journal} {\bibinfo
  {journal} {Phys. Lett. B}\ }\textbf {\bibinfo {volume} {549}},\ \bibinfo
  {pages} {26} (\bibinfo {year} {2002})},\ \Eprint
  {http://arxiv.org/abs/nucl-th/0206026} {arXiv:nucl-th/0206026} \BibitemShut
  {NoStop}%
\bibitem [{\citenamefont {Chen}\ \emph {et~al.}(2003)\citenamefont {Chen},
  \citenamefont {Heeger},\ and\ \citenamefont {Robertson}}]{Chen:2002pv}%
  \BibitemOpen
  \bibfield  {author} {\bibinfo {author} {\bibfnamefont {J.-W.}\ \bibnamefont
  {Chen}}, \bibinfo {author} {\bibfnamefont {K.~M.}\ \bibnamefont {Heeger}}, \
  and\ \bibinfo {author} {\bibfnamefont {R.~G.~H.}\ \bibnamefont {Robertson}},\
  }\href {\doibase 10.1103/PhysRevC.67.025801} {\bibfield  {journal} {\bibinfo
  {journal} {Phys. Rev. C}\ }\textbf {\bibinfo {volume} {67}},\ \bibinfo
  {pages} {025801} (\bibinfo {year} {2003})},\ \Eprint
  {http://arxiv.org/abs/nucl-th/0210073} {arXiv:nucl-th/0210073} \BibitemShut
  {NoStop}%
\bibitem [{\citenamefont {Balantekin}\ and\ \citenamefont
  {Yuksel}(2003)}]{Balantekin:2003ep}%
  \BibitemOpen
  \bibfield  {author} {\bibinfo {author} {\bibfnamefont {A.~B.}\ \bibnamefont
  {Balantekin}}\ and\ \bibinfo {author} {\bibfnamefont {H.}~\bibnamefont
  {Yuksel}},\ }\href {\doibase 10.1103/PhysRevC.68.055801} {\bibfield
  {journal} {\bibinfo  {journal} {Phys. Rev. C}\ }\textbf {\bibinfo {volume}
  {68}},\ \bibinfo {pages} {055801} (\bibinfo {year} {2003})},\ \Eprint
  {http://arxiv.org/abs/hep-ph/0307227} {arXiv:hep-ph/0307227} \BibitemShut
  {NoStop}%
\bibitem [{\citenamefont {Brown}\ \emph {et~al.}(2002)\citenamefont {Brown},
  \citenamefont {Butler},\ and\ \citenamefont {Guenther}}]{Brown:2002ih}%
  \BibitemOpen
  \bibfield  {author} {\bibinfo {author} {\bibfnamefont {K.~I.~T.}\
  \bibnamefont {Brown}}, \bibinfo {author} {\bibfnamefont {M.~N.}\ \bibnamefont
  {Butler}}, \ and\ \bibinfo {author} {\bibfnamefont {D.~B.}\ \bibnamefont
  {Guenther}},\ }\href@noop {} {\  (\bibinfo {year} {2002})},\ \Eprint
  {http://arxiv.org/abs/nucl-th/0207008} {arXiv:nucl-th/0207008} \BibitemShut
  {NoStop}%
\bibitem [{\citenamefont {Savage}\ \emph {et~al.}(2017)\citenamefont {Savage},
  \citenamefont {Shanahan}, \citenamefont {Tiburzi}, \citenamefont {Wagman},
  \citenamefont {Winter}, \citenamefont {Beane}, \citenamefont {Chang},
  \citenamefont {Davoudi}, \citenamefont {Detmold},\ and\ \citenamefont
  {Orginos}}]{Savage:2016kon}%
  \BibitemOpen
  \bibfield  {author} {\bibinfo {author} {\bibfnamefont {M.~J.}\ \bibnamefont
  {Savage}}, \bibinfo {author} {\bibfnamefont {P.~E.}\ \bibnamefont
  {Shanahan}}, \bibinfo {author} {\bibfnamefont {B.~C.}\ \bibnamefont
  {Tiburzi}}, \bibinfo {author} {\bibfnamefont {M.~L.}\ \bibnamefont {Wagman}},
  \bibinfo {author} {\bibfnamefont {F.}~\bibnamefont {Winter}}, \bibinfo
  {author} {\bibfnamefont {S.~R.}\ \bibnamefont {Beane}}, \bibinfo {author}
  {\bibfnamefont {E.}~\bibnamefont {Chang}}, \bibinfo {author} {\bibfnamefont
  {Z.}~\bibnamefont {Davoudi}}, \bibinfo {author} {\bibfnamefont
  {W.}~\bibnamefont {Detmold}}, \ and\ \bibinfo {author} {\bibfnamefont
  {K.}~\bibnamefont {Orginos}},\ }\href {\doibase
  10.1103/PhysRevLett.119.062002} {\bibfield  {journal} {\bibinfo  {journal}
  {Phys. Rev. Lett.}\ }\textbf {\bibinfo {volume} {119}},\ \bibinfo {pages}
  {062002} (\bibinfo {year} {2017})},\ \Eprint
  {http://arxiv.org/abs/1610.04545} {arXiv:1610.04545 [hep-lat]} \BibitemShut
  {NoStop}%
\bibitem [{\citenamefont {Chang}\ \emph {et~al.}(2018)\citenamefont {Chang},
  \citenamefont {Davoudi}, \citenamefont {Detmold}, \citenamefont {Gambhir},
  \citenamefont {Orginos}, \citenamefont {Savage}, \citenamefont {Shanahan},
  \citenamefont {Wagman},\ and\ \citenamefont {Winter}}]{Chang:2017eiq}%
  \BibitemOpen
  \bibfield  {author} {\bibinfo {author} {\bibfnamefont {E.}~\bibnamefont
  {Chang}}, \bibinfo {author} {\bibfnamefont {Z.}~\bibnamefont {Davoudi}},
  \bibinfo {author} {\bibfnamefont {W.}~\bibnamefont {Detmold}}, \bibinfo
  {author} {\bibfnamefont {A.~S.}\ \bibnamefont {Gambhir}}, \bibinfo {author}
  {\bibfnamefont {K.}~\bibnamefont {Orginos}}, \bibinfo {author} {\bibfnamefont
  {M.~J.}\ \bibnamefont {Savage}}, \bibinfo {author} {\bibfnamefont {P.~E.}\
  \bibnamefont {Shanahan}}, \bibinfo {author} {\bibfnamefont {M.~L.}\
  \bibnamefont {Wagman}}, \ and\ \bibinfo {author} {\bibfnamefont
  {F.}~\bibnamefont {Winter}} (\bibinfo {collaboration} {NPLQCD}),\ }\href
  {\doibase 10.1103/PhysRevLett.120.152002} {\bibfield  {journal} {\bibinfo
  {journal} {Phys. Rev. Lett.}\ }\textbf {\bibinfo {volume} {120}},\ \bibinfo
  {pages} {152002} (\bibinfo {year} {2018})},\ \Eprint
  {http://arxiv.org/abs/1712.03221} {arXiv:1712.03221 [hep-lat]} \BibitemShut
  {NoStop}%
\bibitem [{\citenamefont {Parre\~no}\ \emph {et~al.}(2021)\citenamefont
  {Parre\~no}, \citenamefont {Shanahan}, \citenamefont {Wagman}, \citenamefont
  {Winter}, \citenamefont {Chang}, \citenamefont {Detmold},\ and\ \citenamefont
  {Illa}}]{Parreno:2021ovq}%
  \BibitemOpen
  \bibfield  {author} {\bibinfo {author} {\bibfnamefont {A.}~\bibnamefont
  {Parre\~no}}, \bibinfo {author} {\bibfnamefont {P.~E.}\ \bibnamefont
  {Shanahan}}, \bibinfo {author} {\bibfnamefont {M.~L.}\ \bibnamefont
  {Wagman}}, \bibinfo {author} {\bibfnamefont {F.}~\bibnamefont {Winter}},
  \bibinfo {author} {\bibfnamefont {E.}~\bibnamefont {Chang}}, \bibinfo
  {author} {\bibfnamefont {W.}~\bibnamefont {Detmold}}, \ and\ \bibinfo
  {author} {\bibfnamefont {M.}~\bibnamefont {Illa}} (\bibinfo {collaboration}
  {NPLQCD}),\ }\href {\doibase 10.1103/PhysRevD.103.074511} {\bibfield
  {journal} {\bibinfo  {journal} {Phys. Rev. D}\ }\textbf {\bibinfo {volume}
  {103}},\ \bibinfo {pages} {074511} (\bibinfo {year} {2021})},\ \Eprint
  {http://arxiv.org/abs/2102.03805} {arXiv:2102.03805 [hep-lat]} \BibitemShut
  {NoStop}%
\bibitem [{\citenamefont {Barnea}\ \emph {et~al.}(2015)\citenamefont {Barnea},
  \citenamefont {Contessi}, \citenamefont {Gazit}, \citenamefont {Pederiva},\
  and\ \citenamefont {van Kolck}}]{Barnea:2013uqa}%
  \BibitemOpen
  \bibfield  {author} {\bibinfo {author} {\bibfnamefont {N.}~\bibnamefont
  {Barnea}}, \bibinfo {author} {\bibfnamefont {L.}~\bibnamefont {Contessi}},
  \bibinfo {author} {\bibfnamefont {D.}~\bibnamefont {Gazit}}, \bibinfo
  {author} {\bibfnamefont {F.}~\bibnamefont {Pederiva}}, \ and\ \bibinfo
  {author} {\bibfnamefont {U.}~\bibnamefont {van Kolck}},\ }\href {\doibase
  10.1103/PhysRevLett.114.052501} {\bibfield  {journal} {\bibinfo  {journal}
  {Phys. Rev. Lett.}\ }\textbf {\bibinfo {volume} {114}},\ \bibinfo {pages}
  {052501} (\bibinfo {year} {2015})},\ \Eprint {http://arxiv.org/abs/1311.4966}
  {arXiv:1311.4966 [nucl-th]} \BibitemShut {NoStop}%
\bibitem [{\citenamefont {Eliyahu}\ \emph {et~al.}(2020)\citenamefont
  {Eliyahu}, \citenamefont {Bazak},\ and\ \citenamefont
  {Barnea}}]{Eliyahu:2019nkz}%
  \BibitemOpen
  \bibfield  {author} {\bibinfo {author} {\bibfnamefont {M.}~\bibnamefont
  {Eliyahu}}, \bibinfo {author} {\bibfnamefont {B.}~\bibnamefont {Bazak}}, \
  and\ \bibinfo {author} {\bibfnamefont {N.}~\bibnamefont {Barnea}},\ }\href
  {\doibase 10.1103/PhysRevC.102.044003} {\bibfield  {journal} {\bibinfo
  {journal} {Phys. Rev. C}\ }\textbf {\bibinfo {volume} {102}},\ \bibinfo
  {pages} {044003} (\bibinfo {year} {2020})},\ \Eprint
  {http://arxiv.org/abs/1912.07017} {arXiv:1912.07017 [nucl-th]} \BibitemShut
  {NoStop}%
\bibitem [{\citenamefont {Detmold}\ and\ \citenamefont
  {Shanahan}(2021)}]{Detmold:2021oro}%
  \BibitemOpen
  \bibfield  {author} {\bibinfo {author} {\bibfnamefont {W.}~\bibnamefont
  {Detmold}}\ and\ \bibinfo {author} {\bibfnamefont {P.~E.}\ \bibnamefont
  {Shanahan}},\ }\href {\doibase 10.1103/PhysRevD.103.074503} {\bibfield
  {journal} {\bibinfo  {journal} {Phys. Rev. D}\ }\textbf {\bibinfo {volume}
  {103}},\ \bibinfo {pages} {074503} (\bibinfo {year} {2021})},\ \Eprint
  {http://arxiv.org/abs/2102.04329} {arXiv:2102.04329 [nucl-th]} \BibitemShut
  {NoStop}%
\bibitem [{\citenamefont {Francis}\ \emph {et~al.}(2019)\citenamefont
  {Francis}, \citenamefont {Green}, \citenamefont {Junnarkar}, \citenamefont
  {Miao}, \citenamefont {Rae},\ and\ \citenamefont {Wittig}}]{Francis:2018qch}%
  \BibitemOpen
  \bibfield  {author} {\bibinfo {author} {\bibfnamefont {A.}~\bibnamefont
  {Francis}}, \bibinfo {author} {\bibfnamefont {J.}~\bibnamefont {Green}},
  \bibinfo {author} {\bibfnamefont {P.}~\bibnamefont {Junnarkar}}, \bibinfo
  {author} {\bibfnamefont {C.}~\bibnamefont {Miao}}, \bibinfo {author}
  {\bibfnamefont {T.}~\bibnamefont {Rae}}, \ and\ \bibinfo {author}
  {\bibfnamefont {H.}~\bibnamefont {Wittig}},\ }\href {\doibase
  10.1103/PhysRevD.99.074505} {\bibfield  {journal} {\bibinfo  {journal} {Phys.
  Rev. D}\ }\textbf {\bibinfo {volume} {99}},\ \bibinfo {pages} {074505}
  (\bibinfo {year} {2019})},\ \Eprint {http://arxiv.org/abs/1805.03966}
  {arXiv:1805.03966 [hep-lat]} \BibitemShut {NoStop}%
\bibitem [{\citenamefont {H\"orz}\ \emph {et~al.}(2021)\citenamefont {H\"orz}
  \emph {et~al.}}]{Horz:2020zvv}%
  \BibitemOpen
  \bibfield  {author} {\bibinfo {author} {\bibfnamefont {B.}~\bibnamefont
  {H\"orz}} \emph {et~al.},\ }\href {\doibase 10.1103/PhysRevC.103.014003}
  {\bibfield  {journal} {\bibinfo  {journal} {Phys. Rev. C}\ }\textbf {\bibinfo
  {volume} {103}},\ \bibinfo {pages} {014003} (\bibinfo {year} {2021})},\
  \Eprint {http://arxiv.org/abs/2009.11825} {arXiv:2009.11825 [hep-lat]}
  \BibitemShut {NoStop}%
\bibitem [{\citenamefont {Green}\ \emph {et~al.}(2021)\citenamefont {Green},
  \citenamefont {Hanlon}, \citenamefont {Junnarkar},\ and\ \citenamefont
  {Wittig}}]{Green:2021qol}%
  \BibitemOpen
  \bibfield  {author} {\bibinfo {author} {\bibfnamefont {J.~R.}\ \bibnamefont
  {Green}}, \bibinfo {author} {\bibfnamefont {A.~D.}\ \bibnamefont {Hanlon}},
  \bibinfo {author} {\bibfnamefont {P.~M.}\ \bibnamefont {Junnarkar}}, \ and\
  \bibinfo {author} {\bibfnamefont {H.}~\bibnamefont {Wittig}},\ }\href
  {\doibase 10.1103/PhysRevLett.127.242003} {\bibfield  {journal} {\bibinfo
  {journal} {Phys. Rev. Lett.}\ }\textbf {\bibinfo {volume} {127}},\ \bibinfo
  {pages} {242003} (\bibinfo {year} {2021})},\ \Eprint
  {http://arxiv.org/abs/2103.01054} {arXiv:2103.01054 [hep-lat]} \BibitemShut
  {NoStop}%
\bibitem [{\citenamefont {Amarasinghe}\ \emph {et~al.}(2021)\citenamefont
  {Amarasinghe}, \citenamefont {Baghdadi}, \citenamefont {Davoudi},
  \citenamefont {Detmold}, \citenamefont {Illa}, \citenamefont {Parre\~no},
  \citenamefont {Pochinsky}, \citenamefont {Shanahan},\ and\ \citenamefont
  {Wagman}}]{Amarasinghe:2021lqa}%
  \BibitemOpen
  \bibfield  {author} {\bibinfo {author} {\bibfnamefont {S.}~\bibnamefont
  {Amarasinghe}}, \bibinfo {author} {\bibfnamefont {R.}~\bibnamefont
  {Baghdadi}}, \bibinfo {author} {\bibfnamefont {Z.}~\bibnamefont {Davoudi}},
  \bibinfo {author} {\bibfnamefont {W.}~\bibnamefont {Detmold}}, \bibinfo
  {author} {\bibfnamefont {M.}~\bibnamefont {Illa}}, \bibinfo {author}
  {\bibfnamefont {A.}~\bibnamefont {Parre\~no}}, \bibinfo {author}
  {\bibfnamefont {A.~V.}\ \bibnamefont {Pochinsky}}, \bibinfo {author}
  {\bibfnamefont {P.~E.}\ \bibnamefont {Shanahan}}, \ and\ \bibinfo {author}
  {\bibfnamefont {M.~L.}\ \bibnamefont {Wagman}},\ }\href@noop {} {\  (\bibinfo
  {year} {2021})},\ \Eprint {http://arxiv.org/abs/2108.10835} {arXiv:2108.10835
  [hep-lat]} \BibitemShut {NoStop}%
\bibitem [{\citenamefont {Beane}\ \emph
  {et~al.}(2013{\natexlab{a}})\citenamefont {Beane}, \citenamefont {Chang},
  \citenamefont {Cohen}, \citenamefont {Detmold}, \citenamefont {Lin},
  \citenamefont {Luu}, \citenamefont {Orginos}, \citenamefont {Parre\~{n}o},
  \citenamefont {Savage},\ and\ \citenamefont {Walker-Loud}}]{Beane:2012vq}%
  \BibitemOpen
  \bibfield  {author} {\bibinfo {author} {\bibfnamefont {S.~R.}\ \bibnamefont
  {Beane}}, \bibinfo {author} {\bibfnamefont {E.}~\bibnamefont {Chang}},
  \bibinfo {author} {\bibfnamefont {S.~D.}\ \bibnamefont {Cohen}}, \bibinfo
  {author} {\bibfnamefont {W.}~\bibnamefont {Detmold}}, \bibinfo {author}
  {\bibfnamefont {H.~W.}\ \bibnamefont {Lin}}, \bibinfo {author} {\bibfnamefont
  {T.~C.}\ \bibnamefont {Luu}}, \bibinfo {author} {\bibfnamefont
  {K.}~\bibnamefont {Orginos}}, \bibinfo {author} {\bibfnamefont
  {A.}~\bibnamefont {Parre\~{n}o}}, \bibinfo {author} {\bibfnamefont {M.~J.}\
  \bibnamefont {Savage}}, \ and\ \bibinfo {author} {\bibfnamefont
  {A.}~\bibnamefont {Walker-Loud}} (\bibinfo {collaboration} {NPLQCD}),\ }\href
  {\doibase 10.1103/PhysRevD.87.034506} {\bibfield  {journal} {\bibinfo
  {journal} {Phys. Rev. D}\ }\textbf {\bibinfo {volume} {87}},\ \bibinfo
  {pages} {034506} (\bibinfo {year} {2013}{\natexlab{a}})},\ \Eprint
  {http://arxiv.org/abs/1206.5219} {arXiv:1206.5219 [hep-lat]} \BibitemShut
  {NoStop}%
\bibitem [{\citenamefont {Yamazaki}\ \emph {et~al.}(2012)\citenamefont
  {Yamazaki}, \citenamefont {Ishikawa}, \citenamefont {Kuramashi},\ and\
  \citenamefont {Ukawa}}]{Yamazaki:2012hi}%
  \BibitemOpen
  \bibfield  {author} {\bibinfo {author} {\bibfnamefont {T.}~\bibnamefont
  {Yamazaki}}, \bibinfo {author} {\bibfnamefont {K.-i.}\ \bibnamefont
  {Ishikawa}}, \bibinfo {author} {\bibfnamefont {Y.}~\bibnamefont {Kuramashi}},
  \ and\ \bibinfo {author} {\bibfnamefont {A.}~\bibnamefont {Ukawa}},\ }\href
  {\doibase 10.1103/PhysRevD.86.074514} {\bibfield  {journal} {\bibinfo
  {journal} {Phys. Rev. D}\ }\textbf {\bibinfo {volume} {86}},\ \bibinfo
  {pages} {074514} (\bibinfo {year} {2012})},\ \Eprint
  {http://arxiv.org/abs/1207.4277} {arXiv:1207.4277 [hep-lat]} \BibitemShut
  {NoStop}%
\bibitem [{\citenamefont {Beane}\ \emph
  {et~al.}(2013{\natexlab{b}})\citenamefont {Beane}, \citenamefont {Chang},
  \citenamefont {Cohen}, \citenamefont {Detmold}, \citenamefont {Junnarkar},
  \citenamefont {Lin}, \citenamefont {Luu}, \citenamefont {Orginos},
  \citenamefont {Parre\~no}, \citenamefont {Savage},\ and\ \citenamefont
  {Walker-Loud}}]{Beane:2013br}%
  \BibitemOpen
  \bibfield  {author} {\bibinfo {author} {\bibfnamefont {S.~R.}\ \bibnamefont
  {Beane}}, \bibinfo {author} {\bibfnamefont {E.}~\bibnamefont {Chang}},
  \bibinfo {author} {\bibfnamefont {S.~D.}\ \bibnamefont {Cohen}}, \bibinfo
  {author} {\bibfnamefont {W.}~\bibnamefont {Detmold}}, \bibinfo {author}
  {\bibfnamefont {P.}~\bibnamefont {Junnarkar}}, \bibinfo {author}
  {\bibfnamefont {H.~W.}\ \bibnamefont {Lin}}, \bibinfo {author} {\bibfnamefont
  {T.~C.}\ \bibnamefont {Luu}}, \bibinfo {author} {\bibfnamefont
  {K.}~\bibnamefont {Orginos}}, \bibinfo {author} {\bibfnamefont
  {A.}~\bibnamefont {Parre\~no}}, \bibinfo {author} {\bibfnamefont {M.~J.}\
  \bibnamefont {Savage}}, \ and\ \bibinfo {author} {\bibfnamefont
  {A.}~\bibnamefont {Walker-Loud}} (\bibinfo {collaboration} {NPLQCD}),\ }\href
  {\doibase 10.1103/PhysRevC.88.024003} {\bibfield  {journal} {\bibinfo
  {journal} {Phys. Rev. C}\ }\textbf {\bibinfo {volume} {88}},\ \bibinfo
  {pages} {024003} (\bibinfo {year} {2013}{\natexlab{b}})},\ \Eprint
  {http://arxiv.org/abs/1301.5790} {arXiv:1301.5790 [hep-lat]} \BibitemShut
  {NoStop}%
\bibitem [{\citenamefont {Berkowitz}\ \emph {et~al.}(2017)\citenamefont
  {Berkowitz}, \citenamefont {Kurth}, \citenamefont {Nicholson}, \citenamefont
  {Jo\'o}, \citenamefont {Rinaldi}, \citenamefont {Strother}, \citenamefont
  {Vranas},\ and\ \citenamefont {Walker-Loud}}]{Berkowitz:2015eaa}%
  \BibitemOpen
  \bibfield  {author} {\bibinfo {author} {\bibfnamefont {E.}~\bibnamefont
  {Berkowitz}}, \bibinfo {author} {\bibfnamefont {T.}~\bibnamefont {Kurth}},
  \bibinfo {author} {\bibfnamefont {A.}~\bibnamefont {Nicholson}}, \bibinfo
  {author} {\bibfnamefont {B.}~\bibnamefont {Jo\'o}}, \bibinfo {author}
  {\bibfnamefont {E.}~\bibnamefont {Rinaldi}}, \bibinfo {author} {\bibfnamefont
  {M.}~\bibnamefont {Strother}}, \bibinfo {author} {\bibfnamefont {P.~M.}\
  \bibnamefont {Vranas}}, \ and\ \bibinfo {author} {\bibfnamefont
  {A.}~\bibnamefont {Walker-Loud}} (\bibinfo {collaboration} {CalLat}),\ }\href
  {\doibase 10.1016/j.physletb.2016.12.024} {\bibfield  {journal} {\bibinfo
  {journal} {Phys. Lett. B}\ }\textbf {\bibinfo {volume} {765}},\ \bibinfo
  {pages} {285} (\bibinfo {year} {2017})},\ \Eprint
  {http://arxiv.org/abs/1508.00886} {arXiv:1508.00886 [hep-lat]} \BibitemShut
  {NoStop}%
\bibitem [{\citenamefont {Orginos}\ \emph {et~al.}(2015)\citenamefont
  {Orginos}, \citenamefont {Parre\~no}, \citenamefont {Savage}, \citenamefont
  {Beane}, \citenamefont {Chang},\ and\ \citenamefont
  {Detmold}}]{Orginos:2015aya}%
  \BibitemOpen
  \bibfield  {author} {\bibinfo {author} {\bibfnamefont {K.}~\bibnamefont
  {Orginos}}, \bibinfo {author} {\bibfnamefont {A.}~\bibnamefont {Parre\~no}},
  \bibinfo {author} {\bibfnamefont {M.~J.}\ \bibnamefont {Savage}}, \bibinfo
  {author} {\bibfnamefont {S.~R.}\ \bibnamefont {Beane}}, \bibinfo {author}
  {\bibfnamefont {E.}~\bibnamefont {Chang}}, \ and\ \bibinfo {author}
  {\bibfnamefont {W.}~\bibnamefont {Detmold}},\ }\href {\doibase
  10.1103/PhysRevD.92.114512} {\bibfield  {journal} {\bibinfo  {journal} {Phys.
  Rev. D}\ }\textbf {\bibinfo {volume} {92}},\ \bibinfo {pages} {114512}
  (\bibinfo {year} {2015})},\ \bibinfo {note} {[Erratum: Phys. Rev. D {\bf
  102}, 039903 (2020)]},\ \Eprint {http://arxiv.org/abs/1508.07583}
  {arXiv:1508.07583 [hep-lat]} \BibitemShut {NoStop}%
\bibitem [{\citenamefont {Yamazaki}\ \emph {et~al.}(2015)\citenamefont
  {Yamazaki}, \citenamefont {Ishikawa}, \citenamefont {Kuramashi},\ and\
  \citenamefont {Ukawa}}]{Yamazaki:2015asa}%
  \BibitemOpen
  \bibfield  {author} {\bibinfo {author} {\bibfnamefont {T.}~\bibnamefont
  {Yamazaki}}, \bibinfo {author} {\bibfnamefont {K.-i.}\ \bibnamefont
  {Ishikawa}}, \bibinfo {author} {\bibfnamefont {Y.}~\bibnamefont {Kuramashi}},
  \ and\ \bibinfo {author} {\bibfnamefont {A.}~\bibnamefont {Ukawa}},\ }\href
  {\doibase 10.1103/PhysRevD.92.014501} {\bibfield  {journal} {\bibinfo
  {journal} {Phys. Rev. D}\ }\textbf {\bibinfo {volume} {92}},\ \bibinfo
  {pages} {014501} (\bibinfo {year} {2015})},\ \Eprint
  {http://arxiv.org/abs/1502.04182} {arXiv:1502.04182 [hep-lat]} \BibitemShut
  {NoStop}%
\bibitem [{\citenamefont {Wagman}\ \emph {et~al.}(2017)\citenamefont {Wagman},
  \citenamefont {Winter}, \citenamefont {Chang}, \citenamefont {Davoudi},
  \citenamefont {Detmold}, \citenamefont {Orginos}, \citenamefont {Savage},\
  and\ \citenamefont {Shanahan}}]{Wagman:2017tmp}%
  \BibitemOpen
  \bibfield  {author} {\bibinfo {author} {\bibfnamefont {M.~L.}\ \bibnamefont
  {Wagman}}, \bibinfo {author} {\bibfnamefont {F.}~\bibnamefont {Winter}},
  \bibinfo {author} {\bibfnamefont {E.}~\bibnamefont {Chang}}, \bibinfo
  {author} {\bibfnamefont {Z.}~\bibnamefont {Davoudi}}, \bibinfo {author}
  {\bibfnamefont {W.}~\bibnamefont {Detmold}}, \bibinfo {author} {\bibfnamefont
  {K.}~\bibnamefont {Orginos}}, \bibinfo {author} {\bibfnamefont {M.~J.}\
  \bibnamefont {Savage}}, \ and\ \bibinfo {author} {\bibfnamefont {P.~E.}\
  \bibnamefont {Shanahan}} (\bibinfo {collaboration} {NPLQCD}),\ }\href
  {\doibase 10.1103/PhysRevD.96.114510} {\bibfield  {journal} {\bibinfo
  {journal} {Phys. Rev. D}\ }\textbf {\bibinfo {volume} {96}},\ \bibinfo
  {pages} {114510} (\bibinfo {year} {2017})},\ \Eprint
  {http://arxiv.org/abs/1706.06550} {arXiv:1706.06550 [hep-lat]} \BibitemShut
  {NoStop}%
\bibitem [{\citenamefont {Illa}\ \emph {et~al.}(2021)\citenamefont {Illa} \emph
  {et~al.}}]{Illa:2020nsi}%
  \BibitemOpen
  \bibfield  {author} {\bibinfo {author} {\bibfnamefont {M.}~\bibnamefont
  {Illa}} \emph {et~al.} (\bibinfo {collaboration} {NPLQCD}),\ }\href {\doibase
  10.1103/PhysRevD.103.054508} {\bibfield  {journal} {\bibinfo  {journal}
  {Phys. Rev. D}\ }\textbf {\bibinfo {volume} {103}},\ \bibinfo {pages}
  {054508} (\bibinfo {year} {2021})},\ \Eprint
  {http://arxiv.org/abs/2009.12357} {arXiv:2009.12357 [hep-lat]} \BibitemShut
  {NoStop}%
\bibitem [{\citenamefont {Dudek}\ \emph {et~al.}(2013)\citenamefont {Dudek},
  \citenamefont {Edwards},\ and\ \citenamefont {Thomas}}]{Dudek:2012xn}%
  \BibitemOpen
  \bibfield  {author} {\bibinfo {author} {\bibfnamefont {J.~J.}\ \bibnamefont
  {Dudek}}, \bibinfo {author} {\bibfnamefont {R.~G.}\ \bibnamefont {Edwards}},
  \ and\ \bibinfo {author} {\bibfnamefont {C.~E.}\ \bibnamefont {Thomas}}
  (\bibinfo {collaboration} {Hadron Spectrum}),\ }\href {\doibase
  10.1103/PhysRevD.87.034505} {\bibfield  {journal} {\bibinfo  {journal} {Phys.
  Rev. D}\ }\textbf {\bibinfo {volume} {87}},\ \bibinfo {pages} {034505}
  (\bibinfo {year} {2013})},\ \bibinfo {note} {[Erratum: Phys. Rev. D {\bf 90},
  099902 (2014)]},\ \Eprint {http://arxiv.org/abs/1212.0830} {arXiv:1212.0830
  [hep-ph]} \BibitemShut {NoStop}%
\bibitem [{\citenamefont {Wilson}\ \emph {et~al.}(2015)\citenamefont {Wilson},
  \citenamefont {Briceno}, \citenamefont {Dudek}, \citenamefont {Edwards},\
  and\ \citenamefont {Thomas}}]{Wilson:2015dqa}%
  \BibitemOpen
  \bibfield  {author} {\bibinfo {author} {\bibfnamefont {D.~J.}\ \bibnamefont
  {Wilson}}, \bibinfo {author} {\bibfnamefont {R.~A.}\ \bibnamefont {Briceno}},
  \bibinfo {author} {\bibfnamefont {J.~J.}\ \bibnamefont {Dudek}}, \bibinfo
  {author} {\bibfnamefont {R.~G.}\ \bibnamefont {Edwards}}, \ and\ \bibinfo
  {author} {\bibfnamefont {C.~E.}\ \bibnamefont {Thomas}},\ }\href {\doibase
  10.1103/PhysRevD.92.094502} {\bibfield  {journal} {\bibinfo  {journal} {Phys.
  Rev. D}\ }\textbf {\bibinfo {volume} {92}},\ \bibinfo {pages} {094502}
  (\bibinfo {year} {2015})},\ \Eprint {http://arxiv.org/abs/1507.02599}
  {arXiv:1507.02599 [hep-ph]} \BibitemShut {NoStop}%
\bibitem [{\citenamefont {Lang}\ and\ \citenamefont
  {Verduci}(2013)}]{Lang:2012db}%
  \BibitemOpen
  \bibfield  {author} {\bibinfo {author} {\bibfnamefont {C.~B.}\ \bibnamefont
  {Lang}}\ and\ \bibinfo {author} {\bibfnamefont {V.}~\bibnamefont {Verduci}},\
  }\href {\doibase 10.1103/PhysRevD.87.054502} {\bibfield  {journal} {\bibinfo
  {journal} {Phys. Rev. D}\ }\textbf {\bibinfo {volume} {87}},\ \bibinfo
  {pages} {054502} (\bibinfo {year} {2013})},\ \Eprint
  {http://arxiv.org/abs/1212.5055} {arXiv:1212.5055 [hep-lat]} \BibitemShut
  {NoStop}%
\bibitem [{\citenamefont {Kiratidis}\ \emph {et~al.}(2015)\citenamefont
  {Kiratidis}, \citenamefont {Kamleh}, \citenamefont {Leinweber},\ and\
  \citenamefont {Owen}}]{Kiratidis:2015vpa}%
  \BibitemOpen
  \bibfield  {author} {\bibinfo {author} {\bibfnamefont {A.~L.}\ \bibnamefont
  {Kiratidis}}, \bibinfo {author} {\bibfnamefont {W.}~\bibnamefont {Kamleh}},
  \bibinfo {author} {\bibfnamefont {D.~B.}\ \bibnamefont {Leinweber}}, \ and\
  \bibinfo {author} {\bibfnamefont {B.~J.}\ \bibnamefont {Owen}},\ }\href
  {\doibase 10.1103/PhysRevD.91.094509} {\bibfield  {journal} {\bibinfo
  {journal} {Phys. Rev. D}\ }\textbf {\bibinfo {volume} {91}},\ \bibinfo
  {pages} {094509} (\bibinfo {year} {2015})},\ \Eprint
  {http://arxiv.org/abs/1501.07667} {arXiv:1501.07667 [hep-lat]} \BibitemShut
  {NoStop}%
\bibitem [{\citenamefont {Formaggio}\ and\ \citenamefont
  {Zeller}(2012)}]{Formaggio:2013kya}%
  \BibitemOpen
  \bibfield  {author} {\bibinfo {author} {\bibfnamefont {J.}~\bibnamefont
  {Formaggio}}\ and\ \bibinfo {author} {\bibfnamefont {G.}~\bibnamefont
  {Zeller}},\ }\href {\doibase 10.1103/RevModPhys.84.1307} {\bibfield
  {journal} {\bibinfo  {journal} {Rev. Mod. Phys.}\ }\textbf {\bibinfo {volume}
  {84}},\ \bibinfo {pages} {1307} (\bibinfo {year} {2012})},\ \Eprint
  {http://arxiv.org/abs/1305.7513} {arXiv:1305.7513 [hep-ex]} \BibitemShut
  {NoStop}%
\bibitem [{\citenamefont {Strait}\ \emph {et~al.}(2016)\citenamefont {Strait}
  \emph {et~al.}}]{DUNE:2016evb}%
  \BibitemOpen
  \bibfield  {author} {\bibinfo {author} {\bibfnamefont {J.}~\bibnamefont
  {Strait}} \emph {et~al.} (\bibinfo {collaboration} {DUNE}),\ }\href@noop {}
  {\  (\bibinfo {year} {2016})},\ \Eprint {http://arxiv.org/abs/1601.05823}
  {arXiv:1601.05823 [physics.ins-det]} \BibitemShut {NoStop}%
\bibitem [{\citenamefont {Acciarri}\ \emph {et~al.}(2015)\citenamefont
  {Acciarri} \emph {et~al.}}]{DUNE:2015lol}%
  \BibitemOpen
  \bibfield  {author} {\bibinfo {author} {\bibfnamefont {R.}~\bibnamefont
  {Acciarri}} \emph {et~al.} (\bibinfo {collaboration} {DUNE}),\ }\href@noop {}
  {\  (\bibinfo {year} {2015})},\ \Eprint {http://arxiv.org/abs/1512.06148}
  {arXiv:1512.06148 [physics.ins-det]} \BibitemShut {NoStop}%
\bibitem [{\citenamefont {Adler}(1968)}]{Adler:1968tw}%
  \BibitemOpen
  \bibfield  {author} {\bibinfo {author} {\bibfnamefont {S.~L.}\ \bibnamefont
  {Adler}},\ }\href {\doibase 10.1016/0003-4916(68)90278-9} {\bibfield
  {journal} {\bibinfo  {journal} {Annals Phys.}\ }\textbf {\bibinfo {volume}
  {50}},\ \bibinfo {pages} {189} (\bibinfo {year} {1968})}\BibitemShut
  {NoStop}%
\bibitem [{\citenamefont {Hernandez}\ \emph {et~al.}(2007)\citenamefont
  {Hernandez}, \citenamefont {Nieves},\ and\ \citenamefont
  {Valverde}}]{Hernandez:2007qq}%
  \BibitemOpen
  \bibfield  {author} {\bibinfo {author} {\bibfnamefont {E.}~\bibnamefont
  {Hernandez}}, \bibinfo {author} {\bibfnamefont {J.}~\bibnamefont {Nieves}}, \
  and\ \bibinfo {author} {\bibfnamefont {M.}~\bibnamefont {Valverde}},\ }\href
  {\doibase 10.1103/PhysRevD.76.033005} {\bibfield  {journal} {\bibinfo
  {journal} {Phys. Rev. D}\ }\textbf {\bibinfo {volume} {76}},\ \bibinfo
  {pages} {033005} (\bibinfo {year} {2007})},\ \Eprint
  {http://arxiv.org/abs/hep-ph/0701149} {arXiv:hep-ph/0701149} \BibitemShut
  {NoStop}%
\bibitem [{\citenamefont {Hernandez}\ \emph {et~al.}(2010)\citenamefont
  {Hernandez}, \citenamefont {Nieves}, \citenamefont {Valverde},\ and\
  \citenamefont {Vicente~Vacas}}]{Hernandez:2010bx}%
  \BibitemOpen
  \bibfield  {author} {\bibinfo {author} {\bibfnamefont {E.}~\bibnamefont
  {Hernandez}}, \bibinfo {author} {\bibfnamefont {J.}~\bibnamefont {Nieves}},
  \bibinfo {author} {\bibfnamefont {M.}~\bibnamefont {Valverde}}, \ and\
  \bibinfo {author} {\bibfnamefont {M.~J.}\ \bibnamefont {Vicente~Vacas}},\
  }\href {\doibase 10.1103/PhysRevD.81.085046} {\bibfield  {journal} {\bibinfo
  {journal} {Phys. Rev. D}\ }\textbf {\bibinfo {volume} {81}},\ \bibinfo
  {pages} {085046} (\bibinfo {year} {2010})},\ \Eprint
  {http://arxiv.org/abs/1001.4416} {arXiv:1001.4416 [hep-ph]} \BibitemShut
  {NoStop}%
\bibitem [{\citenamefont {Fox}\ \emph {et~al.}(1982)\citenamefont {Fox},
  \citenamefont {Gupta}, \citenamefont {Martin},\ and\ \citenamefont
  {Otto}}]{Fox:1981xz}%
  \BibitemOpen
  \bibfield  {author} {\bibinfo {author} {\bibfnamefont {G.}~\bibnamefont
  {Fox}}, \bibinfo {author} {\bibfnamefont {R.}~\bibnamefont {Gupta}}, \bibinfo
  {author} {\bibfnamefont {O.}~\bibnamefont {Martin}}, \ and\ \bibinfo {author}
  {\bibfnamefont {S.}~\bibnamefont {Otto}},\ }\href {\doibase
  10.1016/0550-3213(82)90384-4} {\bibfield  {journal} {\bibinfo  {journal}
  {Nucl. Phys. B}\ }\textbf {\bibinfo {volume} {205}},\ \bibinfo {pages} {188}
  (\bibinfo {year} {1982})}\BibitemShut {NoStop}%
\bibitem [{\citenamefont {Michael}\ and\ \citenamefont
  {Teasdale}(1983)}]{Michael:1982gb}%
  \BibitemOpen
  \bibfield  {author} {\bibinfo {author} {\bibfnamefont {C.}~\bibnamefont
  {Michael}}\ and\ \bibinfo {author} {\bibfnamefont {I.}~\bibnamefont
  {Teasdale}},\ }\href {\doibase 10.1016/0550-3213(83)90674-0} {\bibfield
  {journal} {\bibinfo  {journal} {Nucl. Phys. B}\ }\textbf {\bibinfo {volume}
  {215}},\ \bibinfo {pages} {433} (\bibinfo {year} {1983})}\BibitemShut
  {NoStop}%
\bibitem [{\citenamefont {L{\"u}scher}\ and\ \citenamefont
  {Wolff}(1990)}]{Luscher:1990ck}%
  \BibitemOpen
  \bibfield  {author} {\bibinfo {author} {\bibfnamefont {M.}~\bibnamefont
  {L{\"u}scher}}\ and\ \bibinfo {author} {\bibfnamefont {U.}~\bibnamefont
  {Wolff}},\ }\href {\doibase 10.1016/0550-3213(90)90540-T} {\bibfield
  {journal} {\bibinfo  {journal} {Nucl. Phys. B}\ }\textbf {\bibinfo {volume}
  {339}},\ \bibinfo {pages} {222} (\bibinfo {year} {1990})}\BibitemShut
  {NoStop}%
\bibitem [{\citenamefont {Peardon}\ \emph {et~al.}(2009)\citenamefont
  {Peardon}, \citenamefont {Bulava}, \citenamefont {Foley}, \citenamefont
  {Morningstar}, \citenamefont {Dudek}, \citenamefont {Edwards}, \citenamefont
  {Joo}, \citenamefont {Lin}, \citenamefont {Richards},\ and\ \citenamefont
  {Juge}}]{HadronSpectrum:2009krc}%
  \BibitemOpen
  \bibfield  {author} {\bibinfo {author} {\bibfnamefont {M.}~\bibnamefont
  {Peardon}}, \bibinfo {author} {\bibfnamefont {J.}~\bibnamefont {Bulava}},
  \bibinfo {author} {\bibfnamefont {J.}~\bibnamefont {Foley}}, \bibinfo
  {author} {\bibfnamefont {C.}~\bibnamefont {Morningstar}}, \bibinfo {author}
  {\bibfnamefont {J.}~\bibnamefont {Dudek}}, \bibinfo {author} {\bibfnamefont
  {R.~G.}\ \bibnamefont {Edwards}}, \bibinfo {author} {\bibfnamefont
  {B.}~\bibnamefont {Joo}}, \bibinfo {author} {\bibfnamefont {H.-W.}\
  \bibnamefont {Lin}}, \bibinfo {author} {\bibfnamefont {D.~G.}\ \bibnamefont
  {Richards}}, \ and\ \bibinfo {author} {\bibfnamefont {K.~J.}\ \bibnamefont
  {Juge}} (\bibinfo {collaboration} {Hadron Spectrum}),\ }\href {\doibase
  10.1103/PhysRevD.80.054506} {\bibfield  {journal} {\bibinfo  {journal} {Phys.
  Rev. D}\ }\textbf {\bibinfo {volume} {80}},\ \bibinfo {pages} {054506}
  (\bibinfo {year} {2009})},\ \Eprint {http://arxiv.org/abs/0905.2160}
  {arXiv:0905.2160 [hep-lat]} \BibitemShut {NoStop}%
\bibitem [{\citenamefont {Morningstar}\ \emph {et~al.}(2011)\citenamefont
  {Morningstar}, \citenamefont {Bulava}, \citenamefont {Foley}, \citenamefont
  {Juge}, \citenamefont {Lenkner}, \citenamefont {Peardon},\ and\ \citenamefont
  {Wong}}]{Morningstar:2011ka}%
  \BibitemOpen
  \bibfield  {author} {\bibinfo {author} {\bibfnamefont {C.}~\bibnamefont
  {Morningstar}}, \bibinfo {author} {\bibfnamefont {J.}~\bibnamefont {Bulava}},
  \bibinfo {author} {\bibfnamefont {J.}~\bibnamefont {Foley}}, \bibinfo
  {author} {\bibfnamefont {K.~J.}\ \bibnamefont {Juge}}, \bibinfo {author}
  {\bibfnamefont {D.}~\bibnamefont {Lenkner}}, \bibinfo {author} {\bibfnamefont
  {M.}~\bibnamefont {Peardon}}, \ and\ \bibinfo {author} {\bibfnamefont
  {C.~H.}\ \bibnamefont {Wong}},\ }\href {\doibase 10.1103/PhysRevD.83.114505}
  {\bibfield  {journal} {\bibinfo  {journal} {Phys. Rev. D}\ }\textbf {\bibinfo
  {volume} {83}},\ \bibinfo {pages} {114505} (\bibinfo {year} {2011})},\
  \Eprint {http://arxiv.org/abs/1104.3870} {arXiv:1104.3870 [hep-lat]}
  \BibitemShut {NoStop}%
\bibitem [{\citenamefont {Detmold}\ \emph
  {et~al.}(2021{\natexlab{a}})\citenamefont {Detmold}, \citenamefont {Murphy},
  \citenamefont {Pochinsky}, \citenamefont {Savage}, \citenamefont {Shanahan},\
  and\ \citenamefont {Wagman}}]{Detmold:2019fbk}%
  \BibitemOpen
  \bibfield  {author} {\bibinfo {author} {\bibfnamefont {W.}~\bibnamefont
  {Detmold}}, \bibinfo {author} {\bibfnamefont {D.~J.}\ \bibnamefont {Murphy}},
  \bibinfo {author} {\bibfnamefont {A.~V.}\ \bibnamefont {Pochinsky}}, \bibinfo
  {author} {\bibfnamefont {M.~J.}\ \bibnamefont {Savage}}, \bibinfo {author}
  {\bibfnamefont {P.~E.}\ \bibnamefont {Shanahan}}, \ and\ \bibinfo {author}
  {\bibfnamefont {M.~L.}\ \bibnamefont {Wagman}},\ }\href {\doibase
  10.1103/PhysRevD.104.034502} {\bibfield  {journal} {\bibinfo  {journal}
  {Phys. Rev. D}\ }\textbf {\bibinfo {volume} {104}},\ \bibinfo {pages}
  {034502} (\bibinfo {year} {2021}{\natexlab{a}})},\ \Eprint
  {http://arxiv.org/abs/1908.07050} {arXiv:1908.07050 [hep-lat]} \BibitemShut
  {NoStop}%
\bibitem [{\citenamefont {Li}\ \emph {et~al.}(2021{\natexlab{b}})\citenamefont
  {Li}, \citenamefont {Xia}, \citenamefont {Feng}, \citenamefont {Jin},\ and\
  \citenamefont {Liu}}]{Li:2020hbj}%
  \BibitemOpen
  \bibfield  {author} {\bibinfo {author} {\bibfnamefont {Y.}~\bibnamefont
  {Li}}, \bibinfo {author} {\bibfnamefont {S.-C.}\ \bibnamefont {Xia}},
  \bibinfo {author} {\bibfnamefont {X.}~\bibnamefont {Feng}}, \bibinfo {author}
  {\bibfnamefont {L.-C.}\ \bibnamefont {Jin}}, \ and\ \bibinfo {author}
  {\bibfnamefont {C.}~\bibnamefont {Liu}},\ }\href {\doibase
  10.1103/PhysRevD.103.014514} {\bibfield  {journal} {\bibinfo  {journal}
  {Phys. Rev. D}\ }\textbf {\bibinfo {volume} {103}},\ \bibinfo {pages}
  {014514} (\bibinfo {year} {2021}{\natexlab{b}})},\ \Eprint
  {http://arxiv.org/abs/2009.01029} {arXiv:2009.01029 [hep-lat]} \BibitemShut
  {NoStop}%
\bibitem [{\citenamefont {L{\"u}scher}(1986)}]{Luscher:1986pf}%
  \BibitemOpen
  \bibfield  {author} {\bibinfo {author} {\bibfnamefont {M.}~\bibnamefont
  {L{\"u}scher}},\ }\href {\doibase 10.1007/BF01211097} {\bibfield  {journal}
  {\bibinfo  {journal} {Commun. Math. Phys.}\ }\textbf {\bibinfo {volume}
  {105}},\ \bibinfo {pages} {153} (\bibinfo {year} {1986})}\BibitemShut
  {NoStop}%
\bibitem [{\citenamefont {L{\"u}scher}(1991{\natexlab{a}})}]{Luscher:1990ux}%
  \BibitemOpen
  \bibfield  {author} {\bibinfo {author} {\bibfnamefont {M.}~\bibnamefont
  {L{\"u}scher}},\ }\href {\doibase 10.1016/0550-3213(91)90366-6} {\bibfield
  {journal} {\bibinfo  {journal} {Nucl. Phys. B}\ }\textbf {\bibinfo {volume}
  {354}},\ \bibinfo {pages} {531} (\bibinfo {year}
  {1991}{\natexlab{a}})}\BibitemShut {NoStop}%
\bibitem [{\citenamefont {L{\"u}scher}(1991{\natexlab{b}})}]{Luscher:1991cf}%
  \BibitemOpen
  \bibfield  {author} {\bibinfo {author} {\bibfnamefont {M.}~\bibnamefont
  {L{\"u}scher}},\ }\href {\doibase 10.1016/0550-3213(91)90584-K} {\bibfield
  {journal} {\bibinfo  {journal} {Nucl. Phys. B}\ }\textbf {\bibinfo {volume}
  {364}},\ \bibinfo {pages} {237} (\bibinfo {year}
  {1991}{\natexlab{b}})}\BibitemShut {NoStop}%
\bibitem [{\citenamefont {Rummukainen}\ and\ \citenamefont
  {Gottlieb}(1995)}]{Rummukainen:1995vs}%
  \BibitemOpen
  \bibfield  {author} {\bibinfo {author} {\bibfnamefont {K.}~\bibnamefont
  {Rummukainen}}\ and\ \bibinfo {author} {\bibfnamefont {S.~A.}\ \bibnamefont
  {Gottlieb}},\ }\href {\doibase 10.1016/0550-3213(95)00313-H} {\bibfield
  {journal} {\bibinfo  {journal} {Nucl. Phys. B}\ }\textbf {\bibinfo {volume}
  {450}},\ \bibinfo {pages} {397} (\bibinfo {year} {1995})},\ \Eprint
  {http://arxiv.org/abs/hep-lat/9503028} {arXiv:hep-lat/9503028 [hep-lat]}
  \BibitemShut {NoStop}%
\bibitem [{\citenamefont {Lellouch}\ and\ \citenamefont
  {L{\"u}scher}(2001)}]{Lellouch:2000pv}%
  \BibitemOpen
  \bibfield  {author} {\bibinfo {author} {\bibfnamefont {L.}~\bibnamefont
  {Lellouch}}\ and\ \bibinfo {author} {\bibfnamefont {M.}~\bibnamefont
  {L{\"u}scher}},\ }\href {\doibase 10.1007/s002200100410} {\bibfield
  {journal} {\bibinfo  {journal} {Commun. Math. Phys.}\ }\textbf {\bibinfo
  {volume} {219}},\ \bibinfo {pages} {31} (\bibinfo {year} {2001})},\ \Eprint
  {http://arxiv.org/abs/hep-lat/0003023} {arXiv:hep-lat/0003023} \BibitemShut
  {NoStop}%
\bibitem [{\citenamefont {Brice\~no}\ \emph {et~al.}(2015)\citenamefont
  {Brice\~no}, \citenamefont {Hansen},\ and\ \citenamefont
  {Walker-Loud}}]{Briceno:2014uqa}%
  \BibitemOpen
  \bibfield  {author} {\bibinfo {author} {\bibfnamefont {R.~A.}\ \bibnamefont
  {Brice\~no}}, \bibinfo {author} {\bibfnamefont {M.~T.}\ \bibnamefont
  {Hansen}}, \ and\ \bibinfo {author} {\bibfnamefont {A.}~\bibnamefont
  {Walker-Loud}},\ }\href {\doibase 10.1103/PhysRevD.91.034501} {\bibfield
  {journal} {\bibinfo  {journal} {Phys. Rev. D}\ }\textbf {\bibinfo {volume}
  {91}},\ \bibinfo {pages} {034501} (\bibinfo {year} {2015})},\ \Eprint
  {http://arxiv.org/abs/1406.5965} {arXiv:1406.5965 [hep-lat]} \BibitemShut
  {NoStop}%
\bibitem [{\citenamefont {Brice\~no}\ and\ \citenamefont
  {Hansen}(2015)}]{Briceno:2015csa}%
  \BibitemOpen
  \bibfield  {author} {\bibinfo {author} {\bibfnamefont {R.~A.}\ \bibnamefont
  {Brice\~no}}\ and\ \bibinfo {author} {\bibfnamefont {M.~T.}\ \bibnamefont
  {Hansen}},\ }\href {\doibase 10.1103/PhysRevD.92.074509} {\bibfield
  {journal} {\bibinfo  {journal} {Phys. Rev. D}\ }\textbf {\bibinfo {volume}
  {92}},\ \bibinfo {pages} {074509} (\bibinfo {year} {2015})},\ \Eprint
  {http://arxiv.org/abs/1502.04314} {arXiv:1502.04314 [hep-lat]} \BibitemShut
  {NoStop}%
\bibitem [{\citenamefont {Brice\~no}\ and\ \citenamefont
  {Hansen}(2016)}]{Briceno:2015tza}%
  \BibitemOpen
  \bibfield  {author} {\bibinfo {author} {\bibfnamefont {R.~A.}\ \bibnamefont
  {Brice\~no}}\ and\ \bibinfo {author} {\bibfnamefont {M.~T.}\ \bibnamefont
  {Hansen}},\ }\href {\doibase 10.1103/PhysRevD.94.013008} {\bibfield
  {journal} {\bibinfo  {journal} {Phys. Rev. D}\ }\textbf {\bibinfo {volume}
  {94}},\ \bibinfo {pages} {013008} (\bibinfo {year} {2016})},\ \Eprint
  {http://arxiv.org/abs/1509.08507} {arXiv:1509.08507 [hep-lat]} \BibitemShut
  {NoStop}%
\bibitem [{\citenamefont {Baroni}\ \emph {et~al.}(2019)\citenamefont {Baroni},
  \citenamefont {Brice\~no}, \citenamefont {Hansen},\ and\ \citenamefont
  {Ortega-Gama}}]{Baroni:2018iau}%
  \BibitemOpen
  \bibfield  {author} {\bibinfo {author} {\bibfnamefont {A.}~\bibnamefont
  {Baroni}}, \bibinfo {author} {\bibfnamefont {R.~A.}\ \bibnamefont
  {Brice\~no}}, \bibinfo {author} {\bibfnamefont {M.~T.}\ \bibnamefont
  {Hansen}}, \ and\ \bibinfo {author} {\bibfnamefont {F.~G.}\ \bibnamefont
  {Ortega-Gama}},\ }\href {\doibase 10.1103/PhysRevD.100.034511} {\bibfield
  {journal} {\bibinfo  {journal} {Phys. Rev. D}\ }\textbf {\bibinfo {volume}
  {100}},\ \bibinfo {pages} {034511} (\bibinfo {year} {2019})},\ \Eprint
  {http://arxiv.org/abs/1812.10504} {arXiv:1812.10504 [hep-lat]} \BibitemShut
  {NoStop}%
\bibitem [{\citenamefont {Alexandrou}\ \emph {et~al.}(2007)\citenamefont
  {Alexandrou}, \citenamefont {Leontiou}, \citenamefont {Negele},\ and\
  \citenamefont {Tsapalis}}]{Alexandrou:2006mc}%
  \BibitemOpen
  \bibfield  {author} {\bibinfo {author} {\bibfnamefont {C.}~\bibnamefont
  {Alexandrou}}, \bibinfo {author} {\bibfnamefont {T.}~\bibnamefont
  {Leontiou}}, \bibinfo {author} {\bibfnamefont {J.~W.}\ \bibnamefont
  {Negele}}, \ and\ \bibinfo {author} {\bibfnamefont {A.}~\bibnamefont
  {Tsapalis}},\ }\href {\doibase 10.1103/PhysRevLett.98.052003} {\bibfield
  {journal} {\bibinfo  {journal} {Phys. Rev. Lett.}\ }\textbf {\bibinfo
  {volume} {98}},\ \bibinfo {pages} {052003} (\bibinfo {year} {2007})},\
  \Eprint {http://arxiv.org/abs/hep-lat/0607030} {arXiv:hep-lat/0607030}
  \BibitemShut {NoStop}%
\bibitem [{\citenamefont {Alexandrou}\ \emph {et~al.}(2008)\citenamefont
  {Alexandrou}, \citenamefont {Koutsou}, \citenamefont {Neff}, \citenamefont
  {Negele}, \citenamefont {Schroers},\ and\ \citenamefont
  {Tsapalis}}]{Alexandrou:2007dt}%
  \BibitemOpen
  \bibfield  {author} {\bibinfo {author} {\bibfnamefont {C.}~\bibnamefont
  {Alexandrou}}, \bibinfo {author} {\bibfnamefont {G.}~\bibnamefont {Koutsou}},
  \bibinfo {author} {\bibfnamefont {H.}~\bibnamefont {Neff}}, \bibinfo {author}
  {\bibfnamefont {J.~W.}\ \bibnamefont {Negele}}, \bibinfo {author}
  {\bibfnamefont {W.}~\bibnamefont {Schroers}}, \ and\ \bibinfo {author}
  {\bibfnamefont {A.}~\bibnamefont {Tsapalis}},\ }\href {\doibase
  10.1103/PhysRevD.77.085012} {\bibfield  {journal} {\bibinfo  {journal} {Phys.
  Rev. D}\ }\textbf {\bibinfo {volume} {77}},\ \bibinfo {pages} {085012}
  (\bibinfo {year} {2008})},\ \Eprint {http://arxiv.org/abs/0710.4621}
  {arXiv:0710.4621 [hep-lat]} \BibitemShut {NoStop}%
\bibitem [{\citenamefont {Alexandrou}\ \emph {et~al.}(2011)\citenamefont
  {Alexandrou}, \citenamefont {Koutsou}, \citenamefont {Negele}, \citenamefont
  {Proestos},\ and\ \citenamefont {Tsapalis}}]{Alexandrou:2010uk}%
  \BibitemOpen
  \bibfield  {author} {\bibinfo {author} {\bibfnamefont {C.}~\bibnamefont
  {Alexandrou}}, \bibinfo {author} {\bibfnamefont {G.}~\bibnamefont {Koutsou}},
  \bibinfo {author} {\bibfnamefont {J.~W.}\ \bibnamefont {Negele}}, \bibinfo
  {author} {\bibfnamefont {Y.}~\bibnamefont {Proestos}}, \ and\ \bibinfo
  {author} {\bibfnamefont {A.}~\bibnamefont {Tsapalis}},\ }\href {\doibase
  10.1103/PhysRevD.83.014501} {\bibfield  {journal} {\bibinfo  {journal} {Phys.
  Rev. D}\ }\textbf {\bibinfo {volume} {83}},\ \bibinfo {pages} {014501}
  (\bibinfo {year} {2011})},\ \Eprint {http://arxiv.org/abs/1011.3233}
  {arXiv:1011.3233 [hep-lat]} \BibitemShut {NoStop}%
\bibitem [{\citenamefont {Alexandrou}\ \emph {et~al.}(2013)\citenamefont
  {Alexandrou}, \citenamefont {Negele}, \citenamefont {Petschlies},
  \citenamefont {Strelchenko},\ and\ \citenamefont
  {Tsapalis}}]{Alexandrou:2013ata}%
  \BibitemOpen
  \bibfield  {author} {\bibinfo {author} {\bibfnamefont {C.}~\bibnamefont
  {Alexandrou}}, \bibinfo {author} {\bibfnamefont {J.~W.}\ \bibnamefont
  {Negele}}, \bibinfo {author} {\bibfnamefont {M.}~\bibnamefont {Petschlies}},
  \bibinfo {author} {\bibfnamefont {A.}~\bibnamefont {Strelchenko}}, \ and\
  \bibinfo {author} {\bibfnamefont {A.}~\bibnamefont {Tsapalis}},\ }\href
  {\doibase 10.1103/PhysRevD.88.031501} {\bibfield  {journal} {\bibinfo
  {journal} {Phys. Rev. D}\ }\textbf {\bibinfo {volume} {88}},\ \bibinfo
  {pages} {031501} (\bibinfo {year} {2013})},\ \Eprint
  {http://arxiv.org/abs/1305.6081} {arXiv:1305.6081 [hep-lat]} \BibitemShut
  {NoStop}%
\bibitem [{\citenamefont {Alexandrou}\ \emph {et~al.}(2016)\citenamefont
  {Alexandrou}, \citenamefont {Negele}, \citenamefont {Petschlies},
  \citenamefont {Pochinsky},\ and\ \citenamefont
  {Syritsyn}}]{Alexandrou:2015hxa}%
  \BibitemOpen
  \bibfield  {author} {\bibinfo {author} {\bibfnamefont {C.}~\bibnamefont
  {Alexandrou}}, \bibinfo {author} {\bibfnamefont {J.~W.}\ \bibnamefont
  {Negele}}, \bibinfo {author} {\bibfnamefont {M.}~\bibnamefont {Petschlies}},
  \bibinfo {author} {\bibfnamefont {A.~V.}\ \bibnamefont {Pochinsky}}, \ and\
  \bibinfo {author} {\bibfnamefont {S.~N.}\ \bibnamefont {Syritsyn}},\ }\href
  {\doibase 10.1103/PhysRevD.93.114515} {\bibfield  {journal} {\bibinfo
  {journal} {Phys. Rev. D}\ }\textbf {\bibinfo {volume} {93}},\ \bibinfo
  {pages} {114515} (\bibinfo {year} {2016})},\ \Eprint
  {http://arxiv.org/abs/1507.02724} {arXiv:1507.02724 [hep-lat]} \BibitemShut
  {NoStop}%
\bibitem [{\citenamefont {Owen}\ \emph {et~al.}(2015)\citenamefont {Owen},
  \citenamefont {Kamleh}, \citenamefont {Leinweber}, \citenamefont {Mahbub},\
  and\ \citenamefont {Menadue}}]{Owen:2015fra}%
  \BibitemOpen
  \bibfield  {author} {\bibinfo {author} {\bibfnamefont {B.~J.}\ \bibnamefont
  {Owen}}, \bibinfo {author} {\bibfnamefont {W.}~\bibnamefont {Kamleh}},
  \bibinfo {author} {\bibfnamefont {D.~B.}\ \bibnamefont {Leinweber}}, \bibinfo
  {author} {\bibfnamefont {M.~S.}\ \bibnamefont {Mahbub}}, \ and\ \bibinfo
  {author} {\bibfnamefont {B.~J.}\ \bibnamefont {Menadue}},\ }\href {\doibase
  10.1103/PhysRevD.92.034513} {\bibfield  {journal} {\bibinfo  {journal} {Phys.
  Rev. D}\ }\textbf {\bibinfo {volume} {92}},\ \bibinfo {pages} {034513}
  (\bibinfo {year} {2015})},\ \Eprint {http://arxiv.org/abs/1505.02876}
  {arXiv:1505.02876 [hep-lat]} \BibitemShut {NoStop}%
\bibitem [{\citenamefont {Lang}\ \emph {et~al.}(2017)\citenamefont {Lang},
  \citenamefont {Leskovec}, \citenamefont {Padmanath},\ and\ \citenamefont
  {Prelovsek}}]{Lang:2016hnn}%
  \BibitemOpen
  \bibfield  {author} {\bibinfo {author} {\bibfnamefont {C.~B.}\ \bibnamefont
  {Lang}}, \bibinfo {author} {\bibfnamefont {L.}~\bibnamefont {Leskovec}},
  \bibinfo {author} {\bibfnamefont {M.}~\bibnamefont {Padmanath}}, \ and\
  \bibinfo {author} {\bibfnamefont {S.}~\bibnamefont {Prelovsek}},\ }\href
  {\doibase 10.1103/PhysRevD.95.014510} {\bibfield  {journal} {\bibinfo
  {journal} {Phys. Rev. D}\ }\textbf {\bibinfo {volume} {95}},\ \bibinfo
  {pages} {014510} (\bibinfo {year} {2017})},\ \Eprint
  {http://arxiv.org/abs/1610.01422} {arXiv:1610.01422 [hep-lat]} \BibitemShut
  {NoStop}%
\bibitem [{\citenamefont {Wu}\ \emph {et~al.}(2018)\citenamefont {Wu},
  \citenamefont {Leinweber}, \citenamefont {Liu},\ and\ \citenamefont
  {Thomas}}]{Wu:2017qve}%
  \BibitemOpen
  \bibfield  {author} {\bibinfo {author} {\bibfnamefont {J.-j.}\ \bibnamefont
  {Wu}}, \bibinfo {author} {\bibfnamefont {D.~B.}\ \bibnamefont {Leinweber}},
  \bibinfo {author} {\bibfnamefont {Z.-w.}\ \bibnamefont {Liu}}, \ and\
  \bibinfo {author} {\bibfnamefont {A.~W.}\ \bibnamefont {Thomas}},\ }\href
  {\doibase 10.1103/PhysRevD.97.094509} {\bibfield  {journal} {\bibinfo
  {journal} {Phys. Rev. D}\ }\textbf {\bibinfo {volume} {97}},\ \bibinfo
  {pages} {094509} (\bibinfo {year} {2018})},\ \Eprint
  {http://arxiv.org/abs/1703.10715} {arXiv:1703.10715 [nucl-th]} \BibitemShut
  {NoStop}%
\bibitem [{\citenamefont {Roberts}\ \emph {et~al.}(2013)\citenamefont
  {Roberts}, \citenamefont {Kamleh},\ and\ \citenamefont
  {Leinweber}}]{Roberts:2013ipa}%
  \BibitemOpen
  \bibfield  {author} {\bibinfo {author} {\bibfnamefont {D.~S.}\ \bibnamefont
  {Roberts}}, \bibinfo {author} {\bibfnamefont {W.}~\bibnamefont {Kamleh}}, \
  and\ \bibinfo {author} {\bibfnamefont {D.~B.}\ \bibnamefont {Leinweber}},\
  }\href {\doibase 10.1016/j.physletb.2013.06.056} {\bibfield  {journal}
  {\bibinfo  {journal} {Phys. Lett. B}\ }\textbf {\bibinfo {volume} {725}},\
  \bibinfo {pages} {164} (\bibinfo {year} {2013})},\ \Eprint
  {http://arxiv.org/abs/1304.0325} {arXiv:1304.0325 [hep-lat]} \BibitemShut
  {NoStop}%
\bibitem [{\citenamefont {Engel}\ \emph {et~al.}(2013)\citenamefont {Engel},
  \citenamefont {Lang}, \citenamefont {Mohler},\ and\ \citenamefont
  {Sch\"afer}}]{Engel:2013ig}%
  \BibitemOpen
  \bibfield  {author} {\bibinfo {author} {\bibfnamefont {G.~P.}\ \bibnamefont
  {Engel}}, \bibinfo {author} {\bibfnamefont {C.~B.}\ \bibnamefont {Lang}},
  \bibinfo {author} {\bibfnamefont {D.}~\bibnamefont {Mohler}}, \ and\ \bibinfo
  {author} {\bibfnamefont {A.}~\bibnamefont {Sch\"afer}} (\bibinfo
  {collaboration} {BGR}),\ }\href {\doibase 10.1103/PhysRevD.87.074504}
  {\bibfield  {journal} {\bibinfo  {journal} {Phys. Rev. D}\ }\textbf {\bibinfo
  {volume} {87}},\ \bibinfo {pages} {074504} (\bibinfo {year} {2013})},\
  \Eprint {http://arxiv.org/abs/1301.4318} {arXiv:1301.4318 [hep-lat]}
  \BibitemShut {NoStop}%
\bibitem [{\citenamefont {Mahbub}\ \emph {et~al.}(2013)\citenamefont {Mahbub},
  \citenamefont {Kamleh}, \citenamefont {Leinweber}, \citenamefont {Moran},\
  and\ \citenamefont {Williams}}]{Mahbub:2013ala}%
  \BibitemOpen
  \bibfield  {author} {\bibinfo {author} {\bibfnamefont {M.~S.}\ \bibnamefont
  {Mahbub}}, \bibinfo {author} {\bibfnamefont {W.}~\bibnamefont {Kamleh}},
  \bibinfo {author} {\bibfnamefont {D.~B.}\ \bibnamefont {Leinweber}}, \bibinfo
  {author} {\bibfnamefont {P.~J.}\ \bibnamefont {Moran}}, \ and\ \bibinfo
  {author} {\bibfnamefont {A.~G.}\ \bibnamefont {Williams}},\ }\href {\doibase
  10.1103/PhysRevD.87.094506} {\bibfield  {journal} {\bibinfo  {journal} {Phys.
  Rev. D}\ }\textbf {\bibinfo {volume} {87}},\ \bibinfo {pages} {094506}
  (\bibinfo {year} {2013})},\ \Eprint {http://arxiv.org/abs/1302.2987}
  {arXiv:1302.2987 [hep-lat]} \BibitemShut {NoStop}%
\bibitem [{\citenamefont {Alexandrou}\ \emph {et~al.}(2014)\citenamefont
  {Alexandrou}, \citenamefont {Korzec}, \citenamefont {Koutsou},\ and\
  \citenamefont {Leontiou}}]{Alexandrou:2013fsu}%
  \BibitemOpen
  \bibfield  {author} {\bibinfo {author} {\bibfnamefont {C.}~\bibnamefont
  {Alexandrou}}, \bibinfo {author} {\bibfnamefont {T.}~\bibnamefont {Korzec}},
  \bibinfo {author} {\bibfnamefont {G.}~\bibnamefont {Koutsou}}, \ and\
  \bibinfo {author} {\bibfnamefont {T.}~\bibnamefont {Leontiou}},\ }\href
  {\doibase 10.1103/PhysRevD.89.034502} {\bibfield  {journal} {\bibinfo
  {journal} {Phys. Rev. D}\ }\textbf {\bibinfo {volume} {89}},\ \bibinfo
  {pages} {034502} (\bibinfo {year} {2014})},\ \Eprint
  {http://arxiv.org/abs/1302.4410} {arXiv:1302.4410 [hep-lat]} \BibitemShut
  {NoStop}%
\bibitem [{\citenamefont {Alexandrou}\ \emph {et~al.}(2015)\citenamefont
  {Alexandrou}, \citenamefont {Leontiou}, \citenamefont {Papanicolas},\ and\
  \citenamefont {Stiliaris}}]{Alexandrou:2014mka}%
  \BibitemOpen
  \bibfield  {author} {\bibinfo {author} {\bibfnamefont {C.}~\bibnamefont
  {Alexandrou}}, \bibinfo {author} {\bibfnamefont {T.}~\bibnamefont
  {Leontiou}}, \bibinfo {author} {\bibfnamefont {C.~N.}\ \bibnamefont
  {Papanicolas}}, \ and\ \bibinfo {author} {\bibfnamefont {E.}~\bibnamefont
  {Stiliaris}},\ }\href {\doibase 10.1103/PhysRevD.91.014506} {\bibfield
  {journal} {\bibinfo  {journal} {Phys. Rev. D}\ }\textbf {\bibinfo {volume}
  {91}},\ \bibinfo {pages} {014506} (\bibinfo {year} {2015})},\ \Eprint
  {http://arxiv.org/abs/1411.6765} {arXiv:1411.6765 [hep-lat]} \BibitemShut
  {NoStop}%
\bibitem [{\citenamefont {Kiratidis}\ \emph {et~al.}(2017)\citenamefont
  {Kiratidis}, \citenamefont {Kamleh}, \citenamefont {Leinweber}, \citenamefont
  {Liu}, \citenamefont {Stokes},\ and\ \citenamefont
  {Thomas}}]{Kiratidis:2016hda}%
  \BibitemOpen
  \bibfield  {author} {\bibinfo {author} {\bibfnamefont {A.~L.}\ \bibnamefont
  {Kiratidis}}, \bibinfo {author} {\bibfnamefont {W.}~\bibnamefont {Kamleh}},
  \bibinfo {author} {\bibfnamefont {D.~B.}\ \bibnamefont {Leinweber}}, \bibinfo
  {author} {\bibfnamefont {Z.-W.}\ \bibnamefont {Liu}}, \bibinfo {author}
  {\bibfnamefont {F.~M.}\ \bibnamefont {Stokes}}, \ and\ \bibinfo {author}
  {\bibfnamefont {A.~W.}\ \bibnamefont {Thomas}},\ }\href {\doibase
  10.1103/PhysRevD.95.074507} {\bibfield  {journal} {\bibinfo  {journal} {Phys.
  Rev. D}\ }\textbf {\bibinfo {volume} {95}},\ \bibinfo {pages} {074507}
  (\bibinfo {year} {2017})},\ \Eprint {http://arxiv.org/abs/1608.03051}
  {arXiv:1608.03051 [hep-lat]} \BibitemShut {NoStop}%
\bibitem [{\citenamefont {Edwards}\ \emph {et~al.}(2011)\citenamefont
  {Edwards}, \citenamefont {Dudek}, \citenamefont {Richards},\ and\
  \citenamefont {Wallace}}]{Edwards:2011jj}%
  \BibitemOpen
  \bibfield  {author} {\bibinfo {author} {\bibfnamefont {R.~G.}\ \bibnamefont
  {Edwards}}, \bibinfo {author} {\bibfnamefont {J.~J.}\ \bibnamefont {Dudek}},
  \bibinfo {author} {\bibfnamefont {D.~G.}\ \bibnamefont {Richards}}, \ and\
  \bibinfo {author} {\bibfnamefont {S.~J.}\ \bibnamefont {Wallace}},\ }\href
  {\doibase 10.1103/PhysRevD.84.074508} {\bibfield  {journal} {\bibinfo
  {journal} {Phys. Rev. D}\ }\textbf {\bibinfo {volume} {84}},\ \bibinfo
  {pages} {074508} (\bibinfo {year} {2011})},\ \Eprint
  {http://arxiv.org/abs/1104.5152} {arXiv:1104.5152 [hep-ph]} \BibitemShut
  {NoStop}%
\bibitem [{\citenamefont {Chen}\ \emph {et~al.}(2004)\citenamefont {Chen},
  \citenamefont {Dong}, \citenamefont {Draper}, \citenamefont {Horvath},
  \citenamefont {Liu}, \citenamefont {Mathur}, \citenamefont {Tamhankar},
  \citenamefont {Srinivasan}, \citenamefont {Lee},\ and\ \citenamefont
  {Zhang}}]{Chen:2004gp}%
  \BibitemOpen
  \bibfield  {author} {\bibinfo {author} {\bibfnamefont {Y.}~\bibnamefont
  {Chen}}, \bibinfo {author} {\bibfnamefont {S.-J.}\ \bibnamefont {Dong}},
  \bibinfo {author} {\bibfnamefont {T.}~\bibnamefont {Draper}}, \bibinfo
  {author} {\bibfnamefont {I.}~\bibnamefont {Horvath}}, \bibinfo {author}
  {\bibfnamefont {K.-F.}\ \bibnamefont {Liu}}, \bibinfo {author} {\bibfnamefont
  {N.}~\bibnamefont {Mathur}}, \bibinfo {author} {\bibfnamefont
  {S.}~\bibnamefont {Tamhankar}}, \bibinfo {author} {\bibfnamefont
  {C.}~\bibnamefont {Srinivasan}}, \bibinfo {author} {\bibfnamefont {F.~X.}\
  \bibnamefont {Lee}}, \ and\ \bibinfo {author} {\bibfnamefont {J.-b.}\
  \bibnamefont {Zhang}},\ }\href@noop {} {\  (\bibinfo {year} {2004})},\
  \Eprint {http://arxiv.org/abs/hep-lat/0405001} {arXiv:hep-lat/0405001}
  \BibitemShut {NoStop}%
\bibitem [{\citenamefont {Sun}\ \emph {et~al.}(2020)\citenamefont {Sun} \emph
  {et~al.}}]{xQCD:2019jke}%
  \BibitemOpen
  \bibfield  {author} {\bibinfo {author} {\bibfnamefont {M.}~\bibnamefont
  {Sun}} \emph {et~al.} (\bibinfo {collaboration} {$\chi$QCD}),\ }\href
  {\doibase 10.1103/PhysRevD.101.054511} {\bibfield  {journal} {\bibinfo
  {journal} {Phys. Rev. D}\ }\textbf {\bibinfo {volume} {101}},\ \bibinfo
  {pages} {054511} (\bibinfo {year} {2020})},\ \Eprint
  {http://arxiv.org/abs/1911.02635} {arXiv:1911.02635 [hep-ph]} \BibitemShut
  {NoStop}%
\bibitem [{\citenamefont {Liu}(2017)}]{Liu:2016rwa}%
  \BibitemOpen
  \bibfield  {author} {\bibinfo {author} {\bibfnamefont {K.-F.}\ \bibnamefont
  {Liu}},\ }\href {\doibase 10.1142/S021830131740016X} {\bibfield  {journal}
  {\bibinfo  {journal} {Int. J. Mod. Phys. E}\ }\textbf {\bibinfo {volume}
  {26}},\ \bibinfo {pages} {1740016} (\bibinfo {year} {2017})},\ \Eprint
  {http://arxiv.org/abs/1609.02572} {arXiv:1609.02572 [hep-ph]} \BibitemShut
  {NoStop}%
\bibitem [{\citenamefont {Verduci}\ and\ \citenamefont
  {Lang}(2014)}]{Verduci:2014csa}%
  \BibitemOpen
  \bibfield  {author} {\bibinfo {author} {\bibfnamefont {V.}~\bibnamefont
  {Verduci}}\ and\ \bibinfo {author} {\bibfnamefont {C.~B.}\ \bibnamefont
  {Lang}},\ }\href {\doibase 10.22323/1.214.0121} {\bibfield  {journal}
  {\bibinfo  {journal} {PoS}\ }\textbf {\bibinfo {volume} {LATTICE2014}},\
  \bibinfo {pages} {121} (\bibinfo {year} {2014})},\ \Eprint
  {http://arxiv.org/abs/1412.0701} {arXiv:1412.0701 [hep-lat]} \BibitemShut
  {NoStop}%
\bibitem [{\citenamefont {Verduci}(2014)}]{Verduci:2014btc}%
  \BibitemOpen
  \bibfield  {author} {\bibinfo {author} {\bibfnamefont {V.}~\bibnamefont
  {Verduci}},\ }\emph {\bibinfo {title} {{Pion-Nucleon Scattering in Lattice
  QCD}}},\ \href@noop {} {Ph.D. thesis},\ \bibinfo  {school} {Graz U.}
  (\bibinfo {year} {2014})\BibitemShut {NoStop}%
\bibitem [{\citenamefont {Benhar}\ \emph {et~al.}(2008)\citenamefont {Benhar},
  \citenamefont {day},\ and\ \citenamefont {Sick}}]{Benhar:2006wy}%
  \BibitemOpen
  \bibfield  {author} {\bibinfo {author} {\bibfnamefont {O.}~\bibnamefont
  {Benhar}}, \bibinfo {author} {\bibfnamefont {D.}~\bibnamefont {day}}, \ and\
  \bibinfo {author} {\bibfnamefont {I.}~\bibnamefont {Sick}},\ }\href {\doibase
  10.1103/RevModPhys.80.189} {\bibfield  {journal} {\bibinfo  {journal} {Rev.
  Mod. Phys.}\ }\textbf {\bibinfo {volume} {80}},\ \bibinfo {pages} {189}
  (\bibinfo {year} {2008})},\ \Eprint {http://arxiv.org/abs/nucl-ex/0603029}
  {arXiv:nucl-ex/0603029} \BibitemShut {NoStop}%
\bibitem [{\citenamefont {Lynn}\ \emph {et~al.}(2019)\citenamefont {Lynn},
  \citenamefont {Tews}, \citenamefont {Gandolfi},\ and\ \citenamefont
  {Lovato}}]{Lynn:2019rdt}%
  \BibitemOpen
  \bibfield  {author} {\bibinfo {author} {\bibfnamefont {J.~E.}\ \bibnamefont
  {Lynn}}, \bibinfo {author} {\bibfnamefont {I.}~\bibnamefont {Tews}}, \bibinfo
  {author} {\bibfnamefont {S.}~\bibnamefont {Gandolfi}}, \ and\ \bibinfo
  {author} {\bibfnamefont {A.}~\bibnamefont {Lovato}},\ }\href {\doibase
  10.1146/annurev-nucl-101918-023600} {\bibfield  {journal} {\bibinfo
  {journal} {Ann. Rev. Nucl. Part. Sci.}\ }\textbf {\bibinfo {volume} {69}},\
  \bibinfo {pages} {279} (\bibinfo {year} {2019})},\ \Eprint
  {http://arxiv.org/abs/1901.04868} {arXiv:1901.04868 [nucl-th]} \BibitemShut
  {NoStop}%
\bibitem [{\citenamefont {Carlson}\ \emph {et~al.}(2015)\citenamefont
  {Carlson}, \citenamefont {Gandolfi}, \citenamefont {Pederiva}, \citenamefont
  {Pieper}, \citenamefont {Schiavilla}, \citenamefont {Schmidt},\ and\
  \citenamefont {Wiringa}}]{Carlson:2014vla}%
  \BibitemOpen
  \bibfield  {author} {\bibinfo {author} {\bibfnamefont {J.}~\bibnamefont
  {Carlson}}, \bibinfo {author} {\bibfnamefont {S.}~\bibnamefont {Gandolfi}},
  \bibinfo {author} {\bibfnamefont {F.}~\bibnamefont {Pederiva}}, \bibinfo
  {author} {\bibfnamefont {S.~C.}\ \bibnamefont {Pieper}}, \bibinfo {author}
  {\bibfnamefont {R.}~\bibnamefont {Schiavilla}}, \bibinfo {author}
  {\bibfnamefont {K.~E.}\ \bibnamefont {Schmidt}}, \ and\ \bibinfo {author}
  {\bibfnamefont {R.~B.}\ \bibnamefont {Wiringa}},\ }\href {\doibase
  10.1103/RevModPhys.87.1067} {\bibfield  {journal} {\bibinfo  {journal} {Rev.
  Mod. Phys.}\ }\textbf {\bibinfo {volume} {87}},\ \bibinfo {pages} {1067}
  (\bibinfo {year} {2015})},\ \Eprint {http://arxiv.org/abs/1412.3081}
  {arXiv:1412.3081 [nucl-th]} \BibitemShut {NoStop}%
\bibitem [{\citenamefont {Benhar}\ \emph {et~al.}(2015)\citenamefont {Benhar},
  \citenamefont {Lovato},\ and\ \citenamefont {Rocco}}]{Benhar:2015ula}%
  \BibitemOpen
  \bibfield  {author} {\bibinfo {author} {\bibfnamefont {O.}~\bibnamefont
  {Benhar}}, \bibinfo {author} {\bibfnamefont {A.}~\bibnamefont {Lovato}}, \
  and\ \bibinfo {author} {\bibfnamefont {N.}~\bibnamefont {Rocco}},\ }\href
  {\doibase 10.1103/PhysRevC.92.024602} {\bibfield  {journal} {\bibinfo
  {journal} {Phys. Rev. C}\ }\textbf {\bibinfo {volume} {92}},\ \bibinfo
  {pages} {024602} (\bibinfo {year} {2015})},\ \Eprint
  {http://arxiv.org/abs/1502.00887} {arXiv:1502.00887 [nucl-th]} \BibitemShut
  {NoStop}%
\bibitem [{\citenamefont {Rocco}\ \emph {et~al.}(2016)\citenamefont {Rocco},
  \citenamefont {Lovato},\ and\ \citenamefont {Benhar}}]{Rocco:2015cil}%
  \BibitemOpen
  \bibfield  {author} {\bibinfo {author} {\bibfnamefont {N.}~\bibnamefont
  {Rocco}}, \bibinfo {author} {\bibfnamefont {A.}~\bibnamefont {Lovato}}, \
  and\ \bibinfo {author} {\bibfnamefont {O.}~\bibnamefont {Benhar}},\ }\href
  {\doibase 10.1103/PhysRevLett.116.192501} {\bibfield  {journal} {\bibinfo
  {journal} {Phys. Rev. Lett.}\ }\textbf {\bibinfo {volume} {116}},\ \bibinfo
  {pages} {192501} (\bibinfo {year} {2016})},\ \Eprint
  {http://arxiv.org/abs/1512.07426} {arXiv:1512.07426 [nucl-th]} \BibitemShut
  {NoStop}%
\bibitem [{\citenamefont {Rocco}\ \emph
  {et~al.}(2019{\natexlab{a}})\citenamefont {Rocco}, \citenamefont {Barbieri},
  \citenamefont {Benhar}, \citenamefont {De~Pace},\ and\ \citenamefont
  {Lovato}}]{Rocco:2018mwt}%
  \BibitemOpen
  \bibfield  {author} {\bibinfo {author} {\bibfnamefont {N.}~\bibnamefont
  {Rocco}}, \bibinfo {author} {\bibfnamefont {C.}~\bibnamefont {Barbieri}},
  \bibinfo {author} {\bibfnamefont {O.}~\bibnamefont {Benhar}}, \bibinfo
  {author} {\bibfnamefont {A.}~\bibnamefont {De~Pace}}, \ and\ \bibinfo
  {author} {\bibfnamefont {A.}~\bibnamefont {Lovato}},\ }\href {\doibase
  10.1103/PhysRevC.99.025502} {\bibfield  {journal} {\bibinfo  {journal} {Phys.
  Rev. C}\ }\textbf {\bibinfo {volume} {99}},\ \bibinfo {pages} {025502}
  (\bibinfo {year} {2019}{\natexlab{a}})},\ \Eprint
  {http://arxiv.org/abs/1810.07647} {arXiv:1810.07647 [nucl-th]} \BibitemShut
  {NoStop}%
\bibitem [{\citenamefont {Rocco}\ \emph
  {et~al.}(2019{\natexlab{b}})\citenamefont {Rocco}, \citenamefont {Nakamura},
  \citenamefont {Lee},\ and\ \citenamefont {Lovato}}]{Rocco:2019gfb}%
  \BibitemOpen
  \bibfield  {author} {\bibinfo {author} {\bibfnamefont {N.}~\bibnamefont
  {Rocco}}, \bibinfo {author} {\bibfnamefont {S.~X.}\ \bibnamefont {Nakamura}},
  \bibinfo {author} {\bibfnamefont {T.~S.~H.}\ \bibnamefont {Lee}}, \ and\
  \bibinfo {author} {\bibfnamefont {A.}~\bibnamefont {Lovato}},\ }\href
  {\doibase 10.1103/PhysRevC.100.045503} {\bibfield  {journal} {\bibinfo
  {journal} {Phys. Rev. C}\ }\textbf {\bibinfo {volume} {100}},\ \bibinfo
  {pages} {045503} (\bibinfo {year} {2019}{\natexlab{b}})},\ \Eprint
  {http://arxiv.org/abs/1907.01093} {arXiv:1907.01093 [nucl-th]} \BibitemShut
  {NoStop}%
\bibitem [{\citenamefont {Liu}\ and\ \citenamefont {Dong}(1994)}]{Liu:1993cv}%
  \BibitemOpen
  \bibfield  {author} {\bibinfo {author} {\bibfnamefont {K.-F.}\ \bibnamefont
  {Liu}}\ and\ \bibinfo {author} {\bibfnamefont {S.-J.}\ \bibnamefont {Dong}},\
  }\href {\doibase 10.1103/PhysRevLett.72.1790} {\bibfield  {journal} {\bibinfo
   {journal} {Phys. Rev. Lett.}\ }\textbf {\bibinfo {volume} {72}},\ \bibinfo
  {pages} {1790} (\bibinfo {year} {1994})},\ \Eprint
  {http://arxiv.org/abs/hep-ph/9306299} {arXiv:hep-ph/9306299} \BibitemShut
  {NoStop}%
\bibitem [{\citenamefont {Liu}(2000)}]{Liu:1999ak}%
  \BibitemOpen
  \bibfield  {author} {\bibinfo {author} {\bibfnamefont {K.-F.}\ \bibnamefont
  {Liu}},\ }\href {\doibase 10.1103/PhysRevD.62.074501} {\bibfield  {journal}
  {\bibinfo  {journal} {Phys. Rev. D}\ }\textbf {\bibinfo {volume} {62}},\
  \bibinfo {pages} {074501} (\bibinfo {year} {2000})},\ \Eprint
  {http://arxiv.org/abs/hep-ph/9910306} {arXiv:hep-ph/9910306} \BibitemShut
  {NoStop}%
\bibitem [{\citenamefont {Aglietti}\ \emph {et~al.}(1998)\citenamefont
  {Aglietti}, \citenamefont {Ciuchini}, \citenamefont {Corbo}, \citenamefont
  {Franco}, \citenamefont {Martinelli},\ and\ \citenamefont
  {Silvestrini}}]{Aglietti:1998mz}%
  \BibitemOpen
  \bibfield  {author} {\bibinfo {author} {\bibfnamefont {U.}~\bibnamefont
  {Aglietti}}, \bibinfo {author} {\bibfnamefont {M.}~\bibnamefont {Ciuchini}},
  \bibinfo {author} {\bibfnamefont {G.}~\bibnamefont {Corbo}}, \bibinfo
  {author} {\bibfnamefont {E.}~\bibnamefont {Franco}}, \bibinfo {author}
  {\bibfnamefont {G.}~\bibnamefont {Martinelli}}, \ and\ \bibinfo {author}
  {\bibfnamefont {L.}~\bibnamefont {Silvestrini}},\ }\href {\doibase
  10.1016/S0370-2693(98)00677-7} {\bibfield  {journal} {\bibinfo  {journal}
  {Phys. Lett. B}\ }\textbf {\bibinfo {volume} {432}},\ \bibinfo {pages} {411}
  (\bibinfo {year} {1998})},\ \Eprint {http://arxiv.org/abs/hep-ph/9804416}
  {arXiv:hep-ph/9804416} \BibitemShut {NoStop}%
\bibitem [{\citenamefont {Detmold}\ and\ \citenamefont
  {Lin}(2006)}]{Detmold:2005gg}%
  \BibitemOpen
  \bibfield  {author} {\bibinfo {author} {\bibfnamefont {W.}~\bibnamefont
  {Detmold}}\ and\ \bibinfo {author} {\bibfnamefont {C.~J.~D.}\ \bibnamefont
  {Lin}},\ }\href {\doibase 10.1103/PhysRevD.73.014501} {\bibfield  {journal}
  {\bibinfo  {journal} {Phys. Rev. D}\ }\textbf {\bibinfo {volume} {73}},\
  \bibinfo {pages} {014501} (\bibinfo {year} {2006})},\ \Eprint
  {http://arxiv.org/abs/hep-lat/0507007} {arXiv:hep-lat/0507007} \BibitemShut
  {NoStop}%
\bibitem [{\citenamefont {Can}\ \emph {et~al.}(2020)\citenamefont {Can} \emph
  {et~al.}}]{Can:2020sxc}%
  \BibitemOpen
  \bibfield  {author} {\bibinfo {author} {\bibfnamefont {K.~U.}\ \bibnamefont
  {Can}} \emph {et~al.},\ }\href {\doibase 10.1103/PhysRevD.102.114505}
  {\bibfield  {journal} {\bibinfo  {journal} {Phys. Rev. D}\ }\textbf {\bibinfo
  {volume} {102}},\ \bibinfo {pages} {114505} (\bibinfo {year} {2020})},\
  \Eprint {http://arxiv.org/abs/2007.01523} {arXiv:2007.01523 [hep-lat]}
  \BibitemShut {NoStop}%
\bibitem [{\citenamefont {Bryan}(1990)}]{Bryan:1990}%
  \BibitemOpen
  \bibfield  {author} {\bibinfo {author} {\bibfnamefont {R.}~\bibnamefont
  {Bryan}},\ }\href {\doibase doi:10.1007/BF02427376} {\bibfield  {journal}
  {\bibinfo  {journal} {Eur. Biophys. J.}\ }\textbf {\bibinfo {volume} {18}},\
  \bibinfo {pages} {165} (\bibinfo {year} {1990})}\BibitemShut {NoStop}%
\bibitem [{\citenamefont {Jarrell}\ and\ \citenamefont
  {Gubernatis}(1996)}]{Jarrell:1996rrw}%
  \BibitemOpen
  \bibfield  {author} {\bibinfo {author} {\bibfnamefont {M.}~\bibnamefont
  {Jarrell}}\ and\ \bibinfo {author} {\bibfnamefont {J.~E.}\ \bibnamefont
  {Gubernatis}},\ }\href {\doibase 10.1016/0370-1573(95)00074-7} {\bibfield
  {journal} {\bibinfo  {journal} {Phys. Rept.}\ }\textbf {\bibinfo {volume}
  {269}},\ \bibinfo {pages} {133} (\bibinfo {year} {1996})}\BibitemShut
  {NoStop}%
\bibitem [{\citenamefont {Burnier}\ and\ \citenamefont
  {Rothkopf}(2013)}]{Burnier:2013nla}%
  \BibitemOpen
  \bibfield  {author} {\bibinfo {author} {\bibfnamefont {Y.}~\bibnamefont
  {Burnier}}\ and\ \bibinfo {author} {\bibfnamefont {A.}~\bibnamefont
  {Rothkopf}},\ }\href {\doibase 10.1103/PhysRevLett.111.182003} {\bibfield
  {journal} {\bibinfo  {journal} {Phys. Rev. Lett.}\ }\textbf {\bibinfo
  {volume} {111}},\ \bibinfo {pages} {182003} (\bibinfo {year} {2013})},\
  \Eprint {http://arxiv.org/abs/1307.6106} {arXiv:1307.6106 [hep-lat]}
  \BibitemShut {NoStop}%
\bibitem [{\citenamefont {Backus}\ and\ \citenamefont
  {Gilbert}(1968)}]{Backus:1968}%
  \BibitemOpen
  \bibfield  {author} {\bibinfo {author} {\bibfnamefont {G.}~\bibnamefont
  {Backus}}\ and\ \bibinfo {author} {\bibfnamefont {F.}~\bibnamefont
  {Gilbert}},\ }\href@noop {} {\bibfield  {journal} {\bibinfo  {journal}
  {Geophysical Journal International}\ }\textbf {\bibinfo {volume} {16}},\
  \bibinfo {pages} {169} (\bibinfo {year} {1968})}\BibitemShut {NoStop}%
\bibitem [{\citenamefont {Hansen}\ \emph {et~al.}(2017)\citenamefont {Hansen},
  \citenamefont {Meyer},\ and\ \citenamefont {Robaina}}]{Hansen:2017mnd}%
  \BibitemOpen
  \bibfield  {author} {\bibinfo {author} {\bibfnamefont {M.~T.}\ \bibnamefont
  {Hansen}}, \bibinfo {author} {\bibfnamefont {H.~B.}\ \bibnamefont {Meyer}}, \
  and\ \bibinfo {author} {\bibfnamefont {D.}~\bibnamefont {Robaina}},\ }\href
  {\doibase 10.1103/PhysRevD.96.094513} {\bibfield  {journal} {\bibinfo
  {journal} {Phys. Rev. D}\ }\textbf {\bibinfo {volume} {96}},\ \bibinfo
  {pages} {094513} (\bibinfo {year} {2017})},\ \Eprint
  {http://arxiv.org/abs/1704.08993} {arXiv:1704.08993 [hep-lat]} \BibitemShut
  {NoStop}%
\bibitem [{\citenamefont {Liang}\ \emph {et~al.}(2020)\citenamefont {Liang},
  \citenamefont {Draper}, \citenamefont {Liu}, \citenamefont {Rothkopf},\ and\
  \citenamefont {Yang}}]{Liang:2019frk}%
  \BibitemOpen
  \bibfield  {author} {\bibinfo {author} {\bibfnamefont {J.}~\bibnamefont
  {Liang}}, \bibinfo {author} {\bibfnamefont {T.}~\bibnamefont {Draper}},
  \bibinfo {author} {\bibfnamefont {K.-F.}\ \bibnamefont {Liu}}, \bibinfo
  {author} {\bibfnamefont {A.}~\bibnamefont {Rothkopf}}, \ and\ \bibinfo
  {author} {\bibfnamefont {Y.-B.}\ \bibnamefont {Yang}} (\bibinfo
  {collaboration} {$\chi$QCD}),\ }\href {\doibase 10.1103/PhysRevD.101.114503}
  {\bibfield  {journal} {\bibinfo  {journal} {Phys. Rev. D}\ }\textbf {\bibinfo
  {volume} {101}},\ \bibinfo {pages} {114503} (\bibinfo {year} {2020})},\
  \Eprint {http://arxiv.org/abs/1906.05312} {arXiv:1906.05312 [hep-ph]}
  \BibitemShut {NoStop}%
\bibitem [{\citenamefont {Fei}\ \emph {et~al.}(2021)\citenamefont {Fei},
  \citenamefont {Yeh},\ and\ \citenamefont {Gull}}]{PhysRevLett.126.056402}%
  \BibitemOpen
  \bibfield  {author} {\bibinfo {author} {\bibfnamefont {J.}~\bibnamefont
  {Fei}}, \bibinfo {author} {\bibfnamefont {C.-N.}\ \bibnamefont {Yeh}}, \ and\
  \bibinfo {author} {\bibfnamefont {E.}~\bibnamefont {Gull}},\ }\href {\doibase
  10.1103/PhysRevLett.126.056402} {\bibfield  {journal} {\bibinfo  {journal}
  {Phys. Rev. Lett.}\ }\textbf {\bibinfo {volume} {126}},\ \bibinfo {pages}
  {056402} (\bibinfo {year} {2021})}\BibitemShut {NoStop}%
\bibitem [{\citenamefont {Fukaya}\ \emph {et~al.}(2020)\citenamefont {Fukaya},
  \citenamefont {Hashimoto}, \citenamefont {Kaneko},\ and\ \citenamefont
  {Ohki}}]{Fukaya:2020wpp}%
  \BibitemOpen
  \bibfield  {author} {\bibinfo {author} {\bibfnamefont {H.}~\bibnamefont
  {Fukaya}}, \bibinfo {author} {\bibfnamefont {S.}~\bibnamefont {Hashimoto}},
  \bibinfo {author} {\bibfnamefont {T.}~\bibnamefont {Kaneko}}, \ and\ \bibinfo
  {author} {\bibfnamefont {H.}~\bibnamefont {Ohki}},\ }\href {\doibase
  10.1103/PhysRevD.102.114516} {\bibfield  {journal} {\bibinfo  {journal}
  {Phys. Rev. D}\ }\textbf {\bibinfo {volume} {102}},\ \bibinfo {pages}
  {114516} (\bibinfo {year} {2020})},\ \Eprint
  {http://arxiv.org/abs/2010.01253} {arXiv:2010.01253 [hep-lat]} \BibitemShut
  {NoStop}%
\bibitem [{\citenamefont {Kronfeld}\ and\ \citenamefont
  {Photiadis}(1985)}]{Kronfeld:1984zv}%
  \BibitemOpen
  \bibfield  {author} {\bibinfo {author} {\bibfnamefont {A.~S.}\ \bibnamefont
  {Kronfeld}}\ and\ \bibinfo {author} {\bibfnamefont {D.~M.}\ \bibnamefont
  {Photiadis}},\ }\href {\doibase 10.1103/PhysRevD.31.2939} {\bibfield
  {journal} {\bibinfo  {journal} {Phys. Rev. D}\ }\textbf {\bibinfo {volume}
  {31}},\ \bibinfo {pages} {2939} (\bibinfo {year} {1985})}\BibitemShut
  {NoStop}%
\bibitem [{\citenamefont {Martinelli}\ and\ \citenamefont
  {Sachrajda}(1989)}]{Martinelli:1988rr}%
  \BibitemOpen
  \bibfield  {author} {\bibinfo {author} {\bibfnamefont {G.}~\bibnamefont
  {Martinelli}}\ and\ \bibinfo {author} {\bibfnamefont {C.~T.}\ \bibnamefont
  {Sachrajda}},\ }\href {\doibase 10.1016/0550-3213(89)90035-7} {\bibfield
  {journal} {\bibinfo  {journal} {Nucl. Phys. B}\ }\textbf {\bibinfo {volume}
  {316}},\ \bibinfo {pages} {355} (\bibinfo {year} {1989})}\BibitemShut
  {NoStop}%
\bibitem [{\citenamefont {Gockeler}\ \emph {et~al.}(1996)\citenamefont
  {Gockeler}, \citenamefont {Horsley}, \citenamefont {Ilgenfritz},
  \citenamefont {Perlt}, \citenamefont {Rakow}, \citenamefont {Schierholz},\
  and\ \citenamefont {Schiller}}]{Gockeler:1995wg}%
  \BibitemOpen
  \bibfield  {author} {\bibinfo {author} {\bibfnamefont {M.}~\bibnamefont
  {Gockeler}}, \bibinfo {author} {\bibfnamefont {R.}~\bibnamefont {Horsley}},
  \bibinfo {author} {\bibfnamefont {E.-M.}\ \bibnamefont {Ilgenfritz}},
  \bibinfo {author} {\bibfnamefont {H.}~\bibnamefont {Perlt}}, \bibinfo
  {author} {\bibfnamefont {P.~E.~L.}\ \bibnamefont {Rakow}}, \bibinfo {author}
  {\bibfnamefont {G.}~\bibnamefont {Schierholz}}, \ and\ \bibinfo {author}
  {\bibfnamefont {A.}~\bibnamefont {Schiller}},\ }\href {\doibase
  10.1103/PhysRevD.53.2317} {\bibfield  {journal} {\bibinfo  {journal} {Phys.
  Rev. D}\ }\textbf {\bibinfo {volume} {53}},\ \bibinfo {pages} {2317}
  (\bibinfo {year} {1996})},\ \Eprint {http://arxiv.org/abs/hep-lat/9508004}
  {arXiv:hep-lat/9508004} \BibitemShut {NoStop}%
\bibitem [{\citenamefont {Alexandrou}\ \emph {et~al.}(2017)\citenamefont
  {Alexandrou}, \citenamefont {Constantinou}, \citenamefont {Hadjiyiannakou},
  \citenamefont {Jansen}, \citenamefont {Kallidonis}, \citenamefont {Koutsou},
  \citenamefont {Vaquero Avil\'es-Casco},\ and\ \citenamefont
  {Wiese}}]{Alexandrou:2017oeh}%
  \BibitemOpen
  \bibfield  {author} {\bibinfo {author} {\bibfnamefont {C.}~\bibnamefont
  {Alexandrou}}, \bibinfo {author} {\bibfnamefont {M.}~\bibnamefont
  {Constantinou}}, \bibinfo {author} {\bibfnamefont {K.}~\bibnamefont
  {Hadjiyiannakou}}, \bibinfo {author} {\bibfnamefont {K.}~\bibnamefont
  {Jansen}}, \bibinfo {author} {\bibfnamefont {C.}~\bibnamefont {Kallidonis}},
  \bibinfo {author} {\bibfnamefont {G.}~\bibnamefont {Koutsou}}, \bibinfo
  {author} {\bibfnamefont {A.}~\bibnamefont {Vaquero Avil\'es-Casco}}, \ and\
  \bibinfo {author} {\bibfnamefont {C.}~\bibnamefont {Wiese}},\ }\href
  {\doibase 10.1103/PhysRevLett.119.142002} {\bibfield  {journal} {\bibinfo
  {journal} {Phys. Rev. Lett.}\ }\textbf {\bibinfo {volume} {119}},\ \bibinfo
  {pages} {142002} (\bibinfo {year} {2017})},\ \Eprint
  {http://arxiv.org/abs/1706.02973} {arXiv:1706.02973 [hep-lat]} \BibitemShut
  {NoStop}%
\bibitem [{\citenamefont {Fan}\ \emph {et~al.}(2018)\citenamefont {Fan},
  \citenamefont {Yang}, \citenamefont {Anthony}, \citenamefont {Lin},\ and\
  \citenamefont {Liu}}]{Fan:2018dxu}%
  \BibitemOpen
  \bibfield  {author} {\bibinfo {author} {\bibfnamefont {Z.-Y.}\ \bibnamefont
  {Fan}}, \bibinfo {author} {\bibfnamefont {Y.-B.}\ \bibnamefont {Yang}},
  \bibinfo {author} {\bibfnamefont {A.}~\bibnamefont {Anthony}}, \bibinfo
  {author} {\bibfnamefont {H.-W.}\ \bibnamefont {Lin}}, \ and\ \bibinfo
  {author} {\bibfnamefont {K.-F.}\ \bibnamefont {Liu}},\ }\href {\doibase
  10.1103/PhysRevLett.121.242001} {\bibfield  {journal} {\bibinfo  {journal}
  {Phys. Rev. Lett.}\ }\textbf {\bibinfo {volume} {121}},\ \bibinfo {pages}
  {242001} (\bibinfo {year} {2018})},\ \Eprint
  {http://arxiv.org/abs/1808.02077} {arXiv:1808.02077 [hep-lat]} \BibitemShut
  {NoStop}%
\bibitem [{\citenamefont {Mondal}\ \emph {et~al.}(2020)\citenamefont {Mondal},
  \citenamefont {Gupta}, \citenamefont {Park}, \citenamefont {Yoon},
  \citenamefont {Bhattacharya},\ and\ \citenamefont {Lin}}]{Mondal:2020cmt}%
  \BibitemOpen
  \bibfield  {author} {\bibinfo {author} {\bibfnamefont {S.}~\bibnamefont
  {Mondal}}, \bibinfo {author} {\bibfnamefont {R.}~\bibnamefont {Gupta}},
  \bibinfo {author} {\bibfnamefont {S.}~\bibnamefont {Park}}, \bibinfo {author}
  {\bibfnamefont {B.}~\bibnamefont {Yoon}}, \bibinfo {author} {\bibfnamefont
  {T.}~\bibnamefont {Bhattacharya}}, \ and\ \bibinfo {author} {\bibfnamefont
  {H.-W.}\ \bibnamefont {Lin}},\ }\href {\doibase 10.1103/PhysRevD.102.054512}
  {\bibfield  {journal} {\bibinfo  {journal} {Phys. Rev. D}\ }\textbf {\bibinfo
  {volume} {102}},\ \bibinfo {pages} {054512} (\bibinfo {year} {2020})},\
  \Eprint {http://arxiv.org/abs/2005.13779} {arXiv:2005.13779 [hep-lat]}
  \BibitemShut {NoStop}%
\bibitem [{\citenamefont {Lin}\ \emph {et~al.}(2018{\natexlab{a}})\citenamefont
  {Lin} \emph {et~al.}}]{Lin:2017snn}%
  \BibitemOpen
  \bibfield  {author} {\bibinfo {author} {\bibfnamefont {H.-W.}\ \bibnamefont
  {Lin}} \emph {et~al.},\ }\href {\doibase 10.1016/j.ppnp.2018.01.007}
  {\bibfield  {journal} {\bibinfo  {journal} {Prog. Part. Nucl. Phys.}\
  }\textbf {\bibinfo {volume} {100}},\ \bibinfo {pages} {107} (\bibinfo {year}
  {2018}{\natexlab{a}})},\ \Eprint {http://arxiv.org/abs/1711.07916}
  {arXiv:1711.07916 [hep-ph]} \BibitemShut {NoStop}%
\bibitem [{\citenamefont {Lin}\ \emph {et~al.}(2018{\natexlab{b}})\citenamefont
  {Lin}, \citenamefont {Melnitchouk}, \citenamefont {Prokudin}, \citenamefont
  {Sato},\ and\ \citenamefont {Shows}}]{Lin:2017stx}%
  \BibitemOpen
  \bibfield  {author} {\bibinfo {author} {\bibfnamefont {H.-W.}\ \bibnamefont
  {Lin}}, \bibinfo {author} {\bibfnamefont {W.}~\bibnamefont {Melnitchouk}},
  \bibinfo {author} {\bibfnamefont {A.}~\bibnamefont {Prokudin}}, \bibinfo
  {author} {\bibfnamefont {N.}~\bibnamefont {Sato}}, \ and\ \bibinfo {author}
  {\bibfnamefont {H.}~\bibnamefont {Shows}},\ }\href {\doibase
  10.1103/PhysRevLett.120.152502} {\bibfield  {journal} {\bibinfo  {journal}
  {Phys. Rev. Lett.}\ }\textbf {\bibinfo {volume} {120}},\ \bibinfo {pages}
  {152502} (\bibinfo {year} {2018}{\natexlab{b}})},\ \Eprint
  {http://arxiv.org/abs/1710.09858} {arXiv:1710.09858 [hep-ph]} \BibitemShut
  {NoStop}%
\bibitem [{\citenamefont {Constantinou}\ \emph {et~al.}(2021)\citenamefont
  {Constantinou} \emph {et~al.}}]{Constantinou:2020hdm}%
  \BibitemOpen
  \bibfield  {author} {\bibinfo {author} {\bibfnamefont {M.}~\bibnamefont
  {Constantinou}} \emph {et~al.},\ }\href {\doibase 10.1016/j.ppnp.2021.103908}
  {\bibfield  {journal} {\bibinfo  {journal} {Prog. Part. Nucl. Phys.}\
  }\textbf {\bibinfo {volume} {121}},\ \bibinfo {pages} {103908} (\bibinfo
  {year} {2021})},\ \Eprint {http://arxiv.org/abs/2006.08636} {arXiv:2006.08636
  [hep-ph]} \BibitemShut {NoStop}%
\bibitem [{\citenamefont {Braun}\ and\ \citenamefont
  {M\"uller}(2008)}]{Braun:2007wv}%
  \BibitemOpen
  \bibfield  {author} {\bibinfo {author} {\bibfnamefont {V.}~\bibnamefont
  {Braun}}\ and\ \bibinfo {author} {\bibfnamefont {D.}~\bibnamefont
  {M\"uller}},\ }\href {\doibase 10.1140/epjc/s10052-008-0608-4} {\bibfield
  {journal} {\bibinfo  {journal} {Eur. Phys. J. C}\ }\textbf {\bibinfo {volume}
  {55}},\ \bibinfo {pages} {349} (\bibinfo {year} {2008})},\ \Eprint
  {http://arxiv.org/abs/0709.1348} {arXiv:0709.1348 [hep-ph]} \BibitemShut
  {NoStop}%
\bibitem [{\citenamefont {Davoudi}\ and\ \citenamefont
  {Savage}(2012)}]{Davoudi:2012ya}%
  \BibitemOpen
  \bibfield  {author} {\bibinfo {author} {\bibfnamefont {Z.}~\bibnamefont
  {Davoudi}}\ and\ \bibinfo {author} {\bibfnamefont {M.~J.}\ \bibnamefont
  {Savage}},\ }\href {\doibase 10.1103/PhysRevD.86.054505} {\bibfield
  {journal} {\bibinfo  {journal} {Phys. Rev. D}\ }\textbf {\bibinfo {volume}
  {86}},\ \bibinfo {pages} {054505} (\bibinfo {year} {2012})},\ \Eprint
  {http://arxiv.org/abs/1204.4146} {arXiv:1204.4146 [hep-lat]} \BibitemShut
  {NoStop}%
\bibitem [{\citenamefont {Monahan}\ and\ \citenamefont
  {Orginos}(2015)}]{Monahan:2015lha}%
  \BibitemOpen
  \bibfield  {author} {\bibinfo {author} {\bibfnamefont {C.}~\bibnamefont
  {Monahan}}\ and\ \bibinfo {author} {\bibfnamefont {K.}~\bibnamefont
  {Orginos}},\ }\href {\doibase 10.1103/PhysRevD.91.074513} {\bibfield
  {journal} {\bibinfo  {journal} {Phys. Rev. D}\ }\textbf {\bibinfo {volume}
  {91}},\ \bibinfo {pages} {074513} (\bibinfo {year} {2015})},\ \Eprint
  {http://arxiv.org/abs/1501.05348} {arXiv:1501.05348 [hep-lat]} \BibitemShut
  {NoStop}%
\bibitem [{\citenamefont {Ji}(2013)}]{Ji:2013dva}%
  \BibitemOpen
  \bibfield  {author} {\bibinfo {author} {\bibfnamefont {X.}~\bibnamefont
  {Ji}},\ }\href {\doibase 10.1103/PhysRevLett.110.262002} {\bibfield
  {journal} {\bibinfo  {journal} {Phys. Rev. Lett.}\ }\textbf {\bibinfo
  {volume} {110}},\ \bibinfo {pages} {262002} (\bibinfo {year} {2013})},\
  \Eprint {http://arxiv.org/abs/1305.1539} {arXiv:1305.1539 [hep-ph]}
  \BibitemShut {NoStop}%
\bibitem [{\citenamefont {Ma}\ and\ \citenamefont {Qiu}(2018)}]{Ma:2014jla}%
  \BibitemOpen
  \bibfield  {author} {\bibinfo {author} {\bibfnamefont {Y.-Q.}\ \bibnamefont
  {Ma}}\ and\ \bibinfo {author} {\bibfnamefont {J.-W.}\ \bibnamefont {Qiu}},\
  }\href {\doibase 10.1103/PhysRevD.98.074021} {\bibfield  {journal} {\bibinfo
  {journal} {Phys. Rev. D}\ }\textbf {\bibinfo {volume} {98}},\ \bibinfo
  {pages} {074021} (\bibinfo {year} {2018})},\ \Eprint
  {http://arxiv.org/abs/1404.6860} {arXiv:1404.6860 [hep-ph]} \BibitemShut
  {NoStop}%
\bibitem [{\citenamefont {Radyushkin}(2017)}]{Radyushkin:2017cyf}%
  \BibitemOpen
  \bibfield  {author} {\bibinfo {author} {\bibfnamefont {A.~V.}\ \bibnamefont
  {Radyushkin}},\ }\href {\doibase 10.1103/PhysRevD.96.034025} {\bibfield
  {journal} {\bibinfo  {journal} {Phys. Rev. D}\ }\textbf {\bibinfo {volume}
  {96}},\ \bibinfo {pages} {034025} (\bibinfo {year} {2017})},\ \Eprint
  {http://arxiv.org/abs/1705.01488} {arXiv:1705.01488 [hep-ph]} \BibitemShut
  {NoStop}%
\bibitem [{\citenamefont {Chambers}\ \emph {et~al.}(2017)\citenamefont
  {Chambers}, \citenamefont {Horsley}, \citenamefont {Nakamura}, \citenamefont
  {Perlt}, \citenamefont {Rakow}, \citenamefont {Schierholz}, \citenamefont
  {Schiller}, \citenamefont {Somfleth}, \citenamefont {Young},\ and\
  \citenamefont {Zanotti}}]{Chambers:2017dov}%
  \BibitemOpen
  \bibfield  {author} {\bibinfo {author} {\bibfnamefont {A.~J.}\ \bibnamefont
  {Chambers}}, \bibinfo {author} {\bibfnamefont {R.}~\bibnamefont {Horsley}},
  \bibinfo {author} {\bibfnamefont {Y.}~\bibnamefont {Nakamura}}, \bibinfo
  {author} {\bibfnamefont {H.}~\bibnamefont {Perlt}}, \bibinfo {author}
  {\bibfnamefont {P.~E.~L.}\ \bibnamefont {Rakow}}, \bibinfo {author}
  {\bibfnamefont {G.}~\bibnamefont {Schierholz}}, \bibinfo {author}
  {\bibfnamefont {A.}~\bibnamefont {Schiller}}, \bibinfo {author}
  {\bibfnamefont {K.}~\bibnamefont {Somfleth}}, \bibinfo {author}
  {\bibfnamefont {R.~D.}\ \bibnamefont {Young}}, \ and\ \bibinfo {author}
  {\bibfnamefont {J.~M.}\ \bibnamefont {Zanotti}},\ }\href {\doibase
  10.1103/PhysRevLett.118.242001} {\bibfield  {journal} {\bibinfo  {journal}
  {Phys. Rev. Lett.}\ }\textbf {\bibinfo {volume} {118}},\ \bibinfo {pages}
  {242001} (\bibinfo {year} {2017})},\ \Eprint
  {http://arxiv.org/abs/1703.01153} {arXiv:1703.01153 [hep-lat]} \BibitemShut
  {NoStop}%
\bibitem [{\citenamefont {Cichy}\ and\ \citenamefont
  {Constantinou}(2019)}]{Cichy:2018mum}%
  \BibitemOpen
  \bibfield  {author} {\bibinfo {author} {\bibfnamefont {K.}~\bibnamefont
  {Cichy}}\ and\ \bibinfo {author} {\bibfnamefont {M.}~\bibnamefont
  {Constantinou}},\ }\href {\doibase 10.1155/2019/3036904} {\bibfield
  {journal} {\bibinfo  {journal} {Adv. High Energy Phys.}\ }\textbf {\bibinfo
  {volume} {2019}},\ \bibinfo {pages} {3036904} (\bibinfo {year} {2019})},\
  \Eprint {http://arxiv.org/abs/1811.07248} {arXiv:1811.07248 [hep-lat]}
  \BibitemShut {NoStop}%
\bibitem [{\citenamefont {Ji}\ \emph {et~al.}(2020{\natexlab{a}})\citenamefont
  {Ji}, \citenamefont {Liu}, \citenamefont {Liu}, \citenamefont {Zhang},\ and\
  \citenamefont {Zhao}}]{Ji:2020ect}%
  \BibitemOpen
  \bibfield  {author} {\bibinfo {author} {\bibfnamefont {X.}~\bibnamefont
  {Ji}}, \bibinfo {author} {\bibfnamefont {Y.}~\bibnamefont {Liu}}, \bibinfo
  {author} {\bibfnamefont {Y.-S.}\ \bibnamefont {Liu}}, \bibinfo {author}
  {\bibfnamefont {J.-H.}\ \bibnamefont {Zhang}}, \ and\ \bibinfo {author}
  {\bibfnamefont {Y.}~\bibnamefont {Zhao}},\ }\href@noop {} {\  (\bibinfo
  {year} {2020}{\natexlab{a}})},\ \Eprint {http://arxiv.org/abs/2004.03543}
  {arXiv:2004.03543 [hep-ph]} \BibitemShut {NoStop}%
\bibitem [{\citenamefont {Lin}\ \emph {et~al.}(2018{\natexlab{c}})\citenamefont
  {Lin}, \citenamefont {Chen}, \citenamefont {Ishikawa},\ and\ \citenamefont
  {Zhang}}]{Lin:2017ani}%
  \BibitemOpen
  \bibfield  {author} {\bibinfo {author} {\bibfnamefont {H.-W.}\ \bibnamefont
  {Lin}}, \bibinfo {author} {\bibfnamefont {J.-W.}\ \bibnamefont {Chen}},
  \bibinfo {author} {\bibfnamefont {T.}~\bibnamefont {Ishikawa}}, \ and\
  \bibinfo {author} {\bibfnamefont {J.-H.}\ \bibnamefont {Zhang}} (\bibinfo
  {collaboration} {LP3}),\ }\href {\doibase 10.1103/PhysRevD.98.054504}
  {\bibfield  {journal} {\bibinfo  {journal} {Phys. Rev. D}\ }\textbf {\bibinfo
  {volume} {98}},\ \bibinfo {pages} {054504} (\bibinfo {year}
  {2018}{\natexlab{c}})},\ \Eprint {http://arxiv.org/abs/1708.05301}
  {arXiv:1708.05301 [hep-lat]} \BibitemShut {NoStop}%
\bibitem [{\citenamefont {Alexandrou}\ \emph
  {et~al.}(2019{\natexlab{a}})\citenamefont {Alexandrou}, \citenamefont
  {Cichy}, \citenamefont {Constantinou}, \citenamefont {Hadjiyiannakou},
  \citenamefont {Jansen}, \citenamefont {Scapellato},\ and\ \citenamefont
  {Steffens}}]{Alexandrou:2019lfo}%
  \BibitemOpen
  \bibfield  {author} {\bibinfo {author} {\bibfnamefont {C.}~\bibnamefont
  {Alexandrou}}, \bibinfo {author} {\bibfnamefont {K.}~\bibnamefont {Cichy}},
  \bibinfo {author} {\bibfnamefont {M.}~\bibnamefont {Constantinou}}, \bibinfo
  {author} {\bibfnamefont {K.}~\bibnamefont {Hadjiyiannakou}}, \bibinfo
  {author} {\bibfnamefont {K.}~\bibnamefont {Jansen}}, \bibinfo {author}
  {\bibfnamefont {A.}~\bibnamefont {Scapellato}}, \ and\ \bibinfo {author}
  {\bibfnamefont {F.}~\bibnamefont {Steffens}},\ }\href {\doibase
  10.1103/PhysRevD.99.114504} {\bibfield  {journal} {\bibinfo  {journal} {Phys.
  Rev. D}\ }\textbf {\bibinfo {volume} {99}},\ \bibinfo {pages} {114504}
  (\bibinfo {year} {2019}{\natexlab{a}})},\ \Eprint
  {http://arxiv.org/abs/1902.00587} {arXiv:1902.00587 [hep-lat]} \BibitemShut
  {NoStop}%
\bibitem [{\citenamefont {Lin}\ and\ \citenamefont
  {Zhang}(2019)}]{Lin:2019ocg}%
  \BibitemOpen
  \bibfield  {author} {\bibinfo {author} {\bibfnamefont {H.-W.}\ \bibnamefont
  {Lin}}\ and\ \bibinfo {author} {\bibfnamefont {R.}~\bibnamefont {Zhang}},\
  }\href {\doibase 10.1103/PhysRevD.100.074502} {\bibfield  {journal} {\bibinfo
   {journal} {Phys. Rev. D}\ }\textbf {\bibinfo {volume} {100}},\ \bibinfo
  {pages} {074502} (\bibinfo {year} {2019})}\BibitemShut {NoStop}%
\bibitem [{\citenamefont {Ji}\ \emph {et~al.}(2021)\citenamefont {Ji},
  \citenamefont {Liu}, \citenamefont {Sch\"afer}, \citenamefont {Wang},
  \citenamefont {Yang}, \citenamefont {Zhang},\ and\ \citenamefont
  {Zhao}}]{Ji:2020brr}%
  \BibitemOpen
  \bibfield  {author} {\bibinfo {author} {\bibfnamefont {X.}~\bibnamefont
  {Ji}}, \bibinfo {author} {\bibfnamefont {Y.}~\bibnamefont {Liu}}, \bibinfo
  {author} {\bibfnamefont {A.}~\bibnamefont {Sch\"afer}}, \bibinfo {author}
  {\bibfnamefont {W.}~\bibnamefont {Wang}}, \bibinfo {author} {\bibfnamefont
  {Y.-B.}\ \bibnamefont {Yang}}, \bibinfo {author} {\bibfnamefont {J.-H.}\
  \bibnamefont {Zhang}}, \ and\ \bibinfo {author} {\bibfnamefont
  {Y.}~\bibnamefont {Zhao}},\ }\href {\doibase 10.1016/j.nuclphysb.2021.115311}
  {\bibfield  {journal} {\bibinfo  {journal} {Nucl. Phys. B}\ }\textbf
  {\bibinfo {volume} {964}},\ \bibinfo {pages} {115311} (\bibinfo {year}
  {2021})},\ \Eprint {http://arxiv.org/abs/2008.03886} {arXiv:2008.03886
  [hep-ph]} \BibitemShut {NoStop}%
\bibitem [{\citenamefont {Huo}\ \emph {et~al.}(2021)\citenamefont {Huo} \emph
  {et~al.}}]{LatticePartonCollaborationLPC:2021xdx}%
  \BibitemOpen
  \bibfield  {author} {\bibinfo {author} {\bibfnamefont {Y.-K.}\ \bibnamefont
  {Huo}} \emph {et~al.} (\bibinfo {collaboration} {Lattice Parton Collaboration
  (LPC)}),\ }\href {\doibase 10.1016/j.nuclphysb.2021.115443} {\bibfield
  {journal} {\bibinfo  {journal} {Nucl. Phys. B}\ }\textbf {\bibinfo {volume}
  {969}},\ \bibinfo {pages} {115443} (\bibinfo {year} {2021})},\ \Eprint
  {http://arxiv.org/abs/2103.02965} {arXiv:2103.02965 [hep-lat]} \BibitemShut
  {NoStop}%
\bibitem [{\citenamefont {Bringewatt}\ \emph {et~al.}(2021)\citenamefont
  {Bringewatt}, \citenamefont {Sato}, \citenamefont {Melnitchouk},
  \citenamefont {Qiu}, \citenamefont {Steffens},\ and\ \citenamefont
  {Constantinou}}]{Bringewatt:2020ixn}%
  \BibitemOpen
  \bibfield  {author} {\bibinfo {author} {\bibfnamefont {J.}~\bibnamefont
  {Bringewatt}}, \bibinfo {author} {\bibfnamefont {N.}~\bibnamefont {Sato}},
  \bibinfo {author} {\bibfnamefont {W.}~\bibnamefont {Melnitchouk}}, \bibinfo
  {author} {\bibfnamefont {J.-W.}\ \bibnamefont {Qiu}}, \bibinfo {author}
  {\bibfnamefont {F.}~\bibnamefont {Steffens}}, \ and\ \bibinfo {author}
  {\bibfnamefont {M.}~\bibnamefont {Constantinou}},\ }\href {\doibase
  10.1103/PhysRevD.103.016003} {\bibfield  {journal} {\bibinfo  {journal}
  {Phys. Rev. D}\ }\textbf {\bibinfo {volume} {103}},\ \bibinfo {pages}
  {016003} (\bibinfo {year} {2021})},\ \Eprint
  {http://arxiv.org/abs/2010.00548} {arXiv:2010.00548 [hep-ph]} \BibitemShut
  {NoStop}%
\bibitem [{\citenamefont {Hobbs}\ \emph {et~al.}(2019)\citenamefont {Hobbs},
  \citenamefont {Wang}, \citenamefont {Nadolsky},\ and\ \citenamefont
  {Olness}}]{Hobbs:2019gob}%
  \BibitemOpen
  \bibfield  {author} {\bibinfo {author} {\bibfnamefont {T.~J.}\ \bibnamefont
  {Hobbs}}, \bibinfo {author} {\bibfnamefont {B.-T.}\ \bibnamefont {Wang}},
  \bibinfo {author} {\bibfnamefont {P.~M.}\ \bibnamefont {Nadolsky}}, \ and\
  \bibinfo {author} {\bibfnamefont {F.~I.}\ \bibnamefont {Olness}},\ }\href
  {\doibase 10.1103/PhysRevD.100.094040} {\bibfield  {journal} {\bibinfo
  {journal} {Phys. Rev. D}\ }\textbf {\bibinfo {volume} {100}},\ \bibinfo
  {pages} {094040} (\bibinfo {year} {2019})},\ \Eprint
  {http://arxiv.org/abs/1904.00022} {arXiv:1904.00022 [hep-ph]} \BibitemShut
  {NoStop}%
\bibitem [{\citenamefont {Davidson}\ \emph {et~al.}(2002)\citenamefont
  {Davidson}, \citenamefont {Forte}, \citenamefont {Gambino}, \citenamefont
  {Rius},\ and\ \citenamefont {Strumia}}]{Davidson:2001ji}%
  \BibitemOpen
  \bibfield  {author} {\bibinfo {author} {\bibfnamefont {S.}~\bibnamefont
  {Davidson}}, \bibinfo {author} {\bibfnamefont {S.}~\bibnamefont {Forte}},
  \bibinfo {author} {\bibfnamefont {P.}~\bibnamefont {Gambino}}, \bibinfo
  {author} {\bibfnamefont {N.}~\bibnamefont {Rius}}, \ and\ \bibinfo {author}
  {\bibfnamefont {A.}~\bibnamefont {Strumia}},\ }\href {\doibase
  10.1088/1126-6708/2002/02/037} {\bibfield  {journal} {\bibinfo  {journal}
  {JHEP}\ }\textbf {\bibinfo {volume} {02}},\ \bibinfo {pages} {037} (\bibinfo
  {year} {2002})},\ \Eprint {http://arxiv.org/abs/hep-ph/0112302}
  {arXiv:hep-ph/0112302} \BibitemShut {NoStop}%
\bibitem [{\citenamefont {Kretzer}\ \emph {et~al.}(2004)\citenamefont
  {Kretzer}, \citenamefont {Olness}, \citenamefont {Pumplin}, \citenamefont
  {Stump}, \citenamefont {Tung},\ and\ \citenamefont {Reno}}]{Kretzer:2003wy}%
  \BibitemOpen
  \bibfield  {author} {\bibinfo {author} {\bibfnamefont {S.}~\bibnamefont
  {Kretzer}}, \bibinfo {author} {\bibfnamefont {F.}~\bibnamefont {Olness}},
  \bibinfo {author} {\bibfnamefont {J.}~\bibnamefont {Pumplin}}, \bibinfo
  {author} {\bibfnamefont {D.}~\bibnamefont {Stump}}, \bibinfo {author}
  {\bibfnamefont {W.-K.}\ \bibnamefont {Tung}}, \ and\ \bibinfo {author}
  {\bibfnamefont {M.~H.}\ \bibnamefont {Reno}},\ }\href {\doibase
  10.1103/PhysRevLett.93.041802} {\bibfield  {journal} {\bibinfo  {journal}
  {Phys. Rev. Lett.}\ }\textbf {\bibinfo {volume} {93}},\ \bibinfo {pages}
  {041802} (\bibinfo {year} {2004})},\ \Eprint
  {http://arxiv.org/abs/hep-ph/0312322} {arXiv:hep-ph/0312322} \BibitemShut
  {NoStop}%
\bibitem [{\citenamefont {Zhang}\ \emph {et~al.}(2021)\citenamefont {Zhang},
  \citenamefont {Lin},\ and\ \citenamefont {Yoon}}]{Zhang:2020dkn}%
  \BibitemOpen
  \bibfield  {author} {\bibinfo {author} {\bibfnamefont {R.}~\bibnamefont
  {Zhang}}, \bibinfo {author} {\bibfnamefont {H.-W.}\ \bibnamefont {Lin}}, \
  and\ \bibinfo {author} {\bibfnamefont {B.}~\bibnamefont {Yoon}},\ }\href
  {\doibase 10.1103/PhysRevD.104.094511} {\bibfield  {journal} {\bibinfo
  {journal} {Phys. Rev. D}\ }\textbf {\bibinfo {volume} {104}},\ \bibinfo
  {pages} {094511} (\bibinfo {year} {2021})},\ \Eprint
  {http://arxiv.org/abs/2005.01124} {arXiv:2005.01124 [hep-lat]} \BibitemShut
  {NoStop}%
\bibitem [{\citenamefont {Ebert}\ \emph {et~al.}(2019)\citenamefont {Ebert},
  \citenamefont {Stewart},\ and\ \citenamefont {Zhao}}]{Ebert:2018gzl}%
  \BibitemOpen
  \bibfield  {author} {\bibinfo {author} {\bibfnamefont {M.~A.}\ \bibnamefont
  {Ebert}}, \bibinfo {author} {\bibfnamefont {I.~W.}\ \bibnamefont {Stewart}},
  \ and\ \bibinfo {author} {\bibfnamefont {Y.}~\bibnamefont {Zhao}},\ }\href
  {\doibase 10.1103/PhysRevD.99.034505} {\bibfield  {journal} {\bibinfo
  {journal} {Phys. Rev. D}\ }\textbf {\bibinfo {volume} {99}},\ \bibinfo
  {pages} {034505} (\bibinfo {year} {2019})},\ \Eprint
  {http://arxiv.org/abs/1811.00026} {arXiv:1811.00026 [hep-ph]} \BibitemShut
  {NoStop}%
\bibitem [{\citenamefont {Shanahan}\ \emph {et~al.}(2020)\citenamefont
  {Shanahan}, \citenamefont {Wagman},\ and\ \citenamefont
  {Zhao}}]{Shanahan:2020zxr}%
  \BibitemOpen
  \bibfield  {author} {\bibinfo {author} {\bibfnamefont {P.}~\bibnamefont
  {Shanahan}}, \bibinfo {author} {\bibfnamefont {M.}~\bibnamefont {Wagman}}, \
  and\ \bibinfo {author} {\bibfnamefont {Y.}~\bibnamefont {Zhao}},\ }\href
  {\doibase 10.1103/PhysRevD.102.014511} {\bibfield  {journal} {\bibinfo
  {journal} {Phys. Rev. D}\ }\textbf {\bibinfo {volume} {102}},\ \bibinfo
  {pages} {014511} (\bibinfo {year} {2020})},\ \Eprint
  {http://arxiv.org/abs/2003.06063} {arXiv:2003.06063 [hep-lat]} \BibitemShut
  {NoStop}%
\bibitem [{\citenamefont {Schlemmer}\ \emph {et~al.}(2021)\citenamefont
  {Schlemmer}, \citenamefont {Vladimirov}, \citenamefont {Zimmermann},
  \citenamefont {Engelhardt},\ and\ \citenamefont
  {Sch\"afer}}]{Schlemmer:2021aij}%
  \BibitemOpen
  \bibfield  {author} {\bibinfo {author} {\bibfnamefont {M.}~\bibnamefont
  {Schlemmer}}, \bibinfo {author} {\bibfnamefont {A.}~\bibnamefont
  {Vladimirov}}, \bibinfo {author} {\bibfnamefont {C.}~\bibnamefont
  {Zimmermann}}, \bibinfo {author} {\bibfnamefont {M.}~\bibnamefont
  {Engelhardt}}, \ and\ \bibinfo {author} {\bibfnamefont {A.}~\bibnamefont
  {Sch\"afer}},\ }\href {\doibase 10.1007/JHEP08(2021)004} {\bibfield
  {journal} {\bibinfo  {journal} {JHEP}\ }\textbf {\bibinfo {volume} {08}},\
  \bibinfo {pages} {004} (\bibinfo {year} {2021})},\ \Eprint
  {http://arxiv.org/abs/2103.16991} {arXiv:2103.16991 [hep-lat]} \BibitemShut
  {NoStop}%
\bibitem [{\citenamefont {Zhang}\ \emph
  {et~al.}(2020{\natexlab{b}})\citenamefont {Zhang} \emph
  {et~al.}}]{LatticeParton:2020uhz}%
  \BibitemOpen
  \bibfield  {author} {\bibinfo {author} {\bibfnamefont {Q.-A.}\ \bibnamefont
  {Zhang}} \emph {et~al.} (\bibinfo {collaboration} {Lattice Parton}),\ }\href
  {\doibase 10.1103/PhysRevLett.125.192001} {\bibfield  {journal} {\bibinfo
  {journal} {Phys. Rev. Lett.}\ }\textbf {\bibinfo {volume} {125}},\ \bibinfo
  {pages} {192001} (\bibinfo {year} {2020}{\natexlab{b}})},\ \Eprint
  {http://arxiv.org/abs/2005.14572} {arXiv:2005.14572 [hep-lat]} \BibitemShut
  {NoStop}%
\bibitem [{\citenamefont {Li}\ \emph {et~al.}(2022)\citenamefont {Li} \emph
  {et~al.}}]{Li:2021wvl}%
  \BibitemOpen
  \bibfield  {author} {\bibinfo {author} {\bibfnamefont {Y.}~\bibnamefont {Li}}
  \emph {et~al.},\ }\href {\doibase 10.1103/PhysRevLett.128.062002} {\bibfield
  {journal} {\bibinfo  {journal} {Phys. Rev. Lett.}\ }\textbf {\bibinfo
  {volume} {128}},\ \bibinfo {pages} {062002} (\bibinfo {year} {2022})},\
  \Eprint {http://arxiv.org/abs/2106.13027} {arXiv:2106.13027 [hep-lat]}
  \BibitemShut {NoStop}%
\bibitem [{\citenamefont {Shanahan}\ \emph {et~al.}(2021)\citenamefont
  {Shanahan}, \citenamefont {Wagman},\ and\ \citenamefont
  {Zhao}}]{Shanahan:2021tst}%
  \BibitemOpen
  \bibfield  {author} {\bibinfo {author} {\bibfnamefont {P.}~\bibnamefont
  {Shanahan}}, \bibinfo {author} {\bibfnamefont {M.}~\bibnamefont {Wagman}}, \
  and\ \bibinfo {author} {\bibfnamefont {Y.}~\bibnamefont {Zhao}},\ }\href
  {\doibase 10.1103/PhysRevD.104.114502} {\bibfield  {journal} {\bibinfo
  {journal} {Phys. Rev. D}\ }\textbf {\bibinfo {volume} {104}},\ \bibinfo
  {pages} {114502} (\bibinfo {year} {2021})},\ \Eprint
  {http://arxiv.org/abs/2107.11930} {arXiv:2107.11930 [hep-lat]} \BibitemShut
  {NoStop}%
\bibitem [{\citenamefont {Ji}\ \emph {et~al.}(2015)\citenamefont {Ji},
  \citenamefont {Sun}, \citenamefont {Xiong},\ and\ \citenamefont
  {Yuan}}]{Ji:2014hxa}%
  \BibitemOpen
  \bibfield  {author} {\bibinfo {author} {\bibfnamefont {X.}~\bibnamefont
  {Ji}}, \bibinfo {author} {\bibfnamefont {P.}~\bibnamefont {Sun}}, \bibinfo
  {author} {\bibfnamefont {X.}~\bibnamefont {Xiong}}, \ and\ \bibinfo {author}
  {\bibfnamefont {F.}~\bibnamefont {Yuan}},\ }\href {\doibase
  10.1103/PhysRevD.91.074009} {\bibfield  {journal} {\bibinfo  {journal} {Phys.
  Rev. D}\ }\textbf {\bibinfo {volume} {91}},\ \bibinfo {pages} {074009}
  (\bibinfo {year} {2015})},\ \Eprint {http://arxiv.org/abs/1405.7640}
  {arXiv:1405.7640 [hep-ph]} \BibitemShut {NoStop}%
\bibitem [{\citenamefont {Ji}\ \emph {et~al.}(2020{\natexlab{b}})\citenamefont
  {Ji}, \citenamefont {Liu},\ and\ \citenamefont {Liu}}]{Ji:2019sxk}%
  \BibitemOpen
  \bibfield  {author} {\bibinfo {author} {\bibfnamefont {X.}~\bibnamefont
  {Ji}}, \bibinfo {author} {\bibfnamefont {Y.}~\bibnamefont {Liu}}, \ and\
  \bibinfo {author} {\bibfnamefont {Y.-S.}\ \bibnamefont {Liu}},\ }\href
  {\doibase 10.1016/j.nuclphysb.2020.115054} {\bibfield  {journal} {\bibinfo
  {journal} {Nucl. Phys. B}\ }\textbf {\bibinfo {volume} {955}},\ \bibinfo
  {pages} {115054} (\bibinfo {year} {2020}{\natexlab{b}})},\ \Eprint
  {http://arxiv.org/abs/1910.11415} {arXiv:1910.11415 [hep-ph]} \BibitemShut
  {NoStop}%
\bibitem [{\citenamefont {Ebert}\ \emph {et~al.}(2022)\citenamefont {Ebert},
  \citenamefont {Schindler}, \citenamefont {Stewart},\ and\ \citenamefont
  {Zhao}}]{Ebert:2022fmh}%
  \BibitemOpen
  \bibfield  {author} {\bibinfo {author} {\bibfnamefont {M.~A.}\ \bibnamefont
  {Ebert}}, \bibinfo {author} {\bibfnamefont {S.~T.}\ \bibnamefont
  {Schindler}}, \bibinfo {author} {\bibfnamefont {I.~W.}\ \bibnamefont
  {Stewart}}, \ and\ \bibinfo {author} {\bibfnamefont {Y.}~\bibnamefont
  {Zhao}},\ }\href@noop {} {\  (\bibinfo {year} {2022})},\ \Eprint
  {http://arxiv.org/abs/2201.08401} {arXiv:2201.08401 [hep-ph]} \BibitemShut
  {NoStop}%
\bibitem [{\citenamefont {Constantinou}\ \emph {et~al.}(2022)\citenamefont
  {Constantinou} \emph {et~al.}}]{Constantinou:2022yye}%
  \BibitemOpen
  \bibfield  {author} {\bibinfo {author} {\bibfnamefont {M.}~\bibnamefont
  {Constantinou}} \emph {et~al.},\ }\href@noop {} {\  (\bibinfo {year}
  {2022})},\ \Eprint {http://arxiv.org/abs/2202.07193} {arXiv:2202.07193
  [hep-lat]} \BibitemShut {NoStop}%
\bibitem [{\citenamefont {Chen}\ and\ \citenamefont
  {Detmold}(2005)}]{Chen:2004zx}%
  \BibitemOpen
  \bibfield  {author} {\bibinfo {author} {\bibfnamefont {J.-W.}\ \bibnamefont
  {Chen}}\ and\ \bibinfo {author} {\bibfnamefont {W.}~\bibnamefont {Detmold}},\
  }\href {\doibase 10.1016/j.physletb.2005.08.041} {\bibfield  {journal}
  {\bibinfo  {journal} {Phys. Lett. B}\ }\textbf {\bibinfo {volume} {625}},\
  \bibinfo {pages} {165} (\bibinfo {year} {2005})},\ \Eprint
  {http://arxiv.org/abs/hep-ph/0412119} {arXiv:hep-ph/0412119} \BibitemShut
  {NoStop}%
\bibitem [{\citenamefont {Segarra}\ \emph {et~al.}(2021)\citenamefont {Segarra}
  \emph {et~al.}}]{Segarra:2020gtj}%
  \BibitemOpen
  \bibfield  {author} {\bibinfo {author} {\bibfnamefont {E.~P.}\ \bibnamefont
  {Segarra}} \emph {et~al.},\ }\href {\doibase 10.1103/PhysRevD.103.114015}
  {\bibfield  {journal} {\bibinfo  {journal} {Phys. Rev. D}\ }\textbf {\bibinfo
  {volume} {103}},\ \bibinfo {pages} {114015} (\bibinfo {year} {2021})},\
  \Eprint {http://arxiv.org/abs/2012.11566} {arXiv:2012.11566 [hep-ph]}
  \BibitemShut {NoStop}%
\bibitem [{\citenamefont {Kovarik}\ \emph {et~al.}(2016)\citenamefont {Kovarik}
  \emph {et~al.}}]{Kovarik:2015cma}%
  \BibitemOpen
  \bibfield  {author} {\bibinfo {author} {\bibfnamefont {K.}~\bibnamefont
  {Kovarik}} \emph {et~al.},\ }\href {\doibase 10.1103/PhysRevD.93.085037}
  {\bibfield  {journal} {\bibinfo  {journal} {Phys. Rev. D}\ }\textbf {\bibinfo
  {volume} {93}},\ \bibinfo {pages} {085037} (\bibinfo {year} {2016})},\
  \Eprint {http://arxiv.org/abs/1509.00792} {arXiv:1509.00792 [hep-ph]}
  \BibitemShut {NoStop}%
\bibitem [{\citenamefont {Abdul~Khalek}\ \emph {et~al.}(2020)\citenamefont
  {Abdul~Khalek}, \citenamefont {Ethier}, \citenamefont {Rojo},\ and\
  \citenamefont {van Weelden}}]{AbdulKhalek:2020yuc}%
  \BibitemOpen
  \bibfield  {author} {\bibinfo {author} {\bibfnamefont {R.}~\bibnamefont
  {Abdul~Khalek}}, \bibinfo {author} {\bibfnamefont {J.~J.}\ \bibnamefont
  {Ethier}}, \bibinfo {author} {\bibfnamefont {J.}~\bibnamefont {Rojo}}, \ and\
  \bibinfo {author} {\bibfnamefont {G.}~\bibnamefont {van Weelden}},\ }\href
  {\doibase 10.1007/JHEP09(2020)183} {\bibfield  {journal} {\bibinfo  {journal}
  {JHEP}\ }\textbf {\bibinfo {volume} {09}},\ \bibinfo {pages} {183} (\bibinfo
  {year} {2020})},\ \Eprint {http://arxiv.org/abs/2006.14629} {arXiv:2006.14629
  [hep-ph]} \BibitemShut {NoStop}%
\bibitem [{\citenamefont {Walt}\ \emph {et~al.}(2019)\citenamefont {Walt},
  \citenamefont {Helenius},\ and\ \citenamefont {Vogelsang}}]{Walt:2019slu}%
  \BibitemOpen
  \bibfield  {author} {\bibinfo {author} {\bibfnamefont {M.}~\bibnamefont
  {Walt}}, \bibinfo {author} {\bibfnamefont {I.}~\bibnamefont {Helenius}}, \
  and\ \bibinfo {author} {\bibfnamefont {W.}~\bibnamefont {Vogelsang}},\ }\href
  {\doibase 10.1103/PhysRevD.100.096015} {\bibfield  {journal} {\bibinfo
  {journal} {Phys. Rev. D}\ }\textbf {\bibinfo {volume} {100}},\ \bibinfo
  {pages} {096015} (\bibinfo {year} {2019})},\ \Eprint
  {http://arxiv.org/abs/1908.03355} {arXiv:1908.03355 [hep-ph]} \BibitemShut
  {NoStop}%
\bibitem [{\citenamefont {Eskola}\ \emph {et~al.}(2017)\citenamefont {Eskola},
  \citenamefont {Paakkinen}, \citenamefont {Paukkunen},\ and\ \citenamefont
  {Salgado}}]{Eskola:2016oht}%
  \BibitemOpen
  \bibfield  {author} {\bibinfo {author} {\bibfnamefont {K.~J.}\ \bibnamefont
  {Eskola}}, \bibinfo {author} {\bibfnamefont {P.}~\bibnamefont {Paakkinen}},
  \bibinfo {author} {\bibfnamefont {H.}~\bibnamefont {Paukkunen}}, \ and\
  \bibinfo {author} {\bibfnamefont {C.~A.}\ \bibnamefont {Salgado}},\ }\href
  {\doibase 10.1140/epjc/s10052-017-4725-9} {\bibfield  {journal} {\bibinfo
  {journal} {Eur. Phys. J. C}\ }\textbf {\bibinfo {volume} {77}},\ \bibinfo
  {pages} {163} (\bibinfo {year} {2017})},\ \Eprint
  {http://arxiv.org/abs/1612.05741} {arXiv:1612.05741 [hep-ph]} \BibitemShut
  {NoStop}%
\bibitem [{\citenamefont {de~Florian}\ \emph {et~al.}(2012)\citenamefont
  {de~Florian}, \citenamefont {Sassot}, \citenamefont {Zurita},\ and\
  \citenamefont {Stratmann}}]{deFlorian:2011fp}%
  \BibitemOpen
  \bibfield  {author} {\bibinfo {author} {\bibfnamefont {D.}~\bibnamefont
  {de~Florian}}, \bibinfo {author} {\bibfnamefont {R.}~\bibnamefont {Sassot}},
  \bibinfo {author} {\bibfnamefont {P.}~\bibnamefont {Zurita}}, \ and\ \bibinfo
  {author} {\bibfnamefont {M.}~\bibnamefont {Stratmann}},\ }\href {\doibase
  10.1103/PhysRevD.85.074028} {\bibfield  {journal} {\bibinfo  {journal} {Phys.
  Rev. D}\ }\textbf {\bibinfo {volume} {85}},\ \bibinfo {pages} {074028}
  (\bibinfo {year} {2012})},\ \Eprint {http://arxiv.org/abs/1112.6324}
  {arXiv:1112.6324 [hep-ph]} \BibitemShut {NoStop}%
\bibitem [{\citenamefont {Hirai}\ \emph {et~al.}(2007)\citenamefont {Hirai},
  \citenamefont {Kumano},\ and\ \citenamefont {Nagai}}]{Hirai:2007sx}%
  \BibitemOpen
  \bibfield  {author} {\bibinfo {author} {\bibfnamefont {M.}~\bibnamefont
  {Hirai}}, \bibinfo {author} {\bibfnamefont {S.}~\bibnamefont {Kumano}}, \
  and\ \bibinfo {author} {\bibfnamefont {T.~H.}\ \bibnamefont {Nagai}},\ }\href
  {\doibase 10.1103/PhysRevC.76.065207} {\bibfield  {journal} {\bibinfo
  {journal} {Phys. Rev. C}\ }\textbf {\bibinfo {volume} {76}},\ \bibinfo
  {pages} {065207} (\bibinfo {year} {2007})},\ \Eprint
  {http://arxiv.org/abs/0709.3038} {arXiv:0709.3038 [hep-ph]} \BibitemShut
  {NoStop}%
\bibitem [{\citenamefont {Cloet}\ \emph {et~al.}(2009)\citenamefont {Cloet},
  \citenamefont {Bentz},\ and\ \citenamefont {Thomas}}]{Cloet:2009qs}%
  \BibitemOpen
  \bibfield  {author} {\bibinfo {author} {\bibfnamefont {I.~C.}\ \bibnamefont
  {Cloet}}, \bibinfo {author} {\bibfnamefont {W.}~\bibnamefont {Bentz}}, \ and\
  \bibinfo {author} {\bibfnamefont {A.~W.}\ \bibnamefont {Thomas}},\ }\href
  {\doibase 10.1103/PhysRevLett.102.252301} {\bibfield  {journal} {\bibinfo
  {journal} {Phys. Rev. Lett.}\ }\textbf {\bibinfo {volume} {102}},\ \bibinfo
  {pages} {252301} (\bibinfo {year} {2009})},\ \Eprint
  {http://arxiv.org/abs/0901.3559} {arXiv:0901.3559 [nucl-th]} \BibitemShut
  {NoStop}%
\bibitem [{\citenamefont {Winter}\ \emph {et~al.}(2017)\citenamefont {Winter},
  \citenamefont {Detmold}, \citenamefont {Gambhir}, \citenamefont {Orginos},
  \citenamefont {Savage}, \citenamefont {Shanahan},\ and\ \citenamefont
  {Wagman}}]{Winter:2017bfs}%
  \BibitemOpen
  \bibfield  {author} {\bibinfo {author} {\bibfnamefont {F.}~\bibnamefont
  {Winter}}, \bibinfo {author} {\bibfnamefont {W.}~\bibnamefont {Detmold}},
  \bibinfo {author} {\bibfnamefont {A.~S.}\ \bibnamefont {Gambhir}}, \bibinfo
  {author} {\bibfnamefont {K.}~\bibnamefont {Orginos}}, \bibinfo {author}
  {\bibfnamefont {M.~J.}\ \bibnamefont {Savage}}, \bibinfo {author}
  {\bibfnamefont {P.~E.}\ \bibnamefont {Shanahan}}, \ and\ \bibinfo {author}
  {\bibfnamefont {M.~L.}\ \bibnamefont {Wagman}},\ }\href {\doibase
  10.1103/PhysRevD.96.094512} {\bibfield  {journal} {\bibinfo  {journal} {Phys.
  Rev. D}\ }\textbf {\bibinfo {volume} {96}},\ \bibinfo {pages} {094512}
  (\bibinfo {year} {2017})},\ \Eprint {http://arxiv.org/abs/1709.00395}
  {arXiv:1709.00395 [hep-lat]} \BibitemShut {NoStop}%
\bibitem [{\citenamefont {Detmold}\ \emph
  {et~al.}(2021{\natexlab{b}})\citenamefont {Detmold}, \citenamefont {Illa},
  \citenamefont {Murphy}, \citenamefont {Oare}, \citenamefont {Orginos},
  \citenamefont {Shanahan}, \citenamefont {Wagman},\ and\ \citenamefont
  {Winter}}]{Detmold:2020snb}%
  \BibitemOpen
  \bibfield  {author} {\bibinfo {author} {\bibfnamefont {W.}~\bibnamefont
  {Detmold}}, \bibinfo {author} {\bibfnamefont {M.}~\bibnamefont {Illa}},
  \bibinfo {author} {\bibfnamefont {D.~J.}\ \bibnamefont {Murphy}}, \bibinfo
  {author} {\bibfnamefont {P.}~\bibnamefont {Oare}}, \bibinfo {author}
  {\bibfnamefont {K.}~\bibnamefont {Orginos}}, \bibinfo {author} {\bibfnamefont
  {P.~E.}\ \bibnamefont {Shanahan}}, \bibinfo {author} {\bibfnamefont {M.~L.}\
  \bibnamefont {Wagman}}, \ and\ \bibinfo {author} {\bibfnamefont
  {F.}~\bibnamefont {Winter}} (\bibinfo {collaboration} {NPLQCD}),\ }\href
  {\doibase 10.1103/PhysRevLett.126.202001} {\bibfield  {journal} {\bibinfo
  {journal} {Phys. Rev. Lett.}\ }\textbf {\bibinfo {volume} {126}},\ \bibinfo
  {pages} {202001} (\bibinfo {year} {2021}{\natexlab{b}})},\ \Eprint
  {http://arxiv.org/abs/2009.05522} {arXiv:2009.05522 [hep-lat]} \BibitemShut
  {NoStop}%
\bibitem [{\citenamefont {Carlson}\ \emph {et~al.}(2002)\citenamefont
  {Carlson}, \citenamefont {Jourdan}, \citenamefont {Schiavilla},\ and\
  \citenamefont {Sick}}]{Carlson:2001mp}%
  \BibitemOpen
  \bibfield  {author} {\bibinfo {author} {\bibfnamefont {J.}~\bibnamefont
  {Carlson}}, \bibinfo {author} {\bibfnamefont {J.}~\bibnamefont {Jourdan}},
  \bibinfo {author} {\bibfnamefont {R.}~\bibnamefont {Schiavilla}}, \ and\
  \bibinfo {author} {\bibfnamefont {I.}~\bibnamefont {Sick}},\ }\href {\doibase
  10.1103/PhysRevC.65.024002} {\bibfield  {journal} {\bibinfo  {journal} {Phys.
  Rev. C}\ }\textbf {\bibinfo {volume} {65}},\ \bibinfo {pages} {024002}
  (\bibinfo {year} {2002})},\ \Eprint {http://arxiv.org/abs/nucl-th/0106047}
  {arXiv:nucl-th/0106047 [nucl-th]} \BibitemShut {NoStop}%
\bibitem [{\citenamefont {Wiringa}\ \emph {et~al.}(1995)\citenamefont
  {Wiringa}, \citenamefont {Stoks},\ and\ \citenamefont
  {Schiavilla}}]{Wiringa:1994wb}%
  \BibitemOpen
  \bibfield  {author} {\bibinfo {author} {\bibfnamefont {R.~B.}\ \bibnamefont
  {Wiringa}}, \bibinfo {author} {\bibfnamefont {V.~G.~J.}\ \bibnamefont
  {Stoks}}, \ and\ \bibinfo {author} {\bibfnamefont {R.}~\bibnamefont
  {Schiavilla}},\ }\href {\doibase 10.1103/PhysRevC.51.38} {\bibfield
  {journal} {\bibinfo  {journal} {Phys. Rev. C}\ }\textbf {\bibinfo {volume}
  {51}},\ \bibinfo {pages} {38} (\bibinfo {year} {1995})},\ \Eprint
  {http://arxiv.org/abs/nucl-th/9408016} {arXiv:nucl-th/9408016 [nucl-th]}
  \BibitemShut {NoStop}%
\bibitem [{\citenamefont {Machleidt}\ \emph {et~al.}(1987)\citenamefont
  {Machleidt}, \citenamefont {Holinde},\ and\ \citenamefont {Elster}}]{Bonnr}%
  \BibitemOpen
  \bibfield  {author} {\bibinfo {author} {\bibfnamefont {R.}~\bibnamefont
  {Machleidt}}, \bibinfo {author} {\bibfnamefont {K.}~\bibnamefont {Holinde}},
  \ and\ \bibinfo {author} {\bibfnamefont {C.}~\bibnamefont {Elster}},\ }\href
  {\doibase http://dx.doi.org/10.1016/S0370-1573(87)80002-9} {\bibfield
  {journal} {\bibinfo  {journal} {Physics Reports}\ }\textbf {\bibinfo {volume}
  {149}},\ \bibinfo {pages} {1 } (\bibinfo {year} {1987})}\BibitemShut
  {NoStop}%
\bibitem [{\citenamefont {Machleidt}(2001)}]{machleidt2001}%
  \BibitemOpen
  \bibfield  {author} {\bibinfo {author} {\bibfnamefont {R.}~\bibnamefont
  {Machleidt}},\ }\href {\doibase 10.1103/PhysRevC.63.024001} {\bibfield
  {journal} {\bibinfo  {journal} {Phys. Rev. C}\ }\textbf {\bibinfo {volume}
  {63}},\ \bibinfo {pages} {024001} (\bibinfo {year} {2001})}\BibitemShut
  {NoStop}%
\bibitem [{\citenamefont {Pieper}(2008)}]{Pieper:2008}%
  \BibitemOpen
  \bibfield  {author} {\bibinfo {author} {\bibfnamefont {S.~C.}\ \bibnamefont
  {Pieper}},\ }\href {\doibase 10.1063/1.2932280} {\bibfield  {journal}
  {\bibinfo  {journal} {AIP Conference Proceedings}\ }\textbf {\bibinfo
  {volume} {1011}},\ \bibinfo {pages} {143} (\bibinfo {year} {2008})},\ \Eprint
  {http://arxiv.org/abs/http://aip.scitation.org/doi/pdf/10.1063/1.2932280}
  {http://aip.scitation.org/doi/pdf/10.1063/1.2932280} \BibitemShut {NoStop}%
\bibitem [{\citenamefont {Villars}(1947)}]{Villars47}%
  \BibitemOpen
  \bibfield  {author} {\bibinfo {author} {\bibfnamefont {F.}~\bibnamefont
  {Villars}},\ }\href {\doibase 10.1103/PhysRev.72.256.2} {\bibfield  {journal}
  {\bibinfo  {journal} {Phys. Rev.}\ }\textbf {\bibinfo {volume} {72}},\
  \bibinfo {pages} {256} (\bibinfo {year} {1947})}\BibitemShut {NoStop}%
\bibitem [{\citenamefont {Chemtob}\ and\ \citenamefont
  {Rho}(1971)}]{Chemtob71}%
  \BibitemOpen
  \bibfield  {author} {\bibinfo {author} {\bibfnamefont {M.}~\bibnamefont
  {Chemtob}}\ and\ \bibinfo {author} {\bibfnamefont {M.}~\bibnamefont {Rho}},\
  }\href {\doibase http://dx.doi.org/10.1016/0375-9474(71)90520-3} {\bibfield
  {journal} {\bibinfo  {journal} {Nucl.~Phys. A}\ }\textbf {\bibinfo {volume}
  {163}},\ \bibinfo {pages} {1} (\bibinfo {year} {1971})}\BibitemShut {NoStop}%
\bibitem [{\citenamefont {Friar}(1977)}]{Friar77}%
  \BibitemOpen
  \bibfield  {author} {\bibinfo {author} {\bibfnamefont {J.~L.}\ \bibnamefont
  {Friar}},\ }\href {\doibase 10.1016/0003-4916(77)90337-2} {\bibfield
  {journal} {\bibinfo  {journal} {Annals Phys.}\ }\textbf {\bibinfo {volume}
  {104}},\ \bibinfo {pages} {380} (\bibinfo {year} {1977})}\BibitemShut
  {NoStop}%
\bibitem [{\citenamefont {Rho}\ and\ \citenamefont {Wilkinson}(1979)}]{rho79}%
  \BibitemOpen
  \bibfield  {author} {\bibinfo {author} {\bibfnamefont {M.}~\bibnamefont
  {Rho}}\ and\ \bibinfo {author} {\bibfnamefont {D.}~\bibnamefont
  {Wilkinson}},\ }\href@noop {} {\emph {\bibinfo {title} {Mesons in nuclei}}},\
  \bibinfo {series} {Mesons in Nuclei}\ No.\ \bibinfo {number} {v. 2}\
  (\bibinfo  {publisher} {North Holland Pub. Co.},\ \bibinfo {year}
  {1979})\BibitemShut {NoStop}%
\bibitem [{\citenamefont {Towner}(1984)}]{Towner84}%
  \BibitemOpen
  \bibfield  {author} {\bibinfo {author} {\bibfnamefont {I.}~\bibnamefont
  {Towner}},\ }\href {\doibase http://dx.doi.org/10.1016/0146-6410(84)90013-9}
  {\bibfield  {journal} {\bibinfo  {journal} {Progress in Particle and
  Nucl.~Phys.}\ }\textbf {\bibinfo {volume} {11}},\ \bibinfo {pages} {91}
  (\bibinfo {year} {1984})}\BibitemShut {NoStop}%
\bibitem [{\citenamefont {Riska}(1984)}]{Riska84}%
  \BibitemOpen
  \bibfield  {author} {\bibinfo {author} {\bibfnamefont {D.~O.}\ \bibnamefont
  {Riska}},\ }\href {\doibase http://dx.doi.org/10.1016/0146-6410(84)90017-6}
  {\bibfield  {journal} {\bibinfo  {journal} {Progress in Particle and
  Nucl.~Phys.}\ }\textbf {\bibinfo {volume} {11}},\ \bibinfo {pages} {199}
  (\bibinfo {year} {1984})}\BibitemShut {NoStop}%
\bibitem [{\citenamefont {Carlson}\ and\ \citenamefont
  {Schiavilla}(1998)}]{Carlson:1997}%
  \BibitemOpen
  \bibfield  {author} {\bibinfo {author} {\bibfnamefont {J.}~\bibnamefont
  {Carlson}}\ and\ \bibinfo {author} {\bibfnamefont {R.}~\bibnamefont
  {Schiavilla}},\ }\href {\doibase 10.1103/RevModPhys.70.743} {\bibfield
  {journal} {\bibinfo  {journal} {Rev. Mod. Phys.}\ }\textbf {\bibinfo {volume}
  {70}},\ \bibinfo {pages} {743} (\bibinfo {year} {1998})}\BibitemShut
  {NoStop}%
\bibitem [{\citenamefont {Marcucci}\ \emph {et~al.}(1998)\citenamefont
  {Marcucci}, \citenamefont {Riska},\ and\ \citenamefont
  {Schiavilla}}]{Marcucci98}%
  \BibitemOpen
  \bibfield  {author} {\bibinfo {author} {\bibfnamefont {L.}~\bibnamefont
  {Marcucci}}, \bibinfo {author} {\bibfnamefont {D.}~\bibnamefont {Riska}}, \
  and\ \bibinfo {author} {\bibfnamefont {R.}~\bibnamefont {Schiavilla}},\
  }\href {\doibase 10.1103/PhysRevC.58.3069} {\bibfield  {journal} {\bibinfo
  {journal} {Phys. Rev. C}\ }\textbf {\bibinfo {volume} {58}},\ \bibinfo
  {pages} {3069} (\bibinfo {year} {1998})},\ \Eprint
  {http://arxiv.org/abs/nucl-th/9805048} {arXiv:nucl-th/9805048} \BibitemShut
  {NoStop}%
\bibitem [{\citenamefont {Marcucci}\ \emph {et~al.}(2005)\citenamefont
  {Marcucci}, \citenamefont {Viviani}, \citenamefont {Schiavilla},
  \citenamefont {Kievsky},\ and\ \citenamefont {Rosati}}]{Marcucci05}%
  \BibitemOpen
  \bibfield  {author} {\bibinfo {author} {\bibfnamefont {L.~E.}\ \bibnamefont
  {Marcucci}}, \bibinfo {author} {\bibfnamefont {M.}~\bibnamefont {Viviani}},
  \bibinfo {author} {\bibfnamefont {R.}~\bibnamefont {Schiavilla}}, \bibinfo
  {author} {\bibfnamefont {A.}~\bibnamefont {Kievsky}}, \ and\ \bibinfo
  {author} {\bibfnamefont {S.}~\bibnamefont {Rosati}},\ }\href {\doibase
  10.1103/PhysRevC.72.014001} {\bibfield  {journal} {\bibinfo  {journal} {Phys.
  Rev. C}\ }\textbf {\bibinfo {volume} {72}},\ \bibinfo {pages} {014001}
  (\bibinfo {year} {2005})},\ \Eprint {http://arxiv.org/abs/nucl-th/0502048}
  {arXiv:nucl-th/0502048} \BibitemShut {NoStop}%
\bibitem [{\citenamefont {Shen}\ \emph {et~al.}(2012)\citenamefont {Shen},
  \citenamefont {Marcucci}, \citenamefont {Carlson}, \citenamefont {Gandolfi},\
  and\ \citenamefont {Schiavilla}}]{Shen:2012xz}%
  \BibitemOpen
  \bibfield  {author} {\bibinfo {author} {\bibfnamefont {G.}~\bibnamefont
  {Shen}}, \bibinfo {author} {\bibfnamefont {L.~E.}\ \bibnamefont {Marcucci}},
  \bibinfo {author} {\bibfnamefont {J.}~\bibnamefont {Carlson}}, \bibinfo
  {author} {\bibfnamefont {S.}~\bibnamefont {Gandolfi}}, \ and\ \bibinfo
  {author} {\bibfnamefont {R.}~\bibnamefont {Schiavilla}},\ }\href {\doibase
  10.1103/PhysRevC.86.035503} {\bibfield  {journal} {\bibinfo  {journal} {Phys.
  Rev. C}\ }\textbf {\bibinfo {volume} {86}},\ \bibinfo {pages} {035503}
  (\bibinfo {year} {2012})},\ \Eprint {http://arxiv.org/abs/1205.4337}
  {arXiv:1205.4337 [nucl-th]} \BibitemShut {NoStop}%
\bibitem [{\citenamefont {Bacca}\ and\ \citenamefont
  {Pastore}(2014)}]{Bacca_Pastore_2014}%
  \BibitemOpen
  \bibfield  {author} {\bibinfo {author} {\bibfnamefont {S.}~\bibnamefont
  {Bacca}}\ and\ \bibinfo {author} {\bibfnamefont {S.}~\bibnamefont
  {Pastore}},\ }\href {\doibase 10.1088/0954-3899/41/12/123002} {\bibfield
  {journal} {\bibinfo  {journal} {J. Phys. G}\ }\textbf {\bibinfo {volume}
  {41}},\ \bibinfo {pages} {123002} (\bibinfo {year} {2014})},\ \Eprint
  {http://arxiv.org/abs/1407.3490} {arXiv:1407.3490 [nucl-th]} \BibitemShut
  {NoStop}%
\bibitem [{\citenamefont {Weinberg}(1979)}]{Weinberg:1979sa}%
  \BibitemOpen
  \bibfield  {author} {\bibinfo {author} {\bibfnamefont {S.}~\bibnamefont
  {Weinberg}},\ }\href {\doibase 10.1103/PhysRevLett.43.1566} {\bibfield
  {journal} {\bibinfo  {journal} {Phys. Rev. Lett.}\ }\textbf {\bibinfo
  {volume} {43}},\ \bibinfo {pages} {1566} (\bibinfo {year}
  {1979})}\BibitemShut {NoStop}%
\bibitem [{\citenamefont {Weinberg}(1990)}]{Weinberg:1990rz}%
  \BibitemOpen
  \bibfield  {author} {\bibinfo {author} {\bibfnamefont {S.}~\bibnamefont
  {Weinberg}},\ }\href {\doibase 10.1016/0370-2693(90)90938-3} {\bibfield
  {journal} {\bibinfo  {journal} {Phys. Lett. B}\ }\textbf {\bibinfo {volume}
  {251}},\ \bibinfo {pages} {288} (\bibinfo {year} {1990})}\BibitemShut
  {NoStop}%
\bibitem [{\citenamefont {Weinberg}(1991)}]{Weinberg:1991um}%
  \BibitemOpen
  \bibfield  {author} {\bibinfo {author} {\bibfnamefont {S.}~\bibnamefont
  {Weinberg}},\ }\href {\doibase 10.1016/0550-3213(91)90231-L} {\bibfield
  {journal} {\bibinfo  {journal} {Nucl. Phys. B}\ }\textbf {\bibinfo {volume}
  {363}},\ \bibinfo {pages} {3} (\bibinfo {year} {1991})}\BibitemShut {NoStop}%
\bibitem [{\citenamefont {van Kolck}(1999)}]{vanKolck93}%
  \BibitemOpen
  \bibfield  {author} {\bibinfo {author} {\bibfnamefont {U.}~\bibnamefont {van
  Kolck}},\ }\href {\doibase 10.1016/S0146-6410(99)00097-6} {\bibfield
  {journal} {\bibinfo  {journal} {Prog. Part. Nucl. Phys.}\ }\textbf {\bibinfo
  {volume} {43}},\ \bibinfo {pages} {337} (\bibinfo {year} {1999})},\ \Eprint
  {http://arxiv.org/abs/nucl-th/9902015} {arXiv:nucl-th/9902015} \BibitemShut
  {NoStop}%
\bibitem [{\citenamefont {Ord{\'o}{\~n}ez}\ and\ \citenamefont {van
  Kolck}(1992)}]{Ordonez:1992xp}%
  \BibitemOpen
  \bibfield  {author} {\bibinfo {author} {\bibfnamefont {C.}~\bibnamefont
  {Ord{\'o}{\~n}ez}}\ and\ \bibinfo {author} {\bibfnamefont {U.}~\bibnamefont
  {van Kolck}},\ }\href {\doibase 10.1016/0370-2693(92)91404-W} {\bibfield
  {journal} {\bibinfo  {journal} {Phys. Lett. B}\ }\textbf {\bibinfo {volume}
  {291}},\ \bibinfo {pages} {459} (\bibinfo {year} {1992})}\BibitemShut
  {NoStop}%
\bibitem [{\citenamefont {Ord{\'o}{\~n}ez}\ \emph {et~al.}(1996)\citenamefont
  {Ord{\'o}{\~n}ez}, \citenamefont {Ray},\ and\ \citenamefont {van
  Kolck}}]{Ordonez96}%
  \BibitemOpen
  \bibfield  {author} {\bibinfo {author} {\bibfnamefont {C.}~\bibnamefont
  {Ord{\'o}{\~n}ez}}, \bibinfo {author} {\bibfnamefont {L.}~\bibnamefont
  {Ray}}, \ and\ \bibinfo {author} {\bibfnamefont {U.}~\bibnamefont {van
  Kolck}},\ }\href {\doibase 10.1103/PhysRevC.53.2086} {\bibfield  {journal}
  {\bibinfo  {journal} {Phys. Rev. C}\ }\textbf {\bibinfo {volume} {53}},\
  \bibinfo {pages} {2086} (\bibinfo {year} {1996})},\ \Eprint
  {http://arxiv.org/abs/hep-ph/9511380} {arXiv:hep-ph/9511380} \BibitemShut
  {NoStop}%
\bibitem [{\citenamefont {Bernard}\ \emph {et~al.}(1995)\citenamefont
  {Bernard}, \citenamefont {Kaiser},\ and\ \citenamefont
  {Mei{\ss}ner}}]{Bernard95}%
  \BibitemOpen
  \bibfield  {author} {\bibinfo {author} {\bibfnamefont {V.}~\bibnamefont
  {Bernard}}, \bibinfo {author} {\bibfnamefont {N.}~\bibnamefont {Kaiser}}, \
  and\ \bibinfo {author} {\bibfnamefont {U.-G.}\ \bibnamefont {Mei{\ss}ner}},\
  }\href {\doibase 10.1142/S0218301395000092} {\bibfield  {journal} {\bibinfo
  {journal} {Int. J. Mod. Phys. E}\ }\textbf {\bibinfo {volume} {4}},\ \bibinfo
  {pages} {193} (\bibinfo {year} {1995})},\ \Eprint
  {http://arxiv.org/abs/hep-ph/9501384} {arXiv:hep-ph/9501384} \BibitemShut
  {NoStop}%
\bibitem [{\citenamefont {Epelbaum}\ \emph {et~al.}(2009)\citenamefont
  {Epelbaum}, \citenamefont {Hammer},\ and\ \citenamefont
  {Mei{\ss}ner}}]{Epelbaum:2008ga}%
  \BibitemOpen
  \bibfield  {author} {\bibinfo {author} {\bibfnamefont {E.}~\bibnamefont
  {Epelbaum}}, \bibinfo {author} {\bibfnamefont {H.-W.}\ \bibnamefont
  {Hammer}}, \ and\ \bibinfo {author} {\bibfnamefont {U.-G.}\ \bibnamefont
  {Mei{\ss}ner}},\ }\href {\doibase 10.1103/RevModPhys.81.1773} {\bibfield
  {journal} {\bibinfo  {journal} {Rev. Mod. Phys.}\ }\textbf {\bibinfo {volume}
  {81}},\ \bibinfo {pages} {1773} (\bibinfo {year} {2009})},\ \Eprint
  {http://arxiv.org/abs/0811.1338} {arXiv:0811.1338 [nucl-th]} \BibitemShut
  {NoStop}%
\bibitem [{\citenamefont {Epelbaum}\ and\ \citenamefont
  {Mei{\ss}ner}(2012)}]{Epelbaum12}%
  \BibitemOpen
  \bibfield  {author} {\bibinfo {author} {\bibfnamefont {E.}~\bibnamefont
  {Epelbaum}}\ and\ \bibinfo {author} {\bibfnamefont {U.-G.}\ \bibnamefont
  {Mei{\ss}ner}},\ }\href {\doibase 10.1146/annurev-nucl-102010-130056}
  {\bibfield  {journal} {\bibinfo  {journal} {Ann. Rev. Nucl. Part. Sci.}\
  }\textbf {\bibinfo {volume} {62}},\ \bibinfo {pages} {159} (\bibinfo {year}
  {2012})},\ \Eprint {http://arxiv.org/abs/1201.2136} {arXiv:1201.2136
  [nucl-th]} \BibitemShut {NoStop}%
\bibitem [{\citenamefont {Epelbaum}\ \emph
  {et~al.}(2015{\natexlab{a}})\citenamefont {Epelbaum}, \citenamefont {Krebs},\
  and\ \citenamefont {Mei{\ss}ner}}]{Epelbaum:2014efa}%
  \BibitemOpen
  \bibfield  {author} {\bibinfo {author} {\bibfnamefont {E.}~\bibnamefont
  {Epelbaum}}, \bibinfo {author} {\bibfnamefont {H.}~\bibnamefont {Krebs}}, \
  and\ \bibinfo {author} {\bibfnamefont {U.-G.}\ \bibnamefont {Mei{\ss}ner}},\
  }\href {\doibase 10.1140/epja/i2015-15053-8} {\bibfield  {journal} {\bibinfo
  {journal} {Eur. Phys. J. A}\ }\textbf {\bibinfo {volume} {51}},\ \bibinfo
  {pages} {53} (\bibinfo {year} {2015}{\natexlab{a}})},\ \Eprint
  {http://arxiv.org/abs/1412.0142} {arXiv:1412.0142 [nucl-th]} \BibitemShut
  {NoStop}%
\bibitem [{\citenamefont {Entem}\ and\ \citenamefont
  {Machleidt}(2003)}]{Entem:2003ft}%
  \BibitemOpen
  \bibfield  {author} {\bibinfo {author} {\bibfnamefont {D.~R.}\ \bibnamefont
  {Entem}}\ and\ \bibinfo {author} {\bibfnamefont {R.}~\bibnamefont
  {Machleidt}},\ }\href {\doibase 10.1103/PhysRevC.68.041001} {\bibfield
  {journal} {\bibinfo  {journal} {Phys. Rev. C}\ }\textbf {\bibinfo {volume}
  {68}},\ \bibinfo {pages} {041001} (\bibinfo {year} {2003})},\ \Eprint
  {http://arxiv.org/abs/nucl-th/0304018} {arXiv:nucl-th/0304018 [nucl-th]}
  \BibitemShut {NoStop}%
\bibitem [{\citenamefont {Machleidt}\ and\ \citenamefont
  {Entem}(2011)}]{Machleidt:2011zz}%
  \BibitemOpen
  \bibfield  {author} {\bibinfo {author} {\bibfnamefont {R.}~\bibnamefont
  {Machleidt}}\ and\ \bibinfo {author} {\bibfnamefont {D.~R.}\ \bibnamefont
  {Entem}},\ }\href {\doibase 10.1016/j.physrep.2011.02.001} {\bibfield
  {journal} {\bibinfo  {journal} {Phys. Rept.}\ }\textbf {\bibinfo {volume}
  {503}},\ \bibinfo {pages} {1} (\bibinfo {year} {2011})},\ \Eprint
  {http://arxiv.org/abs/1105.2919} {arXiv:1105.2919 [nucl-th]} \BibitemShut
  {NoStop}%
\bibitem [{\citenamefont {Ekstr\"om}\ \emph {et~al.}(2015)\citenamefont
  {Ekstr\"om}, \citenamefont {Jansen}, \citenamefont {Wendt}, \citenamefont
  {Hagen}, \citenamefont {Papenbrock}, \citenamefont {Carlsson}, \citenamefont
  {Forss\'en}, \citenamefont {Hjorth-Jensen}, \citenamefont {Navr\'atil},\ and\
  \citenamefont {Nazarewicz}}]{ekstrom2015accurate}%
  \BibitemOpen
  \bibfield  {author} {\bibinfo {author} {\bibfnamefont {A.}~\bibnamefont
  {Ekstr\"om}}, \bibinfo {author} {\bibfnamefont {G.~R.}\ \bibnamefont
  {Jansen}}, \bibinfo {author} {\bibfnamefont {K.~A.}\ \bibnamefont {Wendt}},
  \bibinfo {author} {\bibfnamefont {G.}~\bibnamefont {Hagen}}, \bibinfo
  {author} {\bibfnamefont {T.}~\bibnamefont {Papenbrock}}, \bibinfo {author}
  {\bibfnamefont {B.~D.}\ \bibnamefont {Carlsson}}, \bibinfo {author}
  {\bibfnamefont {C.}~\bibnamefont {Forss\'en}}, \bibinfo {author}
  {\bibfnamefont {M.}~\bibnamefont {Hjorth-Jensen}}, \bibinfo {author}
  {\bibfnamefont {P.}~\bibnamefont {Navr\'atil}}, \ and\ \bibinfo {author}
  {\bibfnamefont {W.}~\bibnamefont {Nazarewicz}},\ }\href {\doibase
  10.1103/PhysRevC.91.051301} {\bibfield  {journal} {\bibinfo  {journal} {Phys.
  Rev. C}\ }\textbf {\bibinfo {volume} {91}},\ \bibinfo {pages} {051301}
  (\bibinfo {year} {2015})},\ \Eprint {http://arxiv.org/abs/1502.04682}
  {arXiv:1502.04682 [nucl-th]} \BibitemShut {NoStop}%
\bibitem [{\citenamefont {Hergert}(2020)}]{10.3389/fphy.2020.00379}%
  \BibitemOpen
  \bibfield  {author} {\bibinfo {author} {\bibfnamefont {H.}~\bibnamefont
  {Hergert}},\ }\href {\doibase 10.3389/fphy.2020.00379} {\bibfield  {journal}
  {\bibinfo  {journal} {Front. in Phys.}\ }\textbf {\bibinfo {volume} {8}},\
  \bibinfo {pages} {379} (\bibinfo {year} {2020})},\ \Eprint
  {http://arxiv.org/abs/2008.05061} {arXiv:2008.05061 [nucl-th]} \BibitemShut
  {NoStop}%
\bibitem [{\citenamefont {Park}\ \emph {et~al.}(1993)\citenamefont {Park},
  \citenamefont {Min},\ and\ \citenamefont {Rho}}]{Park93}%
  \BibitemOpen
  \bibfield  {author} {\bibinfo {author} {\bibfnamefont {T.-S.}\ \bibnamefont
  {Park}}, \bibinfo {author} {\bibfnamefont {D.-P.}\ \bibnamefont {Min}}, \
  and\ \bibinfo {author} {\bibfnamefont {M.}~\bibnamefont {Rho}},\ }\href
  {\doibase 10.1016/0370-1573(93)90099-Y} {\bibfield  {journal} {\bibinfo
  {journal} {Phys. Rept.}\ }\textbf {\bibinfo {volume} {233}},\ \bibinfo
  {pages} {341} (\bibinfo {year} {1993})},\ \Eprint
  {http://arxiv.org/abs/hep-ph/9301295} {arXiv:hep-ph/9301295} \BibitemShut
  {NoStop}%
\bibitem [{\citenamefont {Park}\ \emph {et~al.}(1996)\citenamefont {Park},
  \citenamefont {Min},\ and\ \citenamefont {Rho}}]{Park96}%
  \BibitemOpen
  \bibfield  {author} {\bibinfo {author} {\bibfnamefont {T.-S.}\ \bibnamefont
  {Park}}, \bibinfo {author} {\bibfnamefont {D.-P.}\ \bibnamefont {Min}}, \
  and\ \bibinfo {author} {\bibfnamefont {M.}~\bibnamefont {Rho}},\ }\href
  {\doibase 10.1016/0375-9474(95)00406-8} {\bibfield  {journal} {\bibinfo
  {journal} {Nucl. Phys. A}\ }\textbf {\bibinfo {volume} {596}},\ \bibinfo
  {pages} {515} (\bibinfo {year} {1996})},\ \Eprint
  {http://arxiv.org/abs/nucl-th/9505017} {arXiv:nucl-th/9505017} \BibitemShut
  {NoStop}%
\bibitem [{\citenamefont {Phillips}(2007)}]{Phillips07}%
  \BibitemOpen
  \bibfield  {author} {\bibinfo {author} {\bibfnamefont {D.~R.}\ \bibnamefont
  {Phillips}},\ }\href {\doibase 10.1088/0954-3899/34/2/015} {\bibfield
  {journal} {\bibinfo  {journal} {J. Phys. G}\ }\textbf {\bibinfo {volume}
  {34}},\ \bibinfo {pages} {365} (\bibinfo {year} {2007})},\ \Eprint
  {http://arxiv.org/abs/nucl-th/0608036} {arXiv:nucl-th/0608036} \BibitemShut
  {NoStop}%
\bibitem [{\citenamefont {Pastore}\ \emph {et~al.}(2008)\citenamefont
  {Pastore}, \citenamefont {Schiavilla},\ and\ \citenamefont
  {Goity}}]{Pastore:2008ui}%
  \BibitemOpen
  \bibfield  {author} {\bibinfo {author} {\bibfnamefont {S.}~\bibnamefont
  {Pastore}}, \bibinfo {author} {\bibfnamefont {R.}~\bibnamefont {Schiavilla}},
  \ and\ \bibinfo {author} {\bibfnamefont {J.~L.}\ \bibnamefont {Goity}},\
  }\href {\doibase 10.1103/PhysRevC.78.064002} {\bibfield  {journal} {\bibinfo
  {journal} {Phys. Rev. C}\ }\textbf {\bibinfo {volume} {78}},\ \bibinfo
  {pages} {064002} (\bibinfo {year} {2008})},\ \Eprint
  {http://arxiv.org/abs/0810.1941} {arXiv:0810.1941 [nucl-th]} \BibitemShut
  {NoStop}%
\bibitem [{\citenamefont {Pastore}\ \emph {et~al.}(2009)\citenamefont
  {Pastore}, \citenamefont {Girlanda}, \citenamefont {Schiavilla},
  \citenamefont {Viviani},\ and\ \citenamefont {Wiringa}}]{Pastore:2009is}%
  \BibitemOpen
  \bibfield  {author} {\bibinfo {author} {\bibfnamefont {S.}~\bibnamefont
  {Pastore}}, \bibinfo {author} {\bibfnamefont {L.}~\bibnamefont {Girlanda}},
  \bibinfo {author} {\bibfnamefont {R.}~\bibnamefont {Schiavilla}}, \bibinfo
  {author} {\bibfnamefont {M.}~\bibnamefont {Viviani}}, \ and\ \bibinfo
  {author} {\bibfnamefont {R.~B.}\ \bibnamefont {Wiringa}},\ }\href {\doibase
  10.1103/PhysRevC.80.034004} {\bibfield  {journal} {\bibinfo  {journal} {Phys.
  Rev. C}\ }\textbf {\bibinfo {volume} {80}},\ \bibinfo {pages} {034004}
  (\bibinfo {year} {2009})},\ \Eprint {http://arxiv.org/abs/0906.1800}
  {arXiv:0906.1800 [nucl-th]} \BibitemShut {NoStop}%
\bibitem [{\citenamefont {Pastore}\ \emph {et~al.}(2011)\citenamefont
  {Pastore}, \citenamefont {Girlanda}, \citenamefont {Schiavilla},\ and\
  \citenamefont {Viviani}}]{Pastore:2011ip}%
  \BibitemOpen
  \bibfield  {author} {\bibinfo {author} {\bibfnamefont {S.}~\bibnamefont
  {Pastore}}, \bibinfo {author} {\bibfnamefont {L.}~\bibnamefont {Girlanda}},
  \bibinfo {author} {\bibfnamefont {R.}~\bibnamefont {Schiavilla}}, \ and\
  \bibinfo {author} {\bibfnamefont {M.}~\bibnamefont {Viviani}},\ }\href
  {\doibase 10.1103/PhysRevC.84.024001} {\bibfield  {journal} {\bibinfo
  {journal} {Phys. Rev. C}\ }\textbf {\bibinfo {volume} {84}},\ \bibinfo
  {pages} {024001} (\bibinfo {year} {2011})},\ \Eprint
  {http://arxiv.org/abs/1106.4539} {arXiv:1106.4539 [nucl-th]} \BibitemShut
  {NoStop}%
\bibitem [{\citenamefont {K{\"o}lling}\ \emph {et~al.}(2009)\citenamefont
  {K{\"o}lling}, \citenamefont {Epelbaum}, \citenamefont {Krebs},\ and\
  \citenamefont {Mei{\ss}ner}}]{Kolling:2009iq}%
  \BibitemOpen
  \bibfield  {author} {\bibinfo {author} {\bibfnamefont {S.}~\bibnamefont
  {K{\"o}lling}}, \bibinfo {author} {\bibfnamefont {E.}~\bibnamefont
  {Epelbaum}}, \bibinfo {author} {\bibfnamefont {H.}~\bibnamefont {Krebs}}, \
  and\ \bibinfo {author} {\bibfnamefont {U.-G.}\ \bibnamefont {Mei{\ss}ner}},\
  }\href {\doibase 10.1103/PhysRevC.80.045502} {\bibfield  {journal} {\bibinfo
  {journal} {Phys. Rev. C}\ }\textbf {\bibinfo {volume} {80}},\ \bibinfo
  {pages} {045502} (\bibinfo {year} {2009})},\ \Eprint
  {http://arxiv.org/abs/0907.3437} {arXiv:0907.3437 [nucl-th]} \BibitemShut
  {NoStop}%
\bibitem [{\citenamefont {K{\"o}lling}\ \emph {et~al.}(2011)\citenamefont
  {K{\"o}lling}, \citenamefont {Epelbaum}, \citenamefont {Krebs},\ and\
  \citenamefont {Mei{\ss}ner}}]{Kolling:2011mt}%
  \BibitemOpen
  \bibfield  {author} {\bibinfo {author} {\bibfnamefont {S.}~\bibnamefont
  {K{\"o}lling}}, \bibinfo {author} {\bibfnamefont {E.}~\bibnamefont
  {Epelbaum}}, \bibinfo {author} {\bibfnamefont {H.}~\bibnamefont {Krebs}}, \
  and\ \bibinfo {author} {\bibfnamefont {U.-G.}\ \bibnamefont {Mei{\ss}ner}},\
  }\href {\doibase 10.1103/PhysRevC.84.054008} {\bibfield  {journal} {\bibinfo
  {journal} {Phys. Rev. C}\ }\textbf {\bibinfo {volume} {84}},\ \bibinfo
  {pages} {054008} (\bibinfo {year} {2011})},\ \Eprint
  {http://arxiv.org/abs/1107.0602} {arXiv:1107.0602 [nucl-th]} \BibitemShut
  {NoStop}%
\bibitem [{\citenamefont {K{\"o}lling}\ \emph {et~al.}(2012)\citenamefont
  {K{\"o}lling}, \citenamefont {Epelbaum},\ and\ \citenamefont
  {Phillips}}]{Kolling:2012cs}%
  \BibitemOpen
  \bibfield  {author} {\bibinfo {author} {\bibfnamefont {S.}~\bibnamefont
  {K{\"o}lling}}, \bibinfo {author} {\bibfnamefont {E.}~\bibnamefont
  {Epelbaum}}, \ and\ \bibinfo {author} {\bibfnamefont {D.~R.}\ \bibnamefont
  {Phillips}},\ }\href {\doibase 10.1103/PhysRevC.86.047001} {\bibfield
  {journal} {\bibinfo  {journal} {Phys. Rev. C}\ }\textbf {\bibinfo {volume}
  {86}},\ \bibinfo {pages} {047001} (\bibinfo {year} {2012})},\ \Eprint
  {http://arxiv.org/abs/1209.0837} {arXiv:1209.0837 [nucl-th]} \BibitemShut
  {NoStop}%
\bibitem [{\citenamefont {Krebs}\ \emph {et~al.}(2017)\citenamefont {Krebs},
  \citenamefont {Epelbaum},\ and\ \citenamefont {Mei{\ss}ner}}]{Krebs:2016rqz}%
  \BibitemOpen
  \bibfield  {author} {\bibinfo {author} {\bibfnamefont {H.}~\bibnamefont
  {Krebs}}, \bibinfo {author} {\bibfnamefont {E.}~\bibnamefont {Epelbaum}}, \
  and\ \bibinfo {author} {\bibfnamefont {U.-G.}\ \bibnamefont {Mei{\ss}ner}},\
  }\href {\doibase 10.1016/j.aop.2017.01.021} {\bibfield  {journal} {\bibinfo
  {journal} {Annals Phys.}\ }\textbf {\bibinfo {volume} {378}},\ \bibinfo
  {pages} {317} (\bibinfo {year} {2017})},\ \Eprint
  {http://arxiv.org/abs/1610.03569} {arXiv:1610.03569 [nucl-th]} \BibitemShut
  {NoStop}%
\bibitem [{\citenamefont {Hoferichter}\ \emph {et~al.}(2015)\citenamefont
  {Hoferichter}, \citenamefont {Ruiz~de Elvira}, \citenamefont {Kubis},\ and\
  \citenamefont {Mei\ss{}ner}}]{Hoferichter:2015tha}%
  \BibitemOpen
  \bibfield  {author} {\bibinfo {author} {\bibfnamefont {M.}~\bibnamefont
  {Hoferichter}}, \bibinfo {author} {\bibfnamefont {J.}~\bibnamefont {Ruiz~de
  Elvira}}, \bibinfo {author} {\bibfnamefont {B.}~\bibnamefont {Kubis}}, \ and\
  \bibinfo {author} {\bibfnamefont {U.-G.}\ \bibnamefont {Mei\ss{}ner}},\
  }\href {\doibase 10.1103/PhysRevLett.115.192301} {\bibfield  {journal}
  {\bibinfo  {journal} {Phys. Rev. Lett.}\ }\textbf {\bibinfo {volume} {115}},\
  \bibinfo {pages} {192301} (\bibinfo {year} {2015})},\ \Eprint
  {http://arxiv.org/abs/1507.07552} {arXiv:1507.07552 [nucl-th]} \BibitemShut
  {NoStop}%
\bibitem [{\citenamefont {Siemens}\ \emph {et~al.}(2017)\citenamefont
  {Siemens}, \citenamefont {Ruiz~de Elvira}, \citenamefont {Epelbaum},
  \citenamefont {Hoferichter}, \citenamefont {Krebs}, \citenamefont {Kubis},\
  and\ \citenamefont {Mei\ss{}ner}}]{Siemens:2016jwj}%
  \BibitemOpen
  \bibfield  {author} {\bibinfo {author} {\bibfnamefont {D.}~\bibnamefont
  {Siemens}}, \bibinfo {author} {\bibfnamefont {J.}~\bibnamefont {Ruiz~de
  Elvira}}, \bibinfo {author} {\bibfnamefont {E.}~\bibnamefont {Epelbaum}},
  \bibinfo {author} {\bibfnamefont {M.}~\bibnamefont {Hoferichter}}, \bibinfo
  {author} {\bibfnamefont {H.}~\bibnamefont {Krebs}}, \bibinfo {author}
  {\bibfnamefont {B.}~\bibnamefont {Kubis}}, \ and\ \bibinfo {author}
  {\bibfnamefont {U.~G.}\ \bibnamefont {Mei\ss{}ner}},\ }\href {\doibase
  10.1016/j.physletb.2017.04.039} {\bibfield  {journal} {\bibinfo  {journal}
  {Phys. Lett. B}\ }\textbf {\bibinfo {volume} {770}},\ \bibinfo {pages} {27}
  (\bibinfo {year} {2017})},\ \Eprint {http://arxiv.org/abs/1610.08978}
  {arXiv:1610.08978 [nucl-th]} \BibitemShut {NoStop}%
\bibitem [{\citenamefont {Hoferichter}\ \emph
  {et~al.}(2016{\natexlab{b}})\citenamefont {Hoferichter}, \citenamefont
  {Ruiz~de Elvira}, \citenamefont {Kubis},\ and\ \citenamefont
  {Mei\ss{}ner}}]{Hoferichter:2015hva}%
  \BibitemOpen
  \bibfield  {author} {\bibinfo {author} {\bibfnamefont {M.}~\bibnamefont
  {Hoferichter}}, \bibinfo {author} {\bibfnamefont {J.}~\bibnamefont {Ruiz~de
  Elvira}}, \bibinfo {author} {\bibfnamefont {B.}~\bibnamefont {Kubis}}, \ and\
  \bibinfo {author} {\bibfnamefont {U.-G.}\ \bibnamefont {Mei\ss{}ner}},\
  }\href {\doibase 10.1016/j.physrep.2016.02.002} {\bibfield  {journal}
  {\bibinfo  {journal} {Phys. Rept.}\ }\textbf {\bibinfo {volume} {625}},\
  \bibinfo {pages} {1} (\bibinfo {year} {2016}{\natexlab{b}})},\ \Eprint
  {http://arxiv.org/abs/1510.06039} {arXiv:1510.06039 [hep-ph]} \BibitemShut
  {NoStop}%
\bibitem [{\citenamefont {Entem}\ \emph {et~al.}(2015)\citenamefont {Entem},
  \citenamefont {Kaiser}, \citenamefont {Machleidt},\ and\ \citenamefont
  {Nosyk}}]{entem2015dominant}%
  \BibitemOpen
  \bibfield  {author} {\bibinfo {author} {\bibfnamefont {D.~R.}\ \bibnamefont
  {Entem}}, \bibinfo {author} {\bibfnamefont {N.}~\bibnamefont {Kaiser}},
  \bibinfo {author} {\bibfnamefont {R.}~\bibnamefont {Machleidt}}, \ and\
  \bibinfo {author} {\bibfnamefont {Y.}~\bibnamefont {Nosyk}},\ }\href
  {\doibase 10.1103/PhysRevC.92.064001} {\bibfield  {journal} {\bibinfo
  {journal} {Phys. Rev. C}\ }\textbf {\bibinfo {volume} {92}},\ \bibinfo
  {pages} {064001} (\bibinfo {year} {2015})},\ \Eprint
  {http://arxiv.org/abs/1505.03562} {arXiv:1505.03562 [nucl-th]} \BibitemShut
  {NoStop}%
\bibitem [{\citenamefont {Epelbaum}\ \emph
  {et~al.}(2015{\natexlab{b}})\citenamefont {Epelbaum}, \citenamefont {Krebs},\
  and\ \citenamefont {Mei\ss{}ner}}]{epelbaum2015precision}%
  \BibitemOpen
  \bibfield  {author} {\bibinfo {author} {\bibfnamefont {E.}~\bibnamefont
  {Epelbaum}}, \bibinfo {author} {\bibfnamefont {H.}~\bibnamefont {Krebs}}, \
  and\ \bibinfo {author} {\bibfnamefont {U.-G.}\ \bibnamefont {Mei\ss{}ner}},\
  }\href {\doibase 10.1103/PhysRevLett.115.122301} {\bibfield  {journal}
  {\bibinfo  {journal} {Phys. Rev. Lett.}\ }\textbf {\bibinfo {volume} {115}},\
  \bibinfo {pages} {122301} (\bibinfo {year} {2015}{\natexlab{b}})},\ \Eprint
  {http://arxiv.org/abs/1412.4623} {arXiv:1412.4623 [nucl-th]} \BibitemShut
  {NoStop}%
\bibitem [{\citenamefont {Reinert}\ \emph {et~al.}(2018)\citenamefont
  {Reinert}, \citenamefont {Krebs},\ and\ \citenamefont
  {Epelbaum}}]{reinert2018semilocal}%
  \BibitemOpen
  \bibfield  {author} {\bibinfo {author} {\bibfnamefont {P.}~\bibnamefont
  {Reinert}}, \bibinfo {author} {\bibfnamefont {H.}~\bibnamefont {Krebs}}, \
  and\ \bibinfo {author} {\bibfnamefont {E.}~\bibnamefont {Epelbaum}},\ }\href
  {\doibase 10.1140/epja/i2018-12516-4} {\bibfield  {journal} {\bibinfo
  {journal} {Eur. Phys. J. A}\ }\textbf {\bibinfo {volume} {54}},\ \bibinfo
  {pages} {86} (\bibinfo {year} {2018})},\ \Eprint
  {http://arxiv.org/abs/1711.08821} {arXiv:1711.08821 [nucl-th]} \BibitemShut
  {NoStop}%
\bibitem [{\citenamefont {Barrett}\ \emph {et~al.}(2013)\citenamefont
  {Barrett}, \citenamefont {Navr{\'a}til},\ and\ \citenamefont
  {Vary}}]{barrett2013ab}%
  \BibitemOpen
  \bibfield  {author} {\bibinfo {author} {\bibfnamefont {B.~R.}\ \bibnamefont
  {Barrett}}, \bibinfo {author} {\bibfnamefont {P.}~\bibnamefont
  {Navr{\'a}til}}, \ and\ \bibinfo {author} {\bibfnamefont {J.~P.}\
  \bibnamefont {Vary}},\ }\href {\doibase 10.1016/j.ppnp.2012.10.003}
  {\bibfield  {journal} {\bibinfo  {journal} {Prog. Part. Nucl. Phys.}\
  }\textbf {\bibinfo {volume} {69}},\ \bibinfo {pages} {131} (\bibinfo {year}
  {2013})}\BibitemShut {NoStop}%
\bibitem [{\citenamefont {Jurgenson}\ \emph {et~al.}(2013)\citenamefont
  {Jurgenson}, \citenamefont {Maris}, \citenamefont {Furnstahl}, \citenamefont
  {Navr{\'a}til}, \citenamefont {Ormand},\ and\ \citenamefont
  {Vary}}]{jurgenson2013structure}%
  \BibitemOpen
  \bibfield  {author} {\bibinfo {author} {\bibfnamefont {E.~D.}\ \bibnamefont
  {Jurgenson}}, \bibinfo {author} {\bibfnamefont {P.}~\bibnamefont {Maris}},
  \bibinfo {author} {\bibfnamefont {R.~J.}\ \bibnamefont {Furnstahl}}, \bibinfo
  {author} {\bibfnamefont {P.}~\bibnamefont {Navr{\'a}til}}, \bibinfo {author}
  {\bibfnamefont {W.~E.}\ \bibnamefont {Ormand}}, \ and\ \bibinfo {author}
  {\bibfnamefont {J.~P.}\ \bibnamefont {Vary}},\ }\href {\doibase
  10.1103/PhysRevC.87.054312} {\bibfield  {journal} {\bibinfo  {journal} {Phys.
  Rev. C}\ }\textbf {\bibinfo {volume} {87}},\ \bibinfo {pages} {054312}
  (\bibinfo {year} {2013})},\ \Eprint {http://arxiv.org/abs/1302.5473}
  {arXiv:1302.5473 [nucl-th]} \BibitemShut {NoStop}%
\bibitem [{\citenamefont {Hagen}\ \emph
  {et~al.}(2014{\natexlab{a}})\citenamefont {Hagen}, \citenamefont
  {Papenbrock}, \citenamefont {Ekstr\"om}, \citenamefont {Wendt}, \citenamefont
  {Baardsen}, \citenamefont {Gandolfi}, \citenamefont {Hjorth-Jensen},\ and\
  \citenamefont {Horowitz}}]{hagen2014coupled}%
  \BibitemOpen
  \bibfield  {author} {\bibinfo {author} {\bibfnamefont {G.}~\bibnamefont
  {Hagen}}, \bibinfo {author} {\bibfnamefont {T.}~\bibnamefont {Papenbrock}},
  \bibinfo {author} {\bibfnamefont {A.}~\bibnamefont {Ekstr\"om}}, \bibinfo
  {author} {\bibfnamefont {K.~A.}\ \bibnamefont {Wendt}}, \bibinfo {author}
  {\bibfnamefont {G.}~\bibnamefont {Baardsen}}, \bibinfo {author}
  {\bibfnamefont {S.}~\bibnamefont {Gandolfi}}, \bibinfo {author}
  {\bibfnamefont {M.}~\bibnamefont {Hjorth-Jensen}}, \ and\ \bibinfo {author}
  {\bibfnamefont {C.~J.}\ \bibnamefont {Horowitz}},\ }\href {\doibase
  10.1103/PhysRevC.89.014319} {\bibfield  {journal} {\bibinfo  {journal} {Phys.
  Rev. C}\ }\textbf {\bibinfo {volume} {89}},\ \bibinfo {pages} {014319}
  (\bibinfo {year} {2014}{\natexlab{a}})},\ \Eprint
  {http://arxiv.org/abs/1311.2925} {arXiv:1311.2925 [nucl-th]} \BibitemShut
  {NoStop}%
\bibitem [{\citenamefont {Bogner}\ \emph {et~al.}(2010)\citenamefont {Bogner},
  \citenamefont {Furnstahl},\ and\ \citenamefont {Schwenk}}]{bogner2010low}%
  \BibitemOpen
  \bibfield  {author} {\bibinfo {author} {\bibfnamefont {S.~K.}\ \bibnamefont
  {Bogner}}, \bibinfo {author} {\bibfnamefont {R.~J.}\ \bibnamefont
  {Furnstahl}}, \ and\ \bibinfo {author} {\bibfnamefont {A.}~\bibnamefont
  {Schwenk}},\ }\href {\doibase 10.1016/j.ppnp.2010.03.001} {\bibfield
  {journal} {\bibinfo  {journal} {Prog. Part. Nucl. Phys.}\ }\textbf {\bibinfo
  {volume} {65}},\ \bibinfo {pages} {94} (\bibinfo {year} {2010})},\ \Eprint
  {http://arxiv.org/abs/0912.3688} {arXiv:0912.3688 [nucl-th]} \BibitemShut
  {NoStop}%
\bibitem [{\citenamefont {Hergert}\ \emph {et~al.}(2013)\citenamefont
  {Hergert}, \citenamefont {Bogner}, \citenamefont {Binder}, \citenamefont
  {Calci}, \citenamefont {Langhammer}, \citenamefont {Roth},\ and\
  \citenamefont {Schwenk}}]{hergert2013medium}%
  \BibitemOpen
  \bibfield  {author} {\bibinfo {author} {\bibfnamefont {H.}~\bibnamefont
  {Hergert}}, \bibinfo {author} {\bibfnamefont {S.~K.}\ \bibnamefont {Bogner}},
  \bibinfo {author} {\bibfnamefont {S.}~\bibnamefont {Binder}}, \bibinfo
  {author} {\bibfnamefont {A.}~\bibnamefont {Calci}}, \bibinfo {author}
  {\bibfnamefont {J.}~\bibnamefont {Langhammer}}, \bibinfo {author}
  {\bibfnamefont {R.}~\bibnamefont {Roth}}, \ and\ \bibinfo {author}
  {\bibfnamefont {A.}~\bibnamefont {Schwenk}},\ }\href {\doibase
  10.1103/PhysRevC.87.034307} {\bibfield  {journal} {\bibinfo  {journal} {Phys.
  Rev. C}\ }\textbf {\bibinfo {volume} {87}},\ \bibinfo {pages} {034307}
  (\bibinfo {year} {2013})},\ \Eprint {http://arxiv.org/abs/1212.1190}
  {arXiv:1212.1190 [nucl-th]} \BibitemShut {NoStop}%
\bibitem [{\citenamefont {Piarulli}\ and\ \citenamefont
  {Tews}(2020)}]{10.3389/fphy.2019.00245}%
  \BibitemOpen
  \bibfield  {author} {\bibinfo {author} {\bibfnamefont {M.}~\bibnamefont
  {Piarulli}}\ and\ \bibinfo {author} {\bibfnamefont {I.}~\bibnamefont
  {Tews}},\ }\href {\doibase 10.3389/fphy.2019.00245} {\bibfield  {journal}
  {\bibinfo  {journal} {Front. in Phys.}\ }\textbf {\bibinfo {volume} {7}},\
  \bibinfo {pages} {245} (\bibinfo {year} {2020})},\ \Eprint
  {http://arxiv.org/abs/2002.00032} {arXiv:2002.00032 [nucl-th]} \BibitemShut
  {NoStop}%
\bibitem [{\citenamefont {Krebs}\ \emph {et~al.}(2007)\citenamefont {Krebs},
  \citenamefont {Epelbaum},\ and\ \citenamefont {Meissner}}]{krebs2007nuclear}%
  \BibitemOpen
  \bibfield  {author} {\bibinfo {author} {\bibfnamefont {H.}~\bibnamefont
  {Krebs}}, \bibinfo {author} {\bibfnamefont {E.}~\bibnamefont {Epelbaum}}, \
  and\ \bibinfo {author} {\bibfnamefont {U.-G.}\ \bibnamefont {Meissner}},\
  }\href {\doibase 10.1140/epja/i2007-10372-y} {\bibfield  {journal} {\bibinfo
  {journal} {Eur. Phys. J. A}\ }\textbf {\bibinfo {volume} {32}},\ \bibinfo
  {pages} {127} (\bibinfo {year} {2007})},\ \Eprint
  {http://arxiv.org/abs/nucl-th/0703087} {arXiv:nucl-th/0703087} \BibitemShut
  {NoStop}%
\bibitem [{\citenamefont {Piarulli}\ \emph {et~al.}(2018)\citenamefont
  {Piarulli} \emph {et~al.}}]{piarulli2018light}%
  \BibitemOpen
  \bibfield  {author} {\bibinfo {author} {\bibfnamefont {M.}~\bibnamefont
  {Piarulli}} \emph {et~al.},\ }\href {\doibase 10.1103/PhysRevLett.120.052503}
  {\bibfield  {journal} {\bibinfo  {journal} {Phys. Rev. Lett.}\ }\textbf
  {\bibinfo {volume} {120}},\ \bibinfo {pages} {052503} (\bibinfo {year}
  {2018})},\ \Eprint {http://arxiv.org/abs/1707.02883} {arXiv:1707.02883
  [nucl-th]} \BibitemShut {NoStop}%
\bibitem [{\citenamefont {Piarulli}\ \emph {et~al.}(2015)\citenamefont
  {Piarulli}, \citenamefont {Girlanda}, \citenamefont {Schiavilla},
  \citenamefont {Navarro~P\'erez}, \citenamefont {Amaro},\ and\ \citenamefont
  {Ruiz~Arriola}}]{piarulli2015minimally}%
  \BibitemOpen
  \bibfield  {author} {\bibinfo {author} {\bibfnamefont {M.}~\bibnamefont
  {Piarulli}}, \bibinfo {author} {\bibfnamefont {L.}~\bibnamefont {Girlanda}},
  \bibinfo {author} {\bibfnamefont {R.}~\bibnamefont {Schiavilla}}, \bibinfo
  {author} {\bibfnamefont {R.}~\bibnamefont {Navarro~P\'erez}}, \bibinfo
  {author} {\bibfnamefont {J.~E.}\ \bibnamefont {Amaro}}, \ and\ \bibinfo
  {author} {\bibfnamefont {E.}~\bibnamefont {Ruiz~Arriola}},\ }\href {\doibase
  10.1103/PhysRevC.91.024003} {\bibfield  {journal} {\bibinfo  {journal} {Phys.
  Rev. C}\ }\textbf {\bibinfo {volume} {91}},\ \bibinfo {pages} {024003}
  (\bibinfo {year} {2015})},\ \Eprint {http://arxiv.org/abs/1412.6446}
  {arXiv:1412.6446 [nucl-th]} \BibitemShut {NoStop}%
\bibitem [{\citenamefont {Piarulli}\ \emph {et~al.}(2016)\citenamefont
  {Piarulli}, \citenamefont {Girlanda}, \citenamefont {Schiavilla},
  \citenamefont {Kievsky}, \citenamefont {Lovato}, \citenamefont {Marcucci},
  \citenamefont {Pieper}, \citenamefont {Viviani},\ and\ \citenamefont
  {Wiringa}}]{Piarulli:2016vel}%
  \BibitemOpen
  \bibfield  {author} {\bibinfo {author} {\bibfnamefont {M.}~\bibnamefont
  {Piarulli}}, \bibinfo {author} {\bibfnamefont {L.}~\bibnamefont {Girlanda}},
  \bibinfo {author} {\bibfnamefont {R.}~\bibnamefont {Schiavilla}}, \bibinfo
  {author} {\bibfnamefont {A.}~\bibnamefont {Kievsky}}, \bibinfo {author}
  {\bibfnamefont {A.}~\bibnamefont {Lovato}}, \bibinfo {author} {\bibfnamefont
  {L.~E.}\ \bibnamefont {Marcucci}}, \bibinfo {author} {\bibfnamefont {S.~C.}\
  \bibnamefont {Pieper}}, \bibinfo {author} {\bibfnamefont {M.}~\bibnamefont
  {Viviani}}, \ and\ \bibinfo {author} {\bibfnamefont {R.~B.}\ \bibnamefont
  {Wiringa}},\ }\href {\doibase 10.1103/PhysRevC.94.054007} {\bibfield
  {journal} {\bibinfo  {journal} {Phys. Rev. C}\ }\textbf {\bibinfo {volume}
  {94}},\ \bibinfo {pages} {054007} (\bibinfo {year} {2016})},\ \Eprint
  {http://arxiv.org/abs/1606.06335} {arXiv:1606.06335 [nucl-th]} \BibitemShut
  {NoStop}%
\bibitem [{\citenamefont {Kallidonis}\ \emph {et~al.}(2018)\citenamefont
  {Kallidonis}, \citenamefont {Syritsyn}, \citenamefont {Engelhardt},
  \citenamefont {Green}, \citenamefont {Meinel}, \citenamefont {Negele},\ and\
  \citenamefont {Pochinsky}}]{Kallidonis:2018cas}%
  \BibitemOpen
  \bibfield  {author} {\bibinfo {author} {\bibfnamefont {C.}~\bibnamefont
  {Kallidonis}}, \bibinfo {author} {\bibfnamefont {S.}~\bibnamefont
  {Syritsyn}}, \bibinfo {author} {\bibfnamefont {M.}~\bibnamefont
  {Engelhardt}}, \bibinfo {author} {\bibfnamefont {J.}~\bibnamefont {Green}},
  \bibinfo {author} {\bibfnamefont {S.}~\bibnamefont {Meinel}}, \bibinfo
  {author} {\bibfnamefont {J.}~\bibnamefont {Negele}}, \ and\ \bibinfo {author}
  {\bibfnamefont {A.}~\bibnamefont {Pochinsky}},\ }\href {\doibase
  10.22323/1.334.0125} {\bibfield  {journal} {\bibinfo  {journal} {PoS}\
  }\textbf {\bibinfo {volume} {LATTICE2018}},\ \bibinfo {pages} {125} (\bibinfo
  {year} {2018})},\ \Eprint {http://arxiv.org/abs/1810.04294} {arXiv:1810.04294
  [hep-lat]} \BibitemShut {NoStop}%
\bibitem [{\citenamefont {Alexandrou}\ \emph
  {et~al.}(2019{\natexlab{b}})\citenamefont {Alexandrou}, \citenamefont
  {Bacchio}, \citenamefont {Constantinou}, \citenamefont {Finkenrath},
  \citenamefont {Hadjiyiannakou}, \citenamefont {Jansen}, \citenamefont
  {Koutsou},\ and\ \citenamefont {Vaquero Aviles-Casco}}]{Alexandrou:2018sjm}%
  \BibitemOpen
  \bibfield  {author} {\bibinfo {author} {\bibfnamefont {C.}~\bibnamefont
  {Alexandrou}}, \bibinfo {author} {\bibfnamefont {S.}~\bibnamefont {Bacchio}},
  \bibinfo {author} {\bibfnamefont {M.}~\bibnamefont {Constantinou}}, \bibinfo
  {author} {\bibfnamefont {J.}~\bibnamefont {Finkenrath}}, \bibinfo {author}
  {\bibfnamefont {K.}~\bibnamefont {Hadjiyiannakou}}, \bibinfo {author}
  {\bibfnamefont {K.}~\bibnamefont {Jansen}}, \bibinfo {author} {\bibfnamefont
  {G.}~\bibnamefont {Koutsou}}, \ and\ \bibinfo {author} {\bibfnamefont
  {A.}~\bibnamefont {Vaquero Aviles-Casco}},\ }\href {\doibase
  10.1103/PhysRevD.100.014509} {\bibfield  {journal} {\bibinfo  {journal}
  {Phys. Rev. D}\ }\textbf {\bibinfo {volume} {100}},\ \bibinfo {pages}
  {014509} (\bibinfo {year} {2019}{\natexlab{b}})},\ \Eprint
  {http://arxiv.org/abs/1812.10311} {arXiv:1812.10311 [hep-lat]} \BibitemShut
  {NoStop}%
\bibitem [{\citenamefont {Sufian}\ \emph {et~al.}(2020)\citenamefont {Sufian},
  \citenamefont {Liu},\ and\ \citenamefont {Richards}}]{Sufian:2018qtw}%
  \BibitemOpen
  \bibfield  {author} {\bibinfo {author} {\bibfnamefont {R.~S.}\ \bibnamefont
  {Sufian}}, \bibinfo {author} {\bibfnamefont {K.-F.}\ \bibnamefont {Liu}}, \
  and\ \bibinfo {author} {\bibfnamefont {D.~G.}\ \bibnamefont {Richards}},\
  }\href {\doibase 10.1007/JHEP01(2020)136} {\bibfield  {journal} {\bibinfo
  {journal} {JHEP}\ }\textbf {\bibinfo {volume} {01}},\ \bibinfo {pages} {136}
  (\bibinfo {year} {2020})},\ \Eprint {http://arxiv.org/abs/1809.03509}
  {arXiv:1809.03509 [hep-ph]} \BibitemShut {NoStop}%
\bibitem [{\citenamefont {Jang}\ \emph
  {et~al.}(2020{\natexlab{b}})\citenamefont {Jang}, \citenamefont {Gupta},
  \citenamefont {Lin}, \citenamefont {Yoon},\ and\ \citenamefont
  {Bhattacharya}}]{Jang:2019jkn}%
  \BibitemOpen
  \bibfield  {author} {\bibinfo {author} {\bibfnamefont {Y.-C.}\ \bibnamefont
  {Jang}}, \bibinfo {author} {\bibfnamefont {R.}~\bibnamefont {Gupta}},
  \bibinfo {author} {\bibfnamefont {H.-W.}\ \bibnamefont {Lin}}, \bibinfo
  {author} {\bibfnamefont {B.}~\bibnamefont {Yoon}}, \ and\ \bibinfo {author}
  {\bibfnamefont {T.}~\bibnamefont {Bhattacharya}},\ }\href {\doibase
  10.1103/PhysRevD.101.014507} {\bibfield  {journal} {\bibinfo  {journal}
  {Phys. Rev. D}\ }\textbf {\bibinfo {volume} {101}},\ \bibinfo {pages}
  {014507} (\bibinfo {year} {2020}{\natexlab{b}})},\ \Eprint
  {http://arxiv.org/abs/1906.07217} {arXiv:1906.07217 [hep-lat]} \BibitemShut
  {NoStop}%
\bibitem [{\citenamefont {Briceno}\ \emph {et~al.}(2018)\citenamefont
  {Briceno}, \citenamefont {Dudek},\ and\ \citenamefont
  {Young}}]{Briceno:2017max}%
  \BibitemOpen
  \bibfield  {author} {\bibinfo {author} {\bibfnamefont {R.~A.}\ \bibnamefont
  {Briceno}}, \bibinfo {author} {\bibfnamefont {J.~J.}\ \bibnamefont {Dudek}},
  \ and\ \bibinfo {author} {\bibfnamefont {R.~D.}\ \bibnamefont {Young}},\
  }\href {\doibase 10.1103/RevModPhys.90.025001} {\bibfield  {journal}
  {\bibinfo  {journal} {Rev. Mod. Phys.}\ }\textbf {\bibinfo {volume} {90}},\
  \bibinfo {pages} {025001} (\bibinfo {year} {2018})},\ \Eprint
  {http://arxiv.org/abs/1706.06223} {arXiv:1706.06223 [hep-lat]} \BibitemShut
  {NoStop}%
\bibitem [{\citenamefont {Hansen}\ and\ \citenamefont
  {Sharpe}(2012)}]{Hansen:2012tf}%
  \BibitemOpen
  \bibfield  {author} {\bibinfo {author} {\bibfnamefont {M.~T.}\ \bibnamefont
  {Hansen}}\ and\ \bibinfo {author} {\bibfnamefont {S.~R.}\ \bibnamefont
  {Sharpe}},\ }\href {\doibase 10.1103/PhysRevD.86.016007} {\bibfield
  {journal} {\bibinfo  {journal} {Phys. Rev. D}\ }\textbf {\bibinfo {volume}
  {86}},\ \bibinfo {pages} {016007} (\bibinfo {year} {2012})},\ \Eprint
  {http://arxiv.org/abs/1204.0826} {arXiv:1204.0826 [hep-lat]} \BibitemShut
  {NoStop}%
\bibitem [{\citenamefont {Chen}\ \emph {et~al.}(2017)\citenamefont {Chen},
  \citenamefont {Detmold}, \citenamefont {Lynn},\ and\ \citenamefont
  {Schwenk}}]{Chen:2016bde}%
  \BibitemOpen
  \bibfield  {author} {\bibinfo {author} {\bibfnamefont {J.-W.}\ \bibnamefont
  {Chen}}, \bibinfo {author} {\bibfnamefont {W.}~\bibnamefont {Detmold}},
  \bibinfo {author} {\bibfnamefont {J.~E.}\ \bibnamefont {Lynn}}, \ and\
  \bibinfo {author} {\bibfnamefont {A.}~\bibnamefont {Schwenk}},\ }\href
  {\doibase 10.1103/PhysRevLett.119.262502} {\bibfield  {journal} {\bibinfo
  {journal} {Phys. Rev. Lett.}\ }\textbf {\bibinfo {volume} {119}},\ \bibinfo
  {pages} {262502} (\bibinfo {year} {2017})},\ \Eprint
  {http://arxiv.org/abs/1607.03065} {arXiv:1607.03065 [hep-ph]} \BibitemShut
  {NoStop}%
\bibitem [{\citenamefont {Lynn}\ \emph {et~al.}(2020)\citenamefont {Lynn},
  \citenamefont {Lonardoni}, \citenamefont {Carlson}, \citenamefont {Chen},
  \citenamefont {Detmold}, \citenamefont {Gandolfi},\ and\ \citenamefont
  {Schwenk}}]{Lynn:2019vwp}%
  \BibitemOpen
  \bibfield  {author} {\bibinfo {author} {\bibfnamefont {J.~E.}\ \bibnamefont
  {Lynn}}, \bibinfo {author} {\bibfnamefont {D.}~\bibnamefont {Lonardoni}},
  \bibinfo {author} {\bibfnamefont {J.}~\bibnamefont {Carlson}}, \bibinfo
  {author} {\bibfnamefont {J.~W.}\ \bibnamefont {Chen}}, \bibinfo {author}
  {\bibfnamefont {W.}~\bibnamefont {Detmold}}, \bibinfo {author} {\bibfnamefont
  {S.}~\bibnamefont {Gandolfi}}, \ and\ \bibinfo {author} {\bibfnamefont
  {A.}~\bibnamefont {Schwenk}},\ }\href {\doibase 10.1088/1361-6471/ab6af7}
  {\bibfield  {journal} {\bibinfo  {journal} {J. Phys. G}\ }\textbf {\bibinfo
  {volume} {47}},\ \bibinfo {pages} {045109} (\bibinfo {year} {2020})},\
  \Eprint {http://arxiv.org/abs/1903.12587} {arXiv:1903.12587 [nucl-th]}
  \BibitemShut {NoStop}%
\bibitem [{\citenamefont {King}\ \emph {et~al.}(2020)\citenamefont {King},
  \citenamefont {Andreoli}, \citenamefont {Pastore}, \citenamefont {Piarulli},
  \citenamefont {Schiavilla}, \citenamefont {Wiringa}, \citenamefont
  {Carlson},\ and\ \citenamefont {Gandolfi}}]{King:2020wmp}%
  \BibitemOpen
  \bibfield  {author} {\bibinfo {author} {\bibfnamefont {G.~B.}\ \bibnamefont
  {King}}, \bibinfo {author} {\bibfnamefont {L.}~\bibnamefont {Andreoli}},
  \bibinfo {author} {\bibfnamefont {S.}~\bibnamefont {Pastore}}, \bibinfo
  {author} {\bibfnamefont {M.}~\bibnamefont {Piarulli}}, \bibinfo {author}
  {\bibfnamefont {R.}~\bibnamefont {Schiavilla}}, \bibinfo {author}
  {\bibfnamefont {R.~B.}\ \bibnamefont {Wiringa}}, \bibinfo {author}
  {\bibfnamefont {J.}~\bibnamefont {Carlson}}, \ and\ \bibinfo {author}
  {\bibfnamefont {S.}~\bibnamefont {Gandolfi}},\ }\href {\doibase
  10.1103/PhysRevC.102.025501} {\bibfield  {journal} {\bibinfo  {journal}
  {Phys. Rev. C}\ }\textbf {\bibinfo {volume} {102}},\ \bibinfo {pages}
  {025501} (\bibinfo {year} {2020})},\ \Eprint
  {http://arxiv.org/abs/2004.05263} {arXiv:2004.05263 [nucl-th]} \BibitemShut
  {NoStop}%
\bibitem [{\citenamefont {Gysbers}\ \emph {et~al.}(2019)\citenamefont {Gysbers}
  \emph {et~al.}}]{Gysbers:2019uyb}%
  \BibitemOpen
  \bibfield  {author} {\bibinfo {author} {\bibfnamefont {P.}~\bibnamefont
  {Gysbers}} \emph {et~al.},\ }\href {\doibase 10.1038/s41567-019-0450-7}
  {\bibfield  {journal} {\bibinfo  {journal} {Nature Phys.}\ }\textbf {\bibinfo
  {volume} {15}},\ \bibinfo {pages} {428} (\bibinfo {year} {2019})},\ \Eprint
  {http://arxiv.org/abs/1903.00047} {arXiv:1903.00047 [nucl-th]} \BibitemShut
  {NoStop}%
\bibitem [{\citenamefont {Davoudi}\ \emph {et~al.}()\citenamefont {Davoudi}
  \emph {et~al.}}]{ndbdWP}%
  \BibitemOpen
  \bibfield  {author} {\bibinfo {author} {\bibfnamefont {Z.}~\bibnamefont
  {Davoudi}} \emph {et~al.},\ }\href@noop {} {\enquote {\bibinfo {title}
  {{Neutrinoless Double Beta Decay: A Roadmap for Matching Theory to
  Experiment}},}\ }\bibinfo {note} {{Snowmass 2021 white paper}}\BibitemShut
  {NoStop}%
\bibitem [{\citenamefont {Lovato}\ \emph {et~al.}(2020)\citenamefont {Lovato},
  \citenamefont {Carlson}, \citenamefont {Gandolfi}, \citenamefont {Rocco},\
  and\ \citenamefont {Schiavilla}}]{Lovato:2020kba}%
  \BibitemOpen
  \bibfield  {author} {\bibinfo {author} {\bibfnamefont {A.}~\bibnamefont
  {Lovato}}, \bibinfo {author} {\bibfnamefont {J.}~\bibnamefont {Carlson}},
  \bibinfo {author} {\bibfnamefont {S.}~\bibnamefont {Gandolfi}}, \bibinfo
  {author} {\bibfnamefont {N.}~\bibnamefont {Rocco}}, \ and\ \bibinfo {author}
  {\bibfnamefont {R.}~\bibnamefont {Schiavilla}},\ }\href {\doibase
  10.1103/PhysRevX.10.031068} {\bibfield  {journal} {\bibinfo  {journal} {Phys.
  Rev. X}\ }\textbf {\bibinfo {volume} {10}},\ \bibinfo {pages} {031068}
  (\bibinfo {year} {2020})},\ \Eprint {http://arxiv.org/abs/2003.07710}
  {arXiv:2003.07710 [nucl-th]} \BibitemShut {NoStop}%
\bibitem [{\citenamefont {Leidemann}\ and\ \citenamefont
  {Orlandini}(2013)}]{leidemann2013}%
  \BibitemOpen
  \bibfield  {author} {\bibinfo {author} {\bibfnamefont {W.}~\bibnamefont
  {Leidemann}}\ and\ \bibinfo {author} {\bibfnamefont {G.}~\bibnamefont
  {Orlandini}},\ }\href {\doibase 10.1016/j.ppnp.2012.09.001} {\bibfield
  {journal} {\bibinfo  {journal} {Prog. Part. Nucl. Phys.}\ }\textbf {\bibinfo
  {volume} {68}},\ \bibinfo {pages} {158} (\bibinfo {year} {2013})},\ \Eprint
  {http://arxiv.org/abs/1204.4617} {arXiv:1204.4617 [nucl-th]} \BibitemShut
  {NoStop}%
\bibitem [{\citenamefont {Raghavan}\ \emph {et~al.}(2021)\citenamefont
  {Raghavan}, \citenamefont {Balaprakash}, \citenamefont {Lovato},
  \citenamefont {Rocco},\ and\ \citenamefont {Wild}}]{Raghavan:2020bze}%
  \BibitemOpen
  \bibfield  {author} {\bibinfo {author} {\bibfnamefont {K.}~\bibnamefont
  {Raghavan}}, \bibinfo {author} {\bibfnamefont {P.}~\bibnamefont
  {Balaprakash}}, \bibinfo {author} {\bibfnamefont {A.}~\bibnamefont {Lovato}},
  \bibinfo {author} {\bibfnamefont {N.}~\bibnamefont {Rocco}}, \ and\ \bibinfo
  {author} {\bibfnamefont {S.~M.}\ \bibnamefont {Wild}},\ }\href {\doibase
  10.1103/PhysRevC.103.035502} {\bibfield  {journal} {\bibinfo  {journal}
  {Phys. Rev. C}\ }\textbf {\bibinfo {volume} {103}},\ \bibinfo {pages}
  {035502} (\bibinfo {year} {2021})},\ \Eprint
  {http://arxiv.org/abs/2010.12703} {arXiv:2010.12703 [nucl-th]} \BibitemShut
  {NoStop}%
\bibitem [{\citenamefont {Williamson}\ \emph {et~al.}(1997)\citenamefont
  {Williamson} \emph {et~al.}}]{Williamson:1997zz}%
  \BibitemOpen
  \bibfield  {author} {\bibinfo {author} {\bibfnamefont {C.~F.}\ \bibnamefont
  {Williamson}} \emph {et~al.},\ }\href {\doibase 10.1103/PhysRevC.56.3152}
  {\bibfield  {journal} {\bibinfo  {journal} {Phys. Rev. C}\ }\textbf {\bibinfo
  {volume} {56}},\ \bibinfo {pages} {3152} (\bibinfo {year}
  {1997})}\BibitemShut {NoStop}%
\bibitem [{\citenamefont {Sobczyk}\ \emph {et~al.}(2021)\citenamefont
  {Sobczyk}, \citenamefont {Acharya}, \citenamefont {Bacca},\ and\
  \citenamefont {Hagen}}]{Sobczyk2021}%
  \BibitemOpen
  \bibfield  {author} {\bibinfo {author} {\bibfnamefont {J.~E.}\ \bibnamefont
  {Sobczyk}}, \bibinfo {author} {\bibfnamefont {B.}~\bibnamefont {Acharya}},
  \bibinfo {author} {\bibfnamefont {S.}~\bibnamefont {Bacca}}, \ and\ \bibinfo
  {author} {\bibfnamefont {G.}~\bibnamefont {Hagen}},\ }\href {\doibase
  10.1103/PhysRevLett.127.072501} {\bibfield  {journal} {\bibinfo  {journal}
  {Phys. Rev. Lett.}\ }\textbf {\bibinfo {volume} {127}},\ \bibinfo {pages}
  {072501} (\bibinfo {year} {2021})},\ \Eprint
  {http://arxiv.org/abs/2103.06786} {arXiv:2103.06786 [nucl-th]} \BibitemShut
  {NoStop}%
\bibitem [{\citenamefont {Lovato}\ \emph {et~al.}(2016)\citenamefont {Lovato},
  \citenamefont {Gandolfi}, \citenamefont {Carlson}, \citenamefont {Pieper},\
  and\ \citenamefont {Schiavilla}}]{Lovato:2016gkq}%
  \BibitemOpen
  \bibfield  {author} {\bibinfo {author} {\bibfnamefont {A.}~\bibnamefont
  {Lovato}}, \bibinfo {author} {\bibfnamefont {S.}~\bibnamefont {Gandolfi}},
  \bibinfo {author} {\bibfnamefont {J.}~\bibnamefont {Carlson}}, \bibinfo
  {author} {\bibfnamefont {S.~C.}\ \bibnamefont {Pieper}}, \ and\ \bibinfo
  {author} {\bibfnamefont {R.}~\bibnamefont {Schiavilla}},\ }\href {\doibase
  10.1103/PhysRevLett.117.082501} {\bibfield  {journal} {\bibinfo  {journal}
  {Phys. Rev. Lett.}\ }\textbf {\bibinfo {volume} {117}},\ \bibinfo {pages}
  {082501} (\bibinfo {year} {2016})},\ \Eprint
  {http://arxiv.org/abs/1605.00248} {arXiv:1605.00248 [nucl-th]} \BibitemShut
  {NoStop}%
\bibitem [{\citenamefont {Schmidt}\ and\ \citenamefont
  {Fantoni}(1999)}]{Schmidt:1999lik}%
  \BibitemOpen
  \bibfield  {author} {\bibinfo {author} {\bibfnamefont {K.~E.}\ \bibnamefont
  {Schmidt}}\ and\ \bibinfo {author} {\bibfnamefont {S.}~\bibnamefont
  {Fantoni}},\ }\href {\doibase 10.1016/S0370-2693(98)01522-6} {\bibfield
  {journal} {\bibinfo  {journal} {Phys. Lett. B}\ }\textbf {\bibinfo {volume}
  {446}},\ \bibinfo {pages} {99} (\bibinfo {year} {1999})}\BibitemShut
  {NoStop}%
\bibitem [{\citenamefont {Adams}\ \emph {et~al.}(2021)\citenamefont {Adams},
  \citenamefont {Carleo}, \citenamefont {Lovato},\ and\ \citenamefont
  {Rocco}}]{Adams:2020aax}%
  \BibitemOpen
  \bibfield  {author} {\bibinfo {author} {\bibfnamefont {C.}~\bibnamefont
  {Adams}}, \bibinfo {author} {\bibfnamefont {G.}~\bibnamefont {Carleo}},
  \bibinfo {author} {\bibfnamefont {A.}~\bibnamefont {Lovato}}, \ and\ \bibinfo
  {author} {\bibfnamefont {N.}~\bibnamefont {Rocco}},\ }\href {\doibase
  10.1103/PhysRevLett.127.022502} {\bibfield  {journal} {\bibinfo  {journal}
  {Phys. Rev. Lett.}\ }\textbf {\bibinfo {volume} {127}},\ \bibinfo {pages}
  {022502} (\bibinfo {year} {2021})},\ \Eprint
  {http://arxiv.org/abs/2007.14282} {arXiv:2007.14282 [nucl-th]} \BibitemShut
  {NoStop}%
\bibitem [{\citenamefont {Gnech}\ \emph {et~al.}(2022)\citenamefont {Gnech},
  \citenamefont {Adams}, \citenamefont {Brawand}, \citenamefont {Carleo},
  \citenamefont {Lovato},\ and\ \citenamefont {Rocco}}]{Gnech:2021wfn}%
  \BibitemOpen
  \bibfield  {author} {\bibinfo {author} {\bibfnamefont {A.}~\bibnamefont
  {Gnech}}, \bibinfo {author} {\bibfnamefont {C.}~\bibnamefont {Adams}},
  \bibinfo {author} {\bibfnamefont {N.}~\bibnamefont {Brawand}}, \bibinfo
  {author} {\bibfnamefont {G.}~\bibnamefont {Carleo}}, \bibinfo {author}
  {\bibfnamefont {A.}~\bibnamefont {Lovato}}, \ and\ \bibinfo {author}
  {\bibfnamefont {N.}~\bibnamefont {Rocco}},\ }\href {\doibase
  10.1007/s00601-021-01706-0} {\bibfield  {journal} {\bibinfo  {journal} {Few
  Body Syst.}\ }\textbf {\bibinfo {volume} {63}},\ \bibinfo {pages} {7}
  (\bibinfo {year} {2022})},\ \Eprint {http://arxiv.org/abs/2108.06836}
  {arXiv:2108.06836 [nucl-th]} \BibitemShut {NoStop}%
\bibitem [{\citenamefont {Hagen}\ \emph
  {et~al.}(2014{\natexlab{b}})\citenamefont {Hagen}, \citenamefont
  {Papenbrock}, \citenamefont {Hjorth-Jensen},\ and\ \citenamefont
  {Dean}}]{hagen2013c}%
  \BibitemOpen
  \bibfield  {author} {\bibinfo {author} {\bibfnamefont {G.}~\bibnamefont
  {Hagen}}, \bibinfo {author} {\bibfnamefont {T.}~\bibnamefont {Papenbrock}},
  \bibinfo {author} {\bibfnamefont {M.}~\bibnamefont {Hjorth-Jensen}}, \ and\
  \bibinfo {author} {\bibfnamefont {D.~J.}\ \bibnamefont {Dean}},\ }\href
  {\doibase 10.1088/0034-4885/77/9/096302} {\bibfield  {journal} {\bibinfo
  {journal} {Rept. Prog. Phys.}\ }\textbf {\bibinfo {volume} {77}},\ \bibinfo
  {pages} {096302} (\bibinfo {year} {2014}{\natexlab{b}})},\ \Eprint
  {http://arxiv.org/abs/1312.7872} {arXiv:1312.7872 [nucl-th]} \BibitemShut
  {NoStop}%
\bibitem [{\citenamefont {Efros}\ \emph {et~al.}(1994)\citenamefont {Efros},
  \citenamefont {Leidemann},\ and\ \citenamefont {Orlandini}}]{efros1994}%
  \BibitemOpen
  \bibfield  {author} {\bibinfo {author} {\bibfnamefont {V.~D.}\ \bibnamefont
  {Efros}}, \bibinfo {author} {\bibfnamefont {W.}~\bibnamefont {Leidemann}}, \
  and\ \bibinfo {author} {\bibfnamefont {G.}~\bibnamefont {Orlandini}},\ }\href
  {\doibase 10.1016/0370-2693(94)91355-2} {\bibfield  {journal} {\bibinfo
  {journal} {Phys. Lett. B}\ }\textbf {\bibinfo {volume} {338}},\ \bibinfo
  {pages} {130} (\bibinfo {year} {1994})},\ \Eprint
  {http://arxiv.org/abs/nucl-th/9409004} {arXiv:nucl-th/9409004} \BibitemShut
  {NoStop}%
\bibitem [{\citenamefont {Bacca}\ \emph {et~al.}(2013)\citenamefont {Bacca},
  \citenamefont {Barnea}, \citenamefont {Hagen}, \citenamefont {Orlandini},\
  and\ \citenamefont {Papenbrock}}]{bacca2013}%
  \BibitemOpen
  \bibfield  {author} {\bibinfo {author} {\bibfnamefont {S.}~\bibnamefont
  {Bacca}}, \bibinfo {author} {\bibfnamefont {N.}~\bibnamefont {Barnea}},
  \bibinfo {author} {\bibfnamefont {G.}~\bibnamefont {Hagen}}, \bibinfo
  {author} {\bibfnamefont {G.}~\bibnamefont {Orlandini}}, \ and\ \bibinfo
  {author} {\bibfnamefont {T.}~\bibnamefont {Papenbrock}},\ }\href {\doibase
  10.1103/PhysRevLett.111.122502} {\bibfield  {journal} {\bibinfo  {journal}
  {Phys. Rev. Lett.}\ }\textbf {\bibinfo {volume} {111}},\ \bibinfo {pages}
  {122502} (\bibinfo {year} {2013})},\ \Eprint {http://arxiv.org/abs/1303.7446}
  {arXiv:1303.7446 [nucl-th]} \BibitemShut {NoStop}%
\bibitem [{\citenamefont {Rocco}(2020)}]{Rocco:2020jlx}%
  \BibitemOpen
  \bibfield  {author} {\bibinfo {author} {\bibfnamefont {N.}~\bibnamefont
  {Rocco}},\ }\href {\doibase 10.3389/fphy.2020.00116} {\bibfield  {journal}
  {\bibinfo  {journal} {Front. in Phys.}\ }\textbf {\bibinfo {volume} {8}},\
  \bibinfo {pages} {116} (\bibinfo {year} {2020})}\BibitemShut {NoStop}%
\bibitem [{\citenamefont {Pastore}\ \emph {et~al.}(2020)\citenamefont
  {Pastore}, \citenamefont {Carlson}, \citenamefont {Gandolfi}, \citenamefont
  {Schiavilla},\ and\ \citenamefont {Wiringa}}]{Pastore:2019urn}%
  \BibitemOpen
  \bibfield  {author} {\bibinfo {author} {\bibfnamefont {S.}~\bibnamefont
  {Pastore}}, \bibinfo {author} {\bibfnamefont {J.}~\bibnamefont {Carlson}},
  \bibinfo {author} {\bibfnamefont {S.}~\bibnamefont {Gandolfi}}, \bibinfo
  {author} {\bibfnamefont {R.}~\bibnamefont {Schiavilla}}, \ and\ \bibinfo
  {author} {\bibfnamefont {R.~B.}\ \bibnamefont {Wiringa}},\ }\href {\doibase
  10.1103/PhysRevC.101.044612} {\bibfield  {journal} {\bibinfo  {journal}
  {Phys. Rev. C}\ }\textbf {\bibinfo {volume} {101}},\ \bibinfo {pages}
  {044612} (\bibinfo {year} {2020})},\ \Eprint
  {http://arxiv.org/abs/1909.06400} {arXiv:1909.06400 [nucl-th]} \BibitemShut
  {NoStop}%
\bibitem [{\citenamefont {Barrow}\ \emph {et~al.}(2021)\citenamefont {Barrow},
  \citenamefont {Gardiner}, \citenamefont {Pastore}, \citenamefont
  {Betancourt},\ and\ \citenamefont {Carlson}}]{Barrow:2020mfy}%
  \BibitemOpen
  \bibfield  {author} {\bibinfo {author} {\bibfnamefont {J.~L.}\ \bibnamefont
  {Barrow}}, \bibinfo {author} {\bibfnamefont {S.}~\bibnamefont {Gardiner}},
  \bibinfo {author} {\bibfnamefont {S.}~\bibnamefont {Pastore}}, \bibinfo
  {author} {\bibfnamefont {M.}~\bibnamefont {Betancourt}}, \ and\ \bibinfo
  {author} {\bibfnamefont {J.}~\bibnamefont {Carlson}},\ }\href {\doibase
  10.1103/PhysRevD.103.052001} {\bibfield  {journal} {\bibinfo  {journal}
  {Phys. Rev. D}\ }\textbf {\bibinfo {volume} {103}},\ \bibinfo {pages}
  {052001} (\bibinfo {year} {2021})},\ \Eprint
  {http://arxiv.org/abs/2010.04154} {arXiv:2010.04154 [nucl-th]} \BibitemShut
  {NoStop}%
\bibitem [{\citenamefont {Ivanov}\ \emph {et~al.}(2019)\citenamefont {Ivanov},
  \citenamefont {Antonov}, \citenamefont {Megias}, \citenamefont {Caballero},
  \citenamefont {Barbaro}, \citenamefont {Amaro}, \citenamefont {Ruiz~Simo},
  \citenamefont {Donnelly},\ and\ \citenamefont {Ud\'\i{}as}}]{Ivanov:2018nlm}%
  \BibitemOpen
  \bibfield  {author} {\bibinfo {author} {\bibfnamefont {M.~V.}\ \bibnamefont
  {Ivanov}}, \bibinfo {author} {\bibfnamefont {A.~N.}\ \bibnamefont {Antonov}},
  \bibinfo {author} {\bibfnamefont {G.~D.}\ \bibnamefont {Megias}}, \bibinfo
  {author} {\bibfnamefont {J.~A.}\ \bibnamefont {Caballero}}, \bibinfo {author}
  {\bibfnamefont {M.~B.}\ \bibnamefont {Barbaro}}, \bibinfo {author}
  {\bibfnamefont {J.~E.}\ \bibnamefont {Amaro}}, \bibinfo {author}
  {\bibfnamefont {I.}~\bibnamefont {Ruiz~Simo}}, \bibinfo {author}
  {\bibfnamefont {T.~W.}\ \bibnamefont {Donnelly}}, \ and\ \bibinfo {author}
  {\bibfnamefont {J.~M.}\ \bibnamefont {Ud\'\i{}as}},\ }\href {\doibase
  10.1103/PhysRevC.99.014610} {\bibfield  {journal} {\bibinfo  {journal} {Phys.
  Rev. C}\ }\textbf {\bibinfo {volume} {99}},\ \bibinfo {pages} {014610}
  (\bibinfo {year} {2019})},\ \Eprint {http://arxiv.org/abs/1812.09435}
  {arXiv:1812.09435 [nucl-th]} \BibitemShut {NoStop}%
\bibitem [{\citenamefont {Rocco}\ and\ \citenamefont
  {Barbieri}(2018)}]{Rocco:2018vbf}%
  \BibitemOpen
  \bibfield  {author} {\bibinfo {author} {\bibfnamefont {N.}~\bibnamefont
  {Rocco}}\ and\ \bibinfo {author} {\bibfnamefont {C.}~\bibnamefont
  {Barbieri}},\ }\href {\doibase 10.1103/PhysRevC.98.025501} {\bibfield
  {journal} {\bibinfo  {journal} {Phys. Rev. C}\ }\textbf {\bibinfo {volume}
  {98}},\ \bibinfo {pages} {025501} (\bibinfo {year} {2018})},\ \Eprint
  {http://arxiv.org/abs/1803.00825} {arXiv:1803.00825 [nucl-th]} \BibitemShut
  {NoStop}%
\bibitem [{\citenamefont {Kamano}\ \emph {et~al.}(2013)\citenamefont {Kamano},
  \citenamefont {Nakamura}, \citenamefont {Lee},\ and\ \citenamefont
  {Sato}}]{Kamano:2013iva}%
  \BibitemOpen
  \bibfield  {author} {\bibinfo {author} {\bibfnamefont {H.}~\bibnamefont
  {Kamano}}, \bibinfo {author} {\bibfnamefont {S.}~\bibnamefont {Nakamura}},
  \bibinfo {author} {\bibfnamefont {T.~S.~H.}\ \bibnamefont {Lee}}, \ and\
  \bibinfo {author} {\bibfnamefont {T.}~\bibnamefont {Sato}},\ }\href {\doibase
  10.1103/PhysRevC.88.035209} {\bibfield  {journal} {\bibinfo  {journal} {Phys.
  Rev. C}\ }\textbf {\bibinfo {volume} {88}},\ \bibinfo {pages} {035209}
  (\bibinfo {year} {2013})},\ \Eprint {http://arxiv.org/abs/1305.4351}
  {arXiv:1305.4351 [nucl-th]} \BibitemShut {NoStop}%
\bibitem [{\citenamefont {Nakamura}\ \emph {et~al.}(2015)\citenamefont
  {Nakamura}, \citenamefont {Kamano},\ and\ \citenamefont
  {Sato}}]{Nakamura:2015rta}%
  \BibitemOpen
  \bibfield  {author} {\bibinfo {author} {\bibfnamefont {S.}~\bibnamefont
  {Nakamura}}, \bibinfo {author} {\bibfnamefont {H.}~\bibnamefont {Kamano}}, \
  and\ \bibinfo {author} {\bibfnamefont {T.}~\bibnamefont {Sato}},\ }\href
  {\doibase 10.1103/PhysRevD.92.074024} {\bibfield  {journal} {\bibinfo
  {journal} {Phys. Rev. D}\ }\textbf {\bibinfo {volume} {92}},\ \bibinfo
  {pages} {074024} (\bibinfo {year} {2015})},\ \Eprint
  {http://arxiv.org/abs/1506.03403} {arXiv:1506.03403 [hep-ph]} \BibitemShut
  {NoStop}%
\bibitem [{\citenamefont {Kamano}\ \emph {et~al.}(2016)\citenamefont {Kamano},
  \citenamefont {Nakamura}, \citenamefont {Lee},\ and\ \citenamefont
  {Sato}}]{Kamano:2016bgm}%
  \BibitemOpen
  \bibfield  {author} {\bibinfo {author} {\bibfnamefont {H.}~\bibnamefont
  {Kamano}}, \bibinfo {author} {\bibfnamefont {S.}~\bibnamefont {Nakamura}},
  \bibinfo {author} {\bibfnamefont {T.}~\bibnamefont {Lee}}, \ and\ \bibinfo
  {author} {\bibfnamefont {T.}~\bibnamefont {Sato}},\ }\href {\doibase
  10.1103/PhysRevC.94.015201} {\bibfield  {journal} {\bibinfo  {journal} {Phys.
  Rev. C}\ }\textbf {\bibinfo {volume} {94}},\ \bibinfo {pages} {015201}
  (\bibinfo {year} {2016})},\ \Eprint {http://arxiv.org/abs/1605.00363}
  {arXiv:1605.00363 [nucl-th]} \BibitemShut {NoStop}%
\bibitem [{\citenamefont {Roggero}\ and\ \citenamefont
  {Carlson}(2019)}]{Roggero:2019}%
  \BibitemOpen
  \bibfield  {author} {\bibinfo {author} {\bibfnamefont {A.}~\bibnamefont
  {Roggero}}\ and\ \bibinfo {author} {\bibfnamefont {J.}~\bibnamefont
  {Carlson}},\ }\href {\doibase 10.1103/PhysRevC.100.034610} {\bibfield
  {journal} {\bibinfo  {journal} {Phys. Rev. C}\ }\textbf {\bibinfo {volume}
  {100}},\ \bibinfo {pages} {034610} (\bibinfo {year} {2019})},\ \Eprint
  {http://arxiv.org/abs/1804.01505} {arXiv:1804.01505 [quant-ph]} \BibitemShut
  {NoStop}%
\bibitem [{\citenamefont {Roggero}\ \emph {et~al.}(2020)\citenamefont
  {Roggero}, \citenamefont {Li}, \citenamefont {Carlson}, \citenamefont
  {Gupta},\ and\ \citenamefont {Perdue}}]{Roggero:2020}%
  \BibitemOpen
  \bibfield  {author} {\bibinfo {author} {\bibfnamefont {A.}~\bibnamefont
  {Roggero}}, \bibinfo {author} {\bibfnamefont {A.~C.~Y.}\ \bibnamefont {Li}},
  \bibinfo {author} {\bibfnamefont {J.}~\bibnamefont {Carlson}}, \bibinfo
  {author} {\bibfnamefont {R.}~\bibnamefont {Gupta}}, \ and\ \bibinfo {author}
  {\bibfnamefont {G.~N.}\ \bibnamefont {Perdue}},\ }\href {\doibase
  10.1103/PhysRevD.101.074038} {\bibfield  {journal} {\bibinfo  {journal}
  {Phys. Rev. D}\ }\textbf {\bibinfo {volume} {101}},\ \bibinfo {pages}
  {074038} (\bibinfo {year} {2020})},\ \Eprint
  {http://arxiv.org/abs/1911.06368} {arXiv:1911.06368 [quant-ph]} \BibitemShut
  {NoStop}%
\bibitem [{\citenamefont {Hall}\ \emph {et~al.}(2021)\citenamefont {Hall},
  \citenamefont {Roggero}, \citenamefont {Baroni},\ and\ \citenamefont
  {Carlson}}]{Hall:2021}%
  \BibitemOpen
  \bibfield  {author} {\bibinfo {author} {\bibfnamefont {B.}~\bibnamefont
  {Hall}}, \bibinfo {author} {\bibfnamefont {A.}~\bibnamefont {Roggero}},
  \bibinfo {author} {\bibfnamefont {A.}~\bibnamefont {Baroni}}, \ and\ \bibinfo
  {author} {\bibfnamefont {J.}~\bibnamefont {Carlson}},\ }\href {\doibase
  10.1103/PhysRevD.104.063009} {\bibfield  {journal} {\bibinfo  {journal}
  {Phys. Rev. D}\ }\textbf {\bibinfo {volume} {104}},\ \bibinfo {pages}
  {063009} (\bibinfo {year} {2021})},\ \Eprint
  {http://arxiv.org/abs/2102.12556} {arXiv:2102.12556 [quant-ph]} \BibitemShut
  {NoStop}%
\bibitem [{\citenamefont {Fetter}\ and\ \citenamefont
  {Walecka}(2003)}]{fetterwalecka}%
  \BibitemOpen
  \bibfield  {author} {\bibinfo {author} {\bibfnamefont {A.~L.}\ \bibnamefont
  {Fetter}}\ and\ \bibinfo {author} {\bibfnamefont {J.~D.}\ \bibnamefont
  {Walecka}},\ }\href@noop {} {\emph {\bibinfo {title} {Quantum Theory of
  Many-particle Systems}}}\ (\bibinfo  {publisher} {Dover},\ \bibinfo {year}
  {2003})\BibitemShut {NoStop}%
\bibitem [{\citenamefont {Singh}\ and\ \citenamefont
  {Oset}(1992)}]{Singh:1992dc}%
  \BibitemOpen
  \bibfield  {author} {\bibinfo {author} {\bibfnamefont {S.}~\bibnamefont
  {Singh}}\ and\ \bibinfo {author} {\bibfnamefont {E.}~\bibnamefont {Oset}},\
  }\href {\doibase 10.1016/0375-9474(92)90259-M} {\bibfield  {journal}
  {\bibinfo  {journal} {Nucl. Phys. A}\ }\textbf {\bibinfo {volume} {542}},\
  \bibinfo {pages} {587} (\bibinfo {year} {1992})}\BibitemShut {NoStop}%
\bibitem [{\citenamefont {Kim}\ \emph {et~al.}(1995)\citenamefont {Kim},
  \citenamefont {Piekarewicz},\ and\ \citenamefont {Horowitz}}]{Kim:1994zea}%
  \BibitemOpen
  \bibfield  {author} {\bibinfo {author} {\bibfnamefont {H.-c.}\ \bibnamefont
  {Kim}}, \bibinfo {author} {\bibfnamefont {J.}~\bibnamefont {Piekarewicz}}, \
  and\ \bibinfo {author} {\bibfnamefont {C.}~\bibnamefont {Horowitz}},\ }\href
  {\doibase 10.1103/PhysRevC.51.2739} {\bibfield  {journal} {\bibinfo
  {journal} {Phys. Rev. C}\ }\textbf {\bibinfo {volume} {51}},\ \bibinfo
  {pages} {2739} (\bibinfo {year} {1995})},\ \Eprint
  {http://arxiv.org/abs/nucl-th/9412017} {arXiv:nucl-th/9412017 [nucl-th]}
  \BibitemShut {NoStop}%
\bibitem [{\citenamefont {Nieves}\ \emph {et~al.}(2011)\citenamefont {Nieves},
  \citenamefont {Ruiz~Simo},\ and\ \citenamefont
  {Vicente~Vacas}}]{Nieves:2011pp}%
  \BibitemOpen
  \bibfield  {author} {\bibinfo {author} {\bibfnamefont {J.}~\bibnamefont
  {Nieves}}, \bibinfo {author} {\bibfnamefont {I.}~\bibnamefont {Ruiz~Simo}}, \
  and\ \bibinfo {author} {\bibfnamefont {M.}~\bibnamefont {Vicente~Vacas}},\
  }\href {\doibase 10.1103/PhysRevC.83.045501} {\bibfield  {journal} {\bibinfo
  {journal} {Phys. Rev. C}\ }\textbf {\bibinfo {volume} {83}},\ \bibinfo
  {pages} {045501} (\bibinfo {year} {2011})},\ \Eprint
  {http://arxiv.org/abs/1102.2777} {arXiv:1102.2777 [hep-ph]} \BibitemShut
  {NoStop}%
\bibitem [{\citenamefont {Abe}\ \emph {et~al.}(2016{\natexlab{a}})\citenamefont
  {Abe} \emph {et~al.}}]{T2K:2016jor}%
  \BibitemOpen
  \bibfield  {author} {\bibinfo {author} {\bibfnamefont {K.}~\bibnamefont
  {Abe}} \emph {et~al.} (\bibinfo {collaboration} {T2K}),\ }\href {\doibase
  10.1103/PhysRevD.93.112012} {\bibfield  {journal} {\bibinfo  {journal} {Phys.
  Rev. D}\ }\textbf {\bibinfo {volume} {93}},\ \bibinfo {pages} {112012}
  (\bibinfo {year} {2016}{\natexlab{a}})},\ \Eprint
  {http://arxiv.org/abs/1602.03652} {arXiv:1602.03652 [hep-ex]} \BibitemShut
  {NoStop}%
\bibitem [{\citenamefont {Martini}\ \emph {et~al.}(2011)\citenamefont
  {Martini}, \citenamefont {Ericson},\ and\ \citenamefont
  {Chanfray}}]{Martini:2011wp}%
  \BibitemOpen
  \bibfield  {author} {\bibinfo {author} {\bibfnamefont {M.}~\bibnamefont
  {Martini}}, \bibinfo {author} {\bibfnamefont {M.}~\bibnamefont {Ericson}}, \
  and\ \bibinfo {author} {\bibfnamefont {G.}~\bibnamefont {Chanfray}},\ }\href
  {\doibase 10.1103/PhysRevC.84.055502} {\bibfield  {journal} {\bibinfo
  {journal} {Phys. Rev. C}\ }\textbf {\bibinfo {volume} {84}},\ \bibinfo
  {pages} {055502} (\bibinfo {year} {2011})},\ \Eprint
  {http://arxiv.org/abs/1110.0221} {arXiv:1110.0221 [nucl-th]} \BibitemShut
  {NoStop}%
\bibitem [{\citenamefont {Nieves}\ \emph {et~al.}(2012)\citenamefont {Nieves},
  \citenamefont {Ruiz~Simo},\ and\ \citenamefont
  {Vicente~Vacas}}]{Nieves:2011yp}%
  \BibitemOpen
  \bibfield  {author} {\bibinfo {author} {\bibfnamefont {J.}~\bibnamefont
  {Nieves}}, \bibinfo {author} {\bibfnamefont {I.}~\bibnamefont {Ruiz~Simo}}, \
  and\ \bibinfo {author} {\bibfnamefont {M.~J.}\ \bibnamefont
  {Vicente~Vacas}},\ }\href {\doibase 10.1016/j.physletb.2011.11.061}
  {\bibfield  {journal} {\bibinfo  {journal} {Phys. Lett. B}\ }\textbf
  {\bibinfo {volume} {707}},\ \bibinfo {pages} {72} (\bibinfo {year} {2012})},\
  \Eprint {http://arxiv.org/abs/1106.5374} {arXiv:1106.5374 [hep-ph]}
  \BibitemShut {NoStop}%
\bibitem [{\citenamefont {Aguilar-Arevalo}\ \emph {et~al.}(2010)\citenamefont
  {Aguilar-Arevalo} \emph {et~al.}}]{MiniBooNE:2010bsu}%
  \BibitemOpen
  \bibfield  {author} {\bibinfo {author} {\bibfnamefont {A.~A.}\ \bibnamefont
  {Aguilar-Arevalo}} \emph {et~al.} (\bibinfo {collaboration} {MiniBooNE}),\
  }\href {\doibase 10.1103/PhysRevD.81.092005} {\bibfield  {journal} {\bibinfo
  {journal} {Phys. Rev. D}\ }\textbf {\bibinfo {volume} {81}},\ \bibinfo
  {pages} {092005} (\bibinfo {year} {2010})},\ \Eprint
  {http://arxiv.org/abs/1002.2680} {arXiv:1002.2680 [hep-ex]} \BibitemShut
  {NoStop}%
\bibitem [{\citenamefont {Rodrigues}\ \emph {et~al.}(2016)\citenamefont
  {Rodrigues} \emph {et~al.}}]{MINERvA:2015ydy}%
  \BibitemOpen
  \bibfield  {author} {\bibinfo {author} {\bibfnamefont {P.~A.}\ \bibnamefont
  {Rodrigues}} \emph {et~al.} (\bibinfo {collaboration} {MINERvA}),\ }\href
  {\doibase 10.1103/PhysRevLett.116.071802} {\bibfield  {journal} {\bibinfo
  {journal} {Phys. Rev. Lett.}\ }\textbf {\bibinfo {volume} {116}},\ \bibinfo
  {pages} {071802} (\bibinfo {year} {2016})},\ \bibinfo {note} {[Addendum:
  Phys. Rev. Lett. {\bf 121}, 209902 (2018)]},\ \Eprint
  {http://arxiv.org/abs/1511.05944} {arXiv:1511.05944 [hep-ex]} \BibitemShut
  {NoStop}%
\bibitem [{\citenamefont {Acero}\ \emph {et~al.}(2020)\citenamefont {Acero}
  \emph {et~al.}}]{NOvA:2020rbg}%
  \BibitemOpen
  \bibfield  {author} {\bibinfo {author} {\bibfnamefont {M.~A.}\ \bibnamefont
  {Acero}} \emph {et~al.} (\bibinfo {collaboration} {NOvA, R. Group}),\ }\href
  {\doibase 10.1140/epjc/s10052-020-08577-5} {\bibfield  {journal} {\bibinfo
  {journal} {Eur. Phys. J. C}\ }\textbf {\bibinfo {volume} {80}},\ \bibinfo
  {pages} {1119} (\bibinfo {year} {2020})},\ \Eprint
  {http://arxiv.org/abs/2006.08727} {arXiv:2006.08727 [hep-ex]} \BibitemShut
  {NoStop}%
\bibitem [{\citenamefont {Nieves}\ \emph {et~al.}(2006)\citenamefont {Nieves},
  \citenamefont {Valverde},\ and\ \citenamefont
  {Vicente~Vacas}}]{Nieves:2005rq}%
  \BibitemOpen
  \bibfield  {author} {\bibinfo {author} {\bibfnamefont {J.}~\bibnamefont
  {Nieves}}, \bibinfo {author} {\bibfnamefont {M.}~\bibnamefont {Valverde}}, \
  and\ \bibinfo {author} {\bibfnamefont {M.~J.}\ \bibnamefont
  {Vicente~Vacas}},\ }\href {\doibase 10.1103/PhysRevC.73.025504} {\bibfield
  {journal} {\bibinfo  {journal} {Phys. Rev. C}\ }\textbf {\bibinfo {volume}
  {73}},\ \bibinfo {pages} {025504} (\bibinfo {year} {2006})},\ \Eprint
  {http://arxiv.org/abs/hep-ph/0511204} {arXiv:hep-ph/0511204} \BibitemShut
  {NoStop}%
\bibitem [{\citenamefont {Leitner}\ \emph {et~al.}(2006)\citenamefont
  {Leitner}, \citenamefont {Alvarez-Ruso},\ and\ \citenamefont
  {Mosel}}]{Leitner:2006ww}%
  \BibitemOpen
  \bibfield  {author} {\bibinfo {author} {\bibfnamefont {T.}~\bibnamefont
  {Leitner}}, \bibinfo {author} {\bibfnamefont {L.}~\bibnamefont
  {Alvarez-Ruso}}, \ and\ \bibinfo {author} {\bibfnamefont {U.}~\bibnamefont
  {Mosel}},\ }\href {\doibase 10.1103/PhysRevC.73.065502} {\bibfield  {journal}
  {\bibinfo  {journal} {Phys. Rev. C}\ }\textbf {\bibinfo {volume} {73}},\
  \bibinfo {pages} {065502} (\bibinfo {year} {2006})},\ \Eprint
  {http://arxiv.org/abs/nucl-th/0601103} {arXiv:nucl-th/0601103} \BibitemShut
  {NoStop}%
\bibitem [{\citenamefont {Sobczyk}\ \emph {et~al.}(2020)\citenamefont
  {Sobczyk}, \citenamefont {Nieves},\ and\ \citenamefont
  {S\'anchez}}]{Sobczyk:2020dkn}%
  \BibitemOpen
  \bibfield  {author} {\bibinfo {author} {\bibfnamefont {J.~E.}\ \bibnamefont
  {Sobczyk}}, \bibinfo {author} {\bibfnamefont {J.}~\bibnamefont {Nieves}}, \
  and\ \bibinfo {author} {\bibfnamefont {F.}~\bibnamefont {S\'anchez}},\ }\href
  {\doibase 10.1103/PhysRevC.102.024601} {\bibfield  {journal} {\bibinfo
  {journal} {Phys. Rev. C}\ }\textbf {\bibinfo {volume} {102}},\ \bibinfo
  {pages} {024601} (\bibinfo {year} {2020})},\ \Eprint
  {http://arxiv.org/abs/2002.08302} {arXiv:2002.08302 [nucl-th]} \BibitemShut
  {NoStop}%
\bibitem [{\citenamefont {Lalakulich}\ \emph {et~al.}(2010)\citenamefont
  {Lalakulich}, \citenamefont {Leitner}, \citenamefont {Buss},\ and\
  \citenamefont {Mosel}}]{Lalakulich:2010ss}%
  \BibitemOpen
  \bibfield  {author} {\bibinfo {author} {\bibfnamefont {O.}~\bibnamefont
  {Lalakulich}}, \bibinfo {author} {\bibfnamefont {T.}~\bibnamefont {Leitner}},
  \bibinfo {author} {\bibfnamefont {O.}~\bibnamefont {Buss}}, \ and\ \bibinfo
  {author} {\bibfnamefont {U.}~\bibnamefont {Mosel}},\ }\href {\doibase
  10.1103/PhysRevD.82.093001} {\bibfield  {journal} {\bibinfo  {journal} {Phys.
  Rev. D}\ }\textbf {\bibinfo {volume} {82}},\ \bibinfo {pages} {093001}
  (\bibinfo {year} {2010})},\ \Eprint {http://arxiv.org/abs/1007.0925}
  {arXiv:1007.0925 [hep-ph]} \BibitemShut {NoStop}%
\bibitem [{\citenamefont {Maieron}\ \emph {et~al.}(2003)\citenamefont
  {Maieron}, \citenamefont {Martinez}, \citenamefont {Caballero},\ and\
  \citenamefont {Udias}}]{Maieron03}%
  \BibitemOpen
  \bibfield  {author} {\bibinfo {author} {\bibfnamefont {C.}~\bibnamefont
  {Maieron}}, \bibinfo {author} {\bibfnamefont {M.~C.}\ \bibnamefont
  {Martinez}}, \bibinfo {author} {\bibfnamefont {J.~A.}\ \bibnamefont
  {Caballero}}, \ and\ \bibinfo {author} {\bibfnamefont {J.~M.}\ \bibnamefont
  {Udias}},\ }\href {\doibase 10.1103/PhysRevC.68.048501} {\bibfield  {journal}
  {\bibinfo  {journal} {Phys. Rev. C}\ }\textbf {\bibinfo {volume} {68}},\
  \bibinfo {pages} {048501} (\bibinfo {year} {2003})},\ \Eprint
  {http://arxiv.org/abs/nucl-th/0303075} {arXiv:nucl-th/0303075} \BibitemShut
  {NoStop}%
\bibitem [{\citenamefont {Meucci}\ \emph {et~al.}(2009)\citenamefont {Meucci},
  \citenamefont {Caballero}, \citenamefont {Giusti}, \citenamefont {Pacati},\
  and\ \citenamefont {Udias}}]{Meucci09}%
  \BibitemOpen
  \bibfield  {author} {\bibinfo {author} {\bibfnamefont {A.}~\bibnamefont
  {Meucci}}, \bibinfo {author} {\bibfnamefont {J.~A.}\ \bibnamefont
  {Caballero}}, \bibinfo {author} {\bibfnamefont {C.}~\bibnamefont {Giusti}},
  \bibinfo {author} {\bibfnamefont {F.~D.}\ \bibnamefont {Pacati}}, \ and\
  \bibinfo {author} {\bibfnamefont {J.~M.}\ \bibnamefont {Udias}},\ }\href
  {\doibase 10.1103/PhysRevC.80.024605} {\bibfield  {journal} {\bibinfo
  {journal} {Phys. Rev. C}\ }\textbf {\bibinfo {volume} {80}},\ \bibinfo
  {pages} {024605} (\bibinfo {year} {2009})},\ \Eprint
  {http://arxiv.org/abs/0906.2645} {arXiv:0906.2645 [nucl-th]} \BibitemShut
  {NoStop}%
\bibitem [{\citenamefont {Kim}\ and\ \citenamefont {Wright}(2007)}]{Kim07}%
  \BibitemOpen
  \bibfield  {author} {\bibinfo {author} {\bibfnamefont {K.~S.}\ \bibnamefont
  {Kim}}\ and\ \bibinfo {author} {\bibfnamefont {L.~E.}\ \bibnamefont
  {Wright}},\ }\href {\doibase 10.1103/PhysRevC.76.044613} {\bibfield
  {journal} {\bibinfo  {journal} {Phys. Rev. C}\ }\textbf {\bibinfo {volume}
  {76}},\ \bibinfo {pages} {044613} (\bibinfo {year} {2007})},\ \Eprint
  {http://arxiv.org/abs/0705.0049} {arXiv:0705.0049 [nucl-th]} \BibitemShut
  {NoStop}%
\bibitem [{\citenamefont {Butkevich}\ and\ \citenamefont
  {Kulagin}(2007)}]{Butkevich07}%
  \BibitemOpen
  \bibfield  {author} {\bibinfo {author} {\bibfnamefont {A.~V.}\ \bibnamefont
  {Butkevich}}\ and\ \bibinfo {author} {\bibfnamefont {S.~A.}\ \bibnamefont
  {Kulagin}},\ }\href {\doibase 10.1103/PhysRevC.76.045502} {\bibfield
  {journal} {\bibinfo  {journal} {Phys. Rev. C}\ }\textbf {\bibinfo {volume}
  {76}},\ \bibinfo {pages} {045502} (\bibinfo {year} {2007})},\ \Eprint
  {http://arxiv.org/abs/0705.1051} {arXiv:0705.1051 [nucl-th]} \BibitemShut
  {NoStop}%
\bibitem [{\citenamefont {Pandey}\ \emph {et~al.}(2016)\citenamefont {Pandey},
  \citenamefont {Jachowicz}, \citenamefont {Martini}, \citenamefont
  {Gonz\'alez-Jim\'enez}, \citenamefont {Ryckebusch}, \citenamefont
  {Van~Cuyck},\ and\ \citenamefont {Van~Dessel}}]{Pandey16}%
  \BibitemOpen
  \bibfield  {author} {\bibinfo {author} {\bibfnamefont {V.}~\bibnamefont
  {Pandey}}, \bibinfo {author} {\bibfnamefont {N.}~\bibnamefont {Jachowicz}},
  \bibinfo {author} {\bibfnamefont {M.}~\bibnamefont {Martini}}, \bibinfo
  {author} {\bibfnamefont {R.}~\bibnamefont {Gonz\'alez-Jim\'enez}}, \bibinfo
  {author} {\bibfnamefont {J.}~\bibnamefont {Ryckebusch}}, \bibinfo {author}
  {\bibfnamefont {T.}~\bibnamefont {Van~Cuyck}}, \ and\ \bibinfo {author}
  {\bibfnamefont {N.}~\bibnamefont {Van~Dessel}},\ }\href {\doibase
  10.1103/PhysRevC.94.054609} {\bibfield  {journal} {\bibinfo  {journal} {Phys.
  Rev. C}\ }\textbf {\bibinfo {volume} {94}},\ \bibinfo {pages} {054609}
  (\bibinfo {year} {2016})},\ \Eprint {http://arxiv.org/abs/1607.01216}
  {arXiv:1607.01216 [nucl-th]} \BibitemShut {NoStop}%
\bibitem [{\citenamefont {Gonz\'alez-Jim\'enez}\ \emph
  {et~al.}(2020)\citenamefont {Gonz\'alez-Jim\'enez}, \citenamefont {Barbaro},
  \citenamefont {Caballero}, \citenamefont {Donnelly}, \citenamefont
  {Jachowicz}, \citenamefont {Megias}, \citenamefont {Niewczas}, \citenamefont
  {Nikolakopoulos},\ and\ \citenamefont {Ud\'\i{}as}}]{Gonzalez-Jimenez20}%
  \BibitemOpen
  \bibfield  {author} {\bibinfo {author} {\bibfnamefont {R.}~\bibnamefont
  {Gonz\'alez-Jim\'enez}}, \bibinfo {author} {\bibfnamefont {M.~B.}\
  \bibnamefont {Barbaro}}, \bibinfo {author} {\bibfnamefont {J.~A.}\
  \bibnamefont {Caballero}}, \bibinfo {author} {\bibfnamefont {T.~W.}\
  \bibnamefont {Donnelly}}, \bibinfo {author} {\bibfnamefont {N.}~\bibnamefont
  {Jachowicz}}, \bibinfo {author} {\bibfnamefont {G.~D.}\ \bibnamefont
  {Megias}}, \bibinfo {author} {\bibfnamefont {K.}~\bibnamefont {Niewczas}},
  \bibinfo {author} {\bibfnamefont {A.}~\bibnamefont {Nikolakopoulos}}, \ and\
  \bibinfo {author} {\bibfnamefont {J.~M.}\ \bibnamefont {Ud\'\i{}as}},\ }\href
  {\doibase 10.1103/PhysRevC.101.015503} {\bibfield  {journal} {\bibinfo
  {journal} {Phys. Rev. C}\ }\textbf {\bibinfo {volume} {101}},\ \bibinfo
  {pages} {015503} (\bibinfo {year} {2020})},\ \Eprint
  {http://arxiv.org/abs/1909.07497} {arXiv:1909.07497 [nucl-th]} \BibitemShut
  {NoStop}%
\bibitem [{\citenamefont {Capuzzi}\ \emph {et~al.}(1991)\citenamefont
  {Capuzzi}, \citenamefont {Giusti},\ and\ \citenamefont {Pacati}}]{Capuzzi91}%
  \BibitemOpen
  \bibfield  {author} {\bibinfo {author} {\bibfnamefont {F.}~\bibnamefont
  {Capuzzi}}, \bibinfo {author} {\bibfnamefont {C.}~\bibnamefont {Giusti}}, \
  and\ \bibinfo {author} {\bibfnamefont {F.~D.}\ \bibnamefont {Pacati}},\
  }\href {\doibase 10.1016/0375-9474(91)90269-C} {\bibfield  {journal}
  {\bibinfo  {journal} {Nucl. Phys. A}\ }\textbf {\bibinfo {volume} {524}},\
  \bibinfo {pages} {681} (\bibinfo {year} {1991})}\BibitemShut {NoStop}%
\bibitem [{\citenamefont {Ivanov}\ \emph {et~al.}(2016)\citenamefont {Ivanov},
  \citenamefont {Vignote}, \citenamefont {\'Alvarez-Rodr\'\i{}guez},
  \citenamefont {Meucci}, \citenamefont {Giusti},\ and\ \citenamefont
  {Ud\'\i{}as}}]{Ivanov16b}%
  \BibitemOpen
  \bibfield  {author} {\bibinfo {author} {\bibfnamefont {M.~V.}\ \bibnamefont
  {Ivanov}}, \bibinfo {author} {\bibfnamefont {J.~R.}\ \bibnamefont {Vignote}},
  \bibinfo {author} {\bibfnamefont {R.}~\bibnamefont
  {\'Alvarez-Rodr\'\i{}guez}}, \bibinfo {author} {\bibfnamefont
  {A.}~\bibnamefont {Meucci}}, \bibinfo {author} {\bibfnamefont
  {C.}~\bibnamefont {Giusti}}, \ and\ \bibinfo {author} {\bibfnamefont {J.~M.}\
  \bibnamefont {Ud\'\i{}as}},\ }\href {\doibase 10.1103/PhysRevC.94.014608}
  {\bibfield  {journal} {\bibinfo  {journal} {Phys. Rev. C}\ }\textbf {\bibinfo
  {volume} {94}},\ \bibinfo {pages} {014608} (\bibinfo {year}
  {2016})}\BibitemShut {NoStop}%
\bibitem [{\citenamefont {Udias}\ \emph {et~al.}(1993)\citenamefont {Udias},
  \citenamefont {Sarriguren}, \citenamefont {Moya~de Guerra}, \citenamefont
  {Garrido},\ and\ \citenamefont {Caballero}}]{Udias93}%
  \BibitemOpen
  \bibfield  {author} {\bibinfo {author} {\bibfnamefont {J.~M.}\ \bibnamefont
  {Udias}}, \bibinfo {author} {\bibfnamefont {P.}~\bibnamefont {Sarriguren}},
  \bibinfo {author} {\bibfnamefont {E.}~\bibnamefont {Moya~de Guerra}},
  \bibinfo {author} {\bibfnamefont {E.}~\bibnamefont {Garrido}}, \ and\
  \bibinfo {author} {\bibfnamefont {J.~A.}\ \bibnamefont {Caballero}},\ }\href
  {\doibase 10.1103/PhysRevC.48.2731} {\bibfield  {journal} {\bibinfo
  {journal} {Phys. Rev. C}\ }\textbf {\bibinfo {volume} {48}},\ \bibinfo
  {pages} {2731} (\bibinfo {year} {1993})},\ \Eprint
  {http://arxiv.org/abs/nucl-th/9310004} {arXiv:nucl-th/9310004} \BibitemShut
  {NoStop}%
\bibitem [{\citenamefont {Udias}\ \emph {et~al.}(2001)\citenamefont {Udias},
  \citenamefont {Caballero}, \citenamefont {Moya~de Guerra}, \citenamefont
  {Vignote},\ and\ \citenamefont {Escuderos}}]{Udias01}%
  \BibitemOpen
  \bibfield  {author} {\bibinfo {author} {\bibfnamefont {J.~M.}\ \bibnamefont
  {Udias}}, \bibinfo {author} {\bibfnamefont {J.~A.}\ \bibnamefont
  {Caballero}}, \bibinfo {author} {\bibfnamefont {E.}~\bibnamefont {Moya~de
  Guerra}}, \bibinfo {author} {\bibfnamefont {J.~R.}\ \bibnamefont {Vignote}},
  \ and\ \bibinfo {author} {\bibfnamefont {A.}~\bibnamefont {Escuderos}},\
  }\href {\doibase 10.1103/PhysRevC.64.024614} {\bibfield  {journal} {\bibinfo
  {journal} {Phys. Rev. C}\ }\textbf {\bibinfo {volume} {64}},\ \bibinfo
  {pages} {024614} (\bibinfo {year} {2001})},\ \Eprint
  {http://arxiv.org/abs/nucl-th/0101038} {arXiv:nucl-th/0101038} \BibitemShut
  {NoStop}%
\bibitem [{\citenamefont {Cooper}\ \emph {et~al.}(1993)\citenamefont {Cooper},
  \citenamefont {Hama}, \citenamefont {Clark},\ and\ \citenamefont
  {Mercer}}]{Cooper93}%
  \BibitemOpen
  \bibfield  {author} {\bibinfo {author} {\bibfnamefont {E.~D.}\ \bibnamefont
  {Cooper}}, \bibinfo {author} {\bibfnamefont {S.}~\bibnamefont {Hama}},
  \bibinfo {author} {\bibfnamefont {B.~C.}\ \bibnamefont {Clark}}, \ and\
  \bibinfo {author} {\bibfnamefont {R.~L.}\ \bibnamefont {Mercer}},\ }\href
  {\doibase 10.1103/PhysRevC.47.297} {\bibfield  {journal} {\bibinfo  {journal}
  {Phys. Rev. C}\ }\textbf {\bibinfo {volume} {47}},\ \bibinfo {pages} {297}
  (\bibinfo {year} {1993})}\BibitemShut {NoStop}%
\bibitem [{\citenamefont {Cooper}\ \emph {et~al.}(2009)\citenamefont {Cooper},
  \citenamefont {Hama},\ and\ \citenamefont {Clark}}]{Cooper09}%
  \BibitemOpen
  \bibfield  {author} {\bibinfo {author} {\bibfnamefont {E.~D.}\ \bibnamefont
  {Cooper}}, \bibinfo {author} {\bibfnamefont {S.}~\bibnamefont {Hama}}, \ and\
  \bibinfo {author} {\bibfnamefont {B.~C.}\ \bibnamefont {Clark}},\ }\href
  {\doibase 10.1103/PhysRevC.80.034605} {\bibfield  {journal} {\bibinfo
  {journal} {Phys. Rev. C}\ }\textbf {\bibinfo {volume} {80}},\ \bibinfo
  {pages} {034605} (\bibinfo {year} {2009})}\BibitemShut {NoStop}%
\bibitem [{\citenamefont {Gonz\'alez-Jim\'enez}\ \emph
  {et~al.}(2019)\citenamefont {Gonz\'alez-Jim\'enez}, \citenamefont
  {Nikolakopoulos}, \citenamefont {Jachowicz},\ and\ \citenamefont
  {Ud\'\i{}as}}]{Gonzalez-Jimenez19}%
  \BibitemOpen
  \bibfield  {author} {\bibinfo {author} {\bibfnamefont {R.}~\bibnamefont
  {Gonz\'alez-Jim\'enez}}, \bibinfo {author} {\bibfnamefont {A.}~\bibnamefont
  {Nikolakopoulos}}, \bibinfo {author} {\bibfnamefont {N.}~\bibnamefont
  {Jachowicz}}, \ and\ \bibinfo {author} {\bibfnamefont {J.~M.}\ \bibnamefont
  {Ud\'\i{}as}},\ }\href {\doibase 10.1103/PhysRevC.100.045501} {\bibfield
  {journal} {\bibinfo  {journal} {Phys. Rev. C}\ }\textbf {\bibinfo {volume}
  {100}},\ \bibinfo {pages} {045501} (\bibinfo {year} {2019})},\ \Eprint
  {http://arxiv.org/abs/1904.10696} {arXiv:1904.10696 [nucl-th]} \BibitemShut
  {NoStop}%
\bibitem [{\citenamefont {Nikolakopoulos}\ \emph {et~al.}(2019)\citenamefont
  {Nikolakopoulos}, \citenamefont {Jachowicz}, \citenamefont {Van~Dessel},
  \citenamefont {Niewczas}, \citenamefont {Gonz\'alez-Jim\'enez}, \citenamefont
  {Ud\'\i{}as},\ and\ \citenamefont {Pandey}}]{Nikolakopoulos19}%
  \BibitemOpen
  \bibfield  {author} {\bibinfo {author} {\bibfnamefont {A.}~\bibnamefont
  {Nikolakopoulos}}, \bibinfo {author} {\bibfnamefont {N.}~\bibnamefont
  {Jachowicz}}, \bibinfo {author} {\bibfnamefont {N.}~\bibnamefont
  {Van~Dessel}}, \bibinfo {author} {\bibfnamefont {K.}~\bibnamefont
  {Niewczas}}, \bibinfo {author} {\bibfnamefont {R.}~\bibnamefont
  {Gonz\'alez-Jim\'enez}}, \bibinfo {author} {\bibfnamefont {J.~M.}\
  \bibnamefont {Ud\'\i{}as}}, \ and\ \bibinfo {author} {\bibfnamefont
  {V.}~\bibnamefont {Pandey}},\ }\href {\doibase
  10.1103/PhysRevLett.123.052501} {\bibfield  {journal} {\bibinfo  {journal}
  {Phys. Rev. Lett.}\ }\textbf {\bibinfo {volume} {123}},\ \bibinfo {pages}
  {052501} (\bibinfo {year} {2019})},\ \Eprint
  {http://arxiv.org/abs/1901.08050} {arXiv:1901.08050 [nucl-th]} \BibitemShut
  {NoStop}%
\bibitem [{\citenamefont {Gonz\'alez-Jim\'enez}\ \emph
  {et~al.}(2022)\citenamefont {Gonz\'alez-Jim\'enez}, \citenamefont {Barbaro},
  \citenamefont {Caballero}, \citenamefont {Donnelly}, \citenamefont
  {Jachowicz}, \citenamefont {Megias}, \citenamefont {Niewczas}, \citenamefont
  {Nikolakopoulos}, \citenamefont {Van~Orden},\ and\ \citenamefont
  {Ud\'\i{}as}}]{Gonzalez-Jimenez:2021ohu}%
  \BibitemOpen
  \bibfield  {author} {\bibinfo {author} {\bibfnamefont {R.}~\bibnamefont
  {Gonz\'alez-Jim\'enez}}, \bibinfo {author} {\bibfnamefont {M.~B.}\
  \bibnamefont {Barbaro}}, \bibinfo {author} {\bibfnamefont {J.~A.}\
  \bibnamefont {Caballero}}, \bibinfo {author} {\bibfnamefont {T.~W.}\
  \bibnamefont {Donnelly}}, \bibinfo {author} {\bibfnamefont {N.}~\bibnamefont
  {Jachowicz}}, \bibinfo {author} {\bibfnamefont {G.~D.}\ \bibnamefont
  {Megias}}, \bibinfo {author} {\bibfnamefont {K.}~\bibnamefont {Niewczas}},
  \bibinfo {author} {\bibfnamefont {A.}~\bibnamefont {Nikolakopoulos}},
  \bibinfo {author} {\bibfnamefont {J.~W.}\ \bibnamefont {Van~Orden}}, \ and\
  \bibinfo {author} {\bibfnamefont {J.~M.}\ \bibnamefont {Ud\'\i{}as}},\ }\href
  {\doibase 10.1103/PhysRevC.105.025502} {\bibfield  {journal} {\bibinfo
  {journal} {Phys. Rev. C}\ }\textbf {\bibinfo {volume} {105}},\ \bibinfo
  {pages} {025502} (\bibinfo {year} {2022})},\ \Eprint
  {http://arxiv.org/abs/2104.01701} {arXiv:2104.01701 [nucl-th]} \BibitemShut
  {NoStop}%
\bibitem [{\citenamefont {Nikolakopoulos}\ \emph {et~al.}(2022)\citenamefont
  {Nikolakopoulos}, \citenamefont {Gonz\'alez-Jim\'enez}, \citenamefont
  {Jachowicz}, \citenamefont {Niewczas}, \citenamefont {S\'anchez},\ and\
  \citenamefont {Ud\'\i{}as}}]{Nikolakopoulos22}%
  \BibitemOpen
  \bibfield  {author} {\bibinfo {author} {\bibfnamefont {A.}~\bibnamefont
  {Nikolakopoulos}}, \bibinfo {author} {\bibfnamefont {R.}~\bibnamefont
  {Gonz\'alez-Jim\'enez}}, \bibinfo {author} {\bibfnamefont {N.}~\bibnamefont
  {Jachowicz}}, \bibinfo {author} {\bibfnamefont {K.}~\bibnamefont {Niewczas}},
  \bibinfo {author} {\bibfnamefont {F.}~\bibnamefont {S\'anchez}}, \ and\
  \bibinfo {author} {\bibfnamefont {J.~M.}\ \bibnamefont {Ud\'\i{}as}},\
  }\href@noop {} {\  (\bibinfo {year} {2022})},\ \Eprint
  {http://arxiv.org/abs/2202.01689} {arXiv:2202.01689 [nucl-th]} \BibitemShut
  {NoStop}%
\bibitem [{\citenamefont {Jachowicz}\ and\ \citenamefont
  {Nikolakopoulos}(2021)}]{Jachowicz:2021ieb}%
  \BibitemOpen
  \bibfield  {author} {\bibinfo {author} {\bibfnamefont {N.}~\bibnamefont
  {Jachowicz}}\ and\ \bibinfo {author} {\bibfnamefont {A.}~\bibnamefont
  {Nikolakopoulos}},\ }\href {\doibase 10.1140/epjs/s11734-021-00286-8}
  {\bibfield  {journal} {\bibinfo  {journal} {Eur. Phys. J. ST}\ }\textbf
  {\bibinfo {volume} {230}},\ \bibinfo {pages} {4339} (\bibinfo {year}
  {2021})},\ \Eprint {http://arxiv.org/abs/2110.11321} {arXiv:2110.11321
  [nucl-th]} \BibitemShut {NoStop}%
\bibitem [{\citenamefont {Amaro}\ \emph {et~al.}(2020)\citenamefont {Amaro},
  \citenamefont {Barbaro}, \citenamefont {Caballero}, \citenamefont
  {Gonz\'alez-Jim\'enez}, \citenamefont {Megias},\ and\ \citenamefont
  {Ruiz~Simo}}]{Amaro20}%
  \BibitemOpen
  \bibfield  {author} {\bibinfo {author} {\bibfnamefont {J.~E.}\ \bibnamefont
  {Amaro}}, \bibinfo {author} {\bibfnamefont {M.~B.}\ \bibnamefont {Barbaro}},
  \bibinfo {author} {\bibfnamefont {J.~A.}\ \bibnamefont {Caballero}}, \bibinfo
  {author} {\bibfnamefont {R.}~\bibnamefont {Gonz\'alez-Jim\'enez}}, \bibinfo
  {author} {\bibfnamefont {G.~D.}\ \bibnamefont {Megias}}, \ and\ \bibinfo
  {author} {\bibfnamefont {I.}~\bibnamefont {Ruiz~Simo}},\ }\href {\doibase
  10.1088/1361-6471/abb128} {\bibfield  {journal} {\bibinfo  {journal} {J.
  Phys. G}\ }\textbf {\bibinfo {volume} {47}},\ \bibinfo {pages} {124001}
  (\bibinfo {year} {2020})},\ \Eprint {http://arxiv.org/abs/1912.10612}
  {arXiv:1912.10612 [nucl-th]} \BibitemShut {NoStop}%
\bibitem [{\citenamefont {Gonzal\'ez-Jim\'enez}\ \emph
  {et~al.}(2014)\citenamefont {Gonzal\'ez-Jim\'enez}, \citenamefont {Megias},
  \citenamefont {Barbaro}, \citenamefont {Caballero},\ and\ \citenamefont
  {Donnelly}}]{Gonzalez-Jimenez:2014eqa}%
  \BibitemOpen
  \bibfield  {author} {\bibinfo {author} {\bibfnamefont {R.}~\bibnamefont
  {Gonzal\'ez-Jim\'enez}}, \bibinfo {author} {\bibfnamefont {G.~D.}\
  \bibnamefont {Megias}}, \bibinfo {author} {\bibfnamefont {M.~B.}\
  \bibnamefont {Barbaro}}, \bibinfo {author} {\bibfnamefont {J.~A.}\
  \bibnamefont {Caballero}}, \ and\ \bibinfo {author} {\bibfnamefont {T.~W.}\
  \bibnamefont {Donnelly}},\ }\href {\doibase 10.1103/PhysRevC.90.035501}
  {\bibfield  {journal} {\bibinfo  {journal} {Phys. Rev. C}\ }\textbf {\bibinfo
  {volume} {90}},\ \bibinfo {pages} {035501} (\bibinfo {year} {2014})},\
  \Eprint {http://arxiv.org/abs/1407.8346} {arXiv:1407.8346 [nucl-th]}
  \BibitemShut {NoStop}%
\bibitem [{\citenamefont {Megias}\ \emph {et~al.}(2016)\citenamefont {Megias},
  \citenamefont {Amaro}, \citenamefont {Barbaro}, \citenamefont {Caballero},\
  and\ \citenamefont {Donnelly}}]{Megias16a}%
  \BibitemOpen
  \bibfield  {author} {\bibinfo {author} {\bibfnamefont {G.~D.}\ \bibnamefont
  {Megias}}, \bibinfo {author} {\bibfnamefont {J.~E.}\ \bibnamefont {Amaro}},
  \bibinfo {author} {\bibfnamefont {M.~B.}\ \bibnamefont {Barbaro}}, \bibinfo
  {author} {\bibfnamefont {J.~A.}\ \bibnamefont {Caballero}}, \ and\ \bibinfo
  {author} {\bibfnamefont {T.~W.}\ \bibnamefont {Donnelly}},\ }\href {\doibase
  10.1103/PhysRevD.94.013012} {\bibfield  {journal} {\bibinfo  {journal} {Phys.
  Rev. D}\ }\textbf {\bibinfo {volume} {94}},\ \bibinfo {pages} {013012}
  (\bibinfo {year} {2016})},\ \Eprint {http://arxiv.org/abs/1603.08396}
  {arXiv:1603.08396 [nucl-th]} \BibitemShut {NoStop}%
\bibitem [{\citenamefont {Amaro}\ \emph {et~al.}(2021)\citenamefont {Amaro},
  \citenamefont {Barbaro}, \citenamefont {Caballero}, \citenamefont {Donnelly},
  \citenamefont {González-Jiménez}, \citenamefont {Megias},\ and\
  \citenamefont {Simo}}]{Amaro21}%
  \BibitemOpen
  \bibfield  {author} {\bibinfo {author} {\bibfnamefont {J.~E.}\ \bibnamefont
  {Amaro}}, \bibinfo {author} {\bibfnamefont {M.~B.}\ \bibnamefont {Barbaro}},
  \bibinfo {author} {\bibfnamefont {J.~A.}\ \bibnamefont {Caballero}}, \bibinfo
  {author} {\bibfnamefont {T.~W.}\ \bibnamefont {Donnelly}}, \bibinfo {author}
  {\bibfnamefont {R.}~\bibnamefont {González-Jiménez}}, \bibinfo {author}
  {\bibfnamefont {G.~D.}\ \bibnamefont {Megias}}, \ and\ \bibinfo {author}
  {\bibfnamefont {I.~R.}\ \bibnamefont {Simo}},\ }\href {\doibase
  10.1140/epjs/s11734-021-00289-5} {\bibfield  {journal} {\bibinfo  {journal}
  {Eur. Phys. J. ST}\ }\textbf {\bibinfo {volume} {230}},\ \bibinfo {pages}
  {4321} (\bibinfo {year} {2021})},\ \Eprint {http://arxiv.org/abs/2106.02857}
  {arXiv:2106.02857 [hep-ph]} \BibitemShut {NoStop}%
\bibitem [{\citenamefont {Melendez}\ \emph {et~al.}(2022)\citenamefont
  {Melendez}, \citenamefont {Drischler}, \citenamefont {Furnstahl},
  \citenamefont {Garcia},\ and\ \citenamefont {Zhang}}]{Melendez:2022kid}%
  \BibitemOpen
  \bibfield  {author} {\bibinfo {author} {\bibfnamefont {J.~A.}\ \bibnamefont
  {Melendez}}, \bibinfo {author} {\bibfnamefont {C.}~\bibnamefont {Drischler}},
  \bibinfo {author} {\bibfnamefont {R.~J.}\ \bibnamefont {Furnstahl}}, \bibinfo
  {author} {\bibfnamefont {A.~J.}\ \bibnamefont {Garcia}}, \ and\ \bibinfo
  {author} {\bibfnamefont {X.}~\bibnamefont {Zhang}},\ }\href@noop {} {\
  (\bibinfo {year} {2022})},\ \Eprint {http://arxiv.org/abs/2203.05528}
  {arXiv:2203.05528 [nucl-th]} \BibitemShut {NoStop}%
\bibitem [{\citenamefont {Frame}\ \emph {et~al.}(2018)\citenamefont {Frame},
  \citenamefont {He}, \citenamefont {Ipsen}, \citenamefont {Lee}, \citenamefont
  {Lee},\ and\ \citenamefont {Rrapaj}}]{Frame:2017fah}%
  \BibitemOpen
  \bibfield  {author} {\bibinfo {author} {\bibfnamefont {D.}~\bibnamefont
  {Frame}}, \bibinfo {author} {\bibfnamefont {R.}~\bibnamefont {He}}, \bibinfo
  {author} {\bibfnamefont {I.}~\bibnamefont {Ipsen}}, \bibinfo {author}
  {\bibfnamefont {D.}~\bibnamefont {Lee}}, \bibinfo {author} {\bibfnamefont
  {D.}~\bibnamefont {Lee}}, \ and\ \bibinfo {author} {\bibfnamefont
  {E.}~\bibnamefont {Rrapaj}},\ }\href {\doibase
  10.1103/PhysRevLett.121.032501} {\bibfield  {journal} {\bibinfo  {journal}
  {Phys. Rev. Lett.}\ }\textbf {\bibinfo {volume} {121}},\ \bibinfo {pages}
  {032501} (\bibinfo {year} {2018})},\ \Eprint
  {http://arxiv.org/abs/1711.07090} {arXiv:1711.07090} \BibitemShut {NoStop}%
\bibitem [{\citenamefont {Ekström}\ and\ \citenamefont
  {Hagen}(2019)}]{Ekstrom:2019lss}%
  \BibitemOpen
  \bibfield  {author} {\bibinfo {author} {\bibfnamefont {A.}~\bibnamefont
  {Ekström}}\ and\ \bibinfo {author} {\bibfnamefont {G.}~\bibnamefont
  {Hagen}},\ }\href {\doibase 10.1103/PhysRevLett.123.252501} {\bibfield
  {journal} {\bibinfo  {journal} {Phys. Rev. Lett.}\ }\textbf {\bibinfo
  {volume} {123}},\ \bibinfo {pages} {252501} (\bibinfo {year} {2019})},\
  \Eprint {http://arxiv.org/abs/1910.02922} {arXiv:1910.02922 [nucl-th]}
  \BibitemShut {NoStop}%
\bibitem [{\citenamefont {K\"onig}\ \emph {et~al.}(2020)\citenamefont
  {K\"onig}, \citenamefont {Ekstr\"om}, \citenamefont {Hebeler}, \citenamefont
  {Lee},\ and\ \citenamefont {Schwenk}}]{Konig:2019adq}%
  \BibitemOpen
  \bibfield  {author} {\bibinfo {author} {\bibfnamefont {S.}~\bibnamefont
  {K\"onig}}, \bibinfo {author} {\bibfnamefont {A.}~\bibnamefont {Ekstr\"om}},
  \bibinfo {author} {\bibfnamefont {K.}~\bibnamefont {Hebeler}}, \bibinfo
  {author} {\bibfnamefont {D.}~\bibnamefont {Lee}}, \ and\ \bibinfo {author}
  {\bibfnamefont {A.}~\bibnamefont {Schwenk}},\ }\href {\doibase
  10.1016/j.physletb.2020.135814} {\bibfield  {journal} {\bibinfo  {journal}
  {Phys. Lett. B}\ }\textbf {\bibinfo {volume} {810}},\ \bibinfo {pages}
  {135814} (\bibinfo {year} {2020})},\ \Eprint
  {http://arxiv.org/abs/1909.08446} {arXiv:1909.08446 [nucl-th]} \BibitemShut
  {NoStop}%
\bibitem [{\citenamefont {Yoshida}\ and\ \citenamefont
  {Shimizu}(2021)}]{Yoshida:2021jbl}%
  \BibitemOpen
  \bibfield  {author} {\bibinfo {author} {\bibfnamefont {S.}~\bibnamefont
  {Yoshida}}\ and\ \bibinfo {author} {\bibfnamefont {N.}~\bibnamefont
  {Shimizu}},\ }\href@noop {} {\  (\bibinfo {year} {2021})},\ \Eprint
  {http://arxiv.org/abs/2105.08256} {arXiv:2105.08256 [nucl-th]} \BibitemShut
  {NoStop}%
\bibitem [{\citenamefont {Hu}\ \emph {et~al.}(2021)\citenamefont {Hu} \emph
  {et~al.}}]{Hu:2021trw}%
  \BibitemOpen
  \bibfield  {author} {\bibinfo {author} {\bibfnamefont {B.}~\bibnamefont {Hu}}
  \emph {et~al.},\ }\href@noop {} {\  (\bibinfo {year} {2021})},\ \Eprint
  {http://arxiv.org/abs/2112.01125} {arXiv:2112.01125 [nucl-th]} \BibitemShut
  {NoStop}%
\bibitem [{\citenamefont {Bonilla}\ \emph {et~al.}(2022)\citenamefont
  {Bonilla}, \citenamefont {Giuliani}, \citenamefont {Godbey},\ and\
  \citenamefont {Lee}}]{Bonilla:2022rph}%
  \BibitemOpen
  \bibfield  {author} {\bibinfo {author} {\bibfnamefont {E.}~\bibnamefont
  {Bonilla}}, \bibinfo {author} {\bibfnamefont {P.}~\bibnamefont {Giuliani}},
  \bibinfo {author} {\bibfnamefont {K.}~\bibnamefont {Godbey}}, \ and\ \bibinfo
  {author} {\bibfnamefont {D.}~\bibnamefont {Lee}},\ }\href@noop {} {\
  (\bibinfo {year} {2022})},\ \Eprint {http://arxiv.org/abs/2203.05284}
  {arXiv:2203.05284 [nucl-th]} \BibitemShut {NoStop}%
\bibitem [{\citenamefont {Furnstahl}\ \emph {et~al.}(2020)\citenamefont
  {Furnstahl}, \citenamefont {Garcia}, \citenamefont {Millican},\ and\
  \citenamefont {Zhang}}]{Furnstahl:2020abp}%
  \BibitemOpen
  \bibfield  {author} {\bibinfo {author} {\bibfnamefont {R.~J.}\ \bibnamefont
  {Furnstahl}}, \bibinfo {author} {\bibfnamefont {A.~J.}\ \bibnamefont
  {Garcia}}, \bibinfo {author} {\bibfnamefont {P.~J.}\ \bibnamefont
  {Millican}}, \ and\ \bibinfo {author} {\bibfnamefont {X.}~\bibnamefont
  {Zhang}},\ }\href {\doibase 10.1016/j.physletb.2020.135719} {\bibfield
  {journal} {\bibinfo  {journal} {Phys. Lett. B}\ }\textbf {\bibinfo {volume}
  {809}},\ \bibinfo {pages} {135719} (\bibinfo {year} {2020})},\ \Eprint
  {http://arxiv.org/abs/2007.03635} {arXiv:2007.03635 [nucl-th]} \BibitemShut
  {NoStop}%
\bibitem [{\citenamefont {Bai}\ and\ \citenamefont {Ren}(2021)}]{Bai:2021xok}%
  \BibitemOpen
  \bibfield  {author} {\bibinfo {author} {\bibfnamefont {D.}~\bibnamefont
  {Bai}}\ and\ \bibinfo {author} {\bibfnamefont {Z.}~\bibnamefont {Ren}},\
  }\href {\doibase 10.1103/PhysRevC.103.014612} {\bibfield  {journal} {\bibinfo
   {journal} {Phys. Rev. C}\ }\textbf {\bibinfo {volume} {103}},\ \bibinfo
  {pages} {014612} (\bibinfo {year} {2021})},\ \Eprint
  {http://arxiv.org/abs/2101.06336} {arXiv:2101.06336 [nucl-th]} \BibitemShut
  {NoStop}%
\bibitem [{\citenamefont {Drischler}\ \emph
  {et~al.}(2021{\natexlab{b}})\citenamefont {Drischler}, \citenamefont
  {Quinonez}, \citenamefont {Giuliani}, \citenamefont {Lovell},\ and\
  \citenamefont {Nunes}}]{Drischler:2021qoy}%
  \BibitemOpen
  \bibfield  {author} {\bibinfo {author} {\bibfnamefont {C.}~\bibnamefont
  {Drischler}}, \bibinfo {author} {\bibfnamefont {M.}~\bibnamefont {Quinonez}},
  \bibinfo {author} {\bibfnamefont {P.~G.}\ \bibnamefont {Giuliani}}, \bibinfo
  {author} {\bibfnamefont {A.~E.}\ \bibnamefont {Lovell}}, \ and\ \bibinfo
  {author} {\bibfnamefont {F.~M.}\ \bibnamefont {Nunes}},\ }\href {\doibase
  10.1016/j.physletb.2021.136777} {\bibfield  {journal} {\bibinfo  {journal}
  {Phys. Lett. B}\ }\textbf {\bibinfo {volume} {823}},\ \bibinfo {pages}
  {136777} (\bibinfo {year} {2021}{\natexlab{b}})},\ \Eprint
  {http://arxiv.org/abs/2108.08269} {arXiv:2108.08269 [nucl-th]} \BibitemShut
  {NoStop}%
\bibitem [{\citenamefont {Melendez}\ \emph {et~al.}(2021)\citenamefont
  {Melendez}, \citenamefont {Drischler}, \citenamefont {Garcia}, \citenamefont
  {Furnstahl},\ and\ \citenamefont {Zhang}}]{Melendez:2021lyq}%
  \BibitemOpen
  \bibfield  {author} {\bibinfo {author} {\bibfnamefont {J.}~\bibnamefont
  {Melendez}}, \bibinfo {author} {\bibfnamefont {C.}~\bibnamefont {Drischler}},
  \bibinfo {author} {\bibfnamefont {A.}~\bibnamefont {Garcia}}, \bibinfo
  {author} {\bibfnamefont {R.}~\bibnamefont {Furnstahl}}, \ and\ \bibinfo
  {author} {\bibfnamefont {X.}~\bibnamefont {Zhang}},\ }\href {\doibase
  https://doi.org/10.1016/j.physletb.2021.136608} {\bibfield  {journal}
  {\bibinfo  {journal} {Phys. Lett. B}\ }\textbf {\bibinfo {volume} {821}},\
  \bibinfo {pages} {136608} (\bibinfo {year} {2021})}\BibitemShut {NoStop}%
\bibitem [{\citenamefont {Zhang}\ and\ \citenamefont
  {Furnstahl}(2021)}]{Zhang:2021jmi}%
  \BibitemOpen
  \bibfield  {author} {\bibinfo {author} {\bibfnamefont {X.}~\bibnamefont
  {Zhang}}\ and\ \bibinfo {author} {\bibfnamefont {R.~J.}\ \bibnamefont
  {Furnstahl}},\ }\href@noop {} {\  (\bibinfo {year} {2021})},\ \Eprint
  {http://arxiv.org/abs/2110.04269} {arXiv:2110.04269 [nucl-th]} \BibitemShut
  {NoStop}%
\bibitem [{\citenamefont {Nakamura}\ and\ \citenamefont
  {Zhang}(2022)}]{Nakamura2022}%
  \BibitemOpen
  \bibfield  {author} {\bibinfo {author} {\bibfnamefont {S.}~\bibnamefont
  {Nakamura}}\ and\ \bibinfo {author} {\bibfnamefont {X.}~\bibnamefont
  {Zhang}}\ }(\bibinfo {year} {2022})\ \bibinfo {note} {\emph{in
  preparation}}\BibitemShut {NoStop}%
\bibitem [{\citenamefont {Phillips}\ \emph {et~al.}(2021)\citenamefont
  {Phillips} \emph {et~al.}}]{Phillips:2020dmw}%
  \BibitemOpen
  \bibfield  {author} {\bibinfo {author} {\bibfnamefont {D.~R.}\ \bibnamefont
  {Phillips}} \emph {et~al.},\ }\href {\doibase 10.1088/1361-6471/abf1df}
  {\bibfield  {journal} {\bibinfo  {journal} {J. Phys. G}\ }\textbf {\bibinfo
  {volume} {48}},\ \bibinfo {pages} {072001} (\bibinfo {year} {2021})},\
  \Eprint {http://arxiv.org/abs/2012.07704} {arXiv:2012.07704 [nucl-th]}
  \BibitemShut {NoStop}%
\bibitem [{\citenamefont {{{Bayesian Analysis of Nuclear Dynamics (BAND)
  Framework project}}}(2020)}]{BAND_Framework}%
  \BibitemOpen
  \bibfield  {author} {\bibinfo {author} {\bibnamefont {{{Bayesian Analysis of
  Nuclear Dynamics (BAND) Framework project}}}}\ }(\bibinfo {year} {2020})\
  \bibinfo {note} {\url{https://bandframework.github.io/}}\BibitemShut
  {NoStop}%
\bibitem [{\citenamefont {Yao}\ \emph {et~al.}(2018)\citenamefont {Yao},
  \citenamefont {Alvarez-Ruso}, \citenamefont {Hiller~Blin},\ and\
  \citenamefont {Vicente~Vacas}}]{Yao:2018pzc}%
  \BibitemOpen
  \bibfield  {author} {\bibinfo {author} {\bibfnamefont {D.-L.}\ \bibnamefont
  {Yao}}, \bibinfo {author} {\bibfnamefont {L.}~\bibnamefont {Alvarez-Ruso}},
  \bibinfo {author} {\bibfnamefont {A.~N.}\ \bibnamefont {Hiller~Blin}}, \ and\
  \bibinfo {author} {\bibfnamefont {M.~J.}\ \bibnamefont {Vicente~Vacas}},\
  }\href {\doibase 10.1103/PhysRevD.98.076004} {\bibfield  {journal} {\bibinfo
  {journal} {Phys. Rev. D}\ }\textbf {\bibinfo {volume} {98}},\ \bibinfo
  {pages} {076004} (\bibinfo {year} {2018})},\ \Eprint
  {http://arxiv.org/abs/1806.09364} {arXiv:1806.09364 [hep-ph]} \BibitemShut
  {NoStop}%
\bibitem [{\citenamefont {Yao}\ \emph {et~al.}(2019)\citenamefont {Yao},
  \citenamefont {Alvarez-Ruso},\ and\ \citenamefont
  {Vicente~Vacas}}]{Yao:2019avf}%
  \BibitemOpen
  \bibfield  {author} {\bibinfo {author} {\bibfnamefont {D.-L.}\ \bibnamefont
  {Yao}}, \bibinfo {author} {\bibfnamefont {L.}~\bibnamefont {Alvarez-Ruso}}, \
  and\ \bibinfo {author} {\bibfnamefont {M.~J.}\ \bibnamefont
  {Vicente~Vacas}},\ }\href {\doibase 10.1016/j.physletb.2019.05.036}
  {\bibfield  {journal} {\bibinfo  {journal} {Phys. Lett. B}\ }\textbf
  {\bibinfo {volume} {794}},\ \bibinfo {pages} {109} (\bibinfo {year}
  {2019})},\ \Eprint {http://arxiv.org/abs/1901.00773} {arXiv:1901.00773
  [hep-ph]} \BibitemShut {NoStop}%
\bibitem [{\citenamefont {Guerrero~Navarro}\ and\ \citenamefont
  {Vicente~Vacas}(2020)}]{GuerreroNavarro:2020kwb}%
  \BibitemOpen
  \bibfield  {author} {\bibinfo {author} {\bibfnamefont {G.~H.}\ \bibnamefont
  {Guerrero~Navarro}}\ and\ \bibinfo {author} {\bibfnamefont {M.~J.}\
  \bibnamefont {Vicente~Vacas}},\ }\href {\doibase 10.1103/PhysRevD.102.113016}
  {\bibfield  {journal} {\bibinfo  {journal} {Phys. Rev. D}\ }\textbf {\bibinfo
  {volume} {102}},\ \bibinfo {pages} {113016} (\bibinfo {year} {2020})},\
  \Eprint {http://arxiv.org/abs/2008.04244} {arXiv:2008.04244 [hep-ph]}
  \BibitemShut {NoStop}%
\bibitem [{\citenamefont {Sajjad~Athar}\ and\ \citenamefont
  {Morf\'\i{}n}(2021)}]{SajjadAthar:2020nvy}%
  \BibitemOpen
  \bibfield  {author} {\bibinfo {author} {\bibfnamefont {M.}~\bibnamefont
  {Sajjad~Athar}}\ and\ \bibinfo {author} {\bibfnamefont {J.~G.}\ \bibnamefont
  {Morf\'\i{}n}},\ }\href {\doibase 10.1088/1361-6471/abbb11} {\bibfield
  {journal} {\bibinfo  {journal} {J. Phys. G}\ }\textbf {\bibinfo {volume}
  {48}},\ \bibinfo {pages} {034001} (\bibinfo {year} {2021})},\ \Eprint
  {http://arxiv.org/abs/2006.08603} {arXiv:2006.08603 [hep-ph]} \BibitemShut
  {NoStop}%
\bibitem [{\citenamefont {Andreopoulos}\ \emph {et~al.}(2019)\citenamefont
  {Andreopoulos} \emph {et~al.}}]{Andreopoulos:2019gvw}%
  \BibitemOpen
  \bibfield  {author} {\bibinfo {author} {\bibfnamefont {C.}~\bibnamefont
  {Andreopoulos}} \emph {et~al.} (\bibinfo {collaboration} {NuSTEC}),\ }in\
  \href@noop {} {\emph {\bibinfo {booktitle} {{NuSTEC Workshop on Shallow- and
  Deep-Inelastic Scattering}}}}\ (\bibinfo {year} {2019})\ \bibinfo {note}
  {arXiv:1907.13252},\ \Eprint {http://arxiv.org/abs/1907.13252}
  {arXiv:1907.13252 [hep-ph]} \BibitemShut {NoStop}%
\bibitem [{\citenamefont {Bodek}\ and\ \citenamefont
  {Yang}(2003)}]{Bodek:2002ps}%
  \BibitemOpen
  \bibfield  {author} {\bibinfo {author} {\bibfnamefont {A.}~\bibnamefont
  {Bodek}}\ and\ \bibinfo {author} {\bibfnamefont {U.}~\bibnamefont {Yang}},\
  }\href {\doibase 10.1088/0954-3899/29/8/369} {\bibfield  {journal} {\bibinfo
  {journal} {J. Phys. G}\ }\textbf {\bibinfo {volume} {29}},\ \bibinfo {pages}
  {1899} (\bibinfo {year} {2003})},\ \Eprint
  {http://arxiv.org/abs/hep-ex/0210024} {arXiv:hep-ex/0210024} \BibitemShut
  {NoStop}%
\bibitem [{\citenamefont {Bodek}\ \emph {et~al.}(2005)\citenamefont {Bodek},
  \citenamefont {Park},\ and\ \citenamefont {Yang}}]{Bodek:2004pc}%
  \BibitemOpen
  \bibfield  {author} {\bibinfo {author} {\bibfnamefont {A.}~\bibnamefont
  {Bodek}}, \bibinfo {author} {\bibfnamefont {I.}~\bibnamefont {Park}}, \ and\
  \bibinfo {author} {\bibfnamefont {U.-k.}\ \bibnamefont {Yang}},\ }\href
  {\doibase 10.1016/j.nuclphysbps.2004.11.208} {\bibfield  {journal} {\bibinfo
  {journal} {Nucl. Phys. B Proc. Suppl.}\ }\textbf {\bibinfo {volume} {139}},\
  \bibinfo {pages} {113} (\bibinfo {year} {2005})},\ \Eprint
  {http://arxiv.org/abs/hep-ph/0411202} {arXiv:hep-ph/0411202} \BibitemShut
  {NoStop}%
\bibitem [{\citenamefont {Bodek}\ and\ \citenamefont
  {Yang}(2010)}]{Bodek:2010km}%
  \BibitemOpen
  \bibfield  {author} {\bibinfo {author} {\bibfnamefont {A.}~\bibnamefont
  {Bodek}}\ and\ \bibinfo {author} {\bibfnamefont {U.-k.}\ \bibnamefont
  {Yang}},\ }\href@noop {} {\  (\bibinfo {year} {2010})},\ \Eprint
  {http://arxiv.org/abs/1011.6592} {arXiv:1011.6592 [hep-ph]} \BibitemShut
  {NoStop}%
\bibitem [{\citenamefont {Dasgupta}\ and\ \citenamefont
  {Webber}(1996)}]{Dasgupta:1996hh}%
  \BibitemOpen
  \bibfield  {author} {\bibinfo {author} {\bibfnamefont {M.}~\bibnamefont
  {Dasgupta}}\ and\ \bibinfo {author} {\bibfnamefont {B.}~\bibnamefont
  {Webber}},\ }\href {\doibase 10.1016/0370-2693(96)00674-0} {\bibfield
  {journal} {\bibinfo  {journal} {Phys. Lett. B}\ }\textbf {\bibinfo {volume}
  {382}},\ \bibinfo {pages} {273} (\bibinfo {year} {1996})},\ \Eprint
  {http://arxiv.org/abs/hep-ph/9604388} {arXiv:hep-ph/9604388} \BibitemShut
  {NoStop}%
\bibitem [{\citenamefont {Acero}\ \emph {et~al.}(2019)\citenamefont {Acero}
  \emph {et~al.}}]{Acero:2019ksn}%
  \BibitemOpen
  \bibfield  {author} {\bibinfo {author} {\bibfnamefont {M.}~\bibnamefont
  {Acero}} \emph {et~al.} (\bibinfo {collaboration} {NOvA}),\ }\href {\doibase
  10.1103/PhysRevLett.123.151803} {\bibfield  {journal} {\bibinfo  {journal}
  {Phys. Rev. Lett.}\ }\textbf {\bibinfo {volume} {123}},\ \bibinfo {pages}
  {151803} (\bibinfo {year} {2019})},\ \Eprint
  {http://arxiv.org/abs/1906.04907} {arXiv:1906.04907 [hep-ex]} \BibitemShut
  {NoStop}%
\bibitem [{\citenamefont {Abi}\ \emph {et~al.}(2020)\citenamefont {Abi} \emph
  {et~al.}}]{Abi:2020wmh}%
  \BibitemOpen
  \bibfield  {author} {\bibinfo {author} {\bibfnamefont {B.}~\bibnamefont
  {Abi}} \emph {et~al.} (\bibinfo {collaboration} {DUNE}),\ }\href@noop {} {\
  (\bibinfo {year} {2020})},\ \Eprint {http://arxiv.org/abs/2002.02967}
  {arXiv:2002.02967 [physics.ins-det]} \BibitemShut {NoStop}%
\bibitem [{\citenamefont {Aartsen}\ \emph {et~al.}(2021)\citenamefont {Aartsen}
  \emph {et~al.}}]{IceCube-Gen2:2020qha}%
  \BibitemOpen
  \bibfield  {author} {\bibinfo {author} {\bibfnamefont {M.~G.}\ \bibnamefont
  {Aartsen}} \emph {et~al.} (\bibinfo {collaboration} {IceCube-Gen2}),\ }\href
  {\doibase 10.1088/1361-6471/abbd48} {\bibfield  {journal} {\bibinfo
  {journal} {J. Phys. G}\ }\textbf {\bibinfo {volume} {48}},\ \bibinfo {pages}
  {060501} (\bibinfo {year} {2021})},\ \Eprint
  {http://arxiv.org/abs/2008.04323} {arXiv:2008.04323 [astro-ph.HE]}
  \BibitemShut {NoStop}%
\bibitem [{\citenamefont {Adrian-Martinez}\ \emph {et~al.}(2016)\citenamefont
  {Adrian-Martinez} \emph {et~al.}}]{Adrian-Martinez:2016fdl}%
  \BibitemOpen
  \bibfield  {author} {\bibinfo {author} {\bibfnamefont {S.}~\bibnamefont
  {Adrian-Martinez}} \emph {et~al.} (\bibinfo {collaboration} {KM3Net}),\
  }\href {\doibase 10.1088/0954-3899/43/8/084001} {\bibfield  {journal}
  {\bibinfo  {journal} {J. Phys. G}\ }\textbf {\bibinfo {volume} {43}},\
  \bibinfo {pages} {084001} (\bibinfo {year} {2016})},\ \Eprint
  {http://arxiv.org/abs/1601.07459} {arXiv:1601.07459 [astro-ph.IM]}
  \BibitemShut {NoStop}%
\bibitem [{\citenamefont {Fukuda}\ \emph {et~al.}(2003)\citenamefont {Fukuda}
  \emph {et~al.}}]{Fukuda:2002uc}%
  \BibitemOpen
  \bibfield  {author} {\bibinfo {author} {\bibfnamefont {Y.}~\bibnamefont
  {Fukuda}} \emph {et~al.} (\bibinfo {collaboration} {Super-Kamiokande}),\
  }\href {\doibase 10.1016/S0168-9002(03)00425-X} {\bibfield  {journal}
  {\bibinfo  {journal} {Nucl. Instrum. Meth. A}\ }\textbf {\bibinfo {volume}
  {501}},\ \bibinfo {pages} {418} (\bibinfo {year} {2003})}\BibitemShut
  {NoStop}%
\bibitem [{\citenamefont {Abe}\ \emph {et~al.}(2018)\citenamefont {Abe} \emph
  {et~al.}}]{Abe:2018uyc}%
  \BibitemOpen
  \bibfield  {author} {\bibinfo {author} {\bibfnamefont {K.}~\bibnamefont
  {Abe}} \emph {et~al.} (\bibinfo {collaboration} {Hyper-Kamiokande}),\
  }\href@noop {} {\  (\bibinfo {year} {2018})},\ \Eprint
  {http://arxiv.org/abs/1805.04163} {arXiv:1805.04163 [physics.ins-det]}
  \BibitemShut {NoStop}%
\bibitem [{\citenamefont {Hernandez}\ \emph {et~al.}(2008)\citenamefont
  {Hernandez}, \citenamefont {Nieves}, \citenamefont {Singh}, \citenamefont
  {Valverde},\ and\ \citenamefont {Vicente~Vacas}}]{Hernandez:2007ej}%
  \BibitemOpen
  \bibfield  {author} {\bibinfo {author} {\bibfnamefont {E.}~\bibnamefont
  {Hernandez}}, \bibinfo {author} {\bibfnamefont {J.}~\bibnamefont {Nieves}},
  \bibinfo {author} {\bibfnamefont {S.~K.}\ \bibnamefont {Singh}}, \bibinfo
  {author} {\bibfnamefont {M.}~\bibnamefont {Valverde}}, \ and\ \bibinfo
  {author} {\bibfnamefont {M.~J.}\ \bibnamefont {Vicente~Vacas}},\ }\href
  {\doibase 10.1103/PhysRevD.77.053009} {\bibfield  {journal} {\bibinfo
  {journal} {Phys. Rev. D}\ }\textbf {\bibinfo {volume} {77}},\ \bibinfo
  {pages} {053009} (\bibinfo {year} {2008})},\ \Eprint
  {http://arxiv.org/abs/0710.3562} {arXiv:0710.3562 [hep-ph]} \BibitemShut
  {NoStop}%
\bibitem [{\citenamefont {Rafi~Alam}\ \emph {et~al.}(2010)\citenamefont
  {Rafi~Alam}, \citenamefont {Ruiz~Simo}, \citenamefont {Sajjad~Athar},\ and\
  \citenamefont {Vicente~Vacas}}]{RafiAlam:2010kf}%
  \BibitemOpen
  \bibfield  {author} {\bibinfo {author} {\bibfnamefont {M.}~\bibnamefont
  {Rafi~Alam}}, \bibinfo {author} {\bibfnamefont {I.}~\bibnamefont
  {Ruiz~Simo}}, \bibinfo {author} {\bibfnamefont {M.}~\bibnamefont
  {Sajjad~Athar}}, \ and\ \bibinfo {author} {\bibfnamefont {M.}~\bibnamefont
  {Vicente~Vacas}},\ }\href {\doibase 10.1103/PhysRevD.82.033001} {\bibfield
  {journal} {\bibinfo  {journal} {Phys. Rev. D}\ }\textbf {\bibinfo {volume}
  {82}},\ \bibinfo {pages} {033001} (\bibinfo {year} {2010})},\ \Eprint
  {http://arxiv.org/abs/1004.5484} {arXiv:1004.5484 [hep-ph]} \BibitemShut
  {NoStop}%
\bibitem [{\citenamefont {Alam}\ \emph {et~al.}(2012)\citenamefont {Alam},
  \citenamefont {Simo}, \citenamefont {Athar},\ and\ \citenamefont
  {Vicente~Vacas}}]{Alam:2012zz}%
  \BibitemOpen
  \bibfield  {author} {\bibinfo {author} {\bibfnamefont {M.~R.}\ \bibnamefont
  {Alam}}, \bibinfo {author} {\bibfnamefont {I.~R.}\ \bibnamefont {Simo}},
  \bibinfo {author} {\bibfnamefont {M.~S.}\ \bibnamefont {Athar}}, \ and\
  \bibinfo {author} {\bibfnamefont {M.}~\bibnamefont {Vicente~Vacas}},\ }\href
  {\doibase 10.1103/PhysRevD.85.013014} {\bibfield  {journal} {\bibinfo
  {journal} {Phys. Rev. D}\ }\textbf {\bibinfo {volume} {85}},\ \bibinfo
  {pages} {013014} (\bibinfo {year} {2012})},\ \Eprint
  {http://arxiv.org/abs/1111.0863} {arXiv:1111.0863 [hep-ph]} \BibitemShut
  {NoStop}%
\bibitem [{\citenamefont {Wang}\ \emph {et~al.}(2014)\citenamefont {Wang},
  \citenamefont {Alvarez-Ruso},\ and\ \citenamefont {Nieves}}]{Wang:2013wva}%
  \BibitemOpen
  \bibfield  {author} {\bibinfo {author} {\bibfnamefont {E.}~\bibnamefont
  {Wang}}, \bibinfo {author} {\bibfnamefont {L.}~\bibnamefont {Alvarez-Ruso}},
  \ and\ \bibinfo {author} {\bibfnamefont {J.}~\bibnamefont {Nieves}},\ }\href
  {\doibase 10.1103/PhysRevC.89.015503} {\bibfield  {journal} {\bibinfo
  {journal} {Phys. Rev. C}\ }\textbf {\bibinfo {volume} {89}},\ \bibinfo
  {pages} {015503} (\bibinfo {year} {2014})},\ \Eprint
  {http://arxiv.org/abs/1311.2151} {arXiv:1311.2151 [nucl-th]} \BibitemShut
  {NoStop}%
\bibitem [{\citenamefont {Leitner}\ \emph {et~al.}(2009)\citenamefont
  {Leitner}, \citenamefont {Buss}, \citenamefont {Alvarez-Ruso},\ and\
  \citenamefont {Mosel}}]{Leitner:2008ue}%
  \BibitemOpen
  \bibfield  {author} {\bibinfo {author} {\bibfnamefont {T.}~\bibnamefont
  {Leitner}}, \bibinfo {author} {\bibfnamefont {O.}~\bibnamefont {Buss}},
  \bibinfo {author} {\bibfnamefont {L.}~\bibnamefont {Alvarez-Ruso}}, \ and\
  \bibinfo {author} {\bibfnamefont {U.}~\bibnamefont {Mosel}},\ }\href
  {\doibase 10.1103/PhysRevC.79.034601} {\bibfield  {journal} {\bibinfo
  {journal} {Phys. Rev. C}\ }\textbf {\bibinfo {volume} {79}},\ \bibinfo
  {pages} {034601} (\bibinfo {year} {2009})},\ \Eprint
  {http://arxiv.org/abs/0812.0587} {arXiv:0812.0587 [nucl-th]} \BibitemShut
  {NoStop}%
\bibitem [{\citenamefont {Rafi~Alam}\ \emph {et~al.}(2016)\citenamefont
  {Rafi~Alam}, \citenamefont {Sajjad~Athar}, \citenamefont {Chauhan},\ and\
  \citenamefont {Singh}}]{RafiAlam:2015fcw}%
  \BibitemOpen
  \bibfield  {author} {\bibinfo {author} {\bibfnamefont {M.}~\bibnamefont
  {Rafi~Alam}}, \bibinfo {author} {\bibfnamefont {M.}~\bibnamefont
  {Sajjad~Athar}}, \bibinfo {author} {\bibfnamefont {S.}~\bibnamefont
  {Chauhan}}, \ and\ \bibinfo {author} {\bibfnamefont {S.~K.}\ \bibnamefont
  {Singh}},\ }\href {\doibase 10.1142/S0218301316500105} {\bibfield  {journal}
  {\bibinfo  {journal} {Int. J. Mod. Phys. E}\ }\textbf {\bibinfo {volume}
  {25}},\ \bibinfo {pages} {1650010} (\bibinfo {year} {2016})},\ \Eprint
  {http://arxiv.org/abs/1509.08622} {arXiv:1509.08622 [hep-ph]} \BibitemShut
  {NoStop}%
\bibitem [{\citenamefont {Rein}\ and\ \citenamefont
  {Sehgal}(1981)}]{Rein:1980wg}%
  \BibitemOpen
  \bibfield  {author} {\bibinfo {author} {\bibfnamefont {D.}~\bibnamefont
  {Rein}}\ and\ \bibinfo {author} {\bibfnamefont {L.~M.}\ \bibnamefont
  {Sehgal}},\ }\href {\doibase 10.1016/0003-4916(81)90242-6} {\bibfield
  {journal} {\bibinfo  {journal} {Annals Phys.}\ }\textbf {\bibinfo {volume}
  {133}},\ \bibinfo {pages} {79} (\bibinfo {year} {1981})}\BibitemShut
  {NoStop}%
\bibitem [{\citenamefont {Kabirnezhad}(2018)}]{Kabirnezhad:2017jmf}%
  \BibitemOpen
  \bibfield  {author} {\bibinfo {author} {\bibfnamefont {M.}~\bibnamefont
  {Kabirnezhad}},\ }\href {\doibase 10.1103/PhysRevD.97.013002} {\bibfield
  {journal} {\bibinfo  {journal} {Phys. Rev. D}\ }\textbf {\bibinfo {volume}
  {97}},\ \bibinfo {pages} {013002} (\bibinfo {year} {2018})},\ \Eprint
  {http://arxiv.org/abs/1711.02403} {arXiv:1711.02403 [hep-ph]} \BibitemShut
  {NoStop}%
\bibitem [{\citenamefont {Kabirnezhad}(2020)}]{Kabirnezhad:2020wtp}%
  \BibitemOpen
  \bibfield  {author} {\bibinfo {author} {\bibfnamefont {M.}~\bibnamefont
  {Kabirnezhad}},\ }\href@noop {} {\  (\bibinfo {year} {2020})},\ \Eprint
  {http://arxiv.org/abs/2006.13765} {arXiv:2006.13765 [hep-ph]} \BibitemShut
  {NoStop}%
\bibitem [{\citenamefont {Buss}\ \emph {et~al.}(2012)\citenamefont {Buss},
  \citenamefont {Gaitanos}, \citenamefont {Gallmeister}, \citenamefont {van
  Hees}, \citenamefont {Kaskulov}, \citenamefont {Lalakulich}, \citenamefont
  {Larionov}, \citenamefont {Leitner}, \citenamefont {Weil},\ and\
  \citenamefont {Mosel}}]{Buss:2011mx}%
  \BibitemOpen
  \bibfield  {author} {\bibinfo {author} {\bibfnamefont {O.}~\bibnamefont
  {Buss}}, \bibinfo {author} {\bibfnamefont {T.}~\bibnamefont {Gaitanos}},
  \bibinfo {author} {\bibfnamefont {K.}~\bibnamefont {Gallmeister}}, \bibinfo
  {author} {\bibfnamefont {H.}~\bibnamefont {van Hees}}, \bibinfo {author}
  {\bibfnamefont {M.}~\bibnamefont {Kaskulov}}, \bibinfo {author}
  {\bibfnamefont {O.}~\bibnamefont {Lalakulich}}, \bibinfo {author}
  {\bibfnamefont {A.}~\bibnamefont {Larionov}}, \bibinfo {author}
  {\bibfnamefont {T.}~\bibnamefont {Leitner}}, \bibinfo {author} {\bibfnamefont
  {J.}~\bibnamefont {Weil}}, \ and\ \bibinfo {author} {\bibfnamefont
  {U.}~\bibnamefont {Mosel}},\ }\href {\doibase 10.1016/j.physrep.2011.12.001}
  {\bibfield  {journal} {\bibinfo  {journal} {Phys. Rept.}\ }\textbf {\bibinfo
  {volume} {512}},\ \bibinfo {pages} {1} (\bibinfo {year} {2012})},\ \Eprint
  {http://arxiv.org/abs/1106.1344} {arXiv:1106.1344 [hep-ph]} \BibitemShut
  {NoStop}%
\bibitem [{\citenamefont {Drechsel}\ \emph {et~al.}(2007)\citenamefont
  {Drechsel}, \citenamefont {Kamalov},\ and\ \citenamefont
  {Tiator}}]{Drechsel:2007if}%
  \BibitemOpen
  \bibfield  {author} {\bibinfo {author} {\bibfnamefont {D.}~\bibnamefont
  {Drechsel}}, \bibinfo {author} {\bibfnamefont {S.}~\bibnamefont {Kamalov}}, \
  and\ \bibinfo {author} {\bibfnamefont {L.}~\bibnamefont {Tiator}},\ }\href
  {\doibase 10.1140/epja/i2007-10490-6} {\bibfield  {journal} {\bibinfo
  {journal} {Eur. Phys. J. A}\ }\textbf {\bibinfo {volume} {34}},\ \bibinfo
  {pages} {69} (\bibinfo {year} {2007})},\ \Eprint
  {http://arxiv.org/abs/0710.0306} {arXiv:0710.0306 [nucl-th]} \BibitemShut
  {NoStop}%
\bibitem [{\citenamefont {Alvarez-Ruso}\ \emph {et~al.}(2016)\citenamefont
  {Alvarez-Ruso}, \citenamefont {Hern\'andez}, \citenamefont {Nieves},\ and\
  \citenamefont {Vicente~Vacas}}]{Alvarez-Ruso:2015eva}%
  \BibitemOpen
  \bibfield  {author} {\bibinfo {author} {\bibfnamefont {L.}~\bibnamefont
  {Alvarez-Ruso}}, \bibinfo {author} {\bibfnamefont {E.}~\bibnamefont
  {Hern\'andez}}, \bibinfo {author} {\bibfnamefont {J.}~\bibnamefont {Nieves}},
  \ and\ \bibinfo {author} {\bibfnamefont {M.~J.}\ \bibnamefont
  {Vicente~Vacas}},\ }\href {\doibase 10.1103/PhysRevD.93.014016} {\bibfield
  {journal} {\bibinfo  {journal} {Phys. Rev. D}\ }\textbf {\bibinfo {volume}
  {93}},\ \bibinfo {pages} {014016} (\bibinfo {year} {2016})},\ \Eprint
  {http://arxiv.org/abs/1510.06266} {arXiv:1510.06266 [hep-ph]} \BibitemShut
  {NoStop}%
\bibitem [{\citenamefont {Sa\'ul-Sala}\ \emph {et~al.}(2021)\citenamefont
  {Sa\'ul-Sala}, \citenamefont {Sobczyk}, \citenamefont {Rafi~Alam},
  \citenamefont {Alvarez-Ruso},\ and\ \citenamefont
  {Nieves}}]{Saul-Sala:2021swb}%
  \BibitemOpen
  \bibfield  {author} {\bibinfo {author} {\bibfnamefont {E.}~\bibnamefont
  {Sa\'ul-Sala}}, \bibinfo {author} {\bibfnamefont {J.~E.}\ \bibnamefont
  {Sobczyk}}, \bibinfo {author} {\bibfnamefont {M.}~\bibnamefont {Rafi~Alam}},
  \bibinfo {author} {\bibfnamefont {L.}~\bibnamefont {Alvarez-Ruso}}, \ and\
  \bibinfo {author} {\bibfnamefont {J.}~\bibnamefont {Nieves}},\ }\href
  {\doibase 10.1016/j.physletb.2021.136349} {\bibfield  {journal} {\bibinfo
  {journal} {Phys. Lett. B}\ }\textbf {\bibinfo {volume} {817}},\ \bibinfo
  {pages} {136349} (\bibinfo {year} {2021})},\ \Eprint
  {http://arxiv.org/abs/2101.10749} {arXiv:2101.10749 [hep-ph]} \BibitemShut
  {NoStop}%
\bibitem [{\citenamefont {Ren}\ \emph {et~al.}(2015)\citenamefont {Ren},
  \citenamefont {Oset}, \citenamefont {Alvarez-Ruso},\ and\ \citenamefont
  {Vicente~Vacas}}]{Ren:2015bsa}%
  \BibitemOpen
  \bibfield  {author} {\bibinfo {author} {\bibfnamefont {X.-L.}\ \bibnamefont
  {Ren}}, \bibinfo {author} {\bibfnamefont {E.}~\bibnamefont {Oset}}, \bibinfo
  {author} {\bibfnamefont {L.}~\bibnamefont {Alvarez-Ruso}}, \ and\ \bibinfo
  {author} {\bibfnamefont {M.~J.}\ \bibnamefont {Vicente~Vacas}},\ }\href
  {\doibase 10.1103/PhysRevC.91.045201} {\bibfield  {journal} {\bibinfo
  {journal} {Phys. Rev. C}\ }\textbf {\bibinfo {volume} {91}},\ \bibinfo
  {pages} {045201} (\bibinfo {year} {2015})},\ \Eprint
  {http://arxiv.org/abs/1501.04073} {arXiv:1501.04073 [hep-ph]} \BibitemShut
  {NoStop}%
\bibitem [{\citenamefont {González-Jiménez}\ \emph
  {et~al.}(2017)\citenamefont {González-Jiménez}, \citenamefont {Jachowicz},
  \citenamefont {Niewczas}, \citenamefont {Nys}, \citenamefont {Pandey},
  \citenamefont {Van~Cuyck},\ and\ \citenamefont
  {Van~Dessel}}]{Gonzalez-Jimenez:2016qqq}%
  \BibitemOpen
  \bibfield  {author} {\bibinfo {author} {\bibfnamefont {R.}~\bibnamefont
  {González-Jiménez}}, \bibinfo {author} {\bibfnamefont {N.}~\bibnamefont
  {Jachowicz}}, \bibinfo {author} {\bibfnamefont {K.}~\bibnamefont {Niewczas}},
  \bibinfo {author} {\bibfnamefont {J.}~\bibnamefont {Nys}}, \bibinfo {author}
  {\bibfnamefont {V.}~\bibnamefont {Pandey}}, \bibinfo {author} {\bibfnamefont
  {T.}~\bibnamefont {Van~Cuyck}}, \ and\ \bibinfo {author} {\bibfnamefont
  {N.}~\bibnamefont {Van~Dessel}},\ }\href {\doibase
  10.1103/PhysRevD.95.113007} {\bibfield  {journal} {\bibinfo  {journal} {Phys.
  Rev. D}\ }\textbf {\bibinfo {volume} {95}},\ \bibinfo {pages} {113007}
  (\bibinfo {year} {2017})},\ \Eprint {http://arxiv.org/abs/1612.05511}
  {arXiv:1612.05511 [nucl-th]} \BibitemShut {NoStop}%
\bibitem [{\citenamefont {González-Jiménez}\ \emph
  {et~al.}(2018)\citenamefont {González-Jiménez}, \citenamefont {Niewczas},\
  and\ \citenamefont {Jachowicz}}]{Gonzalez-Jimenez:2017fea}%
  \BibitemOpen
  \bibfield  {author} {\bibinfo {author} {\bibfnamefont {R.}~\bibnamefont
  {González-Jiménez}}, \bibinfo {author} {\bibfnamefont {K.}~\bibnamefont
  {Niewczas}}, \ and\ \bibinfo {author} {\bibfnamefont {N.}~\bibnamefont
  {Jachowicz}},\ }\href {\doibase 10.1103/PhysRevD.97.013004} {\bibfield
  {journal} {\bibinfo  {journal} {Phys. Rev. D}\ }\textbf {\bibinfo {volume}
  {97}},\ \bibinfo {pages} {013004} (\bibinfo {year} {2018})},\ \Eprint
  {http://arxiv.org/abs/1710.08374} {arXiv:1710.08374 [nucl-th]} \BibitemShut
  {NoStop}%
\bibitem [{\citenamefont {Nikolakopoulos}\ \emph {et~al.}(2018)\citenamefont
  {Nikolakopoulos}, \citenamefont {González-Jiménez}, \citenamefont
  {Niewczas}, \citenamefont {Sobczyk},\ and\ \citenamefont
  {Jachowicz}}]{Nikolakopoulos:2018gtf}%
  \BibitemOpen
  \bibfield  {author} {\bibinfo {author} {\bibfnamefont {A.}~\bibnamefont
  {Nikolakopoulos}}, \bibinfo {author} {\bibfnamefont {R.}~\bibnamefont
  {González-Jiménez}}, \bibinfo {author} {\bibfnamefont {K.}~\bibnamefont
  {Niewczas}}, \bibinfo {author} {\bibfnamefont {J.}~\bibnamefont {Sobczyk}}, \
  and\ \bibinfo {author} {\bibfnamefont {N.}~\bibnamefont {Jachowicz}},\ }\href
  {\doibase 10.1103/PhysRevD.97.093008} {\bibfield  {journal} {\bibinfo
  {journal} {Phys. Rev. D}\ }\textbf {\bibinfo {volume} {97}},\ \bibinfo
  {pages} {093008} (\bibinfo {year} {2018})},\ \Eprint
  {http://arxiv.org/abs/1803.03163} {arXiv:1803.03163 [nucl-th]} \BibitemShut
  {NoStop}%
\bibitem [{\citenamefont {Oset}\ and\ \citenamefont
  {Salcedo}(1987)}]{Oset:1987re}%
  \BibitemOpen
  \bibfield  {author} {\bibinfo {author} {\bibfnamefont {E.}~\bibnamefont
  {Oset}}\ and\ \bibinfo {author} {\bibfnamefont {L.~L.}\ \bibnamefont
  {Salcedo}},\ }\href {\doibase 10.1016/0375-9474(87)90185-0} {\bibfield
  {journal} {\bibinfo  {journal} {Nucl. Phys. A}\ }\textbf {\bibinfo {volume}
  {468}},\ \bibinfo {pages} {631} (\bibinfo {year} {1987})}\BibitemShut
  {NoStop}%
\bibitem [{\citenamefont {Hern\'andez}\ \emph {et~al.}(2013)\citenamefont
  {Hern\'andez}, \citenamefont {Nieves},\ and\ \citenamefont
  {Vicente~Vacas}}]{Hernandez:2013jka}%
  \BibitemOpen
  \bibfield  {author} {\bibinfo {author} {\bibfnamefont {E.}~\bibnamefont
  {Hern\'andez}}, \bibinfo {author} {\bibfnamefont {J.}~\bibnamefont {Nieves}},
  \ and\ \bibinfo {author} {\bibfnamefont {M.~J.}\ \bibnamefont
  {Vicente~Vacas}},\ }\href {\doibase 10.1103/PhysRevD.87.113009} {\bibfield
  {journal} {\bibinfo  {journal} {Phys. Rev. D}\ }\textbf {\bibinfo {volume}
  {87}},\ \bibinfo {pages} {113009} (\bibinfo {year} {2013})},\ \Eprint
  {http://arxiv.org/abs/1304.1320} {arXiv:1304.1320 [hep-ph]} \BibitemShut
  {NoStop}%
\bibitem [{\citenamefont {Chanfray}\ and\ \citenamefont
  {Ericson}(2021)}]{Chanfray:2021wie}%
  \BibitemOpen
  \bibfield  {author} {\bibinfo {author} {\bibfnamefont {G.}~\bibnamefont
  {Chanfray}}\ and\ \bibinfo {author} {\bibfnamefont {M.}~\bibnamefont
  {Ericson}},\ }\href {\doibase 10.1103/PhysRevC.104.015203} {\bibfield
  {journal} {\bibinfo  {journal} {Phys. Rev. C}\ }\textbf {\bibinfo {volume}
  {104}},\ \bibinfo {pages} {015203} (\bibinfo {year} {2021})},\ \Eprint
  {http://arxiv.org/abs/2105.02505} {arXiv:2105.02505 [hep-ph]} \BibitemShut
  {NoStop}%
\bibitem [{\citenamefont {Bloom}\ and\ \citenamefont
  {Gilman}(1970)}]{Bloom:1970xb}%
  \BibitemOpen
  \bibfield  {author} {\bibinfo {author} {\bibfnamefont {E.~D.}\ \bibnamefont
  {Bloom}}\ and\ \bibinfo {author} {\bibfnamefont {F.~J.}\ \bibnamefont
  {Gilman}},\ }\href {\doibase 10.1103/PhysRevLett.25.1140} {\bibfield
  {journal} {\bibinfo  {journal} {Phys. Rev. Lett.}\ }\textbf {\bibinfo
  {volume} {25}},\ \bibinfo {pages} {1140} (\bibinfo {year}
  {1970})}\BibitemShut {NoStop}%
\bibitem [{\citenamefont {Lalakulich}\ \emph {et~al.}(2007)\citenamefont
  {Lalakulich}, \citenamefont {Melnitchouk},\ and\ \citenamefont
  {Paschos}}]{Lalakulich:2006yn}%
  \BibitemOpen
  \bibfield  {author} {\bibinfo {author} {\bibfnamefont {O.}~\bibnamefont
  {Lalakulich}}, \bibinfo {author} {\bibfnamefont {W.}~\bibnamefont
  {Melnitchouk}}, \ and\ \bibinfo {author} {\bibfnamefont {E.~A.}\ \bibnamefont
  {Paschos}},\ }\href {\doibase 10.1103/PhysRevC.75.015202} {\bibfield
  {journal} {\bibinfo  {journal} {Phys. Rev. C}\ }\textbf {\bibinfo {volume}
  {75}},\ \bibinfo {pages} {015202} (\bibinfo {year} {2007})},\ \Eprint
  {http://arxiv.org/abs/hep-ph/0608058} {arXiv:hep-ph/0608058} \BibitemShut
  {NoStop}%
\bibitem [{\citenamefont {Lalakulich}\ \emph {et~al.}(2009)\citenamefont
  {Lalakulich}, \citenamefont {Praet}, \citenamefont {Jachowicz}, \citenamefont
  {Ryckebusch}, \citenamefont {Leitner}, \citenamefont {Buss},\ and\
  \citenamefont {Mosel}}]{Lalakulich:2009zza}%
  \BibitemOpen
  \bibfield  {author} {\bibinfo {author} {\bibfnamefont {O.}~\bibnamefont
  {Lalakulich}}, \bibinfo {author} {\bibfnamefont {C.}~\bibnamefont {Praet}},
  \bibinfo {author} {\bibfnamefont {N.}~\bibnamefont {Jachowicz}}, \bibinfo
  {author} {\bibfnamefont {J.}~\bibnamefont {Ryckebusch}}, \bibinfo {author}
  {\bibfnamefont {T.}~\bibnamefont {Leitner}}, \bibinfo {author} {\bibfnamefont
  {O.}~\bibnamefont {Buss}}, \ and\ \bibinfo {author} {\bibfnamefont
  {U.}~\bibnamefont {Mosel}},\ }\href {\doibase 10.1063/1.3274170} {\bibfield
  {journal} {\bibinfo  {journal} {AIP Conf. Proc.}\ }\textbf {\bibinfo {volume}
  {1189}},\ \bibinfo {pages} {276} (\bibinfo {year} {2009})}\BibitemShut
  {NoStop}%
\bibitem [{\citenamefont {Kaskulov}\ and\ \citenamefont
  {Mosel}(2012)}]{Kaskulov:2011pr}%
  \BibitemOpen
  \bibfield  {author} {\bibinfo {author} {\bibfnamefont {M.}~\bibnamefont
  {Kaskulov}}\ and\ \bibinfo {author} {\bibfnamefont {U.}~\bibnamefont
  {Mosel}},\ }\href {\doibase 10.1016/j.ppnp.2011.12.017} {\bibfield  {journal}
  {\bibinfo  {journal} {Prog. Part. Nucl. Phys.}\ }\textbf {\bibinfo {volume}
  {67}},\ \bibinfo {pages} {194} (\bibinfo {year} {2012})},\ \Eprint
  {http://arxiv.org/abs/1111.2610} {arXiv:1111.2610 [nucl-th]} \BibitemShut
  {NoStop}%
\bibitem [{\citenamefont {Praet}\ \emph {et~al.}(2008)\citenamefont {Praet},
  \citenamefont {Lalakulich}, \citenamefont {Jachowicz},\ and\ \citenamefont
  {Ryckebusch}}]{Praet:2008zz}%
  \BibitemOpen
  \bibfield  {author} {\bibinfo {author} {\bibfnamefont {C.}~\bibnamefont
  {Praet}}, \bibinfo {author} {\bibfnamefont {O.}~\bibnamefont {Lalakulich}},
  \bibinfo {author} {\bibfnamefont {N.}~\bibnamefont {Jachowicz}}, \ and\
  \bibinfo {author} {\bibfnamefont {J.}~\bibnamefont {Ryckebusch}},\ }\href
  {\doibase 10.22323/1.074.0138} {\bibfield  {journal} {\bibinfo  {journal}
  {PoS}\ }\textbf {\bibinfo {volume} {NUFACT08}},\ \bibinfo {pages} {138}
  (\bibinfo {year} {2008})}\BibitemShut {NoStop}%
\bibitem [{\citenamefont {Mosel}\ and\ \citenamefont
  {Gallmeister}(2017)}]{Mosel:2017nzk}%
  \BibitemOpen
  \bibfield  {author} {\bibinfo {author} {\bibfnamefont {U.}~\bibnamefont
  {Mosel}}\ and\ \bibinfo {author} {\bibfnamefont {K.}~\bibnamefont
  {Gallmeister}},\ }\href {\doibase 10.1103/PhysRevC.96.015503} {\bibfield
  {journal} {\bibinfo  {journal} {Phys. Rev. C}\ }\textbf {\bibinfo {volume}
  {96}},\ \bibinfo {pages} {015503} (\bibinfo {year} {2017})},\ \bibinfo {note}
  {[Addendum: Phys. Rev. C {\bf 99}, 035502 (2019)]},\ \Eprint
  {http://arxiv.org/abs/1708.04528} {arXiv:1708.04528 [nucl-th]} \BibitemShut
  {NoStop}%
\bibitem [{\citenamefont {Kulagin}\ and\ \citenamefont
  {Petti}(2007)}]{Kulagin:2007ju}%
  \BibitemOpen
  \bibfield  {author} {\bibinfo {author} {\bibfnamefont {S.~A.}\ \bibnamefont
  {Kulagin}}\ and\ \bibinfo {author} {\bibfnamefont {R.}~\bibnamefont
  {Petti}},\ }\href {\doibase 10.1103/PhysRevD.76.094023} {\bibfield  {journal}
  {\bibinfo  {journal} {Phys. Rev. D}\ }\textbf {\bibinfo {volume} {76}},\
  \bibinfo {pages} {094023} (\bibinfo {year} {2007})},\ \Eprint
  {http://arxiv.org/abs/hep-ph/0703033} {arXiv:hep-ph/0703033} \BibitemShut
  {NoStop}%
\bibitem [{\citenamefont {Kulagin}\ and\ \citenamefont
  {Petti}(2006)}]{Kulagin:2004ie}%
  \BibitemOpen
  \bibfield  {author} {\bibinfo {author} {\bibfnamefont {S.~A.}\ \bibnamefont
  {Kulagin}}\ and\ \bibinfo {author} {\bibfnamefont {R.}~\bibnamefont
  {Petti}},\ }\href {\doibase 10.1016/j.nuclphysa.2005.10.011} {\bibfield
  {journal} {\bibinfo  {journal} {Nucl. Phys. A}\ }\textbf {\bibinfo {volume}
  {765}},\ \bibinfo {pages} {126} (\bibinfo {year} {2006})},\ \Eprint
  {http://arxiv.org/abs/hep-ph/0412425} {arXiv:hep-ph/0412425} \BibitemShut
  {NoStop}%
\bibitem [{\citenamefont {Zaidi}\ \emph {et~al.}(2020)\citenamefont {Zaidi},
  \citenamefont {Haider}, \citenamefont {Sajjad~Athar}, \citenamefont {Singh},\
  and\ \citenamefont {Ruiz~Simo}}]{Zaidi:2019asc}%
  \BibitemOpen
  \bibfield  {author} {\bibinfo {author} {\bibfnamefont {F.}~\bibnamefont
  {Zaidi}}, \bibinfo {author} {\bibfnamefont {H.}~\bibnamefont {Haider}},
  \bibinfo {author} {\bibfnamefont {M.}~\bibnamefont {Sajjad~Athar}}, \bibinfo
  {author} {\bibfnamefont {S.~K.}\ \bibnamefont {Singh}}, \ and\ \bibinfo
  {author} {\bibfnamefont {I.}~\bibnamefont {Ruiz~Simo}},\ }\href {\doibase
  10.1103/PhysRevD.101.033001} {\bibfield  {journal} {\bibinfo  {journal}
  {Phys. Rev. D}\ }\textbf {\bibinfo {volume} {101}},\ \bibinfo {pages}
  {033001} (\bibinfo {year} {2020})},\ \Eprint
  {http://arxiv.org/abs/1911.12573} {arXiv:1911.12573 [hep-ph]} \BibitemShut
  {NoStop}%
\bibitem [{\citenamefont {Zaidi}\ \emph {et~al.}(2019)\citenamefont {Zaidi},
  \citenamefont {Haider}, \citenamefont {Sajjad~Athar}, \citenamefont {Singh},\
  and\ \citenamefont {Ruiz~Simo}}]{Zaidi:2019mfd}%
  \BibitemOpen
  \bibfield  {author} {\bibinfo {author} {\bibfnamefont {F.}~\bibnamefont
  {Zaidi}}, \bibinfo {author} {\bibfnamefont {H.}~\bibnamefont {Haider}},
  \bibinfo {author} {\bibfnamefont {M.}~\bibnamefont {Sajjad~Athar}}, \bibinfo
  {author} {\bibfnamefont {S.~K.}\ \bibnamefont {Singh}}, \ and\ \bibinfo
  {author} {\bibfnamefont {I.}~\bibnamefont {Ruiz~Simo}},\ }\href {\doibase
  10.1103/PhysRevD.99.093011} {\bibfield  {journal} {\bibinfo  {journal} {Phys.
  Rev. D}\ }\textbf {\bibinfo {volume} {99}},\ \bibinfo {pages} {093011}
  (\bibinfo {year} {2019})},\ \Eprint {http://arxiv.org/abs/1903.09000}
  {arXiv:1903.09000 [hep-ph]} \BibitemShut {NoStop}%
\bibitem [{\citenamefont {Haider}\ \emph {et~al.}(2016)\citenamefont {Haider},
  \citenamefont {Zaidi}, \citenamefont {Sajjad~Athar}, \citenamefont {Singh},\
  and\ \citenamefont {Ruiz~Simo}}]{Haider:2016zrk}%
  \BibitemOpen
  \bibfield  {author} {\bibinfo {author} {\bibfnamefont {H.}~\bibnamefont
  {Haider}}, \bibinfo {author} {\bibfnamefont {F.}~\bibnamefont {Zaidi}},
  \bibinfo {author} {\bibfnamefont {M.}~\bibnamefont {Sajjad~Athar}}, \bibinfo
  {author} {\bibfnamefont {S.~K.}\ \bibnamefont {Singh}}, \ and\ \bibinfo
  {author} {\bibfnamefont {I.}~\bibnamefont {Ruiz~Simo}},\ }\href {\doibase
  10.1016/j.nuclphysa.2016.06.006} {\bibfield  {journal} {\bibinfo  {journal}
  {Nucl. Phys. A}\ }\textbf {\bibinfo {volume} {955}},\ \bibinfo {pages} {58}
  (\bibinfo {year} {2016})},\ \Eprint {http://arxiv.org/abs/1603.00164}
  {arXiv:1603.00164 [nucl-th]} \BibitemShut {NoStop}%
\bibitem [{\citenamefont {Haider}\ \emph {et~al.}(2015)\citenamefont {Haider},
  \citenamefont {Zaidi}, \citenamefont {Sajjad~Athar}, \citenamefont {Singh},\
  and\ \citenamefont {Ruiz~Simo}}]{Haider:2015vea}%
  \BibitemOpen
  \bibfield  {author} {\bibinfo {author} {\bibfnamefont {H.}~\bibnamefont
  {Haider}}, \bibinfo {author} {\bibfnamefont {F.}~\bibnamefont {Zaidi}},
  \bibinfo {author} {\bibfnamefont {M.}~\bibnamefont {Sajjad~Athar}}, \bibinfo
  {author} {\bibfnamefont {S.~K.}\ \bibnamefont {Singh}}, \ and\ \bibinfo
  {author} {\bibfnamefont {I.}~\bibnamefont {Ruiz~Simo}},\ }\href {\doibase
  10.1016/j.nuclphysa.2015.08.008} {\bibfield  {journal} {\bibinfo  {journal}
  {Nucl. Phys. A}\ }\textbf {\bibinfo {volume} {943}},\ \bibinfo {pages} {58}
  (\bibinfo {year} {2015})},\ \Eprint {http://arxiv.org/abs/1505.08053}
  {arXiv:1505.08053 [nucl-th]} \BibitemShut {NoStop}%
\bibitem [{\citenamefont {Ansari}\ \emph {et~al.}(2021)\citenamefont {Ansari},
  \citenamefont {Athar}, \citenamefont {Haider}, \citenamefont {Simo},
  \citenamefont {Singh},\ and\ \citenamefont {Zaidi}}]{Ansari:2021cao}%
  \BibitemOpen
  \bibfield  {author} {\bibinfo {author} {\bibfnamefont {V.}~\bibnamefont
  {Ansari}}, \bibinfo {author} {\bibfnamefont {M.~S.}\ \bibnamefont {Athar}},
  \bibinfo {author} {\bibfnamefont {H.}~\bibnamefont {Haider}}, \bibinfo
  {author} {\bibfnamefont {I.~R.}\ \bibnamefont {Simo}}, \bibinfo {author}
  {\bibfnamefont {S.~K.}\ \bibnamefont {Singh}}, \ and\ \bibinfo {author}
  {\bibfnamefont {F.}~\bibnamefont {Zaidi}},\ }\href {\doibase
  10.1140/epjs/s11734-021-00277-9} {\bibfield  {journal} {\bibinfo  {journal}
  {Eur. Phys. J. ST}\ }\textbf {\bibinfo {volume} {230}},\ \bibinfo {pages}
  {4433} (\bibinfo {year} {2021})},\ \Eprint {http://arxiv.org/abs/2106.14670}
  {arXiv:2106.14670 [hep-ph]} \BibitemShut {NoStop}%
\bibitem [{\citenamefont {Ansari}\ \emph {et~al.}(2020)\citenamefont {Ansari},
  \citenamefont {Sajjad~Athar}, \citenamefont {Haider}, \citenamefont {Singh},\
  and\ \citenamefont {Zaidi}}]{Ansari:2020xne}%
  \BibitemOpen
  \bibfield  {author} {\bibinfo {author} {\bibfnamefont {V.}~\bibnamefont
  {Ansari}}, \bibinfo {author} {\bibfnamefont {M.}~\bibnamefont
  {Sajjad~Athar}}, \bibinfo {author} {\bibfnamefont {H.}~\bibnamefont
  {Haider}}, \bibinfo {author} {\bibfnamefont {S.~K.}\ \bibnamefont {Singh}}, \
  and\ \bibinfo {author} {\bibfnamefont {F.}~\bibnamefont {Zaidi}},\ }\href
  {\doibase 10.1103/PhysRevD.102.113007} {\bibfield  {journal} {\bibinfo
  {journal} {Phys. Rev. D}\ }\textbf {\bibinfo {volume} {102}},\ \bibinfo
  {pages} {113007} (\bibinfo {year} {2020})},\ \Eprint
  {http://arxiv.org/abs/2010.05538} {arXiv:2010.05538 [hep-ph]} \BibitemShut
  {NoStop}%
\bibitem [{\citenamefont {Brady}\ \emph {et~al.}(2011)\citenamefont {Brady},
  \citenamefont {Accardi}, \citenamefont {Hobbs},\ and\ \citenamefont
  {Melnitchouk}}]{Brady:2011uy}%
  \BibitemOpen
  \bibfield  {author} {\bibinfo {author} {\bibfnamefont {L.~T.}\ \bibnamefont
  {Brady}}, \bibinfo {author} {\bibfnamefont {A.}~\bibnamefont {Accardi}},
  \bibinfo {author} {\bibfnamefont {T.~J.}\ \bibnamefont {Hobbs}}, \ and\
  \bibinfo {author} {\bibfnamefont {W.}~\bibnamefont {Melnitchouk}},\ }\href
  {\doibase 10.1103/PhysRevD.84.074008} {\bibfield  {journal} {\bibinfo
  {journal} {Phys. Rev. D}\ }\textbf {\bibinfo {volume} {84}},\ \bibinfo
  {pages} {074008} (\bibinfo {year} {2011})},\ \bibinfo {note} {[Erratum:
  Phys.Rev.D 85, 039902 (2012)]},\ \Eprint {http://arxiv.org/abs/1108.4734}
  {arXiv:1108.4734 [hep-ph]} \BibitemShut {NoStop}%
\bibitem [{\citenamefont {Schienbein}\ \emph {et~al.}(2008)\citenamefont
  {Schienbein} \emph {et~al.}}]{Schienbein:2007gr}%
  \BibitemOpen
  \bibfield  {author} {\bibinfo {author} {\bibfnamefont {I.}~\bibnamefont
  {Schienbein}} \emph {et~al.},\ }\href {\doibase
  10.1088/0954-3899/35/5/053101} {\bibfield  {journal} {\bibinfo  {journal} {J.
  Phys. G}\ }\textbf {\bibinfo {volume} {35}},\ \bibinfo {pages} {053101}
  (\bibinfo {year} {2008})},\ \Eprint {http://arxiv.org/abs/0709.1775}
  {arXiv:0709.1775 [hep-ph]} \BibitemShut {NoStop}%
\bibitem [{\citenamefont {Georgi}\ and\ \citenamefont
  {Politzer}(1976)}]{Georgi:1976ve}%
  \BibitemOpen
  \bibfield  {author} {\bibinfo {author} {\bibfnamefont {H.}~\bibnamefont
  {Georgi}}\ and\ \bibinfo {author} {\bibfnamefont {H.~D.}\ \bibnamefont
  {Politzer}},\ }\href {\doibase 10.1103/PhysRevD.14.1829} {\bibfield
  {journal} {\bibinfo  {journal} {Phys. Rev. D}\ }\textbf {\bibinfo {volume}
  {14}},\ \bibinfo {pages} {1829} (\bibinfo {year} {1976})}\BibitemShut
  {NoStop}%
\bibitem [{\citenamefont {Ellis}\ \emph {et~al.}(1983)\citenamefont {Ellis},
  \citenamefont {Furmanski},\ and\ \citenamefont {Petronzio}}]{Ellis:1982cd}%
  \BibitemOpen
  \bibfield  {author} {\bibinfo {author} {\bibfnamefont {R.~K.}\ \bibnamefont
  {Ellis}}, \bibinfo {author} {\bibfnamefont {W.}~\bibnamefont {Furmanski}}, \
  and\ \bibinfo {author} {\bibfnamefont {R.}~\bibnamefont {Petronzio}},\ }\href
  {\doibase 10.1016/0550-3213(83)90597-7} {\bibfield  {journal} {\bibinfo
  {journal} {Nucl. Phys. B}\ }\textbf {\bibinfo {volume} {212}},\ \bibinfo
  {pages} {29} (\bibinfo {year} {1983})}\BibitemShut {NoStop}%
\bibitem [{\citenamefont {Accardi}\ \emph {et~al.}(2016)\citenamefont
  {Accardi}, \citenamefont {Brady}, \citenamefont {Melnitchouk}, \citenamefont
  {Owens},\ and\ \citenamefont {Sato}}]{Accardi:2016qay}%
  \BibitemOpen
  \bibfield  {author} {\bibinfo {author} {\bibfnamefont {A.}~\bibnamefont
  {Accardi}}, \bibinfo {author} {\bibfnamefont {L.~T.}\ \bibnamefont {Brady}},
  \bibinfo {author} {\bibfnamefont {W.}~\bibnamefont {Melnitchouk}}, \bibinfo
  {author} {\bibfnamefont {J.~F.}\ \bibnamefont {Owens}}, \ and\ \bibinfo
  {author} {\bibfnamefont {N.}~\bibnamefont {Sato}},\ }\href {\doibase
  10.1103/PhysRevD.93.114017} {\bibfield  {journal} {\bibinfo  {journal} {Phys.
  Rev. D}\ }\textbf {\bibinfo {volume} {93}},\ \bibinfo {pages} {114017}
  (\bibinfo {year} {2016})},\ \Eprint {http://arxiv.org/abs/1602.03154}
  {arXiv:1602.03154 [hep-ph]} \BibitemShut {NoStop}%
\bibitem [{\citenamefont {Gao}\ \emph {et~al.}(2022)\citenamefont {Gao},
  \citenamefont {Hobbs}, \citenamefont {Nadolsky}, \citenamefont {Sun},\ and\
  \citenamefont {Yuan}}]{Gao:2021fle}%
  \BibitemOpen
  \bibfield  {author} {\bibinfo {author} {\bibfnamefont {J.}~\bibnamefont
  {Gao}}, \bibinfo {author} {\bibfnamefont {T.~J.}\ \bibnamefont {Hobbs}},
  \bibinfo {author} {\bibfnamefont {P.~M.}\ \bibnamefont {Nadolsky}}, \bibinfo
  {author} {\bibfnamefont {C.}~\bibnamefont {Sun}}, \ and\ \bibinfo {author}
  {\bibfnamefont {C.~P.}\ \bibnamefont {Yuan}},\ }\href {\doibase
  10.1103/PhysRevD.105.L011503} {\bibfield  {journal} {\bibinfo  {journal}
  {Phys. Rev. D}\ }\textbf {\bibinfo {volume} {105}},\ \bibinfo {pages}
  {L011503} (\bibinfo {year} {2022})},\ \Eprint
  {http://arxiv.org/abs/2107.00460} {arXiv:2107.00460 [hep-ph]} \BibitemShut
  {NoStop}%
\bibitem [{Note1()}]{Note1}%
  \BibitemOpen
  \bibinfo {note} {Refer to \protect \url {https://ncteq.hepforge.org} for
  details of the nCTEQ collaboration}\BibitemShut {NoStop}%
\bibitem [{\citenamefont {Abreu}\ \emph {et~al.}(2020)\citenamefont {Abreu}
  \emph {et~al.}}]{FASER:2019dxq}%
  \BibitemOpen
  \bibfield  {author} {\bibinfo {author} {\bibfnamefont {H.}~\bibnamefont
  {Abreu}} \emph {et~al.} (\bibinfo {collaboration} {FASER}),\ }\href {\doibase
  10.1140/epjc/s10052-020-7631-5} {\bibfield  {journal} {\bibinfo  {journal}
  {Eur. Phys. J. C}\ }\textbf {\bibinfo {volume} {80}},\ \bibinfo {pages} {61}
  (\bibinfo {year} {2020})},\ \Eprint {http://arxiv.org/abs/1908.02310}
  {arXiv:1908.02310 [hep-ex]} \BibitemShut {NoStop}%
\bibitem [{\citenamefont {Ahdida}\ \emph {et~al.}(2020)\citenamefont {Ahdida}
  \emph {et~al.}}]{SHiP:2020sos}%
  \BibitemOpen
  \bibfield  {author} {\bibinfo {author} {\bibfnamefont {C.}~\bibnamefont
  {Ahdida}} \emph {et~al.} (\bibinfo {collaboration} {SHiP}),\ }\href@noop {}
  {\  (\bibinfo {year} {2020})},\ \Eprint {http://arxiv.org/abs/2002.08722}
  {arXiv:2002.08722 [physics.ins-det]} \BibitemShut {NoStop}%
\bibitem [{\citenamefont {Anchordoqui}\ \emph {et~al.}(2021)\citenamefont
  {Anchordoqui} \emph {et~al.}}]{Anchordoqui:2021ghd}%
  \BibitemOpen
  \bibfield  {author} {\bibinfo {author} {\bibfnamefont {L.~A.}\ \bibnamefont
  {Anchordoqui}} \emph {et~al.},\ }\href@noop {} {\  (\bibinfo {year}
  {2021})},\ \Eprint {http://arxiv.org/abs/2109.10905} {arXiv:2109.10905
  [hep-ph]} \BibitemShut {NoStop}%
\bibitem [{\citenamefont {Feng}\ \emph {et~al.}(2022)\citenamefont {Feng} \emph
  {et~al.}}]{Feng:2022inv}%
  \BibitemOpen
  \bibfield  {author} {\bibinfo {author} {\bibfnamefont {J.~L.}\ \bibnamefont
  {Feng}} \emph {et~al.},\ }in\ \href@noop {} {\emph {\bibinfo {booktitle}
  {{2022 Snowmass Summer Study}}}}\ (\bibinfo {year} {2022})\ \Eprint
  {http://arxiv.org/abs/2203.05090} {arXiv:2203.05090 [hep-ex]} \BibitemShut
  {NoStop}%
\bibitem [{\citenamefont {Webber}(1984)}]{Webber:1983if}%
  \BibitemOpen
  \bibfield  {author} {\bibinfo {author} {\bibfnamefont {B.~R.}\ \bibnamefont
  {Webber}},\ }\href {\doibase 10.1016/0550-3213(84)90333-X} {\bibfield
  {journal} {\bibinfo  {journal} {Nucl. Phys. B}\ }\textbf {\bibinfo {volume}
  {238}},\ \bibinfo {pages} {492} (\bibinfo {year} {1984})}\BibitemShut
  {NoStop}%
\bibitem [{\citenamefont {Andersson}\ \emph {et~al.}(1983)\citenamefont
  {Andersson}, \citenamefont {Gustafson}, \citenamefont {Ingelman},\ and\
  \citenamefont {Sjostrand}}]{Andersson:1983ia}%
  \BibitemOpen
  \bibfield  {author} {\bibinfo {author} {\bibfnamefont {B.}~\bibnamefont
  {Andersson}}, \bibinfo {author} {\bibfnamefont {G.}~\bibnamefont
  {Gustafson}}, \bibinfo {author} {\bibfnamefont {G.}~\bibnamefont {Ingelman}},
  \ and\ \bibinfo {author} {\bibfnamefont {T.}~\bibnamefont {Sjostrand}},\
  }\href {\doibase 10.1016/0370-1573(83)90080-7} {\bibfield  {journal}
  {\bibinfo  {journal} {Phys. Rept.}\ }\textbf {\bibinfo {volume} {97}},\
  \bibinfo {pages} {31} (\bibinfo {year} {1983})}\BibitemShut {NoStop}%
\bibitem [{\citenamefont {Chukanov}\ and\ \citenamefont
  {Petti}(2016)}]{Chukanov:2016lra}%
  \BibitemOpen
  \bibfield  {author} {\bibinfo {author} {\bibfnamefont {A.}~\bibnamefont
  {Chukanov}}\ and\ \bibinfo {author} {\bibfnamefont {R.}~\bibnamefont
  {Petti}},\ }\href {\doibase 10.7566/JPSCP.12.010026} {\bibfield  {journal}
  {\bibinfo  {journal} {JPS Conf. Proc.}\ }\textbf {\bibinfo {volume} {12}},\
  \bibinfo {pages} {010026} (\bibinfo {year} {2016})}\BibitemShut {NoStop}%
\bibitem [{\citenamefont {Adams}\ \emph {et~al.}(2019)\citenamefont {Adams}
  \emph {et~al.}}]{Adams:2018fud}%
  \BibitemOpen
  \bibfield  {author} {\bibinfo {author} {\bibfnamefont {C.}~\bibnamefont
  {Adams}} \emph {et~al.} (\bibinfo {collaboration} {MicroBooNE}),\ }\href
  {\doibase 10.1140/epjc/s10052-019-6742-3} {\bibfield  {journal} {\bibinfo
  {journal} {Eur. Phys. J. C}\ }\textbf {\bibinfo {volume} {79}},\ \bibinfo
  {pages} {248} (\bibinfo {year} {2019})},\ \Eprint
  {http://arxiv.org/abs/1805.06887} {arXiv:1805.06887 [hep-ex]} \BibitemShut
  {NoStop}%
\bibitem [{\citenamefont {Hiramoto}\ \emph {et~al.}(2020)\citenamefont
  {Hiramoto} \emph {et~al.}}]{NINJA:2020gbg}%
  \BibitemOpen
  \bibfield  {author} {\bibinfo {author} {\bibfnamefont {A.}~\bibnamefont
  {Hiramoto}} \emph {et~al.} (\bibinfo {collaboration} {NINJA}),\ }\href
  {\doibase 10.1103/PhysRevD.102.072006} {\bibfield  {journal} {\bibinfo
  {journal} {Phys. Rev. D}\ }\textbf {\bibinfo {volume} {102}},\ \bibinfo
  {pages} {072006} (\bibinfo {year} {2020})},\ \Eprint
  {http://arxiv.org/abs/2008.03895} {arXiv:2008.03895 [hep-ex]} \BibitemShut
  {NoStop}%
\bibitem [{\citenamefont {Oshima}\ \emph {et~al.}(2021)\citenamefont {Oshima}
  \emph {et~al.}}]{NINJA:2020bvx}%
  \BibitemOpen
  \bibfield  {author} {\bibinfo {author} {\bibfnamefont {H.}~\bibnamefont
  {Oshima}} \emph {et~al.} (\bibinfo {collaboration} {NINJA}),\ }\href
  {\doibase 10.1093/ptep/ptab027} {\bibfield  {journal} {\bibinfo  {journal}
  {PTEP}\ }\textbf {\bibinfo {volume} {2021}},\ \bibinfo {pages} {033C01}
  (\bibinfo {year} {2021})},\ \Eprint {http://arxiv.org/abs/2012.05221}
  {arXiv:2012.05221 [hep-ex]} \BibitemShut {NoStop}%
\bibitem [{\citenamefont {Sjostrand}\ \emph {et~al.}(2006)\citenamefont
  {Sjostrand}, \citenamefont {Mrenna},\ and\ \citenamefont
  {Skands}}]{Sjostrand:2006za}%
  \BibitemOpen
  \bibfield  {author} {\bibinfo {author} {\bibfnamefont {T.}~\bibnamefont
  {Sjostrand}}, \bibinfo {author} {\bibfnamefont {S.}~\bibnamefont {Mrenna}}, \
  and\ \bibinfo {author} {\bibfnamefont {P.~Z.}\ \bibnamefont {Skands}},\
  }\href {\doibase 10.1088/1126-6708/2006/05/026} {\bibfield  {journal}
  {\bibinfo  {journal} {JHEP}\ }\textbf {\bibinfo {volume} {05}},\ \bibinfo
  {pages} {026} (\bibinfo {year} {2006})},\ \Eprint
  {http://arxiv.org/abs/hep-ph/0603175} {arXiv:hep-ph/0603175} \BibitemShut
  {NoStop}%
\bibitem [{\citenamefont {Sjostrand}\ \emph {et~al.}(2008)\citenamefont
  {Sjostrand}, \citenamefont {Mrenna},\ and\ \citenamefont
  {Skands}}]{Sjostrand:2007gs}%
  \BibitemOpen
  \bibfield  {author} {\bibinfo {author} {\bibfnamefont {T.}~\bibnamefont
  {Sjostrand}}, \bibinfo {author} {\bibfnamefont {S.}~\bibnamefont {Mrenna}}, \
  and\ \bibinfo {author} {\bibfnamefont {P.~Z.}\ \bibnamefont {Skands}},\
  }\href {\doibase 10.1016/j.cpc.2008.01.036} {\bibfield  {journal} {\bibinfo
  {journal} {Comput. Phys. Commun.}\ }\textbf {\bibinfo {volume} {178}},\
  \bibinfo {pages} {852} (\bibinfo {year} {2008})},\ \Eprint
  {http://arxiv.org/abs/0710.3820} {arXiv:0710.3820 [hep-ph]} \BibitemShut
  {NoStop}%
\bibitem [{\citenamefont {Yang}\ \emph {et~al.}(2009)\citenamefont {Yang},
  \citenamefont {Andreopoulos}, \citenamefont {Gallagher}, \citenamefont
  {Hoffmann},\ and\ \citenamefont {Kehayias}}]{Yang:2009zx}%
  \BibitemOpen
  \bibfield  {author} {\bibinfo {author} {\bibfnamefont {T.}~\bibnamefont
  {Yang}}, \bibinfo {author} {\bibfnamefont {C.}~\bibnamefont {Andreopoulos}},
  \bibinfo {author} {\bibfnamefont {H.}~\bibnamefont {Gallagher}}, \bibinfo
  {author} {\bibfnamefont {K.}~\bibnamefont {Hoffmann}}, \ and\ \bibinfo
  {author} {\bibfnamefont {P.}~\bibnamefont {Kehayias}},\ }\href {\doibase
  10.1140/epjc/s10052-009-1094-z} {\bibfield  {journal} {\bibinfo  {journal}
  {Eur. Phys. J. C}\ }\textbf {\bibinfo {volume} {63}},\ \bibinfo {pages} {1}
  (\bibinfo {year} {2009})},\ \Eprint {http://arxiv.org/abs/0904.4043}
  {arXiv:0904.4043 [hep-ph]} \BibitemShut {NoStop}%
\bibitem [{\citenamefont {Bronner}\ and\ \citenamefont
  {Hartz}(2016)}]{Bronner:2016gmz}%
  \BibitemOpen
  \bibfield  {author} {\bibinfo {author} {\bibfnamefont {C.}~\bibnamefont
  {Bronner}}\ and\ \bibinfo {author} {\bibfnamefont {M.}~\bibnamefont
  {Hartz}},\ }\href {\doibase 10.7566/JPSCP.12.010041} {\bibfield  {journal}
  {\bibinfo  {journal} {JPS Conf. Proc.}\ }\textbf {\bibinfo {volume} {12}},\
  \bibinfo {pages} {010041} (\bibinfo {year} {2016})},\ \Eprint
  {http://arxiv.org/abs/1607.06558} {arXiv:1607.06558 [hep-ph]} \BibitemShut
  {NoStop}%
\bibitem [{\citenamefont {Tena-Vidal}\ \emph {et~al.}(2022)\citenamefont
  {Tena-Vidal} \emph {et~al.}}]{GENIE:2021wox}%
  \BibitemOpen
  \bibfield  {author} {\bibinfo {author} {\bibfnamefont {J.}~\bibnamefont
  {Tena-Vidal}} \emph {et~al.} (\bibinfo {collaboration} {GENIE}),\ }\href
  {\doibase 10.1103/PhysRevD.105.012009} {\bibfield  {journal} {\bibinfo
  {journal} {Phys. Rev. D}\ }\textbf {\bibinfo {volume} {105}},\ \bibinfo
  {pages} {012009} (\bibinfo {year} {2022})},\ \Eprint
  {http://arxiv.org/abs/2106.05884} {arXiv:2106.05884 [hep-ph]} \BibitemShut
  {NoStop}%
\bibitem [{\citenamefont {Kuzmin}\ and\ \citenamefont
  {Naumov}(2013)}]{Kuzmin:2013tza}%
  \BibitemOpen
  \bibfield  {author} {\bibinfo {author} {\bibfnamefont {K.~S.}\ \bibnamefont
  {Kuzmin}}\ and\ \bibinfo {author} {\bibfnamefont {V.~A.}\ \bibnamefont
  {Naumov}},\ }\href {\doibase 10.1103/PhysRevC.88.065501} {\bibfield
  {journal} {\bibinfo  {journal} {Phys. Rev. C}\ }\textbf {\bibinfo {volume}
  {88}},\ \bibinfo {pages} {065501} (\bibinfo {year} {2013})},\ \Eprint
  {http://arxiv.org/abs/1311.4047} {arXiv:1311.4047 [hep-ph]} \BibitemShut
  {NoStop}%
\bibitem [{\citenamefont {Katori}\ and\ \citenamefont
  {Mandalia}(2015)}]{Katori:2014fxa}%
  \BibitemOpen
  \bibfield  {author} {\bibinfo {author} {\bibfnamefont {T.}~\bibnamefont
  {Katori}}\ and\ \bibinfo {author} {\bibfnamefont {S.}~\bibnamefont
  {Mandalia}},\ }\href {\doibase 10.1088/0954-3899/42/11/115004} {\bibfield
  {journal} {\bibinfo  {journal} {J. Phys. G}\ }\textbf {\bibinfo {volume}
  {42}},\ \bibinfo {pages} {115004} (\bibinfo {year} {2015})},\ \Eprint
  {http://arxiv.org/abs/1412.4301} {arXiv:1412.4301 [hep-ex]} \BibitemShut
  {NoStop}%
\bibitem [{\citenamefont {Koba}\ \emph {et~al.}(1972)\citenamefont {Koba},
  \citenamefont {Nielsen},\ and\ \citenamefont {Olesen}}]{KOBA1972317}%
  \BibitemOpen
  \bibfield  {author} {\bibinfo {author} {\bibfnamefont {Z.}~\bibnamefont
  {Koba}}, \bibinfo {author} {\bibfnamefont {H.~B.}\ \bibnamefont {Nielsen}}, \
  and\ \bibinfo {author} {\bibfnamefont {P.}~\bibnamefont {Olesen}},\ }\href
  {\doibase 10.1016/0550-3213(72)90551-2} {\bibfield  {journal} {\bibinfo
  {journal} {Nucl. Phys. B}\ }\textbf {\bibinfo {volume} {40}},\ \bibinfo
  {pages} {317} (\bibinfo {year} {1972})}\BibitemShut {NoStop}%
\bibitem [{\citenamefont {Zieminska}\ \emph {et~al.}(1983)\citenamefont
  {Zieminska} \emph {et~al.}}]{Zieminska:1983bs}%
  \BibitemOpen
  \bibfield  {author} {\bibinfo {author} {\bibfnamefont {D.}~\bibnamefont
  {Zieminska}} \emph {et~al.},\ }\href {\doibase 10.1103/PhysRevD.27.47}
  {\bibfield  {journal} {\bibinfo  {journal} {Phys. Rev. D}\ }\textbf {\bibinfo
  {volume} {27}},\ \bibinfo {pages} {47} (\bibinfo {year} {1983})}\BibitemShut
  {NoStop}%
\bibitem [{\citenamefont {Alioli}\ \emph {et~al.}()\citenamefont {Alioli} \emph
  {et~al.}}]{HEPGeneratorWP}%
  \BibitemOpen
  \bibfield  {author} {\bibinfo {author} {\bibfnamefont {S.}~\bibnamefont
  {Alioli}} \emph {et~al.},\ }\href@noop {} {\enquote {\bibinfo {title} {{Event
  Generators for High-Energy Physics Experiments}},}\ }\bibinfo {note}
  {{Snowmass 2021 white paper}}\BibitemShut {NoStop}%
\bibitem [{ECT({\natexlab{a}})}]{ECTStarWorkshop2018}%
  \BibitemOpen
  \href@noop {} {\enquote {\bibinfo {title} {{Modelling neutrino-nucleus
  interactions}},}\ }\bibinfo {howpublished}
  {{\url{https://indico.ectstar.eu/event/19/}}} ({\natexlab{a}}),\ \bibinfo
  {note} {{European Centre for Theoretical Studies in Nuclear Physics (ECT*),
  9--13 July 2018}}\BibitemShut {NoStop}%
\bibitem [{ECT({\natexlab{b}})}]{ECTStarWorkshop2019}%
  \BibitemOpen
  \href@noop {} {\enquote {\bibinfo {title} {{Testing and Improving Models of
  Neutrino Nucleus Interactions in Generators}},}\ }\bibinfo {howpublished}
  {{\url{https://indico.ectstar.eu/event/53/}}} ({\natexlab{b}}),\ \bibinfo
  {note} {{European Centre for Theoretical Studies in Nuclear Physics (ECT*),
  3--7 June 2019}}\BibitemShut {NoStop}%
\bibitem [{FNA()}]{FNALWorkshop2020}%
  \BibitemOpen
  \href@noop {} {\enquote {\bibinfo {title} {{Generator Tools Workshop}},}\
  }\bibinfo {howpublished} {{\url{https://indico.fnal.gov/event/22294/}}},\
  \bibinfo {note} {{Fermi National Accelerator Laboratory, 8--10 January
  2020}}\BibitemShut {NoStop}%
\bibitem [{\citenamefont {Mosel}(2019)}]{Mosel:2019vhx}%
  \BibitemOpen
  \bibfield  {author} {\bibinfo {author} {\bibfnamefont {U.}~\bibnamefont
  {Mosel}},\ }\href {\doibase 10.1088/1361-6471/ab3830} {\bibfield  {journal}
  {\bibinfo  {journal} {J. Phys. G}\ }\textbf {\bibinfo {volume} {46}},\
  \bibinfo {pages} {113001} (\bibinfo {year} {2019})},\ \Eprint
  {http://arxiv.org/abs/1904.11506} {arXiv:1904.11506 [hep-ex]} \BibitemShut
  {NoStop}%
\bibitem [{\citenamefont {Khachatryan}\ \emph {et~al.}(2021)\citenamefont
  {Khachatryan} \emph {et~al.}}]{CLAS:2021neh}%
  \BibitemOpen
  \bibfield  {author} {\bibinfo {author} {\bibfnamefont {M.}~\bibnamefont
  {Khachatryan}} \emph {et~al.} (\bibinfo {collaboration} {CLAS, e4v}),\ }\href
  {\doibase 10.1038/s41586-021-04046-5} {\bibfield  {journal} {\bibinfo
  {journal} {Nature}\ }\textbf {\bibinfo {volume} {599}},\ \bibinfo {pages}
  {565} (\bibinfo {year} {2021})}\BibitemShut {NoStop}%
\bibitem [{\citenamefont {Abratenko}\ \emph {et~al.}(2021)\citenamefont
  {Abratenko} \emph {et~al.}}]{uBooNEGENIE}%
  \BibitemOpen
  \bibfield  {author} {\bibinfo {author} {\bibfnamefont {P.}~\bibnamefont
  {Abratenko}} \emph {et~al.} (\bibinfo {collaboration} {MicroBooNE}),\
  }\href@noop {} {\  (\bibinfo {year} {2021})},\ \Eprint
  {http://arxiv.org/abs/2110.14028} {arXiv:2110.14028 [hep-ex]} \BibitemShut
  {NoStop}%
\bibitem [{\citenamefont {Acero}\ \emph {et~al.}(2021)\citenamefont {Acero}
  \emph {et~al.}}]{NOvA:2021nfi}%
  \BibitemOpen
  \bibfield  {author} {\bibinfo {author} {\bibfnamefont {M.~A.}\ \bibnamefont
  {Acero}} \emph {et~al.} (\bibinfo {collaboration} {{NOvA}}),\ }\href@noop {}
  {\  (\bibinfo {year} {2021})},\ \Eprint {http://arxiv.org/abs/2108.08219}
  {arXiv:2108.08219 [hep-ex]} \BibitemShut {NoStop}%
\bibitem [{\citenamefont {Brdar}\ and\ \citenamefont
  {Kopp}(2021)}]{Brdar:2021cgb}%
  \BibitemOpen
  \bibfield  {author} {\bibinfo {author} {\bibfnamefont {V.}~\bibnamefont
  {Brdar}}\ and\ \bibinfo {author} {\bibfnamefont {J.}~\bibnamefont {Kopp}},\
  }\href@noop {} {\  (\bibinfo {year} {2021})},\ \Eprint
  {http://arxiv.org/abs/2109.08157} {arXiv:2109.08157 [hep-ph]} \BibitemShut
  {NoStop}%
\bibitem [{\citenamefont {Hill}\ \emph {et~al.}(2020)\citenamefont {Hill},
  \citenamefont {Junk} \emph {et~al.}}]{Snowmass2021:nu-H}%
  \BibitemOpen
  \bibfield  {author} {\bibinfo {author} {\bibfnamefont {R.}~\bibnamefont
  {Hill}}, \bibinfo {author} {\bibfnamefont {T.}~\bibnamefont {Junk}},  \emph
  {et~al.},\ }\href@noop {} {\bibfield  {journal} {\bibinfo  {journal}
  {SNOWMASS2021}\ } (\bibinfo {year} {2020})}\BibitemShut {NoStop}%
\bibitem [{\citenamefont {Hill}\ \emph {et~al.}()\citenamefont {Hill} \emph
  {et~al.}}]{HDWP}%
  \BibitemOpen
  \bibfield  {author} {\bibinfo {author} {\bibfnamefont {R.}~\bibnamefont
  {Hill}} \emph {et~al.},\ }\href@noop {} {\enquote {\bibinfo {title} {{H/D
  White paper}},}\ }\bibinfo {note} {{Snowmass 2021 white paper}}\BibitemShut
  {NoStop}%
\bibitem [{\citenamefont {Ankowski}\ and\ \citenamefont
  {Friedland}(2020)}]{Ankowski:2020qbe}%
  \BibitemOpen
  \bibfield  {author} {\bibinfo {author} {\bibfnamefont {A.~M.}\ \bibnamefont
  {Ankowski}}\ and\ \bibinfo {author} {\bibfnamefont {A.}~\bibnamefont
  {Friedland}},\ }\href {\doibase 10.1103/PhysRevD.102.053001} {\bibfield
  {journal} {\bibinfo  {journal} {Phys. Rev. D}\ }\textbf {\bibinfo {volume}
  {102}},\ \bibinfo {pages} {053001} (\bibinfo {year} {2020})},\ \Eprint
  {http://arxiv.org/abs/2006.11944} {arXiv:2006.11944 [hep-ph]} \BibitemShut
  {NoStop}%
\bibitem [{\citenamefont {Benhar}\ \emph {et~al.}(2014)\citenamefont {Benhar}
  \emph {et~al.}}]{Benhar:2014nca}%
  \BibitemOpen
  \bibfield  {author} {\bibinfo {author} {\bibfnamefont {O.}~\bibnamefont
  {Benhar}} \emph {et~al.},\ }\href@noop {} {\enquote {\bibinfo {title}
  {{Measurement of the Spectral Function of $^{40}$Ar through the $(e,e^\prime
  p)$ reaction}},}\ } (\bibinfo {year} {2014}),\ \Eprint
  {http://arxiv.org/abs/1406.4080} {arXiv:1406.4080 [nucl-ex]} \BibitemShut
  {NoStop}%
\bibitem [{\citenamefont {Dai}\ \emph {et~al.}(2018)\citenamefont {Dai} \emph
  {et~al.}}]{JeffersonLabHallA:2018zyx}%
  \BibitemOpen
  \bibfield  {author} {\bibinfo {author} {\bibfnamefont {H.}~\bibnamefont
  {Dai}} \emph {et~al.} (\bibinfo {collaboration} {Jefferson Lab Hall A}),\
  }\href {\doibase 10.1103/PhysRevC.98.014617} {\bibfield  {journal} {\bibinfo
  {journal} {Phys. Rev. C}\ }\textbf {\bibinfo {volume} {98}},\ \bibinfo
  {pages} {014617} (\bibinfo {year} {2018})},\ \Eprint
  {http://arxiv.org/abs/1803.01910} {arXiv:1803.01910 [nucl-ex]} \BibitemShut
  {NoStop}%
\bibitem [{\citenamefont {Dai}\ \emph {et~al.}(2019)\citenamefont {Dai} \emph
  {et~al.}}]{Dai:2018gch}%
  \BibitemOpen
  \bibfield  {author} {\bibinfo {author} {\bibfnamefont {H.}~\bibnamefont
  {Dai}} \emph {et~al.},\ }\href {\doibase 10.1103/PhysRevC.99.054608}
  {\bibfield  {journal} {\bibinfo  {journal} {Phys. Rev. C}\ }\textbf {\bibinfo
  {volume} {99}},\ \bibinfo {pages} {054608} (\bibinfo {year} {2019})},\
  \Eprint {http://arxiv.org/abs/1810.10575} {arXiv:1810.10575 [nucl-ex]}
  \BibitemShut {NoStop}%
\bibitem [{\citenamefont {Murphy}\ \emph {et~al.}(2019)\citenamefont {Murphy}
  \emph {et~al.}}]{Murphy:2019wed}%
  \BibitemOpen
  \bibfield  {author} {\bibinfo {author} {\bibfnamefont {M.}~\bibnamefont
  {Murphy}} \emph {et~al.},\ }\href {\doibase 10.1103/PhysRevC.100.054606}
  {\bibfield  {journal} {\bibinfo  {journal} {Phys. Rev. C}\ }\textbf {\bibinfo
  {volume} {100}},\ \bibinfo {pages} {054606} (\bibinfo {year} {2019})},\
  \Eprint {http://arxiv.org/abs/1908.01802} {arXiv:1908.01802 [hep-ex]}
  \BibitemShut {NoStop}%
\bibitem [{\citenamefont {Gu}\ \emph {et~al.}(2021)\citenamefont {Gu} \emph
  {et~al.}}]{JeffersonLabHallA:2020rcp}%
  \BibitemOpen
  \bibfield  {author} {\bibinfo {author} {\bibfnamefont {L.}~\bibnamefont {Gu}}
  \emph {et~al.} (\bibinfo {collaboration} {Jefferson Lab Hall A}),\ }\href
  {\doibase 10.1103/PhysRevC.103.034604} {\bibfield  {journal} {\bibinfo
  {journal} {Phys. Rev. C}\ }\textbf {\bibinfo {volume} {103}},\ \bibinfo
  {pages} {034604} (\bibinfo {year} {2021})},\ \Eprint
  {http://arxiv.org/abs/2012.11466} {arXiv:2012.11466 [nucl-ex]} \BibitemShut
  {NoStop}%
\bibitem [{\citenamefont {Jiang}\ \emph {et~al.}(2022)\citenamefont {Jiang}
  \emph {et~al.}}]{JeffersonLabHallA:2022cit}%
  \BibitemOpen
  \bibfield  {author} {\bibinfo {author} {\bibfnamefont {L.}~\bibnamefont
  {Jiang}} \emph {et~al.} (\bibinfo {collaboration} {Jefferson Lab Hall A}),\
  }\href@noop {} {\  (\bibinfo {year} {2022})},\ \Eprint
  {http://arxiv.org/abs/2203.01748} {arXiv:2203.01748 [nucl-ex]} \BibitemShut
  {NoStop}%
\bibitem [{\citenamefont {Ashkenazi}(2020)}]{adi_ashkenazi_2020_3959538}%
  \BibitemOpen
  \bibfield  {author} {\bibinfo {author} {\bibfnamefont {A.}~\bibnamefont
  {Ashkenazi}},\ }\href {\doibase 10.5281/zenodo.3959538} {\enquote {\bibinfo
  {title} {{Connections between neutrino and electron scattering}},}\ }
  (\bibinfo {year} {2020})\BibitemShut {NoStop}%
\bibitem [{\citenamefont {Papadopoulou}\ \emph {et~al.}(2021)\citenamefont
  {Papadopoulou} \emph {et~al.}}]{electronsforneutrinos:2020tbf}%
  \BibitemOpen
  \bibfield  {author} {\bibinfo {author} {\bibfnamefont {A.}~\bibnamefont
  {Papadopoulou}} \emph {et~al.} (\bibinfo {collaboration} {electrons for
  neutrinos}),\ }\href {\doibase 10.1103/PhysRevD.103.113003} {\bibfield
  {journal} {\bibinfo  {journal} {Phys. Rev. D}\ }\textbf {\bibinfo {volume}
  {103}},\ \bibinfo {pages} {113003} (\bibinfo {year} {2021})},\ \Eprint
  {http://arxiv.org/abs/2009.07228} {arXiv:2009.07228 [nucl-th]} \BibitemShut
  {NoStop}%
\bibitem [{\citenamefont {Hauenstein}\ \emph {et~al.}(2017)\citenamefont
  {Hauenstein} \emph {et~al.}}]{e4nu17}%
  \BibitemOpen
  \bibfield  {author} {\bibinfo {author} {\bibfnamefont {F.}~\bibnamefont
  {Hauenstein}} \emph {et~al.},\ }\href
  {https://www.jlab.org/exp_prog/proposals/17/PR12-17-006.pdf} {\enquote
  {\bibinfo {title} {{Electrons for Neutrinos: Addressing Critical
  Neutrino-Nucleus Issues. A Proposal to Jefferson Lab PAC 45}},}\ } (\bibinfo
  {year} {2017})\BibitemShut {NoStop}%
\bibitem [{\citenamefont {Ashkenazi}\ \emph {et~al.}(2018)\citenamefont
  {Ashkenazi} \emph {et~al.}}]{e4nu18}%
  \BibitemOpen
  \bibfield  {author} {\bibinfo {author} {\bibfnamefont {A.}~\bibnamefont
  {Ashkenazi}} \emph {et~al.},\ }\href
  {https://www.jlab.org/exp_prog/proposals/18/C12-17-006.pdf} {\enquote
  {\bibinfo {title} {{Electrons for Neutrinos: Addressing Critical
  Neutrino-Nucleus Issues. A Run Group Proposal Resubmission to Jefferson Lab
  PAC 46}},}\ } (\bibinfo {year} {2018})\BibitemShut {NoStop}%
\bibitem [{\citenamefont {Ankowski}\ \emph {et~al.}(2020)\citenamefont
  {Ankowski}, \citenamefont {Friedland}, \citenamefont {Li}, \citenamefont
  {Moreno}, \citenamefont {Schuster}, \citenamefont {Toro},\ and\ \citenamefont
  {Tran}}]{Ankowski:2019mfd}%
  \BibitemOpen
  \bibfield  {author} {\bibinfo {author} {\bibfnamefont {A.~M.}\ \bibnamefont
  {Ankowski}}, \bibinfo {author} {\bibfnamefont {A.}~\bibnamefont {Friedland}},
  \bibinfo {author} {\bibfnamefont {S.~W.}\ \bibnamefont {Li}}, \bibinfo
  {author} {\bibfnamefont {O.}~\bibnamefont {Moreno}}, \bibinfo {author}
  {\bibfnamefont {P.}~\bibnamefont {Schuster}}, \bibinfo {author}
  {\bibfnamefont {N.}~\bibnamefont {Toro}}, \ and\ \bibinfo {author}
  {\bibfnamefont {N.}~\bibnamefont {Tran}},\ }\href {\doibase
  10.1103/PhysRevD.101.053004} {\bibfield  {journal} {\bibinfo  {journal}
  {Phys. Rev. D}\ }\textbf {\bibinfo {volume} {101}},\ \bibinfo {pages}
  {053004} (\bibinfo {year} {2020})},\ \Eprint
  {http://arxiv.org/abs/1912.06140} {arXiv:1912.06140 [hep-ph]} \BibitemShut
  {NoStop}%
\bibitem [{\citenamefont {Tice}\ \emph {et~al.}(2014)\citenamefont {Tice} \emph
  {et~al.}}]{Tice:2014pgu}%
  \BibitemOpen
  \bibfield  {author} {\bibinfo {author} {\bibfnamefont {B.}~\bibnamefont
  {Tice}} \emph {et~al.} (\bibinfo {collaboration} {MINERvA}),\ }\href
  {\doibase 10.1103/PhysRevLett.112.231801} {\bibfield  {journal} {\bibinfo
  {journal} {Phys. Rev. Lett.}\ }\textbf {\bibinfo {volume} {112}},\ \bibinfo
  {pages} {231801} (\bibinfo {year} {2014})},\ \Eprint
  {http://arxiv.org/abs/1403.2103} {arXiv:1403.2103 [hep-ex]} \BibitemShut
  {NoStop}%
\bibitem [{\citenamefont {Mousseau}\ \emph {et~al.}(2016)\citenamefont
  {Mousseau} \emph {et~al.}}]{Mousseau:2016snl}%
  \BibitemOpen
  \bibfield  {author} {\bibinfo {author} {\bibfnamefont {J.}~\bibnamefont
  {Mousseau}} \emph {et~al.} (\bibinfo {collaboration} {MINERvA}),\ }\href
  {\doibase 10.1103/PhysRevD.93.071101} {\bibfield  {journal} {\bibinfo
  {journal} {Phys. Rev. D}\ }\textbf {\bibinfo {volume} {93}},\ \bibinfo
  {pages} {071101} (\bibinfo {year} {2016})},\ \Eprint
  {http://arxiv.org/abs/1601.06313} {arXiv:1601.06313 [hep-ex]} \BibitemShut
  {NoStop}%
\bibitem [{\citenamefont {Betancourt}\ \emph {et~al.}(2017)\citenamefont
  {Betancourt} \emph {et~al.}}]{Betancourt:2017uso}%
  \BibitemOpen
  \bibfield  {author} {\bibinfo {author} {\bibfnamefont {M.}~\bibnamefont
  {Betancourt}} \emph {et~al.} (\bibinfo {collaboration} {MINERvA}),\ }\href
  {\doibase 10.1103/PhysRevLett.119.082001} {\bibfield  {journal} {\bibinfo
  {journal} {Phys. Rev. Lett.}\ }\textbf {\bibinfo {volume} {119}},\ \bibinfo
  {pages} {082001} (\bibinfo {year} {2017})},\ \Eprint
  {http://arxiv.org/abs/1705.03791} {arXiv:1705.03791 [hep-ex]} \BibitemShut
  {NoStop}%
\bibitem [{\citenamefont {Abe}\ \emph {et~al.}(2016{\natexlab{b}})\citenamefont
  {Abe} \emph {et~al.}}]{Abe:2015biq}%
  \BibitemOpen
  \bibfield  {author} {\bibinfo {author} {\bibfnamefont {K.}~\bibnamefont
  {Abe}} \emph {et~al.} (\bibinfo {collaboration} {T2K}),\ }\href {\doibase
  10.1103/PhysRevD.93.072002} {\bibfield  {journal} {\bibinfo  {journal} {Phys.
  Rev. D}\ }\textbf {\bibinfo {volume} {93}},\ \bibinfo {pages} {072002}
  (\bibinfo {year} {2016}{\natexlab{b}})},\ \Eprint
  {http://arxiv.org/abs/1509.06940} {arXiv:1509.06940 [hep-ex]} \BibitemShut
  {NoStop}%
\bibitem [{\citenamefont {Adamson}\ \emph {et~al.}(2016)\citenamefont {Adamson}
  \emph {et~al.}}]{Adamson:2016hyz}%
  \BibitemOpen
  \bibfield  {author} {\bibinfo {author} {\bibfnamefont {P.}~\bibnamefont
  {Adamson}} \emph {et~al.} (\bibinfo {collaboration} {MINOS}),\ }\href
  {\doibase 10.1103/PhysRevD.94.072006} {\bibfield  {journal} {\bibinfo
  {journal} {Phys. Rev. D}\ }\textbf {\bibinfo {volume} {94}},\ \bibinfo
  {pages} {072006} (\bibinfo {year} {2016})},\ \Eprint
  {http://arxiv.org/abs/1608.05702} {arXiv:1608.05702 [hep-ex]} \BibitemShut
  {NoStop}%
\bibitem [{\citenamefont {Wu}\ \emph {et~al.}(2008)\citenamefont {Wu} \emph
  {et~al.}}]{Wu:2007ab}%
  \BibitemOpen
  \bibfield  {author} {\bibinfo {author} {\bibfnamefont {Q.}~\bibnamefont {Wu}}
  \emph {et~al.} (\bibinfo {collaboration} {NOMAD}),\ }\href {\doibase
  10.1016/j.physletb.2007.12.027} {\bibfield  {journal} {\bibinfo  {journal}
  {Phys. Lett. B}\ }\textbf {\bibinfo {volume} {660}},\ \bibinfo {pages} {19}
  (\bibinfo {year} {2008})},\ \Eprint {http://arxiv.org/abs/0711.1183}
  {arXiv:0711.1183 [hep-ex]} \BibitemShut {NoStop}%
\bibitem [{\citenamefont {Adamson}\ \emph {et~al.}(2010)\citenamefont {Adamson}
  \emph {et~al.}}]{Adamson:2009ju}%
  \BibitemOpen
  \bibfield  {author} {\bibinfo {author} {\bibfnamefont {P.}~\bibnamefont
  {Adamson}} \emph {et~al.} (\bibinfo {collaboration} {MINOS}),\ }\href
  {\doibase 10.1103/PhysRevD.81.072002} {\bibfield  {journal} {\bibinfo
  {journal} {Phys. Rev. D}\ }\textbf {\bibinfo {volume} {81}},\ \bibinfo
  {pages} {072002} (\bibinfo {year} {2010})},\ \Eprint
  {http://arxiv.org/abs/0910.2201} {arXiv:0910.2201 [hep-ex]} \BibitemShut
  {NoStop}%
\bibitem [{\citenamefont {Lyubushkin}\ \emph {et~al.}(2009)\citenamefont
  {Lyubushkin} \emph {et~al.}}]{Lyubushkin:2008pe}%
  \BibitemOpen
  \bibfield  {author} {\bibinfo {author} {\bibfnamefont {V.}~\bibnamefont
  {Lyubushkin}} \emph {et~al.} (\bibinfo {collaboration} {NOMAD}),\ }\href
  {\doibase 10.1140/epjc/s10052-009-1113-0} {\bibfield  {journal} {\bibinfo
  {journal} {Eur. Phys. J. C}\ }\textbf {\bibinfo {volume} {63}},\ \bibinfo
  {pages} {355} (\bibinfo {year} {2009})},\ \Eprint
  {http://arxiv.org/abs/0812.4543} {arXiv:0812.4543 [hep-ex]} \BibitemShut
  {NoStop}%
\bibitem [{\citenamefont {Antonello}\ \emph {et~al.}(2015)\citenamefont
  {Antonello} \emph {et~al.}}]{Antonello:2015lea}%
  \BibitemOpen
  \bibfield  {author} {\bibinfo {author} {\bibfnamefont {M.}~\bibnamefont
  {Antonello}} \emph {et~al.} (\bibinfo {collaboration} {MicroBooNE, LAr1-ND,
  ICARUS-WA104}),\ }\href@noop {} {\  (\bibinfo {year} {2015})},\ \Eprint
  {http://arxiv.org/abs/1503.01520} {arXiv:1503.01520 [physics.ins-det]}
  \BibitemShut {NoStop}%
\bibitem [{\citenamefont {Stowell}\ \emph {et~al.}(2019)\citenamefont {Stowell}
  \emph {et~al.}}]{PhysRevD.100.072005}%
  \BibitemOpen
  \bibfield  {author} {\bibinfo {author} {\bibfnamefont {P.}~\bibnamefont
  {Stowell}} \emph {et~al.} (\bibinfo {collaboration} {MINERvA}),\ }\href
  {\doibase 10.1103/PhysRevD.100.072005} {\bibfield  {journal} {\bibinfo
  {journal} {Phys. Rev. D}\ }\textbf {\bibinfo {volume} {100}},\ \bibinfo
  {pages} {072005} (\bibinfo {year} {2019})},\ \Eprint
  {http://arxiv.org/abs/1903.01558} {arXiv:1903.01558 [hep-ex]} \BibitemShut
  {NoStop}%
\bibitem [{\citenamefont {Abreu}\ \emph {et~al.}(2021)\citenamefont {Abreu}
  \emph {et~al.}}]{FASER:2021mtu}%
  \BibitemOpen
  \bibfield  {author} {\bibinfo {author} {\bibfnamefont {H.}~\bibnamefont
  {Abreu}} \emph {et~al.} (\bibinfo {collaboration} {FASER}),\ }\href {\doibase
  10.1103/PhysRevD.104.L091101} {\bibfield  {journal} {\bibinfo  {journal}
  {Phys. Rev. D}\ }\textbf {\bibinfo {volume} {104}},\ \bibinfo {pages}
  {L091101} (\bibinfo {year} {2021})},\ \Eprint
  {http://arxiv.org/abs/2105.06197} {arXiv:2105.06197 [hep-ex]} \BibitemShut
  {NoStop}%
\bibitem [{\citenamefont {Wang}\ \emph {et~al.}(2015)\citenamefont {Wang} \emph
  {et~al.}}]{Wang:2014guo}%
  \BibitemOpen
  \bibfield  {author} {\bibinfo {author} {\bibfnamefont {D.}~\bibnamefont
  {Wang}} \emph {et~al.},\ }\href {\doibase 10.1103/PhysRevC.91.045506}
  {\bibfield  {journal} {\bibinfo  {journal} {Phys. Rev. C}\ }\textbf {\bibinfo
  {volume} {91}},\ \bibinfo {pages} {045506} (\bibinfo {year} {2015})},\
  \Eprint {http://arxiv.org/abs/1411.3200} {arXiv:1411.3200 [nucl-ex]}
  \BibitemShut {NoStop}%
\bibitem [{\citenamefont {Hirata}\ \emph {et~al.}(2002)\citenamefont {Hirata},
  \citenamefont {Katagiri}, \citenamefont {Ochi},\ and\ \citenamefont
  {Takaki}}]{Hirata:2001sw}%
  \BibitemOpen
  \bibfield  {author} {\bibinfo {author} {\bibfnamefont {M.}~\bibnamefont
  {Hirata}}, \bibinfo {author} {\bibfnamefont {N.}~\bibnamefont {Katagiri}},
  \bibinfo {author} {\bibfnamefont {K.}~\bibnamefont {Ochi}}, \ and\ \bibinfo
  {author} {\bibfnamefont {T.}~\bibnamefont {Takaki}},\ }\href {\doibase
  10.1103/PhysRevC.66.014612} {\bibfield  {journal} {\bibinfo  {journal} {Phys.
  Rev. C}\ }\textbf {\bibinfo {volume} {66}},\ \bibinfo {pages} {014612}
  (\bibinfo {year} {2002})},\ \Eprint {http://arxiv.org/abs/nucl-th/0112079}
  {arXiv:nucl-th/0112079} \BibitemShut {NoStop}%
\bibitem [{\citenamefont {Yang}\ and\ \citenamefont
  {Bodek}(1999)}]{Yang:1998zb}%
  \BibitemOpen
  \bibfield  {author} {\bibinfo {author} {\bibfnamefont {U.-K.}\ \bibnamefont
  {Yang}}\ and\ \bibinfo {author} {\bibfnamefont {A.}~\bibnamefont {Bodek}},\
  }\href {\doibase 10.1103/PhysRevLett.82.2467} {\bibfield  {journal} {\bibinfo
   {journal} {Phys. Rev. Lett.}\ }\textbf {\bibinfo {volume} {82}},\ \bibinfo
  {pages} {2467} (\bibinfo {year} {1999})},\ \Eprint
  {http://arxiv.org/abs/hep-ph/9809480} {arXiv:hep-ph/9809480} \BibitemShut
  {NoStop}%
\bibitem [{\citenamefont {Capella}\ \emph {et~al.}(1994)\citenamefont
  {Capella}, \citenamefont {Kaidalov}, \citenamefont {Merino},\ and\
  \citenamefont {Tran Thanh~Van}}]{Capella:1994cr}%
  \BibitemOpen
  \bibfield  {author} {\bibinfo {author} {\bibfnamefont {A.}~\bibnamefont
  {Capella}}, \bibinfo {author} {\bibfnamefont {A.}~\bibnamefont {Kaidalov}},
  \bibinfo {author} {\bibfnamefont {C.}~\bibnamefont {Merino}}, \ and\ \bibinfo
  {author} {\bibfnamefont {J.}~\bibnamefont {Tran Thanh~Van}},\ }\href
  {\doibase 10.1016/0370-2693(94)90988-1} {\bibfield  {journal} {\bibinfo
  {journal} {Phys. Lett. B}\ }\textbf {\bibinfo {volume} {337}},\ \bibinfo
  {pages} {358} (\bibinfo {year} {1994})},\ \Eprint
  {http://arxiv.org/abs/hep-ph/9405338} {arXiv:hep-ph/9405338} \BibitemShut
  {NoStop}%
\bibitem [{\citenamefont {Reno}(2006)}]{Reno:2006hj}%
  \BibitemOpen
  \bibfield  {author} {\bibinfo {author} {\bibfnamefont {M.}~\bibnamefont
  {Reno}},\ }\href {\doibase 10.1103/PhysRevD.74.033001} {\bibfield  {journal}
  {\bibinfo  {journal} {Phys. Rev. D}\ }\textbf {\bibinfo {volume} {74}},\
  \bibinfo {pages} {033001} (\bibinfo {year} {2006})},\ \Eprint
  {http://arxiv.org/abs/hep-ph/0605295} {arXiv:hep-ph/0605295} \BibitemShut
  {NoStop}%
\bibitem [{\citenamefont {Kataev}\ \emph {et~al.}(2000)\citenamefont {Kataev},
  \citenamefont {Parente},\ and\ \citenamefont {Sidorov}}]{Kataev:1999bp}%
  \BibitemOpen
  \bibfield  {author} {\bibinfo {author} {\bibfnamefont {A.}~\bibnamefont
  {Kataev}}, \bibinfo {author} {\bibfnamefont {G.}~\bibnamefont {Parente}}, \
  and\ \bibinfo {author} {\bibfnamefont {A.}~\bibnamefont {Sidorov}},\ }\href
  {\doibase 10.1016/S0550-3213(99)00760-9} {\bibfield  {journal} {\bibinfo
  {journal} {Nucl. Phys. B}\ }\textbf {\bibinfo {volume} {573}},\ \bibinfo
  {pages} {405} (\bibinfo {year} {2000})},\ \Eprint
  {http://arxiv.org/abs/hep-ph/9905310} {arXiv:hep-ph/9905310} \BibitemShut
  {NoStop}%
\bibitem [{\citenamefont {Vermaseren}\ \emph {et~al.}(2005)\citenamefont
  {Vermaseren}, \citenamefont {Vogt},\ and\ \citenamefont
  {Moch}}]{Vermaseren:2005qc}%
  \BibitemOpen
  \bibfield  {author} {\bibinfo {author} {\bibfnamefont {J.}~\bibnamefont
  {Vermaseren}}, \bibinfo {author} {\bibfnamefont {A.}~\bibnamefont {Vogt}}, \
  and\ \bibinfo {author} {\bibfnamefont {S.}~\bibnamefont {Moch}},\ }\href
  {\doibase 10.1016/j.nuclphysb.2005.06.020} {\bibfield  {journal} {\bibinfo
  {journal} {Nucl. Phys. B}\ }\textbf {\bibinfo {volume} {724}},\ \bibinfo
  {pages} {3} (\bibinfo {year} {2005})},\ \Eprint
  {http://arxiv.org/abs/hep-ph/0504242} {arXiv:hep-ph/0504242} \BibitemShut
  {NoStop}%
\bibitem [{\citenamefont {Moch}\ \emph {et~al.}(2005)\citenamefont {Moch},
  \citenamefont {Vermaseren},\ and\ \citenamefont {Vogt}}]{Moch:2004xu}%
  \BibitemOpen
  \bibfield  {author} {\bibinfo {author} {\bibfnamefont {S.}~\bibnamefont
  {Moch}}, \bibinfo {author} {\bibfnamefont {J.}~\bibnamefont {Vermaseren}}, \
  and\ \bibinfo {author} {\bibfnamefont {A.}~\bibnamefont {Vogt}},\ }\href
  {\doibase 10.1016/j.physletb.2004.11.063} {\bibfield  {journal} {\bibinfo
  {journal} {Phys. Lett. B}\ }\textbf {\bibinfo {volume} {606}},\ \bibinfo
  {pages} {123} (\bibinfo {year} {2005})},\ \Eprint
  {http://arxiv.org/abs/hep-ph/0411112} {arXiv:hep-ph/0411112} \BibitemShut
  {NoStop}%
\bibitem [{\citenamefont {Moch}\ \emph {et~al.}(2009)\citenamefont {Moch},
  \citenamefont {Vermaseren},\ and\ \citenamefont {Vogt}}]{Moch:2008fj}%
  \BibitemOpen
  \bibfield  {author} {\bibinfo {author} {\bibfnamefont {S.}~\bibnamefont
  {Moch}}, \bibinfo {author} {\bibfnamefont {J.}~\bibnamefont {Vermaseren}}, \
  and\ \bibinfo {author} {\bibfnamefont {A.}~\bibnamefont {Vogt}},\ }\href
  {\doibase 10.1016/j.nuclphysb.2009.01.001} {\bibfield  {journal} {\bibinfo
  {journal} {Nucl. Phys. B}\ }\textbf {\bibinfo {volume} {813}},\ \bibinfo
  {pages} {220} (\bibinfo {year} {2009})},\ \Eprint
  {http://arxiv.org/abs/0812.4168} {arXiv:0812.4168 [hep-ph]} \BibitemShut
  {NoStop}%
\bibitem [{\citenamefont {Schwehr}\ \emph {et~al.}(2016)\citenamefont
  {Schwehr}, \citenamefont {Cherdack},\ and\ \citenamefont
  {Gran}}]{Schwehr:2016pvn}%
  \BibitemOpen
  \bibfield  {author} {\bibinfo {author} {\bibfnamefont {J.}~\bibnamefont
  {Schwehr}}, \bibinfo {author} {\bibfnamefont {D.}~\bibnamefont {Cherdack}}, \
  and\ \bibinfo {author} {\bibfnamefont {R.}~\bibnamefont {Gran}},\ }\href@noop
  {} {\  (\bibinfo {year} {2016})},\ \Eprint {http://arxiv.org/abs/1601.02038}
  {arXiv:1601.02038 [hep-ph]} \BibitemShut {NoStop}%
\bibitem [{\citenamefont {Dolan}\ \emph {et~al.}(2020)\citenamefont {Dolan},
  \citenamefont {Megias},\ and\ \citenamefont {Bolognesi}}]{Dolan:2019bxf}%
  \BibitemOpen
  \bibfield  {author} {\bibinfo {author} {\bibfnamefont {S.}~\bibnamefont
  {Dolan}}, \bibinfo {author} {\bibfnamefont {G.~D.}\ \bibnamefont {Megias}}, \
  and\ \bibinfo {author} {\bibfnamefont {S.}~\bibnamefont {Bolognesi}},\ }\href
  {\doibase 10.1103/PhysRevD.101.033003} {\bibfield  {journal} {\bibinfo
  {journal} {Phys. Rev. D}\ }\textbf {\bibinfo {volume} {101}},\ \bibinfo
  {pages} {033003} (\bibinfo {year} {2020})},\ \Eprint
  {http://arxiv.org/abs/1905.08556} {arXiv:1905.08556 [hep-ex]} \BibitemShut
  {NoStop}%
\bibitem [{\citenamefont {Dolan}\ \emph {et~al.}(2021)\citenamefont {Dolan},
  \citenamefont {Nikolakopoulos}, \citenamefont {Page}, \citenamefont
  {Gardiner}, \citenamefont {Jachowicz},\ and\ \citenamefont
  {Pandey}}]{Dolan:2021rdd}%
  \BibitemOpen
  \bibfield  {author} {\bibinfo {author} {\bibfnamefont {S.}~\bibnamefont
  {Dolan}}, \bibinfo {author} {\bibfnamefont {A.}~\bibnamefont
  {Nikolakopoulos}}, \bibinfo {author} {\bibfnamefont {O.}~\bibnamefont
  {Page}}, \bibinfo {author} {\bibfnamefont {S.}~\bibnamefont {Gardiner}},
  \bibinfo {author} {\bibfnamefont {N.}~\bibnamefont {Jachowicz}}, \ and\
  \bibinfo {author} {\bibfnamefont {V.}~\bibnamefont {Pandey}},\ }\href@noop {}
  {\  (\bibinfo {year} {2021})},\ \Eprint {http://arxiv.org/abs/2110.14601}
  {arXiv:2110.14601 [hep-ex]} \BibitemShut {NoStop}%
\end{thebibliography}%

\end{document}